\documentclass[a4paper, 11pt, oneside]{Thesis}  
\graphicspath{Figures/} 

\usepackage[square, numbers, comma, sort&compress]{natbib}  
\usepackage{verbatim}  
\hypersetup{urlcolor=blue, colorlinks=true} 
\usepackage{subcaption}
\usepackage{tikz}
\usetikzlibrary{arrows.meta,arrows}

\newcommand{\avg}[1]{\left\langle #1\right\rangle}
\newcommand{\cov}[2]{\mathrm{Cov}\left(#1, #2\right)}
\newcommand{\var}[1]{\mathrm{Var}\,#1}
\newcommand{\h}[1]{\hat{#1}}

\newenvironment{mytabular}[1][1]{
  \renewcommand*{\arraystretch}{#1}
  \tabular
}{
  \endtabular
}

\begin{document}

\frontmatter    
\mainmatter	  
\title  {Relativistic gravity in the inhomogeneous Universe}
\authors  {\texorpdfstring
            {Theodore Jacob Anton}
            {Theodore Jacob Anton}
            }
\addresses  {\groupname\\\deptname\\\univname}  
\date       {\today}
\subject    {}
\keywords   {}

\maketitle

\setstretch{1.3}  

\fancyhead{} 
\rhead{\thepage} 
\lhead{}  

\pagestyle{plain} 
\Declaration{

\addtocontents{toc}{\vspace{1em}}  

I, Theodore Jacob Anton, confirm that the research included within this thesis is my own work or that where it has been carried out in collaboration with, or supported by others, that this is duly acknowledged below and my contribution indicated. Previously published material is also acknowledged below.

I attest that I have exercised reasonable care to ensure that the work is original, and does not to the best of my knowledge break any UK law, infringe any third party’s copyright or other Intellectual Property Right, or contain any confidential material. 
 
I accept that Queen Mary University of London has the right to use plagiarism detection software to check the electronic version of the thesis. 

I confirm that this thesis has not been previously submitted for the award of a degree by this or any other university. 

The copyright of this thesis rests with the author and no quotation from it or information derived from it may be published without the prior written consent of the author.  

\vspace{1cm}

Signed: Theodore Jacob Anton \\
\rule[1em]{25em}{0.5pt}  
 
Date: $5^{\rm th}$ May $2024$ \\
\rule[1em]{25em}{0.5pt} 

\vspace{1cm}

\underline{Details of collaboration and publications}

The original research content of this thesis, which is described in Chapters $5-8$\,, was developed in collaboration with Timothy Clifton (in Chapters $5-8$), Daniel B. Thomas (in Chapters $5-6$) and Philip Bull (in Chapter $6$). It is based upon the following papers\footnote{The \texttt{arXiv} preprint number is provided for all papers. They are listed in chronological order.}:
\begin{itemize}

    \item {\it The momentum constraint equation in Parameterised Post-Newtonian Cosmology}.
    
    Theodore Anton and Timothy Clifton. Classical and Quantum Gravity, 39(9):095005, April 2022.
    \texttt{arXiv:2111.10860} \cite{anton2022momentum}.
    
    \item {\it Scale-dependent gravitational couplings in Parameterised Post-Newtonian Cosmology}. 
    
    Daniel B. Thomas, Timothy Clifton and Theodore Anton. Journal of Cosmology and Astroparticle Physics, 2023(04):016, April 2023.
    \texttt{arXiv:2207.14713} \cite{Thomas_2023}.

    \item {\it Modelling the emergence of cosmic anisotropy from non-linear structures}.

    Theodore Anton and Timothy Clifton. Classical and Quantum Gravity, 40(14):145004, June 2023. 
    \texttt{arXiv:2302.05715} \cite{Anton_2023}.

    \item {\it Hubble Diagrams in Statistically Homogeneous, Anisotropic Universes}.

    Theodore Anton and Timothy Clifton. Journal of Cosmology and Astroparticle Physics, 2024(05):120, May 2024. 
    \texttt{arXiv:2402.16585} \cite{anton2024hubble}.

    \item {\it Constraints on the Time Dependence of Post-Newtonian Parameters from Cosmic Microwave Background Anisotropies}.

    Daniel B. Thomas, Theodore Anton, Timothy Clifton and Philip Bull. Submitted to Journal of Cosmology and Astroparticle Physics for publication, May 2024.
    \texttt{arXiv:2405.20388}
    \cite{Thomas_2024}.
    
\end{itemize}
}

\clearpage  

\pagestyle{empty}  

\addtotoc{Abstract}  
\abstract{
\addtocontents{toc}{\vspace{1em}}  
\vspace{0.5cm}
Cosmology is built on a relativistic understanding of gravity, where the geometry of the Universe is dynamically determined by matter and energy. 
In the cosmological concordance model, gravity is described by General Relativity, and it is assumed that on large scales the Universe is homogeneous and isotropic. These fundamental principles should be tested. In this thesis, we explore the implications of breaking them. 

In order to understand possible modifications to gravity on cosmological scales, we extend the formalism of parameterised post-Newtonian cosmology, an approach for building cosmological tests of gravity that are consistent with tests on astrophysical scales. 
We demonstrate how this approach can be used to construct theory-independent equations for the cosmic expansion and its first-order perturbations.
Then, we apply the framework to observations of the anisotropies in the cosmic microwave background. We use these to place novel cosmological constraints on the evolution of the post-Newtonian parameters.

We investigate the consequences of inhomogeneity and isotropy by developing a new approach to studying anisotropy in the Universe, wherein we consider how an anisotropic cosmology might emerge on large scales as a result of averaging over inhomogeneous structures, and demonstrate how the emergent model is affected by backreaction.
We perform a detailed study of light propagation in a wide class of inhomogeneous and anisotropic spacetimes, exploring the conditions under which the Hubble diagram can be accurately predicted by an anisotropic model constructed using explicit averaging, even in the presence of large inhomogeneities. 
We show that observables calculated in a suitable averaged description closely reproduce the true Hubble diagram on large scales, as long as the spacetime possesses a well-defined homogeneity scale.
}

\clearpage  

\setstretch{1.3}  

\acknowledgements{
\addtocontents{toc}{\vspace{1em}}  
\vspace{0.5cm}

I would like to thank my supervisor, Timothy Clifton, for the fantastic, attentive, patient and thoughtful guidance he has given me throughout the last four years. I am hugely grateful in particular that during the first 18 months of my PhD, which took place during the COVID-$19$ pandemic, he was truly understanding of the difficulties this placed on starting off, and helped to make a very strange situation as normal as possible.
\vspace{0.3cm}

I also wish to express my enormous gratitude to Daniel Thomas, who, aside from being a great friend and scientific collaborator, has also been a true mentor to me, and an unofficial second supervisor in my eyes.
\vspace{0.3cm}

I thank my parents Sylvia Metzer and Richard Anton for their endless love and encouragement, and my brothers Sam, Jonah and Nathan. I am especially indebted to Jonah for proofreading this thesis.
\vspace{0.3cm}

I wish to thank my friends at QMUL who have made this experience so enjoyable. I will genuinely really miss sitting in our windowless lunch room. Likewise, I want to say a massive thank you to all my other friends, whether from London, Oxford or wherever else.
\vspace{0.3cm}

I think in particular of my grandfather, David Anton, who passed away in October 2021. I have so many fond memories of time spent with him.
This thesis is dedicated to him and to my other grandparents, sadly none of whom are with us today, but all of whom I remember with love.


}
\clearpage 

\pagestyle{fancy}

\lhead{\emph{Contents}}  
\tableofcontents 

\lhead{\emph{List of Figures}}  
\listoffigures  

\lhead{\emph{List of Tables}}  

\listoftables  

\setstretch{1.5}  

\clearpage  

\setstretch{1.3}  

\pagestyle{empty}  
\dedicatory{In memory of my grandparents, Louise Taylor, David Taylor, Cynthia Anton and David Anton.}

\addtocontents{toc}{\vspace{2em}}  

\pagestyle{fancy}

\chapter{Introduction}

\lhead{\emph{Introduction}}

Of the four known fundamental interactions, gravity is by far the weakest. Nonetheless, the evolution of our Universe is due entirely to gravitational physics. 
Gravitation is an intrinsically relativistic phenomenon, describing the interaction between matter and energy and the curvature of spacetime itself. In order to model how the Universe behaves on cosmological scales, it is therefore necessary to appreciate the profound consequences of relativistic gravity.
This a rather imposing task, because Einstein's General Theory of Relativity, which remains an extraordinarily successful theory over a century after its publication, is a mathematically subtle, nonlinear theory, and its full implications for cosmological physics are still not entirely known.
Moreover, the Universe is highly inhomogeneous, containing a rich cosmic web of nonlinear structures at late times, which large-scale galaxy surveys (see e.g. Refs. \cite{nesseris2022euclid,adame2024desi}) are probing to ever-higher precision. It would seem, then, that a full accounting for all these observations can only be achieved by understanding how relativistic gravitational fields are sourced by realistic, inhomogeneous matter distributions.

However, enormous progress has been made in the field of cosmology through the use of a very simple model, the $\Lambda$CDM concordance model. It is built on three fundamental principles:
\begin{enumerate}
    \item The cosmological principle, which states that the Universe is statistically spatially homogeneous and isotropic. As a corollary, the large-scale geometry of the Universe is taken as being described by the Friedmann-Lema{\^ i}tre-Robertson-Walker (FLRW) metric.
    \item General Relativity as the underlying theory of gravity, that provides the equations of motion for the metric of spacetime.
    \item The $\Lambda$CDM model - dark energy in the form of Einstein's cosmological constant $\Lambda$\,, plus cold dark matter - for $\sim \,95\%$ of the energy content of the Universe.
\end{enumerate}
The simplicity, and ready observational applicability, of the $\Lambda$CDM concordance model, whose properties we will explore in Chapter 3, allows cosmologists largely to forget about many of the complexities of relativistic gravitation.
Those complexitites are sidestepped by imposing the FLRW geometry and General Relativity {\it a priori}, and then dealing with the inhomogeneities present in the Universe through a combination of perturbation theory on large scales (where the relativistic nature of gravity is incorporated, but its fundamental nonlinearity is ignored), and Newtonian techniques on small scales (where nonlinear overdensities in the matter distribution are accounted for, but relativity is not).
These standard approaches are certainly not without merit, providing an astoundingly good fit to many observations, in particular the anisotropies in the cosmic microwave background \cite{aghanim2020planck} and the large-scale structure of the Universe \cite{amon2022dark}.

However, aside from the long-standing problem that the FLRW cosmology in General Relativity appears to be fit observationally only through the inclusion of dark energy and dark matter species of unknown origin, and the well-known present tensions within the standard model \cite{di2021realm,verde2019tensions,amon2022non}, there is good reason to think that a more careful, covariant, treatment of relativistic gravity, rather than an imposed geometry and gravitational theory (with inhomogeneities accounted for through a mixture of perturbative and Newtonian approaches) might be required in order to model cosmological phenomena correctly, especially in the new era of precision measurements.

Let us begin with the problems associated with assuming General Relativity (GR) as the theory of gravity that governs all cosmological physics. 
Although GR is very well-tested in the Solar System and in other astrophysical settings such as binary pulsars \cite{Will_2014}, it is rather less well supported by cosmological observations. To assume that its validity in the Solar System applies equally to the entire Universe involves an extrapolation over several orders of magnitude in length and time scales.
With this, the dark energy and dark matter problems, and the $\Lambda$CDM tensions, in mind (among other problems, including the ultimate need for a quantum theory of gravity), many cosmologists have proposed alternative theories of gravity. 

Of course, one can simply pick specific modified gravity (MG) theories, and then calculate predictions for observables, using the standard perturbative and quasi-Newtonian methods that are successful in the concordance model.
While straightforward, this approach to model-testing does not really allow us to explore all the possible phenomena that may arise in relativistic theories of gravity. It is certainly conceivable that different theories might provide identical predictions for an FLRW spacetime, for linear perturbations about that model, or in the Solar System, but display fundamentally different behaviour in nonlinear and non-perturbative contexts\footnote{A classic example of this is the notion of screening mechanisms \cite{Vainshtein_1972,Khoury_2004,Khoury_2010,brax2012systematic,brax2013screening}, whereby the nonlinearity of certain theories of gravity (typically scalar-tensor theories) is exploited to force them to be indistinguishable from GR in the Solar System, while exhibiting very different cosmological behaviour.}.
In order to test General Relativity and other candidate theories of gravity, one ought therefore to understand them in their full, relativistic form. Moreover, one should study their behaviour in generic inhomogeneous and anisotropic spacetimes, so that the allowed results are not restricted by the symmetries imposed.
This will be the focus of Chapters 5 and 6, where we will build and carry out tests of gravity without specifying any underlying field content, other than the existence of a metric, and without placing any restrictions on the spacetime geometry, other than the existence of an homogeneity scale that allows an effective large-scale cosmological model to be extracted.

The other central problem we will focus on, associated with the application of General Relativity, or indeed any relativistic (metric) theory of gravity, in our Universe, is that of how one should properly deal with inhomogeneity in cosmology. 
An implication of carrying out such a study is that one can ask whether the standard mixture of perturbative and Newtonian approaches is at all appropriate, or whether it might lead to severely misleading predictions that lead us to make erroneous inferences about the matter and energy content of the Universe.
Because gravity is nonlinear, it is in fact not the case that a statistically homogeneous and isotropic universe can necessarily be described by an exactly homogeneous and isotropic FLRW geometry, or that gravitational physics on the largest scales can in general be separated out from the complicated web of nonlinear structure that we know must exist on smaller scales, especially at late times. 
Instead, one should account for the possibility that inhomogeneities themselves affect the evolution of the large-scale properties of the Universe, with potentially profound observational consequences. This is the problem of cosmological backreaction \cite{Buchert_2012,Buchert_2015, Clarkson_2011_b, clifton2013back,van_den_Hoogen_2010}, to which we will devote considerable attention in this thesis.
It is an intrinsically relativistic problem, which one would simply miss by studying structure formation using a purely Newtonian description of gravitational perturbations, on top of an FLRW background metric imposed by hand.

Appreciating the relativistic nature of gravity, and analysing the effects of inhomogeneity accordingly, becomes particularly important if one has reason to believe that the large-scale Universe might actually be anisotropic, as recent observations may suggest \cite{Aluri_2023}. 
Covariant modelling of exact anisotropic cosmological models shows that, under typical conditions, many of them tend to become isotropic at late times \cite{Ellis_1999,barrow1995universe}.
However, much like the standard conclusions about the energy content of the Universe, which rely implicitly on the assumption of an exactly homogeneous and isotropic geometry, being used to model the dynamics of a spacetime which appears only to exhibit those properties statistically, it is not clear that the usual conclusions about anisotropy in the large-scale Universe should remain true in the presence of inhomogeneities on small scales.
This problem will be central to Chapters 7 and 8, in which we will show, using an inherently relativistic and non-perturbative approach, that the possibility of cosmological backreaction might lead to substantial modifications in the late-time evolution of cosmic anisotropy, with ensuing complications for the interpretation of observations in terms of an homogeneous description.

Overall, then, we make the case in this thesis that in order to test and extend our present understanding of cosmology, it is crucial to apply the ideas of relativity carefully in the complex, inhomogeneous and anisotropic Universe in which we make observations, and to do so without assuming from the start the nature of the gravitational theories and geometrical models which we wish to test. 
Although much of the presentation in this thesis will be focused on the mathematical theory that underpins relativistic cosmological modelling, we will seek wherever possible to make contact with observational probes, and to discuss how the results of those probes can be interpreted in a generic and physically rigorous fashion.

The thesis is structured as follows.
\begin{itemize}

    \item In Chapter 2, we explore the relativistic approach to cosmological modelling. We start from Einstein's equivalence principle which underpins the idea of gravity describing the curvature of spacetime, and use it to construct the General Theory of Relativity in its familiar form.
    Then, we introduce several fruitful covariant approaches that will be central to the analysis presented in much of the thesis.
    Finally, we show how the behaviour of light rays can be studied in curved spacetime, and how fundamental observables such as redshifts and distances are defined and calculated in general.

    \item Chapter 3 describes the $\Lambda$CDM concordance cosmology, that we are seeking to test and extend. 
    We discuss the key properties of the homogeneous and isotropic FLRW universe, and the theory of cosmological perturbations that is used in the standard model to study the growth of inhomogeneous structures. We explain how these ideas can be placed in a covariant context, by making contact with the relativistic modelling approaches developed in Chapter 2.
    We also introduce two crucial observational probes that have been used to constrain the concordance model: the cosmic microwave background (in particular its temperature anisotropies), and the Hubble diagram.

    \item In Chapter 4, we discuss several theoretical and observational probes that point to the incompleteness of the $\Lambda$CDM model, and use these to indicate the need to study alternatives to it. We therefore introduce two main classes of alternatives. 
    The first of these, which motivate the research carried out in Chapters 5 and 6, are modified theories of gravity, and we explain how these are tested and compared to General Relativity on astrophysical and cosmological scales, and why theory-independent parameterised frameworks play an important role.
    The second class, which lead on to the research carried out in Chapters 7 and 8, are alternatives to the homogeneous and isotropic FLRW description of the Universe's geometry. We demonstrate the main theoretical tools that are used to develop anisotropic and inhomogeneous cosmological models. Finally, we discuss the related problems of averaging in curved spacetime and cosmological backreaction.

    \item Chapter 5 is focused on building a theory-independent framework for testing gravity on cosmological scales, based on the highly successful parameterised post-Newtonian (PPN) formalism which is used to constrain gravity in astrophysical settings.
    We show that this parameterised post-Newtonian cosmology (PPNC) can be used to describe the large-scale cosmic expansion, and the evolution of scalar and vector perturbations to the FLRW cosmology across a wide range of scales, all the way down from superhorizon scales to the regime of nonlinear structure formation where standard perturbation theory breaks down. 
    We devote particular attention to how peculiar velocities source the evolution of gravitational fields in general through the momentum constraint, and how the framework can be applied to canonical example theories of gravity, leading us to develop a general understanding of the scale and time dependence of gravitational couplings.

    \item In Chapter 6, we apply the PPNC formalism in order to obtain the first constraints on the time dependence of the PPN parameters, which are used to test gravity in the Solar System, from the anisotropies in the cosmic microwave background (CMB) measured by the Planck satellite \cite{planck_power_spectra}. 
    We use the properties of the CMB anisotropies to explore the PPNC theory space, showing that they indicate an illuminating equality between the post-Newtonian parameters, and that it is crucial to consider the effects of modifying gravity not only on cosmological perturbations, as is done in many standard approaches, but also on the background expansion itself.
    We present observational constraints for a variety of different implementations of the PPNC framework, indicating that the data are consistent with General Relativity. We discuss how the constraints may be improved with the inclusion of other datasets.

    \item Chapter 7 is centred around alternatives to the concordance model, that are built not through deviations in the theory of gravity, but rather by questioning the modelling of the Universe's geometry as isotropic on large scales. 
    We use the idea of emergence, based on Buchert's averaging formalism \cite{Buchert_2000,Buchert_2012}, to show how an anisotropic cosmological model can arise on large scales from a generic, inhomogeneous and anisotropic description on small scales. 
    We derive the full set of equations of motion that determine the evolution of such a universe, and interpret them in terms of a locally rotationally symmetric (LRS) Bianchi cosmology, whose evolution is sourced by backreaction from nonlinear structures. 
    We apply our novel approach explicitly to an instructive class of model spacetimes, in order to demonstrate the importance of foliation dependence, and the additional complications that arise in an emergent anisotropic cosmology compared to the isotropic case.

    \item We develop the notion of emergent anisotropy further in Chapter 8, by demonstrating how it can be used to describe Hubble diagram observations in anisotropic universes. 
    We present a non-perturbative ray-tracing method that can be used to determine the redshifts and luminosity distances that observers would infer for distant sources in a series of informative exact inhomogeneous and anisotropic spacetimes.
    We compare these Hubble diagrams to the observations that would be made by fictitious observers residing in the large-scale averaged cosmological models.
    We show how accurate replication of Hubble diagram observables in these spacetimes is closely related to the existence of a well-defined scale of statistical homogeneity, and how the ``average of the Hubble diagram'' converges to the ``Hubble diagram of the average'' in those cases. 
    In contrast, we verify that this is far from guaranteed in spacetimes without a statistical homogeneity scale, and relate this once again to the key issue of foliation dependence.
    
\end{itemize}

We present our conclusions in Chapter 9.

\section{Conventions and notation}

We adopt the following conventions throughout this thesis.

The speed of light $c$ is set to unity. Newton's constant $G$ is typically retained, but at some points, we also set $8\pi G$ equal to unity. We will state clearly when we are doing so. 

We use the metric signature $(-,+,+,+)$, and the conventions of Misner, Thorne and Wheeler for the Riemann and Ricci tensors \cite{Misner_1974}.

Latin indices $a,b,c,d,...$ from the beginning of the alphabet denote spacetime indices, and Latin indices $i,j,k,...$ from the middle of the alphabet are reserved for purely spatial indices. Repeated indices are summed over, according to Einstein's summation convention, unless otherwise stated.

Round brackets around indices indicate symmetrisation,
\begin{equation*}
    t_{(ab)} = \frac{1}{2!}\left[t_{ab}+t_{ba}\right]\,; \quad t_{(abc)} = \frac{1}{3!}\left[t_{abc}+t_{bac}+t_{cab}+t_{cba}+t_{bca}+t_{acb}\right]\,\quad {\rm etc.}\,,
\end{equation*}
and square brackets around them denote anti-symmetrisation,
\begin{equation*}
    t_{[ab]} = \frac{1}{2!}\left[t_{ab}-t_{ba}\right]\,; \quad t_{[abc]} = \frac{1}{3!}\left[t_{abc}-t_{bac}+t_{cab}-t_{cba}+t_{bca}-t_{acb}\right]\,\quad {\rm etc.}\,.
\end{equation*}
Partial derivatives are denoted by either by $\partial_a$ or by a comma. We use overdots for covariant time derivatives along a preferred timelike congruence, and for partial derivatives with respect to a time coordinate $t$, and primes for partial derivatives with respect to either conformal time or a preferred spatial coordinate, as will be made clear from the context. 

\chapter{Relativistic cosmology}

\lhead{\emph{Relativistic cosmology}} 

The central conceit of this thesis is that generalised, covariant frameworks provide insight into the building and testing of cosmological models, that is not apparent if one uses only simplified approaches built for the $\Lambda$CDM concordance model. To that end, it is our aim in this chapter to introduce the fundamental approaches that will be required for the studies that follow. 
We will first introduce the foundational idea of gravity as the curvature of spacetime, leading to Einstein's General Theory of Relativity. 
Then, we will summarise three important covariant approaches to cosmology, that study the properties of four-dimensional spacetime by decomposing it with respect to some physically preferred objects. 
Finally, we will discuss the behaviour of light rays in curved spacetime, with a viewing to understanding observations in the expanding Universe.

\section{Relativistic gravity}

Here, we provide a brief overview of the principles of spacetime curvature and the General Theory of Relativity (GR) that are required for this thesis. The discussion here is nowhere near exhaustive. For further details, see e.g. Refs. \cite{Misner_1974, wald2010general, gron2007einstein}.

\subsection{Einstein's equivalence principle}

The starting point of all relativistic theories of gravity is Einstein's equivalence principle (EEP). Taken as a whole, the EEP states that the laws of physics in any locally inertial frame of reference are identically those of special relativity. It can be divided into three parts \cite{Will_1993}: the weak equivalence principle (WEP), local Lorentz invariance, and invariance under spacetime translations.

The weak equivalence principle asserts that all freely falling test particles fall identically in the presence of the same gravitational field, no matter their mass or composition. This is a corollary of the equivalence of inertial mass $m_I$ - the resistance of an object to being accelerated - and gravitational mass $m_g$ - the charge of an object with respect to the gravitational field. The WEP has been tested to exquisite accuracy using torsion-balance E{\"o}tv{\"o}s experiments \cite{v1922beitrage,Will_2014,Wagner_2012}, which have verified it to one part in $10^{13}\,$ \cite{Wagner_2012}.

Local Lorentz invariance means that the results of any experiment performed in a locally inertial reference frame are independent of the velocity of the frame with respect to any other frame. It is a generalisation of Galilean invariance to include electromagnetism. Therefore, it implies that the speed of light is invariant under all Lorentz boosts. Like the WEP, it is verified to high precision using laboratory experiments \cite{chung2009atom}. In recent years, it has also been tested using pulsars \cite{shao2014tests} and gravitational waves \cite{kostelecky2016testing}.

Invariance under spacetime translations means that the results of any experiment performed in a locally inertial reference frame are independent of the spacetime location at which that experiment is performed. The laws of physics are the same wherever, and whenever, in the Universe one chooses to measure them. Spacetime translation invariance can be tested using atomic clocks \cite{guena2012improved}.

The Einstein equivalence principle implies that locally, the motion of all massive test particles is determined by extremising the single-particle action
\begin{equation}
    S_{\rm particle} = m \int \mathrm{d}\tau = m \int\sqrt{-\mathrm{d}s^2}\,,
\end{equation}
where $\tau$ is the proper time measured by an observer comoving with that particle.
The infinitesimal $\mathrm{d}s^2$ is defined by
\begin{equation}
    \mathrm{d}s^2 = \eta_{ab}\,\mathrm{d}\hat{x}^a \,\mathrm{d}\hat{x}^b\,,
\end{equation}
where the $\hat{x}^a$ refer to coordinates defined in that local inertial frame. The Minkowski metric $\eta_{ab}$ is given in terms of the usual Euclidean coordinates $\hat{x}^a = \left(t, x, y, z\right)$ by 
\begin{equation}
    \eta_{ab} = {\rm diag}\,\left(-1,1,1,1\right)\,.
\end{equation}
In order to satisfy the EEP, it must always be possible to construct such a set of locally free-falling coordinates. This can only be true everywhere in spacetime if the laws of physics obey the principle of general covariance: invariance under all changes of coordinate frame.
Hence, the motion of free-falling particles must always be determined by the extremisation of the single-particle action, where we can write the spacetime interval in terms of an arbitrary set of spacetime coordinates $x^a$ as
\begin{equation}
    \mathrm{d}s^2 = g_{ab}\, \mathrm{d}x^a\,\mathrm{d}x^b\,.
\end{equation}
The object $g_{ab}$ is the spacetime metric tensor. According to the EEP, it is the central object in gravitational physics. If one knows the metric at all points in spacetime, as a function of a prescribed atlas of coordinate systems, then the trajectories of all test bodies can be calculated.
If the test body is massive, then the worldline the particle follows is referred to as timelike, and the tangent vector $t^a$ to that worldline satisfies $g_{ab} t^a t^b = t_a t^a = -1$\,. However, the metric also determines the worldlines of massless particles, which are null, with $t_a t^a = 0$, and can be used to describe spacelike curves with $t_a t^a = 1$. These would be the worldlines followed by particles of negative mass.

Any theory of gravity that obeys the EEP is referred to as a metric theory. In this thesis, we will only be concerned with such theories.
Metric theories are differentiated from one another by how the distributions of matter and energy throughout the Universe determine the metric tensor.

\subsection{Spacetime curvature}

Because $g_{ab}$ varies as one moves through spacetime, it is necessary to define a notion of parallel transport, that tells us how two quantities defined at different points can be compared. This is achieved through a second object, the connection $\nabla_a\,$, which defines the covariant derivative. It is a generalisation of the coordinate partial derivative that is used in locally inertial frames, and is defined to be metric-compatible so that along any given curve, the metric is parallel-transported,
\begin{equation}\label{eq_metric_compatible_connection}
    \nabla_a g_{bc} = 0\,.
\end{equation}
In terms of a coordinate basis $x^a$, the covariant derivative of some tensor $t_{bc}^{\ \ d}$ is defined
\begin{equation}
    \nabla_a t_{bc}^{\ \ d} = \partial_a t_{bc}^{\ \ d} - \Gamma^e_{\ \ ba}t_{ec}^{\ \ d} - \Gamma^e_{\ \ ca}t_{be}^{\ \ d} + \Gamma^d_{\ ea}t_{bc}^{\ \ e}\,,
\end{equation}
with obvious generalisations to tensors of higher and lower rank. Note that there is a distinction here between contravariant (upper) indices and covariant (lower) indices. The components of a vector are lowered and raised using the metric and its inverse, $v_a = g_{ab}\,v^b$ and $v^a = g^{ab} v_b$ respectively\footnote{It is really more correct to refer only to $v^a$ as the components of a vector, and to $v_a$ as the components of a covector or one-form. However, the existence of the inverse metric $g^{ab}$ means that it always possible to convert a covector $v_a$ into a vector $v^a\,$. Therefore, we will be relaxed with our language, and just refer to vectors in both cases.}.
A connection which satisfies Eq. (\ref{eq_metric_compatible_connection}) is referred to as metric-compatible. In the context of differential geometry, the metric and the connection are entirely separate objects in general. However, it is always possible to choose a connection so that it is compatible with the metric. This is typically what is done in relativistic cosmology. If one demands that the connection is not only metric-compatible as above, but also torsion-free ($T^a_{\ \ bc} = \Gamma^a_{\ \ bc} - \Gamma^a_{\ \ cb} = 0$), then the only possible solution is the Levi-Civita connection. In terms of the metric and its partial derivatives, this is defined by the Christoffel symbols
\begin{eqnarray}
    \left\{
\begin{array}{c}
a \\ 
bc
\end{array}
\right\} \, = \, \frac{g^{ad}}{2}\left(\partial_b g_{cd} + \partial_c g_{bd} - \partial_d g_{bc}\right)\,.
\end{eqnarray}
Throughout this thesis, we will always take the connection $\nabla_a$ to be the Levi-Civita connection, meaning that the connection coefficients $\Gamma^a_{\ \ bc}$ are given by the Christoffel symbols. It should be noted, however, that this is a choice, and equally valid formulations of gravity exist which make use of other connections \cite{Jimenez_2018, Beltran_Jimenez_Heisenberg_2019}.

At any individual point, the Christoffel symbols can be made to vanish by choosing a locally inertial coordinate basis, such that the components of the metric with respect to that local basis are given by $\eta_{ab}$\,. If there exists a basis in which the Christoffel symbols vanish globally, then the spacetime is said to be flat. Otherwise, it is curved. It is this curvature of spacetime which constitutes the relativistic gravitational field.

The spacetime curvature is described entirely by the Riemann tensor, which is defined by the Ricci identity
\begin{equation}\label{eq_Ricci_identity}
        \left(\nabla_a\nabla_b - \nabla_b \nabla_a\right) V_c = R_{abcd} V^d\,,
\end{equation}
for any $V_a\,$.
This implies that the components of the Riemann tensor with respect to a local coordinate basis are
\begin{equation}\label{eq_Riemann_tensor}
    R^a_{\ bcd} = \partial_c \Gamma^a_{\ bd} - \partial_d \Gamma^a_{\ bc} + \Gamma^a_{\ ce}\Gamma^e_{\ db} - \Gamma^a_{\ de}\Gamma^e_{\ cb}\,.
\end{equation}
This is a rank-4 tensor, so it has $256$ components. However, they are not independent. By inspection of the above, one sees that the Riemann tensor satisfies $R_{(ab)cd} = R_{ab(cd)} = 0$\,. It also satisfies the first Bianchi identity,  $R_{[abc]d} = 0$\,. Finally, and crucially, is the second Bianchi identity,
\begin{equation}\label{eq_Bianchi_identity}
    \nabla_{[a}R_{bc]de} = 0\,.
\end{equation}
The importance of this equation will be demonstrated shortly. For now, note that the symmetries of the Riemann tensor tell us that there are in fact only $20$ independent components.
These are further decomposed into two parts, each with $10$ independent components. The Ricci tensor $R_{ab}$ is defined
\begin{equation}\label{eq_Ricci_tensor}
    R_{ab} = R^c_{\ acb}\,,
\end{equation}
and it is symmetric in its two indices. Its trace $R = g^{ab}R_{ab}$ is the Ricci scalar. 
The other $10$ components of the Riemann curvature are contained in the Weyl tensor $C_{abcd}\,$, which is conformally invariant and trace-free,
\begin{equation}\label{eq_Weyl_tensor}
    C_{abcd} = R_{abcd} - g_{a[c}R_{d]b} + g_{b[c}R_{d]a} + \frac{R}{6}\left(g_{ac}g_{bd} - g_{ad}g_{bc}\right)\,.
\end{equation}
It shares the symmetries of the Riemann tensor.
The conformal invariance of $C_{abcd}$ (invariance under transformations $g_{ab} \longrightarrow \Omega^2(x^c)\,g_{ab}$\,, where $\Omega$ is an arbitrary non-zero function of the coordinates, that is defined at all points in the spacetime) implies that the Weyl curvature vanishes for any spacetime whose metric $g_{ab}$ can be related to the Minkowski metric $\eta_{ab}$ by a conformal transformation. 

Before we discuss the physical interpretation of curvature, note that contracting the second Bianchi identity (\ref{eq_Bianchi_identity}) twice with respect to the metric gives
\begin{equation}\label{eq_Einstein_tensor}
    \nabla^b G_{ab} = 0\,, \quad {\rm where} \quad G_{ab} = R_{ab} - \frac{R}{2}\,g_{ab}
\end{equation}
is the Einstein tensor. 

In order to understand the physical effects of curvature, it is necessary first to see how the trajectories of particles are determined entirely by the spacetime metric.
Freely falling test particles follow geodesic curves $x^a(\lambda)$ through the spacetime, where the affine parameter $\lambda$ varies smoothly along the curve, and can equally well describe timelike, spacelike or null geodesics\footnote{By an affine parameter, we mean a parameter $\lambda$ such that the curve described by $x^a(\lambda)$ is invariant under affine transformations $\lambda \longrightarrow a\,\lambda + b\,$. It is possible to construct geodesics that are not affinely parametrised, but this is not necessary for our purposes.}. In the timelike case, $\lambda$ is often taken to be the proper time $\tau$ measured by such a geodesic observer.

The tangent vector $t^a = \dfrac{\mathrm{d}x^a}{\mathrm{d}\lambda}$ defines the Lagrangian derivative $\dfrac{D}{\mathrm{d}\lambda} \equiv t^a \nabla_a\,$, so that the curves $x^a(\lambda)$ are described by the (affinely parametrised) geodesic equation,
\begin{eqnarray}\label{eq_geodesic_equation_coord_free}
    \frac{D^2 x^a}{\mathrm{d}\lambda^2} = t^b \nabla_b t^a = 0\,.
\end{eqnarray}
This is typically written in terms of the Christoffel symbols,
\begin{equation}\label{eq_geodesic_equation}
    \frac{\mathrm{d}^2 x^a}{\mathrm{d}\lambda^2} + \Gamma^a_{\ \ bc} \frac{\mathrm{d}x^b}{\mathrm{d}\lambda}\frac{\mathrm{d}x^c}{\mathrm{d}\lambda}\, = 0\,.
\end{equation}
Therefore, a non-zero Christoffel connection causes the trajectories of geodesics to deviate from a straight line with respect to the local coordinate basis. We reiterate that this does not necessarily mean that the spacetime is curved; that is true only if the combination of Christoffel symbols that defines the Riemann tensor is non-zero.

The physical effect of curvature is demonstrated by considering a family of infinitesimally separated geodesics, with tangents $t^a$\,. We can define $\delta x^a$ to be the infinitesimal separation vector between neighbouring geodesics in that family, such that  $t^a = \dfrac{\mathrm{d}x_1^a}{\mathrm{d}\lambda} = \dfrac{\mathrm{d}x_2^a}{\mathrm{d}\lambda}\,, \quad \, {\rm and} \quad \delta x^a(\lambda) = x_2^a(\lambda) - x_1^a(\lambda)\,$.
Each of these curves satisfies the geodesic equation, $\dfrac{D^2}{\mathrm{d}\lambda^2} x_{1,2}^a = 0\,$. By considering how the separation vector $\delta x^a$ changes under an infinitesimal variation of the affine parameter $\lambda\,$, and inserting the expression (\ref{eq_Riemann_tensor}) for $R_{abcd}$ in terms of the Christoffel symbols $\Gamma^a_{\ \ bc}\,$, one obtains the equation of geodesic deviation,
\begin{equation}\label{eq_geodesic_deviation_equation}
   \frac{D^2}{\mathrm{d}\lambda^2}\,\delta x^a = R^a_{\ \ bcd}\, t^b\, t^d\, \delta x^c\,. 
\end{equation}
Hence, spacetime curvature, encoded by a non-vanishing Riemann tensor, causes the worldlines of neighbouring geodesic observers to accelerate relative to one another. The unambiguous signal of a gravitational field is the presence of non-local tidal forces that would change the separation and orientation of a compass of test particles \cite{Szekeres_1965, pirani1961geometrical}.
Now let us consider timelike geodesics specifically, and decompose the Riemann tensor into its Ricci and Weyl parts. Then the geodesic deviation equation can be written as \cite{Szekeres_1965}
\begin{eqnarray}\label{eq_geodesic_deviation_2}
     \frac{D^2}{\mathrm{d}\lambda^2}\,\delta x^a &=& \mathcal{R} \,\delta x^a + \left(\mathcal{R}^a_{\ \ b} + \mathcal{C}^a_{\ \ b}\right)\,\delta x^b\,, \quad {\rm where}\, \\
     \nonumber \mathcal{R} &=& -\frac{1}{3}R_{bc}\,t^b t^c\,, \\
     \nonumber \mathcal{R}_{ab} &=& \frac{1}{2}\left(h_a^{\ c}h_b^{\ d} - \frac{1}{3}h_{ab}h^{cd}\right)R_{cd}\,, \quad {\rm and}\, \\
     \nonumber \mathcal{C}_{ab} &=& C_{adcb} t^c t^d\,.
\end{eqnarray}
Here we have introduced $h_{ab} = g_{ab} + t_a t_b$, which projects all tensors it acts upon into the spacelike 3-surfaces orthogonal to the timelike vector $t^a\,$.
The interpretation of the above is as follows. The Weyl tensor produces a volume-preserving shear on a compass of test particles, with the symmetric and trace-free $\mathcal{C}_{ab}$ constituting a relativistic generalisation of the Newtonian tidal force gradient. The Ricci tensor can also produce shear, through $\mathcal{R}_{ab}\,$, but it has the important additional effect of expanding or contracting the volume of the entire compass, through the double contraction $\mathcal{R}$ of the Ricci tensor with the timelike tangent vector $t^a\,$.
It remains, however, to determine how the matter fields that are present in the universe actually source the curvature of spacetime, and therefore produce a physical effect through the geodesic deviation equation. Next, we will show how this is done, by defining a suitable action for the gravitational field.

\subsection{General Relativity}

The equations of motion of Einstein's General Theory of Relativity are derived most simply by considering an action for the gravitational field which is consistent with the following principles \cite{Clifton_2012}:
\begin{enumerate}
    \item Spacetime is a four-dimensional manifold $\mathcal{M}$, equipped with a metric $g_{ab}$\,. Its tangent spaces $\mathcal{TM}$ are equipped with the Levi-Civita connection $\nabla_a\,$.
    \item The only gravitational degree of freedom is the metric $g_{ab}$\,.
    \item The equations of motion for the metric are local.
    \item Those local equations contain derivatives of the metric no higher than second order.
\end{enumerate}
Lovelock's theorem \cite{lovelock1969uniqueness, lovelock1971einstein, lovelock1972four} states that given these four postulates, the most general action that can be constructed is the Einstein-Hilbert action
\begin{eqnarray}\label{eq_Einstein_Hilbert_action}
    S_{EH} = \frac{1}{16\pi G}\int_{\mathcal{M}}\mathrm{d}^4x\,\sqrt{-g}\,\left(R - 2\Lambda\right)\,,
\end{eqnarray}
where $g$ denotes the determinant of $g_{ab}$, and $\Lambda$ is Einstein's cosmological constant. In the above, we have assumed that there are no surface terms \cite{gibbons1977action, york1986boundary} on the boundary $\partial\mathcal{M}$ of the spacetime manifold.
The full action is $S_{EH} + S_{\rm matter}\,$, and varying it with respect to the metric, and defining the energy-momentum tensor from the matter action according to
\begin{equation}\label{eq_energy_momentum_tensor}
    T_{ab} = \frac{2}{\sqrt{-g}}\frac{\delta S_{\rm matter}}{\delta g^{ab}}\,,
\end{equation}
gives Einstein's field equations,
\begin{equation}\label{eq_Einstein_field_equation}
    G_{ab} = 8\pi G\, T_{ab} - \Lambda\,g_{ab}\,,
\end{equation}
where $G_{ab}$ is the Einstein tensor we introduced in Eq. (\ref{eq_Einstein_tensor}).
Einstein's equations are equations of motion for the Ricci tensor. They tell us that in vacuum, $R_{ab}$ vanishes, so the free part of the gravitational field is entirely contained within the Weyl tensor. It can be sourced by matter through action at a distance, with its equation of motion given by the second Bianchi identity.
Note that at times throughout this thesis, we will set the prefactor $8\pi G$ equal to unity. 

Finally, we can take the covariant divergence of Einstein's equations. From the double contraction of the second Bianchi identity (\ref{eq_Bianchi_identity}), we have $\nabla^b G_{ab} = 0$\,, as per Eq. (\ref{eq_Einstein_tensor}). Inserting this into Einstein's equations gives the equations of motion for the matter fields,
\begin{equation}\label{eq_energy_momentum_tensor_conservation}
    \nabla^b T_{ab} = 0\,.
\end{equation}
Hence, the local conservation of energy-momentum is guaranteed by Einstein's equations. However, for a general metric there are no globally conserved quantities. Global conservation laws exist in curved spacetime only if that spacetime possesses special symmetries (associated with objects called Killing vectors, which we will introduce in Chapter 3), and obeys certain boundary conditions, such as asymptotic flatness.

Einstein's equations (\ref{eq_Einstein_field_equation}) are a set of ten highly nonlinear coupled partial differential equations. Unsurprisingly, they are very hard to solve. There are two typical approaches that are taken to solving them analytically. 

One of those methods is to impose a form of the metric that satisfies some well-motivated symmetry assumptions, and matter content $T_{ab}$ which is consistent with those symmetries. One then writes down a set of equations of reduced complexity, which it is hoped can be solved in closed form. 
We will show in the next chapter how this approach leads to the equations of motion of an exactly homogeneous and isotropic universe. However, the real Universe contains an enormous degree of inhomogeneous and anisotropic structure, so it satisfies neither of these symmetries exactly. Any model for the Universe's geometry that is based on them should be considered only an approximation that is valid only on certain scales.

The other, related, method, is to adopt the perspective of perturbation theory. One takes some known solution, typically one with a high degree of symmetry such as the Minkowski metric, and refers to it as the background metric $g_{ab}^{\ \ (0)}$. 
Then, a perturbatively small correction $\delta g_{ab} \ll g_{ab}^{\ \ (0)}$ is introduced, and Einstein's equations are expanded and solved order-by-order for this small quantity. 
This is a very fruitful approach in cosmology, leading to the field of cosmological perturbation theory \cite{Malik_2009, Ma_1995}. We will discuss this in some detail in the next chapter. However, General Relativity is fundamentally non-linear and non-perturbative, so perturbation theory is by its nature limited in scope, and will invariably break down in certain regimes. Perturbative methods also exhibit gauge dependence, which causes subtleties in the physical interpretation of $\delta g_{ab}$ \cite{Bardeen_1980, Durrer_1994, Clifton_2020}.

However, alternative approaches have been developed, that are not based directly around solving Einstein's equations for $g_{ab}\,$, as a function of coordinates $x^a$. 
Instead, they work through defining a set of covariant scalar, vector and tensor variables which contain physically equivalent information to the metric, and then calculating the equations of motion for those variables. Those equations are equivalent to Einstein's equations, but take very different forms that make them easier to solve and interpret than the metric picture, in some cases. 
We will now introduce some of these formalisms.

\section{Covariant decompositions}

Covariant descriptions of relativistic gravity theories are formulated in general by decomposing all the equations of motion of that theory with respect to some geometrical object of interest. The dynamics of the spacetime are then described by the covariantly defined properties of that object.
We will discuss four such approaches:
\begin{enumerate}
    \item The $1+3$ decomposition, where the preferred object is a congruence of timelike worldlines.
    \item The $3+1$ decomposition, where the preferred object is the continuous set of level surfaces $\Sigma_t$ of a globally defined time function $t(x^a)\,$.
    \item The $1+1+2$ decomposition, where there are two preferred objects: the congruence of timelike curves as in the $1+3$ decomposition, and a congruence of spacelike curves orthogonal to the timelike congruence.
\end{enumerate}

\subsection{The $1+3$ formalism}\label{subsec:1+3}

Suppose we are considering a spacetime which everywhere admits a preferred timelike vector field $u^a$, with $u_a u^a = -1$. This is very natural in cosmology, where this vector can be taken to correspond to, for example, the average 4-velocity of dark matter on large scales, or the normal vector to the homogeneous three-surfaces of a spatially homogeneous geometry.

For now we reserve a discussion about the appropriate choice or uniqueness of $u^a$, although we will come back to this issue later on in the thesis. Here, we simply assume that at least one such timelike vector exists and is well-defined. One can then perform a covariant decomposition of all tensorial quantities into parts parallel and orthogonal to this vector. 

By using the Ricci identities for $u^a$ and the Bianchi identities, and inserting Einstein's equations to replace $R_{ab}$ where appropriate, one obtains a set of equations that is entirely physically equivalent to Einstein's equations in the coordinate approach. However, the $1+3$ picture is often easier to interpret physically than the coordinate picture, because it deals explicitly with covariantly defined objects. This means that the coordinate dependence that plagues physical interpretation of Einstein's equations can be at least partially circumvented. 
For a full discussion of the $1+3$ decomposition and its uses in cosmology, we refer the reader to e.g. Refs. \cite{Ellis_1999, Hawking_1966, ellis2012relativistic, ellis1989covariant, van1996extensions, van_Elst_1997, maartens1997}.

We will now present the required set of $1+3$-covariant equations. One starts by defining the projection tensor $h_a^{\ b}$, which projects all quantities into the instantaneous 3-spaces orthogonal to $u^a$,
\begin{equation}
    h_a^{\ b} = \delta_a^{\ b} + u_a u^b \quad \Rightarrow \quad h_{ab}u^b = 0\,,\quad h_a^{\ b}h_b^{\ c} = h_a^{\ c}\,, \quad h_a^{\ a} = 3\,.
\end{equation}
If $u^a$ is irrotational, meaning that its vorticity, which we will define shortly, vanishes, then it is also hypersurface-orthogonal. It therefore defines a set of 3-surfaces whose induced metric is $h_{ab}$. Then, the $1+3$ decomposition is equivalent to another approach called the $3+1$ decomposition, which we will define in Section \ref{subsec:3+1}. For a general $u^a$\,, however, this is not the case.

The decomposition with respect to $u^a$ means that instead of the spacetime covariant derivative $\nabla_a$, we work with two new types of covariant derivative:
\begin{enumerate}
    \item the covariant time derivative along $u^a$, which is equivalent to a derivative with respect to the proper time measured by comoving observers,
    \begin{equation}
        \dot{t}_a^{\ bc} \equiv u^d \nabla_d t_a^{\ bc}\,.
    \end{equation}
    \item the spatially projected derivative $D_a\,$, which takes any tensor and projects its covariant derivative fully orthogonally to $u^a$. 
    \begin{equation}
        D_a t_b^{\ cd} \equiv h_a^{\ e} h_b^{\ f} h_g^{\ c} h_h^{\ d} \nabla_e t_f^{\ gh}\,.
    \end{equation}
    If $u_a$ is irrotational, such that $h_{ab}$ is the induced metric on the orthogonal hypersurfaces, it follows that $D$ is indeed a genuine covariant derivative in those surfaces, and is metric-compatible such that $D_a h_{bc} = 0$.
\end{enumerate}

We can also define the 3-volume element $\eta_{abc} = u^d \eta_{dabc}$\,. Here $\eta_{abcd} = \eta_{[abcd]}$ is the spacetime 4-volume element, and is equal to $\sqrt{-g}\,\epsilon_{abcd}$, where $\epsilon_{abcd}$ is the alternating Levi-Civita symbol.

We need to take the scalars, vectors and tensors that are defined in the coordinate approach, and make them compatible with the $1+3$-decomposed picture.
For scalars, this is trivial, because they have no free indices, so they remain unaltered. 
For vectors $V^a$, we define their projection orthogonal to $u^a$ to be $V^{\langle a \rangle} \equiv h^a_{\ b} V^b$.
Finally, for rank-2 tensors $t_{ab}$, their symmetric, trace-free projection is given by
\begin{equation}
    t_{\langle ab\rangle} \equiv \left(h_{(a}^{\ \, c}h_{b)}^{\ d} - \frac{1}{3}h_{ab}h^{cd}\right)t_{cd}\,.
\end{equation}
A key aspect of the $1+3$ formalism is the kinematic decomposition of the covariant derivative of $u^a$. This can be written completely generally as
\begin{equation}\label{eq_nabla_u_decomposition}
    \nabla_a u_b = - u_a \dot{u}_b + D_a u_b = - u_a \dot{u}_b + \frac{1}{3}\Theta\,h_{ab} + \sigma_{ab} + \eta_{cab}\,\omega^c\,,
\end{equation}
where we have defined the expansion $\Theta = \nabla_a u^a$, the shear $\sigma_{ab} = \sigma_{\langle ab\rangle} = D_{\langle a} u_{b\rangle}$, the vorticity $\omega^a = \frac{1}{2}\eta^{abc}D_{[b}u_{c]}$, and the acceleration vector $\dot{u}_a = u^b \nabla_b u_a$. 
The expansion and shear are sometimes grouped together into the expansion tensor, $\Theta_{ab} = \frac{1}{3}\,\Theta \,h_{ab} + \sigma_{ab}\,$.
Let us discuss the physical interpretation of each of these quantities.

\begin{figure}[ht]
    \centering
    \includegraphics[width=0.5\linewidth]{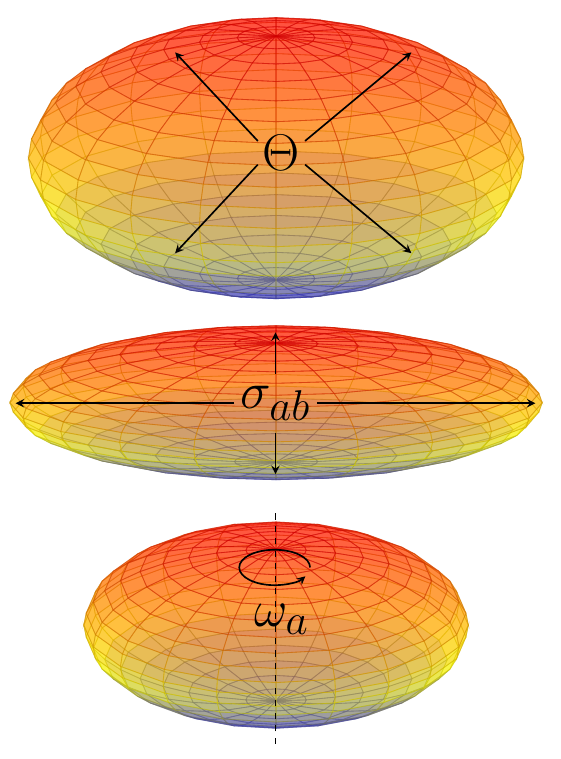}
    \caption{The effect of expansion $\Theta$, shear $\sigma_{ab}$ and vorticity $\omega_a$ on an infinitesimal patch of space. All the points in this patch have worldlines parallel to $u^a\,$.}
    \label{fig_expansion_shear_vorticity}
\end{figure}

The expansion scalar $\Theta$ is so named because it describes the rate at which spatial volumes expand locally. This can be seen by considering an infinitesimally small patch of the instantaneous rest space orthogonal to $u^a$, which has volume $\delta V = \sqrt{h}\,\delta^3 x$. 
By considering an infintesimal variation of the affine parameter along worldlines parallel to $u^a$, one finds that the determinant $h$ of the projection tensor satisfies $\dot{\sqrt{h}} = \Theta\,\sqrt{h}\,$ \cite{Buchert_1996}. Hence, $\dfrac{\dot{\delta V}}{\delta V} = \Theta\,$, so $\Theta$ does indeed tell us about the isotropic local expansion.

The shear tensor $\sigma_{ab}$ is symmetric, trace-free and orthogonal to $u^a\,$. Because it is trace-free, it preserves spatial volumes. However, it causes the congruence to expand at different rates in different spatial directions, as shown in Fig. \ref{fig_expansion_shear_vorticity}, so that some worldlines in the timelike congruence are brought closer together and some are pushed apart.


The vorticity vector $\omega_a$ is formed from the orthogonally projected, antisymmetric part of $\nabla_a \,u_b\,$. It does not affect the distances between neighbouring worldlines in the timelike congruence, but instead causes them to rotate relative to one another.

A simple way to understand the shear and vorticity physically is to take the Newtonian limit. 
Although a Newtonian picture is insufficient, it gives some good intuition which carries over into the relativistic formulation. 
We can identify a Cartesian coordinate basis on Minkowski spacetime, so that $u^a = \left(1, \mathbf{v}\right)\,$, where $\mathbf{v}$ is the 3-velocity associated with the congruence. 
Then $\sigma_{ij} = \partial_{(i} v_{j)} - \frac{1}{3}\,\delta_{ij}\,\nabla \cdot \mathbf{v}\,$, and $\omega_i = -\,\frac{1}{2}\left(\nabla \,\times \, \mathbf{v}\right)_i\,$. 

Here $\nabla \cdot$ and $\nabla \,\times\,$ are the familiar Newtonian divergence and curl operators.
Hence, $\sigma_{ij}$ reduces to the Newtonian shear, which describes tidal deformations of the 3-velocity field, and the vorticity $\omega_i$ reduces to a linear multiple of the Newtonian vorticity vector.

If the vorticity vanishes, $u^a$ is said to be irrotational. One finds that for any general coordinate scalar $S$, 
\begin{equation}
    D_{[a}D_{b]}S = \eta_{abc}\omega^c \dot{S}\,,
\end{equation}
which must vanish if $D_a$ is a covariant derivative. This justifies the earlier claim that if the vorticity $\omega^a$ vanishes, then $h_{ab}$ is the induced metric on the spacelike hypersurfaces orthogonal to $u^a$, and $D_a$ is the covariant derivative in those surfaces.

The acceleration vanishes if $u^a$ is geodesic, i.e. if it is the tangent vector to the worldlines of a freely falling congruence of observers. It therefore measures the effect of non-gravitational forces on $u^a$\,. 
For example, in the presence of an electromagnetic field $F_{ab}\,$, a particle of charge $q$ and mass $m$ will experience an acceleration due to the Lorentz force, given by $m \dot{u}_a = q F_{ab} u^b\,$.

For a complete system we will also require quantities associated with the Ricci and Weyl curvature of the spacetime. The former of these can be related to the matter content of the spacetime using Einstein's equations.
The energy-momentum tensor $T_{ab}$ is decomposed with respect to $u^a$ as
\begin{equation}\label{eq_energy_momentum_tensor_decomposition}
T_{ab} = \rho \, u_a u_b + p\, h_{ab} + 2 \, q_{(a}u_{b)} + \pi_{ab}\,,
\end{equation}
where $\rho = T_{ab}u^a u^b\,$, $p = \frac{1}{3}T_{ab}h^{ab}\,$, $q_a = -T_{bc}u^b h_a^{\ c}\,$, and $\pi_{ab} = T_{\langle ab \rangle}$ are the energy density, isotropic pressure, momentum density and anisotropic stress measured by observers comoving with the $u^a$ congruence. 
For a perfect fluid with 4-velocity $u^a$, the momentum density $q_a$ and the anisotropic stress $\pi_{ab}$ vanish for comoving observers, but an observer with some different 4-velocity $\tilde{u}^a$ would typically measure these quantities to be non-zero. We will frequently deal with pressureless dust, whose energy-momentum tensor is simply $\rho u_a u_b\,$.

Finally, the Weyl tensor can be decomposed into an electric and a magnetic part:
\begin{equation}\label{eq_Weyl_tensor_decomposition}
E_{ab} = C_{acbd}u^c u^d \qquad {\rm and} \qquad H_{ab} =  \frac{1}{2}\eta_{efda}C^{ef}_{\quad bc} u^c u^d \,.
\end{equation}
These are both symmetric and trace-free, and are orthogonal to $u^a$, i.e. $E_{ab}u^b = H_{ab}u^b = 0\,$. Hence, they both have 5 independent components, so together they describe the 10 degrees of freedom in the Weyl tensor. Returning to the geodesic deviation equation split into its Ricci and Weyl parts (\ref{eq_geodesic_deviation_2}), the electric and magnetic Weyl tensors contain all the free gravity that causes a ring or compass of test particles to be sheared.

The electric part can be thought of as a relativistic generalisation of the Newtonian tidal force tensor. The magnetic part is fundamentally non-Newtonian \cite{Sopuerta_1999, clifton2017magnetic, Maartens_1998}. It is sometimes assumed to vanish, leading to the silent cosmological models \cite{van_Elst_1997}. However, in the real Universe, it is known to be non-zero, because gravitational waves contribute to both the electric and magnetic Weyl curvature \cite{Dunsby_1997}. This is very similar to the situation in electromagnetism, where electromagnetic waves are characterised by both the electric and magnetic fields being non-zero in all frames of reference.

The list above defines the full set of kinematic, matter and curvature variables in the $1+3$ formulation. They are grouped into scalars, $\left\lbrace \Theta, \rho, p\right\rbrace\,$, vectors, $\left\lbrace \dot{u}_a, \omega_a, q_a\right\rbrace\,$, and tensors, $\left\lbrace \sigma_{ab}, \pi_{ab}, E_{ab}, H_{ab}\right\rbrace\,$.

Now, we present the equations of motion for these quantities. 
The presentation here follows the approach of Ref. \cite{Ellis_1999}, although we rework the equations somewhat so that what we consider to be the source terms are always on the right hand side of the relevant equation. 
The equations of motion for all of the $1+3$-covariant variables are obtained by various contractions of the following tensors:
\begin{equation}
    S_{abc} \equiv 2\,\nabla_{[a}\nabla_{b]}\,u_c - R_{abcd}\,u^d = 0\,; \qquad W_{abcde} = \nabla_{[a}R_{bc]de} = 0\,,
\end{equation}
where the first vanishes by the Ricci identity (\ref{eq_Ricci_identity}), and the second by the second Bianchi identity (\ref{eq_Bianchi_identity})
Let us first consider the Ricci identities for $u^a$, $S_{abc} = 0$\,. There are six independent equations coming from them.

The first is the Raychaudhuri equation, which is the evolution equation for the isotropic expansion $\Theta$. It is obtained from $u^b S^a_{\ \ ba} = 0\,$:
\begin{equation}\label{eq_Raychaudhuri}
    \dot{\Theta} + \frac{1}{3}\,\Theta^2 + 2\,\sigma^2 - 2\,\omega^2 - \dot{u}_a\dot{u}^a - D_a \dot{u}^a = - 4\pi G\left(\rho + 3p\right) + \Lambda\,.
\end{equation}
On the right hand side, we have replaced the combination $R_{ab} u^a u^b$ by \\
$\left(8\pi G\, T_{ab} - 4\pi G\,T\,g_{ab} + \Lambda g_{ab}\right)u^a u^b\,$, and then used the decomposition (\ref{eq_energy_momentum_tensor_decomposition}) of $T_{ab}\,$. If we were considering a modified gravity theory, rather than GR, then $R_{ab}$ would be replaced according to the field equations of that theory, but the left hand side would be unchanged.

There is a lot of information in this equation. In order to appreciate that information, let us introduce the strong energy condition (SEC), which is the requirement that $\left(T_{ab} - \frac{1}{2}T\,g_{ab}\right) t^a t^b \geq 0$ for all timelike vectors $t^a$\,.
Translated into the language of the $1+3$ decomposition, this means that $\rho + 3p \geq 0\,$. If that is the case, then the effect of matter and energy is to cause timelike congruences to contract, as it drives $\dot{\Theta}\,$, the time derivative of the expansion, towards negative values. 
There are contributions both from the energy density and the pressure, so radiation, which has $p = \frac{1}{3}\rho$, causes more rapid attraction than pressureless dust. 
Phantom fluids, with $p < -\frac{1}{3}\rho\,$, can instead cause gravitational repulsion. This is clearly done by $\Lambda$\,. 
If the total on the RHS is positive, and the shear, vorticity and acceleration are negligible, then the Raychaudhuri equation tells us that the expansion of the Universe, which is described in the isotropic and homogeneous limit by $\Theta = 3H$, will accelerate. It is therefore an equation of fundamental importance for cosmology.

Next we have the evolution equation for the shear. This is obtained from the equation $h_{\langle a}^{\ d}h_{b\rangle}^{\ e} u^c S_{dce} = 0\,$, giving
\begin{equation}\label{eq_shear_evolution}
    \dot{\sigma}_{\langle ab\rangle} + \frac{2}{3}\,\Theta\,\sigma_{ab} - \dot{u}_{\langle a}\dot{u}_{b\rangle} + \sigma_{c\langle a}\sigma_{b\rangle}^{\ c} + \omega_{\langle a}\omega_{b \rangle} - D_{\langle a}\dot{u}_{b \rangle} + E_{ab} = 4\pi G\,\pi_{ab}\,.
\end{equation}
This shows how anisotropy in the expansion of space, which is typically described by the shear tensor, is sourced both by the local matter distribution, through $\pi_{ab}$, and the non-local free gravity, through $E_{ab}\,$.
In the homogeneous and isotropic limit, all the terms in the shear evolution equation are zero.

The evolution equation for the vorticity is given by $h_f^{\ d}h_b^{\ e} u^c S_{dce} \eta^{fb}_{\quad a} = 0\,$,
\begin{equation}\label{eq_vorticity_evolution}
    \dot{\omega}_{\langle a \rangle} + \frac{2}{3}\,\Theta \,\omega_a - \sigma_{ab}\,\omega^b - \frac{1}{2}\,\eta_{abc}D^b\dot{u}^c = 0\,. 
\end{equation}
Unlike the expansion, which is sourced by energy density and pressure, and shear, which is sourced by anisotropic stress, there is no matter term sourcing vorticity. This means that even if they were initially large in the early Universe, vortical flows are usually assumed to have decayed away quickly.

The fourth equation from $S_{abc}$ comes from $h_a^{\ b}S^c_{\ bc} = 0$. It shows that the effect of the momentum density $q_a$ of matter fields is to create gradients in the kinematic variables $\Theta$ and $\sigma_{ab}$\,,
\begin{equation}\label{eq_momentum_constraint_1+3}
    D^b\sigma_{ab} - \frac{2}{3}\,D_a \Theta + \eta_{abc}\left(D^b \omega^c + 2\,\dot{u}^b \omega^c\right) = - 8\pi G\, q_a \,.
\end{equation}
If the vorticity vanishes, so that the rest spaces orthogonal to $u^a$ are level surfaces of some time function $t$, then this equation forms one of the constraints on the initial Cauchy surface. We will discuss this in some more detail when we introduce the $3+1$ formalism.

The next equation is obtained from $\eta^{abc}S_{abc} = 0$\,, and describes the divergence of the vorticity,
\begin{equation}\label{eq_vorticity_divergence}
    D_a \omega^a - \dot{u}_a\omega^a = 0\,.
\end{equation}
If $u^a$ is geodesic, then the vorticity either vanishes or is a pure (divergenceless) vector.

The final equation from the Ricci identity is a somewhat unusual one, because it does not have any time derivatives or spatial divergences. 
Instead, it explicitly identifies the magnetic part of the Weyl tensor with the effects of shear, vorticity and acceleration. It is calculated from $S_{cd\langle a}\eta_{b\rangle}^{\ \: cd} = 0$\,:
\begin{equation}\label{eq_Hweyl_vs_shear_vorticity}
    H_{ab} - \eta_{cd\langle a}D^c \sigma_{b\rangle}^{\ \: d} + D_{\langle a}\omega_{b \rangle} + 2\,\dot{u}_{\langle a}\omega_{b\rangle} = 0\,.
\end{equation}
Therefore, if we are studying a spatially homogeneous cosmology, a geodesic observer, with their worldline orthogonal to the homogeneous surfaces, will measure no magnetic Weyl curvature. There is no equivalent equation for the electric Weyl curvature, which is typically non-zero in anisotropic cosmologies even if they are homogeneous.

The Ricci identities describe how the kinematics of a congruence evolve, but they do not provide equations of motion for the matter or Weyl curvature quantities. These are required in order to have the same amount of information as is contained in the metric, and its first and second derivatives, in the coordinate picture.

In that picture, the Weyl tensor comes out in the wash once the metric has been solved for, and the matter equations of motion come from taking the covariant divergence of Einstein's equations, and using $\nabla^b G_{ab} = 0$.

Let us deal first with the Weyl curvature. We can use the second Bianchi identity to obtain its equations of motion. These tell us how the gravitational field propagates in vacuum, where $R_{ab} = 0$. In particular, they give covariant equations for gravitational waves \cite{Hawking_1966}.

The first is given by $h^{de}u^c W_{cd\langle ab \rangle e} = 0$\,: 
\begin{eqnarray}\label{eq_Eweyl_evolution}
&& \hspace{-1cm} \dot{E}_{\langle ab\rangle} + \Theta\,E_{ab} - 3\,\sigma_{c\langle a}\,E_{b \rangle c} - \eta_{cd\langle a}\,E_{b\rangle}^{\ d}\omega^c - \eta_{cd\langle a}\,D^c\,H_{b\rangle}^{\ d} - 2\,\eta_{cd\langle a}\,H_{b\rangle}^{\ d}\,\dot{u}^c \\
\nonumber &=& -4\pi G\left[\dot{\pi}_{\langle ab\rangle} + \frac{\Theta}{3}\,\pi_{ab} + D_{\langle a}q_{b\rangle} + \left(\rho + p\right)\sigma_{ab} + \sigma_{c\langle a}\pi_{b\rangle c} + 2\,\dot{u}_{\langle a}\,q_{b\rangle} - 2\,\eta_{cd\langle a}\,\pi_{b\rangle}^{\ d}\,\omega^c \right]\,.
\end{eqnarray}
It shows how $H_{ab}$ sources $E_{ab}$ in vacuum, through the term $\eta_{cd\langle a}\,D^c H_{b\rangle}^{\ d}\,$, so that the electric part of the Weyl tensor propagates non-locally.

Similarly, $u^e u^f W_{ecdf\langle a}\eta_{b \rangle}^{\ cd} = 0\,$ gives 
\begin{eqnarray}\label{eq_Hweyl_evolution}
&& \hspace{-1cm} \dot{H}_{\langle ab\rangle} + \Theta\,H_{ab} - 3\,\sigma_{c\langle a}\,H_{b\rangle}^{\ c} - \eta_{cd\langle a}\,H_{b\rangle}^{\ d}\,\omega^c + \eta_{cd\langle a}\,D^c\,E_{b\rangle}^{\ d} + 2\,\eta_{cd\langle a}\,E_{b\rangle}^{\ d}\,\dot{u}^c \\
\nonumber &=& 4\pi G\left[3\,\omega_{\langle a}\,q_{b\rangle} + \eta_{cd\langle a}\,D^c\,\pi_{b\rangle}^{\ d} + \eta_{cd\langle a}\,\sigma_{b\rangle c}\,q^d \right]\,,
\end{eqnarray}
so $E_{ab}$ sources $H_{ab}$ in vacuum through the term $\eta_{cd\langle a}\,D^c E_{b\rangle}^{\ d}\,$, and thus $H_{ab}$ propagates.

A further independent equation comes from $h_a^{\ b}u^c u^d W_{db \ \, ec}^{\ \: \: e} = 0$\,,
\begin{eqnarray}\label{eq_div_E_constraint}
&& \hspace{-1.5cm} D^b E_{ab} - 3\,H_{ab}\,\omega^b - \eta_{abc}\,\sigma^{bd}\,H^c_{\ d} = 4\pi G\left[\frac{2}{3}\,D_a\rho - \frac{2}{3}\,\Theta\,q_a + \sigma_{ab}\,q^b - 3\,\eta_{abc}\,\omega^b\,q^c \right]\,.
\end{eqnarray}
This equation is useful for gravitational waves in vacuum, which are studied in a $1+3$-covariant approach by enforcing the conditions $D^b E_{ab} = D^b H_{ab} = 0$ \cite{Dunsby_1997, Heinesen_2022}, in order to remove all the non-radiative contributions from the Weyl tensor. It tells us that in vacuum, the eigenvectors of the shear $\sigma_{ab}$ and the magnetic Weyl tensor $H_{ab}$ associated with those gravitational waves coincide. These eigenvectors correspond with the $+$ and $\times$ polarisations of the gravitational wave, that cause a ring of test particles to undergo quadrupole (shear) deformation.

The last $1+3$-covariant equation for the Weyl curvature is obtained from $\eta^{bc}_{\quad a}u^d W_{db\ \, ec}^{\ \: \: e} = 0$\,:
\begin{eqnarray}\label{eq_div_H_constraint}
&& \hspace{-1.5cm} D^b H_{ab} + \eta_{abc}\,\sigma^{bd}\,E^c_{\ d} = 4\pi G\left[\pi_{ab}\,\omega^b - 2\,\left(\rho + p\right)\,\omega_a - \eta_{abc}\,D^b q^c - \eta_{abc}\,\sigma^{bd}\,\pi^c_{\ d} \right]\,,
\end{eqnarray}
which means that the eigenvectors of $E_{ab}$ also coincide with the eigenvectors of $\sigma_{ab}$ for gravitational waves in vacuum, as they do for any universe which is silent ($H_{ab}$ = 0) and occupied by a vorticity-free perfect fluid \cite{van_Elst_1997}.

Finally, let us deal with the matter variables. For these, we just have to consider the double-trace of the second Bianchi identity $\nabla^b G_{ab} = W_{abc}^{\quad bc} = 0\,$. Then we project it with $u^a$ and $h_a^{\ b}$, and rewrite everything in $1+3$ language.
The projection $u^a \nabla^b G_{ab} = 0$ gives 
\begin{equation}\label{eq_energy_conservation_equation}
    \dot{\rho} + \Theta\left(\rho+p\right) + D_a q^a + 2\,\dot{u}_a q^a + \sigma_{ab}\pi^{ab} = 0\,. 
\end{equation}
This is a relativistic generalisation of the Newtonian continuity equation, applying equally well to radiation or any other cosmic fluid as it does to non-relativistic matter. It describes all those fluids together, but if they are non-interacting, as is assumed in an FLRW cosmology, then Eq. (\ref{eq_energy_conservation_equation}) provides local conservation equations for each of them independently. It is also important for the covariant study of cosmological perturbations, because it describes how small density perturbations grow into nonlinear structures in the cosmic web.

The projection $h_a^{\ b}\nabla^c G_{bc} = 0$ is 
\begin{equation}\label{eq_momentum_conservation_equation}
\dot{q}_{\langle a \rangle} + D_a p + D^b \pi_{ab} + \frac{4}{3}\,\Theta\, q_a + \sigma_{ab}\,q^b + \eta_{abc}\,\omega^b \,q^c + \left(\rho+p\right)\dot{u}_a + \dot{u}^b\, \pi_{ab} = 0\,,
\end{equation}
which is a relativistic version of the Euler equation. This is our final $1+3$-covariant equation. It does not appear in FLRW cosmology, but it is a vital equation for studying cosmological perturbations in the $1+3$ decomposition, as it tells us how peculiar velocities, which are encoded in $q_a$\,, evolve.

For a general $1+3$ decomposition, Eqs. (\ref{eq_Raychaudhuri}-\ref{eq_momentum_conservation_equation}) form the complete set of equations of motion in an arbitrary spacetime. In most cosmological models, they are far simpler. In addition, we will see shortly that if $u^a$ is irrotational, then the ideas presented in the $3+1$ decomposition provide an extra equation, which is a generalisation of Friedmann's equation, that can be used in the $1+3$ formalism.

The $1+3$-covariant equations are very useful for studying homogeneous cosmologies, for which some $u^a$ always exists which makes the spatially projected derivatives $D_a$ of every quantity vanish. Then one is left with a set of first-order ordinary differential equations in time, plus some algebraic relations in place of the constraint equations\footnote{The situation is complicated somewhat if the homogeneous model is tilted \cite{King_1973,coley1994spacetimes,coley1996new,coley2006fluid}, i.e. the matter 4-velocity is not coincident with the normals to the homogeneous spacelike hypersurfaces. The $1+3$-covariant equations are ODEs in the $t$ variable defined by those normals, but they are not ODEs in the proper time of observers comoving with the matter.}. This makes it possible to analyse the large space of homogeneous models in a general fashion, without having to specify some model for the metric \cite{wainwright1997dynamical}.

It should be noted that we have not indicated here how the preferred timelike congruence that defines the $1+3$ decomposition should be picked out in a cosmological spacetime. This is a complicated and debated issue.
Some authors have suggested that $u^a$ should be taken to correspond to the average 4-velocity of dark matter \cite{Ellis_1999}. However, others have chosen to specify it by geometric means, using for example the properties of the kinematic and matter variables, or the electric and magnetic parts of the Weyl tensor with respect to the congruence \cite{Umeh_2011, Maartens_1998, Barnes_1989, Clifton_2020, Dunsby_1997, Sopuerta_1996, Heinesen_2022, bruni1994dynamics, rendall1996constant}. 
We will return to this issue in Chapters 4 and 7. For now, it suffices to say that in cosmology there is typically at least one well-motivated choice of preferred timelike congruence. Once we know it exists, we can perform the full $1+3$ decomposition with respect to it using the procedure detailed above.

\subsection{The $3+1$ formalism}\label{subsec:3+1}

We will now move on to the $3+1$ formalism, which is similar to the $1+3$ formalism in that it splits up spacetime into three spacelike dimensions and one timelike dimension. It is arguably more mathematically abstract, dealing with ideas in differential geometry, and we will not discuss it in as much detail. However, it introduces two concepts that will be very useful in the chapters that follow: spacetime foliations and gauges, and the initial-value constraints in relativistic gravity.

In the $1+3$ decomposition the fundamental object is a timelike vector $u^a$ that threads the spacetime. 
The properties of the spacetime are described by specifying the kinematics of that $u^a$ at every point. There is no requirement that this $u^a$ forms orthogonal hypersurfaces; it does so only if its vorticity vanishes. 

The $3+1$ decomposition considers the opposite perspective, where the fundamental object is an infinite family of three-dimensional hypersurfaces $\Sigma_t$\,, that are each level surfaces of the single parameter $t$. This parameter is some continuously varying function of the spacetime coordinates. 
The properties of the whole spacetime are obtained by determining the intrinsic properties of each surface $\Sigma_t$, and how those surfaces are related to one another as the function $t$ changes. 

The $3+1$ formalism was originally developed in a series of pioneering papers by Arnowitt, Deser and Misner (ADM) \cite{arnowitt1959dynamical, arnowitt2008republication, arnowitt1960canonical, arnowitt1961coordinate}. For a full discussion, see e.g Refs. \cite{Misner_1974, baumgarte2010numerical}. 
Another form of $3+1$ decomposition, related to the ADM approach, was used in a seminal work by Choquet-Bruhat, to demonstrate that GR has a well-posed initial value problem \cite{Choquet_Bruhat_1969}.
The formalism is an Hamiltonian formulation of General Relativity, allowing energy and momentum to be defined consistently in asymptotically flat spacetime. Perhaps most notably from a current point of view, it is the basis of the entire field of numerical relativity \cite{baumgarte2010numerical}.

The starting point for the ADM formalism is a foliation of spacetime. This means defining some function $t(x^a)$ which slices up the spacetime into the family of hypersurfaces $\Sigma_t$\,. These surfaces are referred to as the leaves of the foliation.
They have normals $n_a = -N\,\nabla_a t$, where $N(x^a)$ is the lapse function. The lapse function specifies the proper time interval between successive leaves $\Sigma_t$ and $\Sigma_{t+\mathrm{d}t}$\,, as measured by observers whose worldlines have tangent $n^a$, such that $\mathrm{d}\tau = N\, \mathrm{d}t$\,.

\begin{figure}
    \centering
    \includegraphics[width=\linewidth]{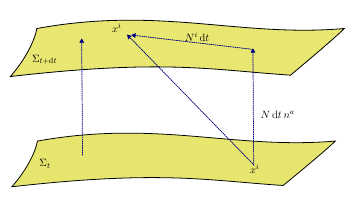}
    \caption{Physical interpretation of lapse function $N$ and shift vector $N^i\,$. The lapse function relates the proper time interval between infinitesimally separated leaves $\Sigma$ of the foliation to the coordinate time interval. The shift vector maps spatial coordinates $x^i$ from one leaf to the next.}
    \label{fig_lapse_and_shift}
\end{figure}

The leaves of the foliation are connected by a time vector $t^a = N n^a + N^a$. In this expression, $N^a$ is known as the shift vector, as it specifies how much the spatial coordinates $x^i$ shift between the hypersurfaces $\Sigma_t$ and $\Sigma_{t+\mathrm{d}t}$\,: $\mathrm{d}x^i = N^i\,\mathrm{d}t$\,. If the shift is non-zero, then curves of constant $x^i$ are not parallel to $n^a\,$.
The physical meanings of the lapse and shift are displayed in Fig. (\ref{fig_lapse_and_shift}).

The choice of the lapse function and shift vector - equivalently the choice of the foliation $\Sigma_t$ - is arbitrary, because of general covariance guaranteeing total coordinate freedom. We can relabel the spatial coordinates $x^i$ as much as we can like, and we are equally free to consider a set of clocks that measure a different time function $t$. A natural example in cosmology is the freedom to switch between using cosmic time and conformal time.
Hence, $N(x^a)$ and $N^i(x^a)$ are gauge variables. They contain 1 and 3 degrees of freedom (DOFs) respectively. These 4 DOFs correspond to our freedom to choose the 4 spacetime coordinates.

The metric is written in general as
\begin{equation}\label{eq_ADM_metric}
\mathrm{d}s^2 = -N^2 \,\mathrm{d}t^2 + f_{ij}\left(\mathrm{d}x^i + N^i \mathrm{d}t\right)\left(\mathrm{d}x^j + N^j \mathrm{d}t\right)\,,
\end{equation}
where $f_{ab}$ is the induced metric on $\Sigma_t$\,, and we have $N^a n_a = 0$, $n^b f_{ab} = 0$, $N^j f_{ij} = N_i\,$. 

One then seeks to identify all the relevant properties of the level surfaces $\Sigma_t$. This is done by determining both the constraint equations that must be satisfied on the initial surface, and the evolution equations which evolve the surfaces forward in time while successfully propagating the constraints.

Before we write down the $3+1$ equations of motion, it is worth clarifying the relationship between the metric (\ref{eq_ADM_metric}) and the quantities present in the $1+3$ formalism.
Consider the case where the preferred timelike vector $u^a$ in that decomposition is irrotational. Then it can also be written in the form $u_a \sim - \nabla_a t\,$, defining its own set of orthogonal hypersurfaces with induced metric $h_{ab}\,$. 
Hence, it must be possible to write the $1+3$ equations of motion in the $3+1$ picture, by mapping $u^a$ on to $n^a$ and $h_{ab}$ on to $f_{ab}$\,.

Let us now return to the $3+1$ situation. The entire spacetime geometry can be described by two symmetric tensors associated with $\Sigma_t$. One is the aforementioned induced metric $f_{ab}\,$, which is also sometimes referred to as the first fundamental form of $\Sigma_t\,$. The other is the extrinsic curvature tensor $K_{ab}\,$, also known as the second fundamental form.
The induced metric $f_{ab}$ and its spatial derivatives give all the information about the intrinsic curvature of the surface. Both $f_{ab}$ and $K_{ab}$ are spatially projected tensors.

The extrinsic curvature $K_{ab}$ gives all the information about how the surface is embedded into the 4-dimensional spacetime manifold. It is equal to the negative expansion tensor of the normal $n^a$. Since $n^a$ is constructed from the time function $t$\,, it is irrotational by definition, with
\begin{equation}
    K_{ab} = -f_a^{\ c}f_b^{\ d}\nabla_c n_d\,
\end{equation}
being manifestly symmetric in its indices.
A simple interpretation of $K_{ab}$ can be provided by introducing the Lie derivative of a tensor $t_{ab}^{\ \ c}$ with respect to some vector $v^a\,$,
\begin{equation}\label{eq_Lie_derivative_def}
    \left(\mathcal{L}_{\bf v} t\right)_{ab}^{\ \ c} = v^d\nabla_d t_{ab}^{\ \ c} + \left(\nabla_a v^d\right) t_{db}^{\ \ c} + \left(\nabla_b v^d\right) t_{ad}^{\ \ c} - \left(\nabla_d v^c\right) t_{ab}^{\ \ d}\,,
\end{equation}
which generalises obviously to tensors of higher and lower rank. The Lie derivative can be interpreted as measuring the rate of change of a covariantly defined object (be that a scalar, vector or tensor), as it is carried along an integral curve of the vector $v^a\,$. 

The extrinsic curvature $K_{ab}$ turns out to be very simply related to the Lie derivative of $f_{ab}$ with respect to the normal vector, by $K_{ab} = -\frac{1}{2}\mathcal{L}_{\bf n} f_{ab}\,$.
Hence, it is a measure of how much the induced metric changes as the spacetime is evolved forward from one leaf of the foliation to the next.
Note that $\mathcal{L}_{\bf n} f_{ab}$ is not generally parallel to the Lie derivative with respect to the time vector, $\mathcal{L}_{\bf t} f_{ab}\,$, because of the presence of the shift vector in $t^a = N\,n^a + N^a\,$. 

To get the equations of motion, one exploits a set of geometrical identities about how 3-dimensional hypersurfaces are embedded in a 4-dimensional manifold. These are the Gauss and Codazzi embedding equations, plus Ricci's equation for the normal derivative of $K_{ab}$\,. The final step is the same as in the $1+3$ decomposition: wherever it appears, we replace $R_{ab}$ with $8\pi G\,\left(T_{ab} - \frac{1}{2}\,T\,g_{ab}\right) + \Lambda g_{ab}$\,.

Gauss' equation relates the spatial curvature of the leaves of the foliation to the full projection of the spacetime curvature,
\begin{equation}\label{eq_Gauss_embedding_equation}
    ^{(3)}R_{abcd} + K_{ac}K_{bd} - K_{ad}K_{bc} = f_a^{\ e}f_b^{\ f}f_c^{\ g}f_d^{\ h}\, ^{(4)}R_{efgh}\,,
\end{equation}
where we have made explicit whether the Riemann tensor in question refers to the Christoffel symbols $^{(4)}\Gamma^a_{\ bc}$ associated with the spacetime 4-metric $g_{ab}$, or those $^{(3)}\Gamma^a_{\ bc}$ associated with the induced 3-metric $f_{ab}$\,.

Codazzi's equation relates the variation in the extrinsic curvature over the leaves to the partial (3 indices projected) projection of the 4-dimensional Riemann tensor,
\begin{equation}\label{eq_Codazzi_equation}
    D_b K_{ac} - D_a K_{bc} = f_a^{\ d}f_b^{\ e}f_c^{\ f}n^g\, ^{(4)}R_{defg}\,.
\end{equation}
Ricci's equation relates the normal Lie derivative of $K_{ab}$ to the double projection of the Riemann tensor in the normal direction,
\begin{equation}\label{eq_Lie_deriv_of_K}
    \mathcal{L}_{\bf n}K_{ab} = n^c n^d f_a^{\ e}f_b^{\ f}\, ^{(4)}R_{cedf} - \frac{1}{N}\,D_a D_b\,N - K_{ac}K_b^{\ c}\,.
\end{equation}
Using these, and the relation $K_{ab} = -\frac{1}{2}\mathcal{L}_{\bf n} f_{ab}\,$, we obtain the full set of equations of motion. There are 4 constraints, one for each of the degrees of freedom in $N$ and $N^i$\,, and 12 evolution equations: 6 for each DOF in $f_{ab}\,$, and 6 for each DOF in $K_{ab}\,$. Like in the $1+3$ formalism, all of these evolution equations are first-order ODEs.

The first initial-value constraint is the Hamiltonian constraint, provided by fully contracting Gauss' equation (\ref{eq_Gauss_embedding_equation}),
\begin{equation}\label{eq_ADM_equation_Hamiltonian_constraint}
   ^{(3)}R + K^2 - K_{ij}K^{ij} = 2\, n^a n^b G_{ab} = 16\pi G\,\rho + 2\,\Lambda\,,
\end{equation}
where in the final equality we have used Einstein's equations, but have retained the middle equality in order to show how the Hamiltonian constraint could be constructed in a modified theory of gravity. We have denoted the trace of the extrinsic curvature by $f^{ij}K_{ij} = K\,$.

The other three constraints are given together by the momentum constraint, which is obtained by contracting Codazzi's equation,
\begin{equation}\label{eq_ADM_equation_momentum_constraint}
   D^j\left(K_{ij}-f_{ij}K\right) = -G_{ab} n^a h_i^{\ b} = 8\pi G\, q_i \,.
\end{equation}

There are also $3+1$ evolution equations. We will not make use of them in this thesis, but include them here for the sake of completeness.

The evolution equations for the 6 independent components of the extrinsic curvature come from taking Eq. (\ref{eq_Lie_deriv_of_K}), and then converting $\mathcal{L}_{\bf n} \longrightarrow \mathcal{L}_t \longrightarrow \partial_t$\,:
\begin{eqnarray}\label{eq_ADM_equation_extrinsic_curvature}
\partial_t K_{ij} &=& -D_i D_j N + N\left(^{(3)}R_{ij} - 2K_{ki}K_j^{\ k} + K K_{ij}\right) \\
\nonumber && - N\left[\pi_{ij} + \frac{1}{2}f_{ij}\left(\rho - p\right)\right] + N^k D_k K_{ij} + 2 K_{k(i}D_{j)}N^k\,.
\end{eqnarray}

Similarly, the evolution equations for the 6 independent components of $f_{ab}$ are provided by $K_{ab} = -\frac{1}{2}\mathcal{L}_{\bf n} f_{ab}\,$, which essentially defines the extrinsic curvature:
\begin{equation}\label{eq_ADM_equation_induced_metric}
\partial_t f_{ij} = -2N\,K_{ij} + 2D_{(i}N_{j)}\,.
\end{equation}
These evolution equations are very much like Hamilton's equations in classical mechanics. Roughly speaking, $f_{ij}$ serve the same role as the canonical coordinates, and their equations of motion (\ref{eq_ADM_equation_induced_metric}) just define the canonical momenta. The canonical momenta, which are related to $K_{ij}$ \cite{Misner_1974}, then have equations of motion (\ref{eq_ADM_equation_extrinsic_curvature}) that contain the source fields directly.

The choice of foliation $t(x^a)$, appearing explicitly in the evolution equations (\ref{eq_ADM_equation_extrinsic_curvature}) and (\ref{eq_ADM_equation_induced_metric}) through the presence of the lapse and shift, is therefore of fundamental importance in the $3+1$ decomposition. This is entirely equivalent to the choice of preferred congruence $u^a$ in the $1+3$ decomposition, as long as that $u^a$ is chosen so as to be irrotational. In the chapters that follow, we will use irrotational congruences wherever possible, so that ideas from the $1+3$ and $3+1$ decompositions can be used interchangeably.

The Hamiltonian constraint is especially illuminating if we translate it into $1+3$ language, for an irrotational $u^a$. Then $K_{ab} = -h_a^{\ c} h_b^{\ d}\nabla_c u_d = - \Theta_{ab} = -\left(\frac{1}{3}\,\Theta\,h_{ab} + \sigma_{ab}\right)\,,$ and Eq. (\ref{eq_ADM_equation_Hamiltonian_constraint}) becomes
\begin{equation}\label{eq_Hamiltonian_constraint_1+3}
\frac{2}{3}\Theta^2 - 2\sigma^2 + \, ^{(3)}R  = 16\pi G\,\rho + 2\Lambda\,, 
\end{equation}
where we have introduced the shear scalar $\sigma^2 = \frac{1}{2}\sigma_{ab}\sigma^{ab}\,$.
In FLRW cosmologies $\Theta = 3H$ and $\sigma_{ab} = 0\,$, so the Hamiltonian constraint just reduces to the Friedmann equation. 
We will usually think of the Hamiltonian constraint as one of the $1+3$ equations, as long as $\omega_a = 0$\,.

Likewise, the momentum constraint equation (\ref{eq_ADM_equation_momentum_constraint}) for a foliation orthogonal to an irrotational $u^a$ gives $\frac{2}{3}D_i\Theta - D^j \sigma_{ij} = -8\pi G\,q_i\,$, which is the equation $h_a^{\ b}S^c_{\ bc} = 0$ in the $1+3$ decomposition, with $\omega_a = 0\,$.
In a later chapter, we will study how an initial value problem can be constructed in a theory-independent way in cosmology. This means that one has to obtain the Hamiltonian and momentum constraints.
The analysis here tells us that as long as we work with an irrotational congruence, we can do this using the $1+3$ equations, rather than having to work with the $3+1$ formalism directly.

\subsection{The $1+1+2$ formalism}\label{subsec:1+1+2}

The $1+3$ decomposition is very well-suited to situations where there is a preferred timelike vector picked out in the spacetime, and the $3+1$ decomposition to the related situations where there is a preferred set of three-dimensional spacelike hypersurfaces. 

If the universe is spatially isotropic, then there will be no further preferred direction. However, many cosmological models, such as the Bianchi universes, are anisotropic in general. 
Often, they have a single preferred spatial direction, along which cosmological observables, such as galaxy number counts or the CMB temperature, display variation which is much stronger than any other direction, and so it might not be ideally treated using perturbation theory about an isotropic background metric. 

In that case, it will be useful to decompose all the physical quantities of interest once more than in the $1+3$ decomposition, with respect to that preferred spacelike vector. We will therefore obtain a set of equations that is, like the $1+3$ equations, physically equivalent to Einstein's equations, but which is written entirely in an even more decomposed form. This is referred to as the $1+1+2$ formalism.
It was first introduced by Greenberg \cite{Greenberg_1970}, developed in Refs. \cite{Tsamparlis_1983, Mason_1985, Zafiris_1997, Clarkson_2003, Betschart_2004}, and worked out in full by Clarkson \cite{Clarkson_2007} and Keresztes et al. \cite{Keresztes_2015}.

Let us therefore suppose that we have a preferred spacelike vector $m^a$, which is orthogonal to the preferred timelike vector $u^a$\,, i.e. $m_a u^a = 0$ and $m_a m^a = 1$.
We will first need to define all the kinematic quantities associated with $m^a$, just like we did for $u^a$ in the $1+3$ formalism. 
Then we will take all of these, and all of the $1+3$ kinematic, curvature and matter variables, and decompose them all with respect to $m^a$ as well as $u^a$. This will give us a new set of $1+1+2$-covariant scalars, vectors and tensors. 

Finally, we will follow the same approach as in the $1+3$ decomposition, to get new equations of motion: write down the Ricci identity for $m^a$, and perform all its projections with respect to both $m^a$ and $u^a$. The rest of the equations come from projecting all the $1+3$ equations with respect to $m^a$, and rewriting all the objects that appear in those equations in their $1+1+2$ forms. 
The full set of equations that results is a full decomposition of Einstein's equations with respect to both a timelike and a spacelike preferred vector. These first-order PDEs are very useful for studying anisotropic cosmological models, because many of the important degrees of freedom are contained in covariantly defined scalars, which are typically much easier to analyse than vectors and tensors. 

The covariant derivative of $m^a$ can be decomposed as
\begin{equation}\label{eq_nabla_a m_b}
\nabla_a m_b = - u_a \dot{m}_b + D_a m_b  + u_b m^c D_a u_c + u_a u_b m^c \dot{u}_c \,.
\end{equation}
A projection tensor onto the two-dimensional spaces orthogonal to both $u^a$ and $m^a$ can be defined as
\begin{equation} \label{eq_bigm}
M_{ab} = h_{ab} - m_a m_b = g_{ab} + u_a u_b - m_a m_b \,.
\end{equation}
We also define $\epsilon_{ab} = \epsilon_{[ab]} = \eta_{abc}m^c$, the 2-dimensional area form on that space.

The projected time and space derivatives of $m^a$ can be written as
\begin{equation}\label{eq_D_a_m_b}
\dot{m}_a = \mathcal{A} \, u_a + \alpha_a \qquad {\rm and} \qquad D_a m_b = m_a a_b + \frac{1}{2}\phi M_{ab} + \xi\epsilon_{ab} + \zeta_{ab}\, ,
\end{equation}
where we have defined a new set of kinematic variables for the spacelike vector $m^a$, that are analogous to the kinematic variables $\Theta$, $\sigma_{ab}$, $\omega_a$ and $\dot{u}_a$ that exist for the timelike vector $u_a\,$:
\begin{itemize}

    \item $\alpha_a = M_{ab} \dot{m}^b$\,, the projection of $\dot{m}^a$ orthogonal to $u^a$.
    
    \item $\mathcal{A} =  -u^a \dot{m}_a = m_a \dot{u}^a$. This can equally well be thought of as the projection of the acceleration of $u^a$ parallel to $m^a$, or the (negative) projection of the acceleration of $m^a$ in the timelike direction $u^a\,$. These are equivalent because $m_a u^a = 0\,$.
    
    \item $a_a = m^b D_b m_a\,$, the non-geodesy of the preferred spacelike vector $m^a\,$. 
    
    \item $\phi = D_a m^a\,$, the expansion of the orthogonal 2-spaces along the direction parallel to $m^a\,$. This is rather like $\Theta = \nabla_a u^a\,,$ but rather than telling us the local rate of volume expansion with respect to some timelike coordinate defined by $u^a$, it tells us the local rate of area expansion with respect to a spacelike coordinate defined by $m^a$\,.
    The argument follows identically to that for $\Theta\,$. Consider a small patch of the orthogonal 2-space, of area $\delta A\,$. Now perform an infinitesimal variation of the affine parameter along a spacelike curve parallel to $m^a\,$, It follows that the determinant of the 2-space projection tensor $M_{ab}$ satisfies $\sqrt{M}' = \phi\,\sqrt{M}\,$, where we have denoted the derivative with respect to the spacelike affine parameter with a prime. Thus $\dfrac{\delta A'}{\delta A} = \phi\,$.
    
    \item $\xi = \frac{1}{2}\,\epsilon^{ab}\, D_a m_b\,$, the twisting of the 2-spaces along $m^a$\,. We can think of this as the equivalent for $m^a$ of the vorticity $\omega_a$ associated with $u^a\,$. It is a scalar rather than a vector, because a rank-2 antisymmetric tensor only has one degree of freedom.
    
    \item $\zeta_{ab} = \left[M^{\ c}_{(a}M^{\ d}_{b)} - \frac{1}{2}M_{ab}M^{cd}\right]D_{\langle c} m_{d \rangle}\,$, the area-preserving shear of the 2-spaces along $m^a$. This can be considered the equivalent of $\sigma_{ab}$ for $m^a\,$. Here we have used $M^{\ c}_{(a}M^{\ d}_{b)} - \frac{1}{2}M_{ab}M^{cd}$ to project orthogonally to $m^a$, symmetrise, and then make the resulting tensor trace-free.
    
\end{itemize}

This completes the set of new kinematic quantities. The rest of the $1+1+2$-decomposed variables come from taking all the vector and tensor quantities defined in the $1+3$ decomposition and then projecting them parallel and orthogonal to $m^a\,$. 

\begin{figure}
    \centering
    \includegraphics[width=0.9\linewidth]{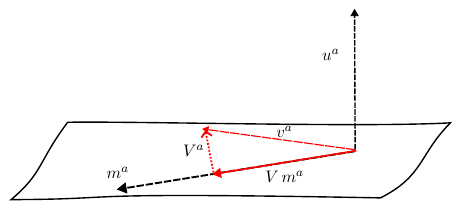}
    \caption{The $1+1+2$ decomposition of a spatial vector $v^a$, with respect to both $u^a$ and $m^a\,$.}
    \label{fig_112_decomposition_u_m}
\end{figure}

Let us start with vectors $v_a = v_{\langle a \rangle} = h_a^{\ b}v_b$. They can be decomposed as
\begin{equation} \label{eq_vdef}
v_a = V \,m_a + V_a \, ,
\end{equation}
where $V = v_b m^b$ is the projection of $v_a$ parallel to $m_a$, and $V_a = M_a^{\ b}\,v_b$ is the projection of $v_a$ into the screen spaces. 
This is displayed in Fig. \ref{fig_112_decomposition_u_m}\,, where we have suppressed the third spatial dimension so that both the timelike direction and the orthogonal rest spaces can be visualised on the same diagram.

For projected, symmetric and trace-free tensors $t_{ab} = t_{\langle ab\rangle}$, we can write
\begin{equation} \label{eq_tdef}
t_{ab}= \tau\left(m_a m_b - \frac{1}{2}M_{ab}\right) + 2 \tau_{(a}m_{b)} + \tau_{ab} \, ,
\end{equation}
where $\tau = t_{ab} m^a m^b$, $\tau_a = M_{ab}m_c t^{bc}$ and $\tau_{ab} = \left[M^{\ c}_{(a}M^{\ d}_{b)} - \frac{1}{2}M_{ab}M^{cd}\right]t_{cd}$
are the projections of $t_{ab}$ twice parallel, once parallel and once orthogonal, and twice orthogonal to $m^a$, respectively.

Using this approach, we can covariantly split up the acceleration and vorticity vectors associated with $u^a$\,,
\begin{eqnarray}
&& \dot{u}_a = \mathcal{A}\,m_a + \mathcal{A}_a \qquad {\rm and} \qquad \omega_a = \Omega\, m_a + \Omega_a \, .
\end{eqnarray}
The expansion $\Theta$ associated with $u^a$ is a scalar, so it is left alone. The final kinematic quantity we need is the shear $\sigma_{ab}$, which can be decomposed as
\begin{eqnarray}
\sigma_{ab} = \Sigma \left(m_a m_b - \frac{1}{2}M_{ab}\right) + 2\Sigma_{(a}m_{b)} + \Sigma_{ab} \,.
\end{eqnarray}
Here the quantities $\mathcal{A}$, $\Omega$ and $\Sigma$ are covariantly defined scalars, obtained using the rules defined by Eqs. (\ref{eq_vdef}) and (\ref{eq_tdef}). Using those same rules, $\mathcal{A}_a$, $\Omega_a$, $\Sigma_a$ and $\Sigma_{ab}$ are covariantly defined vectors and tensors that live in the two-dimensional spacelike surfaces orthogonal to both $u^a$ and $m^a$.

We now have all the kinematic quantities associated with both our preferred vectors. We are not quite done, because we need the objects associated with the Ricci and Weyl curvature of the spacetime. 
As in the $1+3$ decomposition, we assume that gravity is described by General Relativity. Hence, $R_{ab}$ is given by the matter and energy content of the spacetime, via Einstein's equations.

Recall that the energy-momentum tensor was already covariantly decomposed with respect to $u^a$ according to Eq. (\ref{eq_energy_momentum_tensor_decomposition}), into energy density $\rho$, pressure $p$, momentum density $q_a$ and anisotropic stress $\pi_{ab}\,$. To put $T_{ab}$ into $1+1+2$-covariant form, we therefore split $q_a$ and $\pi_{ab}$ with respect to $m^a\,$. 
Defining all quantities according to Eqs. (\ref{eq_vdef}) and (\ref{eq_tdef}), we get
\begin{eqnarray*}
&& q_a = Q m_a + Q_a \quad
{\rm and} \quad \pi_{ab} = \Pi \left(m_a m_b - \frac{1}{2}M_{ab}\right) + 2\Pi_{(a}m_{b)} + \Pi_{ab} \,.
\end{eqnarray*}

Finally, the electric and magnetic parts of the Weyl tensor, defined with respect to $u^a$, are decomposed again by projecting orthogonal and parallel to $m^a$. This gives
\begin{eqnarray*}
E_{ab} &=& \mathcal{E} \left(m_a m_b - \frac{1}{2}M_{ab}\right) + 2\mathcal{E}_{(a}m_{b)} + \mathcal{E}_{ab}\,, \\ 
{\rm and} \quad H_{ab} &=& \mathcal{H} \left(m_a m_b - \frac{1}{2}M_{ab}\right) + 2\mathcal{H}_{(a}m_{b)} + \mathcal{H}_{ab}\, .
\end{eqnarray*}

The full, irreducible set of $1+1+2$-covariant quantities is now complete. They are needed in order to fully specify the spacetime geometry in a way that is equivalent to writing down the metric at every point.
To recap, let us group them together into all the scalars, vectors and tensors.

The scalars are
\begin{equation}\label{eq_112scalars}
\left \lbrace \Theta, \mathcal{A}, \Sigma, \Omega, \phi, \xi, \mathcal{E}, \mathcal{H}, \rho, p, Q, \Pi \right \rbrace\,,
\end{equation}
the vectors are
\begin{equation}
\left \lbrace \mathcal{A}_a, \Sigma_a, \Omega_a, \alpha_a, a_a, \mathcal{E}_a, \mathcal{H}_a, Q_a, \Pi_a \right\rbrace\,,
\end{equation}
and the tensors are
\begin{equation}
\left \lbrace \Sigma_{ab}, \zeta_{ab}, \mathcal{E}_{ab}, \mathcal{H}_{ab}, \Pi_{ab} \right \rbrace\,.
\end{equation}

The full set of $1+1+2$-covariant equations is derived in Refs. \cite{Clarkson_2007, Keresztes_2015}. It is very lengthy, and we will not display it in its entirety. In this thesis, we will only make use of the $1+1+2$-scalar equations. It is these equations that we present here. They are covariant scalar equations, making them ideal for physical interpretation, because there is no possibility of spurious coordinate effects.
Note that for the remainder of this section, we set $8\pi G \equiv 1$ for simplicity. To reinsert it, one need only multiply all matter variables $\left\lbrace \rho, p, Q, \Pi, Q_a, \Pi_a, \Pi_{ab}\right\rbrace$ by $8\pi G$ wherever they appear.

We will split the $1+1+2$-scalar equations into four types, and deal with each in turn.
\begin{enumerate}
    \item $1+3$ scalar equations, written in terms of $1+1+2$-covariant quantities.
    \item $1+3$ vector equations, projected parallel to the preferred spacelike vector $m^a$\,.
    \item $1+3$ tensor equations, projected twice parallel to $m^a\,$.
    \item Novel equations coming from the Ricci identity for $m^a$; $R_{abc} = 2\nabla_{[a}\nabla_{b]}m_c - R_{abcd}m^d = 0\,$.
\end{enumerate}

Type 1: $1+3$ scalar equations. There are three: the Raychaudhuri equation (\ref{eq_Raychaudhuri}), the constraint (\ref{eq_vorticity_divergence}) on the divergence of the vorticity, and the local energy conservation equation (\ref{eq_energy_conservation_equation}). We just need to rewrite the variables that appear in these equations in $1+1+2$-covariant form; e.g. $\omega_a$ is replaced by $\Omega\,m_a + \Omega_a\,$.

The Raychaudhuri equation (\ref{eq_Raychaudhuri}) becomes
\begin{eqnarray}\label{eq_Raychaudhuri_112}
    \dot{\Theta} + \frac{1}{3}\,\Theta^2 + \frac{3}{2}\,\Sigma^2 - 2\,\Omega^2 - \mathcal{A}^2 - \phi\,\mathcal{A} - m^a D_a\mathcal{A} && \\
    \nonumber - M^{ab}\,D_a \mathcal{A}_b + a_a \mathcal{A}^a - \mathcal{A}_a\mathcal{A}^a + 2\,\Sigma_a \Sigma^a - 2\,\Omega_a \Omega^a + \Sigma_{ab}\Sigma^{ab} &=& -\,\frac{1}{2}\,\left(\rho + 3p\right) + \Lambda\,.
\end{eqnarray}
This does not seem like a simplification, compared to the $1+3$ form (\ref{eq_Raychaudhuri}) of this equation. However, note that on the left hand side we have grouped together all the scalar terms, then the (contractions of) derivatives of scalars, then (contractions of) vectors and tensors. The utility of the $1+1+2$ formalism will become clear when we demonstrate the conditions under which all the vectors and tensors in the $1+1+2$ decomposition vanish, leaving us with an equation that is composed purely of scalars.

The vorticity divergence equation (\ref{eq_vorticity_divergence}) becomes
\begin{equation}\label{eq_vorticity_divergence_112}
    \left(\phi - \mathcal{A}\right)\,\Omega + m^a D_a\Omega + M^{ab}\,D_a\Omega_b - a_a\,\Omega^a - \mathcal{A}_a\,\Omega^a = 0\,.
\end{equation}

Finally, the local energy conservation equation (\ref{eq_energy_conservation_equation}) gives
\begin{eqnarray}\label{eq_energy_conservation_112}
    \dot{\rho} + \Theta\,\left(\rho + p\right) + \phi\,Q + 2\,\mathcal{A}\,Q + \frac{3}{2}\,\Sigma\,\Pi + m^a D_a Q + M^{ab}\,D_a Q_b && \\
    \nonumber -a_a\,Q^a + 2\,\mathcal{A}_a\,Q^a + 2\,\Sigma_a\,\Pi^a + \Sigma_{ab}\,\Pi^{ab} &=& 0\,.
\end{eqnarray}

Type 2: $1+3$ vector equations, projected parallel to $m^a\,$. There are five of these, coming from the vorticity evolution equation (\ref{eq_vorticity_evolution}), the $1+3$ momentum constraint (\ref{eq_momentum_constraint_1+3})\footnote{This is only really the momentum constraint if $\omega_a = 0$\,, but it is convenient just to refer to the equation by that name regardless.}, the $E_{ab}$ and $H_{ab}$ divergence equations (\ref{eq_div_E_constraint}) and (\ref{eq_div_H_constraint}), and finally the local momentum conservation equation (\ref{eq_momentum_conservation_equation}).

Projecting Eq. (\ref{eq_vorticity_evolution}) parallel to $m^a$ gives
\begin{eqnarray}\label{eq_vorticity_evolution_112}
    \dot{\Omega} + \left(\frac{2}{3}\,\Theta - \Sigma\right)\,\Omega - \mathcal{A}\,\xi - \Omega_a\,\Sigma^a - \Omega_a\,\alpha^a - \frac{1}{2}\epsilon^{ab}\,D_a\mathcal{A}_b = 0\,.
\end{eqnarray}

The momentum constraint (\ref{eq_momentum_constraint_1+3}) similarly produces the scalar equation
\begin{eqnarray}\label{eq_momentum_constraint_112}
    m^a D_a \Sigma - \frac{2}{3}\,m^a D_a \Theta +\frac{3}{2}\,\phi\,\Sigma + 2\,\xi\,\Omega + M^{ab}\,D_a\Sigma_b + \epsilon^{ab}\,D_a \Omega_b && \\
    \nonumber - 2\,\Sigma_a\,a^a + 2\,\epsilon^{ab}\,\mathcal{A}_a\,\Omega_b - \Sigma_{ab}\,\zeta^{ab} &=& -\,Q\,.
\end{eqnarray}

The equations (\ref{eq_div_E_constraint}) and (\ref{eq_div_H_constraint}) for the divergences of the electric and magnetic Weyl curvature tensors give respectively
\begin{eqnarray}
&& \nonumber \hspace{-1cm} m^a D_a \, \mathcal{E} + \frac{3}{2}\,\phi\,\mathcal{E} - 3\,\Omega\,\mathcal{H} + M^{ab}\,D_a\mathcal{E}_b - 2\,\mathcal{E}_a\,a^a - 3\,\Omega_a\,\mathcal{H}^a - \epsilon^{ab}\,\Sigma_{ac}\,\mathcal{H}^c_{\ b} - \mathcal{E}_{ab}\,\zeta^{ab} \\
\nonumber &=& m^a D_a\left(\frac{1}{3}\,\rho - \frac{1}{2}\,\Pi\right) - \frac{3}{4}\,\phi\,\Pi - \frac{1}{3}\,\Theta\,Q + \frac{1}{2}\,\Sigma Q - \frac{1}{2}\,M^{ab}\,D_a\Pi_b \\
&& \, + \Pi_a\,a^a + \frac{1}{2}\,\Sigma_a\, Q^a - \frac{3}{2}\,\epsilon^{ab}\,\Omega_a\,Q_b + \frac{1}{2}\,\Pi_{ab}\,\zeta^{ab}\,,
\end{eqnarray}
and
\begin{eqnarray}
&& \nonumber \hspace{-1cm} m^a D_a\,\mathcal{H} + \frac{3}{2}\,\phi\mathcal{H} + 3\,\mathcal{E}\,\Omega - 2\,\mathcal{H}_a\,a^a + M^{ab}\,D_a\mathcal{H}_b + 3\,\Omega_a\,\mathcal{E}^a + \epsilon^{ab}\,\Sigma_a^{\ c}\,\mathcal{E}_{bc} - \zeta_{ab}\,\mathcal{H}^{ab} \\
&=& -\left(\rho + p - \frac{1}{2}\,\Pi\right)\,\Omega - Q\,\xi - \frac{1}{2}\,\epsilon^{ab}\,D_a Q_b + \frac{1}{2}\,\Omega_a\,\Pi^a - \frac{1}{2}\,\epsilon^{ab}\,\Sigma_a^{\ c}\,\Pi_{bc}\,.
\end{eqnarray}

The scalar projection of the local momentum conservation equation (\ref{eq_momentum_conservation_equation}) parallel to $m^a$ is 
\begin{eqnarray}\label{eq_momentum_conservation_112}
    \dot{Q} + \frac{4}{3}\,\Theta\,Q + \Sigma\,Q + \left(\frac{3}{2}\,\phi + \mathcal{A}\right)\,\Pi + \left(\rho + p\right)\,\mathcal{A} + m^a D_a\left(p + \Pi\right) && \\
    \nonumber + \, M^{ab}\,D_a\Pi_b - \left(\alpha^a - \Sigma^a + \epsilon^{ab}\,\Omega_b\right)\,Q_a + \left(\mathcal{A}^a - 2\,a^a\right)\,\Pi_a - \zeta_{ab}\,\Pi^{ab} &=& 0\,.
\end{eqnarray}

Next we have type 3, the double projections with $m^a$ of tensor equations from the $1+3$ formalism. There are four of them, coming from the shear evolution equation (\ref{eq_shear_evolution}), the equation (\ref{eq_Hweyl_vs_shear_vorticity}) for $H_{ab}$ in terms of shear and vorticity, and the evolution equations (\ref{eq_Eweyl_evolution}) and (\ref{eq_Hweyl_evolution}) for the electric and magnetic parts of the Weyl tensor.

The evolution equation for the scalar shear $\Sigma$ is 
\begin{eqnarray}\label{eq_shear_evol_eqn_112}
\dot{\Sigma} + \frac{2}{3}\,\Theta\,\Sigma + \frac{1}{2}\,\Sigma^2 + \frac{2}{3}\,\Omega^2 + \frac{1}{3}\,\phi\,\mathcal{A} - \frac{2}{3}\,\mathcal{A}^2 + \mathcal{E} - \frac{2}{3}\,m^a D_a\mathcal{A} + \frac{1}{3}\,M^{ab}D_a\mathcal{A}_b && \\
\nonumber \, - 2\,\alpha_a\,\Sigma^a + \frac{1}{3}\,\Sigma_a\,\Sigma^a - \frac{1}{3}\,\mathcal{A}_a\,\mathcal{A}^a + \frac{2}{3}\,a_a\,\mathcal{A}^a - \frac{1}{3}\,\Omega_a\,\Omega^a - \frac{1}{3}\,\Sigma_{ab}\,\Sigma^{ab} &=& \frac{1}{2}\,\Pi\,.
\end{eqnarray}

The double contraction with $m^a$ of the $H_{ab}$ equation (\ref{eq_Hweyl_vs_shear_vorticity}) gives an equation for the scalar magnetic Weyl curvature $\mathcal{H}\,$ in terms of the vorticity and shear scalars and vectors, and the shear tensor:
\begin{equation}\label{eq_Hweyl_vs_vorticity_112}
    \mathcal{H} - \phi\,\Omega - 3\,\xi\,\Sigma + 2\,\mathcal{A}\,\Omega - M^{ab}D_a\Omega_b - \epsilon^{ab}D_a\Sigma_b + \epsilon^{ab}\,\zeta_{ac}\,\Sigma_b^{\ c} = 0\,.
\end{equation}
The evolution equation for the scalar electric Weyl curvature is
\begin{eqnarray}
    && \dot{\mathcal{E}} + \Theta\,\mathcal{E} - \frac{3}{2}\,\Sigma\,\mathcal{E} - 3\,\xi\,\mathcal{H} - \epsilon^{ab}D_a\mathcal{H}_b - 2\,\epsilon^{ab}\,\mathcal{A}_a\,\mathcal{H}_b  \\
    \nonumber && \, - \left(2\,\alpha^a + \Sigma^a - \epsilon^{ab}\,\Omega_b\right)\,\mathcal{E}_a  - \Sigma^{ab}\,\mathcal{E}_{ab} + \epsilon^{ab}\,\mathcal{H}_{bc}\zeta_a^{\ c} = -\frac{1}{2}\,\dot{\Pi} - \frac{1}{6}\,\Theta\,\Pi  \\
    \nonumber && \, - \frac{1}{4}\left(2\,\rho + 2\,p + \Pi\right)\,\Sigma + \frac{1}{6}\left(\phi - 4\,\mathcal{A}\right)\,Q  - \frac{1}{3}\,m^a D_a Q + \frac{1}{6}\,M^{ab}D_a Q_b \\
    \nonumber && \, + \frac{1}{3}\left(a_a + \mathcal{A}_a\right)\,Q^a + \left(\alpha^a - \frac{1}{6}\,\Sigma^a - \frac{1}{2}\epsilon^{ab}\,\Omega_b\right)\,\Pi_a - \frac{1}{2}\,\Sigma^{ab}\,\Pi_{ab}\,.
\end{eqnarray}
In the above, we have placed all terms involving matter fields on the right hand side. Like the pressure $p$, the anisotropic stress scalar $\Pi$ does not have its own evolution equation. In practice, they are not usually treated as independent. Instead, equations of state $p\,(\rho)$ and $\Pi\,(\rho)$ are supplied.

The evolution equation for $\mathcal{H}$ follows in the same way:
\begin{eqnarray}
    && \hspace{-1cm}  \dot{\mathcal{H}} + \Theta\,\mathcal{H} - \frac{3}{2}\,\Sigma\,\mathcal{H} + 3\,\xi\,\mathcal{E} + \epsilon^{ab}D_a\mathcal{E}_b + 2\,\epsilon^{ab}\,\mathcal{A}_a\,\mathcal{E}_b - \left(2\,\alpha^a + \Sigma^a - \epsilon^{ab}\,\Omega_b\right)\,\mathcal{H}_a \\
    \nonumber && \hspace{-1cm} \, + \, \Sigma^{ab}\,\mathcal{H}_{ab} + \frac{1}{2}\,\epsilon^{ab}\,\mathcal{E}_{bc}\,\zeta_a^{\ c} = \Omega\,Q + \frac{3}{2}\,\xi\,\Pi + \frac{1}{2}\,\epsilon^{ab}D_a\Pi_b - \frac{1}{2}\,\Omega^a\,Q_a - \frac{1}{2}\,\epsilon^{ab}\,Q_a\,\Sigma_b\,.
\end{eqnarray}

Finally we present the equations of type 4, which are the genuinely novel scalar equations arising in the $1+1+2$ decomposition, that cannot be derived using projections of the existing $1+3$ equations. They are obtained through various scalar projections of the Ricci identity for $m^a$, $R_{abc} = 2\,\nabla_{[a}\nabla_{b]} \, m_c - R_{abcd}\,m^d = 0\,$. 

This means that they are evolution and constraint equations for kinematic scalar variables associated with the preferred spacelike congruence. There are two such scalars, the area expansion $\phi$ of the congruence as $m^a$ is parallel-transported along its integral curves, and the twisting $\xi$ of the congruence that occurs when $m^a$ is parallel-transported. 
Therefore, there are four new scalar equations: an evolution equation and a constraint for $\xi$, and an evolution equation and a constraint for $\phi\,$.

The projection $u^a\,\epsilon^{bc}\,R_{abc}$ gives the evolution equation for $\xi$\,,
\begin{eqnarray}\label{xi_evolution_eqn_112}
    \dot{\xi} + \frac{1}{3}\,\Theta\,\xi - \frac{1}{2}\,\Sigma\,\xi + \frac{1}{2}\,\phi\,\Omega - \mathcal{A}\,\Omega - \frac{1}{2}\,\mathcal{H} - \frac{1}{2}\,\epsilon^{ab}D_a\alpha_b && \\
    \nonumber \, - \,\frac{1}{2}\left(a_a + \mathcal{A}_a\right)\,\Omega^a - \frac{1}{2}\,\epsilon^{ab}\left(a_a + \mathcal{A}_a\right)\left(\alpha_b + \Sigma_b\right) + \frac{1}{2}\,\epsilon^{ca}\,\zeta^b_{\ c}\Sigma_{ab} &=& 0\,.
\end{eqnarray}
Like the vorticity $\omega_a$ associated with the timelike congruence, the twisting scalar $\xi$ associated with the spacelike congruence has no source term from the matter fields.

The evolution equation for $\phi$ comes from the projection $u^a\,M^{bc}\,R_{abc}\, = 0\,$,
\begin{eqnarray}\label{eq_phi_evolution_eqn_112}
\dot{\phi} + \frac{1}{3}\,\Theta\,\phi - \frac{1}{2}\,\Sigma\,\phi - \frac{2}{3}\,\Theta\,\mathcal{A} + \Sigma\,\mathcal{A} - 2\,\xi\,\Omega - M^{ab}D_a\alpha_b + a_a\,\mathcal{A}^a && \\
\nonumber \, - \, \alpha_a\,\mathcal{A}^a + \left(\Sigma^a - \epsilon^{ab}\,\Omega_b\right)\,\mathcal{A}_a -  \left(\Sigma^a - \epsilon^{ab}\,\Omega_b\right)\,a_a + \zeta^{ab}\,\Sigma_{ab} &=& \, Q\,.
\end{eqnarray}
This shows how the area of the two-dimensional spacelike surfaces orthogonal to $m^a$ is increased by the presence of a positive momentum density $Q$ in that direction. This spreading out of neighbouring spacelike curves from one another is somewhat like the effect of a negative spatial curvature in an FLRW cosmology. We will see shortly that this is not a coincidence.

We get a constraint on the projected derivative of $\xi$ from $m^a \,\epsilon^{bc}\,R_{abc} = 0\,$,
\begin{equation}\label{eq_phi_xi_constraint_112}
    m^a D_a \xi + \phi\,\xi - \frac{1}{3}\,\Theta\,\Omega - \Sigma\,\Omega - \frac{1}{2}\,\epsilon^{ab}\left(D_a a_b + \Sigma_a\,a_b\right) - \frac{1}{2}\left(a_a + 2\,\alpha_a\right)\,\Omega^a = 0\,,
\end{equation}
showing that if $u^a$ and $m^a$ are both geodesic and $u^a$ is irrotational, then there is no source for gradients in the twist. This demonstrates once again the similarity between $\omega_a$ and $\xi\,$.

Finally, $m^a \,M^{bc}\,R_{abc}$ gives 
\begin{eqnarray}\label{eq_hamiltonian_constraint_112}
    -m^a D_a \phi - \frac{1}{2}\,\phi^2 + \frac{2}{9}\,\Theta^2 + \frac{1}{3}\,\Theta\,\Sigma - \Sigma^2 + 2\,\xi^2 - \mathcal{E} + M^{ab}D_a a_b  && \\
    \nonumber \, - a_a\, a^a - \Sigma_a\,\Sigma^a + \Omega_a\,\Omega^a + 2\,\epsilon^{ab}\,\alpha_a\,\Omega_b - \zeta_{ab}\,\zeta^{ab} &=& \frac{2}{3}\,\rho + \frac{1}{2}\,\Pi + \frac{2}{3}\,\Lambda\,.
\end{eqnarray}
This is a very important equation, because it provides the generalised Friedmann equation in the $1+1+2$ formalism. 

If $u^a$ is irrotational, such that $\Omega $ and $\Omega_a = 0$, then (\ref{eq_hamiltonian_constraint_112}) is equivalent to the Hamiltonian constraint equation. Comparing (\ref{eq_hamiltonian_constraint_112}) to (\ref{eq_Hamiltonian_constraint_1+3}), one finds that the Ricci curvature of spacelike hypersurfaces can be related to the $1+1+2$-covariant variables, by
\begin{eqnarray}\label{eq_ricci_3_112}
    ^{(3)}R &=& -\frac{3}{2}\,\phi^2 - 3\,m^a D_a \phi - \frac{3}{2}\,\Sigma^2 + \Theta\,\Sigma + 6\,\xi^2 - \mathcal{E} - \frac{3}{2}\,\Pi \\
    \nonumber && \, - \, \Sigma_a\,\Sigma^a - 3\,a_a\,a^a + 3\,M^{ab}D_a a_b + \Sigma_{ab}\,\Sigma^{ab} - 3\,\zeta_{ab}\,\zeta^{ab}\,.
\end{eqnarray}
For an homogeneous and isotropic geometry, only the first two terms remain. The projected gradient of $\phi$ is permitted because $\phi$ can contain a single factor of the preferred spatial coordinate (e.g., if one chooses to work in spherical coordinates, and $m^a$ is aligned with the $r$ direction of an FLRW geometry)\,. 
In an FLRW universe in the canonical foliation, $^{(3)}R = \dfrac{6K}{a^2}\,$, as we will show in the next chapter. Hence, in the FLRW limit $\phi$ can be associated with spatial curvature. 

The power of the $1+1+2$ formalism will become apparent in Chapters 7 and 8. There, we will discuss how the preferred timelike vector $u^a$ and preferred spacelike vector $m^a$ might be picked out in either a model geometry or the real Universe. 
For now, consider what happens when the spacelike hypersurfaces are locally rotationally symmetric about the direction defined by $m^a\,$ \cite{Ellis_1967, Stewart_1968}. Then, if any vectors or tensors were non-zero, they would break that symmetry, as they would introduce some directional dependence into the two-dimensional surfaces orthogonal to both $u^a$ and $m^a\,$, at each point in spacetime. 
Therefore, all of the $1+1+2$-covariant vectors and tensors must vanish \cite{Clarkson_2003, vanElst_1996}, so the entire spacetime is characterised by a set of covariantly defined scalars $S$. 

If we further assume spatial homogeneity in the spacelike hypersurfaces orthogonal to $u^a\,$, it follows that all the derivatives $m^a D_a S$ are zero. The $1+1+2$ equations of motion then give us a set of first-order ODEs in time, and algebraic constraints, for a set of covariantly defined scalars. 
Although these equations are still nonlinear and difficult to solve, they are very helpful for systematic analysis of these types of universe. Moreover, the first-order vector and tensor perturbations of that model geometry will automatically be gauge invariant \cite{Clarkson_2007, stewart1974perturbations}. This has obvious advantages for the physical interpretation of perturbations, because there will be no spurious gauge modes in the vector or tensor sectors.

\section{Light propagation in curved spacetime}\label{sec:light_propagation}

So far, we have mainly focused on how cosmological spacetimes can be studied using the properties of timelike congruences and spacelike hypersurfaces. However, almost everything we know about the Universe has been learnt from observations of distant sources. 
That is, it has come from studying electromagnetic waves, arriving at us along the past light cone. These light rays are tangents to null curves through the curved spacetime geometry.
It is therefore crucial to understand how the behaviour of bundles of null rays is determined by the curvature of spacetime. This subject was pioneered by Sachs and Kristian \cite{sachs1961gravitational, Kristian_1966}, and an excellent summary is provided in Refs. \cite{ellis2012relativistic, Fleury:2015hgz}. 
We will not present an exhaustive treatment, but will highlight some key concepts, that will be particularly relevant for the discussions of the cosmic microwave background in Chapters 4 and 6, and the Hubble diagram in Chapters 4 and 8.

Electromagnetism is described covariantly by a four-potential $A_a$\,, which is invariant under gauge transformations $A_a \longrightarrow A_a + \nabla_a \chi$\,. The gauge-invariant action for classical electromagnetism is
\begin{equation}
    S_{EM} = \frac{1}{16\pi}\int\,\mathrm{d}^4x\,\sqrt{-g}\,F_{ab}\,F^{ab}\,,
\end{equation}
where $F_{ab} = 2\nabla_{[a}A_{b]}$ is known as the Faraday tensor. It is an antisymmetric tensor, and therefore it contains all 6 degrees of freedom in the electromagnetic field. 
Given some timelike congruence $u^a$, these can be decomposed into an electric and magnetic part, each with three independent components: $E_a = F_{ab}\,u^b$ and $H_a = \frac{1}{2}\eta_{abcd}\,u^b F^{cd}\,$.

Varying $S_{EM}$ with respect to $A_a$, and choosing it to be in the Lorentz gauge $\nabla_a A^a = 0\,$, we obtain Maxwell's equations in curved spacetime,
\begin{equation}\label{eq_Maxwells_equations_curved_spacetime}
    \nabla^b\,\nabla_b A_a + R_{ab}\,A^b = -j_a\,,
\end{equation}
where the source term $j_a$ is the 4-current density, containing contributions from the charge and current densities as measured by some timelike observer.
In cosmology, we can safely set $j_a$ to zero.

\subsection{Geometric optics}\label{subsec:geometric_optics}

An electromagnetic field is, by definition, a solution to Eq. (\ref{eq_Maxwells_equations_curved_spacetime}). 
One of the simplest such solutions is a plane wave, propagating through the spacetime. The wave has a null tangent vector $k_a$\,, satisfying $k_a k^a = 0$. It is described by an electromagnetic 4-potential of the form $A_a = \mathfrak{A}_a\,e^{i\psi}\,$.

Then, we can introduce a very helpful approximation, which will be the starting point for studying light propagation on cosmological scales. This is called the eikonal, or geometric optics, approximation. It states that the oscillation of the phase $\psi$ is much faster than the variation of either the amplitude $\mathfrak{A}_a$, or of the gravitational fields present in the spacetime. 
It follows that the propagation vector $k_a$ is related to the phase simply by $k_a = \nabla_a \psi\,$.
Taking the covariant derivative of this equation and antisymmetrising, one finds that the vorticity associated with the null congruence vanishes, $ \hat{\omega}_{ab} = \nabla_{[a}k_{b]} = 0\,$\footnote{Note that we have used a hat to distinguish the vorticity of a null congruence from its timelike counterpart.}.
This is a direct consequence of geometric optics \cite{pirani1964introduction}. Generic null congruences can have vorticity, but it has no source term coming from the spacetime curvature, as we will demonstrate shortly. 
Therefore, it is not expected to be relevant in cosmological spacetimes, even in the rare cases, such as strong gravitational lensing, where the geometric optics approximation does not apply.

Consider taking the covariant derivative of the null condition $k_a\,k^a = 0\,$, applying the relation $k_a = \nabla_a\psi$, and using that $\nabla_{[a}k_{b]} = 0\,$.
This shows that the null ray must satisfy
\begin{equation}
    k^b \nabla_b k_a = 0 \quad \Leftrightarrow \quad \frac{D k_a}{d\lambda} = 0\,,
\end{equation}
i.e. under the eikonal approximation, light travels not only on null curves, but on null geodesic curves. Here $\lambda$ is an affine parameter along the null ray.

An important quantity to define is the energy $E$ of a photon, as measured locally by a future-directed timelike observer with 4-velocity $u^a\,$. 
In order for $E$ to be well-defined, the integral curve of $u^a$ must intersect with the null geodesic. Then, the definition of $E$ will depend on whether the null geodesic is future-directed, with $k^a u_a < 0\,$, or past-directed with $k^a u_a > 0\,$. For future-directed null rays, $E = - k^a u_a\,$, and for past-directed null rays, $E = + k^a u_a\,$.
This is an important distinction, because light rays are time-reversible \cite{ellis2009republication}. Depending on the context, it can be useful to consider either future or past directed null curves.

Once we have defined a timelike reference congruence $u^a$, the null tangent vector $k^a$ can be decomposed covariantly as
\begin{equation}\label{eq_null_geodesic_u_e_splitting}
    k^a = E\left(u^a + e^a\right) \quad {\rm (future-dir.)}\,; \quad k^a = -E\left(u^a - e^a\right)\quad {\rm (past-dir.)}\,.
\end{equation}
Here, $e^a = h^a_{\ b}\,k^b$ is the spatial propagation direction of the ray. 
A very important quantity is the redshift $z$, which is defined from the ratio of the energies at the emitting location and the observing location,
\begin{equation}
    1 + z = \frac{E_{\rm emit}}{E_{\rm obs}} = \frac{\pm k^a u_a\vert_{\lambda}}{\pm k^a u_a \vert_{0}}\,,
\end{equation}
where a plus (minus) is for a past (future) directed $k^a\,$, and we have chosen the affine parameter $\lambda$ along the null geodesic to be equal to zero at the observer's location. This is always possible, because of the invariance of null geodesics under affine transformations $\lambda \longrightarrow a\,\lambda + b\,$.

The splitting in Eq. (\ref{eq_null_geodesic_u_e_splitting}) makes it possible to define the rate at which an observer with 4-velocity $u^a$ would find space to be expanding in their immediate vicinity, if they make observations by receiving light rays with tangents $k^a\,$. 
This is a very important quantity in cosmology: unlike the expansion $\Theta = \nabla_a u^a$ of spacelike hypersurfaces, this parameter, which we will call $H^{\parallel}$, can be inferred directly from observations on the past light cone.
By considering the change in the infinitesimal separation $\delta x^a$ between two timelike curves in the $u^a$ congruence as the affine parameter $\lambda$ is varied, one finds that 
\begin{equation}
    H^{\parallel} = \frac{1}{E^2}\,k^a \,k^b \,\nabla_a u_b \, = \,e^a \,e^b \,\nabla_a u_b \, = \, \frac{\Theta}{3} + \sigma_{ab}\,e^a\,e^b\,.
\end{equation}
This is a covariant generalisation of the Hubble parameter, that is familiar from studies of homogeneous and isotropic cosmologies. In a generic expanding spacetime, it depends on the observing direction $e^a\,$, through the presence of the shear term $\sigma_{ab}$ producing a quadrupole signature.
It has been shown that the properties of null geodesics also allow higher-order cosmographic parameters, such as the deceleration parameter, to be defined in a covariant way \cite{Heinesen_2021, Heinesen_2022, Clarkson_2011, Umeh_2011, macpherson2021luminosity, Macpherson:2022eve}.
The parameter $H^{\parallel}$ makes it possible to work with the direct observable $z$ instead of the unobservable $\lambda\,$, because they are related through
\begin{equation}\label{eq_dz_dlambda}
    \frac{\mathrm{d}z}{\mathrm{d}\lambda} = \pm\left(1+z\right)^2 \, H^{\parallel}\left(z\right)\,
\end{equation}
where a plus (minus) is for past (future) directed null geodesics.

If one were only interested in calculating the trajectories of individual null rays, then the geodesic equation is sufficient\footnote{In the eikonal approximation, so that light rays are automatically geodesic.}. However, in practice, we usually want to know about the properties of null geodesic bundles. 
In particular, this is necessary so that covariant notions of flux and the angular size of objects can be defined. These both give rise to sensible definitions of distance that can be studied in curved spacetime.

The basic set-up is as follows. We consider an infinite family of infinitesimally separated null geodesics, sharing a tangent $k^a\,$. Unlike for a timelike congruence, where $u^a$ defined a set of 3-dimensional orthogonal spaces, the null condition $k_a k^a = 0$ tells us that the congruence is orthogonal to a 2-dimensional space, that is itself orthogonal to both $u^a$ and $e^a\,$.
That 2D space is referred to as the screen space, because if the light ray is to be intercepted by a screen with worldline parallel to $u^a\,$, that screen must be set up orthogonal to the direction $e^a\,$.

We can learn about the properties of the null congruence by studying the properties of images projected on to such screens. In particular, the cross-sectional area $\delta A$ of the null bundle will vary as we move along the curve $x^a(\lambda)\,$.  
The tensor $s_{ab} = h_{ab} + e_a\,e_b = g_{ab} + u_a\, u_b + e_a\,e_b$ projects all objects it acts upon into the screen space. Moreover, under the eikonal approximation $k^a$ is vorticity-free, so $s_{ab}$ is the screen space metric.

The screen space is spanned by a pair of real orthonormal spacelike vectors $s_1^{\ a}$ and $s_2^{\ a}$, or equivalently by the complex null vector $s^a = \dfrac{1}{\sqrt{2}}\left(s_1^{\ a} - is_2^{\ a}\right)$ and its complex conjugate $\bar{s}^{a}\,$. These vectors are parallel-transported along the null geodesics. They are related to the screen space projection tensor by
\begin{equation}
    s_{ab} = s^1_{\ a}\,s^1_{\ b} + s^2_{\ a}\,s^2_{\ b}\, = \, s_a\,\bar{s}_b + \bar{s}_a\,s_b\,.
\end{equation}

The separation vector $\delta x^a$ between neighbouring null geodesics is just some real linear combination of $s^a$ and $\bar{s}^a\,$.
Any such separation vector is a solution of the geodesic deviation equation (\ref{eq_geodesic_deviation_equation}). Hence, tracking how separation vectors in the bundle evolve along the null curve allows the Riemann curvature to be probed. 

\begin{figure}
    \centering
    \includegraphics[width=\linewidth]{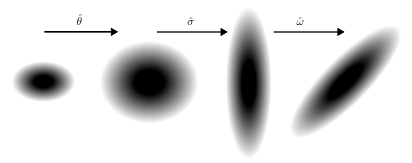}
    \caption{Visualisation of the effects of $\hat{\theta}\,$, $\hat{\sigma}$ and $\hat{\omega}$ on an image formed in the screen space of a null ray.}
    \label{fig_Sachs_formalism_optical_scalars}
\end{figure}

As in the $1+3$ decomposition, this is achieved most conveniently by studying a set of covariant variables that are associated with the properties of the congruence. In fact, for null congruences these can all be written as scalars, which are known as the optical scalars.
We have
\begin{itemize}
    \item The null expansion $\hat{\theta} = \frac{1}{2}\,\nabla_a k^a\,$. This tells us about the expansion or contraction of (images in) the screen space as we move along the null ray. It follows from the definition of $\delta A$ that $\dfrac{\mathrm{d}}{\mathrm{d}\lambda}\ln{\delta A} = 2\,\hat{\theta}\,$.
    \item The null shear $\hat{\sigma}_{ab} = \left(s_{(a}^{\ c}s_{b)}^{\ d} - \frac{1}{2}\,s_{ab}s^{cd}\right)\nabla_c k_d\,$. It can be packaged into a single complex scalar, $\hat{\sigma} = - s^a\,s^b\,\hat{\sigma}_{ab}\,$. This tells us how the screen space can be sheared, so that a circular image would be flattened into an ellipse.
    \item The null vorticity $\hat{\omega}_{ab} = s_{[a}^{\ c}s_{b]}^{\ d} \nabla_c k_d$\,. It can be packaged into a single real scalar, $\hat{\omega} = -i\, s^a\, \bar{s}^b \, \hat{\omega}_{ab}$ \cite{dolan2018geometrical}. It corresponds to twisting of the screen space, so that an image would be rotated. Under the eikonal approximation, $\hat{\omega}$ vanishes.
\end{itemize}
The effects of each of the optical scalars on an image formed in the screen space (similar to an image of an ellipsoidal galaxy), are displayed in Fig. \ref{fig_Sachs_formalism_optical_scalars}.

The evolution equations for the optical scalars are obtained from contractions of the Ricci identity for the null vector $k^a\,$, $2\,\nabla_{[a}\nabla_{b]}\,k_c = R_{abcd}\,k^d\,$. These are the Sachs equations \cite{sachs1961gravitational}, which fully govern the properties of null geodesic congruences in curved spacetime.

The first is the evolution equation for $\hat{\theta}$, which comes from taking the trace of the Ricci identity, projecting it with $k^a\,$, and then using $k_a\,k^a = 0\,$. Doing so, one obtains
\begin{equation}\label{eq_sachs_1}
    \frac{\mathrm{d}}{\mathrm{d}\lambda}\,\hat{\theta} + \hat{\theta}^2 + \bar{\hat{\sigma}}\hat{\sigma} - \hat{\omega}^2 = -\frac{1}{2}\,R_{ab}\,k^a\,k^b\,.
\end{equation}
This equation shows how local Ricci curvature causes light beams to focus. In GR, the null energy condition (NEC) tells us that $8\pi G\, R_{ab}\,k^a\,k^b = T_{ab}\,k^a\,k^b \geq 0\,$ for any $k^a\,$. 
Translating $T_{ab}$ into the energy-momentum tensor of a perfect fluid, we see that the NEC is violated only for exotic fluids with $\rho + p < \,0\,$. 
Therefore, Eq. (\ref{eq_sachs_1}) ensures that under ordinary conditions, matter fields always cause null congruences to converge. The term $-\frac{1}{2}\,R_{ab}\,k^a\,k^b$ can be identified as $\Phi_{00}\,$, one of the real scalars associated with the Ricci tensor in the Newman-Penrose approach \cite{Newman_1961}. 

The next equation is the evolution equation for the complex null shear $\hat{\sigma}\,$. This comes from contracting the Ricci identity with $k^b\,$, then projecting into the screen space and taking the symmetric and trace-free part. Finally, we contract the resulting tensor equation twice with the complex null vector $s^a\,$. This gives
\begin{equation}\label{eq_sachs_2}
    \frac{\mathrm{d}}{\mathrm{d}\lambda}\,\hat{\sigma} + 2\,\hat{\theta}\,\hat{\sigma} = C_{abcd}\,s^a\,k^b\,s^c\,k^d\,.
\end{equation}
The shearing of a null congruence is sourced non-locally, by the Weyl curvature. This is the generic origin of weak gravitational lensing. 
Moreover, because the null shear is coupled to the null expansion in Eq. (\ref{eq_sachs_1}), the Weyl curvature indirectly sources the null expansion, through the evolution of $\hat{\sigma}\,$.
Again, we can identify the right hand side with a Newman-Penrose scalar, in this case the complex scalar $\Psi_0\,$.

The last Sachs equation is an evolution equation for the null vorticity $\hat{\omega}\,$, which we remind the reader vanishes anyway in the eikonal approximation. It arises from following the same procedure as for $\hat{\sigma}\,$, but then taking the antisymmetric part rather than the symmetric and trace-free part. Finally, we contract with $s^a\,\bar{s}^b\,$. The resulting equation is
\begin{equation}\label{eq_sachs_3}
    \frac{\mathrm{d}}{\mathrm{d}\lambda}\,\hat{\omega} + 2\,\hat{\theta}\,\hat{\omega} = 0\,,
\end{equation}
where the vanishing of the right hand side is a consequence of the symmetries of the Weyl tensor.
Eq. (\ref{eq_sachs_3}) shows that the vorticity of null congruences has no source from the spacetime curvature. Hence, it is negligible in the late Universe. 
This result further justifies the approach of geometric optics. For the rest of this thesis, we will work entirely within that approximation when dealing with light propagation.

\subsection{Distance measures}\label{subsec:distance_measures}

In cosmology, it is very informative to calculate the distance to sources of electromagnetic, and recently also gravitational \cite{belgacem2018gravitational}, radiation. However, there is no unique, covariant, definition of distance in General Relativity. Any notion of distance that is built out of a coordinate basis is by its nature not compatible with general covariance, so it cannot be used in curved spacetime.

Instead, cosmologists define notions of distance that are directly related to observable quantities. This means that they are physically well-defined, but that there is no unique choice of how to do this. 
It depends on the observation at hand, whether that is, for example, the angular size of galaxies or clusters of galaxies, the flux of radiation received from a luminous source such as a Type Ia supernova, or the strain associated with a gravitational wave.

Consider a source that we see subtending a solid angle $\delta\Omega_{\rm obs}$ on our local sky. Suppose we also have some information about the intrinsic cross-sectional area of the source, $\delta A_{\rm obs}\,$.
\footnote{It may seem rather confusing that the area of the source is labelled with the subscript ``obs''. The reason we do this is that $\delta A_{\rm obs}$ is the area of an image on the screen space that would be obtained for future-directed null rays converging at the observer (that have travelled to us from some source), or alternatively past-directed null rays emanating from the observer. 
Conversely, $\delta A_{\rm source}$ is the area on the screen space that would be obtained for future-directed null rays emanating from the source.}
This would occur, for example, if it is a galaxy or cosmological feature of a known size. Objects with this property are known as standard rulers.
A promiment example is the scale of baryon acoustic oscillations (BAO), associated with the sound horizon at last scattering \cite{eisenstein2005detection}. We will discuss the BAO scale further in Chapter 3.

We define the angular diameter distance $d_A$ by
\begin{equation}
    \delta A_{\rm obs} = d_A^2\, \delta\Omega_{\rm obs}\,,
\end{equation}
where we should understand $d_A$ as being a property of a congruence of past-directed null geodesics diverging from the observer.
Because the null expansion $\hat{\theta}$ associated with the congruence is related to areas on the screen space by $\dfrac{\mathrm{d}}{\mathrm{d}\lambda}\,\delta A = 2\hat{\theta}\,\delta A\,$, it follows that the angular diameter distance satisfies
\begin{equation}
    \hat{\theta} = \frac{\mathrm{d}}{\mathrm{d}\lambda}\,\ln{d_A}\,.
\end{equation}

Thus, Sachs' equations for the optical scalars can be rewritten in terms of the angular diameter distance,
\begin{eqnarray}
    \label{eq_sachs_dA_1} \frac{\mathrm{d}^2}{\mathrm{d}\lambda^2}\,d_A &=& \left(\Phi_{00} - \bar{\hat{\sigma}}\hat{\sigma}\right)\,d_A \,, \quad {\rm and} \\
    \label{eq_sachs_dA_2} \frac{\mathrm{d}}{\mathrm{d}\lambda}\left(\hat{\sigma}\,d_A^2\right) &=& \Psi_0\, d_A^2\,.
\end{eqnarray}
From a numerical perspective, these are preferable to working with the null expansion scalar directly, because $\hat{\theta} \longrightarrow -\infty$ at the observer ($\lambda = 0$)\,. If it did not, then the past (future)-directed null geodesics would not be emanating from (converging to) the observer's spacetime location.
In contrast, $d_A$ is well behaved at the observer.

Sachs' equations (\ref{eq_sachs_dA_1}) and (\ref{eq_sachs_dA_2}) are solved with the initial conditions
\begin{equation}\label{eq_sachs_initial_conditions}
    d_A(0) = 0\,, \quad \frac{\mathrm{d}}{\mathrm{d}z}\,d_A \vert_0 = \frac{1}{H_0^{\parallel}}\,, \quad \hat{\sigma}(0) = 0\,.
\end{equation}
Hence, knowledge of the spacetime curvature allows one to construct the angular diameter distance as a function of redshift along any given null geodesic in the spacetime.

\begin{figure}[ht]
    \centering
    \includegraphics[width=0.85\linewidth]{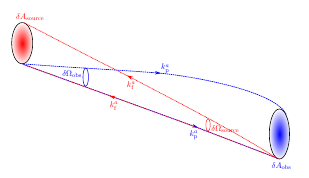}
    \caption{Comparison of future-directed null geodesics (red) with tangents $k^a_{\rm f}$ travelling from the source to the observer, and past-directed null geodesics (blue) $k^a_{\rm p}$ travelling from the observer to the source. The future-directed curves define the luminosity distance, and the past-directed ones the angular diameter distance. Based on a figure in Ref. \cite{ellis2012relativistic}.}
    \label{fig_reciprocity_theorem}
\end{figure}

The other standard distance measure is the luminosity distance $d_L\,$. It is defined by considering a source of known intrinsic luminosity $L_{\rm source}\,$, and measuring the flux of radiation received from it at the observer,
\begin{equation}
    F_{\rm obs} = \frac{L_{\rm source}}{4\pi\,d_L^2}\,.
\end{equation}
This is particularly useful when $L_{\rm source}$ can be inferred directly. Such sources are known as standard, or at least standardisable, candles. The most notable standardisable candles for cosmological observations are Cepheid variable stars, whose intrinsic luminosity is related to the period of pulsations in their brightness \cite{altavilla2004cepheid}, and Type Ia supernovae, which have a roughly constant luminosity associated with the Chandrasekhar limit for white dwarf stars \cite{chandrasekhar1931maximum}. We will come back to these in Chapter 3.

The flux $F$ ultimately comes from the energy density $\rho^{EM} = T_{ab}^{\rm EM}\, u^a \,u^b$ of the radiative electromagnetic field, as measured by a timelike observer with 4-velocity $u^a\,$. 
It is defined by time-averaging $\rho^{EM}$ over many oscillations of the electromagnetic wave, and projecting on to an area of the screen space associated with $k^a\,$.  
The rate of energy transfer to the observer's screen is depreciated relative to the energy transfer rate at the source's spacetime location by two factors of $\left(1+z\right)$:
\begin{enumerate}
    \item The energy of each individual photon is reduced by a factor $1+z$, as it travels from the source at redshift $z$ to the observer at redshift $0\,$.
    \item The rate itself is associated with a time derivative. The proper time interval measured in the observer frame is $\left(1+z\right)\,\times$ the proper time interval measured in the source frame. Hence, there is a factor $1+z$ associated with time dilation of the period of the electromagnetic wave.
\end{enumerate}

This means that $F_{\rm obs}$ is related to $\delta A_{\rm source}\,$, the screen space area for null rays diverging from the source, and $\delta\,\Omega_{\rm source}\,$, the solid angle that the radiation emanating from the source is diverging into, by
\begin{equation}
    F_{\rm obs} = \frac{L}{4\pi\left(\frac{\delta A_{\rm source}}{\delta \Omega_{\rm source}}\right)}\,\frac{1}{1+z}\frac{1}{1+z}\,,
\end{equation}
and so
\begin{equation}
    \delta A_{\rm source} = \frac{1}{\left(1+z\right)^2}\, d_L^2 \,\delta\Omega_{\rm source}\,.
\end{equation}
This demonstrates clearly that $d_L$ is a property of future-directed null geodesics emanating from the source, in contrast to $d_A$\,. The null worldlines are directed forward in time until they intersect with the timelike worldline of the observer. 
This means that the initial conditions for the Sachs equations, solved using a future-directed null geodesic congruence, would be very unclear, because we have no observational access to $\delta \,\Omega_{\rm source}\,$.

However, the distance measures $d_A$ and $d_L$ are not independent. They are related to one another in a very simple fashion, because of the time-reversal symmetry of null geodesics.
The time-reversibility is demonstrated in Fig. \ref{fig_reciprocity_theorem}, where we have shown a past-directed and a future-directed null geodesic coinciding. The only difference between them is that the affine parameter $\lambda$ increases in opposite directions along the curve.

As long as photon number is conserved, Etherington's reciprocity theorem \cite{Etherington_1933, ellis2009republication, ellis2012relativistic} implies that
\begin{equation}\label{eq_reciprocity_theorem}
    d_L = \left(1+z\right)^2\,d_A\,,
\end{equation}
which is also referred to as the distance duality relation. We will not prove this statement here. However, we refer the reader to Ref. \cite{ellis2012relativistic} for a very clearly presented proof.

Etherington's theorem is completely general, irrespective of the spacetime geometry. It remains valid in any metric theory of gravity, as long as that theory does not violate photon number conservation. To date, there is no statistically significant observational evidence for any violations of Eq. (\ref{eq_reciprocity_theorem}) (see e.g. Refs. \cite{liao2016distance, rana2017probing, holanda2010testing, holanda2012test, renzi2022resilience}).
The generality of the reciprocity theorem has occasionally caused some confusion in the literature, where some erroneous claims have been made that the reciprocity theorem relies on the symmetries of the FLRW cosmology.
In Chapter 8 we will make extensive use of the reciprocity theorem in inhomogeneous and anisotropic spacetimes.

The reciprocity theorem is enormously helpful for calculating the relationship between luminosity distance and redshift in generic spacetimes, which is a fundamental observable referred to as the Hubble diagram. 
The theorem circumvents the need to integrate Sachs' equations forward along future-directed null geodesics from the source to the observer, in order to calculate the flux. 
Hence, there is no need to specify any initial conditions at the source location, which would require guessing information to which we have no observational access.
Instead, one specifies the initial conditions (\ref{eq_sachs_initial_conditions}) at the observer location, and integrates Sachs' equations along past-directed null geodesics back to the source. 
This gives $d_A(z)\,$, and then $d_L(z)$ is calculated trivially, using the reciprocity theorem.

\chapter{The concordance cosmological model}

\lhead{\emph{The concordance model}} 

In this chapter, we will introduce the key elements of the $\Lambda$CDM model that has become the concordance model in cosmology over the last three decades.
This is necessary in order to understand how and why alternatives to the $\Lambda$CDM cosmology may be constructed.
We will discuss the evolution of the homogeneous and isotropic FLRW universe, and the theory of cosmological perturbations that underpins the standard interpretation of cosmological inhomogeneities and anisotropies.
Finally, we will review some important observational probes of the FLRW universe and its perturbations, focusing on the cosmic microwave background (CMB) and the Hubble diagram.

\section{Homogeneous and isotropic cosmology}

The fundamental principle upon which the entire concordance model is built is the cosmological principle. Let us introduce this first phenomenologically, and then formally.

A phenomenological definition states that on sufficiently large spatial scales, the Universe is statistically homogeneous (the same at all spatial locations) and isotropic (the same in all spatial directions) \cite{Aluri_2023}. 
This is a generalisation of the Copernican principle, which states that our observing location in the Universe has no special properties, to all observing locations \cite{weinberg1972gravitation}.

The strongest evidence for cosmic isotropy is provided by the cosmic microwave background. The temperature of the CMB has been found consistently to be uniform across the sky to approximately one part in $10^5$ \cite{Planck_2013, Planck_2015, Planck_2018}\footnote{This is a slight oversimplification. The largest anisotropy by far in the CMB is the dipole $(l = 1)$ signal, which is of order $10^{-3}$ rather than $10^{-5}\,$. The dipole is typically attributed to the kinematic motion of the Solar System with respect to the CMB. When the kinematic term is subtracted off, the intrinsic dipole is expected to be of order $10^{-5}\,$. However, this interpretation is the subject of much debate (see e.g. Refs. \cite{Secrest_2021, Secrest_2022,dalang2022kinematic,colin2019evidence}), which we will return to in Chapter 4.}. 

As a two-dimensional map, the CMB does not provide direct evidence for homogeneity. However, let us consider the implications of CMB isotropy in conjunction with the Copernican principle.
The Copernican principle is somewhat difficult to justify observationally, although some tests have been proposed \cite{caldwell2008test, clarkson2008general, valkenburg2013testing}. It can be argued that it is a well-motivated assumption, as there does not appear to be any fundamental reason why we would be privileged observers in the Universe.
If one assumes that the Copernican principle is valid, and measures an exactly isotropic CMB, then it follows that any other observer at some other location would also see an isotropic CMB. 
Hence, the Universe must be spatially homogeneous. This was proved by Ehlers, Geren and Sachs \cite{ehlers1968isotropic}. Although their proof relies on the rather stringent assumption of total isotropy, some near-Ehlers-Geren-Sachs theorems have been proved which generalise their result under weaker assumptions \cite{ellis1983anisotropic,ellis1983exact,stoeger1995proving, rasanen2009relation}.
Similarly, Hasse and Perlick showed that if all observers in a cosmological spacetime measure a Hubble diagram $d_L(z)$ that is isotropic up to $\mathcal{O}(z^3)\,$, then that spacetime is necessarily FLRW \cite{hasse1999spacetime}.

Ultimately, however, the Copernican principle is an assumption, and without overwhelming observational support, it is perhaps safer not to impose it {\it a priori} on philosophical grounds. 
Then, the isotropy of the CMB does not provide evidence for the Universe's statistical homogeneity. Instead, we should test homogeneity directly, using observations of the large-scale distribution of matter. On small scales, the cosmic web is highly inhomogeneous, but statistical homogeneity merely requires that above some homogeneity scale $L_{\rm hom}$, there exist no inhomogeneous structures.
Clearly, if $L_{\rm hom}$ is comparable to the size of the observable Universe, then homogeneity is not a reasonable approximation.

Estimates of $L_{\rm hom}$ are provided by observations of the cosmic large-scale structure (LSS), especially the Sloan Digital Sky Survey (SDSS) \cite{gonccalves2018cosmic,gonccalves2021measuring,hogg2005cosmic,pandey2015testing,pandey2016probing,sarkar2016information,pandey2021testing} and Baryon Oscillation Spectroscopic Survey (BOSS) \cite{laurent201614,ntelis2017exploring}. The estimated value varies substantially depending on the dataset and statistical techniques used, how strictly one defines statistical homogeneity\footnote{Some ultra-large structures have been observed, such as the Sloan Great Wall at around $420\,$\,Mpc \cite{Sloan_Great_Wall}, and even larger ones have been claimed \cite{horvath2014possible,lopez2022giant}. A very stringent interpretation might require that the homogeneity scale be pushed beyond the size of any of these structures.}, and also on the assumed value of $h = H_0/\left(100\,{\rm km}\,{\rm s}^{-1}\,\rm{Mpc}^{-1}\right)\, \approx 0.7 \,$. Estimates typically vary between $L_{\rm hom} = 60\,h^{-1}\,{\rm Mpc}$ and $300\,h^{-1}\,{\rm Mpc}\,$. For comparison, the present-day Hubble horizon is around $3000\,h^{-1}\,{\rm Mpc}\,$.

We have roughly stated the cosmological principle, and summarised its observational support. Let us now return to its precise statement, and the implications thereof.
To do so, it will be necessary to formulate what we mean by homogeneity and isotropy rather more precisely, by introducing the concept of symmetries of spacetime. This concept will also be crucial in later chapters.

When we refer to a spacetime, what we really mean is a four-dimensional Lorentzian manifold $\mathcal{M}$\,, equipped with a metric $g_{ab}$ and its associated Levi-Civita connection. A symmetry of spacetime is some map $\mathcal{M} \longrightarrow \mathcal{M}$ under which the metric is invariant. 
Symmetries can be divided into two classes: continuous and discrete, depending on the character of the group that describes the transformations. 
In the context of curved spacetime, we are only really interested in continuous symmetries.
A simple example of a spacetime possessing continuous symmetries is the Schwarzschild geometry, which is invariant under time translation and spatial rotations.

\begin{figure}
    \centering
    \includegraphics[width=0.9\linewidth]{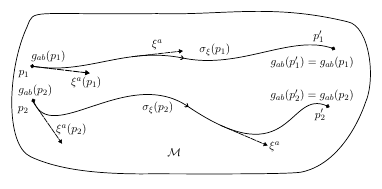}
    \caption{Schematic of the effect of a Killing vector $\xi^a\,$. For all points in the manifold, the metric $g_{ab}$ is invariant along an integral curve of $\xi^a$ from that point.}
    \label{fig_killing_transport}
\end{figure}

Mathematically, we can describe a symmetry transformation by a Killing vector field $\xi^a\,$. It is defined by the equation
\begin{equation}\label{eq_Killing_equation}
    \left(\mathcal{L}_{\xi}\,g\right)_{ab} = 0 \quad \Longrightarrow \quad \nabla_{(a}\xi_{b)} = 0\,,
\end{equation}
where we recall the definition of the Lie derivative $\mathcal{L}$ from Eq. (\ref{eq_Lie_derivative_def}).
The vanishing of the Lie derivative means that the metric is invariant along any integral curve of $\xi^a\,$, so this equation is a precise statement of the definition of a spacetime symmetry. A Killing vector field is the generator of such a symmetry.
The meaning of a Killing vector $\xi^a$ is demonstrated in Fig. \ref{fig_killing_transport}, where we see that for any point $p \in \mathcal{M}\,$, the flow $\sigma_{\xi}(p)$ preserves the metric; i.e. for all points $p'$ on that integral curve, $g_{ab}(p') = g_{ab}(p)\,$.
The symmetries of a spacetime are defined entirely by their Killing vector fields, which are the non-trivial solutions to Eq. (\ref{eq_Killing_equation}). In a generic spacetime which has no symmetries, Killing's equation is over-determined, so it has no non-trivial solutions.

A spacetime is said to be spatially homogeneous if it possesses three non-trivial, linearly independent, spacelike Killing vector fields $\xi^{H}_{1,2,3}$ defined everywhere on the spacetime. This implies that $\xi^{H}_{1,2,3}$ span three-dimensional spacelike hypersurfaces $\Sigma_t\,$, in which any point in the surface can be mapped to any other point under the action of the symmetry group\footnote{There is actually an exception to this implication, called the Kantowski-Sachs cosmology, which we will come back to in Chapter 4. For now, we can ignore it.}.
These $\Sigma_t$ are sometimes called surfaces of transitivity \cite{Stephani_2003, Ellis_1999}. 

A spacetime is isotropic if it possesses three non-trivial Killing vector fields $\xi^{I}_{1,2,3}$ whose associated symmetries map any point in the spacetime on to itself. The spacetime is spatially isotropic if the subgroup formed by those isotropies acts in spacelike hypersurfaces, i.e. the Killing vectors $\xi^{I}_{1,2,3}$ are spacelike. 
Therefore, an homogeneous and isotropic spacetime possesses three-dimensional surfaces of transitivity $\Sigma_t$, with a further three-dimensional group of spacelike symmetries that keep every point in those surfaces fixed. Translations do not keep points fixed, so we are left only with rotations. The only possible three-dimensional group we are left with is $SO(3)\,$, so this is the isotropy group of the spacetime.
Furthermore, the presence of these six Killing vector fields tells us that the fundamental timelike worldlines, orthogonal to the surfaces of transitivity, must have vanishing shear, vorticity and acceleration.

In general, inhomogeneous spacetimes can be isotropic (e.g. Schwarzschild), and homogeneous spacetimes can be anisotropic (e.g. the Bianchi universes).
There is an enormous variety of spatially homogeneous cosmological models, which we will discuss further in Chapter 4.

Let us now introduce the metric tensor $g_{ab}$ that is consistent with spatial homogeneity and isotropy. It is easiest to do this by working in the $3+1$ formalism.
First, we need to fix the gauge. We can introduce a synchronous time coordinate $t$, such that the spacetime can be foliated into homogeneous surfaces of constant $t$, with lapse $N = 1$ and shift $N_i = 0\,$. Those surfaces are covered by some spatial coordinates $x^i$.
Consider again the ADM form of the metric (\ref{eq_ADM_metric}). The assumptions of homogeneity and isotropy mean that there must exist exactly six independent spacelike Killing vector fields - three for homogeneity and three for isotropy. 
Therefore, the induced metric $h_{ij}$ on $\Sigma_t$ must be equal to a maximally symmetric 3-metric $\gamma_{ij}$ which is shared by all the surfaces, up to a conformal factor that depends only on the label $t$ of $\Sigma_t$\,.

We are left with the following form for the spacetime metric,
\begin{equation}\label{eq_FRW_metric}
    \mathrm{d}s^2 = -\,\mathrm{d}t^2 + a^2(t)\,\gamma_{ij}\,\mathrm{d}x^i\,\mathrm{d}x^j\,,
\end{equation}
where the maximally symmetric $\gamma_{ij}$ is most conveniently written in coordinates $\left(r,\theta,\phi\right)$ that are rather like the spherical polar coordinates on three-dimensional Euclidean space,
\begin{equation}\label{eq_Robertson_Walker}
    \gamma_{ij}\,\mathrm{d}x^i\,\mathrm{d}x^j  = \frac{1}{1 - K\,r^2}\,\mathrm{d}r^2 + r^2\,\mathrm{d}\theta^2 + r^2\,\sin^2{\theta}\,\mathrm{d}\phi^2\,.
\end{equation}
This is the Robertson-Walker (RW) class of 3-metrics \cite{robertson1935kinematics,robertson1936kinematics}.
The parameter $K$ describes the spatial curvature associated with $h_{ij} = a^2\,\gamma_{ij}\,$.  This is demonstrated by calculating the Ricci curvature $^{(3)}\,R_{ij}$ associated with $h_{ij}\,$, whereby $^{(3)}R_{ij} = \dfrac{1}{3}\,^{(3)}R\,h_{ij}\,$, with $^{(3)}R = \dfrac{6K}{a^2}\,$.
The Robertson-Walker class can thus be divided into three subclasses, depending on the sign of the spatial curvature parameter $K$: flat ($K = 0$), open ($K < 0$) and closed ($K > 0$).
Eq. (\ref{eq_FRW_metric}) defines the Friedmann-Lema{\^i}tre-Robertson-Walker metric: a Robertson-Walker homogeneous and isotropic induced 3-metric, embedded within a Friedmann-Lema{\^i}tre expanding universe. 

Often, cosmologists choose not to set the lapse equal to unity, but rather to work with a different time variable called conformal time $\tau$, such that $N = a(\tau)\,$. This makes the conformal flatness of the metric apparent,
\begin{equation}\label{eq_FRW_metric_conformal_time}
    \mathrm{d}s^2 = a^2(\tau)\left[-\,\mathrm{d}\tau^2 + \gamma_{ij}\,\mathrm{d}x^i\,\mathrm{d}x^j\right]\,,
\end{equation}
whence it can be deduced that the Weyl tensor $C_{abcd}$ associated with the FLRW metric vanishes identically\footnote{This is immediately obvious in the flat case $K = 0\,$, such that $\gamma_{ij} = \delta_{ij}\,$, but some further analysis shows that it remains true in both the open and closed cases \cite{ellis2012relativistic}.}.

\subsection{The FLRW universe}

In this section, we provide a brief overview of the dynamics of the FLRW universe described by Eq. (\ref{eq_FRW_metric}). We will work first in the standard approach, which uses the coordinate picture of General Relativity. 
However, we will also discuss the results in the $1+3$ formalism. 
We start from Einstein's equations (\ref{eq_Einstein_field_equation}). The FLRW metric has Einstein tensor components
\begin{equation}
    G_{00} = 3 H^2 + \frac{3 K}{a^2}\,, \quad G_{0i} = 0\,,\quad {\rm and} \quad G_{ij} = -a^2\left(3 H^2 + 2 \dot{H} + \frac{K}{a^2}\right)\,\gamma_{ij}\,,
\end{equation}
where we have introduced the Hubble parameter $H(t) = \dfrac{\dot{a}(t)}{a(t)}\,$.
With $C_{abcd} = 0$, the Riemann tensor components are easily constructed according to Eq. (\ref{eq_Weyl_tensor}). 
The equations of motion can also be written in terms of conformal time $\tau$\,, through the conformal Hubble parameter $\mathcal{H} = \dfrac{1}{a}\dfrac{\mathrm{d}a}{\mathrm{d}\tau} = \dfrac{a'}{a} = a H\,$\,.
Using $\tau$ rather than $t$ is a useful simplification in the study of cosmological perturbations.

Let us now consider the right hand side of Einstein's equations. In order for the energy-momentum tensor to respect the Killing symmetries of the metric, all possible matter and radiation fields must have the same 4-velocity $u^a = \delta^a_{\ t}\,$, and the degrees of freedom in $T_{ab}$ must be functions of time alone.
Recall that a general $T_{ab}$ can be decomposed with respect to a timelike vector field $u^a$ into an energy density $\rho$, pressure $p$, momentum density $q_a$ and anisotropic stress $\pi_{ab}\,$.
In this case, $T_{ab}$ cannot have any $1+3$-covariantly defined vector or tensor degrees of freedom, as they would violate the symmetries.
Therefore, we must have
\begin{equation}
    T_{ab} = \rho(t)\,u_a\,u_b + p(t)\,h_{ab}\,.
\end{equation}

Einstein's equations are given entirely by the following pair, the first and second Friedmann equations,
\begin{eqnarray}
    \frac{\mathcal{H}^2}{a^2} + \frac{K}{a^2}  = H^2 + \frac{K}{a^2} &=& \frac{8\pi G}{3}\,\rho + \frac{1}{3}\,\Lambda\,,\label{eq_Fried_1}\\
    \frac{\mathcal{H}'}{a^2} = \dot{H} + H^2 &=& -\frac{4\pi G}{3}\left(\rho + 3\,p\right) + \frac{1}{3}\,\Lambda\, . \label{eq_Fried_2}
\end{eqnarray}
The contracted Bianchi identities tell us that $\dot{\rho} + 3H\left(\rho + p\right) = 0\,$, and so the first Friedmann equation is the first integral of the second.

From the perspective of the $1+3$ formalism, these equations are straightforward consequences of homogeneity and isotropy. 
Because $C_{abcd} = 0\,$, its electric and magnetic parts must vanish. Homogeneity ensures $D_a\,\rho = D_a\,p = 0\,$, and we already have $q_a = 0$ and $\pi_{ab} = 0\,$.
To determine the kinematic variables, we note that any non-vanishing shear, vorticity or acceleration would necessarily give rise to anisotropy, because they would pick out at least one preferred spatial direction. Hence, $\sigma_{ab} = 0\,$ and $\omega_a = \dot{u}_a = 0\,$. 
The isotropic expansion is restricted to being a function of time, i.e. $D_a\,\Theta = 0\,$. As it describes the expansion of volumes, and $H$ the expansion of length scales, it follows that $\Theta$ and $H$ must be related by $\Theta = 3H\,$.

The only non-trivial $1+3$-covariant equations are the Raychaudhuri equation (\ref{eq_Raychaudhuri}), the local energy conservation equation (\ref{eq_energy_conservation_equation}) and the Hamiltonian constraint (\ref{eq_Hamiltonian_constraint_1+3}).
Identifying $^{(3)}R = \dfrac{6K}{a^2}$ as explained above, these three equations are identical to the Friedmann equations and the $\dot{\rho}$ equation.

The pressure and energy density of a given cosmological fluid are related by an equation of state (EOS), $p = p\left(\rho\right)\,$. Although certain exotic fluids may have more complicated equations of state, standard fluids all have linear equations $p = w\rho\,$. 
For non-relativistic matter, which is a combination $\rho$ of baryonic matter $\rho_b$ and cold dark matter $\rho_c$\,, $w = 0\,$, whereas for radiation, $w = \dfrac{1}{3}\,$.
It is also possible to think of the cosmological constant as a constant-density dark energy fluid with EOS parameter $w = -1$: $\rho_{\Lambda} = - p_{\Lambda} = \dfrac{\Lambda}{8\pi G}\,$.

The matter, radiation and dark energy fluids do not interact in the standard cosmology, so they all solve the energy conservation equation individually. Hence, the first Friedmann equation (Hamiltonian constraint) can be simplified by defining the critical density $\rho_{\rm crit}(a) = \dfrac{3 H^2(a)}{8\pi G}$\,, and the fractional density parameters 
\begin{equation*}
    \Omega_m = \frac{\rho_m}{\rho_{\rm crit}} = \frac{\rho_{m0}}{\rho_{\rm crit}\,a^3}\,, \quad \Omega_r = \frac{\rho_r}{\rho_{\rm crit}} = \frac{\rho_{r0}}{\rho_{\rm crit}\,a^4}\,, \quad \Omega_{\Lambda} = \frac{\Lambda}{3H^2(a)}\, \quad {\rm and} \quad \Omega_K = \frac{-K}{a^2 H^2}\,.
\end{equation*}
The Friedmann equation is then conveniently expressed in terms of the present-day density parameters and the present-day Hubble parameter $H_0\,$,
\begin{equation}
    H^2\left(a\right) = H_0^2\left[\frac{\Omega_{m0}}{a^3} + \frac{\Omega_{r0}}{a^4} + \Omega_{\Lambda 0} + \frac{\Omega_{K0}}{a^2}\right]\,,
\end{equation}
so that $\Omega_{m0} + \Omega_{r0} + \Omega_{\Lambda 0} + \Omega_{K0} = 1\,$.
In the absence of any more complicated matter fields such as a novel dynamical scalar\footnote{A period of inflationary expansion in the early Universe is expected to have been driven by at least one such scalar field (see e.g. Refs. \cite{liddle1998introduction,brandenberger2000inflationary}).}, an FLRW cosmology is described by specifying three of these four parameters, plus the integration constant $H_0\,$.

The radiation density parameter $\Omega_{r0}$ is constrained by the monopole temperature of the CMB, which is measured to be a perfect black body to within 1 part in $10^{5}\,$; $T_{\rm CMB} = 2.72548 \, \pm \, 0.00057 \, {\rm K}\,$ \cite{fixsen2009temperature}. 
The energy density of black body radiation is given by Planck's law, so $\rho_r(T) = C T^4\,$, where $C$ is a collection of fundamental constants. This tells us that $\Omega_{r0} \sim 5\, \times\, 10^{-5}\,$, so radiation makes a negligible contribution to the energy density of the late Universe. However, because $\Omega_r(a) \sim a^{-4}\,$, radiation dominated the energy content of the Universe before matter did (after the end of inflation).

Observational constraints on the spatial curvature parameter are consistent with zero, i.e. a spatially flat Universe. The tighest constraint is obtained by combining Planck measurements of the temperature and polarisation anisotropies in the CMB with BAO (baryon acoustic oscillation) measurements \cite{aghanim2020planck}, yielding $\Omega_{K0} = 0.0007 \pm 0.0019\,$. If one only uses CMB data, then the constraint is much weaker, and mildly favours a closed Universe: $\Omega_{K0} = -0.044^{+ 0.018}_{- 0.015}\,$ \cite{di2020planck}. However, this result is still consistent with zero to within $3\sigma\,$.
A near-flat universe is a concrete prediction of the theory of cosmic inflation in the early Universe \cite{brandenberger2000inflationary}.

The $\Lambda$CDM concordance model arises, then, from the negligible amount of radiation in the late Universe, and from assuming flatness. The reduced parameter space can then be strongly constrained using the temperature, polarisation and lensing anisotropies in the CMB. 
Hence, the strongest constraints on $H_0$, $\Omega_{m0}$ and $\Omega_{\Lambda 0}$ come from the Planck data \cite{aghanim2020planck}. They are
\begin{equation*}
    H_0 = 67.37 \pm 0.54\, {\rm km\, s}^{-1}\,{\rm Mpc}^{-1}\,, \quad \Omega_{m0} = 0.3147 \pm 0.0074\,, \quad {\rm and} \quad \Omega_{\Lambda 0} = 0.6889 \pm 0.0056\,.
\end{equation*}

With these parameters in hand, one can integrate the Friedmann equation to obtain $a(t)\,$. As a result, we are provided with an estimate for the age of the Universe assuming a flat $\Lambda$CDM cosmology,
\begin{equation}
    t_0 = \frac{1}{H_0}\,\int_0^1 \frac{\mathrm{d}a}{\sqrt{\Omega_{m0}a^{-3} + \left(1-\Omega_{m0}\right)}} \, \approx \, 13.8 \, {\rm Gyr}\,.
\end{equation}

Before we go onto discuss the history of the Universe in some more detail, let us define two extra quantities.

The first is the physical Hubble horizon, which is just defined to be $r^{\rm phys}_H(t) = \dfrac{1}{H(t)}\,$, in units where $c = 1\,$. Note that $r^{\rm phys}_H(t_0) \approx 4100 \,{\rm Mpc}\,$, which tells us the characteristic length scale of the present Universe. It is related to the comoving Hubble horizon $r_H^{\rm com}$ by $r_H^{\rm com}(a) = \dfrac{r_H^{\rm phys}(a)}{a}\,$. By a comoving distance, we mean simply the separation in the $r$ coordinate in Eq. (\ref{eq_FRW_metric}) between two points on a spacelike surface $\Sigma_t\,$. 
When we refer to the Hubble horizon $r_H$ in this thesis, we will always mean the comoving Hubble horizon $r_H^{\rm com}\,$.

The second is the particle horizon. It is defined by considering null geodesics in the FLRW spacetime, and calculating the largest possible comoving coordinate distance a light ray can have travelled through the Universe to reach an observer at some time $t$. 
Although this can be calculated by solving the geodesic equation, it is easiest just to exploit the symmetries of the metric. As null geodesics are conformally invariant, let us use the conformal-time form (\ref{eq_FRW_metric_conformal_time}) of the FLRW metric.
Now, setting the interval equal to zero at each point along the null curve, we must have $r(t) - r(0) = \tau(t) = \int_0^t\,\dfrac{\mathrm{d}t'}{a(t')}\,$.
The particle horizon $r_P(t)$ is thus equal to the conformal time that has elapsed between the hot Big Bang, which corresponds to $\tau = 0\,$\footnote{By the hot Big Bang, we really mean the end of inflation. The conformal time coordinate can be analytically extended into the inflationary epoch by considering negative values of $\tau\,$.}, and some later cosmic time $t\,$.
Written in terms of the scale factor, it is
\begin{equation}\label{eq_particle_horizon}
    r_P(a) = \int_0^a \frac{\mathrm{d}\tilde{a}}{\tilde{a}^2 H(\tilde{a})}\,.
\end{equation}
The particle horizon gives us a concrete definition for the coordinate extent of the observable Universe.

\subsection{Thermal history of the Universe}

The $\Lambda$CDM model provides a standard model for the history of the Universe, going back from the present day until before the hot Big Bang. We will now present a brief summary of this history. It is conveniently divided into a series of epochs that are distinguished by the Universe's thermal properties in each of them. 

\begin{enumerate}
    \item The pre-inflationary epoch: $t \lesssim 10^{-36} \, {\rm s}$. For $t \lesssim t_{\rm Planck} = \sqrt{\frac{\hbar c}{G^5}} \sim 10^{-43}\, {\rm s}$, classical gravity breaks down and a quantum theory of gravity (QG) is required, with all four fundamental forces being unified.
    After $t_{\rm Planck}\,$, the QG symmetry is presumed to break down to a Grand Unified Theory (GUT), which later breaks into the electroweak and strong interactions.
    
    \item The inflationary epoch: $10^{-36}\,{\rm s}
 \lesssim 10^{-32}\,{\rm s}$.
    In this period, the universe undergoes an ultra-rapid exponential expansion, with the scale factor believed to expand by at least 60 $e$-folds, i.e. $\dfrac{a_{\rm end}}{a_{\rm start}} = e^N \sim e^{60} \,$.
    The inflationary expansion is typically explained by the existence of a novel, massive scalar field $\phi$, slowly rolling down its potential $V(\phi)\,$. This field is referred to as the inflaton.
    Quantum fluctuations in the inflaton field generated before inflation are stretched out over enormous length scales, from deep within the horizon to far outside it.
    This leads to them becoming classical, after which fluctuations can be described using the general-relativistic theory of cosmological perturbations, which we will introduce shortly.
    
    The theory of inflation is consistent with a small value for $\Omega_K$, as the curvature scale can be pushed far beyond the horizon by the rapid expansion. It also provides a good explanation for the well-known horizon problem: apparently causally disconnected regions of the sky are observed to have the same CMB temperature. Since the comoving Hubble horizon shrinks by a factor of order $e^N$ during inflation, such regions would have been in causal contact in the pre-inflationary epoch.

    \item The Hot Big Bang/reheating epoch: $10^{-32}\,{\rm s}\lesssim t \lesssim 10^{-10}\,{\rm s}$.
    The Hot Big Bang (HBB) does not usually refer to $t = 0\,$, but rather to the end of inflation. At this point, the inflaton rolls down to the minimum of $V(\phi)$, and decays into the standard model particles.
    This causes the Universe to become filled with a very hot plasma: $t \sim 10^{-32}\,{\rm s}$ corresponds to temperatures of order $10^{13}\, {\rm GeV}\,$.
    The expansion causes that plasma to cool as $T \sim a^{-1}$.
    
    \item Radiation domination: $10^{-10}\,{\rm s} \lesssim t \leq t_{\rm eq}\, \sim 5\,\times 10^4 \, {\rm yr}\,$.
     The electroweak symmetry breaks once the characteristic temperature of the Universe is of order ${\rm TeV}\,$, producing the electromagnetic and weak interactions.
     The Universe is dominated by a thermal bath of photons.
     Around $20\,{\rm s}$ after the HBB, the process of Big Bang Nucleosynthesis (BBN) begins, with free neutrons and protons fusing into deuterium. 
     BBN continues until $t \sim 300 \, {\rm s}\,$. The stable nuclei produced are almost entirely hydrogen $^1{\rm H}$, deuterium $^{2}{\rm H}\,$, helium $^3{\rm He}$ and $^4{\rm He}$, and lithium $^7{\rm Li}$ \cite{cyburt2016big}.

     At the beginning of radiation domination, $\Omega_m$ is negligible compared to $\Omega_r\,$. However, $\Omega_m \sim a^{-3}\,$ and $\Omega_r \sim a^{-4}\,$, so eventually matter comes to dominate over radiation. This occurs at the matter-radiation equality scale, $a_{\rm eq} = \dfrac{\Omega_{r0}}{\Omega_{m0}} \approx \dfrac{1}{3400}\,$. For the best-fit Planck values of the cosmological parameters, this corresponds to around $5 \times 10^4 \, {\rm yr}$ after the HBB.
     Even before $a_{\rm eq}$, a series of acoustic waves occur in the density of the subdominant baryonic matter fluid, caused by the competition between gravitational perturbations and radiative pressure from the photon field. These waves are known as the baryon acoustic oscillations (BAO). 

    \item Matter domination: $t_{\rm eq} < t \leq t_{\rm DE} \sim 10\,{\rm Gyr}\,$.
    The energy density of the Universe is dominated by non-relativistic matter during this epoch.
    Initially, although radiation is subdominant after $t_{\rm eq}\,$, the Universe is still too hot for neutral atoms to form. Hence, there is an abundance of free electrons, causing continual Compton scattering of photons which means their mean free path is very small.
    The baryon acoustic oscillations also continue.
    
    Once the temperature drops to around $0.3\,{\rm eV}\,$, at $a \sim \dfrac{1}{1090}$, electrons and protons rapidly combine into neutral hydrogen. This is referred to as the process of recombination. 
    As there are no longer free electrons, the Universe becomes transparent to radiation, with the mean free path of photons becoming essentially infinite. This is the related process of decoupling. The scale factor at which it occurs is referred to as the last scattering surface $a_{\rm LS}$. 
    These photons are observed today as the Cosmic Microwave Background (CMB), which has been mapped to extraordinary detail. Fluctuations in the CMB temperature provide the initial conditions for gravitational collapse.
    Moreover, the baryon acoustic oscilations end at decoupling. The horizon associated with those sound waves leaves an observable signature in the present-day matter distribution, the BAO scale.

    After recombination and decoupling, the next several hundred ${\rm Myr}$ see dark matter overdensities, seeded in the CMB, collapse into halos under the influence of gravity, beginning the process of structure formation that leads to the observed cosmic web of large-scale structure (LSS). Neutral baryonic matter falls into these potential wells. In the strongest, most nonlinear, overdensities, baryonic matter collapses to form the first stars. 
    
    \item Accelerating era: $t > t_{\rm DE}\,$.
    The expansion of the Universe goes from deceleration to acceleration at $t_{\rm DE}\,$. In the $\Lambda$CDM model, this occurs at $a_{\rm DE} = \left(\dfrac{\Omega_{m0}}{\Omega_{\Lambda 0}}\right)^{1/3} \approx 0.75\,$. The accelerated expansion is driven by the appearance of Einstein's cosmological constant in the FLRW case (\ref{eq_Fried_2}) of the Raychaudhuri equation (\ref{eq_Raychaudhuri}).
    In the $\Lambda$CDM model, $\Lambda$ will eventually come to dominate the cosmological energy content entirely. Hence, the FLRW universe will tend towards an exponentially expanding de Sitter space, as $t \longrightarrow \infty\,$.
    
\end{enumerate}

Now, let us focus on the gravitational physics that produces the CMB anisotropies and later the cosmic large scale structure.
To do this, we will introduce the theory of cosmological perturbations.

\section{Perturbations in cosmology}\label{subsec:CPT}

The theory of relativistic cosmological perturbations is a very well-studied field. We will provide an overview of some key elements of the linear theory, focusing especially on how the standard coordinate approach to cosmological perturbations can be understood using the covariant approaches introduced in Chapter 2. For a full discussion of cosmological perturbation theory (CPT), we refer the reader to Refs. \cite{mukhanov1992theory,Durrer_1994,Ma_1995,Malik_2009,peter2009primordial,durrer2020cosmic}.

We note that first-order perturbations are dominant on large scales. Second-order perturbations can usually be safely ignored when modelling large-scale phenomena such as the cosmic microwave background, with the notable exception of studies of primordial non-Gaussianity. 
Hence, we will not calculate any second-order quantities, because for the contents of this thesis the linear theory is sufficient.

We assume that the spacetime metric is perturbatively close to some FLRW metric $\bar{g}_{ab}$ on all the scales we are interested in, so that $g_{ab} = \bar{g}_{ab} + \delta g_{ab}\,$, with $\vert \delta g_{ab} \vert \ll \vert \bar{g}_{ab} \vert\,$.
For a strict application of CPT, the first and second derivatives of $\delta g_{ab}$ must also be small. 
It is possible to write generalised versions of the CPT equations where $\delta g_{ab}$ remains small but its derivatives do not have to \cite{Adamek_2015,Milillo_2015,Goldberg_2017a,Goldberg_2017b}. 
We will come back to this point in Chapter 5, but for now we assume that the metric perturbations and their derivatives are sufficiently small that only terms that are linear in $\delta g_{ab}$ and its derivatives need be retained.

The metric can be written in general as
\begin{equation}
    \mathrm{d}s^2 = a^2(\tau)\left[-\left(1-2\phi\right)\,\mathrm{d}\tau^2 + 2\tilde{H}_i\,\mathrm{d}\tau\,\mathrm{d}x^i + \left(\gamma_{ij} + 2\tilde{H}_{ij}\right)\,\mathrm{d}x^i\,\mathrm{d}x^j\,\right]\,,
\end{equation}
where $\gamma_{ij}$ is once again a maximally symmetric 3-metric, and we have used conformal time $\tau$ rather than cosmic time $t$, as we we will do throughout this section.
The perturbation $\phi$ gives us the lapse function, $N(\tau, x^i) = a(\tau)\left(1-\phi(\tau,x^i)\right)\,$, and $\tilde{H}_i$ defines the shift vector $N_i = a^2\,\tilde{H}_i$\,. The perturbation $\tilde{H}_{ij}$ describes perturbations to the curvature of constant-$\tau$ spatial sections\footnote{The tilde on $\tilde{H}_{ij}$ is to avoid confusion with the magnetic part $H_{ab}$ of the Weyl tensor, and on $\tilde{H}_i$ to avoid confusion with the magnetic part $H_a$ of the Faraday tensor.}.

The metric perturbations $\tilde{H}_i$ and $\tilde{H}_{ij}$ are conveniently split according to the scalar-vector-tensor (SVT) decomposition. As $\tilde{H}_i$ is a spatial 3-vector, and the Christoffel symbols associated with $\gamma_{ij}$ define a covariant derivative $\tilde{D}_i$\footnote{We have avoided using $D_i$ here, to prevent confusion introduced by the scale factor in front of the induced metric.}, it follows that $\tilde{H}_i$ can be expressed without loss of generality as
\begin{equation}
    \tilde{H}_i = \tilde{D}_i B + B_i\,,
\end{equation}
where $\tilde{D}_i B^i = \gamma^{ij}\tilde{D}_i B_j = 0\,$. 
Note that if $K = 0$ (i.e. the FLRW background is spatially flat), then $\gamma_{ij} = \delta_{ij}$, and so $\tilde{D}_i$ is just $\partial_i\,$, the partial derivative with respect to the coordinates $x^i$ on the $\tau = {\rm cst.}$ hypersurfaces\footnote{We offer the cautionary note that a non-zero $B_i$ introduces vorticity which means that the $\tau = {\rm cst}$ spaces are not true hypersurfaces. However, this effect is small, so it is conventional to use the language of the $1+3$ and $3+1$ decompositions interchangeably in CPT.}. 
Likewise, $\tilde{H}_{ij}$ is decomposed as
\begin{equation}
    \tilde{H}_{ij} = \psi\,\gamma_{ij} + \tilde{D}_i\tilde{D}_j F + \tilde{D}_{(i}F_{j)} + \frac{1}{2}F_{ij}\,,
\end{equation}
where $\tilde{D}_iF^i = \gamma^{ij}F_{ij} = 0\,$, and $\tilde{D}^j F_{ij} = 0\,$.

\subsection{The gauge problem}\label{subsec:gauge}

The ten degrees of freedom in the metric perturbation $\delta g_{ab}$ are separated into the four scalar degrees of freedom $\left\lbrace \phi,\psi,B,F\right\rbrace$, four vector DOFs described by the two divergenceless vectors $\left\lbrace B_i, F_i\right\rbrace$ (each containing two DOFs), and two tensor DOFs described by the transverse trace-free tensor $F_{ij}$. 
The scalar perturbations can be thought of as generalisations of the Newtonian gravitational potential, the vector perturbations as post-Newtonian frame-dragging terms, and the tensors as the $+$ and $\times$ gravitational wave polarisations.
The dynamics of the perturbations are obtained by expanding Einstein's equations to linear order in the metric perturbations, and similarly in perturbations $\delta T_{ab}$ to the energy-momentum tensor, so that $\delta G_{ab} = 8\pi G\, \delta T_{ab}\,$. The scalar, vector and tensor equations decouple in the linear theory, making the SVT decomposition very helpful.
Note that because of the freedom in choosing the lapse and the shift in the $3+1$ decomposition, the number of true physical degrees of freedom is not 10 but 6: two each in the scalar, vector and tensor sectors. 

There is a serious issue regarding the physical interpretation of $\delta g_{ab}\,$, because of the non-covariant nature of the splitting of $g_{ab}$ into $\bar{g}_{ab}$ and $\delta g_{ab}\,$.
This is the gauge problem in cosmological perturbation theory \cite{Bardeen_1980,Malik_2009}. In writing down $\delta g_{ab}$ in a coordinate basis $(\tau, x^i)$, we are performing a map from the true spacetime manifold $\mathcal{M}$ to some fictitious background manifold $\bar{\mathcal{M}}$\,, on which our coordinates are defined. 
There is no unique choice of $\bar{\mathcal{M}}$, so we could change to some other background $\bar{\mathcal{M}}'$, and then map points in that manifold on to points in $\mathcal{M}\,$. Hence, the same set of coordinates $(\tau, x^i)$ would now refer to a different point in the true spacetime $\mathcal{M}\,$.
The change in mapping from $\bar{\mathcal{M}} \longrightarrow \mathcal{M}$ to $\bar{\mathcal{M}}' \longrightarrow \mathcal{M}$ is an active gauge transformation: we hold the coordinates fixed, and consider the change in the points $p \in \mathcal{M} \longrightarrow p' \in \mathcal{M}$ under the action of the gauge generators $\chi^a\,$.

\begin{figure}
    \centering
    \includegraphics[width=0.8\linewidth]{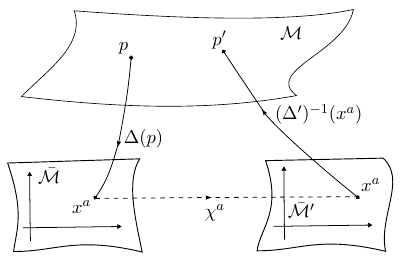}
    \caption{The effect of an active gauge transformation $\chi^a\,$. By changing the choice of ``background'' manifold $\bar{\mathcal{M}}$ on to which the spacetime manifold $\mathcal{M}$ is mapped, $\chi^a$ changes the spacetime point $p$ to which the coordinate point $x^a$ refers.}
    \label{fig_gauge_transformation}
\end{figure}

An active gauge transformation is depicted in Fig. \ref{fig_gauge_transformation}. The map $\Delta: \mathcal{M} \longrightarrow \bar{\mathcal{M}}$ maps points $p$ on the spacetime manifold $\mathcal{M}$ on to coordinate points (in the comoving coordinate chart of) $x^a$ on some background $\bar{\mathcal{M}}$. 
The metric is split as $g_{ab} \longrightarrow \bar{g}_{ab} + \delta g_{ab}\,$. 
The inverse map $\Delta^{-1}$ maps coordinate points $x^a$ on $\bar{\mathcal{M}}$ on to points $p$ on $\mathcal{M}\,$.
The alternative map $\Delta': \mathcal{M} \longrightarrow \bar{\mathcal{M}}'$ is such that $g_{ab} \longrightarrow \bar{g}_{ab}' + \delta g_{ab}'\,$.
Consider the following process. We first perform the map $\Delta(p)$ to $x^a$ on $\bar{\mathcal{M}}\,$.
We perform an active gauge transformation $\bar{\mathcal{M}} \longrightarrow \bar{\mathcal{M}}'\,$, by a generator $\chi^a\,$. This holds coordinate points fixed such that $x^a$ on $\bar{\mathcal{M}}$ is mapped to $x^a$ on $\bar{\mathcal{M}}'\,$.
Finally, we map from $x^a$ on $\bar{\mathcal{M}}'$ back on to the true spacetime $\mathcal{M}\,$, using $\left(\Delta'\right)^{-1}\,$. This takes us to a point $p' \neq p\,$. Hence, the active gauge transformation has changed the spacetime point to which the coordinates $x^a$ refer.

One can also study passive gauge transformations, in which the spacetime points $p$ are held fixed and the coordinates that label them change \cite{Malik_2008}.
We will focus on the active approach, as it will allow us to write down a set of gauge choices directly in terms of the perturbations $\left\lbrace \phi, \psi, B, F, B_i, F_i, F_{ij}\right\rbrace\,$.

Under an active gauge transformation by a generator $\chi^a$, a covariantly defined object $A$, evaluated at some fixed coordinate location $x^a = (\tau, x^i)$, is transformed to $A'$ under the exponential map, $A \longrightarrow A' = \exp{\left(\mathcal{L}_{\chi}\right)}\,A$. Writing $A = \bar{A} + \delta A$, and linearising in both $\delta A$ and $\chi^a\,$, one sees that $\bar{A}' = \bar{A}$, and $\delta A' = \delta A + \mathcal{L}_{\chi}\bar{A}\,$.
The gauge generator is a 4-vector like any other, so its components can be decomposed into two scalars and a divergenceless vector: $\chi^a = \left(\chi^0, \tilde{D}^i\zeta + \zeta^i\right)\,$, with $\tilde{D}_i \zeta^i = 0\,$.

Acting on the metric with $\exp{\left(\mathcal{L}_{\chi}\right)}$ gives the transformations of the metric perturbations,
\begin{eqnarray*}
    \phi \longrightarrow \phi - \mathcal{H}\chi^0 - {\chi^0}'\,; \quad && \quad \psi \longrightarrow \psi + \mathcal{H}\chi^0\,; \\
    B \longrightarrow B + \zeta' - \chi^0\,; \quad && \quad F \longrightarrow F + \zeta\,; \\
    B_i \longrightarrow B_i + \zeta_i'\,; \quad && \quad F_i \longrightarrow F_i + \zeta_i\,; \\
    F_{ij} \longrightarrow F_{ij}\,, \quad &&
\end{eqnarray*}
from which we see that the gravitational wave sector $F_{ij}$ is gauge-independent.
All other metric perturbations are dependent on the choice of gauge, i.e on the fictitious background manifold with respect to which our coordinate grid is defined. 
However, we can package the 6 physical degrees of freedom in $\delta g_{ab}$ into gauge-invariant variables: two gauge-invariant scalar perturbations, a gauge-invariant divergenceless vector perturbation, and a gauge-invariant tensor perturbation,
\begin{eqnarray}\label{eq_Bardeen_potentials}
    \Phi &=& \phi + \mathcal{H}\left(F' - B\right) + \left(F'-B\right)'\,, \\
    \nonumber \Psi &=& \psi + \mathcal{H}\left(F'-B\right)\,, \\
    \nonumber \beta_i &=& B_i - F_i'\,, \quad {\rm and} \,\\
    \nonumber F_{ij} &=& F_{ij}\,.
\end{eqnarray}
The scalars $\Phi$ and $\Psi$ are the celebrated Bardeen potentials \cite{Bardeen_1980}, and $\beta_i$ provides a gauge-invariant definition of the frame-dragging potential \cite{Thomas_2015,bruni2014computing}. 

Hence, gauge-invariant results can be obtained in CPT by either (a) working exclusively with gauge-invariant quantities or (b) fixing the gauge, by choosing which four of the eight gauge-dependent DOFs in the metric to set to zero.
This is similar, but not identical, to fixing the lapse and shift in the $3+1$-covariant formalism. Moreover, we stress that this similarity is only true perturbatively.
Gauge choice in the $3+1$ formalism means a choice of coordinate basis $(t, x^i)$, with respect to which the real spacetime $\mathcal{M}$ is covariantly foliated into surfaces of constant $t$. Thus, a change of gauge in that context means a passive transformation, with the coordinates themselves being changed.
In contrast, in perturbation theory it means a choice of the fictitious $\bar{\mathcal{M}}$, which is changed by an active transformation, with the coordinate basis held fixed.

The perturbation $\delta T_{ab}$ to the energy-momentum tensor is
\begin{equation*}
    \delta T_{ab} = \left(\delta\rho + \delta p\right) \bar{u}_a \bar{u}_b + 2\left(\bar{\rho}+\bar{p}\right)\bar{u}_{(a}\delta u_{b)} + \bar{p}\,\delta g_{ab} + \delta p\, \bar{g}_{ab} + 2\bar{u}_{(a}q_{b)} + \pi_{ab}\,.
\end{equation*}
The momentum density and anisotropic stress are covariantly defined objects which vanish at the level of the background, so at first order they are gauge-independent, by the Stewart-Walker lemma \cite{stewart1974perturbations}. However, the density and pressure perturbations are gauge dependent. Specifically, under the same gauge transformation as above,
\begin{eqnarray*}
    \delta \rho \longrightarrow \delta\rho + \chi^0\bar{\rho}'\,; \quad && \quad \delta p \longrightarrow \delta p + \chi^0 \bar{p}'\,.
\end{eqnarray*}
The 4-velocity $u^a$ can be written in general as $u^a = \left(u^0, \dfrac{1}{a}v^i\right)\,$, where we now have some additional freedom, corresponding to the freedom in the choice of $u^a$ in the $1+3$ decomposition. 
We can choose $u^a$ to track the matter field such that $v^i$ vanishes, in which case the momentum density $q^a$ measured by observers with worldlines tangent to $u^a$ will be non-zero in general. Alternatively, we can define $u^a$ such that $q^a$ vanishes, in which case $v^i$ is non-zero and is considered a peculiar velocity. 
Adopting the latter perspective, we can split $v_i$ into a scalar and divergenceless vector part, $v_i = \tilde{D}_i v + \mathfrak{v}_i\,$, where $\tilde{D}_i \mathfrak{v}^i = 0\,$.
These have the gauge transformation properties
\begin{eqnarray*}
    v \longrightarrow v - \zeta'\,; \quad && \quad \mathfrak{v}_i \longrightarrow \mathfrak{v}_i - \zeta_i'\,.
\end{eqnarray*}
Hence, we can identify gauge-invariant quantities associated with the matter perturbations,
\begin{eqnarray*}
    \delta \rho^N = \delta\rho - \bar{\rho}'\left(F' - B\right)\,, \quad \delta p^N = \delta p - \bar{p}'\left(F' - B\right)\,, \quad v^N = v + F'\,, \quad \mathfrak{v}^N_i = \mathfrak{v}_i + F'_i\,.
\end{eqnarray*}
The notation N here refers to the Newtonian gauge\footnote{Note that $\pi^N_{ij} 
= \pi_{ij}$ trivially, due to its gauge independence.}, which is the gauge choice we will make for all CPT calculations. 
It corresponds to the conditions $B = F = B_i = F_i = 0\,$. The metric perturbations are identically equal to the gauge-invariant perturbations in Eq. (\ref{eq_Bardeen_potentials}), and the matter perturbations are equal to the gauge-invariants we just defined. 
Changing notation slightly, we can write the line element as
\begin{equation}\label{eq_metric_Newtonian_gauge}
    \mathrm{d}s^2 = a^2(\tau)\left[-\left(1-2\Phi\right)\mathrm{d}\tau^2 + 2B_i\,\mathrm{d}\tau\mathrm{d}x^i + \left\lbrace\left(1+2\Psi\right) + F_{ij}\right\rbrace\delta_{ij}\mathrm{d}x^i\mathrm{d}x^j \right]\,.
\end{equation}
where it should be noted that we have assumed that the background FLRW cosmology is spatially flat, motivated by the tight observational bounds on $\Omega_K$, so that $\gamma_{ij} = \delta_{ij}$.
This is also sometimes called the Poisson or longitudinal gauge. Some authors use the phrase ``Newtonian gauge'' to refer only to the purely scalar version of the above, with just the $\Phi$ and $\Psi$ perturbations. 
As we only consider first-order perturbations in this thesis, the scalar, vector and tensor terms are entirely decoupled from one another, so it is possible to use the terms Newtonian and longitudinal gauge interchangeably.

\subsection{Cosmological perturbations in the Newtonian gauge}\label{subsec:newtonian_gauge}

Let us now focus on the properties of linear cosmological perturbations in the Newtonian gauge.

The components of the perturbed Einstein tensor are
\begin{eqnarray}\label{eq_CPT_Einstein_tensor_components}
    \delta G_0^{\ \ 0} &=&\frac{1}{a^2}\left[-2\nabla^2\Psi + 6\mathcal{H}\left(\Psi'+\mathcal{H}\Phi\right)\right], \\
    \nonumber \delta G_0^{\ \ i} &=& \frac{\delta^{ij}}{a^2}\left[-2\left(\Psi'+\mathcal{H}\Phi\right)_{,j} - \frac{1}{2}\nabla^2 B_j + 2\left(\mathcal{H}^2-\mathcal{H}'\right)B_j\right], \quad {\rm and} \\
    \nonumber \delta G_i^{\ \ j} &=& -\frac{2\delta_i^{\ j}}{a^2}\left[\left(\mathcal{H}^2 + 2\mathcal{H}'\right)\Phi + \mathcal{H}\Phi' + 2\mathcal{H}\Psi' + \Psi'' + \frac{1}{3}\nabla^2\left(\Phi - \Psi\right)\right] \\
    \nonumber && + \frac{\delta^{jk}}{a^2}\left(\Phi-\Psi\right)_{,\langle ik\rangle} - \frac{\delta^{jk}}{a^2}\partial_{(i}\left[B_{k)}'+2\mathcal{H}B_{k)}\right] \\
    \nonumber && + \frac{\delta^{jk}}{a^2}\left(F_{ik}'' + 2\mathcal{H}F_{ik}' - \nabla^2 F_{ik}\right)\,.
\end{eqnarray}
Here, we have written all the components with one index raised and one lowered, which turns out to make calculations simpler. We have also replaced the spatially projected derivatives $\tilde{D}_i$ with partial derivatives, as we are dealing with a spatially flat background.

Hence, the scalars $\Phi$ and $\Psi$ satisfy
\begin{eqnarray}\label{eq_CPT_newtonian_gauge_GR}
    \Psi' + \mathcal{H}\Phi &=& 4\pi G\,\left(\bar{\rho} + \bar{p}\right) a^2\, v\, \\
    \nonumber \nabla^2 \Psi &=& -4\pi G\,a^2\left(\delta\rho + 3\left(\bar{\rho}+\bar{p}\right)\mathcal{H}v\right)\,, \\
    \nonumber \frac{1}{3}\nabla^2\Phi + \Psi'' + 2\mathcal{H}\Psi' + \mathcal{H}\Phi' + \left(\mathcal{H}^2 + 2\mathcal{H}'\right)\Phi &=& -\frac{4\pi G a^2}{3}\left[\delta \rho - \delta p + 3\left(\bar{\rho}+\bar{p}\right)\mathcal{H}v\right]\,\\
    \nonumber \Phi - \Psi &=& 8\pi G a^2\, \Pi\,,
\end{eqnarray}
where $\Pi$ is the scalar part of the anisotropic stress tensor, $\pi_{ij} = \Pi_{,ij} + \Pi_{(i,j)} + \Pi_{ij}\,$.
These equations explain the characterisation of this gauge as Newtonian: on scales much smaller than the Hubble horizon, terms involving time derivatives can be neglected, and so $\Phi$ and $\Psi$ both satisfy equations very similar to the Poisson equation of Newtonian gravity.
Anisotropic stress is typically negligible at linear order, as it is primarily sourced by neutrinos, which are very subdominant. Hence, we get $\Phi = \Psi\,$. The equality of the two Newtonian gauge potentials is generically broken in modified gravity theories, as we will demonstrate in Chapter 5.

The vector perturbations satisfy
\begin{eqnarray}
    2\left(\mathcal{H}' - \mathcal{H}^2\right)B_i + \frac{1}{2}\nabla^2 B_i &=& 8\pi G \left(\bar{\rho} + \bar{p}\right)\, a^2 \, \mathfrak{v}_i\,, \\
    \nonumber B_i' +2\mathcal{H}B_i &=& -8\pi G\, a^2 \,\Pi_i\,,
\end{eqnarray}
showing that in the expected absence of a substantial anisotropic stress vector, vectors $B_i$ decay as the Universe expands.

Finally, the tensor perturbations satisfy
\begin{equation}
    F_{ij}'' + 2\mathcal{H}F_{ij}' - \nabla^2 F_{ij} = 8\pi G\,a^2\,\Pi_{ij}\,.
\end{equation}
As they describe the free propagation of gravitons, the tensor modes do not have a constraint equation.

Let us now see how the CPT equations fit into a $1+3$-covariant description of gravity. This will be useful when we derive theory-agnostic versions of some of these equations in Chapter 5.
At the level of the background, there was a single canonical choice of preferred timelike vector, $u^a = \dfrac{1}{a}\delta^a_{\ \tau}\,$.
In the presence of perturbations, the normalisation $u_a u^a = -1$ gives $u^0 = \dfrac{1}{a}\left(1+\Phi\right)\,$.
Suppose we still want $u^a$ to be orthogonal to constant-$\tau$ hypersurfaces, such that $u^i = 0\,$. The kinematic variables associated with this congruence are
\begin{equation*}
    \Theta = \frac{3}{a}\left(\mathcal{H} + \Psi' + \mathcal{H}\Phi\right)\,, \quad \sigma_{ij} = a\left(F'_{ij} + 2\mathcal{H}F_{ij}\right)\,, \quad \omega_{ij} = -a B_{[i,j]}\,, \quad \dot{u}_i = -\Phi_{,i} + \mathcal{H}B_i + B_i'\,.
\end{equation*}
Unless there are gravitational waves present, the Newtonian gauge describes a shear-free congruence, and unless there are frame-dragging vector potentials present, that congruence is irrotational.
The electric and magnetic parts of the Weyl tensor are respectively
\begin{eqnarray*}
    E_{ij} &=& -\frac{1}{2}\left[\left(\Phi+\Psi\right)_{,\langle ij\rangle} - B'_{(i,j)} + F_{ij}'' + \nabla^2 F_{ij}\right]\,, \\
    H_{ij} &=& \frac{1}{2}\,\eta_{lk(i}\,\partial^l \left(2{F_{j)}^{\ k}}' - \partial_{j)}B^k\right)\,.
\end{eqnarray*}
Therefore, if we restrict ourselves purely to scalar perturbations, then the Weyl curvature is purely electric. It turns out that this is also true at second order in perturbation theory \cite{Umeh_2011}. This $1+3$ perspective highlights why the Newtonian gauge really is ``Newtonian'', since the entirely non-Newtonian gravity is contained in $H_{ab}$ \cite{Dunsby_1997,Maartens_1998,clifton2017magnetic}. 
On the other hand, the combination $\Phi + \Psi$ enters into the electric part of the Weyl curvature, which we recall from our study of the Sachs equations in Section \ref{subsec:geometric_optics} sources weak gravitational lensing in the geometric optics approximation.
As expected from the discussion of the $1+3$ decomposition, gravitational waves $F_{ij}$ cause both the electric and magnetic Weyl curvature to be non-zero.

The upshot of these calculations is that we could have got to the perturbation theory equations in a covariant way \cite{Hawking_1966}, using the $1+3$ equations of motion for the comoving congruence $u^a$ we just defined:
\begin{itemize}
    \item The Hamiltonian constraint (\ref{eq_Hamiltonian_constraint_1+3}) gives the $\nabla^2 \Psi$ equation.
    \item The momentum constraint (\ref{eq_momentum_constraint_1+3}) gives the $\Psi' + \mathcal{H}\Phi$ and $\nabla^2 B_i$ equations.
    \item The Raychaudhuri equation (\ref{eq_Raychaudhuri}) gives the equation for $\Psi''\,$.
    \item The shear evolution equation (\ref{eq_shear_evolution}) produces the slip relation between $\Psi$ and $\Phi\,$.
    \item The equations of motion for the electric and magnetic parts of the Weyl curvature, Eqs. (\ref{eq_Eweyl_evolution}-\ref{eq_div_H_constraint}), give rise to the evolution equations for $B_i$ and $F_{ij}\,$.
\end{itemize}

This gives us a set of retrospective justifications for the Newtonian gauge choice.
\begin{enumerate}
    \item In the absence of vector and tensor perturbations, it provides a silent foliation \cite{Dunsby_1997,van_Elst_1997} of the perturbed FLRW spacetime, because $H_{ab}$ vanishes at both first and second order if we select $u^a = \dfrac{1}{a}\left(1 + \Phi\right)\delta^a_{\ \tau}$ and only consider the scalar perturbations $\Phi$ and $\Psi$\,. This silent property is particularly advantageous for calculating spatial averages \cite{Umeh_2011,Adamek_2018}.
    \item The scalar metric perturbations $\Phi$ and $\Psi$ are gauge-invariant, so using them circumvents the problem of gauge ambiguities that is a serious issue for e.g. the synchronous gauge \cite{Bardeen_1980,bruni1997perturbations}.
    \item The equations of motion for the scalar perturbations are similar to the Newton-Poisson equation, in the small-scale (aka. quasi-static \cite{bakerbull,lagos2018general}) limit.
    \item The Newtonian correspondence turns out to be rather deep, with significant consequences for small-scale modelling. Unlike the vast majority of other standard CPT gauge choices, the Netwonian gauge remains valid in the presence of nonlinear density contrasts, so it can be safely extended into a post-Newtonian expansion in the weak-field regime \cite{Clifton_2020}. We will make use of this property in Chapter 4. Its validity on small scales makes the Newtonian gauge well-suited to relativistic N-body simulations \cite{adamek2014distance,Adamek_2015}.
\end{enumerate} 

The set of CPT equations is completed by the contracted Bianchi identities, producing the relativistic continuity and Euler equations. These are
\begin{eqnarray}
    \delta \rho' + 3\mathcal{H}\left(\delta\rho + \delta p \right) + \left(\bar{\rho} + \bar{p}\right)\left(3\Psi' + \nabla^2 v\right) &=& 0\,, \label{eq_CPT_continuity_eqn}\\
    v' + \mathcal{H}v - 3c_s^2\mathcal{H}v + \frac{\delta p + \nabla^2 \Pi}{\bar{\rho} + \bar{p}} - \Phi &=& 0\,, \label{eq_CPT_Euler_eqn}\\
    \left(\mathfrak{v}_i + B_i\right)' + \mathcal{H}\left(\mathfrak{v}_i + B_i\right) - 3c_s^2 \mathcal{H}\mathfrak{v}_i + \frac{\nabla^2 \Pi_i}{\bar{\rho} + \bar{p}} &=& 0\,.
\end{eqnarray}
In the second and third equations, we have defined the sound speed $c_s^2 = \dfrac{\partial p}{\partial \rho}\,$, in order to evaluate $\bar{p}' = -3c_s^2\mathcal{H}\,\left(\bar{\rho} + \bar{p}\right)\,$.
In the late Universe, we can specialise to cold dark matter with $\bar{p} = 0$ and $c_s^2 = 0$\,. The above equations tell us that pure vector velocity perturbations decay with the vector metric perturbations. By contrast, density and scalar velocity perturbations can grow, sourced by the potential $\Phi$. 

As an example, consider $\delta_m = \dfrac{\delta \rho_m}{\bar{\rho}_m}\,$, evaluated during the matter-dominated era that is relevant for most structure formation. By combining the continuity, Raychaudhuri and Euler equations and taking the small-scale limit so that the time derivatives of $\Psi$ can be neglected, one obtains \cite{Dodelson_2003}
\begin{equation}
    \delta_m'' + \mathcal{H}\delta_m' - \frac{3}{2}\Omega_m(a)\mathcal{H}^2(a) \delta_m = 0\,,
\end{equation}
where during matter domination $a \sim \tau^2$ so $\mathcal{H} = \dfrac{2}{\tau}\,$. This has a growing solution $\delta_m \sim \tau^2$, demonstrating how structure forms in linear theory.

In reality, first-order CPT breaks down on small scales where $\delta_m$ becomes comparable to unity. Then, one must use alternative techniques, in order to determine the behaviour of $\delta_m$\,. For larger but sub-unity $\delta_m$, second-order perturbation theory can be used \cite{malik2004evolution}. However, for $\delta_m > 1$, the entire premise of CPT no longer makes sense. Instead, nonlinear Newtonian methods are usually employed, especially N-body simulations under Newtonian gravity (see e.g. \cite{efstathiou1985numerical,bertschinger1991cosmological,springel2005simulations,pillepich2018simulating,vogelsberger2020cosmological}).

This concludes our discussion of the theory of cosmological perturbations. We will next introduce some key observables, that have provided evidence for the $\Lambda$CDM concordance model.

\section{Standard observations}

The first key observational probe in cosmology that we wish to introduce is the cosmic microwave background. We will focus in particular on its temperature anisotropies. The second is the Hubble diagram, which is the relation between redshift and luminosity distance for directly observed distant astrophysical sources.
There are many other observations that are made in cosmology, such as surveys of the large scale structure of galaxies, the baryon acoustic oscillations, and gravitational wave measurements.
These are rich sources of cosmological information, but we do not make calculations relating to them in this thesis, so we will not discuss them here. 

\subsection{The cosmic microwave background}\label{subsec:CMB}

It is not much of an exaggeration to say that cosmology became a science when the cosmic microwave background (CMB) was first observed \cite{penzias1979measurement}, and a precision science when the anisotropies in its temperature were first mapped by COBE in the 1990s \cite{smoot1991preliminary,efstathiou1992cobe}. Since then, the CMB has been measured ever more precisely by first WMAP \cite{spergel2003first,komatsu2009five,bennett2013nine} and then Planck \cite{Planck_2013,Planck_2015,Planck_2018,Planck_2020}. 
The CMB is an extraordinarily uniform black body at $\bar{T} = 2.72548 \pm 0.00057 \, {\rm K}\,$ \cite{fixsen2009temperature}, which is redshifted by a factor of $z_{\rm recom} = 1089.88 \pm 0.22$ \cite{aghanim2020planck} from its emission temperature. It contains fluctuations $\delta T$ of order tens of $\mu{\rm K}\,$, whose statistics convey a wealth of information about the Universe, as do the statistics of its polarisations and gravitational lensing \cite{aghanim2016planck,aghanim2020planck,durrer2020cosmic}. 
Here, we will focus primarily on the temperature anisotropies, and will very briefly discuss polarisations.

The CMB is a fundamentally two-dimensional observable: we observe at essentially a single spacetime point, and as it has one redshift $z_{\rm recom}$ we do not have access to any information in the radial direction (placing ourselves at the origin of a spherical coordinate system). 
The temperature field can therefore be written as a single scalar function of the polar coordinates $\theta$ and $\phi$ on our celestial sphere.
It is convenient to decompose this function as
\begin{equation}
    \frac{\delta T(\theta, \phi)}{\bar{T}} = \sum_{l=1}^{\infty} \sum_{m=-l}^l a_{lm} Y_{lm}(\theta, \phi)\,,
\end{equation}
where the spherical harmonic functions $Y_{lm}(\theta,\phi)$ form an orthonormal basis on the two-sphere, $\int_{S^2}\mathrm{d}\theta\mathrm{d}\phi\sin{\theta}\,Y^{*}_{l'm'}(\theta,\phi)\,Y_{lm}(\theta, \phi) = \delta_{ll'}\,\delta_{mm'}\,$. We have removed the monopole $l = 0$, as it refers by definition to the all-sky average $\bar{T}\,$.

The central object in the study of the CMB is the 2-point angular correlation function of the temperature fluctuations. Given two observing directions $\hat{\mathbf{e}}_1$ and $\hat{\mathbf{e}}_2$, the angle between them is defined $\cos{\theta} = \hat{\mathbf{e}}_1 {\bf\cdot} \hat{\mathbf{e}}_2\,$. Then, the 2-point correlator is
\begin{equation}
    \left\langle \frac{\delta T}{\bar{T}}(\hat{\mathbf{e}}_1)\,\frac{\delta T}{\bar{T}}(\hat{\mathbf{e}}_2)\right\rangle = \frac{1}{4\pi}\sum_{l = 1}^{\infty}\left(2l + 1\right)\,\mathcal{C}_l\,P_l(\cos{\theta})\,,
\end{equation}
where $\mathcal{C}_l$ is the angular power spectrum (really $\mathcal{C}_l^{TT}$), defined such that $\left\langle a^{*}_{l'm'}a_{lm}\right\rangle = \mathcal{C}_l\delta_{ll'}\delta_{mm'}\,$\footnote{In this expression, it has been assumed that $\delta T$ is a Gaussian random field, so that $\left\langle a_{lm} \right\rangle = 0\,$.}. It is often expressed in terms of $\mathcal{D}_l = \left(2\pi\right)^{-1}\,l\,(l+1)\,\mathcal{C}_l\,$, in order to remove the scaling of the moments with the multipole value $l\,$.

As most inflationary models indicate that the CMB temperature should be very close to a Gaussian random field, its 2-point correlation function, or equivalently the angular power spectrum, is sufficient to describe all the statistical information contained in the field, by Wick's theorem \cite{durrer2001theory}. 
Higher-point correlators can be studied to test for possible deviations from Gaussianity \cite{komatsu2005measuring,liguori2006testing,emami2015probing}, but they are not relevant for this thesis. 

\begin{figure}
    \centering
    \hspace{-0.32cm}
    \includegraphics[width=0.7\linewidth]{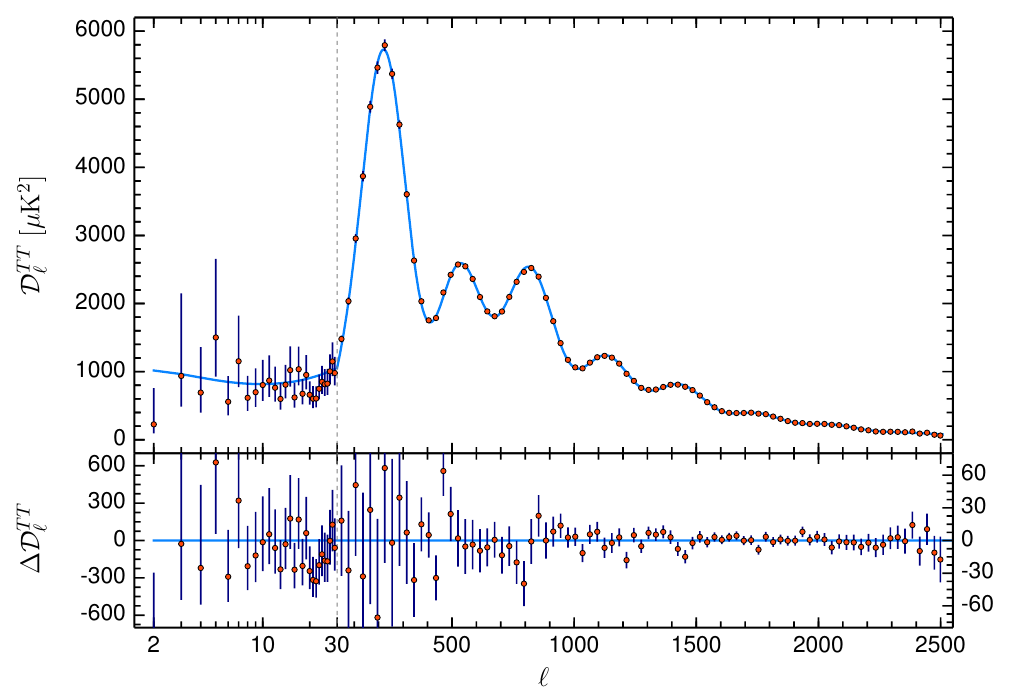}
    \includegraphics[width=0.68\linewidth]{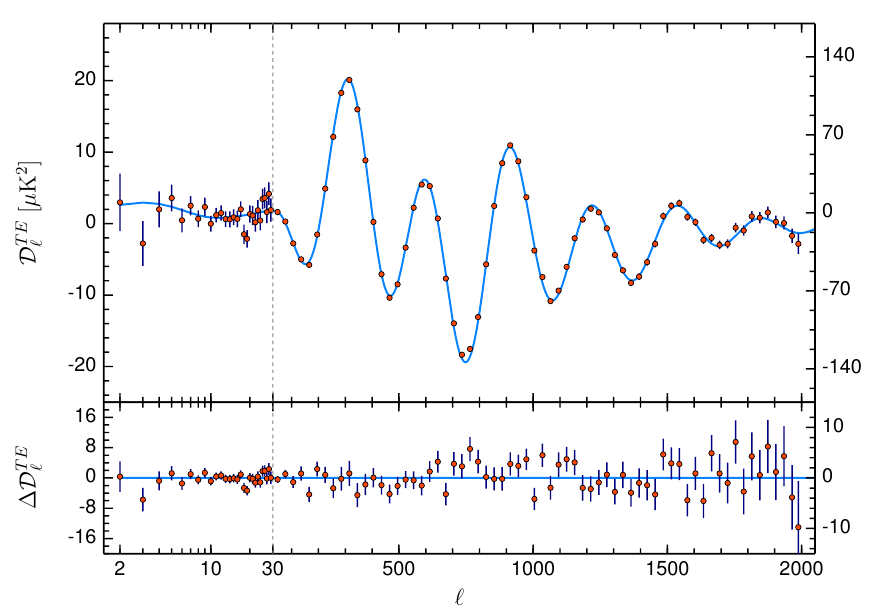}
    \hspace{-0.8cm}
    \includegraphics[width=0.7\linewidth]{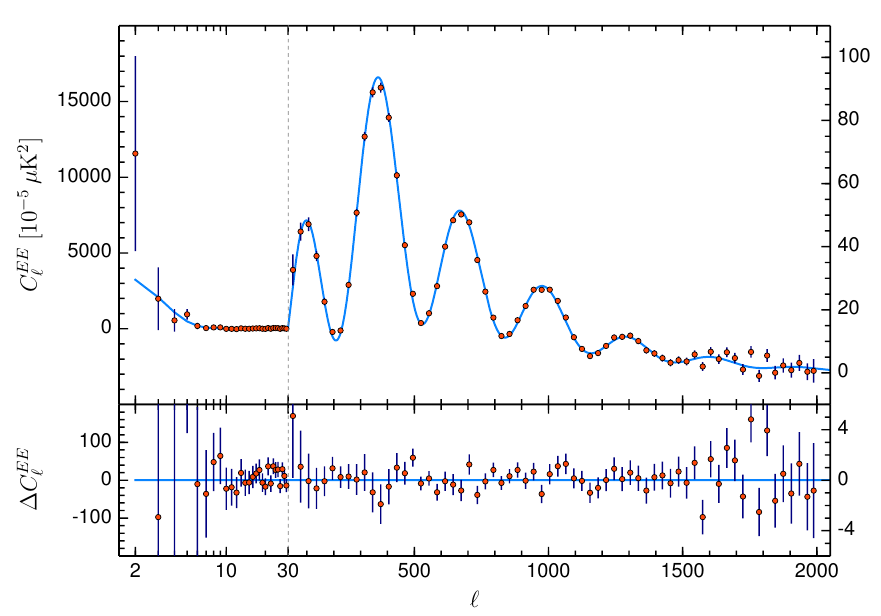}
    \caption{CMB temperature and E-mode polarisation angular power spectra, as measured by the Planck satellite, with their $1\sigma$ confidence intervals. From Ref. \cite{aghanim2020planck}.}
    \label{fig_Planck_2018_TT_TE_EE}
\end{figure}

In addition to the temperature correlations, let us also briefly mention the polarisation correlations present in the CMB. 
The polarisation signal has two irreducible parts: an irrotational $E$ part (rather like the electric field of classical electromagnetism, or indeed the electric part of the Weyl tensor), and a divergenceless $B$ part (rather like the magnetic field, or the magnetic part of the Weyl tensor).
Tensor perturbations produced in the early Universe can source $B$-mode polarisations \cite{polnarev1985polarization,seljak1997signature}, but these have not yet been observed. 
However, the $E$-mode polarisation signal has been mapped \cite{kovac2002detection}, and its 2-point correlation $\mathcal{C}_l^{EE}$ and cross-correlation with temperature fluctuations, $\mathcal{C}_l^{TE}$, have been measured much like $\mathcal{C}_l^{TT}$ \cite{planck_power_spectra,aghanim2020planck}.
The 2018 Planck $TT$, $TE$ and $EE$ spectra are displayed in Fig. \ref{fig_Planck_2018_TT_TE_EE}.
Note that the dipole moment ($l = 1$) has been removed. It is around $100 \times$ larger than any of the $l \geq 2$ multipoles. The exact physical interpretation of the dipole is controversial \cite{Aluri_2023}. For the remainder of this chapter, we will ignore it. 

In order to understand the physics behind the $TT$, $TE$ and $EE$ spectra, which are used for the analysis in Chapter 6, one need only consider the scalar metric and energy-momentum perturbations. Most of the information comes from the $TT$ spectrum, so it is there that we will focus our attention. We will not provide a derivation of the terms that contribute to $\mathcal{C}_l^{TT}$, but will merely summarise some key results. For a full discussion, see e.g. Refs \cite{durrer2001theory,durrer2020cosmic,hu2002cosmic}.
The power spectrum at a given multipole is obtained by integrating the source contribution to that multipole over all wavenumbers,
\begin{equation}\label{eq_source_terms_integrate}
    \mathcal{C}_l = \frac{2}{\pi}\int \mathrm{d}\ln{k}\, k^3 \, \left\langle \vert S_l^2(k) \vert \right\rangle \,.
\end{equation}
The source $S_l(k)$ comes from several physical effects. Ignoring polarisation terms, which are irrelevant for our purposes, it can be written as 
\begin{equation}
    S_l(k) = S_l^{\rm density}(k) + S_l^{\rm SW}(k) + S_l^{\rm Dop}(k) + S_l^{\rm ISW}(k)\,,
\end{equation}
where
\begin{eqnarray}
    S_l^{\rm density}(k) &=& \frac{1}{4}\,\frac{\delta\rho_r(k,\tau_{\rm recom})}{\bar{\rho}_r(\tau_{\rm recom})}\,j_l\left(k\left(\tau_0 - \tau_{\rm recom}\right)\right)\,, \label{eq_source_term_density} \\
    S_l^{\rm SW}(k) &=& - \Phi(k,\tau_{\rm recom})\,j_l\left(k\left(\tau_0 - \tau_{\rm recom}\right)\right)\,, \label{eq_source_term_Sachs_Wolfe} \\
    S_l^{\rm Dop}(k) &=& v_b(k,\tau_{\rm recom})\,j_l'\left(k\left(\tau_0-\tau_{\rm recom}\right)\right)\,, \quad {\rm and} \, \label{eq_source_term_Doppler} \\
    S_l^{\rm ISW}(k) &=& -\int_{\tau_{\rm recom}}^{\tau_0}\mathrm{d}\tau\,\left(\Phi'(k,\tau)+\Psi'(k,\tau)\right)\,j_l\left(k\left(\tau_0 - \tau\right)\right)\,. \label{eq_source_term_ISW}
\end{eqnarray}
Here $j_l(x)$ is the $l^{\rm th}$ spherical Bessel function, $j_l'$ denotes a derivative of that function with respect to its argument, and $v_b$ is the scalar part of the baryon 3-velocity, $v_b^i = \delta^{ij}\partial_j v_b + \mathfrak{v}_b^i$\,.
Because the Sachs-Wolfe and photon density perturbation source functions have the same coefficient $j_l(k(\tau_0 - \tau_{\rm recom}))$, it is not possible to distinguish them in $\mathcal{C}_l^{TT}\,$. They are therefore usually thought of as being one single term, which is loosely referred to as the ``temperature/Sachs-Wolfe'' term, or indeed just Sachs-Wolfe.

\begin{figure}
    \centering
    \includegraphics[width=\linewidth]{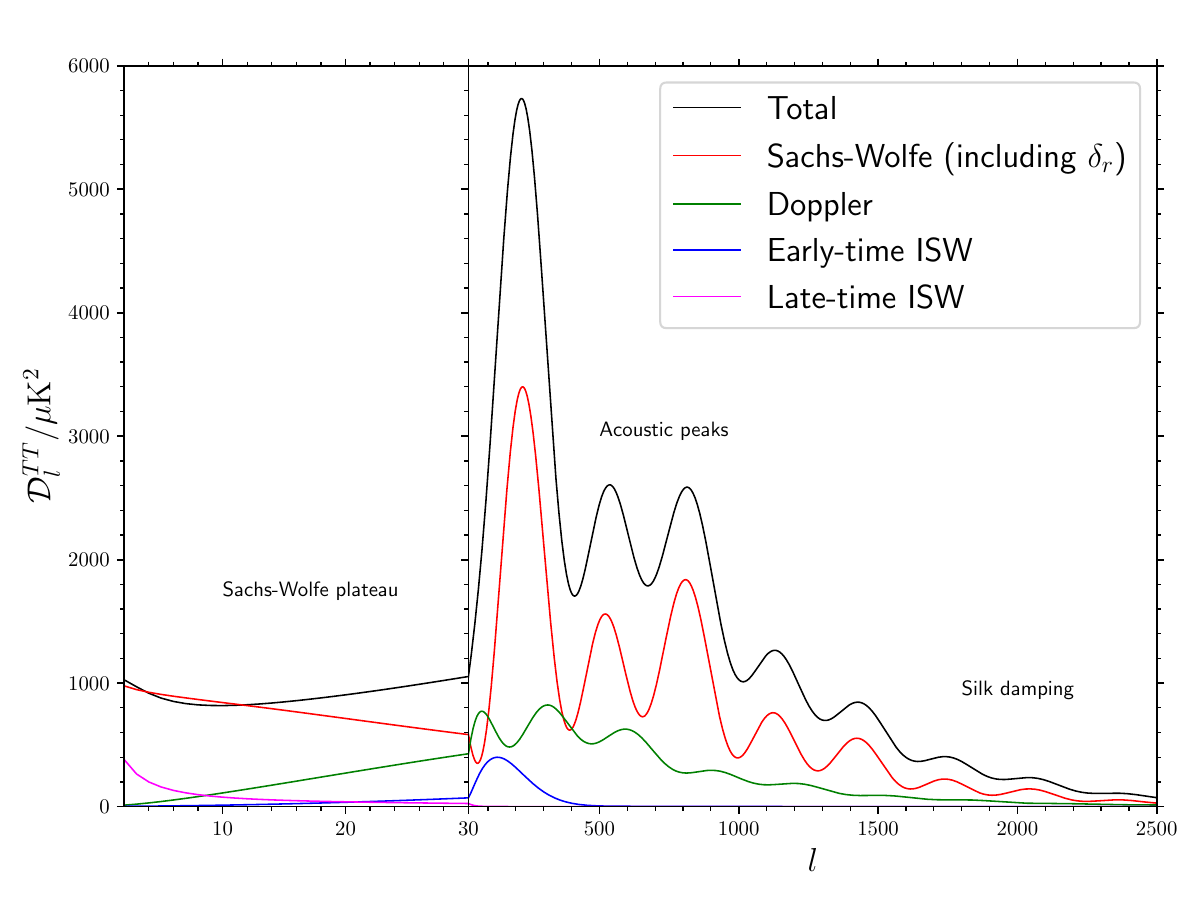}
    \caption{Contributions of the Sachs-Wolfe (including the intrinsic photon density perturbation), Doppler, and integrated Sachs-Wolfe source terms to CMB temperature anisotropies, for a $\Lambda$CDM cosmology with its parameters set to the Planck 2018 best-fit values. Lensing and polarisation effects have been ignored, and the ISW term is separated out into its early-time and late-time parts.}
    \label{fig_Cl_TT_sources_LCDM}
\end{figure}
The contributions of each of the source functions, for a fiducial $\Lambda$CDM cosmology, are shown in Fig. \ref{fig_Cl_TT_sources_LCDM}.
Let us now explain each of these contributions in turn.

First we have the Sachs-Wolfe term $\dfrac{\delta\rho_r}{4\bar{\rho}_r} - \Phi\, = \dfrac{1}{4}\delta_r - \Phi\,$. The intrinsic photon density perturbation directly corresponds to a temperature perturbation, by the black-body relation $\rho_r \sim T^4\,$. 
The perturbation $\Phi$ to the lapse function, $N = a\left(1-\Phi\right)\,$, corresponds to a gravitational time dilation, or equivalently gravitational redshift. 

The behaviour of the Sachs-Wolfe term can be split into two regimes: perturbations that were super-horizon at recombination, and perturbations that were sub-horizon. Calculating the particle horizon $r_P$ at $a_{\rm recom}$, and dividing this by $d_A(a_{\rm recom})$, one finds that the causal sphere for CMB photons corresponds to an angular scale $l \sim 100\,$. 
    
Hence, low-$l$ multipoles, $l \lesssim 100$, result from physics on super-horizon scales. 
On these scales, it can be shown that during matter domination, $\Phi \approx \dfrac{3}{8}\delta_r\,$ \cite{Dodelson_2003,durrer2001theory}.
Assuming that the initial conditions for the scalar perturbations are adiabatic \cite{gordon2000adiabatic}, the perturbation $\Phi$ is ultimately related to the gauge-invariant curvature perturbation $\mathcal{R}$\footnote{See e.g. Ref. \cite{Malik_2009} for a precise definition of $\mathcal{R}\,$.} laid down during inflation, $\Phi = \dfrac{3}{5}\mathcal{R}\,$. Therefore, on super-horizon scales we have \cite{sachs1967perturbations}
\begin{eqnarray}
    \mathcal{C}_l^{\rm SW} \approx \frac{2}{25 \pi}\int\mathrm{d}\ln{k}\,k^3 \,\left\langle \vert \mathcal{R}^2(k)\vert\right\rangle \, \vert j_l\left(k(\tau_0-\tau_{\rm recom})\right)\vert^2\,,
\end{eqnarray}
where we have dropped the time dependence of $\mathcal{R}$ because it is conserved on super-horizon scales.
The term $\left\langle \vert \mathcal{R}^2(k)\vert\right\rangle$ is nothing but the power spectrum of inflationary curvature perturbations. This is very nearly scale-invariant: $P_{\mathcal{R}}(k) \sim k^{n_s - 1}\,$, with $n_s = 0.9652 \pm 0.0042$ \cite{aghanim2020planck}.
It follows from the properties of the spherical harmonic functions $j_l$ that $\mathcal{C}_l^{\rm SW} \sim \dfrac{1}{l\left(l+1\right)}\,$, i.e. $\mathcal{D}_l^{\rm SW}$ is approximately constant with $l\,$. 
In other words, the $\Lambda$CDM model, with an inflationary paradigm for the initial curvature perturbations, predicts a nearly flat $D_l^{\rm SW}$ at low multipoles, referred to as the Sachs-Wolfe plateau (it would be a genuinely flat plateau if $n_s$ were equal to unity). This is displayed by the red curve in the left part of Fig. \ref{fig_Cl_TT_sources_LCDM}\,. Fig. \ref{fig_Planck_2018_TT_TE_EE} shows that it is borne out in the observed Planck spectrum.
    
The multipoles $l \gtrsim 100\,$, which are governed by sub-horizon adiabatic perturbations, behave rather differently. On sub-horizon scales, scalar perturbations are not frozen in time, but rather undergo oscillations. 
These occur due to acoustic oscillations in the photon-baryon fluid, which is tightly coupled by Thomson scattering. 
The Sachs-Wolfe term on sub-horizon scales is found to obey \cite{hu1996small,Dodelson_2003}
\begin{equation*}
\left(\frac{1}{4}\delta_r - \Phi\right)(k, \tau_{\rm LS}) \approx \left(\frac{1}{4}\delta_r - \Phi\right)(k, 0)\,\cos{\left(kr_s(a_{\rm LS})\right)}\,.
\end{equation*} 
Here $\tau = 0$ refers as usual to the end of the inflationary epoch.
The scale $r_s(a_{\rm LS})$ is the horizon associated with the sound waves, evaluated at last scattering:
\begin{equation}\label{eq_sound_horizon}
    r_s(a) = \int_0^{a} \frac{\mathrm{d}a\,c_s(a)}{a^2 H(a)}\,, \quad {\rm where} \quad c_s^2(a) = \frac{1}{3}\left(1-\frac{3\bar{\rho}_b(a)}{4\bar{\rho}_r(a)}\right)\,.
\end{equation}
The form of its sub-horizon solution shows that there will be acoustic peaks in the Sachs-Wolfe term for $k r_s(a_{\rm LS}) = n\pi\,$, where $n \in \mathbb{Z}^{+}\,$. In $\mathcal{C}_l^{TT}$, the peaks show up at multipoles $l_{\rm peak} = \dfrac{\pi}{\theta_{\rm peak}}$, with the first at the characteristic angular scale $\theta_{*} = \dfrac{r_s(a_{\rm LS})}{d_A(a_{\rm LS})}$\,. The Planck experiment has measured seven such peaks \cite{aghanim2020planck}. The location of the first is tightly constrained to be $100\,\theta_{*} = 1.04110 \pm 0.00031\,$.
The acoustic peaks are the dominant effect in the CMB $TT$ power spectrum for $l \gtrsim 100\,$.

An increase in the angular diameter distance to last scattering causes an decrease in $\theta_{*}\,$, shifting the locations of the acoustic peaks to the right. The only fundamental distance scale available from the cosmological background is $H_0^{-1}\,$. Unsurprisingly, then, a lower $H_0$ increases $d_A(a_{\rm LS})$, so $H_0$ is negatively correlated with $l_{\rm peak}\,$. 
The fact that $H_0$ is the key cosmological parameter in setting the locations of the acoustic peaks, by determining the angular diameter distance to the last scattering surface, will be important in Chapter 6. There, we will see that in some non-standard cosmological models (i.e. modifications to gravity), another parameter competes with $H_0$ in its effect on $d_A(a_{\rm LS})$ and thus the peak locations.

The heights of the acoustic peaks are primarily sensitive to the time interval\footnote{For the argument here, it does not matter whether one considers conformal time $\tau$ or cosmic time $t$ as the time coordinate.} that elapses between matter-radiation equality and last scattering. 
As the latter is fixed by the CMB monopole temperature, the peak heights are really determined by $\tau_{\rm eq}\,$. 
An earlier matter-radiation equality time means that dark matter is becoming the dominant component of $T_{ab}$ sooner. Acoustic oscillations in the photon-baryon fluid have less time to drive acoustic anisotropies in the CMB, before dark matter takes over and suppresses the oscillations. Thus, the acoustic peaks are smaller.
This shows that the dark matter density parameter $\omega_c = \Omega_{c 0}h^2$ is negatively correlated with the peak heights. By contrast, $\omega_b = \Omega_{b 0}h^2$ is positively correlated with them, for the simple reason that baryon acoustic oscillations are enhanced if there are more baryons present in the Universe. 
In reality, however, $\omega_b$ is tightly controlled by BBN abundances \cite{copi1995big,schramm1998big,burles2001big}. Therefore, it is mostly the free parameter $\omega_c$ that one is free to vary in order to change the peak heights.

\begin{figure}
    \centering
    \includegraphics[width=0.9\linewidth]{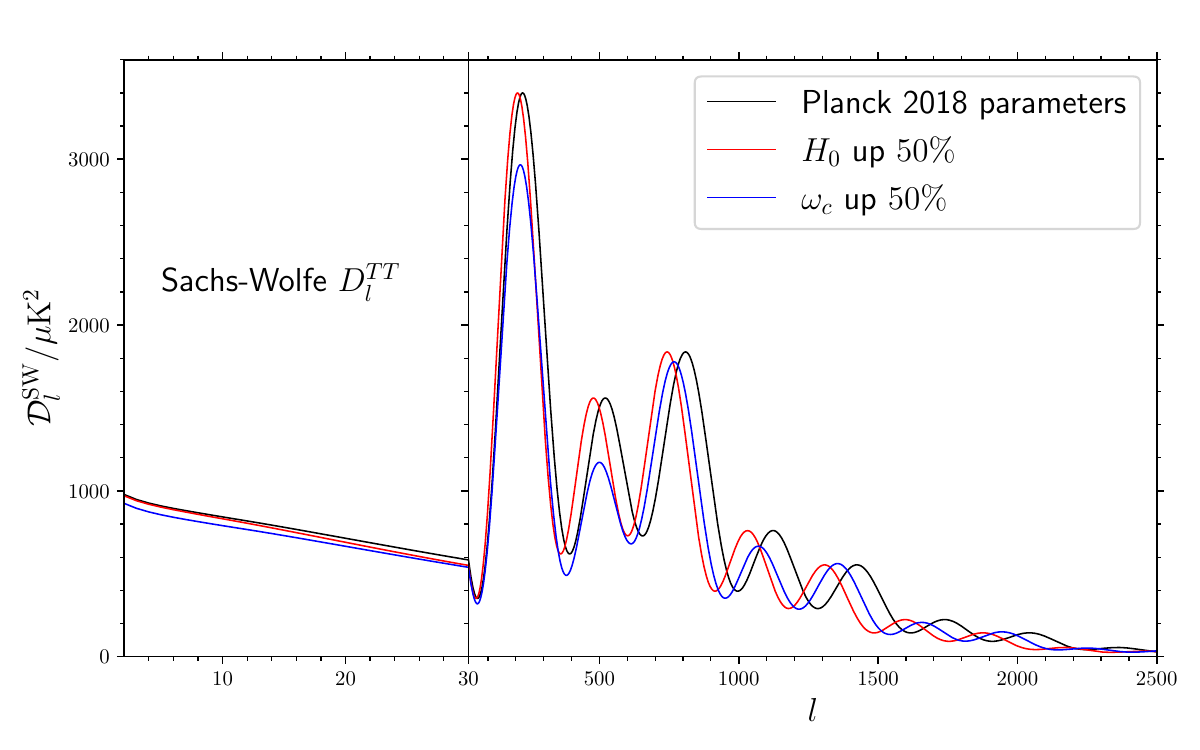}
    \caption{The effects of varying $H_0$ and $\omega_c$ on the Sachs-Wolfe contribution to the CMB $TT$ angular power spectrum, for a flat $\Lambda$CDM cosmology.}
    \label{fig_sachs_wolfe_var_H0_omega_c}
\end{figure}

The effects of $H_0$ and $\omega_c$ on the Sachs-Wolfe term are displayed in Fig. \ref{fig_sachs_wolfe_var_H0_omega_c}, where we have considered the effect of increasing each of those parameters by $50\%$ from their Planck 2018 best-fit value, while holding all other $\Lambda$CDM cosmological parameters constant. 
The effect of varying $H_0\,$, shifting the locations in $l$ of the acoustic peaks, is clearly visible. Varying $\omega_c$ also causes a slight horizontal shift, but its primary effect is to alter the heights of the peaks, especially the first peak.

The final principal feature of the Sachs-Wolfe source function is the damping tail, where CMB power is washed out on small angular scales, $l \gtrsim l_{\rm damping} \sim 1600\,$, by diffusive Silk damping \cite{silk1968cosmic}. 
This refers to the process where, on small scales, photons diffuse from hot overdensities to cold underdensities, which reduces the size of density and temperature perturbations \cite{hu1997physics}. Further details of diffusion damping are not relevant for this thesis. 

Let us now move on to the Doppler source function, $v_b(k,\tau_{\rm recom})\,j_l'\left(k\left(\tau_0-\tau_{\rm recom}\right)\right)\,$. 
It is so named because it arises precisely due to the Doppler shift in the energy of a photon that is emitted from a patch of the photon-baryon fluid which has a special-relativistic boost velocity $v_b^i = \partial^i v_b$ with respect to the CMB rest frame (the frame in which the kinematic dipole vanishes).
Like the Sachs-Wolfe term, it describes effects in the recombination epoch, rather than integrated effects over cosmic history. 

It follows that the physical effects driving the Doppler term are essentially the same ones that drive the Sachs-Wolfe term, as the baryon velocity potential $v_b$ is sourced by the lapse perturbation $\Phi$ through the relativistic Euler equation (\ref{eq_CPT_Euler_eqn}). 
However, the baryon acoustic oscillations in $v_b$ are out of phase with the oscillations in $\delta_r$ and $\Phi$. Under the same approximation scheme on small scales, the solution for $v_b$ at last scattering is \cite{hu1996small,Dodelson_2003}
\begin{equation*}
    v_b(k,\tau_{\rm LS}) \approx \sqrt{3}\left(\frac{1}{4}\delta_r - \Phi\right)(k,0)\,\sin{\left(kr_s(a_{\rm LS})\right)}.
\end{equation*}
The out-of-phase nature of the acoustic oscillations in the Doppler term is demonstrated by the green curve in Fig. \ref{fig_Cl_TT_sources_LCDM}.

Finally, consider the integrated Sachs-Wolfe (ISW) term (\ref{eq_source_term_ISW}).
It describes the integrated effect of photons being redshifted by evolving gravitational wells they encounter as they travel through the Universe. 
For a typical, $\Lambda$CDM-like cosmology, the ISW term is smaller than the Sachs-Wolfe and Doppler terms that arise directly from the recombination epoch. It only receives substantial contributions from super-horizon perturbations. 
Those contributions are non-zero only when the Weyl potential $\psi_W = \Phi + \Psi$\footnote{The Weyl potential is so named because $\psi_W$ is precisely what enters into the Weyl tensor at linear order in scalar perturbations in the Newtonian gauge.} has a non-zero conformal time derivative. 
During matter domination in GR, the perturbations $\Phi$ and $\Psi$ are frozen in time, which means that there is no ISW effect at all for most of cosmic history. 

However, there are two contributions that survive. One is the early ISW term that comes from just after recombination, when matter is dominant but there is still a non-negligible contribution from radiation to the overall cosmic energy density. 
This means that the perturbations can evolve on super-horizon scales, which are then brought inside the horizon as the Universe expands into the fully matter-dominated era. The early ISW effect therefore arises in modes that are around, and just outside, the horizon at recombination. As $r_P(a_{\rm recom})$ corresponds to $l \sim 100$, the early-time ISW is seen in $l$ of this order. This is displayed by the blue curve in Fig. \ref{fig_Cl_TT_sources_LCDM}.

The other contribution is the late ISW term, that arises after $a_{\rm DE}\,$, once dark energy dominates over matter. Then, large-scale perturbations begin to evolve again. 
The late-time ISW effect occurs, therefore, in multipoles that correspond to scales comparable to $r_P$ at late times. The $\Lambda$CDM model thus predicts an ISW rise in $\mathcal{D}_l^{TT}$ above the Sachs-Wolfe plateau, at the lowest multipoles $l \lesssim 10\,$. This rise is shown on the very left end of the magenta curve in Fig. \ref{fig_Cl_TT_sources_LCDM}.
Because they are directly sensitive to the time evolution of the scalar metric perturbations, both the early and late time ISW effects have been suggested as ways to test both the $\Lambda$CDM paradigm \cite{vagnozzi2021consistency} and gravity more generally \cite{zhang2006testing,cai2014integrated,frusciante2021signatures}.

This concludes our discussion on the cosmic microwave background, for now. In Chapter 4, we will look again at the CMB observations, in order to discuss some of the anomalies and tensions that have arisen in the data. These may point to a failure of the $\Lambda$CDM model, or even the breakdown of the homogeneous and isotropic FLRW paradigm itself.

\subsection{The Hubble diagram}\label{subsec:Hubble_diagram}

Possibly the most basic observations one can perform in cosmology are measurements of the fluxes $F$ received from, and redshifts $z$ of, distant astrophysical sources, such as radio galaxies, supernovae and quasars. As discussed in Section \ref{subsec:distance_measures}, if the intrinsic luminosities of a given set of sources are known, or at least can be estimated, then these flux measurements allow one to calculate the luminosity distances $d_L$ to those sources.
Then, one can construct a relation $d_L(z)\,$, or equivalently $\mu(z)$, where $\mu$ is the distance modulus. This is the Hubble diagram. 

Unlike the CMB, where observations of the temperature anisotropies require a vast amount of forward modelling in order to produce $\mathcal{C}_l^{TT}$, the construction of the Hubble diagram is essentially independent of the underlying cosmological model. One simply collects a set of data points $(z, d_L)$, and fits them to a curve.
This curve then tells us cosmological information directly, as it corresponds to a solution to the Sachs equations (\ref{eq_sachs_dA_1}-\ref{eq_sachs_dA_2}). A Taylor series expansion of the Sachs equations at low redshift tells us that at linear order in $z$, $d_L \simeq \dfrac{z}{H_0^{\parallel}} \simeq \dfrac{z}{\left[\left(\Theta/3\right) + \sigma_{ab}e^ae^b\right]\vert_{\rm obs}}\,$. Therefore, the monopole of the linear Hubble diagram contains direct information about the local expansion of space. 

In a general spacetime, the curve $d_L(z)$ depends on 
\begin{enumerate}
\item The time at which the observer is making measurements. More formally, this means the leaf $\Sigma_t$ of the foliation of the spacetime into constant-$t$ surfaces, on which the observer is placed.
\item The spatial location of the observer, within that leaf $\Sigma_t\,$.
\item The spatial direction $e^a$ in which they choose to observe.
\end{enumerate} 
However, in an FLRW cosmology, the Killing symmetries of the metric ensure that $d_L(z)$ depends only on the observing time. At low redshift, it is given by
\begin{equation}
    d_L(z) = \frac{z}{H_0} + \frac{\left(1-q_0\right)}{2H_0}z^2 + \mathcal{O}\left(z^3\right) + \: ... \: ,
\end{equation}
where $q_0$ is the deceleration parameter, $q_0 = -\dfrac{a\ddot{a}}{\dot{a}^2}\Big\vert_{t_0}\,$. Higher order terms in the Taylor series expansion in $z$ contain additional information about the FLRW universe; for example, the cubic term depends on the spatial curvature parameter $K\,$ \cite{visser2004jerk}. 

\begin{figure}
    \centering
    \includegraphics[width=0.9\linewidth]{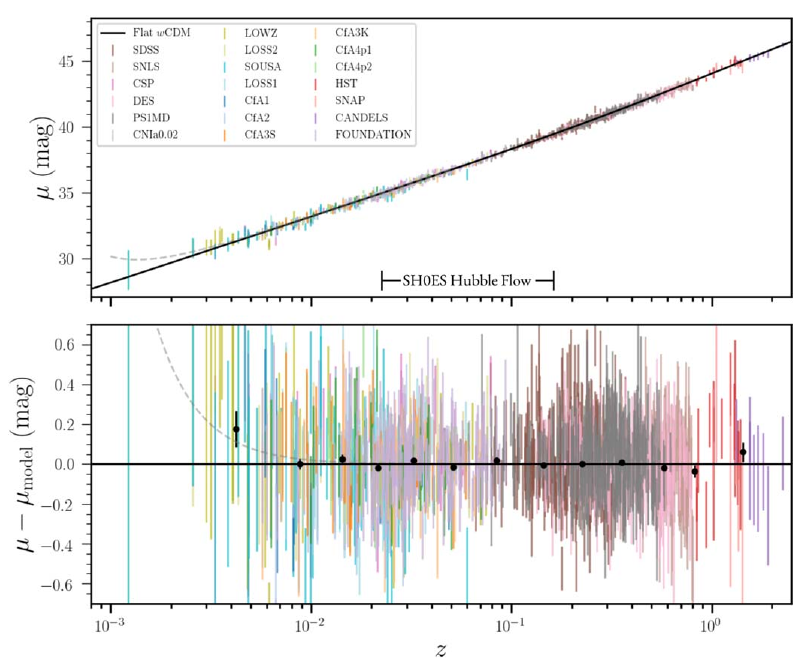}
    \caption{Hubble diagram $\mu(z)$ for the Pantheon+ catalogue of $1701$ Type Ia supernovae. The colours refer to the 18 different surveys that make up the catalogue. The black curve is what would be obtained in a flat FLRW universe, with the Pantheon+/SH0ES best-fit flat $w$CDM cosmological parameters. From Ref. \cite{brout2022pantheon+}.}
    \label{fig_Pantheon+_hubble_diagram}
\end{figure}

In a $\Lambda$CDM cosmology, $q_0 = \dfrac{1}{2}\Omega_{m0} - \Omega_{\Lambda 0}\,$, which under the assumption of flatness becomes $q_0 = -\dfrac{3}{2}\left(\Omega_{\Lambda 0}-\dfrac{1}{3}\right)\,$.
Therefore, a negative deceleration parameter, corresponding to cosmic acceleration, means that $\Omega_{\Lambda 0} > \dfrac{1}{2}\Omega_{m0}\,$, and thus constitutes direct evidence for dark energy, if one interprets the observed Hubble diagram in terms of an exactly homogeneous and isotropic cosmology. 
It is precisely this measurement, interpreted under that assumption, that led in the late 1990s to the acceptance of the cosmological constant into the standard model \cite{Riess_1998}. 
In that flat $\Lambda$CDM model, the full curve $d_L(z)$ can be calculated exactly:
\begin{equation}
    d_L(z) = \frac{\left(1+z\right)}{H_0}\int_0^z \frac{\mathrm{d}\tilde{z}}{\sqrt{\Omega_{\Lambda 0} + \Omega_{m0}\left(1+\tilde{z}\right)^3 + \Omega_{r0}\left(1+\tilde{z}\right)^4 }}\,,
\end{equation}
where we have included the radiation contribution for the sake of completeness, but it is negligible for the redshift ranges, $0 < z \lesssim 3$\,, over which the Hubble diagram is constructed.
The most complete Hubble diagram to date is shown in Fig. \ref{fig_Pantheon+_hubble_diagram}. 
In the lower panel, the observed distance modulus $\mu$ is compared to $\mu_{\rm model}$, the distance modulus that one would find in an exact FLRW universe with the best-fit set of cosmological parameters inferred from the survey\footnote{In the figure, the baseline cosmological model that has been adopted is flat $w$CDM, rather than flat $\Lambda$CDM. This refers to the equation of state $w$ of dark energy, which is assumed to be a constant throughout cosmic history but which is not assumed to be equal to $-1$, as it would be for $\Lambda$. The best-fit equation of state is found to be $w = -0.90 \pm 0.14$\,.}.

In reality, there are many observational challenges that must be met, in order to construct the Hubble diagram over a sufficient range of redshifts. The most complete Hubble diagram to date is obtained using the cosmic distance ladder. 
The ultimate aim is to make use of the intrinsic luminosity of Type Ia supernovae. This luminosity is set very roughly by the Chandrasekhar limit \cite{chandrasekhar1931maximum}, which puts the maximum mass of a white dwarf (WD) at around $1.4 M_{\odot}$ (the exact value of $M_{\rm Chandrasekhar}$ is sensitive to the chemical composition of the WD).
A Type Ia supernova is believed to occur when a WD, accreting mass from a binary companion, grows beyond the Chandrasekhar mass and thus explodes as it can no longer be supported by electron degeneracy pressure. 
The absolute magnitude of such an event therefore does not change much across different Type Ia supernovae \cite{nomoto1997type}. Hence, although it is not quite correct to call SNEIa standard candles \cite{branch1992type}, they are standardisable, as the details of their B-band light curves (the observed flux in the B colour band as a function of time) allow the absolute magnitude, or equivalently the intrinsic luminosity, of the events to be estimated \cite{tripp1998two,wood2008type}.
An important property of these observations, that will be relevant in Chapter 8, is that the absolute B-band magnitude $M_B$ of SNEIa cannot be accurately determined independently of the cosmological parameters. Instead, it is a nuisance parameter, that is fit to the data $m_B$ under the assumption of an FLRW cosmology:
\begin{equation}\label{eq_SNE1a_nuisance_parameters}
    \mu = 5\log{\left(\frac{d_L}{\rm Mpc}\right)} + 25 = m_B - M_B + \  {\rm other \: nuisance \: parameters}\,.
\end{equation}
Hence, an overall shift in magnitudes is fundamentally unobservable. We can only measure magnitude differences, and so it is not inconceivable that the distance moduli of Type Ia supernovae contain a theoretical systematic from incorrectly applying an FLRW cosmological model, and thereby obtaining an incorrect estimate for $M_B$ or the other nuisance parameters (whose details \cite{tripp1998two} we will not concern ourselves with here).

Type Ia supernovae are used to construct the Hubble diagram on redshift scales $z \gtrsim 0.05$ at which matter might be thought to be comoving with the Hubble flow (the fictitious worldlines orthogonal to the homogeneous hypersurfaces of an FLRW metric). In order to obtain accurate distance estimates on those scales, it is necessary to calibrate large-scale observations using smaller-scale ones. 
This process of calibration is the cosmic distance ladder. The most important step in the standard SH0ES calibration makes use of Cepheid variable stars \cite{altavilla2004cepheid}, although the tip of the red giant branch can also be used as an independent calibrator, with slightly different implications for $H_0\,$ estimation \cite{freedman2019carnegie,freedman2020calibration,freedman2021measurements}.

The cosmic distance ladder, used to construct the Hubble diagram, is the basis of late-time estimates of $H_0$, most recently the SH0ES collaboration's constraint $H_0 = 73.04 \pm 1.04\,{\rm km}\,{\rm s}^{-1}\,{\rm Mpc}^{-1}$\,\cite{riess2022comprehensive}. This is in tension, at around $5\sigma$, with the Planck $\Lambda$CDM estimate of $67.37 \pm 0.54 \,{\rm km}\,{\rm s}^{-1}\,{\rm Mpc}^{-1}$ \cite{aghanim2020planck}.
It is sometimes said that unlike the Planck constraint, which, being obtained from the sound horizon at last scattering, is intrinsically dependent on the assumption of a $\Lambda$CDM cosmology, the SH0ES estimate is model-independent.
This is probably a slight oversimplification, because the calibration of the cosmic distance ladder requires cosmological parameters to be estimated concurrently with the nuisance parameters associated with SNEIa, Cepheids and so on. 
However, it remains the consensus that at present there is a substantial Hubble tension in the $\Lambda$CDM concordance model, between indirect early-time and quasi-direct late-time estimates of $H_0\,$, although some authors have questioned whether the extent of the discrepancy truly constitutes a tension \cite{rameez2021there,freedman2021measurements}.
As yet there is no clear indication of any observational systematics in either the CMB or cosmic distance ladder measurements that could reconcile the tension. 
This is considered by many to be the greatest crisis in the field of cosmology at present, with no compelling theoretical explanation appearing to resolve it \cite{di2021realm}. It may be that the $H_0$ tension, which we will explore further in the next chapter, points to a fundamental failing in our understanding of physics in the Universe.

\chapter{Alternatives to the concordance cosmology}

\lhead{\emph{Alternative models}} 

Having introduced the $\Lambda$CDM concordance model, and the standard techniques used to study it, we will now explain why the model may be incomplete. We will review some underlying theoretical issues in the model, as well as some observational problems that it faces at present.

We will then explore two ways that one might generically alter the standard model. The first is by changing the laws of gravity itself, so that General Relativity might be replaced by some modified theory of gravity. This requires the development of generalised frameworks to test gravity. We will review the parameterised post-Newtonian (PPN) formalism that is used to constrain deviations from GR on astrophysical scales. We will also discuss some approaches that have been used on cosmological scales.

The second way of breaking the standard model that we will study is removing the assumption of homogeneity and isotropy that underpins the FLRW paradigm. We will introduce some anisotropic and inhomogeneous cosmological models that have been proposed as alternatives to FLRW. Finally, we will discuss the fundamental problem of averaging in curved spacetime, that it is necessary to understand if one wishes to build an homogeneous description of the Universe on large scales explicitly. This will lead us to the concept of cosmological backreaction.

\section{Problems with the standard model}\label{subsec:LCDM_issues}

Despite the many successes of the $\Lambda$CDM cosmological model, there are several reasons to think it may not be a complete description of our Universe. These can be split into two broad camps: theoretical and observational problems. Let us outline them in turn.

\subsection{Theoretical issues}\label{subsec:theoretical_issues}

The cosmological constant problem is considered by many to be the most fundamental in cosmology. This refers to the phenomenally small value of Einstein's cosmological constant $\Lambda$, when compared to a na{\" i}ve expectation from quantum field theory (QFT). 
Suppose that, in accordance with QFT principles, one interprets $\Lambda$ in terms of the zero-point energy density associated with the vacuum. Then one should sum all the way up to a cut-off scale where the quantisation of spacetime becomes significant: the Planck scale.
It would then follow that $\rho_{\Lambda} \sim M_{\rm Planck}^4\,$. However, the observed value of $\Lambda$, although no longer widely believed to be zero as it was previously, is around $120$ orders of magnitude smaller than $M_{\rm Planck}^4$.
This problem exists independently of the more recently discovered problem of explaining the Universe's accelerated expansion \cite{weinberg1972gravitation,carroll1992cosmological}. It can thus be thought of as the old cosmological constant problem.  

The new cosmological constant problem, then, is the need to account for the accelerated expansion of the Universe, that is inferred from the Hubble diagram of Type Ia supernovae. Although Einstein's cosmological constant evidently provides a phenomenological explanation for the acceleration, the enormous discrepancy between $\rho_{\Lambda}$ and $M_{\rm Planck}^4$ means that thinking of $\Lambda$ on such simple terms might be problematic. 
Instead, it is really safer to refer to the $\Lambda$ component of $\Lambda$CDM just as dark energy (DE). That is, it is some degree of freedom that appears to have phenomenologically indistinguishable properties from $\Lambda$ at the level of Einstein's equations, but whose underlying physical nature is unknown \cite{copeland2006dynamics}.

An enormous number of models of dark energy have been proposed to explain the accelerated expansion, in order to circumvent the problems associated with identifying the accelerating component with  $\Lambda$\,. 
Of course, these are only solutions to the new cosmological constant problem. A solution to the full problem would require explaining not only why these new fields cause the observed acceleration, but also how they explain the old problem, of why the associated energy density is so extraordinarily small compared to the QFT cut-off. 
Alternative models to $\Lambda$ often attribute the accelerated expansion to new fundamental fields in the Universe (particularly novel scalars; see e.g. \cite{frusciante2020effective,padmanabhan2002accelerated,Gubitosi_2013}), or deviations from GR in the laws of gravity \cite{Clifton_2012,Joyce_2016,Battye_2012,Bloomfield_2013}. Some also suggest that it is seen due to the effects of inhomogeneities \cite{Buchert_2007,Adamek_2018,Buchert_2012,Rasanen_2004,Li_2008,Brown_2009}, as we will discuss later in this chapter.

According to the best Planck constraints \cite{aghanim2020planck}, not only does around $69\%$ of the Universe's present-day energy budget exist in the form of some unknown dark energy, but most of the remaining $31\%$ (around $26\%$ of the total) is in the form of cold dark matter (DM\footnote{Or CDM for cold dark matter}), which is also unaccounted for in the standard model of particle physics. Whereas the phenomenological evidence for dark energy is purely cosmological, dark matter has ample astrophysical evidence. It was originally proposed to explain the anomalous rotation curves of galaxies \cite{zwicky1979masses,rubin1980rotational}, and since then has also found astrophysical support from the X-ray emission properties of galaxy clusters \cite{dietrich2012filament} and strong gravitational lensing \cite{massey2010dark}.

Like DE, the fundamental nature of DM is the subject of a great deal of research. A plethora of DM models have been proposed (see e.g. \cite{arun2017dark} for a review). These include, but are not limited to, weakly interacting massive particles (WIMPS) \cite{roszkowski2018wimp,schumann2019direct}, axions or axion-like particles \cite{duffy2009axions,niemeyer2020small,chadha2022axion}, wave/fuzzy/ultra-light dark matter \cite{hu2000fuzzy,lee2018brief,hui2021wave}, primordial black holes \cite{carr2016primordial,carr2020primordial,green2021primordial}, and Kaluza-Klein particles \cite{cheng2002kaluza}.

It has been suggested in some quarters that the apparent requirement for dark matter to exist on astrophysical scales could be explained by a modification to gravity in the ultra-low-acceleration Newtonian regime \cite{bekenstein1984does}, a fundamentally phenomenological and non-covariant proposal known as modified Newtonian dynamics (MOND) \cite{famaey2012modified}. If correct, MOND could do away entirely with the need for dark matter in galaxies \cite{sanders2002modified}. 
General-covariant relativistic theories that incorporate MOND have been developed \cite{bekenstein2004relativistic,zlosnik2006vector,zlosnik2007modifying}, although they are typically very complicated. 
In order to constitute valid theories of gravity and dark matter they must also account for CDM-like behaviour in large-scale structure and the cosmic microwave background.
Some relativistic completions of MOND indeed appear to be ruled out by cosmological datasets \cite{zuntz2010vector}, although novel theories of this kind have since been developed that may be compatible with both astrophysical and cosmological observations \cite{skordis2021new,thomas2023consistent}. 

Another fundamental theoretical issue that is not addressed in the cosmological standard model is the ultimate need for an ultraviolet (UV) completion of General Relativity - in other words, a quantum theory of gravity. 
In most contexts, the UV incompleteness of GR is rendered irrelevant by the characteristic energy scale being vastly smaller than $M_{\rm Planck}\,$. However, it is apparent that a classical description of gravity must break down in the Planck epoch $t \lesssim 10^{-43}\,{\rm s}$, if not before. 
In order to probe physics at that very early time, it is therefore necessary to develop candidates theories of quantum gravity. Although we will not discuss the details of any such theories here, it is worth noting some general features that they appear to display at lower energies (i.e. in the IR), that may be relevant in cosmology or astrophysics.
\begin{enumerate}

    \item There are typically higher-order effective field theory contributions from curvature in the gravitational action, e.g. $R^2$, $R_{abcd}R^{abcd}\,$. These are suppressed by inverse powers of $M_{\rm Planck}\,$, so they are typically irrelevant at low curvature. 
    They might be important in spacetime regions of high curvature, for instance in the immediate vicinity of black holes, or in the early Universe; in fact, a strong candidate inflationary model, Starobinsky inflation, is driven by an $R^2$ action \cite{starobinsky1980new, mukhanov1981quantum}.
    
    \item UV completions of gravity generically give rise to novel degrees of freedom in the IR, particularly scalar fields. Scalars are interesting from the perspective of both late and early time cosmology, because in the early Universe they can be studied as candidates for the inflaton field, and in the late Universe they may provide compelling models of dark energy that could replace $\Lambda\,$ \cite{copeland2006dynamics}. 

    \item Many quantum gravity frameworks, most famously string theory, feature extra spacetime dimensions beyond the four of GR. These extra dimensions are mostly bundled into very small scales below the Planck length $\sim 10^{-35}\,{\rm m}\,$. However, these models can manifest themselves in the IR through 5-dimensional gravity theories, in which our 4-dimensional brane is considered a hypersurface embedded in a 5-dimensional bulk. 
    Thus, there would be corrections to 4-dimensional GR, through the gravitational effects of the bulk on our brane, that may be cosmologically observable. Theories of this kind have attracted considerable interest, particularly the Dvali-Gabadadze-Porrati (DGP) \cite{dvali2000metastable}, Kaluza-Klein \cite{overduin1997kaluza} and Randall-Sundrum \cite{randall1999out} models.

\end{enumerate}
These considerations have led many cosmologists to study classical theories of gravity that incorporate higher-order curvature terms, novel degrees of freedom (especially scalars) and higher spacetime dimensions, as testing grounds for some of the ideas that may be important in the UV completion of gravity. We will discuss some such theories in more detail later in this chapter.

Finally, let us consider the Copernican and cosmological principles. Even putting aside the considerable challenge of verifying or falsifying them observationally, from a theoretical perspective it is not clear that an FLRW geometry should provide an accurate description of our Universe. 
After all, we know that we live in a complex web of structure. Hence, the real Universe has no Killing vectors, not six. If we are to draw strong conclusions about our Universe from FLRW solutions to Einstein's equations, it is important to show that these conclusions are not overly dependent on idealised symmetry properties, so that we can trust the cosmological ``fitting'' model we construct \cite{Ellis_fitting_1987}.

Some crucial insights were made in this regard by Hawking, whose used Penrose's singularity theorems in order to show that an early-time singularity is generic in relativistic cosmology, under certain assumptions \cite{hawking1970singularities}. 
For the late Universe, though, the situation is much more complicated, and there may be considerable insight to be gained from studying spacetime models with a lower degree of symmetry than FLRW. 
The problem of interpreting cosmological observations in terms of an essentially fictional homogeneous and isotropic ``background'' spacetime is made more severe by the nonlinearity of General Relativity. A corollary of that nonlinearity is that there is no clean hierarchy of scales in cosmology: we should not automatically expect that gravitational physics on large scales is not affected by gravitational physics on small scales \cite{korzynski2015nonlinear}. 
Thus, some cosmologists have conjectured that the evolution of the Universe on large scales might be substantially affected by a phenomenon known as backreaction (see Ref. \cite{Buchert_2012} for an overview). We will introduce this concept in more detail at the end of the chapter.

\subsection{Observational issues}\label{subsec:observational_issues}

Aside from the theoretical concerns that suggest that the $\Lambda$CDM concordance model is incomplete, and merely a placeholder for a more fundamental description, there are several observational tensions and anomalies that may point to the model breaking down. 
These have mostly emerged over the last decade, as measurements of the cosmic microwave background, large-scale structure and cosmic distance ladder sources have become sufficiently precise to reveal statistically significant inconsistencies between datasets.
For the purposes of the discussion here, we will divide these observational problems into two categories:
\begin{enumerate}
    \item Tensions within the $\Lambda$CDM model itself.
    \item Anomalous signals of anisotropy and inhomogeneity, that may be contradictory to the FLRW paradigm. 
\end{enumerate}
Note that we have drawn a distinction between {\it tensions} and {\it anomalies}. By tensions, we refer to contradictory inferences of the same parameter, separated by several standard deviations, between two different types of observation. Those inferences are made within the assumption of a certain cosmological model (typically spatially flat $\Lambda$CDM). 
By anomalies, we mean observed phenomena that are different from expected in the standard cosmology, but which are (a) less statistically significant than a tension and/or (b) direct observations (e.g. of a very large correlated structure) rather than parameter values inferred using a specific model.

Let us first deal with the $\Lambda$CDM tensions. The most famous is the Hubble tension \cite{di2021realm}, in inferred values of $H_0$.
As discussed in the previous chapter, the Hubble constant is one of the six fitting parameters $\left\lbrace H_0, \omega_c, \omega_b, A_s, n_s, \tau_{\rm reio} \right\rbrace$ for the CMB anisotropies, in the flat $\Lambda$CDM model. Thus, although it cannot be directly measured from early-time observables, $H_0$ can be inferred from the CMB, by assuming the concordance model and adopting standard Bayesian inference techniques. This yields a best-fit $H_0$ value from the Planck data of $67.4 \pm 0.5\,\,{\rm km}\,{\rm s}^{-1}\,{\rm Mpc}^{-1}\,$. Another early-time, indirect, measurement of $H_0$ can be obtained using BAO data with a BBN prior \cite{cuceu2019baryon}. This gives a similar value, indicating a consistent, low, value of $H_0$ from early-time probes.

\begin{figure}
    \centering
    \includegraphics[width=0.8\linewidth]{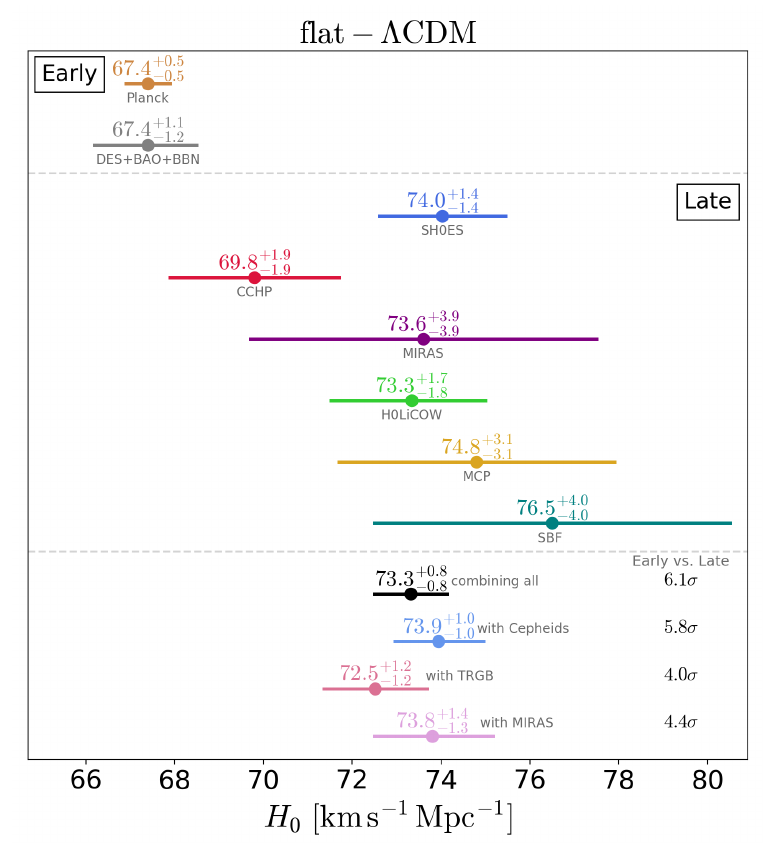}
    \caption{Summary of the tension between early and late-time measurements of $H_0$, inferred assuming a flat $\Lambda$CDM cosmology. From Ref. \cite{verde2019tensions}, courtsey of Vivien Bonvin.}
    \label{fig_hubble_tension}
\end{figure}

In contrast, direct, late-time measurements of $H_0$, obtained by constructing a Hubble diagram from Type Ia supernovae and fitting it to $d_L = \dfrac{z}{H_0} + ...$ at low redshift, lead to a value of $H_0$ around $73 \pm 1\,{\rm km}\,{\rm s}^{-1}\,{\rm Mpc}^{-1}\,$. The exact value and error depends somewhat on the calibration method used for the absolute SNEIa magnitude and nuisance parameters. 
In particular, calibrating the supernovae using Cepheids yields a higher value of $H_0$, that is over $5\sigma$ in tension with Planck \cite{riess2022comprehensive}, whereas using the tip of the Red Giant branch gives a lower value that is only around $4\sigma$ in tension \cite{freedman2021measurements}, and measurements of this kind from James Webb Space Telescope (JWST) data might conceivably give rise in the near future to a Hubble diagram $H_0$ that is consistent with the CMB \cite{freedman2023progress}.
However, other late time observables also give high $H_0$ values (although with rather larger confidence intervals) \cite{birrer2019h0licow,pesce2020megamaser,schombert2020using}. 
Thus it appears, at this stage at least, that late-time measurements point to a consistent picture for $H_0$, at high values around $73 \,{\rm km}\,{\rm s}^{-1}\,{\rm Mpc}^{-1}\,$.
A summary of the tension, grouped into early-time and late-time inferences, is displayed in Fig. \ref{fig_hubble_tension}. 

As yet, no consensus has been reached as to how the $H_0$ tension might be resolved. Many proposed solutions that modify physics at late times appear in fact to make the tension worse \cite{di2021realm}. Some early-time mechanisms have been suggested as strong candidates to resolve the tension, most notably early dark energy \cite{poulin2019early}, although this is hotly debated \cite{kamionkowski2023hubble,hill2020early}. However, the fundamental physical motivation for these mechanisms is often unclear. 
Of course, it may be that $H_0$ itself should not be regarded as a single parameter, but rather a function of observing direction on the sky \cite{Heinesen_2021,Heinesen_2022,macpherson2021luminosity,Macpherson:2022eve,kalbouneh2024cosmography,maartens2024covariant}, if we live in a Universe that cannot appropriately be described by an FLRW model at the percent level of precision that is relevant for present and upcoming observations.

Another notable tension in the concordance cosmology has emerged over the last few years in the inferred value of $\sigma_8$, the clustering amplitude $\sqrt{\left\langle\delta_k \delta_k\right\rangle}$ of matter on scales $k^{-1} = 8 h^{-1}\,{\rm Mpc}\,$. This is also sometimes expressed in terms of $S_8 = \sigma_8 \sqrt{\dfrac{\Omega_{m0}}{0.3}}$\,. 
Like the $H_0$ tension, the $\sigma_8$ tension arises between late-time measurements, in this case using weak lensing \cite{asgari2021kids,amon2022dark}, and indirect inferences that are obtained by forward-modelling from cosmic microwave background constraints \cite{efstathiou2019detailed}.

\begin{figure}
    \centering
    \includegraphics[width=0.6\linewidth]{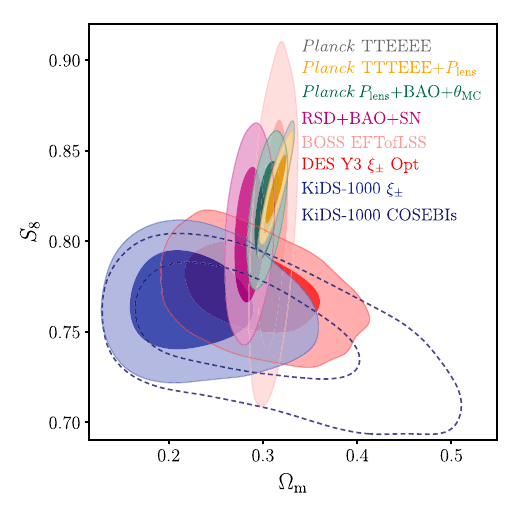}
    \caption{Summary of the tension between early and late-time measurements of $S_8$\,, inferred assuming a flat $\Lambda$CDM cosmology. From Ref. \cite{amon2022non}.}
    \label{fig_s8_tension}
\end{figure}

The tension is weaker than in $H_0$, at around $2.5\, \sigma\,$. Specifically, the best-fit values for $S_8$ from cosmic shear are $0.759^{+0.024}_{-0.021}$ from KiDS-1000 \cite{asgari2021kids} and $0.772^{+0.018}_{-0.017}$ from DES \cite{amon2022dark}, whereas the Planck $TT$, $TE$ and $EE$ data predict $S_8 = 0.828 \pm 0.016$ \cite{efstathiou2019detailed}.
However, the relative paucity of observational systematics in weak lensing \cite{amon2022dark} means that the clustering amplitude discrepancy is considered by many cosmologists to be likely to evolve into a statistically significant, robust tension. Moreover, it is often the case that models that alleviate the $H_0$ tension worsen the $\sigma_8$ tension, and vice versa \cite{hill2020early,secco2023,vagnozzi2023seven,poulin2018implications}. 
This potentially spells further trouble for the entire paradigm itself: it is not easy to solve the observational problems with simple extensions of the concordance model. If observational systematics are not to save $\Lambda$CDM, then a more drastic alteration to our understanding may be required.

Let us now move on to a rather different set of observational problems, that are not concerned with tensions within the standard model. Instead, they are a collection of anomalies that are difficult to reconcile with the cosmological principle. 
Although many of these have individually low statistical significance, and may yet all be explained by systematics, taken together they give reason at least to consider alternatives to the FLRW geometry.

The most prominent such anomaly is found in measurements of the cosmic dipole, to the extent that some cosmologists have argued that it ought to be considered a tension on the same level as $H_0$ and $\sigma_8\,$ \cite{peebles2022anomalies,Secrest_2021,Secrest_2022}, although others have pointed out theoretical systematics that might plague the interpretation of the dipole anomaly as a genuine tension \cite{dalang2022kinematic,guandalin2023theoretical}. 
We will not argue for either position here. Instead, we will just provide a brief description of the current state of play. 

\begin{figure}
    \centering
    \includegraphics[width=0.8\linewidth]{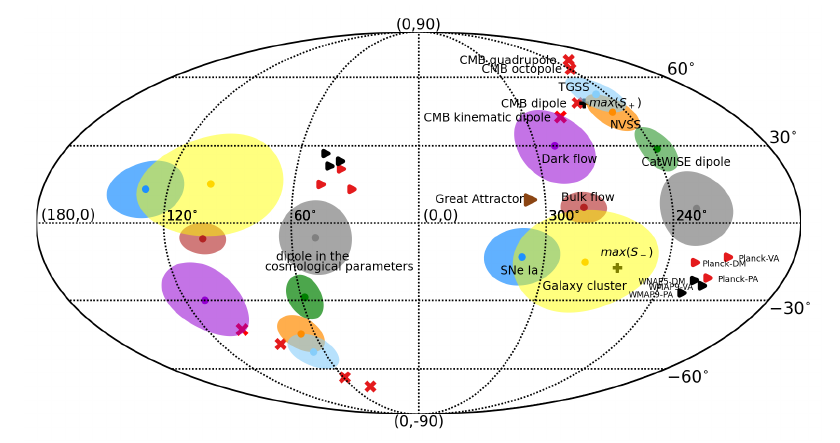}
    \caption{Visualisation of various dipoles, bulk flows and other anisotropic anomalies, as a Mollweide projection in galactic coordinates. From Ref. \cite{Aluri_2023}.}
    \label{fig_dipole_tension}
\end{figure}

The dipole anomaly refers to a discrepancy in the inferred magnitudes of the Sun's kinematic 3-velocity with respect to the supposed cosmic rest frame. This rest frame refers to the supposed comoving 4-velocity $u^a$ which all matter and radiation fields should have on large scales, in an FLRW geometry, as per the discussion in the previous chapter. 
It follows from the FLRW assumption that the dipole that is seen in any large-scale observable, whether that is the CMB temperature\footnote{Apart from the expected dipole $\mathcal{O}(10^{-5})$ from fluctuations in the early Universe.}, the galaxy distribution, or the value of cosmological parameters such as $H_0$ or $q_0$ that are inferred from Hubble diagram observations, should be entirely attributed to a Doppler effect. That Doppler effect arises due to the non-relativistic 3-velocity $v^a$ of the Solar System with respect to $u^a$\, (i.e. $u^a_{\rm SS} = \gamma\left(u^a + v^a\right) \approx u^a + v^a\,$).
All observations, then, should produce the same magnitude and direction of $v^a\,$. It turns out that they do roughly agree on the direction of the dipole, as shown in Fig. \ref{fig_dipole_tension}.
The discrepancy arises between the value obtained from the CMB dipole $\mathcal{C}_1^{TT}$, and the value obtained from the observed dipole in the number counts of astrophysical sources (primarily quasars, radio galaxies and SNEIa). Observations of the cosmic microwave background \cite{Planck_2020} suggest that 
\begin{equation*}
    v =  369.82 \pm 0.11 \, {\rm km}\,{\rm s}^{-1}\,, \quad {\rm in \ the \ direction} \quad \left(l, b\right) = \left(264^{\circ}, 48^{\circ}\right)\,,
\end{equation*}
with respect to the CMB rest frame.

For number counts, the dipole that one would expect if its origin were purely due to our kinematic motion was derived in a seminal paper by Ellis \& Baldwin \cite{Ellis_1984}. They showed that for a survey of sources (which are assumed to be distributed isotropically in their rest frame) with a spectral index $\alpha$ and a magnification bias $x$, the kinematic dipole in the observed number count would be
\begin{equation}
    \hspace{-1cm} \mathcal{D}_{\rm kinematic} = \left[2 + x\left(1+\alpha\right)\right]\, v\,,
\end{equation}
pointing in the direction of $v^a\,$. Here, we have defined $\mathcal{D}_{\rm kinematic}$ as the deviation in the observed number count $N$ per unit solid angle (integrated over all source comoving distances), compared to what would be expected in the source rest frame, i.e.
\begin{equation*}
    \mathcal{D}_{\rm kinematic} \, \cos{\theta} = \frac{\left(\frac{\mathrm{d}N}{\mathrm{d}\Omega}\right)_{\rm obs}-\left(\frac{\mathrm{d}N}{\mathrm{d}\Omega}\right)_{\rm source}}{\left(\frac{\mathrm{d}N}{\mathrm{d}\Omega}\right)_{\rm source}}\,,
\end{equation*}
where $\theta$ is the angle between the observing direction $\hat{e}$ and the observer's direction of motion $\hat{v}$ with respect to the source frame, such that $\cos{\theta} = \hat{e}\mathbf{\cdot}\hat{v}\,$ \cite{Ellis_1984,dalang2022kinematic}.

Hence, a sufficiently precise measurement of the dipole in astrophysical number counts (assuming the sources are distributed in an homogeneous and isotropic fashion in the Universe) provides a way of measuring our kinematic motion, with respect to the average rest frame of matter fields.
If the cosmological principle holds, then both the CMB and number count dipoles should correspond entirely to the same kinematic peculiar velocity, with respect to the same rest frame, the canonical congruence of the FLRW geometry. That is, they must agree on both the magnitude and direction of $v^a\,$. 
Most estimates to date of the matter dipole, particularly quasars \cite{Secrest_2021}, report roughly the same direction as the CMB. However, the inferred magnitude of the dipole is anomalously high. Applying the Ellis-Baldwin formula would suggest that $\mathcal{D} \approx 0.007\,$. However, Secrest et al. \cite{Secrest_2021}, using the CATWISE catalogue of around $1.36$ million quasars, find
$\mathcal{D} \approx  0.01554\,$, i.e. over twice as large as predicted from the CMB. With their reported confidence intervals, this constitutes a tension at around $4.9\,\sigma$ compared to Planck.

The anomalous dipole measurement is supported by some other anomalies that indicate anisotropy in the Universe pointing in the same direction as the CMB and quasar dipoles \cite{Aluri_2023}, as per Fig. \ref{fig_dipole_tension}, but with a larger magnitude.
Galaxy clustering observations have been claimed to indicate a dipole that is around $5\sigma$ above the Ellis-Baldwin expectation \cite{Migkas_2021}, and there also appears to be a larger-than-expected dipole in the distribution of SNEIa sources \cite{antoniou2010searching,sorrenti2023dipole}, as well as possible directional dependence of cosmological parameters \cite{Yeung_2022}.
Taken together, these measurements may give reason to believe that interpreting the cosmic dipole purely in terms of our kinematic motion is erroneous. Instead, one could conclude that there is an intrinsic dipole arising from large-scale anisotropy in the Universe itself. Then, there is no need for the magnitudes of the matter and CMB dipoles to be consistent.

Finally, let us briefly mention some curious signals that are found in cosmological datasets, and which are not easily explained in an homogeneous and isotropic Universe with $\Lambda$CDM energy-momentum content.
There are several anomalous features of the cosmic microwave background, especially at low multipoles $l$ \cite{schwarz2016cmb,muir2018covariance}\footnote{Note that issues of cosmic variance \cite{scott2016information} and galactic masking \cite{rassat2014planck} make the error bars on $\mathcal{D}_l^{TT}$ at low $l$ larger than they are on small angular scales, so conclusions made using low-$l$ phenomena should be taken with a pinch of salt.}.
These include the parity asymmetry \cite{land2005parity,Ben-David_2012}, referring to an apparent excess of power in odd multipoles compared to even ones, the hemispherical power asymmetry \cite{Hansen_2004,Hansen_2009}, the apparent aligment of the CMB quadrupole and octupole \cite{de2004significance}, and the overall suppression of power on large angular scales compared to the $\Lambda$CDM expectation \cite{Efstathiou_2010,schwarz2016cmb}.

In the late Universe, there are some notable observations that are anomalous with respect to the FLRW paradigm. As discussed in the previous chapter, it is assumed that above some homogeneity scale $L_{\rm hom}$ there ought to be no inhomogeneous structures.
Observations of coherent bulk flows of matter on scales of hundreds of ${\rm Mpc}$ \cite{Kashlinsky_2008, Magoulas_2014, hoffman2015cosmic}, and even larger ultra-large correlated structures \cite{Sloan_Great_Wall,horvath2020clustering,clowes2013structure,balazs2015giant} of claimed sizes up to $\sim 2000-3000\,{\rm Mpc}$\,, place doubt on the existence of an homogeneity scale that is much smaller than the present day Hubble-horizon at $\sim 3000\,h^{-1}\,{\rm Mpc}\,$. 
These claimed observations are controversial, and we do not wish to comment on their veracity. However, as cosmological LSS surveys become ever more precise \cite{nesseris2022euclid,maartens2015overview}, it may be that at least some of these anomalies are supported to high enough significance to constitute a genuine cosmological tension, like the $H_0$ tension in the $\Lambda$CDM model. 
Then, an homogeneous and isotropic model for our Universe on large-scales may no longer suffice, and it will be necessary to use anisotropic and/or inhomogeneous models to accurately predict observables.

Now that we have discussed a variety of theoretical and observational problems that indicate that the concordance model may be an incomplete description of our Universe, let us now introduce some key approaches that have been taken to replacing it. This will be the focus of the remainder of the present chapter.
The three central tenets of the concordance model are (a) General Relativity as the theory of gravity (b) the FLRW metric (plus small perturbations) as the geometrical model and (c) $\Lambda$CDM energy-momentum content that sources the evolution of the FLRW spacetime according to Einstein's equations.
Therefore, we will divide the alternative models into two generic classes: alternatives to General Relativity, and alternatives to the homogeneous and isotropic cosmology. There is of course also a third class, where the GR and FLRW tenets are retained but novel fields are introduced into $T_{ab}\,$. We will not focus on these in this thesis, although certain models we will come across in our discussion will essentially fall into the third camp.

\section{Alternatives to General Relativity}

Theories of gravity that modify or extend Einstein's theory of General Relativity have existed almost as long as GR itself, such as the Kaluza-Klein theory first introduced in the 1920s \cite{overduin1997kaluza}. In the field of cosmology, they originally attracted interest in the 1950s and 1960s \cite{brans1961mach}, due to the idea, first proposed by Dirac, that the value of Newton's constant could have evolved through cosmic time \cite{dirac1974cosmological}. 
Later, considerations of UV completions of gravity, as well as simply the availability of astrophysical tests of gravity, led to a plethora of modified theories of gravity being proposed, as well as generalised frameworks that could be used to compare their predictions.
Most recently, the observed accelerated expansion of the Universe has led cosmologists to seek out gravitational theories that could account for that acceleration without recourse to a mysterious dark energy field such as $\Lambda\,$.

\subsection{Modified theories of gravity}

\begin{figure}
   \centering
    \includegraphics[width=0.8\linewidth]{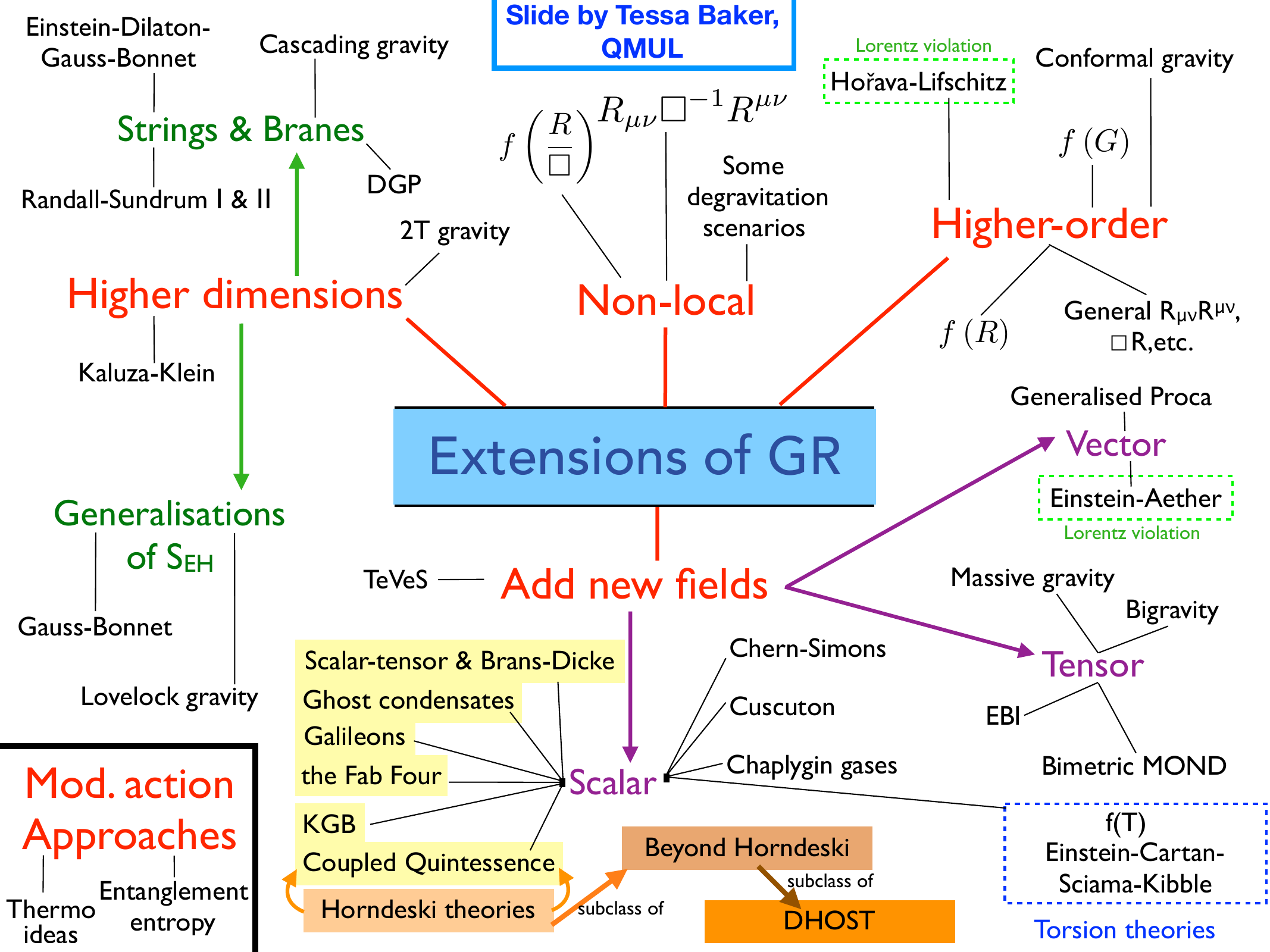}
    \caption{Visualisation of the landscape of modified theories of gravity, organised by the ways in which they break the assumptions of Lovelock's theorem. Courtesy of Tessa Baker.}
    \label{fig_MG_landscape_baker}
\end{figure}

In this section, we will discuss some specific modified theories of gravity of particular interest. The landscape of modified gravity (MG) theories is vast, with an enormous number of different theories having been proposed, as visualised in Fig. \ref{fig_MG_landscape_baker}.

In order to understand how one can build alternatives to GR, it is useful to return to Lovelock's theorem \cite{lovelock1969uniqueness,lovelock1971einstein,lovelock1972four}, which we first introduced in Chapter 2. It told us that GR is the unique metric theory of gravity one can construct in four spacetime dimensions that is local and second-order, and contains no novel gravitational degrees of freedom.
We will not consider non-metric theories, as they violate the WEP, so in order to obtain a distinct theory we must either introduce non-locality, more spacetime dimensions, new fields, or derivatives of $g_{ab}$ beyond second order.
For an overview of MG theories and their application to cosmology, we refer the reader to Ref. \cite{Clifton_2012}.

The oldest, and most well-studied, MG theories are scalar-tensor theories, that introduce a novel scalar field $\phi$ that is non-minimally coupled to the metric $g_{ab}\,$\footnote{There exist scalar field theories in which $\phi$ is minimally coupled, such as the quintessence theories \cite{zlatev1999quintessence}, that are sometimes also referred to as modified gravity theories. However, these do not affect the coupling strength of matter to the metric, so we will refer to such models as scalar field models of dark energy, not scalar-tensor theories of gravity.}. 
By the EEP, the scalar $\phi$ is not coupled to matter fields, but affects them indirectly, by sourcing the dynamics of the metric, whose geodesics determine the motion of particles.
The most general set of second-order theories of this kind is the Horndeski class \cite{horndeski1974second}.
The full Horndeski action is \cite{koyama2016cosmological}
\begin{eqnarray}\label{eq_Horndeski_action}
    S_{\rm Horndeski} &=& \int\mathrm{d}^4x\,\sqrt{-g}\,\left[K\left(\phi, X\right) + \mathcal{L}_3 + \mathcal{L}_4 + \mathcal{L}_5 \right]\,, \quad {\rm where} \\
    \nonumber \mathcal{L}_3 &=& - G_3\left(\phi, X\right)\Box{\phi}\,, \\
    \nonumber \mathcal{L}_4 &=& G_4\left(\phi, X\right)R + G_{4X}\left(\phi, X\right)\left[\left(\Box{\phi}\right)^2 - \nabla^a \nabla^b \phi\nabla_a \nabla_b \phi\right]\,, \quad {\rm and} \\
    \nonumber \mathcal{L}_5 &=& G_5\left(\phi, X\right) G^{ab}\nabla_a\nabla_b\phi \\
    \nonumber && - \frac{1}{6}G_{5X}\left(\phi, X\right)\left[\left(\Box{\phi}\right)^3 - 3\nabla^a\nabla^b\phi\,\nabla_a\nabla_b\phi \,\Box{\phi} + 2\nabla_a\nabla_b\phi\,\nabla_c\nabla^a\phi\, \nabla^c\nabla^b\phi\right]\,.
\end{eqnarray}
Here $X = -\frac{1}{2}\nabla_a \phi \nabla^a \phi$ is the canonical kinetic term associated with the scalar field, $\Box = \nabla^a \nabla_a$ is the covariant d'Alembertian operator, and $K$, $\mathcal{L}_3$, $\mathcal{L}_4$ and $\mathcal{L}_5$ are arbitrary functions of their arguments. The notation $F_{X}$ denotes $\dfrac{\mathrm{d}F}{\mathrm{d}X}\,$.
Note that we are working in the so-called Jordan frame\footnote{The terminology ``frame'' here is somewhat misleading, because it does not refer to a frame of reference in the standard relativistic sense. It means instead a choice of definition of the metric, where the possible field definitions are related by conformal transformations. Nonetheless, we will adapt the use of ``frame'' to refer to the two standard field definitions, in accordance with the standard practice in the literature.}, in which the matter action $S_{\rm matter} \equiv S_{\rm matter}\left(\psi, g_{ab}\right)\,$, for some collection of matter fields $\psi$; i.e. the matter action is independent of $\phi$. 
An alternative formulation of scalar-tensor theories is provided by the Einstein frame, in which the scalar field is minimally coupled to the Ricci scalar associated with the metric $\tilde{g}_{ab}\,$, but $S_{\rm matter} \equiv S_{\rm matter}\left(\psi, \tilde{g}_{ab}, \phi\right)\,$ is dependent on $\phi\,$. This means that particles do not follow geodesics of $\tilde{g}_{ab}\,$, which is related to the Jordan frame metric $g_{ab}$ through a conformal transformation.

The Horndeski action is very phenomenologically rich, containing several special cases of physical interest. Cosmologists are often interested in scalar-tensor theories that contain screening mechanisms \cite{Vainshtein_1972,Khoury_2004,Khoury_2010,brax2013screening}. 
These are phenomena whereby the nonlinear terms in $\mathcal{L}_3$ and $\mathcal{L}_4$ produce self-coupling of the scalar field that suppresses its effects in certain spacetime regions, such as regions of high energy density and/or spacetime curvature, or simply on length scales that are small compared to some characteristic Yukawa scale. 
The advantage of screened scalar-tensor theories is that on cosmological scales, the scalar field has non-trivial dynamics, and can affect the large-scale expansion (for example driving acceleration under certain conditions), but in the astrophysical regime, screening mechanisms kick in and force the theory essentially to revert to GR. This allows the theory to evade tight Solar System constraints on deviations from General Relativity.

Measurements of the propagation speed $c_T$ of gravitational radiation, from the gravitational wave event GW170817 and its electromagnetic counterpart GRB170817A \cite{abbott2017gw170817}, tightly constrain $c_T$ to be equal to unity to within 1 part in $10^{15}\,$. If one is interested in modifications to GR in the strong-field regime, then this does not particularly restrict the allowed space in the Horndeski class \cite{clifton2020observational}.
Suppose, however, that one wishes to consider scalar-tensor theories that are phenomenologically distinct from GR in late-time cosmology and in the weak-field regime. Then, the $c_T = 1$ requirement restricts the class substantially \cite{baker2017strong}, by removing $\mathcal{L}_5$, and requiring that the part of the action proportional to $G_{4X}$ must vanish, such that $\mathcal{L}_4 \longrightarrow G_4(\phi)\,R\,$.
The canonical scalar-tensor theory is then obtained by setting the self-interaction function $G_3$ to zero. Finally, one may redefine the field so that $G_4(\phi) = \dfrac{\phi}{16\pi G}\,$, and set the kinetic term $K(\phi, X) = -\dfrac{\omega(\phi)}{16\pi G}\dfrac{2X}{\phi}\,$, where $\omega(\phi)$ is some arbitrary coupling function of the scalar field alone\footnote{The factor of $\omega(\phi)/\phi$ in the kinetic term, rather than just using the canonical kinetic term $X$\,, is a result of us working in the Jordan frame. If we were to work in the Einstein frame, then the kinetic term would be canonical.}
This produces the well-known Bergmann-Wagoner scalar-tensor action \cite{bergmann1968comments,wagoner1970scalar}
\begin{equation}\label{eq_bergmann_wagoner_action}
    S_{\rm ST} = \frac{1}{16\pi G}\int\mathrm{d}^4x\,\sqrt{-g}\,\left[\phi R - \frac{\omega(\phi)}{\phi}\nabla_a \phi \nabla^a \phi 
 - 2\Lambda(\phi)\,\right]\,,
\end{equation}
where $\Lambda$ is like $\omega$ an arbitrary function of $\phi\,$. Throughout the remainder of this thesis, we will only consider scalar-tensor theories of the form $S_{\rm ST}\,$. 
Varying the full action, including $S_{\rm matter}$, with respect to the metric gives the equivalent of Einstein's equations,
\begin{equation}\label{eq_EOM_metric_scalar_tensor}
    G_{ab} + \left[\Box{\phi} + \frac{\omega}{2\phi}\nabla^c\phi\nabla_c\phi + \Lambda\right]\frac{g_{ab}}{\phi} - \frac{1}{\phi}\nabla_a\nabla_b \phi - \frac{\omega}{\phi^2}\nabla_a\phi\nabla_b\phi = \frac{8\pi G}{\phi}T_{ab}\,,
\end{equation}
Varying $S_{\rm ST}$ with respect to $\phi$ gives a Klein-Gordon equation for the scalar field, as $S_{\rm matter}$ is independent of $\phi$\,:
\begin{equation}\label{eq_EOM_KleinGordon_scalar_tensor}
    \Box{\phi} + \frac{1}{2\omega}\left(\frac{\mathrm{d}\omega}{\mathrm{d}\phi} - \frac{\omega}{\phi}\right)\nabla^a \phi \nabla_a \phi - \frac{\phi}{\omega}\frac{\mathrm{d}\Lambda}{\mathrm{d}\phi} + \frac{\phi}{2\omega}R = 0\,.
\end{equation}
We will explore cosmological solutions to these equations in detail in Chapter 5. In the limit $\omega(\phi) \longrightarrow \infty$ and $\Lambda(\phi) \longrightarrow \Lambda = {\rm cst.}\,$, they are simply Einstein's equations with a cosmological constant, as $\omega \longrightarrow \infty$ makes the Klein-Gordon equation (\ref{eq_EOM_KleinGordon_scalar_tensor}) trivial.

The simplest novel case is the Brans-Dicke theory \cite{brans1961mach}, which is defined by $\Lambda = 0$ and $\omega(\phi) \longrightarrow \omega = {\rm cst.}\,$.
The coupling parameter $\omega$ of the Brans-Dicke theory is very tightly constrained by Cassini observations  \cite{bertotti2003test} of the Shapiro time delay \cite{shapiro1964fourth} in the Solar System. These give $\omega \gtrsim 4\,\times 10^4$\, \cite{Clifton_2012}, essentially rendering the Brans-Dicke theory phenomenologically indistinct from GR.
However, as arguably the simplest MG theory, the Brans-Dicke theory retains considerable interest. It is an excellent testing ground for many of the conceptual challenges that cosmologists are interested in when studying deviations from General Relativity. 

Another type of gravity theory containing new degrees of freedom is the vector-tensor class of theories, where in addition to the metric one has a timelike vector field $A^a\,$. These theories are less well-explored than scalar-tensor theories, due primarily to their mathematical complexity, and also because novel vectors are arguably less well-motivated from EFT considerations than novel scalars. 
However, they introduce some very interesting physical effects. In particular, the field $A^a$ defines a preferred congruence in the Universe \cite{jacobson2001gravity,eling2004static}. Hence, these theories exhibit Lorentz violation, and so they are simple candidate theories that violate the EEP. 

Vector-tensor theories are divided into two distinct subclasses. The first is the set of constrained theories, for which $A_a A^a = -1$\,. These are also known as Einstein-\AE ther (EA) theories. 
The Einstein-\AE ther action is 
\begin{eqnarray}\label{eq_einstein_aether}
S_{\rm EA} &=& \frac{1}{16\pi G}\int \mathrm{d}^4 x\,\sqrt{-g}\,\left[R + \mathcal{K} +\lambda\left(A_a A^a + 1\right)\right]\,, \quad {\rm where} \\
\mathcal{K} &=& \left(c_1 g^{ab}g_{cd} + c_2 \delta^a_{\ c}\delta^b_{\ d} + c_3 \delta^a_{\ d}\delta^b_{\ c} - c_4 A^a A^b g_{cd}\right) \nabla_a A^c \, \nabla_b A^d\,,
\end{eqnarray}
and the Lagrange multiplier $\lambda$ enforces the normalisation of $A_a\,$. EA theories can be generalised by replacing $\mathcal{K} \longrightarrow f(\mathcal{K})$ \cite{bonvin2008generalized}. EA and generalised EA theories have been used to study gravitational alternatives to dark matter, as certain functional forms of $f(\mathcal{K})$ can exhibit MOND-like behaviour at low acceleration scales \cite{zlosnik2006vector,zlosnik2007modifying,thomas2023consistent}.

The other subclass of vector-tensor MG is the set of unconstrained theories, where $A_a A^a$ is free to vary in the range $-1 < A_a A^a < 0\,$. Unconstrained vector-tensor theories have some problematic properties, for example concerning their stability and the appearance of ghosts \cite{jimenez2009viability}. However, they provide a tractable way to study the properties of Lorentz-violating gravity theories, which led to them being developed as theoretical test cases \cite{nordtvedt1972conservation,Hellings_1973}.
The general action for these theories is
\begin{eqnarray}\label{eq_vector_tensor_action_unconstrained}
    && S_{\rm VT} = \frac{1}{16\pi G}\int\mathrm{d}^4 x \,\sqrt{-g}\,[\left(1 + \omega A_a A^a\right) R + \eta\, A^a A^b R_{ab} \\
    \nonumber &&  \hspace{3cm} + \left(\tau - 4\epsilon\right)\,\nabla_{[a}A_{b]}\,\nabla^{[a}A^{b]} + \tau \,\nabla_{(a}A_{b)}\,\nabla^{(a}A^{b)}]\,,
\end{eqnarray}
where we have explicitly separated out the gauge-violating term $\nabla_{(a}A_{b)}\,\nabla^{(a}A^{b)}\,$\,, and $\left\lbrace \omega, \eta, \tau, \epsilon \right\rbrace$ are constants.
We will make use of a special case of $S_{\rm VT}$ in Chapter 5, for which we will state the equivalents of the Einstein and Proca equations of motion. We do not produce the full equations here, as the metric EOM in particular is very lengthy and not especially insightful. 
Although vector-tensor theories are not typically considered to be well suited for cosmological applications, scalar-vector-tensor theories, where both $\phi$ and $A_a$ are introduced as novel gravitational DOFs, have found some interest in recent years \cite{skordis2008generalizing,skordis2009tensor,heisenberg2018scalar,heisenberg2018cosmology,skordis2021new,skordis2022aether}. 
It has been suggested that they can give rise to MOND-like physics on galactic scales, while still providing a good fit to CMB and LSS observations on cosmological scales \cite{skordis2021new}.

Let us briefly discuss theories of gravity that introduce higher derivatives of the metric, beyond second order. This is equivalent to introducing higher powers of curvature invariants, so the simplest approach is to define an $f(R)$ theory,
\begin{equation}
    S = \frac{1}{16\pi G}\int \mathrm{d}^4 x \,\sqrt{-g}\, f(R)\,.
\end{equation}
In a low-energy effective field theory of gravity, it is entirely expected that we would have an EFT expansion of the form $f(R) = R + \alpha R^2 + \mathcal{O}(R^3)\,$. Terminating this power series at the $R^2$ term produces a theory that is both renormalisable at one-loop order \cite{stelle1977renormalization} and a viable inflationary model \cite{starobinsky1980new}.
The $f(R)$ gravity theories are very well-studied in cosmology, with sophisticated N-body simulations having been developed in this theory \cite{zhao2011n,li2012haloes,achitouv2016imprint} that make it the benchmark for LSS constraints on deviations from General Relativity. 
Much of this work has focused on the Hu-Sawicki model \cite{hu2007models}, for which the free function in the action takes the form \cite{hu2016testing}
\begin{equation}
f_{\rm HS}(R) = R - m^2 \frac{c_1 \left(\frac{R}{m^2}\right)^n}{\left(\frac{R}{m^2}\right)^n + 1}\,,
\end{equation}
where $m^2 = H_0^2 \Omega_{m0}$\,. This model has the notable property that, in the standard (FLRW + perturbations) approach to cosmology, the background expansion is entirely equivalent to $\Lambda$CDM, with MG effects arising only in the perturbation theory.

Despite the popularity of $f(R)$ gravity, it faces a number of challenges. Power-law functional forms of $f$ are highly restricted by Solar System measurements \cite{clifton2005power}, and other simple forms are constrained by the integrated Sachs-Wolfe effect in the CMB \cite{cai2014integrated}, LSS observations \cite{lombriser2012constraints} and cross-correlations between the two \cite{giannantonio2010new}. 
Although certain forms like Hu-Sawicki remain viable, and continue to be studied in LSS surveys, they are often explicitly designed to avoid any modifications to the FLRW cosmology, rather than being derived from fundamental theory considerations. Hence, one might argue that these models are essentially phenomenological testing grounds rather than well-motivated extensions to GR.

If we are willing to accept higher derivatives of $g_{ab}$, then $R$ is no longer the unique choice of curvature scalar that can enter the action. It is particularly illuminating to consider the introduction of the Gauss-Bonnet scalar \cite{fernandes20224d},
\begin{eqnarray}
\mathcal{G} &=& R^2 - 4R_{ab}R^{ab} + R_{abcd}R^{abcd}\,,  \quad {\rm so \ that} \\
\nonumber S &=& \frac{1}{16\pi G} \int \mathrm{d}^4 x \,\sqrt{-g}\,\left(R - 2\Lambda + \alpha \mathcal{G}\right)\,.
\end{eqnarray}
It turns out that $\mathcal{G}$ is related to a topological invariant in $\leq 4$ spacetime dimensions \cite{lovelock1970dimensionally}. Thus, varying the action above yields Einstein's equations identically, with the Gauss-Bonnet term having no effect.
However, research in recent years has shown that a non-trivial MG theory in four spacetime dimensions can be built using $\mathcal{G}$\, \cite{glavan2020einstein}. This is the four-dimensional Einstein-Gauss-Bonnet (4DEGB) theory \cite{fernandes2020derivation,clifton2020observational,fernandes20224d}, which is obtained by letting the number of spacetime dimensions $D$ vary as a free parameter, and rescaling the coefficient $\alpha \longrightarrow \dfrac{\alpha}{D-4}\,$. 
Finally, one takes the limit $D \longrightarrow 4$, and adds a counterterm to the action in order to regularise the theory. The dimensional regularisation procedure is controversial \cite{gurses2020there}, although the fact that 4DEGB can be mapped on to an Horndeski scalar-tensor theory might give one confidence that the results of the procedure are valid \cite{fernandes20224d}.
Supposing it is valid, it has been shown that the 4DEGB theory contains interesting novel phenomena in the context of black hole spacetimes, potentially giving rise to a novel dark matter candidate \cite{fernandes2021black}.

We will not discuss higher-dimensional or non-local alternatives to General Relativity in this thesis, although there are certainly strong motivations to consider them (especially higher spacetime dimensions) from the perspective of the UV completion of gravity.
Instead, we now turn our attention to the ways in which metric theories of gravity can be tested in astrophysics and cosmology. We will focus on generalised frameworks for testing gravitational theories, starting with the gold standard: the parameterised post-Newtonian (PPN) formalism \cite{Will_1993}.

\subsection{Parameterised post-Newtonian formalism}\label{subsec:PPN}

The parameterised post-Newtonian (PPN) formalism is the standard approach that is used to test theories of gravity in isolated, weakly gravitating systems. It is ideally suited to studying gravitational phenomena in the Solar System and binary systems\footnote{For binary systems of compact objects (neutron stars and black holes), the weak gravity assumption breaks down only when the objects are very close together (i.e. $\sim 100$ Schwarzschild radii or fewer). Post-Newtonian techniques can therefore be used to study compact binary mergers in all but the final stages of inspiral.}, where most astrophysical tests of gravity have been performed.
The formalism is genuinely theory-agnostic, making no assumptions on the underlying gravitational field content other than assuming the validity of the WEP, so that test particles move on geodesics of the spacetime metric $g_{ab}\,$.
It was developed in the late 1960s and early 1970s by Will, Nordtvedt and Thorne
\cite{nordtvedt1968equivalence,will1971theoretical_1,will1971theoretical_2,thorne1971theoretical}, following pioneering work by Chandrasekhar \cite{chandrasekhar1965post} on post-Newtonian methods in General Relativity.

The central principle behind the PPN formalism is that in most astrophysical systems, three key conditions are satisfied:
\begin{enumerate}
    \item The characteristic velocity scale $v$ is small compared to the speed of light.
    \item Gravitational fields are weak, i.e. $g_{ab}$ can locally be considered perturbatively close to the Minkowski metric $\eta_{ab}\,$.
    \item The cosmological expansion happens sufficiently slowly that it is irrelevant in studies of isolated systems. Hence, spacetime can be modelled as asymptotically flat, with perturbations to the Minkowski metric vanishing at spatial infinity.
\end{enumerate} 
Thus, $v \ll 1$ can be used as a perturbative\footnote{The PPN formalism is not strictly a form of perturbation theory, but it is helpful to think of it on those terms, in order to make the comparison to cosmological perturbation theory apparent.} parameter, with the relativistic equations of motion solved order-by-order. The Virial theorem ensures that the Newtonian potential $U \sim v^2\,$, and hence $\rho \sim v^2\,$ as $\nabla^2 U = -4\pi G \rho\,$. 
A crucial consequence of the slow-motion condition is that, unlike in CPT where all derivatives enter the equations of motion on equal footing, time derivatives of all quantities are suppressed by $v$ relative to spatial derivatives,
\begin{equation}\label{eq_PPN_hierarchy_time_vs_space_derivatives}
    \frac{\frac{\partial}{\partial t}}{\frac{\partial}{\partial x}} \sim v \ll 1\,.
\end{equation}
The full PPN procedure is as follows \cite{Will_1993}:
\begin{itemize}

    \item Write down the metric as $g_{ab} = \eta_{ab} + h_{ab}\,$, with $ \vert h_{ab} \vert \ll 1$ and $h_{ab} \longrightarrow 0$ at spatial infinity. Specify an order in $v$ to which the components of $h_{ab}$ must be determined. It turns out that in order to calculate the leading-order post-Newtonian corrections to the geodesic equation for both null and timelike geodesics, one needs to know $h_{ij}$ to $\mathcal{O}(v^2)$, $h_{0i}$ to $\mathcal{O}(v^3)\,$, and $h_{00}$ to $\mathcal{O}(v^4)\,$. 
    
    \item Write down all the possible contributions to the energy-momentum tensor that can appear up to the highest required post-Newtonian order. From now on, we will take that order to be $v^4\,$.

    \item Identify the complete set of gravitational potentials that can be sourced by those contributions to $T_{ab}$, while remaining consistent with asymptotic flatness.

    \item Express $h_{00}^{(2)}$, $h_{ij}^{(2)}$, $h_{0i}^{(3)}$ and $h_{00}^{(4)}$ as combinations of those potentials. This involves fixing the gauge, in order to avoid spurious coordinate dependence.

    \item Write the coefficients of those potentials\footnote{And relevant monomials of them, such as $U^2\,$.} in terms of a complete set of constant parameters, where those parameters are defined with a clear physical meaning in mind. These are the PPN parameters.

    \item Constrain the PPN parameters using astrophysical observations.

    \item Given a specific theory of gravity, calculate the PPN parameters in terms of the underlying parameters of the theory. Hence, the observational constraints on the PPN parameters can be used to test individual MG theories of interest.
    
\end{itemize} 

Let us now carry out the first five steps above. The energy-momentum tensor of matter can contain the following scalar contributions: mass density $\rho \sim v^2\,$, pressure $p \sim c_s^2 \rho \sim v^2 v^2 \sim v^4\,$, internal energy density $u = \rho \Pi \sim v^4\,$, and gravitational potential energy density $-\rho h_{00}^{(2)}\, \sim v^4\,$. 
The only possible 3-vector contribution is the momentum density $\rho v_i \sim v^3\,$. Finally, the allowed 3-tensors are $p\,\delta_{ij} \sim v^4$\,, and $\rho v_i v_j \sim v^4\,$. 
Hence, the components of $T_{ab}$ are 
\begin{eqnarray}
    T_{00} &=& \rho\left[1 + \Pi + v^2 - h_{00}^{(2)}\right]\, + \, \mathcal{O}\left(v^6\right)\,, \\
    \nonumber T_{0i} &=& - \rho v_i \, + \, \mathcal{O}\left(v^5\right)\,, \quad {\rm and} \\
    \nonumber T_{ij} &=& \rho v_i v_j + p \delta_{ij} \, + \, \mathcal{O}\left(v^6\right)\,.
\end{eqnarray}
Note that in this section, we will raise and lower spatial indices with the Kronecker delta, so there is no difference between e.g. $v_i$ and $v^i\,$.

The gravitational potentials that can appear at $\mathcal{O}(v^2)$ are the Newtonian potential $U(t, \mathbf{x})$ and the tensor potential $U_{ij}(t, \mathbf{x})\,$, defined by
\begin{equation*}
    U = G\int\frac{\mathrm{d}^3 x' \, \rho\left(t, \mathbf{x}'\right)}{\left\vert \mathbf{x}-\mathbf{x}'\right\vert}\,, \quad U_{ij} = G \int \frac{\mathrm{d}^3 x' \delta_{ik}\delta_{jl}\, \left(x-x'\right)^j \left(x-x'\right)^l\,\rho\left(t,\mathbf{x}'\right)}{\left\vert \mathbf{x}-\mathbf{x}'\right\vert^3}\,.
\end{equation*}
It is useful to define the superpotential $\chi$ by $\nabla^2 \chi = -2U\,$. Then, $U_{ij} = U \delta_{ij} + \chi_{,ij}\,$, so at $\mathcal{O}(v^2)$ the only allowed potential terms are $U$ and $\chi_{,ij}\,$.

At $\mathcal{O}(v^3)$ there are two vector potentials $V_i$ and $W_i$\,:
\begin{equation*}
    V_i = G\int\frac{\mathrm{d}^3 x' \, \rho\left(t, \mathbf{x}'\right) v_i\left(t, \mathbf{x}'\right)}{\left\vert \mathbf{x}-\mathbf{x}'\right\vert}\,, \quad W_i = G\int\frac{\mathrm{d}^3 x'\,\left(x-x'\right)^j\delta_{ij}\,\left(\mathbf{x}-\mathbf{x}'\right){\bf \cdot}\mathbf{v}(t,\mathbf{x}')\rho(t,\mathbf{x}')}{\left\vert \mathbf{x}-\mathbf{x}'\right\vert^3}\,,
\end{equation*}
where $V_i$ satisfies a Poisson equation, $\nabla^2 V_i = -4\pi G \rho v_i\,$\,, and the continuity equation implies the additional identity $V_{i,i} = - U_{,t}\,$. 
The potential $W_i$ is rather unwieldy, but it turns out it can be related to derivatives of the superpotential, by $W_i = V_i - \chi_{,ti}\,$, such that $\nabla^2 W_i = -4\pi G\rho v_i + 2U_{,ti}\,$.
Therefore, in the absence of preferred frames in the Universe, the only source terms with one spatial index are $V_i$ and $\chi_{,ti}\,$.
If there is a cosmologically preferred frame, which might be defined, for example, by the rest frame of the timelike vector field $A^a$ in a vector-tensor theory of gravity, then there are two additional contributions that can appear. 
Defining the non-relativistic 3-velocity $w^i$ of the astrophysical system of interest with respect to the preferred frame, these are $U w_i$ and $w^j \chi_{,ij}\,$.

Now, let us consider the possible gravitational potentials at $\mathcal{O}(v^4)\,$. Because we only need $h_{00}^{(4)}$ in the metric, it is only necessary to calculate the scalar potentials at this order. Clearly, a contribution $U^2$ is allowed.
The new required potentials are the following six:
\begin{eqnarray*}
    \Phi_1 &=& G\int\frac{\mathrm{d}^3 x' \, \rho\left(t, \mathbf{x}'\right)\,v^2\left(t,\mathbf{x}'\right)}{\left\vert \mathbf{x}-\mathbf{x}'\right\vert}  \quad \Rightarrow \quad \nabla^2 \Phi_1 = -4\pi G \rho v^2\,, \\
    \Phi_2 &=& G\int\frac{\mathrm{d}^3 x' \, \rho\left(t, \mathbf{x}'\right)\,U\left(t,\mathbf{x}'\right)}{\left\vert \mathbf{x}-\mathbf{x}'\right\vert}  \quad \Rightarrow \quad \nabla^2 \Phi_2 = -4\pi G \rho U\,, \\
    \Phi_3 &=& G\int\frac{\mathrm{d}^3 x' \, \rho\left(t, \mathbf{x}'\right)\,\Pi\left(t,\mathbf{x}'\right)}{\left\vert \mathbf{x}-\mathbf{x}'\right\vert}  \quad \Rightarrow \quad \nabla^2 \Phi_3 = -4\pi G \rho \Pi\,, \\
    \Phi_4 &=& G\int\frac{\mathrm{d}^3 x' \,p\left(t,\mathbf{x}'\right)}{\left\vert \mathbf{x}-\mathbf{x}'\right\vert}  \quad \Rightarrow \quad \nabla^2 \Phi_4 = -4\pi G p\,, \\
    \Phi_6 &=& G\int\frac{\mathrm{d}^3 x' \, \rho\left(t, \mathbf{x}'\right)\,\left(\mathbf{v}{\bf \cdot}\left(\mathbf{x}-\mathbf{x}'\right)\right)^2}{\left\vert \mathbf{x}-\mathbf{x}'\right\vert}\,, \quad {\rm and} \\
    \Phi_W &=& G^2 \int\frac{\mathrm{d}^3x'\,\mathrm{d}^3x''\,\rho(t,x')\,\rho(t,x'')\, \left(x-x'\right)^i\,\delta_{ij}}{\left\vert\mathbf{x}-\mathbf{x}'\right\vert^3}\,\left[\frac{\left(x'-x''\right)^j}{\left\vert\mathbf{x}-\mathbf{x}''\right\vert} - \frac{\left(x-x''\right)^j}{\left\vert\mathbf{x}'-\mathbf{x}''\right\vert}\right]\,.
\end{eqnarray*}
There is also a term $\Phi_5$, but it turns out not to be independent, satisfying $\Phi_5 = \chi_{,tt} + \Phi_1 +2\Phi_4 - \Phi_6\,$, so it can be ignored.
The Whitehead potential $\Phi_W$ encodes the possibility that there are gravitationally preferred locations in the Universe. In the absence of these rather exotic preferred-location effects, $\Phi_W$ makes no contribution to $h_{ab}\,$.
Preferred-frame effects can also make a contribution at this order, through the combinations $w^2 U$, $w^i V_i$ and $w^i w^j \chi_{,ij}\,$.

The final step we need in order to write down the form of the metric is to fix the gauge. The term $h_{00}^{(2)}$ is invariant under post-Newtonian gauge transformations \cite{Clifton_2020}. 
The remaining gauge conditions are fixed by demanding that
\begin{itemize}
    \item There are no $\mathcal{O}(v)$ metric perturbations, i.e. $h_{0i}^{(1)} = 0\,$.
    \item The space-space part of the metric $h_{ij}
    $ is diagonal, so there are no contributions of the form $\chi_{,ij}\,$, only $U\delta_{ij}\,$.
    \item There is no explicit time dependence in $h_{00}$, so $\Phi_5\,$, which contains a contribution from $\chi_{,tt}\,$, is absent.
\end{itemize}
Taken together, these conditions define the PPN gauge, which is closely related to the general-relativistic post-Newtonian gauge originally used by Chandrasekhar \cite{chandrasekhar1965post}. 

With all the above considerations in mind, the PPN test metric can be written down. The metric perturbations take the form \cite{Will_1993}
\begin{eqnarray}
    h_{00}^{(2)} &=& 2\alpha U\,, \label{eq_PPN_h00_2} \\
    h_{ij}^{(2)} &=& 2\gamma U\,\delta_{ij}\,, \label{eq_PPN_hij_2} \\
    h_{0i}^{(3)} &=& -2\left(\alpha + \gamma + \frac{\alpha_1}{4}\right)V_i + \frac{1}{2}\left(\alpha + \alpha_2 - \zeta_1 + 2\xi\right)\chi_{,ti} + \varphi_i^{\rm PF}\,, \quad {\rm where} \label{eq_PPN_h0i_3} \\
    \varphi_i^{\rm PF} &=& -\frac{1}{2}\alpha_1 w_i U - \alpha_2 w^j \chi_{,ij}\,, \quad {\rm and} \label{eq_PPN_PF_3} \\
    h_{00}^{(4)} &=& -2\beta U^2 + \left(2\alpha + 2\gamma + \alpha_3 + \zeta_1 - 2\xi\right)\Phi_1 \label{eq_PPN_h00_4} \\
    \nonumber && + 2\left(\alpha + 3\gamma - 2\beta + \zeta_2 + \xi\right)\Phi_2 + 2\left(\alpha + \zeta_3\right)\Phi_3 + 2\left(3\gamma + 3\zeta_4 - 2\xi\right)\Phi_4  \\
    \nonumber && - \left(\zeta_1 - 2\xi\right)\Phi_6 - 2\xi\Phi_W + \varphi^{\rm PF}\,, \quad {\rm where} \\
    \varphi^{\rm PF} &=& \left(\alpha_3 - \alpha_1\right) w^2 U + \left(2\alpha_3 - \alpha_1\right) w^i V_i + \alpha_2 w^i w^j \chi_{,ij}\,. \label{eq_PPN_PF_4}
\end{eqnarray}
The set $\left\lbrace \alpha, \gamma, \beta, \xi, \alpha_{1,2,3}, \zeta_{1,2,3,4}\right\rbrace$ are the 11 PPN parameters. In GR, the only non-zero PPN parameters are $\alpha = \gamma = \beta = 1\,$.
Ten of the PPN parameters are independent, with the exception of $\zeta_4$, for which consideration of the properties of perfect fluids generates the algebraic relation $\zeta_4 = \frac{1}{6}\left(3\alpha_3 + 2\zeta_1 - 3\zeta_3\right)\,$ \cite{will1976active}.

The parameter $\alpha$ is typically set to unity by absorbing it into the definition of Newton's constant $G\,$, as the form $2\alpha U$ of the purely Newtonian term $h_{00}^{(2)}$ follows directly from the weak equivalence principle.
However, as we will be interested in Chapters 5 and 6 in cosmological applications where $G_{\rm eff} \sim \alpha G$ can evolve over cosmic time, we will retain $\alpha$ in the PPN equations. 

\begin{table}
\centering
\hspace{-1.5cm}
\begin{tabular}{|c|c|c|c|} 
\hline 
Physical effect  & Parameter & Constraint & Origin of constraint \\ 
\hline 
Effective $G$ & $\alpha$ & $1$ & By definition \\
\hline
Spatial curvature & $\gamma$ & $1 + \left(2.1 \pm 2.3\right)\times 10^{-5}$ & Shapiro delay \cite{bertotti2003test} \\
\hline
Nonlinearity & $\beta$ & $1 + \left(-2.7 \pm 3.9\right)\times 10^{-5}$ & Mercury's perihelion \cite{park2017precession} \\
\hline 
Preferred locations & $\xi$ & $0 \pm 3.9 \times 10^{-9}$ & Pulsar precession \cite{Shao_2013} \\
\hline
Preferred frames & $\alpha_1$ & $\left(-0.7 \pm 1.8\right)\times 10^{-4}$ & Pulsar binaries \cite{Shao_2012} \\
\hline
Preferred frames & $\alpha_2$ & $\left(1.8 \pm 5.0\right)\times 10^{-5}$ & Pulsar torque \cite{Shao_2012} \\
\hline
Preferred frames & $\alpha_3$ & $0 \pm 4 \times 10^{-20}$ & Pulsar timing \cite{stairs2005discovery} \\
\hline
Momentum conservation & $\zeta_1$ & $0 \pm 2 \times 10^{-2}$ & Lunar laser ranging \cite{williams2004progress} \\
\hline
Momentum conservation & $\zeta_2$ & $0 \pm 4 \times 10^{-5}$ & Binary acceleration \cite{Will_1992} \\
\hline
Momentum conservation & $\zeta_3$ & $0 \pm 10^{-8}$ & Lunar acceleration \cite{Will_2014} \\
\hline
Momentum conservation & $\zeta_4$ & -------------- & Not independent \cite{will1976active} \\
\hline
\end{tabular} \\ 
\caption{The full set of coupling parameters in the PPN test metric.}
\label{table_PPN_parameters}
\end{table}

The form of the coefficients of each post-Newtonian potential has been chosen so that the PPN parameters all have a clear physical interpretation, as shown in Table \ref{table_PPN_parameters}. 
Specific experiments constrain certain combinations of the PPN parameters. For example, the bending of light around the Sun is sensitive to $\alpha + \gamma $ \cite{lambert2009determining}, and Lense-Thirring precession depends on $\alpha + \gamma + \dfrac{1}{4}\alpha_1\,$ \cite{Gravity_Probe_B}.
Fully conservative theories of gravity, in which total momentum and total angular momentum are both globally conserved in asymptotically flat spacetime, satisfy $\alpha_1 = \alpha_2 = \alpha_3 = \zeta_1 = \zeta_2 = \zeta_3 = \zeta_4 = 0\,$.
A corollary of global angular momentum conservation in asymptotically flat spacetime is that there are no preferred-frame effects in the theory of gravity.
Semi-conservative theories drop the requirement of angular momentum conservation, but retain momentum conservation. For these theories, the parameters $\zeta_{1,2,3,4}$ still vanish, but $\alpha_{1,2,3}$ can be non-zero, with preferred-frame effects allowed.

Finally, let us discuss the interpretation of the remaining three parameters $\left\lbrace \gamma, \beta, \xi\right\rbrace$ that may appear in fully conservative theories and which cannot, unlike $\alpha$, be normalised to unity at the present day by a rescaling of Newton's constant.
The leading-order post-Newtonian parameter $\gamma$ can be interpreted as setting the amount of spatial curvature that is produced by non-relativistic matter. To see this, consider foliating the perturbed Minkowski spacetime into constant-time surfaces. Then, one can calculate the Ricci curvature scalar associated with the induced metric on the three-dimensional spacelike hypersurfaces. Evaluating this to $\mathcal{O}(v^2)$ gives $^{(3)}R = - 2\nabla^2 h_{00}^{(2)} = 16\pi G\, \gamma \,\rho\,$, thereby justifying the ``spatial curvature'' interpretation.
The interpretation of $\beta$ as the parameter describing the degree of nonlinearity in the gravitational field follows from it multiplying $U^2$ in Eq. (\ref{eq_PPN_h00_4}).
Similarly, the notion that $\xi$ describes preferred-location effects is shown by its direct coupling to $\Phi_W$ in Eq. (\ref{eq_PPN_h00_4}).

The classical parameterised post-Newtonian formalism is the most successful framework for testing astrophysical gravity in a theory-agnostic fashion. However, as it is based around small perturbations of Minkowski spacetime, it is clearly not suitable for use on cosmological scales. We will come back to this issue in Chapter 5, where we will show how the PPN formalism can be generalised into a cosmologically viable approach.
To conclude the present discussion of testing alternatives to General Relativity, let us next provide a short overview of some generalised approaches that have been used to model deviations from GR in cosmology.

\subsection{Cosmological approaches to testing gravity}\label{subsec:PPF}

Theory-agnostic frameworks are harder to construct on cosmological scales than they are on astrophysical scales. There are a number of reasons for this, each of which introduces a new level of complexity:
\begin{enumerate}
    \item The evolution in time of the background expansion cannot be ignored. Therefore, any parametrisation of modified gravity we construct should be time-dependent.
    \item We deal with a vast range of length scales, ranging from deep within the horizon to well beyond it. It is not clear that any treatment on small scales would be valid on large scales, and vice versa.
    \item Our understanding of gravity in isolated astrophysical systems relies on the notion of asymptotically flat spacetime. This is not a viable assumption in cosmology.
    \item Any cosmological approach to testing gravity is likely to involve cosmological perturbation theory. One will invariably encounter the gauge problem that makes interpreting the results of CPT calculations far from straightforward.
    \item In order to make predictions for cosmological observables, particularly the CMB, one needs to model not just the matter-dominated epoch but also the radiation and dark energy epochs.
\end{enumerate}

Despite these challenges, several helpful approaches have been developed. One of the most widely applicable is the Parameterised Post-Friedmann (PPF) formalism \cite{hu2007models,Hu_2008,amin2008subhorizon,Skordis_2009,Baker_2011,Baker_2013,bakerbull}. It is built on linear cosmological perturbation theory in Newtonian gauge, analysed in Fourier space. Deviations from General Relativity are described by two free functions of time and scale, through the equations
\begin{eqnarray}
    \label{eq_PPF_1} k^2 \Psi &=& 4\pi G \, a^2 \,\mu\left(\tau, k\right)\,\left[\delta \rho + 3\left(\bar{\rho}+\bar{p}\right)\mathcal{H}v \right]\,, \quad {\rm and} \\
    \label{eq_PPF_2} \Phi - \Psi &=& \Sigma\left(\tau, k\right)\,\Psi\,.
\end{eqnarray}
The GR equivalents of these are the Fourier transforms of the second and fourth equations in the set (\ref{eq_CPT_newtonian_gauge_GR})\footnote{With the scalar anisotropic stress $\Pi$ assumed to vanish.}, whence one sees that the free functions $\mu(\tau, k)$ and $\Sigma(\tau, k)$ are respectively equal to $1$ and $0$ in GR, at all times and on all scales. 
The function $\mu(\tau, k)$ can be thought of as an effective Newton's constant, at least in the context of linear scalar perturbations. 
Note that it has been assumed that the effects of density and velocity perturbations are modulated by the same factor $\mu\,$. This is equivalent to stating that the scalar part of the CPT momentum constraint - the first equation in (\ref{eq_CPT_newtonian_gauge_GR}) - is given by
\begin{equation}
    \Psi' + \mathcal{H}\Phi = 4\pi G\,a^2\,\mu\left(\tau, k\right)\,\left(\bar{\rho} + \bar{p}\right)\,v\,.
\end{equation}
The second free function $\Sigma\left(\tau,k\right)$ is referred to as the gravitational slip \cite{ppnvscosmo}. In modified gravity theories, $\Sigma$ is generically non-zero. For example, in the Bergman-Wagoner scalar-tensor theories, it is directly related to the scalar field perturbation, by $\Sigma = \dfrac{\delta \phi}{\bar{\phi}\Psi}\,$. 

It has been argued that the equality of the Bardeen potentials is the defining special feature of GR in the context of CPT \cite{Bertschinger_2006}. Constraining the slip therefore provides a way to use the PPF framework to test GR.
This requires identifying observables that depend only on the properties of timelike curves (i.e. redshift space distortions), which are sensitive only to $\Phi$, and comparing them to observables that depend on the properties of null curves (i.e. weak lensing), which are sensitive to $\psi_W = \Phi + \Psi\,$. Combining this measurement would allow $\Sigma$ to be constrained \cite{song2011complementarity,bonvin2023modified}, at least on the scales where linear CPT can safely be applied.

The PPF framework is a very good, deliberately simple, starting point. However, like all theory-agnostic approaches, it does have some limitations. It does not consistently modify the FLRW expansion, which is often set to be equal to the $\Lambda$CDM $a(t)$ solution throughout, along with the perturbations.
Moreover, it is not easy to make contact with approaches used in regimes of nonlinear densities, such as the PPN formalism, so astrophysical tests of gravity might not be able to constrain $\mu$ and $\Sigma$ on small scales.
From a mathematical point of view, it is also notable that the PPF equations are incomplete, as they do not contain the Raychaudhuri equation, and the momentum constraint is assumed to be described by the same coupling function $\mu$ as the Hamiltonian constraint. We will show in Chapter 5 that this assumption is not always a safe one, with potentially significant consequences. 

To finish the discussion of alternatives to GR in this chapter, let us very briefly summarise two other methods that have been adopted by cosmologists to test the MG landscape. 
One of them is built on the Horndeski class of gravity theories, described by the action (\ref{eq_Horndeski_action}). As scalar-tensor theories are, along with $f(R)$ and nDGP (the normal branch of the DGP brane-world theory \cite{dvali2000metastable}), the most widely-studied MG theories in most observational applications, there is substantial benefit in developing a parameterised framework to constrain the vast space of theories contained within $S_{\rm Horndeski}\,$.

The most well-known such framework is referred to as the $\alpha$ parametrisation. Like the PPF formalism, the $\alpha$ functions are defined through the scalar sector of the linear CPT equations. 
In that context, it has been shown that deviations from GR can be characterised through four functions of time \cite{bellini2016constraints}: $\alpha_M$\,, $\alpha_K$\,, $\alpha_B$ and $\alpha_T\,$, all of which vanish in GR with standard $\Lambda$CDM energy-momentum content.
They each have distinct physical interpretations.
\begin{itemize}

    \item $\alpha_M$\,: the evolution rate in cosmic time of the effective Planck mass $M_{\rm Planck}^{\rm eff} = \left(8\pi G_{\rm eff}\right)^{-1/2}\,$. Here $G_{\rm eff}$ refers purely to the effective Newton's constant associated with scalar perturbations to FLRW, with no obvious relation to the $G_{\rm eff}$ that is measured by $\alpha_{\rm PPN}$ in astrophysical settings.
    
    \item $\alpha_K$\,: the kineticity of the scalar field. Note that a minimally coupled quintessence scalar, which is not really modified gravity, has $\alpha_K \neq 0$.
    
    \item $\alpha_B$\,: the amount of braiding between the kinetic terms associated with the scalar field and the metric. Kinetic braiding produces a characteristic scale in the evolution of perturbations, in addition to the usual Hubble horizon scale \cite{Bellini_2014}.
    
    \item $\alpha_T$\,: the excess propagation speed of tensor modes. The GW speed has been constrained to $\left\vert \alpha_T \right\vert \lesssim 10^{-15}$ at LIGO frequencies by the dual messenger event GW170817/GRB170817A \cite{baker2017strong}, although it has been suggested that $\alpha_T$ could deviate from zero at lower frequencies, such as in the LISA band \cite{de2018gravitational,baker2022measuring}\,.
    
\end{itemize}

Like the PPF formalism, the Horndeski EFT described by the $\alpha_i(t)$ contains the feature that the background expansion must be specified independently. The $\alpha_i$ are rather complicated functions of both $H(t)$ and the various terms in the Horndeski action.
It is not clear how one should model their evolution, with many authors simply making the ansatz that $H(t)$ is identically that of $\Lambda$CDM, and then $\alpha_i = \Omega_{\rm DE}(t)\, c_i\,$ \cite{bellini2016constraints}. Although not without merit, this assumption is clearly restrictive.

The other approach we wish to mention is the effective field theory of dark energy (EFTofDE) \cite{Gubitosi_2013} (see Ref. \cite{frusciante2020effective} for a review). This describes a larger range of theories than the Horndeski EFT, covering for example the Ho{\v r}ava vector-tensor theory, by using a large set of free functions of time\footnote{In the formulation adopted by Ref. \cite{frusciante2020effective} there are nine EFT functions.} in the EFT action.
It also includes directly the possibility of modifications to the cosmic expansion. 
However, the EFTofDE has the same problem as the Horndeski $\alpha_i(t)$ in that it is rather unclear what the functional forms of all the EFT functions of time should be, and whether a simple form such as a proportionality to $\Omega_{\rm DE}$ is appropriate \cite{linder2016effective}.

In the remainder of this thesis, we will not make use of EFT approaches such as these two, although they are certainly useful, because they can be applied directly to cosmological datasets that probe the dynamics of scalar perturbations. Instead, our studies of cosmological tests of gravity in Chapters 5 and 6 will make use of a theory-agnostic framework that is closely related to the PPN formalism.

This concludes our review of alternatives theories of gravity, and the ways they are studied in astrophysical and cosmological settings. 
For the remainder of this chapter, we will focus on another set of alternatives to the cosmological concordance model. These will be cosmological models that break the cosmological principle of homogeneity and isotropy.

\section{Inhomogeneity and anisotropy}\label{sec:inhomogeneity_anisotropy}

Although the homogeneous and isotropic FLRW universe is the most well-known cosmological model, there is a wide variety of cosmological spacetimes with fewer symmetries.
These can be used both as genuine alternatives to FLRW to be tested observationally, and as useful mathematical models to understand the problems at hand in anisotropic and/or inhomogeneous cosmologies.
First, we will describe the general properties of homogeneous, anisotropic cosmological models. In doing so, we will demonstrate the utility of the $1+3$ and $1+1+2$ decompositions for describing these geometries. We will also introduce some significant inhomogeneous models.
Finally, we will discuss the important problems of cosmological averaging and backreaction, and review some possible solutions to them. In particular, we will introduce the averaging formalism developed by Buchert, which has potentially profound implications for the interpretation of large-scale cosmological observations.

\subsection{Anisotropic cosmological models}\label{subsec:anisotropic_models}

The simplest generalisations of the homogeneous and isotropic FLRW geometry are geometries which retain spatial homogeneity, but break spatial isotropy. To understand the properties of these models, we recall the introduction of continuous symmetries of spacetime in Chapter 3. 
Let us now expand further on the notion of a spacetime symmetry group\footnote{These are also called isometry groups.} that we introduced. The discussion here is based on summaries by Gr\o n \& Hervik \cite{gron2007einstein}, and Ellis \& van Elst \cite{Ellis_1999}.

In Chapter 3, we explained that a spatially homogeneous spacetime can be foliated into three-dimensional spaces $\Sigma_t\,$, which are level surfaces of some time function $t$\,. The induced metric on $\Sigma_t$ possesses three spacelike Killing vectors $\xi^H_{\ A}\,$, $A = 1,2,3\,$, which typically form a transitive subgroup of the full symmetry group $G$.
It turns out that there is actually an exception, where the spacetime is spatially homogeneous, with ${\rm dim}(G) = 4\,$, but does not have such a transitive subgroup. This is called the Kantowski-Sachs (KS) cosmology. We not study the KS model in this thesis, so we can now just ignore it and focus on the usual situation with a three-dimensional subgroup acting transitively on the surfaces $\Sigma_t\,$.
The spacetime may also possess additional KVFs $\xi^I$\,, the generators of spatial isotropies, although this is not a necessity. 

The Lie group $G$ of symmetries is described by a Lie algebra, which is the tangent space to $G$ at the identity. That Lie algebra is precisely the $\geq 3$-dimensional vector space spanned by the $\xi_A$ (where we have dropped the $H$ and $I$ superscripts for ease of notation). These Killing vector fields (KVFs) generate metric-preserving flows $\sigma_{\xi}$ on the constant-$t$ hypersurfaces through the map $\exp{\left(\mathcal{L}_{\xi}\right)}$\,, as displayed in Fig. \ref{fig_killing_transport}\,.
The Lie algebra is equipped with the commutator
\begin{equation}
    \left[\mathbf{A},\mathbf{B}\right]\,, \ {\rm s.t.} \ \left[\mathbf{A},\mathbf{B}\right]^a = \left(\mathcal{L}_{\mathbf{A}}\mathbf{B}\right)^a = A^b \partial_b B^a - B^b \partial_b A^a\,.
\end{equation}
By the group closure property, we can write the commutator of any two KVFs as
\begin{equation}
    \left[\xi_A, \xi_B\right] = D^C_{\ AB}\,\xi_C\,,
\end{equation}
where the quantities $D^C_{\ AB}\,$, which we will discuss very shortly, are manifestly anti-symmetric in $A$ and $B$\,.
In the above, we have assumed that the surfaces of transitivity $\Sigma_t$ are simply transitive. This means that the dimension of the isotropy group is $0$, so the dimension of the total symmetry group is $3\,$. That is not the case in general, because we can have isotropy groups of dimension $1$ and $3$, making the surfaces multiply transitive. 
The latter are the FLRW geometries, and we will come to the former case soon. These special cases do not substantially change the analysis that follows.

The KVFs $\xi_A$ span the surfaces of transitivity, so they do form a basis on $\Sigma_t$. However, this basis has the undesirable property that in general the objects $D^C_{\ AB}$ are dependent on the spatial coordinates, rather than being constant on $\Sigma_t$ \cite{gron2007einstein}. 
Instead, it is helpful to define the adjoint basis $\mathbf{e}_A$\,, which is invariant under parallel transport along integral curves of all the $\xi_A\,$. That is,
\begin{equation}
    \left[\xi_A, \mathbf{e}_B \right] = \mathcal{L}_{\xi_A} \mathbf{e}_B = 0 \ \ \forall \ \  A,B = 1,2,3\,.
\end{equation}
The basis vectors $\mathbf{e}_A$ provide the adjoint representation of the exact same Lie algebra. Because of the property of invariance under Killing transport, the structure constants $C^C_{\ AB}\,$ in this new representation, defined by the commutator\footnote{It is often useful to change basis yet again, to obtain an orthonormal tetrad in which the physical interpretation of the commutators is clearer \cite{Ellis_1999,wainwright1997dynamical}, but this is not required for the brief overview here.}
\begin{equation}
    \left[\mathbf{e}_A, \mathbf{e}_B \right] = C^C_{\ AB}\, \mathbf{e}_C\,,
\end{equation}
are genuinely constants on each $\Sigma_t\,$.
The implication of this result is rather profound. It means that all the possible spatially homogeneous geometries (except for the Kantowski-Sachs special case) are defined by their structure constants $C^C_{\ AB}\,$. 

Once the structure constants in the adjoint representation have been specified, the metric can be constructed directly in line with the spacetime symmetries. The $\mathbf{e}_A$ are dual to covectors $\mathbf{\omega}_A$\,, defined by $\mathbf{\omega}^A\left(\mathbf{e}_B\right) = \delta^A_{\ B}\,$. These $\mathbf{\omega}_A$ are themselves invariant under Killing transport. 
Finally, the line element can be written down as \cite{gron2007einstein,taub1951empty}
\begin{equation}
    \mathrm{d}s^2 = -\mathrm{d}t^2 + g_{AB}(t)\, \mathbf{\omega}^A \otimes \mathbf{\omega}^B\,,
\end{equation}
where $\otimes$ denotes the tensor product\footnote{For a formal definition of $\otimes$, see e.g. Ref. \cite{nakahara2018geometry}.}.

Homogeneous cosmological models which have three-dimensional surfaces $\Sigma_t$ of transitivity (i.e. all homogeneous cosmologies except for Kantowski-Sachs) can be classified into nine different types, according to the structure constants of their Lie group of symmetries. This is the Bianchi classification, with each of the nine types giving rise to the Bianchi cosmological models of that type. A full exposition of these spacetimes is provided by Ref. \cite{Stephani_2003}. 
We will not provide the full classification, but we will describe the subset of the Bianchi types that contain FLRW geometries as special cases. These are
\begin{itemize}

    \item Type {\it I}. Their structure constants are very simple: all $C^C_{\ AB} = 0\,$. The type {\it I} models are essentially anisotropic generalisations of the spatially flat FLRW cosmology. 
    
    \item Type {\it V}. The structure constants are $C^1_{\ 13} = C^2_{\ 23} = 1\,$, and $C^3_{\ 13} = C^2_{\ 13} = C^3_{\ 23} = C^1_{\ 23} = C^1_{\ 12} = C^2_{\ 12} = C^3_{\ 12} = 0\,$. The spatially open FLRW geometry is contained as a special case.

    \item Type {\it IX}. The structure constants are $C^C_{\ AB} = \delta^{CD}\epsilon_{ABD}\,$, where $\epsilon$ is the Levi-Civita alternating symbol. The spatially closed FLRW geometry is a special case.

    \item Type {\it VII}. It is best understood by splitting it further into two subclasses, ${\it VII}_0$ and ${\it VII}_h\,$. 
    \begin{itemize}
        \item ${\it VII}_0\,$: $C^2_{\ 13} = C^1_{\ 32} = 1\,$, $C^1_{\ 13} = C^1_{\ 12} = C^2_{\ 21} = C^3_{\ 13} = C^3_{\ 23} = C^3_{\ 12} = 0\,$, and $C^2_{\ 23} = 0\,$. Like type {\it I}, Bianchi type ${\it VII}_0$ contains the spatially flat FLRW cosmology.
        \item ${\it VII}_h\,$: ${\it VII}_0\,$: $C^2_{\ 13} = C^1_{\ 32} = 1\,$, $C^1_{\ 13} = C^1_{\ 12} = C^2_{\ 21} = C^3_{\ 13} = C^3_{\ 23} = C^3_{\ 12} = 0\,$, and $C^2_{\ 23} = \kappa > 0\,$. The details of $\kappa$ are not relevant for the present summary. Like type {\it V}, Bianchi type ${\it VII}_h$ contains the spatially open FLRW cosmology.
    \end{itemize}

\end{itemize}

All the homogeneous cosmological models we will study in this thesis belong to the Bianchi classification. Let us now turn to the issue of isotropy. We know that the largest possible isotropy group acting in $\Sigma_t$ is $SO(3)\,$. This has no subgroups of order $2$, so the next largest allowed isotropy group is $SO(2) \cong U(1)$, the one-parameter group of continuous rotations. 
Cosmological models that exhibit three-dimensional surfaces of transitivity, and a $U(1)$ isotropy group acting within those surfaces such that they are multiply transitive, are referred to as the locally rotationally symmetric (LRS) Bianchi cosmologies \cite{Ellis_1967,Stewart_1968,vanElst_1996}.  LRS solutions exist in the Bianchi types {\it I}, {\it II}, {\it III}, {\it V}, {\it VII} and {\it IX}\,.
For these models, the entire dynamics can be described by first-order scalar ODEs in time, in the $1+1+2$ decomposition, by choosing the preferred timelike vector to be orthogonal to the transitive surfaces $\Sigma_t\,$, and the preferred spacelike vector $m^a$ to be parallel to the rotational symmetry axis. 
For example, an LRS Bianchi type {\it I} model is characterised by the metric
\begin{equation}\label{eq_LRS_Bianchi_I}
    \mathrm{d}s^2 = -\mathrm{d}t^2 + A^2(t)\,\mathrm{d}r^2 + B^2(t)\left(\mathrm{d}y^2 + \mathrm{d}z^2\right)\,,
\end{equation}
and an LRS Bianchi type {\it V} model by
\begin{equation}\label{eq_LRS_Bianchi_V}
    \mathrm{d}s^2 = -\mathrm{d}t^2 + A^2(t)\,\mathrm{d}r^2 + B^2(t)\,e^{-2\beta r}\left(\mathrm{d}y^2 + \mathrm{d}z^2\right)\,,
\end{equation}
where the constant parameter $\beta$ can be thought of a generalisation of the spatial curvature of an open FLRW model. In the isotropic limit of Eq. (\ref{eq_LRS_Bianchi_V}), the Ricci scalar associated with the constant-$t$ hypersurfaces is $^{(3)}R = -6\beta/A^2 = -6\beta/B^2\,$.

Let us now consider the $1+1+2$ equations of motion for the simple, instructive, LRS Bianchi {\it I} model. This will allow us to make an important point about the late-time behaviour of the Bianchi cosmologies.
Picking the preferred timelike and spacelike vectors to point in the $t$ and $r$ coordinate directions respectively, the local rotational symmetry means that the only non-zero $1+1+2$ variables are purely time-dependent scalars. The kinematic scalars are
\begin{eqnarray}
    \Theta &=& \frac{\dot{A}}{A} + \frac{2\dot{B}}{B}\,, \\
    \Sigma &=& \frac{2}{3}\left(\frac{\dot{A}}{A} - \frac{\dot{B}}{B}\right)\,, \quad {\rm and} \\
    \Omega &=& \mathcal{A} \ = \ 0\,,
\end{eqnarray}
from which it is clear that $\Sigma(t)$ entirely encodes the anisotropy in the Bianchi {\it I} cosmology. 
In order to understand the anisotropic expansion of this universe, one can thus consider Eq. (\ref{eq_shear_evol_eqn_112}), the evolution equation for the scalar shear in the $1+1+2$ formalism. For an LRS Bianchi {\it I} model, this equation reduces to 
\begin{equation}
    \frac{\mathrm{d}}{\mathrm{d}t}\Sigma + \frac{2}{3}\Theta\Sigma + \frac{1}{2}\Sigma^2 + \mathcal{E} = \frac{1}{2}\Pi\,,
\end{equation}
where we have used that the lapse function is equal to unity to replace the covariant time derivative with a total derivative with respect to $t$\,.
The electric Weyl curvature scalar is
\begin{eqnarray}
    \mathcal{E} &=& -\frac{1}{3AB^2}\left[A\dot{B}^2 + B^2\ddot{A} - B\dot{A}\dot{B} - AB\ddot{B}\right]\,.
\end{eqnarray}
Hence, the shear evolution equation reduces to 
\begin{equation}
    \frac{1}{3AB^2}\left[B\dot{A}\dot{B} - A\dot{B}^2 + B^2\ddot{A} - AB\ddot{B}\right] = \frac{1}{2}\Pi\,.
\end{equation}
We can define the volume scale factor $S(t)$ for this universe by $S = \left(AB^2\right)^{1/3}\,$. Assuming that the scalar anisotropic stress $\Pi$ associated with the matter fields is negligible, it follows that
\begin{equation}
    \frac{\mathrm{d}}{\mathrm{d}t}\left(S^3 \,\Sigma\right) = 0\,.
\end{equation}
This means that as space expands, the shear decays as $S^{-3}\,$, and so space isotropises\footnote{We have verified this explicitly for the special LRS case where the scale factors in the $y$ and $z$ coordinate directions are equal for all $t\,$. However, it turns out that the same result is true in the general case \cite{Ellis_1999}.}. During matter domination one finds that at late times the ratio $\Sigma(t)/\Theta(t) \sim t^{-1}$: anisotropy is power-law suppressed.
For a $\Lambda$-dominated Bianchi {\it I} universe, the situation is even more severe, as $\Sigma/\Theta$ is exponentially suppressed \cite{gron2007einstein}.

Thus, in order to have significant late-time anisotropy in a Bianchi {\it I} cosmology, the shear must have been extraordinarily large in the early Universe. However, this is ruled out by CMB observations which indicate a highly isotropic Universe at last scattering \cite{Planck_2018}\,.
Therefore, a Bianchi {\it I} model does not really make any predictions for the present-day Universe that are meaningfully different from FLRW.

A natural follow-up question is to ask how generic a feature isotropisation is in the Bianchi cosmologies. If there is a positive $\Lambda$, as appears to be the case observationally, then Bianchi cosmologies tend to isotropise at late times towards an asymptotic de Sitter state \cite{wald1983asymptotic}. 
For a generic matter content, isotropisation is more complicated, and there remains the possibility that a Bianchi model could have a long-lived intermediate state of near-perfect isotropy, but then become highly anisotropic again in the far future \cite{lim2004asymptotic,ellis2006bianchi}. The Bianchi type ${\it VII}_h$ cosmologies can exhibit this behaviour \cite{wainwright1998isotropy}. Thus, type ${\it VII}_h$ models have attracted substantial attention, and have been adopted as canonical alternatives to FLRW in the fitting of CMB data \cite{jaffe2005evidence,jaffe2006viability,bridges2007markov,mcewen2013bayesian,saadeh2016framework,Saadeh_2016}. 
However, the more recent of these studies (particularly Ref. \cite{Saadeh_2016}) have found that Bianchi ${\it VII}_h$ cosmologies must be very nearly isotropic in the present day. 
Although there are exceptions, it certainly appears that the typical behaviour of the Bianchi cosmological spacetimes is to become isotropic at late times. Therefore, the type {\it I}, {\it V}, {\it VII} and {\it IX} models that have FLRW limits must essentially become FLRW cosmologies at late times, even though their early-time behaviour can be very different, exhibiting for example Kasner dynamics \cite{kasner1925solutions} or chaotic Mixmaster oscillations \cite{misner1969mixmaster}.

A very important subdivision of the Bianchi models is between tilted and untilted models. A Bianchi model is referred to as untilted, or orthogonal, if the 4-velocity $u^a$ of matter coincides with the normal $n^a$ to the homogeneous spacelike hypersurfaces (surfaces of transitivity). 
In a tilted cosmology \cite{King_1973,coley1996new,coley2006fluid}, this is not the case, and so an observer who is comoving with the matter field will not find their instantaneous rest spaces to be homogeneous. 
This implies that even if the matter is a perfect fluid with $T_{ab} = \rho u_a u_b + p h_{ab}\,$, observers with 4-velocity $n^a$ will generically measure non-zero momentum density $\tilde{q}_a = -T_{bc}n^b f^c_{\ a}$ (where $f_{ab}$ is the induced metric on the homogeneous surfaces) and anisotropic stress $\tilde{\pi}_{ab} = \left(f_{(a}^{\ c}f_{b)}^{\ d} - \frac{1}{3}f_{ab}f^{cd}\right)T_{cd}\,$, as well as different values for the energy density $\tilde{\rho}$ and pressure $\tilde{p}\,$.

Tilted solutions are generic to the Bianchi classification, except for type {\it I} spacetimes which are necessarily untilted \cite{King_1973}. 
In cosmology, we are often interested in solutions that admit an irrotational timelike congruence. Tilted models with $\omega_a = 0$ exist for Bianchi types {\it II}, {\it IV}, {\it V}, {\it VI} and {\it VII}, within types {\it V} and ${\it VII}_h$ admit tilted LRS solutions. We will examine a specific tilted, irrotational, type {\it V} model in detail in Chapters 7 and 8. 
Tilted models have attracted some theoretical attention in recent years, motivated at least in part by the claimed observations of anomalously large bulk flows of matter in the Universe, as described earlier in this chapter. 
Specifically, tilted type {\it V} and ${\it VII}_h$ Bianchi spacetimes have recently been proposed as viable cosmological models that naturally admit a large-scale dipole in cosmic observables, brought about by a tilted flow that can grow at late times \cite{krishnan2023dipole,krishnan2023tilt}.

In fact, near-FLRW tilted cosmologies have even been explored as possible alternatives to dark energy for explaining the observed accelerated expansion, at least within our local spatial patch of the Universe. This attention has arisen due to the fact that a large-scale, contracting, tilted flow can give rise to an apparent negative deceleration parameter within the flow \cite{Tsagas_2009,tsagas2015deceleration,tsagas2022deceleration,Santiago_2022,asvesta2022observational}, although it could be argued that the assumed existence of a canonical FLRW reference foliation makes the physical interpretation of the result unclear.

We will devote considerable study to the Bianchi cosmologies in Chapters 7 and 8, especially their LRS subclass. 
Now, we turn our attention to inhomogeneous cosmological models.

\subsection{Inhomogeneous cosmological models}\label{subsec:inhomogeneous_models}

Inhomogeneous models are of obvious interest to cosmologists, because the Universe we live in is ultimately spatially inhomogeneous, containing the vast cosmic web of structures.
However, compared to the Bianchi cosmologies which can be studied using systematic approaches (see e.g. Ref. \cite{ellis1969class}), inhomogeneous cosmologies present an even greater theoretical challenge. 
They are typically mathematically complicated, because of the low degree of symmetry in the metric, which can have at most three Killing vector fields. For a full review of inhomogeneous cosmologies, we refer the reader to Ref. \cite{krasinski1997inhomogeneous}.

The class of inhomogeneous cosmological models with the largest number of symmetries is the well-studied Lema{\^ i}tre-Tolman-Bondi (LTB) family of spacetimes  \cite{redlich2014probing,garcia2008confronting}. 
The LTB metric is
\begin{equation}\label{eq_LTB_metric}
    \mathrm{d}s^2 = -\mathrm{d}t^2 + \frac{R'(t,r)^2}{1+2E(r)}\,\mathrm{d}r^2 + R^2(t,r)\,\mathrm{d}\Omega^2\,,
\end{equation}
where a prime denotes $\partial_r$ and $\mathrm{d}\Omega^2$ is the usual line element on $S^2\,$. The metric possesses three KVFs: two for the two-dimensional homogeneous surfaces on each constant-$r$ spherical shell, and one for the local rotational symmetry. The LTB models, exhibiting spherical symmetry in their spatial sections in the canonical foliation, are closely related to spacetimes with plane symmetry, which we will study in detail in Chapter 8.

The LTB geometry is often studied as a canonical example of a cosmological model that violates the Copernican principle. Such a violation would happen if we were placed at the centre of an LTB metric, perhaps embedded within a larger-scale FLRW cosmology through an Einstein-Strauss Swiss cheese model \cite{Marra:2007pm,Vanderveld:2008vi,Biswas:2007gi,Fleury:2014gha,Koksbang:2017arw}.
This possibility caused excitement in the past due to the conjecture of the Hubble bubble \cite{zehavi1998local,conley2007there,sinclair2010residual}, where the Local Group would be situated deep within a large void. 
This could lead to a value of the Hubble constant inferred from low-redshift observations that would be lower than the CMB value, or even to a negative inferred deceleration parameter without the need for dark energy. 
However, the Hubble bubble is now believed to be largely discredited by observations \cite{camarena2022void}, especially of the kinematic Sunyaev-Zel'dovich effect \cite{bull2012kinematic}.
Despite its apparent failure to account fully for those observational problems, the LTB and LTB-Swiss Cheese families retain a great deal of interest in the cosmological community. For example, they have been used to study the effects of inhomogeneities on the Hubble diagram \cite{clifton2009hubble,Fleury:2014gha,Koksbang:2017arw,Koksbang_2020_a,Koksbang_2020_b}, structure formation \cite{Koksbang:2015ima,adamek2016spherically,marra2022behomo}, and the properties of cosmic voids \cite{alonso2010large,biswas2010testing}.

In an ideal world, we would really like to study cosmology without making any symmetry assumptions. Of course, this is an onerous task. Arguably the most tractable exact cosmological spacetimes that contain no Killing symmetries at all are the dust models developed by Szekeres \cite{szekeres1975quasispherical}, which are split into classes {\it I} and {\it II}. As an example, the metric for the Szekeres type {\it I} models can be written as \cite{pizana2022gravitational}
\begin{equation}
    \mathrm{d}s^2 = -\mathrm{d}t^2 + \frac{\left(R'(t,r) - \frac{R(t,r)\mathcal{E}'(r,y,z)}{\mathcal{E}(r,y,z)}\right)^2}{1+2E(r)}\,\mathrm{d}r^2 + \frac{R^2(t,r)}{\mathcal{E}^2(r,y,z)}\,\left(\mathrm{d}y^2 + \mathrm{d}z^2\right)\,.
\end{equation}
This metric is essentially a generalisation of the LTB metric (\ref{eq_LTB_metric}) that removes all the symmetries. 
Despite their mathematical complexity, the Szekeres dust models have the attractive property that their lack of symmetry means they can be used to model gravitational collapse and structure formation in a relativistic and non-perturbative fashion \cite{sussman2016coarse}.
We will not study the LTB or Szekeres models in this thesis, but we have described them here because they serve as a useful introduction to inhomogeneous exact solutions to Einstein's equations, which are central to the analysis carried out in Chapter 8.

For the remainder of this chapter, we will focus on the fundamental problem of how one should construct an average cosmological model on large scales, given the fundamental nonlinearity of General Relativity. This will lead us on to the notion of cosmological backreaction.
                        
\subsection{Averaging and backreaction}\label{subsec:averaging_problem}

On small scales, the Universe is very inhomogeneous. Therefore, we know that the spacetime geometry is described by some complicated inhomogeneous metric tensor $g_{ab}$. From it, we define the Einstein tensor $G_{ab}(g)$ at every point in spacetime\footnote{We have dropped the indices on the metric here, for ease of notation.}, which is manifestly nonlinear in $g_{ab}$ and its derivatives. This Einstein tensor satisfies of course $G_{ab}(g) = 8\pi G\, T_{ab}\,$.
However, in order to construct a cosmological model, one assumes that on sufficiently large scales (i.e. above the homogeneity scale $L_{\rm hom}$), the Universe can be modelled using some homogeneous model metric $\bar{g}_{ab}\,$. 
The evolution of this model geometry is assumed to be sourced by a perfect fluid energy-momentum tensor $\avg{T_{ab}}$, that arises from averaging over all the inhomogeneities in the matter distribution within $L_{\rm hom}\,$. Therefore, it is supposed that 
\begin{equation}
    G_{ab}\left(\bar{g}\right) \overset{!}{=} \avg{G_{ab}(g)} =  8\pi G\, \avg{T_{ab}}\,,
\end{equation}
where $\avg{\cdot}$ denotes some averaging procedure, of which possible definitions will be discussed in what follows.
That is, one makes the implicit assumption that the procedures of averaging the spacetime metric, and calculating the evolution of the metric through Einstein's equations, commute.
This is not true, because General Relativity is a nonlinear theory, so $G_{ab}\left(\bar{g}\right) \neq \avg{G_{ab}(g)}\,$.
Therefore, as pointed out by van den Hoogen in Ref. \cite{van_den_Hoogen_2010}, one should say instead that
\begin{eqnarray}
    G_{ab}\left(\bar{g}\right) &=& 8\pi G\,\avg{T_{ab}} + C_{ab}\,, \quad {\rm where} \\
    C_{ab} &=& G_{ab}\left(\bar{g}\right) - \avg{G_{ab}(g)}\,,
\end{eqnarray}
which specifically encodes the non-commutativity of averaging and evolution, can be thought of as a gravitational correlation tensor. 
An immediate implication of non-vanishing $C_{ab}$ is that the smoothed energy-momentum tensor $\bar{T}_{ab}(g) \overset{!}{=} \avg{T_{ab}}$ has $\bar{\nabla}^b \bar{T}_{ab} \neq 0\,$, where $\bar{\nabla}_a$ denotes the Levi-Civita connection associated with $\bar{g}_{ab}\,$ \cite{coley2010averaging}.
The non-conservation of $\bar{T}_{ab}$\,, with respect to the fictitious large-scale metric $\bar{g}_{ab}$\,, is a direct consequence of the effect of the inhomogeneities in the spacetime geometry below $L_{\rm hom}\,$ on the dynamics above $L_{\rm hom}\,$. 
This coupling between physics on different scales is reflective of the nonlinearity of full GR. One may contrast it with the situation in linear cosmological perturbation theory we saw in Chapter 3, where the evolution of modes on different scales $k$ is entirely decoupled.

Even more pertinently, the correlation tensor $C_{ab}$ tells us that the expansion of the Universe on cosmic scales, which we would typically describe with some homogeneous model $\bar{g}_{ab}$ (i.e. an FLRW or Bianchi metric), can be affected by the growth of inhomogeneities on small scales. 
This is the problem of cosmological backreaction \cite{Buchert_2012,Buchert_2015, Clarkson_2011_b, clifton2013back,van_den_Hoogen_2010}. It may be particularly crucial in the late Universe, in which we know that highly nonlinear structures form in the cosmic web. 
As we will explicitly show shortly, backreaction from inhomogeneities has been proposed as a means of generating accelerated expansion at late times, thereby playing the role of dark energy without the need to introduce any new physics \cite{Buchert_2012,Li_2008}. Indeed, the large-scale Hubble diagram in certain cosmological models can indicate backreaction effects that make the inferred deceleration parameter negative, even though space is locally decelerating \cite{Bull_2012}.
Cosmological backreaction has been studied extensively over the last 30 years, especially with regard to the accelerated cosmic expansion. Methods to investigate it have included exact GR solutions \cite{clifton2012exact,Clifton_2019,Koksbang_2019,Koksbang_2020_a,Koksbang_2020_b}, second-order CPT \cite{Clarkson_2011,Umeh_2011,Kolb_2008,Behrend_2008,Brown_2009_b}, the post-Newtonian formalism \cite{Sanghai_2016}, relativistic simulations \cite{Adamek_2018,macpherson2017inhomogeneous,macpherson2019einstein} and approaches based on direct observables \cite{larena2009testing,Heinesen_2021,Heinesen_2022,macpherson2021luminosity,Macpherson:2022eve}.

The backreaction problem is closely related to two others: the fitting problem in cosmological modelling \cite{Ellis_fitting_1987,kolb2010cosmological}, and the mathematical problem of averaging in curved spacetime \cite{van_den_Hoogen_2010}.
Let us now summarise these interrelated issues. For a more detailed exposition, see e.g. Ref. \cite{ellis2012relativistic}. The fitting problem, as defined by Ellis \& Stoeger, refers to the issue that in the real, inhomogeneous Universe, there is no unique procedure to determine the best-fit FLRW background model\footnote{The fitting problem naturally extends to other homogeneous models such as an anisotropic Bianchi cosmology.} from observations on our past light cone. 
The fitting problem has some overlap with the gauge issue in cosmological perturbation theory, because the supposed best fit model is assumed to be the fictitious, gauge-dependent, background manifold $\bar{\mathcal{M}}$ with respect to which perturbations are defined, as per Fig. \ref{fig_gauge_transformation}.

Ultimately, the fitting problem arises in the standard approach to cosmology because we take a top-down approach to cosmological model-building. We prescribe some form of $\bar{g}_{ab}$ from the start, forward-model to make predictions for observations as a function of a set of cosmological parameters, and then use our observational data to fit those parameters.
A more satisfying approach might be to adopt a bottom-up perspective, wherein we would average the spacetime geometry explicitly. Then, there is no fitting ambiguity, because no specific form of the metric has been prescribed.

However, in order to do this, one must define a sensible procedure for averaging tensor fields, such as the metric $g_{ab}$, in curved spacetime, in order to calculate $C_{ab}$. 
Tensor fields cannot be straightforwardly compared at different points in a curved spacetime \cite{van_den_Hoogen_2010}, so quantities that exist at different points in an averaging domain first have to be transported to the same point for comparison, which is a non-unique and path-dependent process. Only then can averages be defined.
Moreover, the non-commutativity of averaging and evolution, as discussed above, means that the evolution of an averaged quantity cannot be guaranteed to be the same as an average taken on some future time slice of the true evolution.
This is the notoriously difficult problem of averaging in GR, for which no covariant, exact and unique solution has been found.

Approaches to cosmological averaging can be divided into three camps:
\begin{enumerate}
    \item Procedures that average on three-dimensional spacelike hypersurfaces.
    \item Procedures that average on the past light cone directly, with a sole view to describing observables, rather than trying to model the overall cosmic expansion.
    \item Procedures that average on four-dimensional spacetime itself.
\end{enumerate}
With some exceptions, notably Zalaletdinov's macroscopic gravity approach which falls into the third camp and to which we will come soon \cite{Zalaletdinov_1997}, most successful averaging approaches have sidestepped the difficulties associated with averaging tensor fields by focusing entirely on scalars. 
Because of their coordinate invariance, covariantly defined scalars can be directly compared between spacetime points. Thus, they are much easier to average. For this reason, it has been suggested that the averaging and backreaction problems could be studied using only the properties of scalar curvature invariants \cite{coley2010averaging}.

We will devote most of our attention to the first camp, because it is perhaps most closely related to what one does in the concordance cosmology, where one studies the properties of space, evolved from one instant in time to the next. Of course, in a framework of this kind there is invariably dependence on the chosen foliation of spacetime \cite{Buchert_2018}. 
The extent of foliation dependence of spatial averaging in cosmology is debated, with some authors suggesting that as long as all possible foliations are related by non-relativistic 3-velocities, then it should not be too severe \cite{mourier2024splitting}, but others finding strong foliation dependence in the extent of backreaction inferred from numerical simulations \cite{Adamek_2018}.

The main approach from the first camp, and probably the most well-known approach to cosmological backreaction overall, is the averaging formalism developed by Buchert in an influential series of papers \cite{Buchert_1996,Buchert_2000,Buchert_2000_b,Buchert_2001}, the main results of which are collected together in Ref. \cite{Buchert_2012}. 
The central idea of Buchert's scheme is that covariantly defined scalars, calculated in the $1+3$ decomposition, can be averaged over domains $\mathcal{D}$ of spacelike hypersurfaces orthogonal to an irrotational timelike congruence. 
In Buchert's original equations, this is assumed to be the geodesic 4-velocity $u_a = -\nabla_a t$ of non-relativistic matter. The spacetime is foliated into constant-$t$ hypersurfaces, with the lapse and shift equal to unity and zero respectively. However, the equations can be generalised beyond the synchronous unit-lapse case, as we will discuss in Chapter 7.

In Buchert's dust foliation, which we will assume is valid for the remainder of this section, scalar averages are calculated on orthogonal hypersurfaces according to \cite{Buchert_2000}
\begin{equation} \label{eq_Buchert_averaging_procedure}
\avg{S}_{\mathcal{D}}(t) \equiv \frac{1}{V_{\mathcal{D}}(t)}\int_{\mathcal{D}} \mathrm{d}V \: S(t,x^i) \: = \: \frac{\int_{\mathcal{D}}\mathrm{d}^3 x\,\sqrt{^{(3)}g(t,x^i)} \, S(t,x^i)}{\int_{\mathcal{D}}\mathrm{d}^3 x\,\sqrt{^{(3)}g(t,x^i)}} \, ,
\end{equation}
where $S$ is the scalar being averaged over the spatial domain $\mathcal{D}$, which has volume $V_{\mathcal{D}}=\int_{\mathcal{D}}\mathrm{d} V$, and where $\mathrm{d} V =\mathrm{d}^3 x\,\sqrt{^{(3)}g}$ is the spatial volume element. Here the $x^i$ are coordinates on the spacelike hypersurfaces. The object $^{(3)}g$ is ${\rm det}(^{(3)}g_{ab})$, where $^{(3)}g_{ab}=g_{ab}+u_a u_b$\,.
Using the volume $V_{\mathcal{D}}(t)$, an effective scale factor $a_{\mathcal{D}}(t)$ for the domain can be defined as
\begin{equation}
a_{\mathcal{D}}(t) \equiv \left(\frac{V_{\mathcal{D}}(t)}{V_{\mathcal{D}}(t_0)}\right)^{1/3}\,.
\end{equation}
As we stated in Chapter 2, assuming that the spatial coordinates are fixed by the flow of $u^a$ gives that 
\begin{equation}\label{eq_partial_t_det3g}
    \frac{\mathrm{d}}{\mathrm{d}t}\,\sqrt{^{(3)}g} = \Theta\, \sqrt{^{(3)}g}\,,
\end{equation}
whence it follows that 
\begin{equation}
    \frac{1}{a_{\mathcal{D}}(t)}\frac{\mathrm{d}a_{\mathcal{D}}(t)}{\mathrm{d}t} = \avg{\Theta}(t)\,.
\end{equation}
Let us now apply Eq. (\ref{eq_partial_t_det3g}) to the time derivative of Eq. (\ref{eq_Buchert_averaging_procedure}), with the averaging domain $\mathcal{D}$ being propagated by the flow of $u^a\,$.
Doing so, we obtain Buchert's commutation rule for time derivative and averaging operations \cite{Buchert_2000},
\begin{equation} \label{eq_Buchert_commutation_rule}
\partial_t\avg{S}_{\mathcal{D}}(t) - \avg{\partial_t S(t,x^i)}_{\mathcal{D}}  = \avg{S\Theta}_{\mathcal{D}} - \avg{S}_{\mathcal{D}}\avg{\Theta}_{\mathcal{D}} = {\rm Cov}_{\mathcal{D}}(\Theta,S) \, ,
\end{equation}
where we have defined the covariance as ${\rm Cov}_{\mathcal{D}}(S_1, S_2) \equiv  \avg{S_1 S_2}_{\mathcal{D}} - \avg{S_1}_{\mathcal{D}} \avg{S_2}_{\mathcal{D}}\,$. We can also introduce the variance of a scalar over the domain, ${\rm Var}_{\mathcal{D}}(S) \equiv {\rm Cov}_{\mathcal{D}}(S,S)\,$.

The $1+3$-covariant scalars that would define a matter-dominated FLRW universe are the energy density $\rho\,$, the expansion scalar $\Theta\,$, and the Ricci 3-curvature $^{(3)}R$ associated with $^{(3)}g_{ab}\,$.
The scalar equations of motion for those variables are the Raychaudhuri equation (\ref{eq_Raychaudhuri}), the Hamiltonian constraint equation (\ref{eq_Hamiltonian_constraint_1+3}), and the contracted Bianchi identity/local energy conservation equation (\ref{eq_energy_conservation_equation}).
Averaging these according to Buchert's procedure gives the Friedmann-like equations
\begin{eqnarray}\label{eq_Buchert_1}
    3\left(\frac{\dot{a}_{\mathcal{D}}}{a_{\mathcal{D}}}\right)^2 &=& 8\pi G\,\avg{\rho} - \frac{1}{2}\avg{^{(3)}R} - \frac{1}{2}\mathcal{Q}_{\mathcal{D}}\,,\\
    \label{eq_Buchert_2} \frac{3\,\ddot{a}_{\mathcal{D}}}{a_{\mathcal{D}}} &=& -4\pi G\,\avg{\rho} + \mathcal{Q}_{\mathcal{D}}\,.
\end{eqnarray}
The cosmological constant has been set to zero for simplicity, and we have dropped the $\mathcal{D}$ subscripts on the averages for the sake of simplicity of presentation.
The term $\mathcal{Q}_{\mathcal{D}}$ is referred to as the kinematical backreaction. It is given by
\begin{equation}
    \mathcal{Q}_{\mathcal{D}} := \frac{2}{3}\left(\avg{\Theta^2} - \avg{\Theta}^2\right) - 2\avg{\sigma^2}\,,
\end{equation}
where $\sigma^2$ is the shear scalar associated with $u^a\,$.
The backreaction scalar can be readily interpreted physically: it encodes directly the effect of inhomogeneities on the large-scale isotropic expansion.

In particular, one sees from Eq. (\ref{eq_Buchert_2}) that if $\mathcal{Q}_{\mathcal{D}}$ is sufficiently large and positive, then it can drive apparent acceleration, $\ddot{a}_{\mathcal{D}} > 0$, in the expansion of some large domain of the universe \cite{Buchert_2012, Bull_2012}, and this could happen without a cosmological constant. 
This demonstrates why properly accounting for the effects of backreaction on the emergent cosmic expansion might, according to some authors, circumvent the need for dark energy \cite{Rasanen_2004,Rasanen_2006,Buchert_2007,kolb2011backreaction}.
In Chapter 7 we will develop a novel averaging scheme, based on Buchert's formalism, that can be used to study backreaction in anisotropic universes, rather than ones that are necessarily assumed to be phenomenologically close to FLRW on large scales. We will explore some observational consequences of that scheme in Chapter 8.

Although Buchert's averaging procedure is very powerful, it has three main drawbacks:
\begin{itemize}
    \item The averaged equations are not closed.
    \item The results are foliation-dependent.
    \item The covariant averages $\avg{S}(t)$ are defined on spacelike hypersurfaces, rather than on the past light cone, so they are not observable.
\end{itemize}
The first of these three drawbacks is unavoidable in any averaging formalism, because we necessarily lose information about small-scale physics. Moreover, we deliberately remove any knowledge of the vector and tensor degrees of freedom, but these clearly affect the large-scale dynamics, through e.g. the shear tensor $\sigma_{ab}$ that enters into the kinematical backreaction scalar.

An attempt to resolve the second drawback is provided by Gasperini and collaborators \cite{gasperini2009gauge,gasperini2010covariant}. They average covariantly defined scalars in a similar way to Buchert, but generalise the procedure to four spacetime dimensions.
The idea is to define suitable spacetime window functions $W_{\Omega}(x^a)$ that pick out submanifolds $\Omega$ embedded in the spacetime manifold $\mathcal{M}$, through integrals of the kind
\begin{equation}
    I(S, \Omega) = \int_{\mathcal{M}}\mathrm{d}^4x\,\sqrt{-g(x^a)}\,W_{\Omega}(x^a)\,S(x^a)\,.
\end{equation}
In order for $\Omega$ to be a patch of a spacelike hypersurface that is the level surface of some function $A(x^a)$ with timelike gradient $n_a = -\dfrac{\nabla_a A}{\sqrt{-\nabla_b A \nabla^b A}}$\,, the window function is chosen to be
\begin{equation}
    W_{\Omega}(x^a) = \left(n^b \nabla_b\, \theta(A(x^a) - A_0)\right)\,\theta(r_0 - B(x^a))\,,
\end{equation}
where $B(x^a)$ is a function with spacelike gradient that defines the spatial extent of the patch on the $A = A_0$ surface, and $\theta(x)$ is the Heaviside step function.
Thus, averages may be computed over level surfaces of any timelike foliation $A(x^a)$\,, by
\begin{eqnarray}
    \avg{S}_{A_0} &=& \frac{I(S, A_0)}{I(1, A_0)}\,, \quad {\rm where} \\
    \nonumber I(S, A_0) &=& \int_{\mathcal{M}}\mathrm{d}^4x\,\sqrt{-g}\,\sqrt{-\nabla_b A \nabla^b A}\,\delta\left(A - A_0\right)\,\theta\left(r_0 - B\right)\, S\,.
\end{eqnarray}
The results are not restricted to any particular foliation, and allow different foliation choices to be compared systematically \cite{mourier2024splitting}.
However, this approach is less directly applicable than Buchert's, as the averages that one obtains are hard to interpret physically.

The third drawback of the Buchert formalism leads us on to the second camp of approaches to cosmological averaging, where one restricts all attention to the past light cone (PLC), in order to model the observables that would be associated with some large-scale average description of the Universe. 
The most direct averaging formalism of this kind is very closely related to Gasperini's window-function approach, with the difference that rather than the window function picking out spacelike hypersurfaces, it picks out spacetime regions directly related to the properties of the PLC \cite{gasperini2011light,fanizza2020generalized}. 
Natural choices for that region of interest might include either the entire volume of the PLC, truncated on some spacelike hypersurface in the past, or a 2-sphere within the PLC, with obvious utility for Hubble diagram \cite{ben2012backreaction} and CMB observations.
Other approaches to averaging on the past light cone have included adapting Buchert's averaging procedure to averages on the screen space of null geodesic congruences \cite{buchert2023averaging}, and proposals that work directly with the statistics of the Hubble diagram \cite{Bull_2012}.

Next, let us consider the third camp of cosmological averaging procedures, which attempt to average the 4-dimensional spacetime geometry itself. The hope is that doing this would allow one to actually construct the gravitational correlation $C_{ab}\,$, rather than just inferring its effect indirectly through proxies such as Buchert's $\mathcal{Q}_{\mathcal{D}}\,$.
The first covariant procedure of this kind was developed by Isaacson to study gravitational radiation \cite{isaacson1968gravitational_1,isaacson1968gravitational_2}, rather than cosmology. As one might expect for an analysis of gravitational waves, Isaacson's treatment is fundamentally perturbative. 
However, its ideas were successfully adapted into a non-perturbative spacetime averaging scheme, known as the theory of macroscopic gravity, by Zalaletdinov \cite{Zalaletdinov_1997,Mars_1997,Coley_2005}.
In this approach, one defines the average of a tensor field $t^a_{\ b}\,$, at a point $x$ which is associated with an averaging domain $\Sigma_x$ in its neighbourhood, by
\begin{equation}
    \avg{t^a_{\ b}(x)} = \frac{\int_{\Sigma_x}\mathrm{d}^4x'\,\sqrt{-g(x')}\, \mathcal{A}^a_{\ c}\left(x,x'\right)\,\mathcal{A}^d_{\ b}\left(x,x'\right)\, t^c_{\ d}(x')}{\int_{\Sigma_x}\mathrm{d}^4x'\,\sqrt{-g(x')}}\,,
\end{equation}
with natural generalisations to tensors of other rank. 
Here the $\mathcal{A}^a_{\ b}(x,x')$ are called bilocal operators, and their role is to transport the tensor being considered from the spacetime point $x'$ to the point $x$, where the average is defined. 
Note that this averaging procedure does not involve a foliation of spacetime at any stage.
The equations that result from this approach are known as macroscopic field equations, as they describe the emergent dynamics of an average metric at every point in spacetime. Those dynamics are sourced both by the average energy-momentum tensor and by a series of bilocally-averaged objects that together play the role of $C_{ab}$.
The macroscopic Zalaletdinov equations are genuine tensor equations, rather than being restricted to scalars. However, they are very mathematically complex, and are harder to interpret physically than Buchert's equations.

Finally, let us briefly consider an approach to the cosmological backreaction problem that is not based on averaging, but instead studies non-perturbatively the effects of arbitrarily nonlinear small-scale inhomogeneities on a large-scale ``background''. 
This formalism was developed by Green \& Wald \cite{Green_2011,Green_2013,Green_2014}. It is based on the reformulation by Burnett \cite{burnett1989high} of Isaacson's short-wavelength gravitational-wave averaging scheme.

Green \& Wald's proposal is based on a splitting of the metric into a background and a deviation from the background, $g_{ab}(x,\lambda) = g^{(0)}_{ab}(x) + \gamma_{ab}(x,\lambda)\,$. The parameter $\lambda$ is defined so that $g_{ab}^{(0)}(x) = \underset{\lambda \rightarrow 0}{\rm lim} g_{ab}(x,\lambda)\,$.
Then, given a set of four postulates that we will not state explicitly here (but which essentially demand that $\gamma_{ab}$ and its first covariant derivatives are bounded) they show that the backreaction effect of $\gamma_{ab}$ on the evolution of $g^{(0)}_{ab}$ can be summarised through the tensor
\begin{equation}\label{eq_green_wald_mutensor}
    \mu_{abcdef} = \underset{\lambda \rightarrow 0}{\rm wlim}\left[\nabla_a \gamma_{cd}(x,\lambda)\, \nabla_b \gamma_{ef}(x,\lambda)\right]\,,
\end{equation}
where we refer the reader to Ref. \cite{Green_2011} for a precise definition of the weak limit.
The effect of $\mu_{abcdef}$ on the cosmological background is communicated through the effective gravitational energy-momentum tensor,
\begin{eqnarray}\label{eq_green_wald_tab_eff}
    t^{(0)}_{ab} &=& \frac{1}{8}\left[2\mu^{cd \ \ e}_{\ \ \ c \ \ de} -\mu^{c \ \ de}_{\ c \ \ \  de} - \mu^{c \ \ d \ \ e}_{\ c \ \ d \ \ e}\right]g^{(0)}_{ab} + \frac{1}{2}\mu^{cd}_{\ \ \ acbd} - \frac{1}{2}\mu^{c \quad \,\, d}_{\ \ ca \ \ bd} \\
    \nonumber &&  + \frac{1}{4}\mu_{ab \ \ \ cd}^{\ \ \ cd} - \frac{1}{2}\mu^{c \qquad \, d}_{\ \ (ab)c \ \ d} + \frac{3}{4}\mu^{c \quad \ \, \, d}_{\ \ cab \ \ d} - \frac{1}{2}\mu^{cd}_{\ \ \ abcd}\,,
\end{eqnarray}
which is trace-free (so it behaves like gravitational radiation), and, if interpreted as an energy-momentum tensor, satisfies the weak energy condition. 
Then, the Einstein equations for $g^{(0)}_{ab}$ are 
\begin{equation}\label{eq_green_wald_einstein_eq}
    G_{ab}(g^{(0)}) = 8\pi G\,T_{ab}^{(0)} + 8\pi t^{(0)}_{ab}\,.
\end{equation}
Using this formalism, Green \& Wald argued that backreaction is negligible in cosmology \cite{green2015comments,green2016simple}, because of the trace-free nature of $t^{(0)}_{ab}\,$. 
This claim was challenged by Buchert et al. \cite{Buchert_2015}, who constructed an effective gravitational energy-momentum tensor arising from backreaction on an FLRW cosmology with non-zero trace. They argued that a local formulation of the backreaction problem, as in Eq. (\ref{eq_green_wald_einstein_eq}), is unsuitable, because the backreaction problem is inherently non-local.
Ostrowski suggested that Green \& Wald's use of weak limits is problematic, because it does not define a coarse-grained description of the spacetime, which makes its interpretation for cosmological modelling unclear \cite{ostrowski2018green}.

Moreover, Clifton \& Sussman \cite{Clifton_2019} showed that in a variety of spherically-symmetric and plane-symmetric cosmological models, $t_{ab}^{(0)}$ vanishes irresepectively of the choice of background metric for the same underlying spacetime, even if the ``backgrounds'' can have very different physical characteristics (e.g. either an FLRW or an LTB background).
In contrast, Buchert's averaging-based formalism provided a well-defined prescription for the behaviour of the large-scale cosmology. They concluded that the Green \& Wald formalism, although useful for studying nonlinear perturbations to a known background, does not really study the same cosmological backreaction problem as the averaging-based procedures, because there is no clear way to identify what the cosmological background should be.
For these reasons, we will not study local approaches to the cosmological backreaction problem further. Instead, we will focus on non-local techniques that make use of explicit averaging on spacelike hypersurfaces, according to Buchert's scheme (\ref{eq_Buchert_averaging_procedure}). These ideas will be central to Chapters 7 and 8.

This concludes our discussion of alternative cosmological models, and indeed the pedagogical first part of the thesis. The four chapters that follow will be centred around research into alternatives to the concordance model. 

\chapter{Parameterised post-Newtonian cosmology}

\lhead{\emph{Parameterised post-Newtonian cosmology}}

We will now introduce and extend the framework of parameterised post-Newtonian cosmology (PPNC), first developed in Refs. \cite{Sanghai_2015,Sanghai_2017,Sanghai_2019}. 
Section \ref{sec:basic_PPN} is an introduction to the PPNC formalism, as devised by the authors of those papers. Sections \ref{sec:momentum_constraint} and \ref{sec:scale_dependence} are based on research by myself and my direct collaborators Timothy Clifton and Daniel B. Thomas, as presented in Refs. \cite{anton2022momentum} and \cite{Thomas_2023} respectively.

The central idea behind the approach is that the parameterised post-Newtonian (PPN) formalism is by far the most successful theory-independent framework for studying tests of gravity in astrophysical settings. 
The formalism is very simple, expressing the entire phenomenology of relativistic gravity in terms of the set of 11 parameters we introduced in Section \ref{subsec:PPN}.
Its insensitivity to the finer details of gravitational theories means that observational constraints on the PPN parameters can be used to constrain the underlying theory parameters of a wide array of different modified theories of gravity.

To date, no similarly successful theory-independent framework exists in cosmology.
Therefore, it is desirable to use the ideas from the PPN formalism to build such an approach that is valid over cosmological length and time scales, rather than just in isolated astrophysical systems.
Then, we would have a unified version of the PPN formalism that can be used to test gravity on cosmological scales, while reducing to standard PPN on small scales, so that the equations remain valid in the regime of nonlinear density contrasts.
For the reasons laid out in Section \ref{subsec:PPN}, this is a difficult task, particularly because of the enormous range of scales involved in the problem.
However, such an approach, the PPNC formalism, was formulated by Clifton \& Sanghai \cite{Sanghai_2015,Sanghai_2017,Sanghai_2019}. We will first summarise its key features, as of the beginning of my PhD research. 
Then, we will discuss why it must be extended \cite{anton2022momentum}, to include the gravitational effects of the momentum of matter fields. This will naturally lead us on to considerations of vector perturbations and preferred-frame effects. 
Finally, we will study the scale and time dependence of the coupling functions in the PPNC formalism \cite{Thomas_2023}, which are intimately related to the familiar PPN parameters. This will involve a detailed application of the formalism to Bergmann-Wagoner scalar-tensor theories of gravity, as a fiducial test case. 

\section{Adapting the PPN formalism for cosmology}\label{sec:basic_PPN}

The PPNC formalism is a bottom-up approach to cosmology that explicitly allows for the presence of nonlinear, inhomogeneously distributed matter, while carefully discarding the assumptions of asymptotically flat spacetime and negligible time-variation of the cosmological background that exist within the classical PPN formalism \cite{Will_1993}. 
It generalises in a more physically transparent fashion the effective Newton's constant $\mu(\tau, k)$ and gravitational slip $\Sigma(\tau, k)$ from the parameterised post-Friedmann formalism \cite{hu2007models,Hu_2008,amin2008subhorizon,Skordis_2009,Baker_2011,Baker_2013,bakerbull}, whose drawbacks we discussed in Section \ref{subsec:PPF}. 
It shows that the generalisations of these quantities, which in the PPF approach are explicitly tied only to sub-horizon linear scalar perturbations, are closely linked to parameters very much like the classical PPN parameters, which we call PPNC parameters. 
Moreover, the PPNC parameters appear directly in the governing equations for both the FLRW background expansion and the large-scale, time-dependendent parts of the linear perturbation equations.

The basic starting point for the PPNC framework is that both the post-Newtonian regime and the regime of linear cosmological perturbations are examples of weak-field gravity.
This means that there exist coordinate systems in which the metric can be written as
\begin{equation}
g_{ab} = g^{(0)}_{ab} + \delta g_{ab}\, , \qquad {\rm where} \qquad g^{(0)}_{ab} \sim 1 \quad {\rm and} \quad \delta g_{ab} \ll 1. 
\end{equation}
In the classical PPN formalism, $g_{ab}^{(0)}$ is the Minkowski metric, whereas in cosmological perturbation theory it is an FLRW metric. 

The weak-field treatment, common to both CPT and post-Newtonian expansions, is justified by the leading-order part of the gravitational fields of all astrophysical systems except black holes and neutron stars being $\ll 1$\,. 
However, CPT and post-Newtonian expansions differ not only in the geometry of the assumed background. They have a fundamentally different perturbative hierarchy, with the power-counting smallness parameter $\epsilon$ associated with $\delta g_{ab}$ being the density contrast $\delta$ for CPT, but the non-relativistic velocity scale $v$ for post-Newtonian gravity.

Let us first clarify exactly why an approach purely based on cosmological perturbation theory is unsuitable if one wishes to make contact with the post-Newtonian regime. 
Expressed in Newtonian gauge, we recall that the CPT metric is
\begin{equation} \label{eq_pertrw}
\mathrm{d}s^2 = a(\tau)^2 \left[ - (1-2 \h{\Phi}) \mathrm{d}\tau^2 +\left( (1+2 \h{\Psi}) \delta_{ij} + \h{F}_{ij} \right) \mathrm{d}\h{x}^i \mathrm{d}\h{x}^j  + 2 \h{B}_i d\h{\tau} d\hat{x}^i \right] \, ,
\end{equation}
where $\h{B}_i$ is divergenceless and $\h{F}_{ij}$ is transverse and tracefree. We will come back shortly to the suitability of the Newtonian gauge for describing gravitational physics in the presence of nonlinear structures. 
Note that we have used hats on all the quantities, in order to distinguish quantities associated with a perturbed FLRW geometry from those associated with a perturbed Minkowski geometry, which will be unhatted. 
The only exception to this rule is the conformal time coordinate $\tau$\,, which refers to the FLRW cosmology but which we have left unhatted, because it has no analogue in Minkowski spacetime, and therefore there is no ambiguity associated with it. 
We have also assumed that the FLRW background is spatially flat, motivated by the constraints on $\Omega_K$ discussed in Chapter 3.

The perturbative orders of smallness of all fields in this approach are taken to be similar, including the fluctuations in the density contrast, $\delta$, and the 3-velocities of matter fields, $\h{v}^i$, such that
\begin{equation}\label{eq_CPT_order_of_smallness}
\h{\Phi} \sim \h{\Psi} \sim \h{B}_i \sim \h{F}_{ij} \sim \delta \sim \h{v}^i  \ll 1 \,.
\end{equation}
The field equations of any theory of gravity can then be used to find constraint and evolution equations for the background quantities, and subsequently those of all first and higher-order perturbations, as we showed in Section \ref{subsec:CPT}.

However, the equations of motion for the perturbations, derived according to the CPT scheme in Eq. (\ref{eq_CPT_order_of_smallness}), invariably break down on small scales, for the following reasons:
\begin{itemize}
    \item The density contrast is required to be perturbatively small.
    \item The 3-velocity of matter fields is expected to remain as small as the amplitude of gravitational potentials.
\end{itemize} 
Neither of these things is true when we consider scales $\lesssim 100 \, {\rm Mpc}$ in the real Universe, where we can observe density contrasts $\delta\sim 1$ or greater on scales $\lesssim 10 \, {\rm Mpc}$, and where we typically have $\h{v}^2 \sim \h{\Phi}$, due to the Virial theorem. 
This failure means that we cannot use linear cosmological perturbation theory to reliably model gravitational interactions on these scales. Consequently, we face a challenge if we wish to try to use it to relate any parameterised framework for gravity in cosmology to results that we might obtain, for example, from experiments in the Solar System. 
Thus, it is very hard to conceive of a theory-independent parameterised framework for constraining gravity on all scales using cosmological perturbation theory alone.

Post-Newtonian approaches, on the other hand, are perfectly valid for arbitrarily large densities, as long as gravitational fields are weak. However, they are underpinned by the slow-motion requirement (\ref{eq_PPN_hierarchy_time_vs_space_derivatives}),
which is clearly problematic for cosmology, as the Hubble flow velocity increases in proportion to distance, and approaches $\sim 1$ on the scale of the horizon. 

We recall from Section \ref{subsec:PPN}, with a slight change of notation, that the post-Newtonian metric can be written to $\mathcal{O}(v^3)$ as 
\begin{equation}\label{eq_pert_m}
\mathrm{d}s^2 = -(1-2 \Phi)\, \mathrm{d}t^2 + (1+2 \Psi)\,\delta_{ij} \,\mathrm{d}x^i \mathrm{d}x^j + 2 B_{i} \, \mathrm{d}t \mathrm{d}x^i \, ,
\end{equation}
where in this expression we have $\Phi\sim \Psi \sim v^2$ and $\vert B_{i} \vert \sim v^3$, such that the vector gravitational potentials are smaller in magnitude than their scalar counterparts. The transverse and tracefree tensor perturbations $h_{ij}$ are smaller still, entering the post-Newtonian expansion at $\mathcal{O}(v^4)$ and only becoming dynamical at $\mathcal{O}(v^6)\,$, so they have been neglected. 
With the additional hierarchy (\ref{eq_PPN_hierarchy_time_vs_space_derivatives}) between temporal and spatial derivatives, we can see that the structure of post-Newtonian theory is rather different from that of cosmological perturbation theory.

While the line element given in Eq. (\ref{eq_pert_m}) cannot be used to directly describe an entire cosmology, it can be safely applied within a region of spacetime that is small compared to the cosmological horizon, so long as the Hubble flow velocity within that region is of order $v \ll 1$ (if this is not the case, then the slow motion requirement is violated). 
By considering many such regions next to each other, one can then construct a viable cosmological model \cite{Sanghai_2015, Sanghai_2017}. This requires applying appropriate boundary conditions (the Israel junction conditions \cite{Israel_1966}) between each of the regions. Carrying out this procedure, one finds that the large-scale cosmological dynamics emerge from the post-Newtonian gravitational fields that reside within each constituent region. 
This is a construction known as post-Newtonian cosmology, and has been investigated thoroughly in the context of Einstein's equations \cite{Sanghai_2015, Sanghai_2016}. By specifying the governing equations for the perturbations to Minkowski spacetime in terms of the PPN parameters, the post-Newtonian cosmology construction can be used to derive equations for a parameterised post-Newtonian cosmology, without specifying a theory of gravity.

The basic, seemingly na{\" i}ve observation one can make is that the line elements in Eqs. (\ref{eq_pertrw}) and (\ref{eq_pert_m}) appear rather similar, and ought to be described by similar equations on small scales deep within the Hubble horizon.
This link can be made explicit by the coordinate transformation
\begin{eqnarray}\label{eq_statictoexpandingtransformation1}
t &= \hat{t}+\frac{1}{2}\,a^2 H\, \hat{r}^2 + T(\hat{t},\hat{\mathbf{x}}) + \mathcal{O}(v^5) \\
\label{eq_statictoexpandingtransformation2} x^i &= a\,\hat{x}^i\left[1+\frac{1}{4}\,a^2 H^2 \,\hat{r}^2\right] + \mathcal{O}(v^4)\,,
\end{eqnarray}
where $T$ is an as-yet-unspecified gauge function of order $v^3$, $\hat{r}^2 \equiv \delta_{ij} \h{x}^i \h{x}^j$, and we have used cosmic time $\h{t}$ rather than conformal time $\h{\tau}$, for the sake of mathematical simplicity. 
This coordinate transformation demonstrates explicitly that a perturbed Minkowski geometry and a perturbed FLRW geometry are locally isometric, with the isometry breaking down once the $v \ll 1$ assumption no longer applies.

Under such a transformation, the line element (\ref{eq_pert_m}) can be directly transformed into the form of the perturbed FLRW geometry (\ref{eq_pertrw}), as long as we take \cite{Sanghai_2015}
\begin{eqnarray}\label{expandingvsstatic}
{\Phi} &=& \h{\Phi}+\frac{\ddot{a} \, a}{2}\hat{r}^2 \\ \label{expandingvsstatic2}
\Psi &=& \hat{\Psi} - \frac{\dot{a}^2}{4}\hat{r}^2 \\ \label{expandingvsstatic3} B_i &=& \hat{B}_i - 2 \dot{a} \, \hat{x}^j\delta_{ij}\left(\hat{\Phi}+\hat{\Psi}\right) - a \, \dot{a} \, \ddot{a} \, \hat{r}^2\hat{x}^j\delta_{ij} + \frac{1}{a}T_{,i}\, ,
\end{eqnarray}
and $\h{F}_{ij}=0$, and subsequently transform to conformal time. 
The function $T$ will be used in the next section to enforce the Newtonian gauge condition that $\h{B}_i$ is divergenceless. 
We must use the Newtonian gauge throughout our treatment of parameterised post-Newtonian cosmology, because it remains valid under gauge transformations in both the cosmological and post-Newtonian perturbative hierarchies. 
This makes it rather unique, as most standard gauge choices in CPT, such as the synchronous, comoving orthogonal, spatially flat and uniform density gauges, cannot possibly be implemented by a post-Newtonian gauge transformation \cite{Clifton_2020, goldberg2017cosmology,goldberg2017perturbation}.

Eqs. (\ref{expandingvsstatic}--\ref{expandingvsstatic3}) allow us to rewrite the PPN equations of motion for the Minkowski metric perturbations on small scales in terms of cosmologically suitable quantities, namely perturbations to an FLRW metric. 
These equations are valid in small regions where the expansion of the Universe is locally insignificant. However, using the PPNC construction we just described, we can stitch many such regions together in order to build equations of motion for $\h{\Phi}\,$, $\h{\Psi}$ and $\h{B}_i$\,.
As long as the coordinate patches of neighbouring regions overlap, which can be arranged by a suitable choice of $a(\tau)$\,, we can then consider the $(\tau,\h{x}^i)$ coordinate system to span the entire spacetime. Thus, they provide a coordinate system for the cosmological background.
This formulation of post-Newtonian gravity allows the gravitational fields of highly nonlinear density contrasts to be consistently modelled, and simultaneously allows the Friedmann equations to be extracted from them. 
It is therefore ideal for creating a unified framework for testing gravity in both isolated astrophysical systems, and in cosmology on the very largest scales.

Let us now introduce the basic equations of the PPNC framework, which require one to consider the PPN equations only to $\mathcal{O}(v^2)$. We will show, however, that these equations are insufficient to fully characterise the evolution of weak-field perturbations. 
It is necessary to study higher-order terms in the post-Newtonian expansion. These introduce additional complexity, which we will handle in Section \ref{sec:momentum_constraint}.

\subsection{The basic PPNC equations}

The foundational equations of parameterised post-Newtonian cosmology are obtained by considering the PPN equations of motion for the leading-order corrections $\sim v^2$ to the Minkowski metric,
\begin{eqnarray} \label{eq_classic_PPN} 
    \nabla^2 \Phi(t, x^i) = -4\pi G \,\alpha\,\rho(t, x^i)\,, \quad {\rm and} \quad \nabla^2 \Psi(t, x^i) = -4\pi G\,\gamma\,\rho(t,x^i)\,.
\end{eqnarray}
We need to modify these in two ways:
\begin{enumerate}
    \item An extra, spatially constant, contribution, that does not appear in the classic PPN parametrisation, should be added linearly to the RHS of each of these equations. This is required in order to be able to consistently include dark energy, and the gravitational effects of the time variation of any extra degrees of freedom in a theory. 
    We know that the contribution must only be time-dependent, because this means that the only additional contributions to $\Phi$ and $\Psi$ that we can get will be of the form $c(t)\, \h{r}^2$\, (any solution to the Laplace equation, that could otherwise be added linearly to solutions to the Poisson equation, is removed by the periodic boundary conditions in the PPNC stitching procedure). Inspecting Eqs. (\ref{expandingvsstatic}-\ref{expandingvsstatic2}), one sees that these contributions will then simply give $\Phi = \h{\Phi} + \left(\dfrac{\ddot{a}a}{2} + c_1(t)\right)\h{r}^2\,$, and $\Psi = \h{\Psi} - \left(\dfrac{\dot{a}^2}{4} - c_2(t)\right)\h{r}^2\,$. Hence, these novel contributions will be spatially homogeneous, and appear directly where we expect them to in the Friedmann equations, so they behave exactly as desired for a purely cosmological, dark energy-like field.
    \item The PPN parameters should be functions of time, in order for the cosmological model that we construct to be self-consistent \cite{Sanghai_2017}. The requirement that $\alpha \longrightarrow \alpha(t)$ and $\gamma \longrightarrow \gamma(t)$  is justified by considering that in the classical PPN formalism, any variation in these parameters is assumed to be much slower than the characteristic timescale associated with the astrophysical system in question. However, when the characteristic timescale is taken to be the Hubble time, as is the case in cosmology, then this assumption cannot be made {\it a priori}. For example, in a scalar-tensor theory of gravity, the PPN parameters are related to the slowly varying background value of the scalar field, whose time derivatives are expected to comparable to $H\,$. 
\end{enumerate}
The equations in (\ref{eq_classic_PPN}) become therefore
\begin{eqnarray}
    \nabla^2 \Phi(t, x^i) &=& - 4\pi G \,\alpha(t)\,\rho(t,x^i) + \alpha_c(t)\,, \quad {\rm and} \\
    \nabla^2 \Psi(t, x^i) &=& - 4\pi G\,\gamma(t)\,\rho(t,x^i) + \gamma_c(t)\,.
\end{eqnarray}

It should be noted that the pressure of matter fields need not be neglected at leading order when considering the cosmological context, but that on the spatial scales on which post-Newtonian expansions can be applied (i.e. $\lesssim 100 \, {\rm Mpc}$) it must be effectively spatially constant \cite{goldberg2017perturbation, Sanghai_2016}. 
This is because, if one has a fluid with pressure $p$ of the same order of magnitude as its energy density, i.e. $p \sim \rho \sim v^2\,$, then a post-Newtonian expansion of the relativistic Euler equation (\ref{eq_momentum_conservation_equation}) shows that $D_i\, p = p_{,i} = 0$ at $\mathcal{O}(v^2)\,$, as all other terms in Eq. (\ref{eq_momentum_conservation_equation}) are of order $v^4$ or higher. Hence, for a relativistic fluid, $p \equiv p(t)$ only\footnote{Given an equation of state $p \equiv p(\rho)\,$, it follows that the fluid also has $\rho \equiv \rho(t)$ only.}. 
The gravitational effect of isotropic pressure is important if we wish to include radiation fields (with $p = \dfrac{1}{3}\,\rho\,$) in our cosmological model, as is necessary in order to study observations of the cosmic microwave background in the PPNC context.

Let us now apply the transformations (\ref{eq_statictoexpandingtransformation1}) and (\ref{eq_statictoexpandingtransformation2}), identify the small-scale Bardeen potentials via Eqs. (\ref{expandingvsstatic}) and (\ref{expandingvsstatic2}), and finally take the spatial Laplacian $\nabla^2$ with respect to the non-expanding coordinates $x^i\,$\,. 
At leading PN order, this is related to the spatial Laplacian $\h{\nabla}^2$ with respect to the expanding coordinates simply by $\h{\nabla}^2 = a^2 \nabla^2\,$.
We get the equations
\begin{eqnarray}
    -4\pi G \alpha \rho + \alpha_c &=& \frac{1}{a^2}\h{\nabla}^2\h{\Phi} + \frac{3\ddot{a}}{a}\,, {\rm and} \label{eq_integrated_PPN_1} \\
    -4\pi G \gamma\rho + \gamma_c &=& \frac{1}{a^2}\h{\nabla}^2\h{\Psi} - \frac{3H^2}{2a^2}\, \label{eq_integrated_PPN_2}.
\end{eqnarray}
Now, consider integrating these over some domain $\mathcal{D}$ of a constant-$\h{t}$ spacelike hypersurface, with boundary $\partial\mathcal{D}\,$. We can split the density field into $\rho\left(\h{t}, \h{x}^i\right) = \bar{\rho}\left(\h{t}\right) + \delta\rho\left(\h{t},\h{x}^i\right)\,$, where 
\begin{equation}
    \bar{\rho} = \frac{1}{V_{\mathcal{D}}}\int_{\mathcal{D}}\mathrm{d}V\,\rho\,, \quad {\rm and} \quad \int_{\mathcal{D}}\mathrm{d}V\,\delta\rho = 0\,.
\end{equation}
Here the overall size of the domain is such that it is well below the horizon, but above the homogeneity scale. Note that we have never required that the inhomogeneous part $\delta\rho$ of the energy density be small, only that it averages to zero over $\mathcal{D}\,$.

It follows that the equations that describe the expansion of the domain must precisely be the emergent Friedmann equations of the cosmic expansion. 
To get these, we apply periodic boundary conditions, as required by the symmetry of the situation we are considering. These tell us that
\begin{equation}
    \int_{\mathcal{D}}\mathrm{d}V\,\h{\nabla}^2\h{\Phi} = \int_{\partial\mathcal{D}}\mathbf{\mathrm{d}A}{\bf \cdot}\h{\nabla}\h{\Phi} = 0\,,
\end{equation}
and similarly for $\h{\Psi}\,$.
It follows that the appropriate Friedmann equations describing the emergent scale factor $a(\h{t})$ are \cite{Sanghai_2016}
\begin{eqnarray}
\mathcal{H}^2 &=& \frac{8\pi G \gamma}{3} \, \bar{\rho}a^2 - \frac{2\gamma_c a^2}{3}\, \quad {\rm and} \label{eq_parametrisedfriedmanneqns1} \\ 
\mathcal{H}' &=& -\frac{4\pi G \alpha}{3} \, \bar{\rho}a^2 + \frac{\alpha_c a^2}{3} \, ,\label{eq_parametrisedfriedmanneqns2}
\end{eqnarray}
where it should be noted that the modifications to the coupling of matter density $\bar{\rho}$ in the first and second Friedmann equations are different: the (generalisation of the) PPN light-bending parameter $\gamma$ in the first, and the PPN effective Newton's constant $\alpha$ in the second.
We see that a purely phenomenological approach, where one just multiplies $\bar{\rho}$ everywhere by some $G_{\rm eff}\,$, would give an incorrect set of equations for the cosmic expansion. This demonstrates the importance of considering the full PPN parameterisation.

Spatially averaging the continuity equation in the same way gives that for non-relativistic matter,
\begin{equation}
    \bar{\rho}' + 3\mathcal{H}\bar{\rho} = 0\,.
\end{equation}
The time dependences of the PPNC parameters are not independent. In order for the background contribution (\ref{eq_parametrisedfriedmanneqns1}) to the Hamiltonian constraint to be the first integral of the background Raychaudhuri equation (\ref{eq_parametrisedfriedmanneqns2}), the PPNC parameters must satisfy the integrability condition
\begin{equation}\label{eq_integrabilitycondition}
4 \pi G \, \bar{\rho} \left(\alpha-\gamma +\frac{{\rm d}  \gamma}{{\rm d}  \ln a}  \right) = 
\alpha_c+2 \gamma_c +\frac{{\rm d}  \gamma_c}{{\rm d}  \ln a} \,.
\end{equation}
Finally, subtracting off the spatially constant parts of Eqs. (\ref{eq_integrated_PPN_1}) and (\ref{eq_integrated_PPN_2}) gives Poisson equations for the scalar perturbations on small scales,
\begin{equation}\label{eq_PPNC_small_scale_poisson}
    \h{\nabla}^2\h{\Phi} = -4\pi G \,\alpha\,a^2\,\delta\rho\,, \quad {\rm and} \quad \h{\nabla}^2\h{\Psi} = -4\pi G \,\gamma\,a^2\,\delta\rho\,.
\end{equation}
These can be made to look rather like the PPF equations (\ref{eq_PPF_1}) and (\ref{eq_PPF_2}) that use the free functions $\mu$ and $\Sigma$ to multiply the density perturbation. 
However, unlike in those equations, the coefficients are purely functions of time, and are directly related to the PPN parameters studied in astrophysical experiments. In addition, the equations in (\ref{eq_PPNC_small_scale_poisson}) are valid for arbitrarily nonlinear densities, rather than being restricted to the linear regime.
They do have the drawback, though, that they only apply to theories of gravity that can be incorporated into the PPN formalism using only the standard post-Newtonian potentials defined in Section \ref{subsec:PPN}. This condition is satisfied by many modified theories of gravity, but it excludes theories that exhibit nonlinear screening mechanisms on small scales. The PPF framework does not make that restriction.

Eqs. (\ref{eq_parametrisedfriedmanneqns1}--\ref{eq_PPNC_small_scale_poisson}) provide a fully consistent set of equations for the homogeneous and isotropic emergent expansion, and for scalar Newtonian-gauge perturbations deep within the horizon. The equations are explicitly dependent on the generalised PPN parameters $\left\lbrace \alpha(a), \gamma(a), \alpha_c(a), \gamma_c(a)\right\rbrace\,$.
In order to obtain an understanding of scalar perturbations on all spatial scales, it is necessary to consider the other limit, of perturbations beyond the Hubble horizon. Then, one can smoothly interpolate between these regimes, in order to construct all-scales equations for $\h{\Phi}$ and $\h{\Psi}\,$.

\subsection{All-scales parameterisation}

Let us now focus on the evolution of linear perturbations on large scales, beyond the Hubble horizon $\mathcal{H}^{-1}\,$. On these scales, the spatial dependence of perturbations can be safely neglected. A purely time-dependent perturbation to the FLRW geometry is entirely equivalent to an FLRW geometry with perturbed coordinates, and therefore a different scale factor and spatial curvature parameter.
Thus, super-horizon perturbations essentially correspond to a separate universe, which can be studied as a small deviation from the original universe. The separate-universe approach allows one to obtain evolution and constraint equations that are specified purely by the contents of the Friedmann equations, and are therefore agnostic to the underlying theory of gravity that determines the exact form of those equations \cite{wands2000new}.

The structure of these super-horizon equations was calculated explicitly by Bertschinger \cite{Bertschinger_2006}, for both adiabatic and isocurvature modes. For the present discussion, we will focus on the adiabatic perturbations only, which are relevant in the late Universe and for the CMB anisotropies.
The approach used by Bertschinger assumes that the initial FLRW geometry has non-zero curvature, and that this can be written as a second FLRW geometry with a spatial curvature that is equal to the first up to a factor that is perturbatively close to unity. 
In the present discussion, we are interested in spatially flat cosmologies, so let us work instead under the assumption that the second geometry, which we think of as being ``our'' perturbed FLRW cosmology, is spatially flat.
Therefore, we start off in the first, perturbed, geometry, with a spatially curved FLRW metric,
\begin{equation}
\mathrm{d}s^2 = a^2(\tau_*)\left[-\mathrm{d}\tau_*^2 + \frac{{\bf \mathrm{d}\hat{x}}_*^2}{1+\frac{K}{4} {\bf \hat{x}}_*^2}\right].
\end{equation}
Now let us assume the spatial curvature is small such that $K=\delta K \ll \mathcal{H}^2(\tau)$ for all $ \tau\,$, and perturb the coordinates such that $\tau \longrightarrow \tau_* = \tau + A(\tau)$ and $\h{x}^i \longrightarrow \h{x}_*^i = \h{x}^i\left(1+\beta(\tau)\right)$, where $A(\tau) \ll1$ and $\beta(\tau) \ll 1\,$. We then obtain
\begin{eqnarray}
\label{metricA}
\hspace{-1.0cm}\mathrm{d}s^2 &=& a^2(\tau) \Bigg[-\left(1+2A'+2\mathcal{H}A + 2 \frac{\delta K}{a} \frac{\partial a}{\partial K}\right)\mathrm{d}\tau^2 \\
\nonumber \hspace{-1.0cm}&&\qquad + 2\beta' \,x^j\,\delta_{ij} \,\mathrm{d}\tau\mathrm{d}\h{x}^i + \left(1+2\beta + 2\mathcal{H}A + 2 \frac{\delta K}{a} \frac{\partial a}{\partial K} -\frac{1}{4} \delta K \, {\bf x}^2   \right)\delta_{ij}\mathrm{d}\h{x}^i\mathrm{d}\h{x}^j \Bigg]\,.
\end{eqnarray}
Comparing this to a spatially flat geometry with spatially-homogeneous scalar perturbations in Newtonian gauge,
\begin{equation}
\mathrm{d}s^2 = a^2(\tau) \left[ -(1-2 \hat{\Phi}) \,\mathrm{d}\tau^2 + (1+2 \hat{\Psi}) \,\delta_{ij}\mathrm{d}\h{x}^i\mathrm{d}\h{x}^j\right] \, ,
\end{equation}
we see that we must require $\beta' = \delta K = 0$ (i.e. $\beta = \, {\rm cst.}$), and 
\begin{eqnarray}
\hat{\Phi} = -A' - \mathcal{H}A\,; \qquad 
\hat{\Psi} = \beta + \mathcal{H}A \, . 
\end{eqnarray}
The metric in Eq. (\ref{metricA}) can therefore be thought of as an FLRW metric with the super-horizon scalar perturbations $\hat{\Phi}$ and $\hat{\Psi}$ given above. 

Let us now derive the Hamiltonian constraint and Raychaudhuri equations for those large-scale perturbations. At this stage, we cannot calculate any other equations, because we have not yet considered the effects of matter contributions besides energy density and pressure, i.e. momentum density and anisotropic stress.
The Einstein tensor components in Eq. (\ref{eq_CPT_Einstein_tensor_components}) tell us the combinations of $\h{\Phi}$\,, $\h{\Psi}$\,, and their conformal time derivatives, that we need to construct. 
For the Hamiltonian constraint, we have
\begin{equation}
    \mathcal{H}\left(\h{\Psi}'+\mathcal{H}\h{\Phi}\right) = -\left(\mathcal{H}^2-\mathcal{H}'\right)\mathcal{H}A\,.
\end{equation}
Considering the effect of the coordinate perturbation $\delta\tau = A(\tau)$ on the Friedmann equations gives further that on super-horizon scales, the total energy density perturbation $\delta = -3\mathcal{H}A\,$. 
It follows that the Hamiltonian constraint can be written in the ultra-large-scale limit as
\begin{equation}
    -\mathcal{H}^2\h{\Phi} - \mathcal{H}\h{\Psi}' = -\frac{\delta}{3}\left(\mathcal{H}^2-\mathcal{H}'\right)\,.
\end{equation}
Upon inserting the PPNC generalisations (\ref{eq_parametrisedfriedmanneqns1}) and (\ref{eq_parametrisedfriedmanneqns2}) of the Friedmann equations, and the integrability condition (\ref{eq_integrabilitycondition}), this becomes
\begin{equation}\label{eq_PPNC_superhorizon_Hamiltonian_constraint}
    -\mathcal{H}^2\h{\Phi} - \mathcal{H}\h{\Psi}' = -\frac{4\pi G}{3}\,\delta\rho\,a^2\left[\gamma - \frac{1}{3}\frac{\mathrm{d}\gamma}{\mathrm{d}\ln{a}} + \frac{1}{12\pi G \bar{\rho}}\frac{\mathrm{d}\gamma_c}{\mathrm{d}\ln{a}}\right]\,.
\end{equation}
Similarly, the perturbed Raychaudhuri equation, according to the Bertschinger separate-universe approach, is
\begin{equation}
    2\mathcal{H}'\h{\Phi} + \mathcal{H}\h{\Phi}' + \mathcal{H}\h{\Psi}' + \h{\Psi}'' = \frac{\delta}{3}\left[\frac{2\mathcal{H}\mathcal{H}' - \mathcal{H}''}{\mathcal{H}}\right]\,,
\end{equation}
from which we get 
\begin{equation}\label{eq_PPNC_superhorizon_Raychaudhuri}
    2\mathcal{H}'\h{\Phi} + \mathcal{H}\h{\Phi}' + \mathcal{H}\h{\Psi}' + \h{\Psi}'' = - \frac{4\pi G}{3}\,\delta\rho\,a^2\left[\alpha - \frac{1}{3}\frac{\mathrm{d}\alpha}{\mathrm{d}\ln{a}} + \frac{1}{12\pi G \bar{\rho}}\frac{\mathrm{d}\alpha_c}{\mathrm{d}\ln{a}}\right]\,.
\end{equation}

Combining the large-scale results (\ref{eq_PPNC_superhorizon_Hamiltonian_constraint}) and (\ref{eq_PPNC_superhorizon_Raychaudhuri}) with the small-scale results in Eq. (\ref{eq_PPNC_small_scale_poisson}), one finds \cite{Sanghai_2019} that the leading-order perturbations to the Hamiltonian constraint and Raychaudhuri equations can be written on all scales as
\begin{eqnarray}
&\hspace{-25pt}
\dfrac{1}{3}\hat{\nabla}^2 \hat{\Psi} - \mathcal{H}^2\hat{\Phi} - \mathcal{H} \hat{\Psi}'   = -\dfrac{4\pi G}{3} \, \mu(\tau, L) \, \delta\! \rho \, a^2   \label{eq_PPNC_allscales_Hamiltonian} \\[3pt]
&\hspace{-25pt}
\dfrac{1}{3}\hat{\nabla}^2 \hat{\Phi} + 2{\mathcal{H}'}\hat{\Phi} +  \mathcal{H}\hat{\Phi}' + \hat{\Psi}'' +\mathcal{H} \hat{\Psi}'   = -\dfrac{4\pi G}{3} \, \nu(\tau, L) \, \delta\! \rho \, a^2 \, ,
 \label{eq_PPNC_allscales_Raychaudhuri}
\end{eqnarray}
where we have explicitly written the coefficients $\mu$ and $\nu$ of the overall energy density perturbation $\delta\rho$ as functions of $\tau$ and spatial scale $L = k^{-1}$ in general. 
These functions are, however, far from arbitrary: they have small-scale and large-scale limits precisely defined in terms of the PPNC parameters.
Moreover, $\delta\rho$ need not be small compared to $\bar{\rho}\,$. The only requirement we make is that $\h{\Phi}$ and $\h{\Psi}$ remain $\ll 1$ on all scales.

On small scales $\lesssim 100 \, {\rm Mpc}$ we have \cite{Sanghai_2019}
\begin{eqnarray} \label{eq_PPNC_scalarperts_small}
\underset{L \rightarrow 0}{\rm lim} \mu &= \gamma \qquad {\rm and} \qquad \underset{L \rightarrow 0}{\rm lim} \nu = \alpha \, .
\end{eqnarray}
The separate-universe approach that we just employed tells us that on the very largest scales \cite{Sanghai_2019}
\begin{eqnarray} \label{eq_PPNC_scalarperts_large1}
\underset{L \rightarrow \infty}{\rm lim} \mu &=& \gamma - \dfrac{1}{3} \dfrac{{\rm d} \gamma}{{\rm d}  \ln a}+ \dfrac{1}{12 \pi G \bar{\rho}} \dfrac{{\rm d} \gamma_c}{{\rm d} \ln a} \\ \label{eq_PPNC_scalarperts_large2}
\underset{L \rightarrow \infty}{\rm lim} \nu &=& \alpha - \dfrac{1}{3} \dfrac{{\rm d}  \alpha}{{\rm d}  \ln a}+ \dfrac{1}{12 \pi G \bar{\rho}} \dfrac{{\rm d}  \alpha_c}{{\rm d}  \ln a} \, .
\end{eqnarray}
The limiting behaviour given in Eqs. (\ref{eq_PPNC_scalarperts_small}-\ref{eq_PPNC_scalarperts_large2}) shows that the couplings $\mu$ and $\nu$ tend towards a scale-invariant form on both small and large spatial scales, that is entirely a function of the PPNC parameters $\{\alpha(\tau), \gamma(\tau), \alpha_c(\tau), \gamma_c(\tau)\}\,$.
In general, the small-scale and large-scale limits of $\mu$ and $\nu$ are different, as we will verify explicitly later in this chapter for both scalar-tensor and vector-tensor theories of gravity.
Of course, in order to actually use these equations, one must have a viable, physically well-motivated prescription for how to interpolate between the deep sub-horizon and super-horizon regimes. That will be a principle subject of Section \ref{sec:scale_dependence}.

However, the equations we have presented so far, which were first derived in Refs. \cite{Sanghai_2017} and \cite{Sanghai_2019}, do not provide a complete framework to describe both the cosmic expansion and the linear scalar perturbations. In particular, in order to evolve the system in a well-defined way, we require the PPNC equations to constitute an initial value problem. 
From our earlier discussion of the $3+1$ formalism, this is equivalent to having (a) a complete set of constraint equations, i.e. both a Hamiltonian and a momentum constraint, that allow the initial data to be specified on some constant-$\tau$ leaf, and (b) a complete set of evolution equations that preserve the constraints as the equations of motion are evolved over successive leaves of the foliation. 

In the context of CPT, the momentum constraint means understanding how the scalar metric perturbations $\h{\Phi}$ and $\h{\Psi}$, and the divergenceless vector $\h{B}_i\,$, are sourced by the momentum density of matter, i.e. by peculiar velocity perturbations. 
The other non-trivial evolution equation, beyond the Raychaudhuri equation, turns out at leading order to be equivalent to a scalar relation for the slip $\h{\Phi} - \h{\Psi}$, which comes from the shear evolution equation.
Aside from being necessary from a theoretical perspective, these equations are also very important computationally, as it is only once they are derived that we can write down a consistent set of equations that can be solved numerically using an Einstein-Boltzmann code. This will be crucial in Chapter 6, when we come to studying the cosmic microwave background using the PPNC formalism.

\section{Peculiar velocities and the momentum constraint}\label{sec:momentum_constraint}

In this section, we will focus on the momentum constraint in the PPNC framework. The presentation here is based on Ref. \cite{anton2022momentum}.
We recall from Chapter 2 that the $1+3$-covariant equation to which the momentum constraint (\ref{eq_ADM_equation_momentum_constraint}), defined in the $3+1$ ADM formalism, is equivalent, is the equation (\ref{eq_momentum_constraint_1+3}) that connects the gradients of expansion, shear and vorticity to the momentum density of matter $q_a$\,. This is defined implicitly through the Einstein tensor, so we can write the momentum constraint in a theory-independent form as
\begin{equation}\label{eq_1+3momentumconstraint_theory_indep}
D^{b}\sigma_{ab} - \frac{2}{3}D_{a}\Theta + g_{ac}\eta^{cbd}\left[D_{b}\omega_{d} + 2\dot{u}_{b}\omega_{d}\right] = G_b^{\ c} u^b h_{ca}\,,
\end{equation}
which we stress must be satisfied in any metric theory of gravity, so that a complete set of initial data exists on some spacelike hypersurface.

In the perturbed-FLRW context we are interested in here, we can choose coordinates such that $u^a = u^0 \delta^a_{\ \tau}\,$, in which case Eq. (\ref{eq_1+3momentumconstraint_theory_indep}) takes the more familiar form
\begin{equation}\label{eq_cosmology1+3momentumconstraint}
-\frac{1}{a}\left[2\left(\mathcal{H}'-\mathcal{H}^2\right)\hat{B}_i + \frac{1}{2}\hat{\nabla}^2\hat{B}_i + 2\left(\hat{\Psi}'+\mathcal{H}\hat{\Phi}\right)_{,i}\right] = G_0^{\ c}u^0 h_{ci}\, .
\end{equation}
Because of the SVT decomposition, this equation is thought of in cosmology as being comprised of two separate parts: a scalar equation that adds a constraint to the equations (\ref{eq_PPNC_allscales_Hamiltonian}) and (\ref{eq_PPNC_allscales_Raychaudhuri}) we presented in the previous section, and a divergenceless vector equation that contains the frame-dragging potential $\hat{B}_i$\,. 

Our aim in this section is therefore to derive scalar and divergenceless vector equations from Eq. (\ref{eq_cosmology1+3momentumconstraint}), that are valid on all cosmological scales where a weak-field expansion can be applied, whether that is an expansion in the cosmological perturbation theory parameter $\sim \delta$ or the post-Newtonian slow-motion parameter $\sim v\,$.
Following the approach of Section \ref{sec:basic_PPN}, we will study this problem first on small, post-Newtonian, scales - which will require extending the PPNC formalism from $\mathcal{O}(v^2)$ to $\mathcal{O}(v^3)$ - and then on very large scales where the separate universe approach to super-horizon cosmological perturbations can be used.

\subsection{Small scales}

Let us first revisit the momentum constraint in the standard PPN formalism, so that we can go about generalising it to cosmological settings, by deriving the analogue equation for parameterised post-Newtonian perturbations about an FLRW background.
Momentum density terms of the form $\rho v$ are of order $v^3$ in the post-Newtonian perturbative hierarchy, compared to the pure density terms $\rho \sim v^2$ that source the Poisson equations (\ref{eq_PPNC_small_scale_poisson}) for the small-scale scalar perturbations.
A consequence of the need to consider higher-order post-Newtonian potentials is that preferred frame effects must be considered. The gauge-fixing process is also more complicated at this order.

The substantive point making the application of post-Newtonian theory non-trivial in this context is that the perturbation $B_i = h_{0i}^{(3)}$ to the Minkowski metric cannot be written directly in terms of the integral expressions for the post-Newtonian potentials $V_i$ and $\chi_{,ti}$ we introduced in Section \ref{subsec:PPN}, because we do not have asymptotic flatness that allows us to define these quantities in terms of the Green's function $\left\vert\mathbf{x}-\mathbf{x}'\right\vert^{-1}\,$.
Without asymptotic flatness, the best the PPN formalism can do in this regime is to write
\begin{equation}\label{eq_B_i_PPN}
    \nabla^2 B_i = 8\pi G \left[\alpha + \gamma + \frac{1}{4}\alpha_1\right]\rho v_i - \left[\alpha + \alpha_2 - \zeta_1 + 2\xi\right]U_{,ti} + \nabla^2 \varphi^{\rm PF}_i\,,
\end{equation}
where the Newtonian gravitational potential is defined implicitly by $\nabla^2 U \equiv - 4 \pi G \, \rho$, and
\begin{eqnarray}\label{eq_preferred_frame_potential_1}
\nabla^2 {\varphi}^{\rm PF}_i &=& 2 \pi {\alpha_1} {w}_i \, \rho +2 \alpha_2 {w}^j U_{,ij}\,,
\end{eqnarray}
where ${w}^j$ is the velocity of the PPN system with respect to the preferred frame of the theory, if one exists. Recall that in theories, such as scalar-tensor theories of gravity, that have $\alpha_{1,2,3} = \zeta_{1,2,3,4} = 0\,$, global momentum and angular momentum are conserved in asymptotically flat spacetime to high order in the post-Newtonian expansion \cite{Will_1993}. 

Let us now consider how first to upgrade Eq. (\ref{eq_B_i_PPN}) for cosmological treatments, and then how to translate it into a momentum constraint for the small-scale perturbations $\h{\Phi}$, $\h{\Psi}$ and $\h{B}_i$ to an FLRW spacetime.
In addition to making $\alpha$ and $\gamma$ functions of cosmic time, we need to do the same to $\alpha_1$, $\alpha_2$, $\xi$ and $\zeta_1\,$. 
Note that there are very rough constraints on the time variation of $\alpha$, $\gamma$, $\alpha_c$ and $\gamma_c$ from lunar laser ranging \cite{Uzan_2003}, weak lensing \cite{Joudaki_2017} and the CMB \cite{Planck_2020} respectively, but none on the rest of the parameters that enter Eq. (\ref{eq_B_i_PPN}). 
These time-derivative constraints are summarised in Table \ref{table_PPN_time_derivs}, where it should be noted that we have expressed derivatives not with respect to cosmic time but with respect to the logarithm of $a$ (i.e. the number of $e$-foldings $N$ of the scale factor).
For completeness, we note also that inferred estimates of $\Omega_{\Lambda 0}$ from CMB observations give at the present day ($a = 1$), $\alpha_c = \left(2.07 \pm 0.03\right)H_0^2$ and $\gamma_c = \left(-1.04 \pm 0.02\right)H_0^2$ \cite{Planck_2020}, as the $\Lambda$CDM limit of the PPNC Friedmann equations tells us that in that context, $\alpha_c = - 2\gamma_c = \Lambda\,$.
We will perform a much more detailed and rigorous Markov Chain Monte Carlo (MCMC) analysis of the time variation of the PPNC parameters in Chapter 6, using Planck measurements of the cosmic microwave background anisotropies.

\begin{table}[ht]
\begin{center}
\begin{tabular}{|c|c|c|}
\hline
Physical effect & Parameter & Constraint on derivative \\
\hline
Effective $G$  & $\alpha$ & $0 \pm 0.01$ \cite{Uzan_2003} \\
\hline
Spatial curvature & $\gamma$ & $0 \pm 0.1$ \cite{Joudaki_2017} \\
\hline
Preferred locations & $\xi$ & -------- \\
\hline
Momentum conservation & $\zeta_1$ & -------- \\
\hline
Preferred frames & $\alpha_1$ & -------- \\
\hline
Preferred frames & $\alpha_2$ & -------- \\
\hline
Cosmological effects & $\alpha_c$ & $\left(0.12 \pm 0.25\right)H_0^2$ \cite{Planck_2020} \\
\hline 
Cosmological effects & $\gamma_c$ & $\left(-0.06\pm 0.12\right)H_0^2$ \cite{Planck_2020} \\
\hline
\end{tabular}
\caption{Rough observational estimates of the present-day derivatives with respect to $\ln{a}$ of the coupling parameters that appear in the PPNC test metric.} 
\label{table_PPN_time_derivs}
\end{center}
\end{table}

We also require a contribution $B_i^{\rm extra}$ to take care of any novel vector contributions that might need to be added to the RHS of Eq. (\ref{eq_B_i_PPN}) due to purely cosmological vector fields, such as vector dark energy (as in e.g. \cite{jimenez2008cosmic}). One can think of this term as playing a role equivalent to $\alpha_c$ and $\gamma_c$, but for the vector rather than scalar sector.

We will use the transformations from Eqs. (\ref{eq_statictoexpandingtransformation1}) and (\ref{eq_statictoexpandingtransformation2}) to get from the equation (\ref{eq_B_i_PPN}) for $\nabla^2 B_i$ (where it is important to stress that $B_i$ has generically non-vanishing divergence) to a divergenceless vector equation for $\h{\nabla}^2 \h{B}_i$, and a scalar equation for the first time derivatives of $\h{\Phi}$ and $\h{\Psi}\,$. 
As the post-Newtonian expansion on which these equations are based is expected to be valid on scales $\lesssim 100 \, {\rm Mpc}$, this will give us the ``small scale'' limit of the general parameterised momentum constraint equation. 
We will start by considering this equation in conservative theories of gravity, so that we do not need to worry about preferred frame effects yet. Later, we will generalise the result to non-conservative theories by carefully examining the preferred-frame effects that can arise on cosmological scales.

The first step is to calculate $\h{B}_i$ in Newtonian gauge, in terms of $B_i$, $\Phi$ and $\Psi\,$.
Therefore, we must enforce that $\h{B}_{i, i}=0$, where the derivative is taken with respect to the coordinates $\hat{x}^i$ that define the canonical basis on the expanding FLRW system. 
Taking the spatial divergence of Eq. (\ref{expandingvsstatic3}), and enforcing the Newtonian gauge condition, gives that the $\mathcal{O}(v^3)$ gauge function in Eq. (\ref{eq_statictoexpandingtransformation1}) must satisfy
\begin{equation}
\hat{\nabla}^2 T = a^2 B_{i,i} + 6a \dot{a} \left(\hat{\Phi}+\hat{\Psi}\right) + 2a \dot{a} \left(\hat{\Phi}+\hat{\Psi}\right)_{,i} \hat{x}^i+ 5a^2 \dot{a} \ddot{a}\, \hat{r}^2 \,,
\end{equation}
where we have also used the transformations (\ref{expandingvsstatic}) and (\ref{expandingvsstatic2}) to rewrite $\Phi$ and $\Psi$ in terms of $\h{\Phi}$ and $\h{\Psi}\,$.
In this equation, and henceforth, we use the convention that spatial derivatives on a quantity are with respect to the set of coordinates with which that quantity is defined (i.e. whether that object is defined with respect to a Minkowski or FLRW background metric $g_{ab}^{(0)}\,$). 
This means that ``spatial'' derivatives of hatted quantities are taken with respect to $\h{x}^i$, and unhatted quantities are differentiated spatially with respect to $x^i$. 

Next, we operate on Eq. \eqref{expandingvsstatic3} with the Laplacian $\hat{\nabla}^2$, and substitute in the expression above for $T$. This gives us the following expression for the left-hand side of the momentum constraint\footnote{By this, we mean the desired combination of expansion, shear, vorticity and acceleration that appears on the LHS of Eq. (\ref{eq_1+3momentumconstraint_theory_indep}).} on small scales:
\begin{eqnarray} \label{eq_small_scale_momentum_1}
\frac{1}{2a}\h{\nabla}^2\h{B}_i + 2\left(\dot{\h{\Psi}} + H\h{\Phi}\right)_{,i} &=& \frac{a}{2}\,\left(\nabla^2 B_i - B_{j,ji}\right) + H\,\delta_{ij}\,\h{x}^j\,\h{\nabla}^2\left(\h{\Phi}+\h{\Psi}\right) \\
\nonumber && - H\,\h{x}^j\,\left(\h{\Phi}+\h{\Psi}\right)_{,ij} + 2\dot{\h{\Psi}}_{,i} - 2H\,\h{\Psi}_{,i}\,.
\end{eqnarray}

First, we need to deal with the $B_i$ terms on the RHS. We cannot really solve for $B_i$ in closed form without the key PPN assumption of asymptotic flatness. 
However, we can essentially define the solution, by writing new implicit definitions for the superpotential $\chi$ and post-Newtonian vector potential $V_i\,$ as $\nabla^2 \chi = - 2U$ and $\nabla^2 V_i = -4\pi\rho v_i\,$.
Then, upon setting $\alpha_1 = \alpha_2 = \zeta_1 = 0\,$, as per the assumption of fully conservative gravity, Eq. (\ref{eq_B_i_PPN}) can be integrated to
\begin{equation}\label{eq_B_i_PPNC_implicit}
    B_i = -2\left(\alpha + \gamma\right)V_i + \frac{1}{2}\,\left(\alpha + 2\xi\right)\,\dot{\chi}_{,i} + B_i^{\rm extra}\,,
\end{equation}
which shows how the additional cosmological vector term $B_i^{\rm extra}$, whose need we described earlier, enters into the formalism.

Taking the Laplacian of Eq. (\ref{eq_B_i_PPNC_implicit}) with respect to the expanding coordinates, and substituting the result into Eq. (\ref{eq_small_scale_momentum_1}), we get our first equation for the FLRW perturbations directly in terms of the momentum density $\rho v_i$ of matter fields,
\begin{eqnarray} \label{eq_small_scale_momentum_2}
\frac{1}{2a} \h{\nabla}^2 \h{B}_i + 2 \left( \dot{\h{{\Psi}}}+{H} \h{\Phi} \right)_{,i} &=& 4 \pi G\, a (\alpha+ \gamma) \rho v_i + a (\alpha+\gamma) V_{j,ji} \\
\nonumber && +  {H} \hat{x}^j\left[\delta_{ij}\hat{\nabla}^2 (\hat{\Phi} + \hat{\Psi}) - \left(\h{\Phi} + \h{\Psi}\right)_{,ij} \right] \\
\nonumber && + 2 \dot{\h{{\Psi}}}_{,i} - 2 {H} \hat{\Psi}_{, i} + \frac{a}{2}\, \left( \nabla^2 B^{\rm extra}_{i} - B^{\rm extra}_{j ,ji} \right) \, ,
\end{eqnarray}
which at this stage is not especially illuminating.

To proceed further it is useful to rewrite the term containing the factor $V_{j,ji}\,$. We do this by splitting the velocity $v^i$ into a background part $H x^i$ that is purely due to the Hubble flow and a peculiar velocity $\delta v^i$, such that $v^i = H {x}^i + \delta v^i$. 
It is crucial to emphasise that unlike in cosmological perturbation theory, where the inhomogeneous term $\delta v^i$ is much smaller than the homogeneous Hubble flow term, in the post-Newtonian regime the contributions $H x^i$ and $\delta v^i$ are of the same post-Newtonian order, because they both contain one power of the characteristic slow-motion scale $v\,$.
Having made this splitting of the velocity field, we can split $V^i$ similarly, into components due to the background and peculiar velocities, such that $V_i = \bar{V}_i + \delta {V}_i$, where
\begin{equation}\label{eq_Vhat}
\nabla^2 \bar{V}_i \equiv -4\pi G \, \rho H x^j\delta_{ij} \qquad {\rm and} \qquad \nabla^2 \delta {V}_i \equiv -4\pi G \, \rho\, \delta{v}_i \, .
\end{equation}
The former of these implicit definitions allows for the solution $\bar{V}^{i} = Hx^j\delta_{ij} U + H \chi_{,i}\,$, which can easily be verified by explicit differentiation. In the second term here, the spatial derivative should be understood to be taken with respect to the Minkowski $x^i$ coordinates, not the FLRW $\h{x}^i$ coordinates.

Making a similar split of the Newtonian potential $U$ into contributions from the homogeneous and inhomogeneous parts, i.e. taking $\rho= \bar{\rho} + \delta \rho$, allows us to write
\begin{equation}\label{eq_Ubar_and_deltaU}
\nabla^2 U = -4 \pi G \rho = - 4 \pi G \bar{\rho} - 4 \pi G \delta \rho \equiv \nabla^2 \bar{U} + \nabla^2 \delta U \, ,
\end{equation}
where the last equality provides implicit definitions for $\bar{U}$ and $\delta U$. 
One can then trivially write down the solution for the homogeneous part of the Newtonian potential as $\bar{U} = -{2 \pi} \bar{\rho} r^2/3$\,.
Subtracting $\nabla^2\bar{U}$ off from $\nabla^2 U$, one then has immediately that the cosmological perturbations to the FLRW geometry are linearly proportional $\delta U\,$, namely $\hat{\Phi} = \alpha \,\delta U$ and $\hat{\Psi} = \gamma \,\delta U$. 
These expressions, together with the continuity equation $\dot{\rho} + (\rho v^i)_{,i} = 0$, allow us to derive the useful identities 
\begin{eqnarray}\label{eq_PPNC_V_i_identity_1}
\alpha \, \delta V_{i,i} = -\dot{\hat{\Phi}}-{H}\hat{\Phi}+\frac{\dot{\alpha}}{\alpha}\hat{\Phi}
\quad {\rm and} \quad
\gamma \, \delta V_{i,i} = -\dot{\hat{\Psi}}-{H}\hat{\Psi}+\frac{\dot{\gamma}}{\gamma}\hat{\Psi} \, ,
\end{eqnarray}
which can be compared with the identity $V_{i,i} = - \dot{U}$ that exists in the classic PPN formalism \cite{Will_1993}, as we mentioned in Section \ref{subsec:PPN}. That identity does not hold in our approach, because it relies on the integral form of the post-Newtonian potentials that is a consequence of asymptotic flatness for isolated astrophysical systems. 

The part $\bar{V}_i = H x^j \delta_{ij} U + H \chi_{,i}$ of the post-Newtonian vector potential that depends on the Hubble flow is then split further as
\begin{equation}\label{eq_PPNC_V_i_identity_2}
    \bar{V}_i = H x^j \delta_{ij} \bar{U} + H x^j \delta_{ij} \delta U + H\bar{\chi}_{,i} + H\delta\chi_{,i}\,,
\end{equation}
where $\nabla^2 \bar{\chi} = -2\bar{U}$ and $\nabla^2 \delta \chi = -2\delta U\,$.
This gives
\begin{equation}\label{eq_PPNC_V_i_identity_3}
    \bar{V}_{j,j} = 3H \bar{U} + H\delta U + H a \hat{x}^j \delta U_{,j}\,,
\end{equation}
whence
\begin{equation}\label{eq_PPNC_V_i_identity_4}
    \bar{V}_{j,ji} = -4\pi G \bar{\rho} H a \h{x}^j \delta_{ij} + 2H\delta U_{,i} + Ha\h{x}^j\delta U_{,ij}\,.
\end{equation}
Putting the results (\ref{eq_PPNC_V_i_identity_1}--\ref{eq_PPNC_V_i_identity_4}) together on the RHS of Eq. (\ref{eq_small_scale_momentum_2}), we find that all the terms involving the Hubble flow $H a \h{x}^i$ cancel out from the equations of motion. This is a relief, because the origin of the spatial coordinate system for the homogeneous and isotropic FLRW background metric should be arbitrary.

Specifically, we find that the momentum constraint for the post-Newtonian perturbations to FLRW spacetime is
\begin{eqnarray}\label{eq_small_scale_momentum_3}
\nonumber \frac{1}{2a} \h{\nabla}^2 \h{B}_i + 2 \left(\dot{\h{{\Psi}}}+ {H} \h{\Phi} \right)_{,i} &=& 4\pi G \left(\alpha+\gamma \right)\,a\rho \,\delta{v}_i - \Big[\Big(\dot{\hat{\Phi}}-\dot{\hat{\Psi}}\Big) - {H}\left(\hat{\Phi}-\hat{\Psi}\right) \\
&& - \frac{\dot{\alpha}}{\alpha}\hat{\Phi}-\frac{\dot{\gamma}}{\gamma}\hat{\Psi}\Big]_{,i}  
+ \frac{1}{2} a\left[\nabla^2 B^{\mathrm{extra}}_{i} - B^{\mathrm{extra}}_{j,ji}\right] \,,
\end{eqnarray}
where we stress that the coefficient $\rho$ of the peculiar velocity $\delta v_i$ is the entire nonlinear energy density of non-relativistic matter, not just its homogeneous part $\bar{\rho}\,$. 
This can be straightforwardly split into a scalar part
\begin{eqnarray}\label{eq_small_scale_momentum_scalar_1}
2  {\h{{\Psi}}}_{,i}'+2 \mathcal{H} \h{\Phi}_{,i}  &=& 4\pi G \left(\alpha+\gamma \right) \left[ \rho \hat{v}_i \right]^{\rm S} \, a^2 \\
&& \nonumber - \Big[\Big({\hat{\Phi}}-{\hat{\Psi}}\Big)' - \mathcal{H}\left(\hat{\Phi}-\hat{\Psi}\right) - \frac{{\alpha}'}{\alpha}\hat{\Phi}-\frac{{\gamma}'}{\gamma}\hat{\Psi}\Big]_{,i}\,,
\end{eqnarray}
and a divergenceless vector part
\begin{eqnarray}\label{eq_small_scale_momentum_vector}
\hat{\nabla}^2 \hat{B}_i = 8\pi G \left(\alpha+\gamma \right) \left[\rho \hat{v}_i\right]^{\rm V} \, a^2 + \h{\nabla}^2 \hat{B}_i^{\rm extra} \, ,
\end{eqnarray}
where $\h{\nabla}^2 \hat{B}_i^{\rm extra}\equiv a^2(\nabla^2 B^{\mathrm{extra}}_{i} - B^{\mathrm{extra}}_{j,ji})$ is manifestly a divergence-free vector, and where $\h{v}^i = \delta v^i$ is exactly what we typically think of in cosmology as a peculiar velocity: the velocity of the matter fluid with respect to the comoving coordinate basis, given a particular choice of gauge, which in our case is the Newtonian gauge. 
In each of these two expressions we have used the superscripts S and V to indicate that we intend them to correspond respectively to the scalar or divergenceless-vector part of the object within the square brackets. 
We have also converted all time derivatives in these expressions into conformal time, so that the equations can be compared readily to the results obtained from linear cosmological perturbation theory in General Relativity. 
On linear scales, one could replace $\rho \h{v}_i$ with $\bar{\rho} \h{v}_i\,$, and then use a scalar velocity potential to remove the indices in the scalar part of the equation. However, our results do not make perturbative assumptions about $\delta \rho$, with $\rho$ receiving inhomogeneous contributions, so this should not be done in general.

Before we go on, let us note that the scalar equation (\ref{eq_small_scale_momentum_scalar_1}) can be rewritten in a form that will be especially useful for unifying our small-scale result with the large-scale result that we will derive in the next section.
To achieve this, we recall that the function $\mu$ we defined in the all-scales Hamiltonian constraint (\ref{eq_PPNC_allscales_Hamiltonian}) is equal to $\gamma$ on small scales, where $\h{\Phi}$ and $\h{\Psi}$ are respectively given by $\alpha\, \delta U$ and $\gamma \,\delta U\,$.
Thus, Eq. (\ref{eq_small_scale_momentum_scalar_1}) can be expressed as 
\begin{eqnarray}
    \h{\Psi}'_{,i} + \mathcal{H}\h{\Phi}_{,i} &=& 4\pi G \,a^2\,\mu\,\left[\rho \h{v}_i\right]^{\rm S} + \left( \frac{\alpha - \gamma}{\gamma} + \frac{\mathrm{d}\ln{\gamma}}{\mathrm{d}\ln{a}}\right)\,\mathcal{H}\,\h{\Psi}_{,i} \\
    \nonumber && - \left(\frac{\alpha - \gamma}{2}\right)\left[\delta U'_{,i} + \mathcal{H}\delta U_{,i} - 4\pi G \,a^2\,\left[\rho \h{v}_i\right]^{\rm S}\right]\,.
\end{eqnarray}
Taking the conformal time derivative of the equation $\h{\nabla}^2 \delta U = -4\pi G \,a^2\,\delta\rho\,$, and using the continuity equation to deal with the term $\delta\rho'\,$, one sees that the combination in square brackets on the second line vanishes identically.
Hence, the scalar part of the momentum constraint on small scales can be written more compactly as 
\begin{equation}\label{eq_small_scale_momentum_scalar_2}
\h{\Psi}'_{,i} + \mathcal{H}\h{\Phi}_{,i} = 4\pi\,a^2\,\mu\,\left[\rho \h{v}_i\right]^{\rm S} + \mathcal{G}\,\mathcal{H}\,\h{\Psi}_{,i}\,,   
\end{equation}
where we see that the term 
\begin{equation}
    \mathcal{G} = \frac{\alpha - \gamma}{\gamma} + \frac{\mathrm{d}\ln{\gamma}}{\mathrm{d}\ln{a}} 
\end{equation}
is exactly zero in GR, and is therefore an entirely novel term that can arise in MG theories due to a difference between the PPN parameters $\alpha$ and $\gamma\,$, and/or the evolution of $\gamma$ over cosmic history.

The equations (\ref{eq_small_scale_momentum_scalar_2}) and (\ref{eq_small_scale_momentum_vector}) provide the full, theory-independent, momentum constraint equation on small scales, for conservative theories of gravity. 
They are valid in regions of arbitrarily high densities, as long as the gravitational fields are weak enough that a post-Newtonian analysis does not break down (which in cosmology is a very safe assumption, unless one is considering compact objects).
Note that although the PPN parameter $\xi$ is allowed to be non-zero in this class of theories, it appears in neither the scalar nor the vector part of our parameterised momentum constraint equation.

This is our first step towards the momentum constraint on small cosmological scales. 
Let us now consider how the results we have obtained are generalised in non-conservative theories of gravity, which can exhibit violation of global conservation laws at post-Newtonian order.
Moreover, such theories can display preferred-frame effects. These effects do not matter for the scalar perturbations $\h{\Phi}$ and $\h{\Psi}$ to FLRW spacetime, because they are both invariant under local Lorentz boosts at post-Newtonian order, but they are important for the frame-dragging divergenceless vector perturbation $\h{B}_i\,$, which is not boost-invariant.

\subsection{Incorporating preferred-frame effects}\label{subsec:preferred_frame_effects}

Let us now consider how the PPNC vector equation (\ref{eq_small_scale_momentum_vector}) changes when we go from conservative to non-conservative theories of gravity, in which the PPN parameters $\{\alpha_1, \alpha_2, \zeta_1 \}$ are allowed to be non-zero. 
To do this, let us suppose that the perturbed Minkowski coordinate system $\left(t, x^i\right)$, and its perturbed FLRW counterpart $\left(\h{t}, \h{x}^i\right)\,$, are comoving with the preferred frame of the theory, so that the preferred-frame potential $\varphi_i^{\rm PF}$ vanishes.
Then, Eq. (\ref{eq_B_i_PPNC_implicit}) generalises to 
\begin{equation}\label{eq_B_i_preferred_frame_PPNC_implicit}
    B_i = -2\left(\alpha + \gamma\right)V_i + \frac{1}{2}\,\left(\alpha + \alpha_2 - \zeta_1 + 2\xi\right)\,\dot{\chi}_{,i} + B_i^{\rm extra}\,.
\end{equation}
Performing the transformation $B_i \longrightarrow \h{B}_i$ according to Eq. (\ref{expandingvsstatic3}), and again using the gauge function $T$ to ensure that $\h{B}_i$ satisfies the Newtonian gauge condition $\h{B}_{i,i} = 0\,$, we get that 
\begin{eqnarray}\label{eq_B_i_PPNC_in_preferred_frame}
\hat{\nabla}^2 \hat{B}_i = 8\pi G \left(\alpha+\gamma+\frac{\alpha_1}{4} \right) \left[\rho \hat{v}_i\right]^{\rm V} \, a^2 
+ 2 \pi G \alpha_1 a^2 \mathcal{H} \left[\rho\, \hat{x}^i\right]^{\rm V}
+ \h{\nabla}^2 \hat{B}_i^{\rm extra} \, ,
\end{eqnarray}
where it is notable that only $\alpha_1$ is retained in the cosmological version of this equation, with $\alpha_2$ and $\zeta_1$ both being removed by the imposition that $\h{B}_i$ is divergenceless. 

The term $ 2 \pi \alpha_1 a^2 \mathcal{H} \rho\, \hat{x}^i$ on the RHS appears physically problematic as it introduces a spurious dependence of $\h{B}_i$ on the chosen location of the origin of the spatial coordinates $\h{x}^i\,$.
As we have not specified a particular configuration of matter, or any symmetries beyond those of the background, it is hard to see how such a term could possibly be permitted, even in theories with preferred frames. 
Indeed, in the canonical example gravity theories of this kind, which are vector-tensor theories, the preferred frame is usually thought of as corresponding to the frame in which the cosmological vector field has no spatial component, i.e. $A^a = \left(A^0, 0, 0, 0\right)\,$ \cite{jimenez2008cosmic,thomas2023consistent}. It ought to be true by definition that an observer congruence comoving with the cosmic expansion should not measure preferred-frame effects resulting precisely from that expansion.
We will therefore remove the anomalous coordinate-dependent term from Eq. (\ref{eq_B_i_PPNC_in_preferred_frame}), by altering the original PPN equation (\ref{eq_B_i_PPN}) to subtract off the Hubble flow from the Lorentz-violating contribution proportional to $\alpha_1\,$, so that the Lorentz-violating term in the equation is sourced only by peculiar velocities. That is, we have 
\begin{equation}
    \nabla^2 B_i = 8\pi G \,\left(\alpha + \gamma\right)\,\rho \, v_i + 2\pi G \,\alpha_1\,\rho\,\delta v_i - \left(\alpha + \alpha_2 - \zeta_1 + 2\xi\right)U_{,ti} + \nabla^2 B_i^{\rm extra}\,,
\end{equation}
where $\delta v_i = v_i - \delta_{ij}H x^j\,$, and where we have again taken $\varphi_i^{\rm PF}$ to be zero for now. In the classic PPN situation familiar from astrophysical tests of gravity, the Hubble flow term $\delta_{ij} H x^j$ is negligible compared to $v_i\,$, and so the need to subtract it off does not arise.
Transforming again into the cosmological coordinate system, this adjustment to the PPN system results in
\begin{eqnarray}\label{eq_B_i_PPNC_in_preferred_frame_adjusted}
\hat{\nabla}^2 \hat{B}_i = 8\pi G \left(\alpha+\gamma+\frac{\alpha_1}{4} \right) \left[\rho \hat{v}_i\right]^{\rm V} \, a^2 
+ \h{\nabla}^2 \hat{B}_i^{\rm extra} \, .
\end{eqnarray}
It is conceivable that similar changes may need to be made for the term that couples with $\alpha_2$, as this is also a preferred-frame parameter, but as this parameter does not appear in the cosmological equations at the order we are studying we will not concern ourselves with it here.

Now, let us turn our attention to the form this equation takes if we transform away from the preferred frame, so that the local post-Newtonian coordinate system is in motion with respect to the preferred frame. This, of course, will be the general situation - there is no particular reason for any given small-scale astrophysical system to be comoving with a globally defined frame that is ``preferred'' by the underlying theory of gravity.
We will generate the new form of this equation by performing a Lorentz boost in the perturbed Minkowski description of the spacetime, from the preferred frame with 4-velocity $u^a_{\rm PF}$ to some arbitrary $u^a\,$. 
The corresponding perturbed FLRW descriptions, before and after the boost, can then be determined by using the transformations from Eqs. \eqref{eq_statictoexpandingtransformation1} and \eqref{eq_statictoexpandingtransformation2}. 

\begin{figure}[ht]
\begin{tikzpicture}
\coordinate (A) at (0,0);
\coordinate (B) at (0,4);
\coordinate (C) at (4,4);
\coordinate (D) at (4,0);
\coordinate (E) at (0,2);
\coordinate (F) at (2,4);
\coordinate (G) at (4,2);
\coordinate (H) at (2,0);
\coordinate (I) at (-1,2);
\coordinate (J) at (2,5);
\coordinate (K) at (5,2);
\coordinate (L) at (2,-1);
\draw[very thick,{Latex[length=5mm, width=2mm]}-](A)--(B);
\draw[very thick,{Latex[length=5mm, width=2mm]}-](B)--(C);
\draw[very thick,{Latex[length=5mm, width=2mm]}-](D)--(C);
\draw[very thick,{Latex[length=5mm, width=2mm]}-](D)--(A);
\draw[-{Latex[length=4mm, width=1.6mm]},color=red](I)--(A);
\draw[-{Latex[length=4mm, width=1.6mm]},color=red](I)--(B);
\draw[-{Latex[length=4mm, width=1.6mm]},color=red](J)--(B);
\draw[-{Latex[length=4mm, width=1.6mm]},color=red](J)--(C);
\draw[-{Latex[length=4mm, width=1.6mm]},color=red](K)--(C);
\draw[-{Latex[length=4mm, width=1.6mm]},color=red](K)--(D);
\draw[-{Latex[length=4mm, width=1.6mm]},color=red](L)--(D);
\draw[-{Latex[length=4mm, width=1.6mm]},color=red](L)--(A);
\draw[-{Latex[length=4mm, width=1.6mm]}, very thick, color=blue, rounded corners=10pt] (3.5,3.5) -- (0.5,3.5) -- (0.5,0.5) -- (3.5,0.5);
\node[left] at (E) {L};
\node[above] at (F) {E};
\node[right] at (G) {L};
\node[below] at (H) {E};
\node[left] at (I) {Non-expanding coordinates};
\node[above] at (J) {Preferred frame};
\node[right] at (K) {Expanding coordinates};
\node[below] at (L) {General frame};

\end{tikzpicture}
\caption{A schematic of the transformations between perturbed Minkowski and perturbed FLRW geometries, and transformations between the preferred frame and general frames. Lorentz transformations are labelled L, and E denotes transformations between non-expanding and expanding backgrounds.}
\label{fig_coord_systems_PF}
\end{figure}
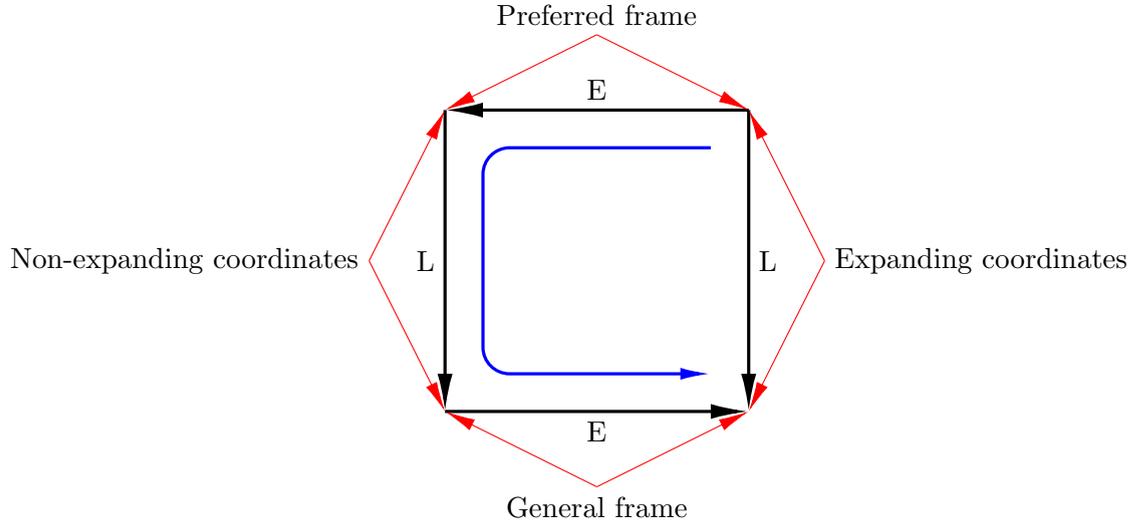

This process is displayed schematically in Fig. \ref{fig_coord_systems_PF}, where the transformation from a perturbed FLRW geometry in the preferred frame to the general frame is indicated by the black arrow from the top-right corner to the bottom-right corner, and which is equivalent to the three transformations around the other sides of the square, collectively denoted by the blue arrow. 
That is exactly what we are trying to do:
\begin{enumerate}
    \item Start off with the form (\ref{eq_small_scale_momentum_vector}) of the perturbed-FLRW equation for $\h{B}_i$ in conservative theories.
    \item Transform back to the perturbed Minkowski system, where we can define the preferred-frame effects that are introduced in non-conservative theories.
    \item Perform a local Lorentz boost in perturbed Minkowski spacetime, to identify how the perturbations to Minkowski spacetime are affected under some boost by the coordinate velocity $w^i$ with respect to the preferred frame.
    \item Transform into expanding coordinates to obtain the new equation for $\h{B}_i$ in non-conservative theories.
\end{enumerate}

We have already discussed the transformations between expanding and static backgrounds, corresponding to the top and bottom of the square in Fig. \ref{fig_coord_systems_PF}, in Section \ref{sec:basic_PPN}. The Lorentz boost between two different coordinate systems covering Minkowski spacetime, corresponding to the left side of the square, are given by the standard expressions:
\begin{eqnarray}
t &\rightarrow& {t} \left( 1+ \frac{1}{2} w^2 + \frac{3}{8} w^4 \right) + \left( 1+ \frac{1}{2} w^2 \right) {x}^i w_i +O(w^5)\,, \\
x^i &\rightarrow& {x}^i +\left( 1+ \frac{1}{2} w^2 \right)  {t}\,  w^i+\frac{1}{2} {x}^j w_j w^i +O(w^4) \, .
\end{eqnarray}
The effects of these on the perturbations to Minkowski spacetime are
\begin{eqnarray}
\Phi \rightarrow \Phi \, , \qquad \Psi \rightarrow \Psi \, ,  \qquad {\rm and} \qquad
B_i \rightarrow B_i - 2 w_i (\Phi+\Psi) \,,
\end{eqnarray}
which justifies our earlier claim that $\Phi$ and $\Psi$ are boost-invariant at post-Newtonian order, but $B_i$ is not.

Transforming back to expanding coordinates after performing this boost, and ensuring yet again that the gauge function $T$ is chosen to preserve the Newtonian gauge condition, we find that 
\begin{equation}\label{eq_small_scale_momentum_vector_including_preferred_frames}
\hat{\nabla}^2 \hat{B}_i = 8\pi G \left(\alpha+\gamma\right)\left[\rho \hat{v}_i\right]^{\rm V}a^2 + 2\pi G \alpha_1\left[\rho\left(\hat{v}_i+\hat{w}_i\right)\right]^{\rm V}a^2 + \h{\nabla}^2 \hat{B}_i^{\rm extra} \,  ,
\end{equation}
where $\hat{w}^i$ is the coordinate velocity of the comoving coordinate system relative to the preferred frame (for example, if the preferred frame is set up to coincide with the Hubble flow, then $\h{w}^i$ vanishes). 
This is the general form of the vector part of the parameterised momentum constraint equation on small scales, written entirely in terms of quantities defined in a perturbed FLRW cosmology in Newtonian gauge. 

As one would expect from the usual interpretation of each of the PPN parameters in the standard PPN formalism, $\hat{w}^i$ does not enter into the terms that couple with $\alpha$ and $\gamma$\,. This follows from the fact that $\alpha$ and $\gamma$ are the $G_{\rm eff}$ and ``spatial curvature'' parameters that are present in standard, conservative, theories of gravity (e.g. $\alpha = \gamma = 1$ in GR), rather than being associated with any exotic preferred-frame effects that may arise in non-conservative gravity theories. 

It should be noted that in order to avoid the presence of another unphysical term of the form $\delta_{ij}\mathcal{H}\h{x}^j$, we have adjusted the preferred frame potential $\varphi_i^{\rm PF}$ so that it too depends only on the velocity of the preferred frame relative to the Hubble flow, $\delta {w}_i = w_i - \delta_{ij} H x^j$\,, just like we did for the $\alpha_1$ term that arose within the preferred frame. That is, we have redefined $\varphi^{\rm PF}_i$ to satisfy
\begin{equation}\label{eq_preferred_frame_potential_2}
    \nabla^2 \varphi^{\rm PF}_i = 2\pi G \,\alpha_1\,\rho\,\delta w_i + 2\alpha_2\, \delta w^j\, U_{,ij}\,,
\end{equation}
which should be compared directly with the classic PPN definition given by Eq. (\ref{eq_preferred_frame_potential_1}). 

This concludes our discussion of the gravitational effects of momentum densities on small-scale perturbations to an homogeneous and isotropic cosmology, in the PPNC approach.
In the next section, we will consider those effects on very large scales, above the Hubble horizon.

\subsection{Super-horizon scales}

We now focus on extending our parameterised equations up to super-horizon scales. Much of this will rely on the separate universe approach to cosmological perturbations that we discussed in Section \ref{sec:basic_PPN}, and which was pioneered in a theory-independent way by Bertschinger \cite{Bertschinger_2006}. 
Recall that using this approach, one could derive the Raychaudhuri and Hamiltonian constraint equations, Eqs. (\ref{eq_PPNC_superhorizon_Raychaudhuri}) and (\ref{eq_PPNC_superhorizon_Hamiltonian_constraint}) respectively, for super-horizon perturbations in the PPNC formalism \cite{Sanghai_2019}. 
Now, we consider how it can be applied to the momentum constraint equation on those scales. In order to do this, we will need to consider the effects of super-horizon peculiar velocity perturbations. 
We can split a peculiar velocity $\h{v}_i$ into scalar and vector parts, $\h{v}_i = \h{v}_i^{\rm S} + \h{\mathfrak{v}}_i\,$, where $\h{\mathfrak{v}}_{i,i} = 0$ and $\h{v}_i^{\rm S} = V_{,i}\,$. Let us deal with the scalar and vector parts, that source the scalar and vector perturbations separately at linear order according to the SVT decomposition, in turn.

Let us start with the scalars. These can be dealt with by considering the effect of a boost in the coordinates $(\tau, \h{x}^i)$ of the expanding FLRW spacetime, such that we transform to new coordinates
\begin{eqnarray}
\tau_* &= \gamma(\hat{v})\left[\tau - \hat{v} \hat{x}\right] = \tau - \hat{v}\hat{x} + \frac{1}{2}\hat{v}^2\tau + ... \\
\hat{x}_* &= \gamma(\hat{v})\left[\hat{x} - \hat{v} \hat{\tau}\right] = \hat{x} - \hat{v}\tau + \frac{1}{2}\hat{v}^2\hat{x} + ... 
\, ,
\end{eqnarray}
and $\hat{y}_* = \hat{y}$ and $\hat{z}_* = \hat{z}$, where $\hat{v}$ is the velocity of the boost in the $\hat{x}$-direction. The conformal part of the metric is unchanged by this transformation, while the scale factor becomes
\begin{equation}
a^2(\tau_*) \simeq a^2(\tau)\left[1 - 2\hat{v}\mathcal{H}\hat{x}\right].
\end{equation}
Hence, the line-element for our geometry becomes
\begin{equation}
\mathrm{d}s^2 = a^2(\tau)\left[-\left(1-2\mathcal{H}V\right)\mathrm{d}\tau^2+ \left(1-2\mathcal{H}V\right)\left\lbrace\mathrm{d}\hat{x}^2 + \mathrm{d}\hat{y}^2 + \mathrm{d}\hat{z}^2\right\rbrace \right],
\end{equation}
where $V = \hat{v}\hat{x}$ is the scalar velocity potential. This is equivalent to the following pair of scalar perturbations:
\begin{equation}\label{scalarsvstheta}
\hat{\Phi} = \mathcal{H}V \qquad {\rm and} \qquad \hat{\Psi} = -\mathcal{H}V \, . 
\end{equation}
Constructing the left-hand side of the momentum constraint equation therefore gives that on super-horizon scales we must have
\begin{equation}\label{eq_superhorizonscalar}
 \hat{\Psi}' +  \mathcal{H}\hat{\Phi} = \left(\mathcal{H}^2 - \mathcal{H}'\right) V \, .
\end{equation}
Inserting the parameterised Friedmann equations (\ref{eq_parametrisedfriedmanneqns1}) and (\ref{eq_parametrisedfriedmanneqns2}), this can equivalently be expressed as 
\begin{equation}\label{eq_superhorizonscalar2}
\hat{\Psi}'_{,i} + \mathcal{H}\hat{\Phi}_{,i} = \frac{4 \pi G}{3} (\alpha + 2 \gamma) \bar{\rho} \hat{v}^{\rm S}_i \,a^2- \frac{1}{3} (\alpha_c+ 2 \gamma_c) \hat{v}^{\rm S}_i\, a^2 \,,
\end{equation}
where we have generalised this expression to an arbitrary boost direction by taking $V_{,i} = \hat{v}^{\rm S}_i$\,. This result can be seen to be consistent with the scalar Hamiltonian constraint and Raychaudhuri equations, derived using Bertschinger's separate universe treatment, and the large-scale momentum conservation equation, which we take as evidence of the validity of our approach.

Using the integrability condition (\ref{eq_integrabilitycondition}), Eq. (\ref{eq_superhorizonscalar2}) gives finally that on super-horizon scales, 
\begin{eqnarray}\label{eq_PPNC_superhorizon_scalar_momentum}
    \left(\h{\Psi}' + \mathcal{H}\h{\Phi}\right)_{,i} &=& 4\pi G\,a^2 \,\left[\gamma - \frac{1}{3}\frac{\mathrm{d}\gamma}{\mathrm{d}\ln{a}} + \frac{1}{12\pi G \bar{\rho}}\frac{\mathrm{d} \gamma_c}{\mathrm{d}\ln{a}}\right]\,\bar{\rho}\,\h{v}_i^{\rm S} \\
    \nonumber &=& 4\pi G a^2 \mu \,\bar{\rho}\,\h{v}_i^{\rm S}\,,
\end{eqnarray}
where we have identified the combination of PPNC parameters in the first line as being precisely the same combination that gives the large-scale ($L \longrightarrow \infty$) limit of the coupling function $\mu(\tau, L)$ that enters into the all-scales Hamiltonian constraint (\ref{eq_PPNC_allscales_Hamiltonian}).

We can put together Eq. (\ref{eq_PPNC_superhorizon_scalar_momentum}) and the small-scale momentum constraint (\ref{eq_small_scale_momentum_scalar_2}) into a single equation that is valid on all cosmological scales,
\begin{equation}\label{eq_generalparametrisedscalareqn}
\hat{\Psi}'_{,i} + \mathcal{H}\hat{\Phi}_{,i} = 4\pi G \mu \left[\rho \hat{v}_i\right]^{\rm S} a^2 + \mathcal{G} \mathcal{H} \hat{\Psi}_{,i} \phantom{\Big]} \,,
\end{equation}
where $\mathcal{G}=\mathcal{G}\left(\tau, L\right)$ is assumed to be a smooth function of time and spatial scale, with limits
\begin{eqnarray}
\label{eq_G1largek} \lim_{L \rightarrow 0}\mathcal{G} &=& \frac{\alpha-\gamma}{\gamma} + \frac{{\rm d}\ln \gamma}{{\rm d}\ln a}\,,
\qquad {\rm and} \qquad
\label{eq_G1smallk} \lim_{L \rightarrow \infty}\mathcal{G} = 0 \,.
\end{eqnarray}
We remind the reader that $\mu$ is a smooth function of $\tau$ and $L$ that varies between small and large scale limits given by Eqs. (\ref{eq_PPNC_scalarperts_small}) and (\ref{eq_PPNC_scalarperts_large1}). 
It is intended that in the super-horizon limit the combination $\rho \hat{v}_i$ should be understood as approaching $\bar{\rho} \hat{v}_i$, as in this limit the density contrast is assumed to be perturbatively small. 

It is worth remembering that the function $\mathcal{G}$ vanishes at all times and on all scales for the case of GR with a cosmological constant, in which case $\alpha = \gamma = 1$\,. This will not be the case in general though, and for modified theories of gravity it is expected that $\mathcal{G} \neq 0$ on small scales.
Na{\" i}vely, one can think of a positive $\mathcal{G}\mathcal{H}\h{\Psi}_{,i}$ term on the RHS as effectively reducing the damping term $\mathcal{H}\h{\Phi}_{,i}$ on the LHS of the momentum constraint, and therefore driving faster growth in $\h{\Psi}\,$.
The importance of the $\mathcal{G}$ coupling function will be discussed at length in Chapter 6, where we will find that it can introduce phenomenology in the CMB temperature anisotropies that deviates substantially from the behaviour expected in the $\Lambda$CDM concordance model. 

The use of a single new modifying function $\mathcal{G}$, with no other terms appearing on the RHS other than the momentum density term and the novel term $\mathcal{G}\mathcal{H}\h{\Psi}_{,i}\,$, is justified by the fact that no other terms could be produced in either the post-Newtonian regime or the regime of linear cosmological perturbations. 
In a parameterised framework such as PPNC, we do not change anything about the potentials on the LHS: the form of the LHS of the equation is entirely prescribed by the Einstein tensor at the relevant orders in $v$ and $\delta\,$, depending on the context.
In the post-Newtonian regime, power-counting in $v$ tells us that the only scalar objects that can possibly arise on the RHS at $\mathcal{O}(v^3)$ are the scalar part of the momentum density $\rho v_i^{\rm S}\,$, time derivatives of scalar perturbations, and scalar perturbations multiplied by $\mathcal{H}$\,. Because $\h{\Phi}$ and $\h{\Psi}$ are related on these scales simply by the ratio $\alpha/\gamma\,$, there is no need for any additional terms of the form $\h{\Phi}'$ or $\mathcal{H}\h{\Phi}\,$ on the right hand side.
In the CPT regime, it is again only these terms that can arise at linear order in the smallness parameter $\epsilon \sim \delta$\,, and in fact the situation is even simpler because $\left[\rho\h{v}_i\right]^{\rm S}$ just becomes $\bar{\rho}\h{v}_i^{\rm S}\,$.
Therefore, it is safe to assume that, given we expect the presence of an intermediate regime in which both the post-Newtonian and linear CPT expansions hold, the form of Eq. (\ref{eq_generalparametrisedscalareqn}) is entirely sufficient to describe the momentum constraint for scalar perturbations on all scales in parameterised post-Newtonian cosmology.

\begin{figure}[ht]
    \centering
    \includegraphics[width=0.8\linewidth]{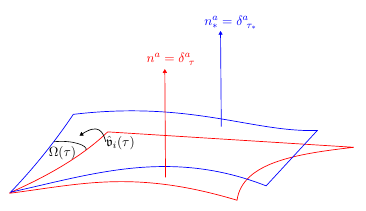}
    \caption{Rotating coordinates by a time-dependent angle $\Omega$ induces the divergenceless vector 3-velocity perturbation $\h{\mathfrak{v}}_i$ in the cosmological fluid, which can be ascribed to a divergenceless vector perturbation $\h{B}_i$ to the FLRW metric. Depicted here are spacelike hypersurfaces of constant $\tau$ and $\tau_*\,$.}
    \label{fig_horizon_scalerotation}
\end{figure}

Let us now move on to the constraint equation for the divergenceless vector perturbation $\h{B}_i$ on very large scales.
In order to construct a divergenceless vector version of the momentum constraint equation for super-horizon scales, let us now consider the case where we rotate our spatial coordinates, rather than boosting. This will produce an apparent vortical motion in the fluid that fills the spacetime, as illustrated in Fig. \ref{fig_horizon_scalerotation}. 

In order to induce this perturbation we rotate coordinates by the angle $\Omega=\Omega(\tau)$, such that
\begin{eqnarray}
\hat{x}_* = \h{x} \cos{\Omega} +\h{y} \sin{\Omega} \,, \qquad \hat{y}_* = \h{y} \cos{\Omega}-\h{x} \sin{\Omega} \, ,
\end{eqnarray}
with $\h{z}_*=\h{z}$ and $\tau_*=\tau$\,, and we interpret the ${x_*}^a$ coordinates as the canonical coordinate basis (which is perturbatively close to $x^a$) on a different FLRW cosmology. Putting this into the line element of a spatially flat FLRW geometry gives us
\begin{eqnarray}
\mathrm{d}s^2 &=& a^2(\tau_*)\left[-\mathrm{d}{\tau_*}^2+\mathrm{d}{\hat{x}_*}^2+\mathrm{d}{\hat{y}_*}^2+\mathrm{d}{\hat{z}_*}^2\right]\\
\nonumber &=& a^2(\tau)\left[-\mathrm{d}\tau^2+\mathrm{d}\hat{x}^2+\mathrm{d}\hat{y}^2+\mathrm{d}\hat{z}^2\right]  +2a^2(\tau)\frac{\mathrm{d}\Omega}{\mathrm{d}\tau}\left(\hat{y}\mathrm{d}\hat{x}-\hat{x}\mathrm{d}\hat{y}\right)\mathrm{d}\tau\, ,
\end{eqnarray}
where we have taken $\Omega = \Omega (\tau)$ and expanded to leading order in $\mathrm{d}\Omega/\mathrm{d}\tau\,$. Comparing this to a linearly perturbed FLRW geometry allows us to identify that the metric in the rotated coordinate system is automatically in Newtonian gauge, and that we have induced a divergenceless vector perturbation
\begin{equation}\label{eq_generalDVhorizonscale}
\hat{B}_i = \frac{{\rm d}\Omega}{{\rm d}\tau} \left(\hat{y}, -\hat{x},0\right) = -\hat{\mathfrak{v}}_i\, ,
\end{equation}
where $\hat{\mathfrak{v}}_i$ is the divergenceless vector part of the fluid's 3-velocity in the rotating coordinates.
Clearly there is nothing special about the direction of the axis of rotation in this example, so we expect the result $\hat{B}_i = - \hat{\mathfrak{v}}_i$ to be valid in general.

Combining our result with the relevant prefactor allows us to write down a straightforward equation for the divergenceless vector part of the super-horizon momentum constraint:
\begin{equation}\label{eq_DV_horizonfullconstraint}
2 \left(\mathcal{H}'-\mathcal{H}^2\right)\hat{B}_i = - 2 \left(\mathcal{H}'-\mathcal{H}^2\right) \hat{\mathfrak{v}}_i \, , 
\end{equation}
or, equivalently, using Eqs. (\ref{eq_parametrisedfriedmanneqns1}) and (\ref{eq_parametrisedfriedmanneqns2}), 
\begin{equation} \label{eq_superhorizon_divergenceless_vector}
2 \left(\mathcal{H}'-\mathcal{H}^2\right)\hat{B}_i = \frac{8 \pi G }{3} (\alpha + 2 \gamma) \bar{\rho}\, \hat{\mathfrak{v}}_i \,a^2- \frac{2}{3} (\alpha_c+ 2 \gamma_c) \hat{\mathfrak{v}}_i \, a^2 \, .
\end{equation}
The right hand side here is of exactly the same form as the scalar part of the super-horizon momentum constraint equation, Eq. (\ref{eq_superhorizonscalar2}), which gives us confidence in its validity. Just like in the scalar case, we can group the PPN parameters to give that on super-horizon scales
\begin{equation}
    2 \left(\mathcal{H}'-\mathcal{H}^2\right)\hat{B}_i = 4\pi G a^2 \, \mu\, \bar{\rho}\, \mathfrak{v}_i\,,
\end{equation}
with $\mu$ equal to its $L \longrightarrow \infty$ value given by Eq. (\ref{eq_PPNC_scalarperts_large1}).

We can now see that it is possible to write Eq. (\ref{eq_superhorizon_divergenceless_vector}) together with the small-scale vector equation (\ref{eq_small_scale_momentum_vector_including_preferred_frames}) in the unified form
\begin{equation}\label{eq_generalparametrisedvectoreqn}
 2 \left(\mathcal{H}'-\mathcal{H}^2\right)\hat{B}_i+\frac{1}{2} \hat{\nabla}^2\hat{B}_i  = 8\pi G (\mu+\mathcal{Q}) \left[\rho \hat{v}_i\right]^{\rm V} a^2  + \alpha_1 \, \pi G \left[\rho \hat{w}_i\right]^{\rm V} a^2\, ,
\end{equation}
where $\mu$ is again given by Eqs. (\ref{eq_PPNC_scalarperts_small}) and (\ref{eq_PPNC_scalarperts_large1}), and we have introduced a new smooth coupling function $\mathcal{Q}(\tau, L)$, which has the limits
\begin{eqnarray}
\label{eq_G2largek} \lim_{L \rightarrow 0}\mathcal{Q} &=& \frac{\alpha - \gamma}{2} + \frac{\alpha_1}{8} 
\qquad {\rm and} \qquad
\label{eq_ G2smallk} \lim_{L \rightarrow \infty}\mathcal{Q} = 0\,.
\end{eqnarray}
Again, the quantity $\left[\rho \hat{v}_i\right]^{\rm V}$ should be understood to reduce to $\bar{\rho} \hat{\mathfrak{v}}_i$ on large scales, when the density contrast becomes perturbatively small. 
The novel coupling function $\mathcal{Q}$ vanishes identically in GR, when $\alpha=\gamma=1$ and $\alpha_1=0$, but is not expected to be zero on small scales in generic modified theories of gravity. 
We have set $\hat{B}_i^{\rm extra}=0$, as it is not required for any of the MG theories we consider in this thesis, as we will demonstrate in the next section.
Note that there are no preferred-frame effects on super-horizon scales, with $\mathcal{Q}$ vanishing for $L \longrightarrow 0$ just as $\mathcal{G}$ does. This is entirely expected, because on these scales the evolution of perturbations is prescribed by the background (hence the validity of the separate-universe approach \cite{wands2000new,Bertschinger_2006}).

As it was for the scalar equation (\ref{eq_generalparametrisedscalareqn}), it is straightforward to see why the form of the vector equation (\ref{eq_generalparametrisedvectoreqn}) is exhaustive\footnote{With the notable exception of $B_i^{\rm extra}\,$, which we have removed as discussed.}. 
On small scales, the only divergenceless vectors that can possibly appear will be sourced by either $\left[\rho \h{v}_i\right]^{\rm V}$ or $\left[\rho \h{w}_i\right]^{\rm V}\,$. On large scales, the only vector momentum term that is allowed is $\bar{\rho} \h{\mathfrak{v}}_i$\,. 
According to the entire ethos of a parameterised framework, we do not change the form of the LHS, which is the most general form of the relevant Einstein tensor component $G_0^{\ i}$ that can appear in both CPT and the post-Newtonian regime. The only things that we can change are the couplings to matter fields on the RHS, and there are no other objects that could be coupled to that have the right post-Newtonian or perturbative order.

The scalar and vector PPNC equations (\ref{eq_generalparametrisedscalareqn}) and (\ref{eq_generalparametrisedvectoreqn}) are the central results of this section. Now, we will demonstrate their applicability, using two example modified theories of gravity and one other model of dark energy.

\subsection{Application to example theories}

Having derived our parameterised momentum constraint equations, we will now show how they work for some example theories of modified gravity and dark energy. This requires determining the PPNC parameters for each theory, and then demonstrating that inputting these into our equations results in the correct small and large-scale limits of their weak-field theory. 

Before we proceed with the worked examples, it will prove useful to collect together some general results about super-horizon adiabatic perturbations, that are true in any metric theory of gravity.
The salient point is the definition of an adiabatic perturbation, which states that the gauge-invariant entropy perturbation $S_{XY}$ between any two scalars $X$ and $Y$ vanishes:
\begin{equation}\label{eq_adiabaticcondition}
S_{XY} := \mathcal{H}\left(\frac{\delta X}{\bar{X}'} - \frac{\delta Y}{\bar{Y}'}\right) = 0.
\end{equation}
Moreover, the separate-universe approach tells us that for adiabatic perturbations on super-horizon scales, we have  $\delta \rho = -3\mathcal{H}V \bar{\rho}\,$.
As a consequence, 
\begin{equation}\label{eq_largescaledeltarho3}
\frac{\delta\rho}{\bar{\rho}'} = \frac{\hat{\Psi}}{\mathcal{H}} \, , \qquad {\rm and \;\; therefore} \qquad
\frac{\delta X}{\bar{X}'} = \frac{\hat{\Psi}}{\mathcal{H}},
\end{equation}
for any scalars $X$ that appear in the theory.
It is also useful to recall that the scalar part of the relativistic Euler equation in linear CPT is
\begin{equation}\label{eq_superhorizon_Eulerequation}
V' + \mathcal{H}V - \hat{\Phi} = 0 \,.
\end{equation}
Let us now consider our example theories of gravity and dark energy, in increasing order of mathematical complexity.

\subsubsection*{Quintessence}

Let us start with quintessence, which refers to a scalar field, $\phi$, that couples minimally to gravity. Because of this minimal coupling, quintessence should be thought of as a model for dark energy, rather than a modified theory of gravity.
The full action for a quintessence model is \cite{copeland2006dynamics}
\begin{eqnarray}\label{eq_quintessence_action}
S = \int \mathrm{d}^4 x \sqrt{-g}\left[\frac{R}{16\pi G} -\frac{1}{2}g^{ab}\nabla_{a}\phi\nabla_{b}\phi - V(\phi)\right] + S_m\left(\psi,g_{ab}\right) \,,
\end{eqnarray}
where $\psi$ denotes matter fields. The field equations are
\begin{equation}\label{eq_quintessence_einstein_equations}
G_{ab} = 8\pi G\, T_{ab} + 8 \pi G \left( g_{ab}\left[-\frac{1}{2}g^{cd}\nabla_{c}\phi\nabla_{d}\phi - V(\phi)\right] + \nabla_{a}\phi \nabla_{b}\phi  \right) \,,
\end{equation}
and the Klein-Gordon equation
\begin{eqnarray}\label{eq_quintessence_Klein_Gordon}
    \Box{\phi} = \frac{\mathrm{d}V}{\mathrm{d}\phi}\,.
\end{eqnarray}
Performing a post-Newtonian expansion about Minkowski spacetime gives the PPNC parameters,
\begin{eqnarray}
\alpha &=& \gamma = 1\,, \\
\nonumber
\gamma_c &=& -4\pi G \left(\frac{\left(\bar{\phi}'\right)^2}{2a^2} + V(\bar{\phi})\right)\,, \quad {\rm and} \quad \alpha_c = -8\pi G \left(\frac{\left(\bar{\phi}'\right)^2}{a^2} - V(\bar{\phi})\right),
\end{eqnarray}
where $\bar{\phi}$ is the time-dependent background value of $\phi$, and $\alpha_1 = \alpha_2 = \xi = \zeta_1 = 0$.
The fact that the standard PPN parameters are all equal to their GR values follows from the fact that quintessence theories are minimally coupled. 
Thus, in astrophysical settings, where the time variation of $\bar{\phi}$ is irrelevant, the results of a post-Newtonian expansion are equivalent to those for the standard Einstein equations $G_{ab} = 8\pi G\,T_{ab}\,$. 
The cosmological parameters $\alpha_c$ and $\gamma_c$ can be seen easily upon inserting an FLRW metric ansatz into Eq. (\ref{eq_quintessence_einstein_equations})\,. 
Let us now examine the CPT equations in quintessence models, in order to show that they are correctly described by the PPNC framework.

\begin{itemize}

    \item Small scales: applying the parameter values above to Eqs. (\ref{eq_generalparametrisedscalareqn}) and (\ref{eq_generalparametrisedvectoreqn}), we find
    \begin{equation}\label{eq_quintessencesmallscale0i}
    \frac{1}{2}\hat{\nabla}^2\hat{B}_i + 2\left[\hat{\Psi}+\mathcal{H}\hat{\Phi}\right]_{,i} = 8\pi G \rho\hat{v}_i a^2 \,,
    \end{equation}
    where the scalar and vector parts can be extracted trivially. This is just the standard GR result, with no effect from either the homogeneous $\bar{\phi}(t)$ or the inhomogeneous perturbation $\delta \phi(\tau, \h{x}^i) \sim v^2\,$.
    It now remains to show that this is the same equation one would obtain from performing a direct post-Newtonian expansion of Eq. (\ref{eq_quintessence_einstein_equations}). For this, we can note that the Klein-Gordon equation can be expanded to give
    \begin{equation}
    \frac{1}{a^2}\left(\bar{\phi}'' + 3 \mathcal{H} \bar{\phi}'\right) = - \frac{dV (\bar{\phi})}{d\phi}  \qquad {\rm and} \qquad \hat{\nabla}^2 \delta \phi = 0 \, ,
    \end{equation}
    where we have separated out the leading-order part of this equation into its background and inhomogeneous parts, but it should be noted that these parts appear at the same perturbative order in the post-Newtonian hierarchy, unlike in cosmological perturbation theory.
    
    As the inhomogeneous equation has no source terms, this implies that the leading-order part of the quintessence field in the post-Newtonian expansion must be homogeneous, which in turn means that all contributions from the scalar field to the leading-order part of the $G_{0i}$ field equation must vanish on small scales \footnote{See Ref. \cite{goldberg2017perturbation} for a more detailed discussion of this phenomenon.}. We are therefore led to an equation that is identical to Eq. (\ref{eq_quintessencesmallscale0i}), from our direct analysis of the field equations (\ref{eq_quintessence_einstein_equations}), which verifies our parameterised equation for this example.

    \item Large scales: for super-horizon scales, the parameterised scalar equation (\ref{eq_generalparametrisedscalareqn}) becomes
    \begin{equation}
    \hat{\Psi}'+\mathcal{H}\hat{\Phi} = 4\pi G\,\bar{\rho}\,a^2\,V + 4\pi G \,\bar{\phi}^{\prime 2}\, V \,.
    \end{equation}
    This can be compared to the equation for scalar super-horizon perturbations, derived directly from the modified Einstein equation (\ref{eq_quintessence_einstein_equations}),
    \begin{equation}
    \hat{\Psi}'+\mathcal{H}\hat{\Phi} = 4\pi G\,\bar{\rho}\,a^2\,V - 4\pi G \,\bar{\phi}'\,\delta\phi.
    \end{equation}
    It can be seen that these two equations are identical provided that $\delta\phi = -V\,\bar{\phi}'$\,, which is guaranteed by the combination of the adiabatic condition (\ref{eq_largescaledeltarho3}) and the Euler equation (\ref{eq_superhorizon_Eulerequation}). The parameterised scalar momentum constraint on super-horizon scales is therefore identical to what is obtained from directly expanding the field equations (\ref{eq_quintessence_einstein_equations}).
    
    The divergenceless vector part is even simpler: the super-horizon limit of our general result \eqref{eq_generalparametrisedvectoreqn} follows immediately from the field equations. 
    Hence, we have shown that our parameterised momentum constraint equation correctly reproduces all of the results that one would obtain from directly dealing with the quintessence model of dark energy, in both the scalar and divergenceless vector parts of the theory, and on both large and small limits. 
    We therefore have our first explicit verification of its validity\footnote{The validity of the PPNC Hamiltonian constraint and Raychaudhuri equations for both quintessence and the Brans-Dicke scalar-tensor theory was demonstrated by Sanghai \& Clifton in Ref. \cite{Sanghai_2017}.}.
    
\end{itemize}

\subsubsection*{Brans-Dicke theory}

Let us now consider Brans-Dicke theory \cite{brans1961mach}, which is a scalar-tensor theory of the Bergmann-Wagoner \cite{bergmann1968comments,wagoner1970scalar} type (\ref{eq_bergmann_wagoner_action}), with the potential $\Lambda(\phi) = 0$ and the coupling function $\omega(\phi)$ just equal to a constant $\omega\,$. 
The field equations are given by Eqs. (\ref{eq_EOM_metric_scalar_tensor}) and (\ref{eq_EOM_KleinGordon_scalar_tensor}). It is worth noting that for the Brans-Dicke case, the generalised Klein-Gordon equation (\ref{eq_EOM_KleinGordon_scalar_tensor}) can be written in a simplified form,
\begin{equation}\label{eq_BDT_klein_gordon}
    \Box{\phi} = \dfrac{8\pi G}{3 + 2\omega}\,T\,,
\end{equation}
where $T$ is the trace of the energy-momentum tensor.

The relevant post-Newtonian parameters for this theory are 
\begin{eqnarray}\label{eq_BransDicke_PPNparameters}
\alpha = \dfrac{4+2\omega}{3+2\omega} \, \dfrac{1}{\bar{\phi}} \,, \quad
\gamma = \dfrac{2+2\omega}{3+2\omega} \, \dfrac{1}{\bar{\phi}} \, ,  \quad {\rm and} \quad
\alpha_1 = \alpha_2 = \xi = \zeta_1 = 0 \, ,
\end{eqnarray}
with additional cosmological parameters
\begin{eqnarray}
\alpha_c =\frac{1}{a^2} \left[ -\frac{\bar{\phi}^{''}}{\bar{\phi}} + \mathcal{H} \frac{\bar{\phi}'}{\bar{\phi}} - \omega\left(\frac{\bar{\phi}'}{\bar{\phi}}\right)^2 \right]\ \ {\rm and} \ \ \gamma_c  = -\frac{1}{2 a^2}\left[\frac{\bar{\phi}^{''}}{\bar{\phi}}-\mathcal{H}\frac{\bar{\phi}'}{\bar{\phi}} + \frac{\omega}{2}\left(\frac{\bar{\phi}'}{\bar{\phi}}\right)^2\right] \,.
\end{eqnarray}
The Friedmann equations can be obtained by applying these PPNC parameter values to Eqs. (\ref{eq_parametrisedfriedmanneqns1}) and (\ref{eq_parametrisedfriedmanneqns2}). 
Let us now carry out for the Brans-Dicke theory the same exercise we did for the quintessence model of dark energy, in order to verify that the PPNC formalism produces the correct results, that can be obtained from the equations of motion of the theory themselves.

\begin{itemize}

    \item Small scales: using the Brans-Dicke PPNC parameters, we can immediately write down the scalar part of the PPNC momentum constraint (\ref{eq_generalparametrisedscalareqn}) on small scales as
    \begin{equation}\label{eq_RHSBDTscalars}
    2\left(\h{\Psi}'+\mathcal{H}\h{\Phi}\right)_{,i} = 8\pi G\,\frac{1}{\bar{\phi}}\frac{2+2\omega}{3+2\omega}\,\left[\rho \h{v}_i\right]^{\rm S}a^2 + 2\left(\frac{1}{1+\omega} - \frac{\bar{\phi}'}{\mathcal{H}\bar{\phi}}\right)\,\mathcal{H}\h{\Psi}_{,i}\,,
    \end{equation}
    and the divergenceless vector part  (\ref{eq_generalparametrisedvectoreqn}) as
    \begin{equation} \label{eq_bdvec}
    \hat{\nabla}^2\hat{B}_i = \frac{16\pi G\, a^2}{\bar{\phi}}\left[\rho \hat{v}_i\right]^{\rm V}\,.
    \end{equation}
    Let us now show that a direct post-Newtonian expansion of the $G_{0i}$ component of the field equations (\ref{eq_EOM_metric_scalar_tensor}) generates the same results.  

    Focusing on the scalar part of Eq. (\ref{eq_EOM_metric_scalar_tensor}) gives
    \begin{equation}\label{eq_BDTmomentumconstraintscalar}
    2\left[\hat{\Psi}'+\mathcal{H}\hat{\Phi}\right]_{,i} = \frac{8\pi G\, a^2}{\bar{\phi}}\left[\rho \hat{v}_i\right]^{\rm S} + \frac{1}{\bar{\phi}}\left[\mathcal{H}\delta\phi-\hat{\Phi}\bar{\phi}'-\delta\phi'\right]_{,i} - \frac{\omega \bar{\phi}'}{\bar{\phi}^2}\delta\phi_{,i}.
    \end{equation}
    To deal with the terms involving $\delta \phi$, let us note that a post-Newtonian expansion of the scalar field equation (\ref{eq_BDT_klein_gordon}) tells us that
    \begin{equation}
    \hat{\nabla}^2 \delta \phi = -\frac{8\pi G}{3+2\omega}\, \delta\rho\, a^2 \, , \qquad {\rm which \;\; implies} \qquad \delta \phi = \frac{\bar{\phi}}{1+\omega} \,\hat{\Psi}\, .
    \end{equation}
    Using this result, we get that in Brans-Dicke theory, we have on small scales
    \begin{eqnarray}\label{eq_BDTmomentumconstraintscalar_2}
        2\left(\h{\Psi}'+\mathcal{H}\h{\Phi}\right)_{,i} &=& \frac{8\pi G\,a^2}{\bar{\phi}}\left[\rho \h{v}_i\right]^{\rm S} - \frac{1}{1+\omega}\h{\Psi}'_{,i} \\
        \nonumber && + \left[\frac{1}{1+\omega} - \frac{3+2\omega}{1+\omega}\frac{\bar{\phi}'}{\mathcal{H}\bar{\phi}}\right]\mathcal{H}\h{\Psi}_{,i}\,.
    \end{eqnarray}
    Finally, we can use the Poisson equation for $\h{\Psi}$ and the continuity equation to obtain
    \begin{equation}
        \h{\Psi}'_{,i} = \frac{4\pi G\,a^2}{\bar{\phi}}\frac{2+2\omega}{3+2\omega}\left[\rho \h{v}_i\right]^{\rm S} - \frac{\bar{\phi}'}{\bar{\phi}}\h{\Psi}_{,i} - \mathcal{H}\h{\Psi}_{,i}\,,
    \end{equation}
    whence it can be seen that replacing $\h{\Psi}'_{,i}$ in  Eq. (\ref{eq_BDTmomentumconstraintscalar_2}) leads it to reduce to Eq. (\ref{eq_RHSBDTscalars}). 
    
    Hence, we have verified our parameterised equation (\ref{eq_generalparametrisedscalareqn}) in this case. 
    We have also verified that the divergenceless vector part of Eq. (\ref{eq_EOM_metric_scalar_tensor}) correctly reproduces Eq. (\ref{eq_bdvec}), which follows straightforwardly as there are no direct contributions from the scalar field to the divergenceless vector part of the $G_{0i}$ equation: it enters only through the combination $\alpha + \gamma=2/\bar{\phi}\,$.

    \item Large scales: On super-horizon scales, the parameterised scalar equation (\ref{eq_generalparametrisedscalareqn}) can be written
    \begin{equation}\label{eq_superhorizonscalarBDTPPNC}
    2\left[\hat{\Psi}'+\mathcal{H}\hat{\Phi}\right] = \frac{8\pi G\,\bar{\rho}\,a^2}{\bar{\phi}} \, V +  \left[-2\mathcal{H} \frac{\bar{\phi}'}{\bar{\phi}} + \frac{\bar{\phi}^{''}}{\bar{\phi}} + \omega\left(\frac{\bar{\phi}'}{\bar{\phi}}\right)^2\right]\, V \, ,
    \end{equation}
    where we have made use of the background equation for the scalar field (\ref{eq_BDT_klein_gordon}), which reads
    \begin{equation}
    \bar{\phi}^{''} + 2\mathcal{H}\bar{\phi} = \frac{8\pi G \bar{\rho}a^2}{3+2\omega} \, .
    \end{equation}

    For adiabatic perturbations, Eqs. (\ref{eq_largescaledeltarho3}) and (\ref{eq_CPT_Euler_eqn}) give
    \begin{equation}\label{eq_BDT_adiabaticscalarfield}
    \frac{\delta\phi}{\bar{\phi}'} = -V \, , \qquad {\rm and} \qquad
    \frac{\bar{\phi}^{''}}{\bar{\phi}} \, V = -\frac{\delta\phi'}{\bar{\phi}} \, .
    \end{equation}
    Substituting these results into Eq. (\ref{eq_superhorizonscalarBDTPPNC}), we get
    \begin{equation}\label{eq_BDTsuperhorizonscalars}
    2\left[\hat{\Psi}'+\mathcal{H}\hat{\Phi}\right] = \frac{8\pi G\,\bar{\rho}\,a^2}{\bar{\phi}}\, V + \frac{1}{\bar{\phi}}\left[\mathcal{H}\delta\phi - \hat{\Phi}\bar{\phi}' - \delta\phi'\right] - \frac{\omega\bar{\phi}'}{\bar{\phi}^2}\delta\phi \,,
    \end{equation}
    which is precisely what is obtained by directly expanding the scalar part of the $G_{0i}$ field equation (\ref{eq_EOM_metric_scalar_tensor}), as seen by linearising Eq. (\ref{eq_BDTmomentumconstraintscalar_2}). 
    The divergenceless vector part again agrees immediately with the relevant limit of Eq. (\ref{eq_generalparametrisedvectoreqn}), which verifies that our parameterised equations reproduce the results of Brans-Dicke theory exactly on both small and large scales, and in both the scalar and divergenceless-vector sectors of the theory.
    
\end{itemize}

\subsubsection*{Vector-tensor theory}

Let us now show our parameterised momentum constraint also works in theories that contain a timelike vector field, $A^a$, as well as the metric. We consider unconstrained theories of the form $S_{\rm VT}$ in Eq. (\ref{eq_vector_tensor_action_unconstrained}), restricting ourselves to a simplified subclass, with $\epsilon = 0$ and $\eta = -2\omega$. 
Nevertheless, the simplified theories will still display all the gravitational phenomena that we are interested in here, the primary novel ones being preferred-frame effects.
The field equations for the metric are \cite{Will_1993}
\begin{equation}
    G_{ab} + \tau \,\Theta^{(\tau)}_{ab} + \omega\, \Theta^{(\omega)}_{ab} - 2\omega\,\Theta^{(\eta)}_{ab} = 8\pi G\,T_{ab}\,,
\end{equation}
where
\begin{eqnarray}
    \Theta^{(\tau)}_{ab} &=& \nabla_c A_a\,\nabla^c A_b + \nabla_a A_c\,\nabla_b A^c - \frac{1}{2}\,g_{ab}\,\nabla_c A_d\,\nabla^c A^d \\
    \nonumber && + \nabla_c\left[A^c \nabla_{(a}A_{b)} - A_{(a}\nabla_{b)}A^c - A_{(a}\nabla^c A_{b)} \right]\,, \\
    \nonumber \Theta^{(\omega)}_{ab} &=& R\,A_a A_b + A_c A^c\, G_{ab} - \nabla_a\nabla_b\left(A_c A^c\right) - g_{ab}\,\Box{\left(A_c A^c\right)}\,, \quad {\rm and} \\
    \nonumber \Theta^{(\eta)}_{ab} &=& 2 A^c A_{(a}R_{b)}c + \frac{1}{2}\,g_{ab}\left[\nabla_c \nabla_d\left(A^c A^d\right) - A^c A^d\,R_{cd}\right] \\
    \nonumber && - \nabla_c\nabla_{(a}\left(A_{b)}A^c\right) + \frac{1}{2}\,\Box{\left(A_a A_b\right)}\,.
\end{eqnarray}
The equation of motion for the vector field is a generalised Proca equation \cite{Hellings_1973},
\begin{equation}\label{VTvectorEOM}
2\omega \,A^{b}\,G_{ab} + \tau\,\Box{A_a} = 0\,.
\end{equation}

The PPN parameters for this theory are
\begin{eqnarray}\label{eq_hellingsPPNCparamsalpha}
\alpha &=& \frac{2\left[\tau +\omega\bar{A}^2\left(8\omega-\tau\right)\right]}{\tau\left[2+\bar{A}^2\left(\tau-4\omega\right)-\omega\bar{A}^4\left(\tau-10\omega\right)\right]}\,, \\
\nonumber \label{eq_hellingsPPNCparamsgamma} \gamma &=& \frac{2\left(1-\omega\bar{A}^2\right)}{2+\bar{A}^2\left(\tau-4\omega\right)-\omega\bar{A}^4\left(\tau-10\omega\right)}\,,\\ 
\nonumber \label{eq_hellingsPFparameter} \alpha_1 &=& \frac{16\tau}{2\tau+\bar{A}^2\left(2\tau\left(\tau+\omega\right)-\left(\tau+2\omega\right)^2\right)}-\frac{16\tau-2\omega\bar{A}^2\left(\tau-4\omega\right)}{2\tau+\tau\bar{A}^2\left(\tau-4\omega\right)-\omega\tau\bar{A}^4\left(\tau-10\omega\right)}\,,
\end{eqnarray} 
and $\zeta_1 = \xi = 0$\,.
The cosmological parameters are
\begin{eqnarray}
\nonumber \alpha_c &=& \frac{1}{a^2} \Bigg[ \left(\alpha-\frac{2}{2-\bar{A}^2\left(\tau-2\omega\right)}\right)\frac{\tau\left(2-\bar{A}^2\left(\tau-2\omega\right)\right)\left(\bar{A}^{''}+2\mathcal{H}\bar{A}'\right)}{\bar{A}^2\left(\tau-2\omega\right)} \\
\label{eq_hellingsgammac} && - \left(\alpha+\frac{6}{2-\bar{A}^2\left(\tau-2\omega\right)}\right)\frac{\tau\bar{A}^{'\,2}}{4}+ \frac{6\mathcal{H}\bar{A}\bar{A}'\left(\tau-2\omega\right)}{2-\left(\tau-2\omega\right)\bar{A}^2} \Bigg]\,, \quad {\rm and} \\
\nonumber \gamma_c &=& \frac{1}{a^2} \Bigg[ \left(\gamma - \frac{2}{2-\bar{A}^2 \left(\tau-2\omega\right)}\right)\frac{\tau\left(2-\bar{A}^2\left(\tau-2\omega\right)\right)\left(\bar{A}^{''}+2\mathcal{H}\bar{A}'\right)}{4\bar{A}^2\left(\tau-2\omega\right)} - \frac{\gamma \tau\bar{A}^{'\,2}}{4} \Bigg] \,,
\end{eqnarray}
where $\bar{A}$ is the background value of $A_0$. 
Note that we have used conformal time for the derivatives $\bar{A}'\,$, but this should not be confused with the unrelated $\tau$ parameter in the vector-tensor field theory action (\ref{eq_vector_tensor_action_unconstrained}). 

Because the PPN parameter $\alpha_1$ is generically non-zero, we need to account carefully for preferred-frame effects, by working out how the velocity $\h{w}_i$ of a generic expanding coordinate system is related to the spatial part of $A_a\,$. 
We will also need to check whether there is some additional cosmological contribution $\h{B}_i^{\rm extra}$ to the small-scale divergenceless vector perturbations. Let us now work through that procedure, so that the all-scales PPNC result can be constructed.

\begin{itemize}
    \item Small scales: Let us focus first on the post-Newtonian regime. In this case we can write $$A_{\mu} = \left(\bar{A} + \delta A_0^{(2)}, \delta A_i^{(1)} + \delta A_i^{(3)}\right)\,,$$ where $\bar{A}\sim \mathcal{O}(v)$, and superscripts indicate the perturbative order in $v$\,.
    The vector field equation of motion gives immediately $\hat{\nabla}^2\delta A^{(1)}_i = 0$\,, which with suitable boundary conditions implies $\delta A^{(1)}_i=\delta A^{(1)}_i(\tau)$\,. As $\delta A^{(1)}_i$ is spatially constant, it must necessarily be the derivative of a scalar, i.e. $\delta A^{(1)}_i = \delta A^{(1)\, {\rm S}}_i\,$.
    
    These theories have a preferred frame, which is picked out by the direction of the timelike vector field $A_a\,$.
    To complete the full set of ingredients required to compute the momentum constraint on small scales in the PPNC framework, we need to add to the PPNC parameters the velocity $\hat{w}_i$ of expansion-comoving observers with respect to the preferred frame, and the cosmological divergenceless vector $\hat{B}^{\rm extra}_i$. We will now find those in turn. 

    To determine the preferred-frame 3-velocity $\hat{w}_i$ that couples to $\alpha_1$, consider a local Lorentz boost from the preferred frame, in which $\hat{w}_i$ vanishes, to a generic frame, in which it does not. 
    The ``preferred frame'' refers in the case of these vector-tensor theories to the frame picked out by a preferred time direction that is aligned with the timelike vector field, i.e. a frame constructed using the coordinates $\left(\tau_*, \mathbf{x}_*\right)$ in which $\delta A^{(1)}_i = 0$.

    We can now perform the Lorentz transformation to the generic frame $\left(\tau,\mathbf{x}\right)$, which for ease of calculation we present as the inverse transformation:
    \begin{eqnarray}
    \tau_* = \gamma_w\left(\tau+\hat{w}_j x^j\right)\,, \quad {\rm and} \quad {x_*}^j = \gamma_w \left(x^j+\hat{w}^j\tau\right)\,.
    \end{eqnarray}
    Computing the transformation of the vector field components in the usual way, we find that the preferred-frame velocity $\hat{w}_i$ is directly related to the local perturbation to the vector field by 
    \begin{equation}
    \hat{w}_i = \frac{\delta A^{(1)}_i}{\bar{A}}\,,
    \end{equation}
    which we recall from Section \ref{subsec:preferred_frame_effects} has no divergenceless vector part. 

    To determine the extra cosmological contribution to the local vector perturbation, as in Eq. (\ref{eq_small_scale_momentum_vector}), we expand the vector part of the $G_{0i}$ field equation about Minkowski spacetime, allowing for the time evolution of $\bar{A}$ (which is negligible in the classic PPN formalism), and look for the relevant additional term. This gives
    \begin{equation}\label{eq_B_i_extra_VT}
    \hat{\nabla}^2\hat{B}^{\rm extra}_i = -\frac{2\tau}{1+\left(\tau+\omega\right)\bar{A}^2}\left[\bar{A}^{''}+2\mathcal{H}\bar{A}'\right]\delta A^{(1)\,{\rm V}}_i = 0\,,
    \end{equation}
    because $\delta A^{(1)}_i$ has no divergenceless vector part. Thus, we do not need to add any $\h{B}_i^{\rm extra}$ term.
    
    With all the required ingredients obtained, we can substitute back into the small-scale momentum constraint Eqs. (\ref{eq_small_scale_momentum_scalar_2}) and (\ref{eq_small_scale_momentum_vector_including_preferred_frames}), which reconstitutes the full momentum constraint for the small-scale metric perturbations in these theories.

    \item Large scales: on super-horizon scales, we know from our earlier analysis that the parameterised momentum constraint must be given by
    \begin{equation}
    2\left(\hat{\Psi}'+\mathcal{H}\hat{\Phi}\right)_{,i} - 2\left(\mathcal{H}^2-\mathcal{H}'\right)\hat{B}_i = 2\left(\mathcal{H}^2-\mathcal{H}'\right)\hat{v}_i.
    \end{equation}
    Using the parameter values from Eqs. (\ref{eq_hellingsPPNCparamsalpha}--\ref{eq_hellingsgammac}), and simplifying them using the equation of motion for the vector field, we have that the coefficient of $\hat{v}_i$ on the right hand side is
    \begin{equation}
    2\left(\mathcal{H}^2-\mathcal{H}'\right) = \frac{8\pi G \bar{\rho}a^2 + \tau \bar{A}'^{\, 2} - 2\mathcal{H}\bar{A}\bar{A}'\left(\tau-2\omega\right)}{1-\frac{1}{2}\,\bar{A}^2\,\left(\tau-2\omega\right)}.
    \end{equation}
    With this quantity in hand, we can thus explicitly reconstruct the momentum constraint on large scales. This is a vast simplification compared to a direct cosmological perturbation theory expansion of the field equations of these theories. 
    
\end{itemize}

Before we move on to studying the time and scale dependence of the gravitational couplings that appear in our generalised Friedmann and perturbation equations, let us recap what we have found in this section.
We have extended the PPNC formalism, that we described in Section \ref{sec:basic_PPN}, and which was introduced in Refs. \cite{Sanghai_2017,Sanghai_2019}, by deriving a parameterised momentum constraint equation. This study required us to consider gravitational physics at $\mathcal{O}(v^3)$ in the post-Newtonian expansion. 
To do this, we have had to introduce several new physical phenomena, in particular the gravitational potentials that result from theories with preferred frame effects, and considerable care had to be taken to ensure that the perturbations to FLRW remain in Newtonian gauge in all cases, so that they are valid on small scales \cite{Clifton_2020}. 
Those final results are the parameterised scalar equation (\ref{eq_generalparametrisedscalareqn}) and divergenceless-vector equation (\ref{eq_generalparametrisedvectoreqn}). 

The PPNC equations - the ones we constructed in this section, as well as the Hamiltonian constraint and Raychaudhuri equations (\ref{eq_PPNC_allscales_Hamiltonian}) and (\ref{eq_PPNC_allscales_Raychaudhuri}) -  are valid on scales where the density contrast is highly nonlinear, as well as on super-horizon scales where terms in the field equations with time derivatives dominate. 
They require the introduction of only one new parameter: $\alpha_1(\tau)$, which is expected to be non-zero in non-conservative theories of gravity only, and which at the present time coincides with the PPN parameter of the same name. 
The additional parameters $\alpha_2$, $\zeta_1$ and $\xi$, which are present in the PPN equation for vector gravitational potentials, are not necessary for the cosmological version of the momentum constraint equation. 

This concludes our discussion of the construction of the fundamental equations of motion in the parameterised post-Newtonian cosmology framework.
In the next section, we will move on to how the various coupling functions $\mu$, $\nu$, $\mathcal{G}$ in the scalar sector, which have small-scale and large-scale limits given by combinations of PPNC parameters, can be interpolated between those extremes\footnote{In the divergenceless vector sector, there is also $\mathcal{Q}\,$, but we will not focus on the vector perturbations further.}. We will also study how their time variation might be modelled.

There are two caveats we wish to provide to the discussion above.
Firstly, it should be noted that the existing scalar perturbation equations do not quite form a closed set. One must generically supply a slip equation to relate $\h{\Phi}$ and $\h{\Psi}$ to one another. 
On small scales in PPNC, this is given by $\h{\Phi} - \h{\Psi} = \left(\dfrac{\alpha-\gamma}{\gamma}\right)\,\h{\Psi}\,$, but a super-horizon slip relation in parameterised frameworks is rather elusive, because it cannot be obtained within the standard theory-independent Bertschinger \cite{Bertschinger_2006} approach to super-horizon adiabatic perturbations, that we have used to derive all the previous large-scale results. 
We have also ignored transverse-tracefree tensor perturbations. These are problematic to include in standard post-Newtonian expansions as they only couple to matter at relatively high orders, but they are obviously present in cosmology, as they correspond to gravitational waves.

\section{Scale and time dependence of the PPNC couplings}\label{sec:scale_dependence}

The parameterised post-Newtonian cosmology framework is remarkably compact.
The generalised Friedmann equations (\ref{eq_parametrisedfriedmanneqns1}-\ref{eq_parametrisedfriedmanneqns2}) and the perturbation equations (\ref{eq_PPNC_allscales_Hamiltonian}), (\ref{eq_PPNC_allscales_Raychaudhuri}), (\ref{eq_generalparametrisedscalareqn}) and (\ref{eq_generalparametrisedvectoreqn}) describe the cosmological behaviour of any theory of gravity that fits into the PPN framework, in terms of direct generalisations of the PPN parameters. 
However, in order to construct the equations of motion for the scalar perturbations we needed to introduce the functions $\mu(\tau, k)$, $\nu(\tau, k)$ and $\mathcal{G}(\tau, k)$\footnote{Here we have switched the second argument from length scale $L = k^{-1}$ to wavenumber $k$\,.}, and for the vector perturbations we required $\mathcal{Q}(\tau, k)$. 
These interpolate between large and small-scale limits which are given by precise combinations of the PPNC parameters, but we have not yet determined how the transition between the deep sub-horizon and the super-horizon limits behaves. 
It is crucial to do so, because we require a prescription for $\mu(\tau, k)$ etc., in order to construct and integrate fully defined parameterised CPT equations. The solutions to these equations can then be used directly for observational predictions, such as the CMB temperature anisotropies, calculated using an Einstein-Boltzmann code.

In this section, we will study this interpolation problem using the Bergmann-Wagoner scalar-tensor theories of gravity \cite{bergmann1968comments,wagoner1970scalar} as a canonical example class, focusing on the scalar perturbations which are relevant for most cosmological observables.
We will demonstrate that simple elementary functions provide a good approximation to the full phenomenology of the theory, with most deviations occurring around the Hubble horizon, which roughly sets the transition scale between the two extremal regimes. 
This section is based on Ref. \cite{Thomas_2023}, where some additional detail is provided which we omit here for the sake of brevity. Note that these investigations were led by Daniel B. Thomas.

Although the PPNC equations are valid in the regime of nonlinear density contrasts, the transition between the deep sub-horizon and super-horizon limits occurs on large enough scales that linear cosmological perturbation theory should provide an excellent approximation. Indeed, in most theories of gravity, it is expected that there should be scales where the regions of validity of CPT and the post-Newtonian approximation overlap \cite{goldberg2017perturbation, thomas2020cosmological,Sankar}.
Hence, in this section we will assume that linear CPT can be safely applied, with density contrasts $\delta \ll 1\,$. The smallest scales we will consider, at around $20-100\,{\rm Mpc}$, will be precisely in that overlap region of validity, so that we can linearise in the standard way while also having the functions $\mu$, $\nu$ and $\mathcal{G}$ equal to their small-scale $L \longrightarrow 0$ limits.

A particular advantage of the applicability of linear perturbation theory is that we can perform a Fourier transform, as discussed in Section \ref{subsec:CPT}, thereby simplifying the scalar sector of the (Fourier space) PPNC perturbation equations to
\begin{eqnarray}
    -\mathcal{H}^2\Phi - \mathcal{H}\Psi' - \frac{1}{3}k^2 \Psi &=& -\frac{4\pi G \, a^2}{3}\,\mu\, \delta\rho\,, \label{eq_PPNC_Fourier_Hamiltonian} \\
    \Psi'' + \mathcal{H}\Phi' + \mathcal{H}\Psi' + 2\mathcal{H}'\Phi - \frac{1}{3}k^2\Phi &=& -\frac{4\pi G \,a^2}{3}\,\nu \,\delta\rho\,, \quad {\rm and} \label{eq_PPNC_Fourier_Raychaudhuri} \\
    \Psi' + \mathcal{H}\Phi &=& 4\pi G\,a^2\,\mu\,\bar{\rho} v + \mathcal{G}\mathcal{H}\Psi\,, \label{eq_PPNC_Fourier_momentum}
\end{eqnarray}
where it should be noted that we have dropped the hats on the perturbations to FLRW spacetime, as we will work exclusively with those perturbations from now on, and will no longer concern ourselves with the perturbed Minkowski picture.
We will now apply these equations to scalar-tensor theories as a test case, in order to determine well-behaved prescriptions for the couplings $\mu(\tau, k)\,$, $\nu(\tau, k)\,$, and $\mathcal{G}(\tau,k)\,$.

\subsection{Application to scalar-tensor theories of gravity}

Let us first examine some key features of the Bergmann-Wagoner scalar tensor theory, focusing in particular on \begin{enumerate}
    \item The equations of motion for a spatially flat FLRW cosmology.
    \item The equations that govern the evolution of linear, scalar, perturbations in Newtonian gauge.
    \item The post-Newtonian parameters of the theory that allow the equivalents of these equations to be constructed in the PPNC framework.
\end{enumerate}
The action (\ref{eq_bergmann_wagoner_action}) gives rise to equations of motion (\ref{eq_EOM_metric_scalar_tensor}) for the metric and (\ref{eq_EOM_KleinGordon_scalar_tensor}) for the novel scalar degree of freedom $\phi\,$. 
The theory is fully conservative, with no preferred-frame effects, and in many ways represents the simplest possible class of deviations from General Relativity. This makes it an ideal, tractable, example that we can use to study the behaviour of the PPNC equations. 
Initially, we will not make any assumptions on the funcitonal form of $\omega(\phi)$ and $\Lambda(\phi)\,$. 

Assuming that we are studying sufficiently late times that radiation can be ignored, the equations that fully describe the evolution of a spatially flat FLRW background are the Friedmann equation
\begin{equation}\label{eq_scalar_tensor_Friedmann}
    \mathcal{H}^2 = \frac{8\pi G\,\bar{\rho}\,a^2}{3\bar{\phi}} - \frac{\mathcal{H}\bar{\phi}'}{\bar{\phi}} + \frac{\omega\,\bar{\phi}'^2}{6\bar{\phi}^2} + \frac{\Lambda a^2}{3\bar{\phi}}\,,
\end{equation}
and the background part of the Klein-Gordon equation,
\begin{equation}\label{eq_scalar_tensor_KG_background}
    \frac{\bar{\phi}''}{\bar{\phi}} = \frac{8\pi G\,\bar{\rho}\,a^2}{\left(3+2\omega\right)\bar{\phi}} - \frac{2\mathcal{H}\bar{\phi}'}{\bar{\phi}} - \frac{\mathrm{d}\omega}{\mathrm{d}\bar{\phi}}\frac{\bar{\phi}'^2}{\left(3+2\omega\right)\bar{\phi}} + \frac{4\Lambda\,a^2}{\left(3+2\omega\right)\bar{\phi}} - \frac{\mathrm{d}\Lambda}{\mathrm{d}\bar{\phi}}\frac{2a^2}{3+2\omega}\,.
\end{equation}
Here we have decomposed $\phi = \bar{\phi}(\tau) + \delta\phi(\tau,x^i)$\,, and written $\omega(\bar{\phi})$ and $\Lambda(\bar{\phi})$ as just $\omega$ and $\Lambda$ for simplicity. As usual, we have also $\bar{\rho}' + 3\mathcal{H}\bar{\rho} = 0\,$ for non-relativistic matter.

The perturbation to the Hamiltonian constraint is
\begin{eqnarray}\label{eq_scalar_tensor_CPT_hamiltonian}
    3\mathcal{H}^2\delta\phi + \frac{\omega}{2}\left(\frac{\bar{\phi}'}{\bar{\phi}}\right)^2 \delta\phi + 3\mathcal{H}\delta\phi' - \frac{\omega\bar{\phi}'}{\bar{\phi}}\delta\phi' + 6\mathcal{H}\bar{\phi}\Psi' + 3\bar{\phi}'\Psi' + 6\mathcal{H}^2\bar{\phi}\Phi && \\
    \nonumber + 6\mathcal{H}\bar{\phi}'\Phi - \frac{\omega\bar{\phi}'^2}{\bar{\phi}}\Phi - \frac{\mathrm{d}\omega}{\mathrm{d}\bar{\phi}}\frac{\bar{\phi}'^2}{2\bar{\phi}}\delta \phi - a^2 \frac{\mathrm{d}\Lambda}{\mathrm{d}\bar{\phi}} \delta\phi + k^2\delta\phi + 2\bar{\phi}k^2 \Psi &=& 8\pi G\,a^2\,\delta\rho\,,
\end{eqnarray}
and the momentum constraint gives
\begin{eqnarray}\label{eq_CPT_scalar_tensor_momentum}
    -\mathcal{H}\delta\phi + \frac{\omega\bar{\phi}'}{\bar{\phi}}\delta\phi + \delta\phi' + \bar{\phi}'\Phi + 2\bar{\phi}\Psi' + 2\bar{\phi}\mathcal{H}\Phi = 8\pi G\,a^2\,\bar{\rho}v\,.
\end{eqnarray}
The shear evolution equation (\ref{eq_shear_evolution}) reduces to a constraint on the gravitational slip,
\begin{equation}
    \Phi - \Psi = \frac{\delta \phi}{\bar{\phi}}\,.
\end{equation}\label{eq_CPT_scalar_tensor_slip}
The perturbations to the evolution equations are the perturbed Raychaudhuri equation,
\begin{eqnarray}\label{eq_CPT_scalar_tensor_raychaudhuri}
    2\bar{\phi}\mathcal{H}\Phi' + 4\mathcal{H}'\bar{\phi}\Phi + 2\mathcal{H}^2\bar{\phi}\Phi + 2\bar{\phi}\Psi'' + 4\bar{\phi}\mathcal{H}\Psi' + 2\mathcal{H}'\delta\phi + \mathcal{H}^2\delta\phi + 2\bar{\phi}''\Phi && \\
    \nonumber + \delta\phi'' + \mathcal{H}\delta\phi' - \frac{\omega\bar{\phi}'^2}{2\bar{\phi}^2}\delta\phi + \frac{\omega\bar{\phi}'}{\bar{\phi}}\delta\phi' + \bar{\phi}'\Phi' + 2\bar{\phi}'\Psi' + 2\bar{\phi}'\mathcal{H}\Phi && \\
    \nonumber - a^2\frac{\mathrm{d}\Lambda}{\mathrm{d}\bar{\phi}}\delta\phi + \frac{\mathrm{d}\omega}{\mathrm{d}\bar{\phi}}\frac{\bar{\phi}'^2}{2\bar{\phi}}\delta\phi + \frac{\omega\bar{\phi}'^2}{\bar{\phi}}\Phi + \frac{2k^2}{3}\left[\delta\phi - \bar{\phi}\left(\Phi - \Psi\right)\right] &=& 0\,,
\end{eqnarray}
and the Klein-Gordon equation
\begin{eqnarray}\label{eq_CPT_scalar_tensor_KG}
    \left(3+2\omega\right)\left[\delta\phi'' + 2\mathcal{H}\delta\phi' + 4\mathcal{H}\bar{\phi}'\Phi + 2\bar{\phi}''\Phi + \bar{\phi}'\Phi' + 3\bar{\phi}'\Psi' + k^2 \delta\phi \right]  && \\
    \nonumber - 2a^2\frac{\mathrm{d}\Lambda}{\mathrm{d}\bar{\phi}}\delta\phi + 2a^2\frac{\mathrm{d}^2\Lambda}{\mathrm{d}\bar{\phi}^2}\bar{\phi}\delta\phi' + \frac{\mathrm{d}^2\omega}{\mathrm{d}\bar{\phi}^2}\bar{\phi}'^2\delta\phi && \\
    \nonumber + \frac{\mathrm{d}\omega}{\mathrm{d}\bar{\phi}}\left(4\mathcal{H}\bar{\phi}'\delta\phi + 2\bar{\phi}''\delta\phi + 2\bar{\phi}'\delta\phi' + 2\bar{\phi}'^2 \Phi \right) &=& 8\pi G\,a^2\,\delta\rho\,.
\end{eqnarray}
Finally, we have the standard perturbation equations for the matter fields coming from the Bianchi identities, 
\begin{eqnarray}\label{eq_CPT_Fourier_conservation}
    \delta\rho' + 3\mathcal{H}\delta\rho + 3\bar{\rho}\Psi' - k^2 \bar{\rho} v = 0 \quad {\rm and} \quad v' + \mathcal{H}v - \Phi = 0\,.
\end{eqnarray}

The PPNC parameters can be calculated by considering instead a post-Newtonian expansion about Minkowski spacetime on small scales, in the usual fashion described in Section \ref{subsec:PPN}.
One gets
\begin{eqnarray}\label{eq_scalar_tensor_PPNC_parameters}
    \alpha &=& \frac{1}{\bar{\phi}}\frac{4 + 2\omega}{3+2\omega}\,, \qquad \gamma = \frac{1}{\bar{\phi}}\frac{2+2\omega}{3+2\omega}\,, \\
    \nonumber a^2 \alpha_c &=& -\frac{\omega\bar{\phi}'^2}{\bar{\phi}^2} - \frac{\bar{\phi}''}{\bar{\phi}} + \frac{\mathcal{H}\bar{\phi}'}{\bar{\phi}} + \frac{\mathrm{d}\omega}{\mathrm{d}\bar{\phi}}\frac{\bar{\phi}'^2}{2\bar{\phi}\left(3+2\omega\right)} + \frac{1+2\omega}{3+2\omega}\frac{\Lambda a^2}{\bar{\phi}} + \frac{1}{3+2\omega}\frac{\mathrm{d}\Lambda}{\mathrm{d}\bar{\phi}}\,, \quad {\rm and} \\
    \nonumber a^2 \gamma_c &=& -\frac{\omega\bar{\phi}'^2}{4\bar{\phi}^2} - \frac{\bar{\phi}''}{2\bar{\phi}} + \frac{\mathcal{H}\bar{\phi}'}{2\bar{\phi}} - \frac{\mathrm{d}\omega}{\mathrm{d}\bar{\phi}}\frac{\bar{\phi}'^2}{2\bar{\phi}\left(3+2\omega\right)} + \frac{1-2\omega}{3+2\omega}\frac{\Lambda a^2}{2\bar{\phi}} + \frac{2}{3+2\omega}\frac{\mathrm{d}\Lambda}{\mathrm{d}\bar{\phi}}\,,
\end{eqnarray}
which can be shown to satisfy the PPNC integrability condition Eq. (\ref{eq_integrabilitycondition}).
In the standard PPN formalism, $\alpha$ would be set to unity by the definition of $G$\,, but this is clearly not possible at all cosmic times, as $\bar{\phi}$ evolves. Instead, one sets the present-day value of $\bar{\phi}$ by demanding that $\alpha(\tau_0) = 1\,$, but $\alpha(\tau)$ will generically evolve away from unity.
In what follows, we will consider the special case where $\omega$ and $\Lambda$ are both constants. This corresponds to the Brans-Dicke theory, plus a cosmological constant.

First, we can specify the constants $\omega$ and $\Lambda$ (or equivalently $\Omega_{\Lambda}$\,, where throughout this section we will be rather slack with our notation and simply take $\Omega_{\Lambda}$ to mean $\Omega_{\Lambda 0}$), and then integrate the background equations (\ref{eq_scalar_tensor_Friedmann}) and (\ref{eq_scalar_tensor_KG_background}), to determine the evolution of $\alpha(\tau)\,$, $\gamma(\tau)\,$, $\alpha_c(\tau)$ and $\gamma_c(\tau)$ at all $a$ being considered. In this case, that is $a \in \left[a_{\rm LS}, 1\right]\,$, so that we are safe to neglect radiation.
The evolution of $a(\tau)$ and $\bar{\phi}(\tau)$ in these theories is very well-studied \cite{nariai1968green}. As long as $\omega > 0\,$, we have $\bar{\phi}' > 0\,$, so at earlier times than the present day, $\bar{\phi}$ was smaller and hence $\alpha$ and $\gamma$ were larger than their values at $\tau_0\,$.

\begin{figure}
    \centering
    \includegraphics[width=0.95\linewidth]{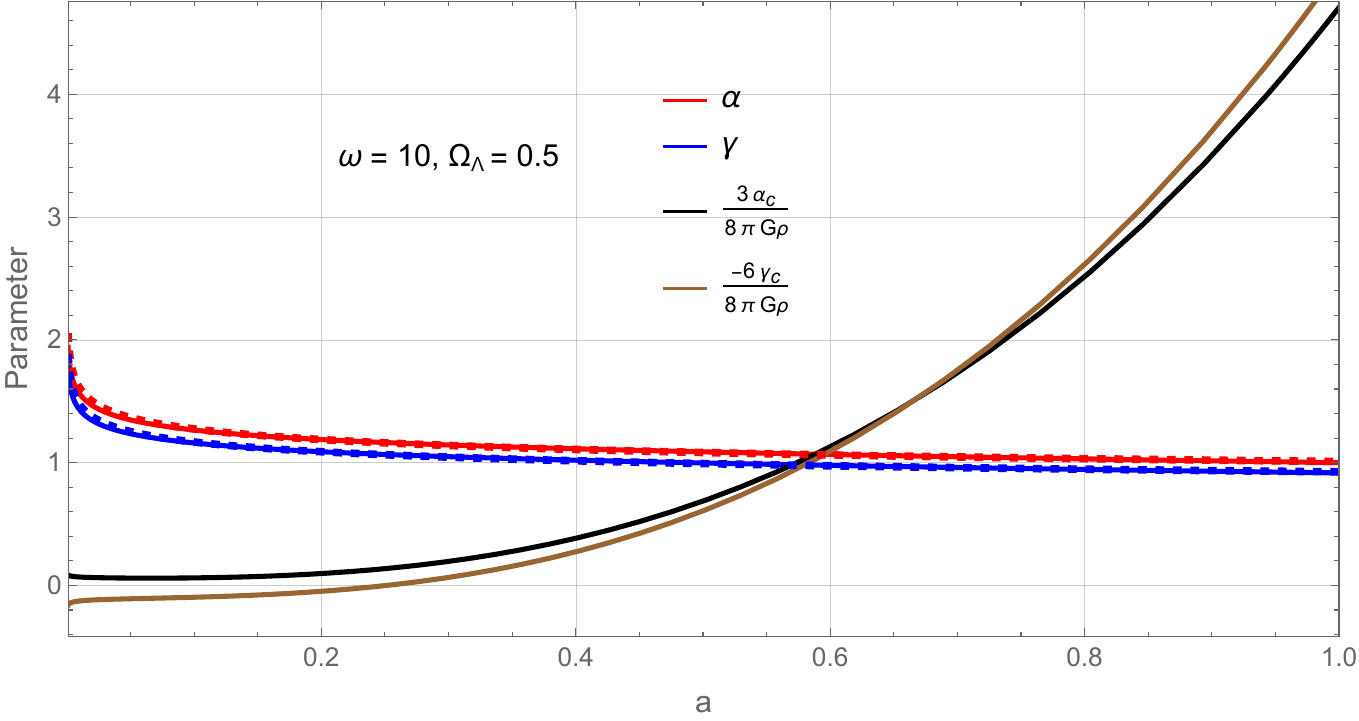}
    \caption{Evolution with scale factor of the PPNC parameters $\alpha(a)$\,, $\gamma(a)\,$, $\alpha_c(a)$ and $\gamma_c(a)\,$, for the Brans-Dicke theory with a cosmological constant. Here, $\alpha_c(a)$ and $\gamma_c(a)$ have been normalised by the competing factors $\dfrac{8\pi G \bar{\rho}(a)}{3}$ and $\dfrac{8\pi G \bar{\rho}(a)}{-6}$ respectively.}
    \label{fig_ppnc_parameters_BDT}
\end{figure}

The evolution of the PPNC parameters is shown in Fig. \ref{fig_ppnc_parameters_BDT}, for the parameter values $\omega = 10$ and $\Omega_{\Lambda} = 0.5\,$. 
The dashed curves are obtained by approximating $\bar{\phi}(a)$, and hence $\alpha(a)$ and $\gamma(a)$\,, by a power law $c_1 a^n + c_2$ in the scale factor. Motivated by the fact that during matter domination, the attractor solution is $\bar{\phi} \sim a^{1/\left(1+\omega\right)}\,$ \cite{nariai1968green}, we set the power law index $n$ to $0.1\,$. The constant $c_1$ is set by demanding $\alpha(\tau_0) = 1\,$, and then $c_2$ can just be fixed by eye in this toy example.
The simple power law prescription for the time evolution of $\alpha$ and $\gamma$ works very well, with the dashed curves almost on top of the corresponding solid curves, that are obtained by calculating the functions $\alpha(a)$ and $\gamma(a)$ from the full evolution of $\bar{\phi}(a)$ that arises by solving the Klein-Gordon equation (\ref{eq_scalar_tensor_KG_background}). 
Of course, in this case, we expected this from the analytic attractor solution. However, this result tells us that a power law should be a good prescription for the generalised PPN parameters $\alpha(a)$ and $\gamma(a)\,$. That observation will be useful in Chapter 6.

\begin{figure}
    \centering
    \includegraphics[width=0.9\linewidth]{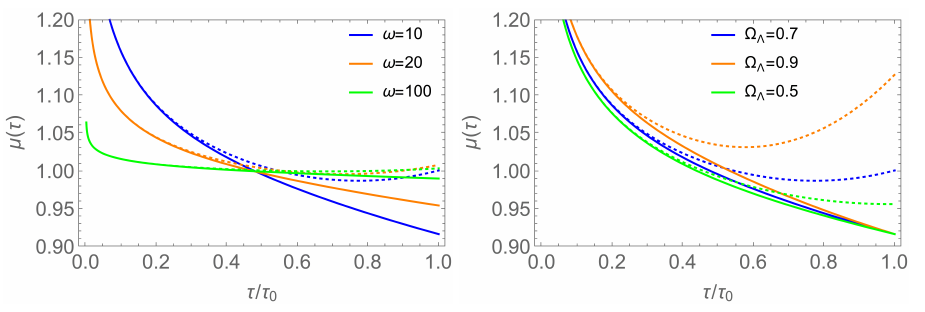}
    \caption{Small-scale (solid) and large-scale (dashed) limits of $\mu\,$, as a function of $\tau\,$, in the Brans-Dicke theory with a cosmological constant. In the left-hand plot, all curves have $\Omega_{\Lambda} = 0.7$\,, and $\omega$ is varied. In the right-hand plot all curves have $\omega = 10\,$, and $\Omega_{\Lambda}$ is varied. From Ref. \cite{Thomas_2023}, courtesy of Daniel B. Thomas.}
    \label{fig_mu_large_small_BDT}
\end{figure}

From the functions $\alpha(\tau)$, $\gamma(\tau)$, $\alpha_c(\tau)$ and $\gamma_c(\tau)$, the dependence in time of the small-scale and large-scale limits of $\mu\,$, $\nu\,$, and $\mathcal{G}$ is determined, according to Eqs. (\ref{eq_PPNC_scalarperts_small}-\ref{eq_PPNC_scalarperts_large2}) and (\ref{eq_G1largek}).
As an example, we display the $k \longrightarrow \infty$ (small-scale) and $k \longrightarrow 0$ (large-scale) limits of $\mu(\tau, k)\,$, for different values of the parameters $\omega$ and $\Omega_{\Lambda}\,$, in Fig. \ref{fig_mu_large_small_BDT}.
We see that at late times the sub-horizon and super-horizon behaviour of the coupling function $\mu(\tau, k)$ is very different, with the difference increasing, as it must, with decreasing $\omega$ (recall that $\omega \longrightarrow \infty$ recovers GR). Equivalent plots can of course be made for $\nu$ and $\mathcal{G}\,$.

However, there are many scales of interest for cosmological observables (such as the CMB and weak lensing), which do not fit into either the small-scale or large-scale limits, but rather describe physics on scales that are in between the limits, such as on scales below but comparable to $r_H(a)\,$.
We therefore need to understand how the coupling functions in our generalised framework behave in between the two extremal regimes, for the Bergmann-Wagoner theory being considered.
In order to understand this, solutions to the background equations (\ref{eq_scalar_tensor_Friedmann}) and (\ref{eq_scalar_tensor_KG_background}) will not be sufficient. One must also solve the linear CPT equations (\ref{eq_scalar_tensor_CPT_hamiltonian}-\ref{eq_CPT_Fourier_conservation}).
These can be integrated numerically using adiabatic initial conditions: one assumes that all the Fourier modes that are physically important are outside the horizon at the starting time, which is taken to be at redshift $1100\,$, roughly corresponding to the last scattering surface $a_{\rm LS}\,$. 
Then, one uses the comoving curvature perturbation on constant-density hypersurfaces to set the initial conditions for $\Phi$\,, $\Psi$ and $\delta\phi$. For further details, see Ref. \cite{Thomas_2023}. 
Once the initial conditions are introduced, it is a straightforward numerical exercise to calculate $\Phi(\tau, k)\,$, $\Psi(\tau, k)$ and $\delta\phi(\tau, k)\,$. In the next section, we will use these solutions to test the validity of a simple interpolated construction of $\mu\,$, $\nu$ and $\mathcal{G}$ from the PPNC parameters.

\subsection{Interpolation between small and large scales}

\begin{figure}
    \centering
    \includegraphics[width=\linewidth]{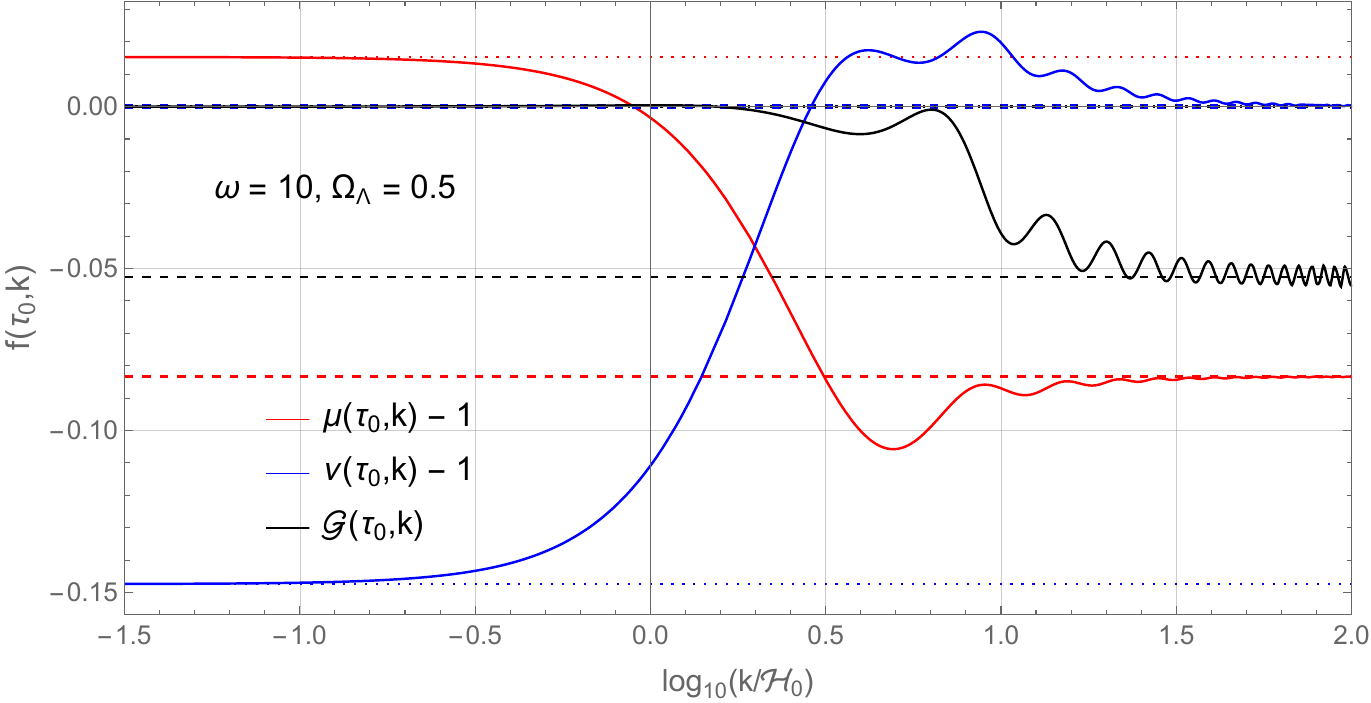}
    \caption{The functions $\mu_{\rm theory}(k)\,$, $\nu_{\rm theory}(k)$ and $\mathcal{G}_{\rm theory}(k)$ calculated in the Brans-Dicke theory with a cosmological constant, evaluated at the present day $\tau_0\,$. The dotted and dashed lines show respectively the $k \longrightarrow 0$ and $k \longrightarrow \infty$ limits that are predicted for each function from the parameters of the PPNC formalism.}
    \label{fig_mu_nu_g_BDT}
\end{figure}

Given the Fourier-space solutions for $\Phi$\,, $\Psi\,$, $\delta\rho$ and $v$ in our scalar-tensor test theory, the parameterised couplings can be reconstructed as
\begin{eqnarray}\label{eq_mu_nu_G_theory}
    \mu_{\rm theory} &=& \frac{k^2\Psi + 3\mathcal{H}^2\Phi + 3\mathcal{H}\Psi'}{4\pi G\,a^2\,\delta\rho}\,, \\
    \nonumber \nu_{\rm theory} &=& \frac{k^2\Phi - 6\mathcal{H}'\Phi - 3\mathcal{H}\Phi'-3\Psi'' + 3\mathcal{H}\Psi'}{4\pi G\,a^2\,\delta\rho}\,, \quad {\rm and} \\
    \nonumber \mathcal{G}_{\rm theory} &=& \frac{\Psi' + \mathcal{H}\Phi - 4\pi G\,a^2\,\mu_{\rm theory}\,\bar{\rho}v}{\mathcal{H}\Psi} \ = \ \frac{\Psi' + \mathcal{H}\Phi}{\mathcal{H}\Psi} - \frac{\bar{\rho}v}{\delta\rho}\frac{\left(k^2\Psi + 3\mathcal{H}^2\Phi + 3\mathcal{H}\Psi'\right)}{\mathcal{H}\Psi}\,.
\end{eqnarray}
We expect that in the limits $k \ll \mathcal{H}$ and $k \gg \mathcal{H}\,$, the functions $\mu_{\rm theory}$\,, $\nu_{\rm theory}$ and $\mathcal{G}_{\rm theory}$ should approach asymptotic values given by the appropriate combinations of PPNC parameters, as per Eqs (\ref{eq_PPNC_scalarperts_small}-\ref{eq_PPNC_scalarperts_large2}) and (\ref{eq_G1largek}).
Let us now verify that prediction, by calculating solutions to the linear perturbation equations for values of $k$ in the range $\log_{10}{k/\mathcal{H}_0} \in \left[-1.5, 2\right]\,$, and then constructing the desired functions according to Eq. (\ref{eq_mu_nu_G_theory}). 
The results are shown in Fig. \ref{fig_mu_nu_g_BDT}, where we have evaluated the solutions at $\tau = \tau_0$\,. Once again, we have used the parameter values $\omega = 10$ and $\Omega_{\Lambda} = 0.5\,$.
For each of the functions $f(\tau_0, k)$ ($\mu\,$ (red), $\nu$ (blue) and $\mathcal{G}\,$ (black)), the dashed lines of the same colour indicate the expected small-scale limit $f(\tau_0, k \longrightarrow \infty)$\,, and the dotted lines of the same colour indicate the expected large-scale limit $f(\tau_0, k \longrightarrow 0)\,$.

We see that the PPNC limits are correctly reproduced in all three functions. Furthermore, the transition between the limits occurs as expected for wavenumbers $k$ that are of the same order of magnitude as the horizon scale $k_H(\tau_0) = 1/r_H(\tau_0) = \mathcal{H}_0$\,.
Those transitions are smooth but non-monotonic. The oscillatory behaviour of the solutions for $k$ values that are in neither the fully post-Newtonian or fully super-horizon regimes is a direct result of the scalar field $\phi$ satisfying its Klein-Gordon equation of motion (\ref{eq_EOM_KleinGordon_scalar_tensor}), resulting in wavelike solutions on scales where neither the time or space derivatives of $\delta\phi$ can be neglected. 
This phenomenon has been investigated by Brando et al. in Ref. \cite{gao} in a slightly different context. They referred to the apparent scale-dependent oscillations in the effective Newton's constant as Gravity Acoustic Oscillations (GAO).

Let us suppose now that we did not have access to the solutions to the underlying equations of motion, as would be the case in the generic theory-independent scenario where we wish to constrain the properties of the PPNC functions $\mu$, $\nu$ and $\mathcal{G}$ on all relevant scales, at many points in cosmic history.
Then, we need a functional form for them that transitions smoothly between the small-scale and large-scale limits. Motivated by the results we have just calculated, one expects the characteristic scale of the transition to be given by the Hubble horizon wavenumber $\mathcal{H}$\,.
The simplest choices for an interpolating function one can make that satisfies these requirements are zero-parameter interpolations that do not introduce any additional scale dependence beyond $\mathcal{H}\,$. The following form, which was suggested by Clifton \& Sanghai in Ref. \cite{Sanghai_2019}, appears to perform well:
\begin{equation}\label{eq_scale_dependent_coupling}
    f(\tau, k) = \frac{S(\tau) + L(\tau)}{2} + \frac{S(\tau) - L(\tau)}{2}\,\tanh{\ln{\left(\frac{k}{\mathcal{H}(\tau)}\right)}}\,, 
\end{equation}
where $S(\tau)$ and $L(\tau)$ are the small-scale and large-scale limits of the coupling function $f\,$, as given by the relevant combinations of PPNC parameters, and we have made the time dependence of both them and $\mathcal{H}$ explicit\footnote{Previous attempts to parameterise the scale dependence of gravitational couplings have used functional forms that are similar to Eq. (\ref{eq_scale_dependent_coupling}) \cite{amin2008subhorizon,bakerbull}.}.

\begin{figure}
    \centering
    \includegraphics[width=\linewidth]{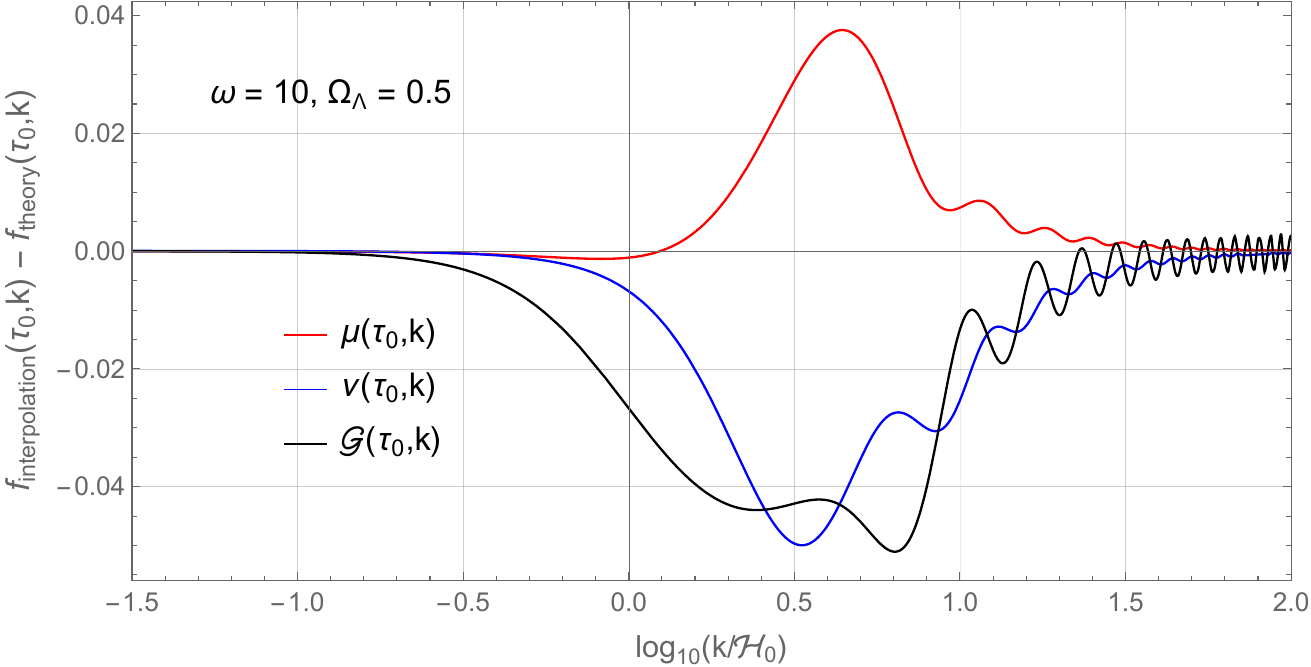}
    \caption{The difference, evaluated at $\tau = \tau_0$\,, between the zero-parameter interpolation (\ref{eq_scale_dependent_coupling}) of the PPNC functions $\mu\,$, $\nu\,$ and $\mathcal{G}$\,, and the form of each function that is obtained by a full numerical integration of the scalar-tensor equations of motion, displayed as a function of $k/\mathcal{H}_0\,$.}
    \label{fig_interpolation_tau0_BDT}
\end{figure}

The interpolation functional form in Eq. (\ref{eq_scale_dependent_coupling}) gives a value of $f$ that is halfway between $S$ and $L$ precisely at the Hubble horizon. As a first approximation it is passable, but it will inevitably miss more complicated features in $\mu$\, $\nu$ and $\mathcal{G}$\,, such as the oscillations that are found in the Bergmann-Wagoner scalar-tensor case.
The relative accuracy of the interpolation for that theory is shown in Fig. \ref{fig_interpolation_tau0_BDT}, where we display the difference at $\tau = \tau_0$ between the interpolated function $f_{\rm interpolation}(k)$ and the function $f_{\rm theory}(k)$ obtained from a full numerical integration of the linearly perturbed scalar-tensor equations of motion (\ref{eq_scalar_tensor_CPT_hamiltonian}-\ref{eq_CPT_scalar_tensor_KG}), as per the definitions in Eq. (\ref{eq_mu_nu_G_theory}).
The interpolation appears to work well on most scales, with the greatest failure in all three cases occurring around the scale $\log_{10}\left(k/\mathcal{H}_0\right) \sim 0.5\,$, where the scalar field oscillations are particularly pronounced. We can be confident, then, that at $\tau_0$ (or equivalently, at redshift zero), the discrepancy between the interpolation and the ``true'' theory result is maximised at some scale that essentially corresponds to an order unity fraction of the present-day Hubble horizon.

\begin{figure}
    \centering
    \includegraphics[width=0.8\linewidth]{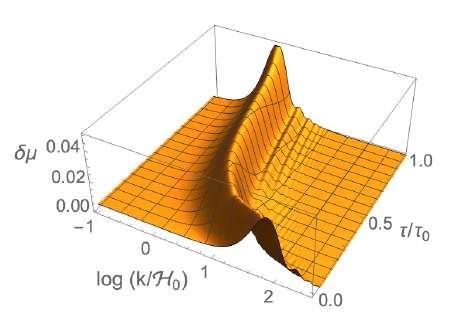}
    \caption{The absolute fractional error in $\mu$ between the true $\mu_{\rm theory}(\tau, k)$ in the scalar-tensor theory being considered ($\omega = 10$ and $\Omega_{\Lambda} = 0.5$) and the function $\mu_{\rm interpolation}(\tau, k)$ that would be obtained from a simple zero-parameter interpolation between the small-scale and large-scale PPNC limits, displayed as a function of both $k/\mathcal{H}_0$ and $\tau/\tau_0$\,. From Ref \cite{Thomas_2023}, courtesy of Daniel B. Thomas.}
    \label{fig_delta_mu_interp}
\end{figure}

This story continues as we consider earlier times (higher redshifts), except for some slight complications in $\nu$ for large values of $\Omega_{\Lambda}$ that we will not concern ourselves with here, but which are explained in Ref. \cite{Thomas_2023}.
This is demonstrated by Fig. \ref{fig_delta_mu_interp}, in which the the (absolute) fractional difference $\delta \mu (\tau, k) = \left\vert\dfrac{\mu_{\rm interpolation} - \mu_{\rm theory}}{\mu_{\rm theory}}\right\vert$ has been calculated as a function of both $k$ and $\tau\,$, for the parameter values $\omega = 10$ and $\Omega_{\Lambda} = 0.7\,$.
It can be seen that $\delta\mu$ is dominated throughout by a large peak at a scale that essentially tracks the evolution of the Hubble horizon. Moreover, that peak is never more than $4\%$ for these parameter values. 
The interpolating function produces at all times, and at all scales except that peak, a value for $\mu$ that is accurate roughly at the percent level or better. This gives us confidence that we can use the simple parameterisation in Eq. (\ref{eq_scale_dependent_coupling}), and obtain sensible dynamics that captures most of the physics we are interested in over virtually all linear scales in cosmology.

This is really our purpose in a theory-agnostic framework: we want to be able to write down a well-motivated, consistent set of gravitational equations that allow us to model simply the effects we might see in cosmology. In general, we will not want to impose any given underlying theory, but instead just use our parameterised Friedmann and perturbation equations to evolve the relevant metric and matter perturbations. 
Of course, if the interpolated couplings we are using are tied to the PPNC parameters of a known theory, then the predictions the parameterised equations make for various phenomena should closely match the predictions obtained using a full integration of that theory's equations of motion.
In the next, final, section of this chapter, we will verify this for our canonical Bergmann-Wagoner class.

\subsection{Using the parameterised equations for cosmological phenomena}

Given the PPNC parameters $\alpha(\tau)\,$, $\gamma(\tau)\,$, $\alpha_c(\tau)$ and $\gamma_c(\tau)\,$ (thus accordingly a solution for $a(\tau)$), and the simple prescription (\ref{eq_scale_dependent_coupling}) for each of the couplings in terms of those parameters, we are nearly in a position to numerically integrate the parameterised scalar perturbation equations (\ref{eq_PPNC_Fourier_Hamiltonian}-\ref{eq_PPNC_Fourier_momentum}), along with the necessary equations (\ref{eq_CPT_Fourier_conservation}) for the matter fields. 
The solutions $\left\lbrace \Phi(\tau, k), \Psi(\tau, k), \delta\rho(\tau, k), v(\tau, k)\right\rbrace$ can then be used to make predictions for physical phenomena of interest, and we can forget entirely about the underlying field content of the theory that gave rise to the evolving PPNC parameters.

The only remaining thing we need is a slip relation, $\Phi(\tau, k) - \Psi(\tau, k) = \Sigma(\tau, k)\Psi(\tau, k)\,$. The gravitational slip is a crucial equation, because it allows the second Bardeen potential (say $\Phi$) to be obtained from the first (say $\Psi$) as a constraint, and therefore reduces by one the number of fields that must be evolved numerically \cite{caldwell2007constraints, amendola2008measuring}.
On small scales, we know that $\Sigma \longrightarrow \dfrac{\alpha - \gamma}{\gamma}$ from the earlier discussion of the PPNC formalism.
However, it is not clear what the prescription for the large-scale limit of the slip should be, in order for the interpolating function $\Sigma(\tau, k)$ to be constructed according to Eq. (\ref{eq_scale_dependent_coupling}).

A theory-independent result for the large-scale limit of the slip has so far not been obtained within the PPNC formalism, although it is the topic of ongoing research \cite{Clifton_2024}. As mentioned in Section \ref{sec:basic_PPN}, it is elusive because it cannot be calculated using the standard separate-universe approach on super-horizon scales \cite{Bertschinger_2006}. 
In this section, and indeed in Chapter 6 where we will continue the discussion of the parameterised post-Newtonian cosmology formalism, we will simply take the $k \longrightarrow 0$ limit of $\Sigma$ to be an additional free parameter. 
In fact, we will take this free parameter to be precisely zero, so on ultra-large scales the Bardeen potentials are equal. Once we have specified $\Sigma\,$, the parameterised equations can be integrated straightforwardly, starting from the initial constant-time hypersurface at $\tau = \tau_{\rm LS}\,$.

We will focus on calculating the density contrast at $\tau_0$\,, which serves as a na{\" i}ve proxy for large scale structure statistics. It is certainly not a replacement for a proper analysis of these observables, but will give us an initial understanding of how the PPNC framework performs when it is used to calculate objects of interest in cosmology in a theory-independent way.
The presentation here is just a brief paraphrasing of the analysis carried out by Daniel B. Thomas in Ref. \cite{Thomas_2023}, where weak lensing and the integrated Sachs-Wolfe effect are also discussed.

\begin{figure}
    \centering
    \includegraphics[width=0.75\linewidth]{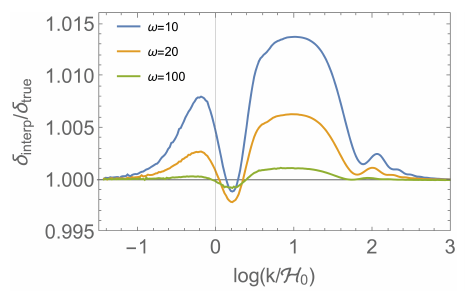}
    \caption{Present-day density contrast $\delta_{\rm interp}(\tau_0, k)$ obtained using the PPNC equations with the interpolation (\ref{eq_scale_dependent_coupling}), as a ratio of the value $\delta_{\rm true}$ that is found by directly integrating the scalar-tensor perturbation equations with $\Omega_{\Lambda} = 0.7\,$. From Ref. \cite{Thomas_2023}, courtesy of Daniel B. Thomas.}
    \label{fig_density_contrast_interp}
\end{figure}

Calculating the present-day Fourier space density contrast $\delta(\tau_0, k)$ using the parameterised equations with our simple interpolation function (\ref{eq_scale_dependent_coupling}), one finds excellent agreement with the results that are obtained if one integrates the linearised scalar-tensor theory equations directly.
The interpolation leads to an error in $\delta(\tau_0,k)$ that is never more than $\sim 1\%\,$, which is for $\log_{10}{\left(k/\mathcal{H}_0\right)} \sim 1\,$.  As expected, the error is greater for smaller $\omega\,$, wherein the scalar-tensor theory being considered is more strongly discrepant from General Relativity, as shown in Fig. \ref{fig_density_contrast_interp}.

\begin{figure}[ht]
    \centering
    \includegraphics[width=\linewidth]{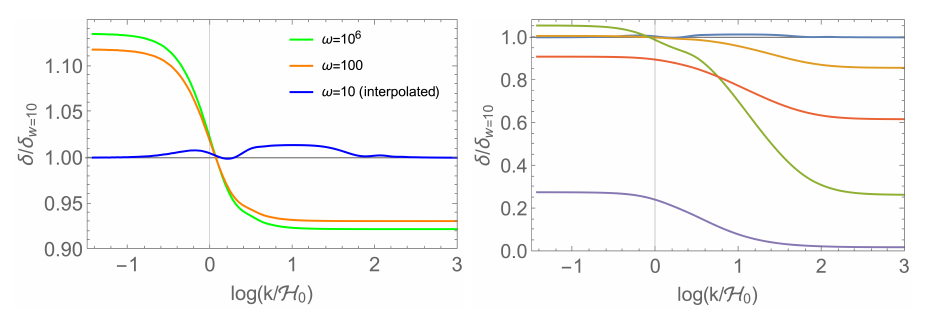}
    \caption{Left: Ratio of $\delta(\tau_0, k)$ to the value obtained in the scalar-tensor theory with $\omega = 10\,$. The effect of varying the $\omega$ parameter to $100$ (orange) and $10^6$ (green) is compared with the effect of using the parameterised equations for the PPNC parameters calculated in the $\omega = 10$ theory (blue). 
    Right: the ratio of $\delta(\tau_0, k)$ to what would be obtained if one assumed a $\Lambda$CDM background evolution (purple and red curves for $\omega = 10$ and $100$ respectively) or the $\Lambda$CDM linear CPT equations (orange and green for $\omega = 10$ and $100$ respectively), instead of using the full set of equations of motion for both the background and perturbations. The blue curve has the same meaning as in the left plot.
    From Ref. \cite{Thomas_2023}, courtesy of Daniel B. Thomas.}
    \label{fig_theory_vs_lcdm_interp}
\end{figure}

Moreover, the left plot of Fig. \ref{fig_theory_vs_lcdm_interp} shows that the error in $\delta(\tau_0,k)$ that arises due to using the parameterised and interpolated equations for a specific theory (shown in blue) \footnote{In this case, the underlying theory is the Brans-Dicke + cosmological constant theory with $\omega = 10$ and $\Omega_{\Lambda} = 0.7$.} is much smaller than the change that is induced by actually changing the coupling parameter $\omega$ of the underlying theory to $100$ (orange) or $10^6$ (green).
As the limit $\omega \longrightarrow \infty$ recovers General Relativity with a cosmological constant, we can take the green curve as corresponding to the GR result.
Hence, the effect of using a flawed, oversimplified interpolating function, that, for example, fails to capture the scalar field oscillations on scales comparable to the Hubble horizon, would not present a large obstacle when trying to constrain the theory parameter $\omega\,$.
This is because the bias it introduces would be much smaller than the deviation from GR, if we were to detect the imprints on $\delta$ of a scalar-tensor theory of this kind with $\omega = 10\,$.

Finally, let us examine how much the matter density field would change if instead of using our full set of equations of motion, we applied modifications to the $\Lambda$CDM equations only for the perturbations or only for the background. 
Of course, in our formalism, to do so would be fundamentally erroneous, because the background and perturbations are governed by the same set of underlying PPNC parameters.
However, in many cosmological frameworks for testing gravity, modifications to GR are often only introduced at the level of perturbations, with the background expansion prescribed to be equivalent to that of the concordance $\Lambda$CDM cosmology \cite{hu2007models,Hu_2008,amin2008subhorizon,Skordis_2009,Baker_2011,Baker_2013,bakerbull}.
It is interesting to consider, therefore, what would happen if we enforce this situation, or its converse, by hand.
The results are shown in the right plot of Fig. \ref{fig_theory_vs_lcdm_interp}. The errors induced by introducing a fictitious $\Lambda$CDM background (e.g. shown in purple for the $\omega = 10$ theory) are in fact larger than those induced by using $\Lambda$CDM perturbation equations instead of Eqs. (\ref{eq_scalar_tensor_CPT_hamiltonian}-\ref{eq_CPT_scalar_tensor_KG}) (shown in orange for $\omega = 10$). 
Both of these errors are much larger than the error introduced by using the simple parametrisation rather than the true scalar-tensor equations (\ref{eq_scalar_tensor_CPT_hamiltonian}-\ref{eq_CPT_scalar_tensor_KG}).
These results indicate the importance of evolving both the FLRW background and the linear perturbations in terms of the same consistent set of parameters. In cosmology, this feature of the background and perturbations being so intimately related is fairly unique to PPNC.
In Chapter 6, we will use the PPNC evolution for both the background and perturbations, to avoid erroneous constraints that would be obtained if, for example, we assumed that the background cosmology was phenomenologically equivalent to $\Lambda$CDM, and then focused only on deviations from GR at the level of linear perturbation theory.

This concludes our discussion of the time dependence of the PPNC parameters $\alpha(\tau)\,$, $\gamma(\tau)\,$, $\alpha_c(\tau)$ and $\gamma_c(\tau)\,$, and the scale dependence of the couplings $\mu(\tau, k)\,$, $\nu(\tau, k)$ and $\mathcal{G}(\tau, k)\,$. 
The detailed study in this section of scalar-tensor theories of gravity as a canonical test case has suggested that power laws in $a$\,, and a zero-parameter interpolation of the form (\ref{eq_scale_dependent_coupling}) in $k\,$, provide viable proposals for the time and scale dependences of these functions which we derived in Sections \ref{sec:basic_PPN} and \ref{sec:momentum_constraint}.

With our prescriptions for the PPNC couplings in hand, it is possible to construct and numerically evolve the fully theory-independent parameterised Friemdann and scalar CPT equations. 
By doing so, we will be able to compute accurate predictions for the anisotropies in the cosmic microwave background, and therefore constrain the space of PPNC parameters using observations of the CMB. This will be the topic of the next chapter.

\section{Discussion}

In this chapter, we have introduced the framework of parameterised post-Newtonian cosmology, which is designed for performing theory-agnostic tests of gravity on a wide range of cosmological scales, while being explicitly compatible with precision observational tests that are made in the astrophysical regime of post-Newtonian gravity. 
We have shown that this formalism gives rise to a compact set of generalised equations for the cosmological expansion, and for scalar and vector perturbations to the FLRW metric. 
We have also verified that it reproduces an accurate description of the phenomenology for a canonical class of scalar-tensor theories of gravity, suggesting that a simple parametrisation of the time and scale dependence of the gravitational couplings that arise can be used to extract viable predictions for cosmological observables.

Of course, there are limitations to the analysis we have presented here. We have not yet determined the scale dependence of the function $\mathcal{Q}(\tau, k)$ that enters the divergenceless vector part (\ref{eq_generalparametrisedvectoreqn}) of the momentum constraint. This would be necessary if we want to study deviations from GR in cosmological vector perturbations, which arise very naturally within the post-Newtonian approach \cite{bruni2014computing,Thomas_2015}.
We have also used a linearised, fluid description of matter fields, which will break down on small scales. Then, it would be necessary to develop N-body simulations that incorporate the formalism, that would need to be relativistic rather than purely Newtonian \cite{adamek2013general,Adamek_2015,Adamek_2018}.

Even with those limitations in place, the existence of a consistent set of theory-independent equations, and viable interpolations across length and time scales that allows them to be implemented computationally, provides us with ample space to explore the consequences of parameterised post-Newtonian cosmology for real observations. 
Indeed, we ultimately expect the principal utility of parameterised frameworks in cosmology to come from their comparison to observational data. Although the present-day values of the PPN parameters are mostly very tightly constrained (see Table \ref{table_PPN_parameters}), their time evolution is not, as per Table \ref{table_PPN_time_derivs}.
The most powerful probe of the overall time evolution in cosmic history of the parameters that enter the scalar sector of the PPNC formalism, is likely to be the anisotropies in the cosmic microwave background, which as we mentioned above will be the focus of Chapter 6.

However, let us first discuss some other probes that might be useful for studying the time dependence of the PPNC parameters purely in the late Universe. 
A full analysis of these late-time probes is a serious challenge, and it will be not be attempted in this thesis, but we will provide an overview of them in order to suggest how a detailed picture of evolving gravitational couplings in cosmology might be built up using the framework we have discussed.

\begin{enumerate}

    \item Because non-relativistic matter couples only to $\alpha$ in the leading-order Newtonian limit, its cosmological time evolution might be most tightly constrained by large-scale-structure observations, particularly the matter power spectrum. 

    \item On the other hand, light couples at leading order to both $\alpha$ and $\gamma$\,. 
    Hence, the parameter $\gamma(\tau)$ may be best constrained with observations that depend on the properties of null geodesics. In cosmology, this would likely mean using weak lensing data.

    \item Beyond leading order, the time dependence of the preferred-frame parameter $\alpha_1(\tau)$, and the scale dependence of $\mathcal{Q}(\tau, k)$\,, through which $\alpha_1$ couples to the vector perturbation $\h{B}_i\,$, could potentially also be constrained by weak gravitational lensing measurements \cite{Thomas_2015,Milillo_2015}, especially observations of lensing $B$ modes \cite{thomas2009rotation,bruni2014computing}.
    Observations of gravitational waves emitted from compact binary inspirals, which can be detected from systems at intermediate cosmological redshifts, may also be helpful for studying cosmological preferred-frame effects, because it is expected that gravitational wave signals should be sensitive to the preferred-frame PPN parameters $\alpha_1$ and $\alpha_2$ \cite{Sampson_2013}. 

\end{enumerate}

All of these parameters are well constrained at the present time by astrophysical tests (with $\alpha(t_0) = 1$ by definition), but cosmological observations should provide the best opportunity to study their time dependence.
In the next chapter, we will derive the first cosmological constraints on the PPNC parameters.

\chapter{Constraining PPNC with the cosmic microwave background}

\lhead{\emph{CMB tests of parameterised post-Newtonian cosmology}}

In Chapter 5, we discussed at length the development of parameterised post-Newtonian cosmology (PPNC), a theory-independent framework based on the parameterised post-Newtonian (PPN) formalism, that makes it possible to construct consistent, well-defined cosmological models without specifying an underlying theory of gravity \cite{Sanghai_2017, Sanghai_2019, anton2022momentum, Thomas_2023, Clifton_2024}. 
The framework leads to generalised Friedmann equations (\ref{eq_parametrisedfriedmanneqns1}-\ref{eq_parametrisedfriedmanneqns2}) in which the PPN parameters $\alpha$ and $\gamma\,$, as well as the minimal additions $\alpha_c$ and $\gamma_c$\,, appear explicitly. They are upgraded from constants to functions of time, or equivalently scale factor $a$\,.
It also gives rise to theory-independent equations of motion (\ref{eq_PPNC_allscales_Hamiltonian}, \ref{eq_PPNC_allscales_Raychaudhuri}, \ref{eq_generalparametrisedscalareqn}) for scalar perturbations to an FLRW cosmology, and the equation (\ref{eq_generalparametrisedvectoreqn}) for divergenceless vector perturbations, into which the PPNC parameters and their time derivatives enter through the gravitational coupling functions $\mu(\tau, k)$\,, $\nu(\tau, k)$\,, $\mathcal{G}(\tau, k)$ and $\mathcal{Q}(\tau, k)\,$.
We showed in Section \ref{sec:scale_dependence} that the time dependence of the PPN parameters is described well by a power law in $a$ for canonical scalar-tensor theories of gravity, and that the scale dependence of the couplings $\mu\,$, $\nu$ and $\mathcal{G}$ can be approximated well by a simple zero-parameter interpolation (\ref{eq_scale_dependent_coupling}) between the sub-horizon and super-horizon PPNC limits\footnote{We did not consider the scale dependence of $\mathcal{Q}$ as it affects only vector perturbations. Vector modes are irrelevant for the temperature, E-mode polarisation and lensing anisotropies in the cosmic microwave background that we will consider in this chapter, so $\mathcal{Q}$ will again not be considered.}.

Our aim now is to constrain the time-dependent PPNC functions $\alpha(t)\,$, $\gamma(t)\,$, $\alpha_c(t)$ and $\gamma_c(t)\,$, and therefore the gravitational couplings $\mu\,$, $\nu$ and $\mathcal{G}\,$. This would allow us to extend the astrophysical constraints on post-Newtonian gravity, summarised by Tables \ref{table_PPN_parameters} and \ref{table_PPN_time_derivs}, to cosmological scales.
Studying the time dependence of the PPN parameters over cosmic history constitutes a direct, physically well-defined, extension of the long-standing idea, going back to Dirac's ``large numbers'' hypothesis \cite{dirac1974cosmological}, that Newton's constant (or indeed other fundamental constants) might vary in time \cite{Uzan_2003,uzan2011varying}. 
For example, Dirac's hypothesis is naturally incorporated by the Brans-Dicke scalar-tensor theory \cite{brans1961mach}, where $G_{\rm eff} (t) \sim \alpha(t) \sim 1/\bar{\phi}(t)\,$ evolves as a power law in the scale factor during matter domination \cite{nariai1968green}. 

Inferred measurements we might make within the PPNC framework of, for example, $\dfrac{\dot{\alpha}}{\alpha}$\,, using cosmological datasets, could therefore be compared unambiguously to measurements made in the Solar System through vastly different means \cite{Will_2014}, such as lunar laser ranging \cite{williams2004progress}, the ephemeris of Mars \cite{konopliv2011mars} and helioseismology \cite{guenther1998testing}. 
Hence, we would be able to complement our understanding of the time variation (or indeed the lack thereof) of $G$ in that setting with a novel set of constraints in an entirely different regime, while retaining the physical clarity and interpretability that makes the post-Newtonian regime the gold standard for testing gravity in a theory-agnostic fashion \cite{Will_1993}.

It is worth noting, before we go on, that there do exist cosmological constraints on an evolving Newton's constant.
The strongest bounds are provided by the abundances of light elements from Big Bang Nucleosynthesis (BBN) \cite{bambi2005response}, although they are model-dependent \cite{Will_2014}. Perhaps more troubling is that it is not clear in general that the ``Newton's constant'' inferred from observations of this kind is equivalent to what we measure in weak-field astrophysical gravity, given that BBN occurs during radiation domination, whereas the $G$ that we measure astrophysically is associated with non-relativistic matter. 
To be specific, in GR, the coefficients of $4\pi \rho_m$ in the Newton-Poisson equation and $\dfrac{8\pi \bar{\rho}_r}{3}$ in the first Friedmann equation are the same, but that equality is, more often than not, broken in modified theories of gravity.

Thus, although the BBN result is certainly very useful, and should not be disregarded, it would be satisfying to compare it to the results of an approach in which the notion of a time-varying Newton's constant is formalised into a different well-motivated picture (and using a different cosmological dataset). 
As we have introduced such an approach in the previous chapter, it is now our purpose in this chapter to carry out an observational investigation into the time dependence of the post-Newtonian parameters in cosmology. 

In order to do this, we wish to use an accurately measured observational dataset, that is sensitive to phenomena across a wide range of cosmological scales, from deep within the Hubble horizon to well beyond it, over large swathes of cosmic history. The anisotropies in the cosmic microwave background, which we introduced in Section \ref{subsec:CMB}, are therefore a natural choice. 
These have the additional property of linearity ($\dfrac{\Delta T}{T} \sim 10^{-5}$ is comfortably within the linear regime), and the angular power spectra are dependent only on the scalar sector of linear cosmological perturbation theory.
They therefore provide a very clean probe of the parameters that enter into the Friedmann equations and the scalar sector of the perturbation theory equations.

Thus, in this chapter we will use observations of the $TT$\,, $TE$\,, $EE$ and $\phi\phi$ (lensing) CMB anisotropies from the Planck satellite \cite{aghanim2020planck,planck_power_spectra}, in order to constrain the degrees of freedom of the PPNC formalism, and hence to constrain the time evolution of the PPN parameters all the way from the time of last scattering through to the present day. 
We will see that the CMB anisotropies alone already provide a wealth of information about the landscape of parameterised post-Newtonian cosmology, and therefore about what kinds of historic and evolving deviations from General Relativity might be allowed in our Universe, for any metric theory of gravity that is compatible with the principles of the PPN formalism\footnote{Broadly speaking, this means theories that do not contain nonlinear screening mechanisms, although attempts to incorporate screened modified gravity theories into the post-Newtonian framework do exist \cite{Avilez_Lopez_2015}.}.
It should be considered as a complementary alternative to previous works that have obtained cosmic microwave background constraints on specific classes of modified gravity theories (especially scalar-tensor models), using effective field theory and similar approaches
\cite{brax2012systematic,frusciante2020effective,Joudaki_2022,Bellini_2018,Peirone_2019}. 

We reiterate once again that parameterised post-Newtonian cosmology is a rather conservative framework: it makes no assumptions about any additional gravitational field content of the underlying theory, and does not allow for more complicated deviations from GR that could be induced, for example, by a theory that contains highly non-trivial small-scale phenomenology such as a screening mechanism or modified Newtonian dynamics (MOND). 
The conservative nature of the formalism means that our constraints will be rather general, and thus constitute something approaching a genuine null test of GR. 

Furthermore, in the PPNC framework, the FLRW background expansion and the scalar perturbations are evolved consistently within the same formalism. This contrasts with several approaches that are typically taken in cosmology, where the cosmological background is fixed {\it a priori} to be equivalent to a $\Lambda$CDM cosmology \cite{Hu_2008,amin2008subhorizon}. 
We will find that this is a crucial feature of our results, showing the importance of a complete, self-consistent framework, as opposed to a na{\" i}ve test of GR that focuses only on the evolution of cosmological scalar perturbations (unless one is certain that the cosmological background expansion is equivalent to $\Lambda$CDM, which is the case, for example, in certain $f(R)$ theories \cite{hu2007models}).

The chapter, which is based on Ref. \cite{Thomas_2024}, will be structured as follows. 
We will first explain in Section \ref{sec:ppnc_setup} how parameterised post-Newtonian tests of cosmological gravity with CMB anisotropies can be performed, by introducing a modified Einstein-Boltzmann code that incorporates the PPNC framework to calculate predictions for CMB observables, and a Markov Chain Monte Carlo approach for obtaining posteriors on the PPNC and standard cosmological parameters from the Planck data.
In Section \ref{sec:ppnc_cmb} we will examine the most prominent effects of PPNC deviations from GR on the cosmic microwave background temperature anisotropies.
Finally, we will present the results of our analysis of the PPNC system using the Planck data in Section \ref{sec:planck_results}, focusing especially on the posteriors on the PPNC parameters $\alpha$ and $\gamma$\,, and their degeneracies with both each other and the various standard cosmological parameters. These degeneracies will have direct physical interpretations in terms of the phenomenology we describe in Section \ref{sec:ppnc_cmb}, and we will discuss some possibilities for alleviating the degeneracies in the future. 

\section{Setting up CMB tests}\label{sec:ppnc_setup}

In order to use the measured anisotropies in the cosmic microwave background to test the PPNC framework, it is necessary to introduce two important tools: a modified Einstein-Boltzmann code to calculate the metric and energy-momentum perturbations that determine the CMB anisotropies, and a Markov Chain Monte Carlo (MCMC) analysis to obtain our constraints.

\subsection{The PPNC CLASS code}\label{subsec:ppnc_class}

The parameterised, theory-independent, equations of motion, for the cosmological background expansion and the (Fourier space) linear scalar perturbations, must be integrated numerically in order to compute predictions for $\mathcal{C}_l^{TT}\,$, $\mathcal{C}_l^{TE}$\,, $\mathcal{C}_l^{EE}$ and $\mathcal{C}_l^{\phi\phi}$\,.
We do this by suitably modifying the Cosmic Linear Anisotropy Solving System (CLASS) code \cite{lesgourgues2011cosmic}, which solves the Einstein-Boltzmann system described in Appendix \ref{subsec:einstein_boltzmann}\footnote{The modified CLASS code was written by Daniel B. Thomas.}. 
For this purpose, one must make some choices of both mathematical parametrisations and physical couplings of species, which are not uniquely determined by the underlying principles of the PPNC framework.

We need to prescribe
\begin{enumerate}
    \item How radiation, and other ultrarelativistic degrees of freedom, should be incorporated. 
    \item The functional form of the gravitational slip $\Sigma = \dfrac{\Phi - \Psi}{\Psi}$ on super-horizon scales $k \longrightarrow 0\,$.
    \item The forms of the time dependence of the post-Newtonian parameters, and the scale dependence of the resultant coupling functions.
\end{enumerate}

Let us first deal with the issue of how to include ultrarelativistic species (i.e. radiation and neutrinos) in the PPNC governing equations. This is obviously necessary in order to calculate the properties of the cosmic microwave background radiation, but so far we have dealt only with non-relativistic matter fields (and dark energy). 
This is really an artefact of the way the PPN formalism is built in the first place. It is designed to describe the gravitational fields of astrophysical systems, for which isotropic pressure $p$ is suppressed by a factor of $v^2$ relative to mass density $\rho$ in the post-Newtonian hierarchy. 
For ultrarelativistic degrees of freedom with $v \sim 1$, pressure contributes at leading order just like energy density; for example, in the case of radiation one has $\bar{p}_r = \dfrac{1}{3}\bar{\rho}_r\,$. 

Of course, one could go back to first principles, and reconstruct the PPN formalism with a different power-counting hierarchy, so that contributions from the generic energy-momentum tensor of ultrarelativistic species are explicitly accounted for.
However, generalising the post-Newtonian framework in this way may lead to it losing its simplicity and predictive power. A first attempt to understand the effects of radiation and neutrinos in parameterised post-Newtonian cosmology is provided in Ref. \cite{Sanghai_2016}. 
In this chapter, we will make a simple, conservative choice: in the absence of any information from the classical PPN formalism on how the coupling of ultrarelativistic species to the metric tensor should be described, we will just retain all these couplings at their standard GR value. 
This choice is not unique, but provides us with a definite means to calculate the effect of the change in gravitational coupling strengths of pressureless matter without having to make significant changes to the radiation era in the early Universe.

Thus, the Friedmann equation that is evolved in the Einstein-Boltzmann code is
\begin{equation}\label{eq_ppnc_CLASS_Friedmann}
    H = \sqrt{\frac{8\pi G}{3}\gamma\bar{\rho}_m + \frac{8 \pi G}{3}\bar{\rho}_{\rm ur} - \frac{2\gamma_c}{3}}\,,
\end{equation}
where $\bar{\rho}_m = \bar{\rho}_c + \bar{\rho}_b\,$ is the total energy density of cold dark matter and baryons, and $\bar{\rho}_{\rm ur}$ is the total energy density in ultrarelativistic species (so in the standard case, it is just $\bar{\rho}_r + \bar{\rho}_{\nu}$).

The equations of motion that are evolved by the code for the Fourier space Newtonian gauge scalar perturbations $\Phi$ and $\Psi$ are
\begin{eqnarray}
    \Phi &=& \left(1+\Sigma\right)\Psi + \frac{12\pi G a^2}{k^2} \sum_i \left(\bar{\rho}_i + \bar{p}_i\right)\sigma_i\,, \quad {\rm and} \label{eq_PPNC_CLASS_slip} \\
    \Psi' &=& - \mathcal{H}\Phi - \frac{4\pi G a^2}{k^2}\left[\left(\mu-1\right)\left(\bar{\rho}_b \theta_b + \bar{\rho}_c \theta_c\right) + \sum_i\left(\bar{\rho}_i + \bar{p}_i\right)\theta_i\right] + \mathcal{G}\mathcal{H}\Psi\,. \label{eq_PPNC_CLASS_momentum}
\end{eqnarray}
Here the labels $i$ refer to the species being considered, $\sigma_i$ is the linear shear stress associated with that species, so that $\left(\bar{\rho}_i + \bar{p}_i\right)\sigma_i$ is the scalar part of the anisotropic stress contribution $\Pi_i$ from the species\footnote{The only sizeable anisotropic stress is from neutrinos, which are suppressed at the linear level.}, and $\theta_i = - k^2 v_i$ is the velocity divergence associated with the species (in previous equations, we have used the velocity potential $v$ rather than $\theta$).
As in the Friedmann equation, in the momentum constraint (\ref{eq_PPNC_CLASS_momentum}) the ultrarelativistic species have been explicitly separated out from non-relativistic matter, so that they do not carry with them a factor of the PPNC coupling $\mu\,$. 
It should also be noted that in the CLASS code, the metric perturbations $\Phi$ and $\Psi$ are replaced by the pair $\phi$ and $\psi$\,, defined by $\phi = -\Psi$ and $\psi = -\Phi\,$. In addition, we note that instead of the slip $\Sigma$\,, the code uses $\eta = 1 + \Sigma\,$, so that in the absence of anisotropic stress $\psi = \eta \phi\,$.

Eq. (\ref{eq_PPNC_CLASS_slip}) leads us naturally on to the second problem we mentioned above: how to prescribe the slip function $\Sigma\,$. On small scales $k \longrightarrow \infty$ we have $\Sigma = \dfrac{\alpha - \gamma}{\gamma}\,$, but there is no known link at this stage between the PPN parameters and the value of $\Sigma$ on ultra-large scales.
As discussed in the previous chapter, we will therefore make the simple prescription that for $k \longrightarrow 0$\,, the slip $\Sigma$ reverts to its GR value of zero. This is just chosen for convenience, and can be revisited if and when a more physically well-motivated prescription for its value is obtained. 
If we wanted to, there is nothing stopping us just including the large-scale limit of the slip of an additional free parameter, and then constraining it alongside our other PPNC parameters. 
However, to do so would be somewhat antithetical to the ethos of our formalism where the coupling functions are decidedly not arbitrary, and so we will not carry out such an analysis here.

Finally, one needs to specify the free functions that enter the governing equations. For the scale dependence of the couplings $\mu\,$, $\mathcal{G}$ and $\Sigma$\,, we will use the zero-parameter interpolation (\ref{eq_scale_dependent_coupling}) introduced and stress-tested for canonical scalar-tensor gravity theories in Section \ref{sec:scale_dependence}. 
Although this choice might miss out on a small amount of more complicated phenomenology on scales similar to the horizon, it has the desirable property from the perspective of parameter constraints that it does not introduce any new free parameters, and so we can focus our constraining power on the time-dependent PPNC parameters themselves.

Therefore, let us turn our attention to the functional form of the PPNC parameters $\{\alpha, \gamma, \alpha_c, \gamma_c\} = \{\alpha(\tau), \gamma(\tau), \alpha_c(\tau), \gamma_c(\tau)\}$, which must be input directly into the code.
First, consider the time dependence of the standard post-Newtonian parameters $\alpha$ and $\gamma$\,. We choose them to be functions of the scale factor $a$ of the form
\begin{eqnarray} \label{eq_ppnc_power_law_n}
\alpha(a)=A\left(\frac{a_1}{a}\right)^n+B \qquad {\rm and} \qquad    \gamma(a)=C\left(\frac{a_1}{a}\right)^n+D\,.
\end{eqnarray} 
Here $a_1$ is the initial scale factor of the numerical integration. In all the results presented, we have made the arbitrary choice $a_1 = 10^{-10}\,$, but its exact value does not matter as long as it is so deep into the radiation-dominated era that matter fields were entirely irrelevant at that time.
The PPNC modifications to gravity are taken not to be present when the initial conditions are set, which is a safe choice because these initial conditions are laid down so deep into the radiation era that the coupling of gravity to the entirely negligible non-relativistic matter content of the Universe is irrelevant. The evolving parameters then ``switch on'' at $a_1\,$, and begin to affect the cosmological dynamics only once there is a non-negligible $\Omega_m$ in the Universe.

The power law form for $\alpha$ and $\gamma$ is chosen for its simplicity, and the smoothness of the interpolated coupling functions $\mu(\tau, k)\,, \mathcal{G}(\tau, k)$ and $\Sigma(\tau, k)$ that result. 
In other fruitful approaches, couplings of this kind have been studied as piecewise functions in redshift bins, as in Ref. \cite{Sankar}. However, this is not a viable possibility in the PPNC framework, because the time derivatives of $\alpha$\,, $\gamma$\,, $\alpha_c$ and $\gamma_c$ enter explicitly into the equations of motion, and so the time derivatives of the parameters must always be well-defined and continuous.
The power law functional form is further motivated by its validity in the scalar-tensor theory class that was considered in Section \ref{sec:scale_dependence}, as shown by Fig. \ref{fig_ppnc_parameters_BDT}. In that case, the same power law index for $\alpha$ and $\gamma$ was found to be a good description, with $n \approx 0.1$ for the Brans-Dicke theory with $\omega = 10$ that was most strongly deviant from GR. 

The constants $\{A, B, C, D\}$ are calculated from 
\begin{itemize}

    \item The present-day values of $\alpha$ and $\gamma\,$, which we denote by $\alpha_0$ and $\gamma_0$. By the definition of Newton's constant, which we measure in the Solar System at $t_0\,$, $\alpha_0 = 1\,$. The parameter $\gamma_0$ is constrained by Solar System experiments to be unity, within 1 part in $10^5$ \cite{bertotti2003test}, and so in our analyses we will always choose $\gamma_0 = 1\,$.
    
    \item Their initial values $\alpha(a_1)$ and $\gamma(a_1)$\,. As these are strongly degenerate with the power law index $n$\,, it is more helpful to specify instead the ``average'' values $\bar{\alpha}$ and $\bar{\gamma}$ over the range of scale factors $\left[a_1, 1\right]\,$\,, defined by
    \begin{equation}\label{eq_mean_alpha_mean_gamma}
    \bar{\alpha}\equiv \frac{\int^{0}_{\ln a_1}{\mathrm{d}\ln{a}\ \alpha(a)}}{\int^{0}_{\ln a_1}{\mathrm{d} \ln a}} \qquad {\rm and} \qquad \bar{\gamma}\equiv \frac{\int^{0}_{\ln a_1}{\mathrm{d} \ln{a}\ \gamma(a)}}{\int^{0}_{\ln a_1}{\mathrm{d} \ln a}}\,.
    \end{equation}
    After these averages are specified, the initial condition on $\gamma(a)$ is then obtained by
    \begin{equation}\label{eq_mean_gamma_to_gamma_a1}
        \gamma(a_1) = \frac{\bar{\gamma} - \left(\frac{1}{1-a_1^n} - \frac{1}{n\ln{a}_1}\right)\gamma_0}{1 - \frac{1}{1 - a_1^n} - \frac{1}{n\ln{a}_1}}\,,
    \end{equation}
    and equivalently for $\alpha(a)\,$\,.
    
\end{itemize}

\begin{figure}
    \centering
    \includegraphics[width=0.95\linewidth]{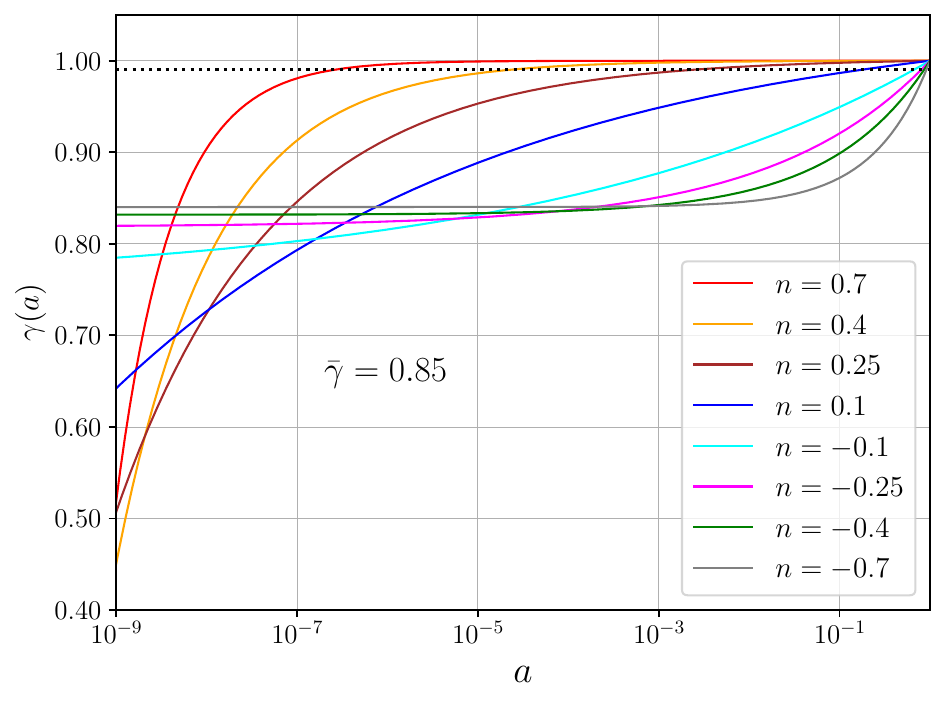}
    \caption{Evolution of the PPN parameter $\gamma(a)$ for different values of the power-law index $n$, shown as a function of the scale factor $a$. In each case shown, $\bar{\gamma}$ is equal to $0.85\,$. The dotted line is at $\gamma = 0.99\,$.}
    \label{fig_gamma_of_a_power_laws}
\end{figure}

The value of $n$ tells us about the rate of change of the parameters in time, and therefore indicates at what points in cosmic history they are evolving most rapidly.
The behaviour of $\gamma(a)$ for different values of the power-law index $n$\,, but the same $\bar{\gamma}\,$, is displayed in Fig. \ref{fig_gamma_of_a_power_laws}. 
For large positive $n$\,, any possible modifications in GR in that parameter are pushed to very early times, when its derivative is very large, whereas for large negative $n$\,, the parameter varies mostly in the late Universe, before which it is roughly a constant $\neq 1\,$.
For the sake of simplicity, we have chosen the power law index for both $\alpha$ and $\gamma$ to be the same, which would indeed be the case if they were both controlled by the evolution of the same underlying degree of freedom. This condition could of course be removed, at the expense of constraining power and computational efficiency.

Let us finally consider the dark energy-like fields present in the Universe, which are communicated in the PPNC formalism entirely through the parameters $\alpha_c(\tau)$ and $\gamma_c(\tau)$\,. Recall that in GR with $\Lambda$CDM energy-momentum content, $\alpha_c = -2\gamma_c = \Lambda\,$.
In order that $\gamma_c$\,, which sources the expansion through the first Friedmann equation (\ref{eq_parametrisedfriedmanneqns1}), behaves as similarly as possible to a cosmological constant, we will demand that $\gamma_c$ is constant. It is determined by its fractional contribution to $H$ at the present day, much like how in the $\Lambda$CDM picture $\Lambda$ is implicitly set by specifying $\Omega_{m0}\,$.
Thus, we have
\begin{equation}
    \gamma_c = \gamma_{c0} = \frac{3}{2}\Omega_{m0}H_0^2\left(\gamma_0 - 1\right) - \frac{\Lambda}{2} \,.
\end{equation}

The integrability condition (\ref{eq_integrabilitycondition}) is then used to specify $\alpha_c$\,, so that
\begin{equation}
    \alpha_c(a) = \frac{3\,\Omega_{m0}\, H_0^2}{2a^3}\left(\alpha(a) - \gamma(a) + \frac{\mathrm{d}\gamma(a)}{\mathrm{d}\ln{a}}\right) - 2\gamma_{c0}\,,
\end{equation}
where we have dropped the term $\dfrac{\mathrm{d}\gamma_c}{\mathrm{d}\ln{a}}$ that usually appears in Eq. (\ref{eq_integrabilitycondition}), as it is manifestly zero for a constant $\gamma_c$\,, and we have also made use of the first Friedmann equation (\ref{eq_parametrisedfriedmanneqns1}), evaluated at the present day.

Note that there is a gauge-related subtlety involved in integrating our equations using the CLASS Einstein-Boltzmann code. The initial conditions in CLASS are always specified in the synchronous gauge \cite{lesgourgues2011cosmic}, and then a gauge transformation is made if one desires the system to be evolved in Newtonian gauge.
However, the entire PPNC framework is explicitly constructed with perturbations defined in the Newtonian gauge, as it is only with this gauge choice that an explicit link with the post-Newtonian regime can be made safely \cite{Clifton_2020}, and so it would appear problematic to have to specify initial conditions in a gauge in which our formalism is ill-defined. 
The issue is sidestepped by setting the PPNC modifications to gravity to be zero at the initial scale factor $a_1$, and only switching them on afterwards, for $a > a_1\,$.
Therefore, we can safely make the gauge transformation from synchronous to Newtonian gauge at $a_1$\, assuming GR equations, in order to set the initial conditions. Then, we evolve forward in Newtonian gauge using our full PPNC equations. 

In order to highlight the relative effects of PPNC modifications to the background expansion and the evolution of scalar perturbations, relative to $\Lambda$CDM, flags can be switched on in the initialisation file of the PPNC CLASS code that allow it to be run with either the background or perturbation equations modified from their $\Lambda$CDM forms, or both. 
Of course, the case where they are both modified constitutes the physically correct set of equations, and the cases where modifications can be switched off at either the background or perturbation level only are considered only for the purposes of understanding how modifying each sector affects observations.

We will find in our analysis that the term $\mathcal{G}\mathcal{H}\Psi$ in the momentum constraint (\ref{eq_PPNC_Fourier_momentum}), which evolves the Bardeen potentials \footnote{Strictly speaking, only $\Psi$ is evolved by the code, and then $\Phi$ is set at each timestep by the slip relation (\ref{eq_PPNC_CLASS_slip}).} via Eq. (\ref{eq_PPNC_CLASS_momentum}), plays an important role. 
Therefore, the code also contains a flag that allows this term, which vanishes identically for all $a$ and $k$ in GR, to be artificially removed from Eq. (\ref{eq_PPNC_CLASS_momentum}), in order to highlight its effect.
At present, the PPNC CLASS code is not public, but it will be made public in the near future, with an official code release describing in detail all the modifications to the standard CLASS setup, and the initialisations that can be used to explore the PPNC theory space.

\subsection{Markov Chain Monte Carlo approach}\label{subsec:mcmc}

Our constraints are obtained using a Markov Chain Monte Carlo (MCMC) approach\footnote{MCMC methods are a standard tool in cosmology, as they are ideally suited to the problem of determining cosmological parameters from data using Bayesian inference \cite{christensen2001bayesian}.}, using the Metropolis-Hasting algorithm within the open-source MontePython package \cite{audren2013conservative,audren2013monte}. 
We use a jumping factor of $2.1\,$, and for each run with a fixed power law we start ten chains, which continue until the Gelman-Rubin $1-R$ \cite{gelman1992inference} convergence criterion is smaller than 0.01 for all parameters\footnote{The plots presented in this chapter are my own. However, they are made using data from the MCMC analyses, which were implemented by Daniel B. Thomas.}. 
For the runs where the power law index $n$ is a free parameter, we start 120 chains, and use the same convergence criterion.

In our MCMC analyses, we vary seven standard cosmological parameters: the usual six $H_0$\,, $\omega_c$\,, $\omega_b$\,, $\tau_{\rm reio}\,$, $\ln{\left(10^{10} A_s\right)}$ and $n_s$ in the canonical $\Lambda$CDM cosmology (where the first three parameters are as explained in Section \ref{subsec:CMB}, and the final three are respectively the optical depth to reionisation, and the amplitude and spectral tilt of primordial scalar perturbations), plus the helium fraction $Y_{\rm P}$ from Big Bang nucleosynthesis.
Although $Y_{\rm P}$ is not independent of the first six parameters in a $\Lambda$CDM context, that is no longer true in the present case, because the background expansion can be altered in the radiation era by the modified PPNC Friedmann equations (\ref{eq_parametrisedfriedmanneqns1}-\ref{eq_parametrisedfriedmanneqns2}). Hence, we must vary $Y_{\rm P}$ as an independent parameter.
The FLRW spatial curvature is set to zero, as is $\Omega_\Lambda$\,, because $\alpha_c$ and $\gamma_c$ already account for dark energy. Isocurvature perturbations are ignored.

Let us now move on to the novel parameters that are being varied in the MCMC study.
In all of our runs we have at least two new parameters: the ``average'' PPN parameters $\bar{\alpha}$ and $\bar{\gamma}$\,, defined as in Eq. (\ref{eq_mean_alpha_mean_gamma}). From these, we can also display the derived parameters $\alpha(a_1)$ and $\gamma(a_1)$\,, calculated according to Eq. (\ref{eq_mean_gamma_to_gamma_a1}), and the present-time derivatives $\dot{\alpha}_0$ and $\dot{\gamma}_0\,$. 
In all cases, we set the present-day values of $\alpha$ and $\gamma$ to unity. The latter is not truly guaranteed, but merely suggested by Solar System experiments \cite{bertotti2003test}, and so technically one ought to include the Shapiro time delay constraint (displayed in Table \ref{table_PPN_parameters}) as a Gaussian prior. However, the variance in $\bar{\gamma}$ that would arise from doing this is negligibly small compared to the variance in the CMB constraint. Thus, for the sake of simplicity we are safe to ignore this, and set $\gamma(t_0) = 1\,$.

The only thing left to consider is the power law index $n$ that specifies the evolution of the post-Newtonian parameters through Eq. (\ref{eq_ppnc_power_law_n}).
We consider two different sets of chains:
\begin{enumerate}
    \item Runs in which the value of $n$ is fixed. We consider runs with several different indices $n$\,, the results of which are presented in Section \ref{subsec:fixed_power_law}. These runs therefore have nine independent parameters (as well as the usual Planck nuisance parameters): the six standard cosmological parameters present in $\Lambda$CDM, plus $Y_{\rm P}$, $\bar{\alpha}$ and $\bar{\gamma}\,$.
    
    \item Runs with $n$ as a free parameter, such that there are ten independent parameters. The results of these runs are presented in Section \ref{subsec:varying_power_law}. This is rather closer than the fixed-$n$ cases to what one would really like to do from a Bayesian perspective, because there is no particular reason to choose any given $n$\,, but of course the inclusion of an additional parameter comes at the cost of constraining power for the rest of the PPNC parameter space.
    
\end{enumerate} 

Constraints are obtained from the Planck 2018 dataset \cite{Planck_2020, planck_power_spectra}, which comprises the low-$l$ likelihood, the full TT, EE and TE high-$l$ likelihood with the complete ``not-lite'' set of nuisance parameters, and the lensing potential likelihood.
For the $20$ nuisance parameters, we adopt Gaussian priors, with means and variances fixed at the same values used by the Planck collaboration for the above set of likelihoods \cite{planck_power_spectra}.

The set of cosmological parameters constrained is summarised in Table \ref{table_mcmc_priors}. We have also displayed the ranges of their flat priors, as well as the derived parameters we constrain. Finally, we list the parameters which are held fixed - the PPNC boundary conditions already discussed, plus the number of massive neutrinos $M_{\rm ncdm}$\,, and their associated mass $m_{\rm ncdm}$ and temperature $T_{\rm ncdm}$\,, the effective number of ultrarelativistic (massless neutrino) species $N_{\rm ur}$\,. Taken together, these parameters fix the effective number of neutrino species to be its $\Lambda$CDM value $N_{\rm eff} = 3.046\,$.

\begin{table}
\centering
\begin{mytabular}[1.2]{|c|c|c|} 
\hline 
Cosmological parameters &  Derived parameters & Fixed parameters \\
\hline
$100 \, \omega_b \, \in \left[0, \infty\right)$ & $z_{\rm reio}$ & $N_{\rm ncdm} = 1$ \\
\hline
$\omega_c \, \in \left[0, \infty\right)$ & $\sigma_8$ & $m_{\rm ncdm}/{\rm eV} = 0.06$ \\
\hline
$\ln\left(10^{10} \, A_s\right)\, \in \left(-\infty, \infty\right)$ & $100\,\theta_{*}$ & $N_{\rm ur} = 2.0328$ \\
\hline
$n_s \, \in \left[0, \infty\right)$ & $\gamma(a_1)$ & $T_{\rm ncdm}/{\rm K} = 0.7161$ \\
\hline
$\tau_{\rm reio}\, \in \left[0.004, 1\right]$ & $\alpha(a_1)$ & $a_1 = 10^{-10}$ \\
\hline
$Y_{\rm P} \, \in \left[0, 1\right]$ & $\dot{\gamma}_0/H_0$ & $\gamma(t_0) = 1$ \\
\hline
$100\,h\, \in \left(0, \infty\right)$ & $\dot{\alpha}_0/H_0$ & $\alpha(t_0) = 1$ \\
\hline
$\bar{\gamma}\, \in \left[0, 50\right]$ & ----- & ----- \\
\hline
$\bar{\alpha} \,\in \left[0, 50\right]$ & ----- & ----- \\
\hline
$n\, \in \left(-15, 0.25\right]$ & ----- & ----- \\
\hline
\end{mytabular} 
\caption{Left column: full list of cosmological parameters constrained in the MCMC analysis, with the bounds on their prior distributions (which are flat in between the bounds). Centre: full list of derived parameters constrained as a result. Right: full list of parameters which are held fixed in the analysis. Note that in several of the chains, the PPNC power law index $n$ is held fixed, rather than being varied as a model parameter.}
\label{table_mcmc_priors}
\end{table}


For the chains in which the power law index $n$ is varied, we apply a flat prior with a lower bound of $-15$\,. The upper bound of the prior is more complicated, and the results are invariably sensitive to it. 
The reason for this is that increasing the value of $n$ moves the consequences of modifying gravity to earlier and earlier times, as displayed in Fig. \ref{fig_gamma_of_a_power_laws}. Thus, all deviations from GR are pushed deeper and deeper into the radiation era, where the matter contribution and thus the effect of the modified gravitational coupling parameters is increasingly negligible. 
For example, one could set the upper bound of $n$ to $1\,$. This is very conservative, allowing all the effects of modifying gravity to be confined to very early times well before matter-radiation equality. 
In this case, the posterior is driven to the range of values of $n$ up against the upper bound of the prior, where the altered gravitational parameters have little effect. Therefore, the resulting constraints on $\bar{\alpha}$ and $\bar{\gamma}$ lack physical meaning and are rather misleading.

Let us therefore determine a less conservative prior, in order to obtain some more meaningful results.
We choose the upper bound on the prior for $n$ to be $0.25$\,. This is motivated by requiring that deviations from GR in $\gamma(a)$ and $\alpha(a)\,$, if they exist, should persist non-negligibly into the matter era, rather than being confined entirely to the radiation era where they have little physical effect, since the coupling of gravity to matter is irrelevant if matter is negligible compared to radiation.
To visualise what this means for the evolution of $\alpha(a)$ and $\gamma(a)$\,, consider again Fig. \ref{fig_gamma_of_a_power_laws}. 
For the large positive power laws  with $n = 0.4$ and $0.7$, the curve $\gamma(a)$ only drops below the dotted line at $\gamma = 0.99$ for $a < 10^{-4}$. Thus, almost all the modifications to GR are appearing well before matter-radiation equality in these cases. This means that they have very little effect, because the PPNC parameters have been assumed not to couple to radiation.
The choice of $n = 0.25$ as our borderline cutoff case is suggested by the corresponding brown curve in Fig. \ref{fig_gamma_of_a_power_laws}, for which $\gamma$ is below $0.99$ for $a \lesssim 10^{-3}\,$, i.e. until around the recombination epoch.

\begin{table}
    \centering
\begin{tabular}{|c|c|c|c|c|c|c|} 
 \hline 
Power law index $n$ & $\bar{\gamma}$ & $\bar{\alpha}$ & $\gamma_{\rm eq}$ & $\alpha_{\rm eq}$ & $\gamma_{\rm LS}$ & $\alpha_{\rm LS}$ \\ \hline 
$0.1$ & $0.87$ & $1.13$ & $0.9439$ & $1.0562$ & $0.9548$ & $1.0452$ \\
\hline
$-0.4$ & $0.87$ & $1.13$ & $0.8598$ & $1.1402$ & $0.8631$ & $1.1369$ \\
\hline
$0.4$ & $0.87$ & $1.13$ & $0.9970$ & $1.0030$ & $0.9982$ & $1.0019$ \\
\hline
 \end{tabular} \\ 
    \caption{Values of $\gamma$ and $\alpha$ at matter-radiation equality and last scattering, for three different PPNC power laws with the same values of $\bar{\gamma}$ and $\bar{\alpha}\,$.}
    \label{table_equality_LS}
\end{table}

To see the differences between the power laws as they pertain to the CMB, consider Table \ref{table_equality_LS}, in which we have computed the values of the PPN parameters at matter-radiation equality and last scattering, for the power laws $n = 0.1$\,, $-0.4$ and $0.4\,$, with the same values of $\bar{\gamma}$ and $\bar{\alpha}\,$. 
It shows that for the $0.1$ and $-0.4$ power laws, the functional form of $\gamma(a)$ and $\alpha(a)$ allows for non-negligible modifications to gravity in the matter-dominated era. In contrast, for the $n = 0.4$ power law, virtually all the evolution in $\alpha(a)$ and $\gamma(a)$ happens for $a \ll a_{\rm LS}$\,.
Hence, there is very little constraining power on $\bar{\alpha}$ and $\bar{\gamma}$\,, because by the time of matter-radiation equality, $\alpha$ and $\gamma$ have already relaxed to within $0.3\%$ of unity. We take this to be justification for our intuition about the prior on $n$ for the varying-$n$ chains.

This concludes our discussion of how our CMB tests of the PPNC landscape are conducted. 
In the next section, we will consider in some detail what the main physical effects of PPNC deviations from General Relativity are, and how they imprint themselves on the cosmic microwave background anisotropies. With those in mind, we will be able to understand the key MCMC results. These are presented in Section \ref{sec:planck_results}.

\section{CMB phenomenology in parameterised post-Newtonian cosmology}\label{sec:ppnc_cmb}

In this section, we will investigate the main features of the PPNC theory space, as they pertain to the CMB temperature anisotropies. 
We choose to do this before we present the results of the Planck MCMC analyses in Section \ref{sec:planck_results}, so that when we come to those we will have a physical intuition for what the results might mean.

We will focus first on the relationship between the PPNC parameters (especially their weighted averages $\bar{\alpha}$ and $\bar{\gamma}$).
We will then explore the modifications to the FLRW background expansion that can be induced in our framework, which are manifested primarily through the acoustic peaks in the CMB. These will allow us to predict degeneracies with the cosmological parameters, most notably $H_0$ and $\omega_c\,$. 

\subsection{Relationship between the PPN parameters}\label{subsec:alpha_gamma}

\begin{figure}
    \centering
    \includegraphics[width=0.75\linewidth]{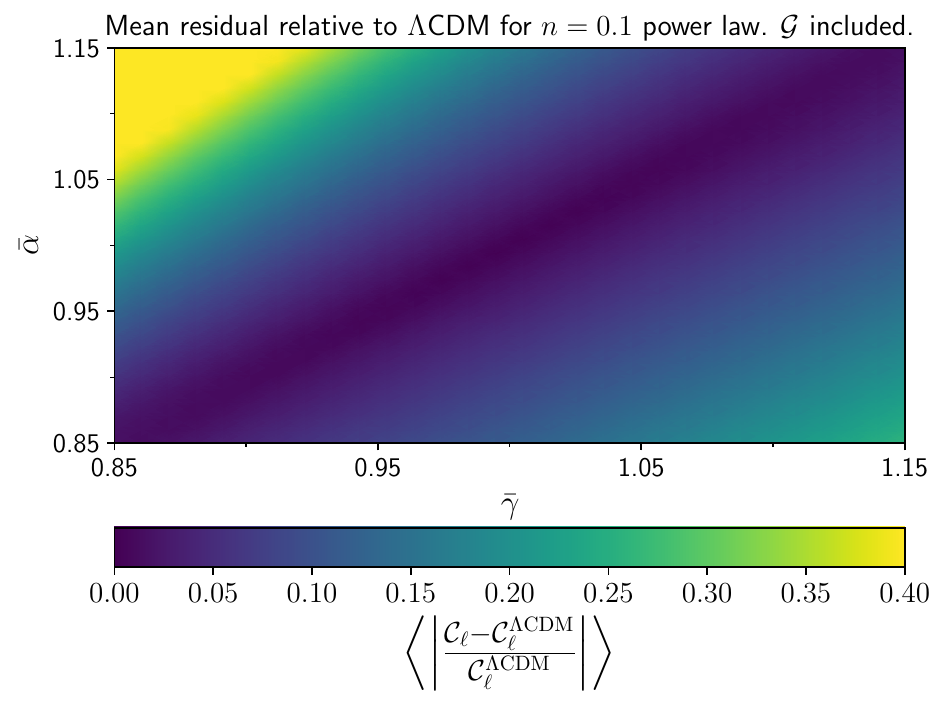}
    \caption{Colour map of the mean absolute residual of $\mathcal{C}_l$ relative to $\Lambda$CDM, averaged over multipoles $l \in \left[2,1001\right]$. Results are displayed as a function of $\bar{\alpha}$ and $\bar{\gamma}$, for the $n = 0.1$ PPNC power law. We have evolved the cosmological background and scalar perturbations using the full set of PPNC equations.}
    \label{fig_mean_residual_0pt1_with_G}
\end{figure}

Perhaps the simplest thing one can do is simply select a power law for $\gamma(a)$ and $\alpha(a)\,$, hold the standard cosmological parameters fixed at their Planck 2018 best-fit values, and then na{\" i}vely compute CMB statistics, for a large set of pairs $\left(\bar{\gamma},\bar{\alpha}\right)\,$. 
Nevertheless, this simple exercise can provide a good deal of insight. The first results of carrying it out are shown by Fig. \ref{fig_mean_residual_0pt1_with_G}, which displays the mean absolute residual, averaged over $l \in \left[2, 1001\right]$, in $\mathcal{C}_l^{TT}$ relative to the best-fit $\Lambda$CDM, for a PPNC power law with $n = 0.1$\,.
It indicates a very clear preference for $\alpha$ and $\gamma$ to be roughly equal\footnote{Note that $\bar{\alpha}$ and $\bar{\gamma}$ being equal means that $\alpha(a) = \gamma(a)$ for all $a$\,, as they have the same power law index $n$, and we are implementing the condition $\gamma(a = 1) = 1$, while $\alpha(a = 1) = 1$ by definition.}, with $\bar{\alpha} \neq \bar{\gamma}$ disfavoured. 
For example, the mean absolute residual is  $80.2\,\%$ for $\left(\bar{\alpha}, \bar{\gamma}\right) = \left(1.15, 0.85\right)$, whereas it is $1.67\,\%$ for $\left(\bar{\alpha}, \bar{\gamma}\right) = \left(0.85, 0.85\right)$, and $1.61\,\%$ for $\left(1.15, 1.15\right)$\,, even though in all of these cases $\vert \bar{\alpha} -1 \vert$ and $\vert \bar{\gamma} - 1 \vert$ are the same.
We find that this behaviour, pushing $\alpha$ and $\gamma$ to be very close to one another for the resultant CMB temperature anisotropies to be observationally viable, persists generically for other choices of $n\,$.

The positive degeneracy between $\bar{\alpha}$ and $\bar{\gamma}$, or equivalently the strong preference for $\bar{\alpha} \approx \bar{\gamma}$, can be explained with reference to the PPNC Friedmann equation Eq. (\ref{eq_ppnc_CLASS_Friedmann}), and the perturbation equations that are used in the PPNC CLASS code, Eqs. (\ref{eq_PPNC_CLASS_slip}-\ref{eq_PPNC_CLASS_momentum}). 
Eq. (\ref {eq_ppnc_CLASS_Friedmann}) is used in the PPNC-CLASS code to evolve the FLRW background. It contains only $\gamma$, and not $\alpha$, which is irrelevant to the background expansion for our choice of $\gamma_c$ time evolution \footnote{In principle, $\alpha$ could enter into the background if one chose to specify the time evolution of $\alpha_c$ rather than $\gamma_c$, and then calculate the evolution of $\gamma_c$ from the integrability condition Eq. (\ref{eq_integrabilitycondition}), which contains $\alpha$.}.
However, the perturbation equations involve both $\alpha$ and $\gamma$, through the interpolating functions $\mu(a, k)$\,, $\mathcal{G}(a, k)$ and $\Sigma(a, k)\,$. These are explicitly dependent on the PPNC parameters and their time derivatives, via their functional forms given by Eq. (\ref{eq_scale_dependent_coupling}).

As the CMB observables depend on the evolution of both the background cosmology, which is sensitive only to $\gamma(a)$, and the metric perturbations, which are sensitive to both $\gamma(a)$ and $\alpha(a)$, it follows that if $\bar{\alpha}$ and $\bar{\gamma}$ are equal, then there must be a cancellation between the effects of the PPNC system's departures from $\Lambda$CDM on the background, and their effects on the perturbations. 
We will show shortly that this is indeed the case. This cancellation is increasingly broken as $\bar{\alpha} - \bar{\gamma}$ is taken further away from zero. 

The combination $\alpha - \gamma$ appears explicitly at only one place in the evolution equations\footnote{It appears in the slip relation, Eq. (\ref{eq_PPNC_CLASS_slip}), but this is just an algebraic constraint, and does not tell us anything about how the perturbations should evolve.} for the perturbations $\Phi$ and $\Psi\,$.
This is in the term $\mathcal{G}\mathcal{H}\Psi$ in the momentum constraint (\ref{eq_PPNC_CLASS_momentum}), which determines the evolution of $\Psi$.
Recall that with our choice of interpolating function, $\mathcal{G}$ has the form
\begin{eqnarray}
    \mathcal{G}(a, k) = \frac{1}{\gamma}\left(\alpha - \gamma + \frac{\mathrm{d}\gamma}{\mathrm{d}\ln{a}}\right)\left[\frac{1}{2} + \frac{1}{2}\,\tanh{\ln{\left(\frac{k}{\mathcal{H}(a)}\right)}} \right]\,.
\end{eqnarray}
In the limit $k \longrightarrow \infty$, corresponding to modes deep inside the horizon, $\mathcal{G}$ is given by $\dfrac{\alpha - \gamma}{\gamma} + \dfrac{\mathrm{d}\ln{\gamma}}{\mathrm{d}\ln{a}}\,$. 
The function $\mathcal{G}$ vanishes at all times and for all perturbation wavenumbers in $\Lambda$CDM. If $\alpha = \gamma$, then $\mathcal{G}$ remains small outside of GR, with contributions to it coming only from $\dfrac{\mathrm{d}\ln{\gamma}}{\mathrm{d}\ln{a}}\,$. 
Therefore, the explicitly non-GR contributions to the perturbation dynamics are suppressed, even if $\gamma = \alpha$ is deviant from unity, and so a $\Lambda$CDM-like CMB power spectrum can be obtained with a suitable set of cosmological parameters.
By contrast, if $\bar{\alpha} \neq \bar{\gamma}$, then the non-GR term $\mathcal{G}\mathcal{H}\Psi$ drives distinct phenomenology in the CMB, which could push it well outside the Planck error bars no matter how much the basic $\Lambda$CDM parameters are adjusted.

\begin{figure}
    \centering
    \includegraphics[width=0.75\linewidth]{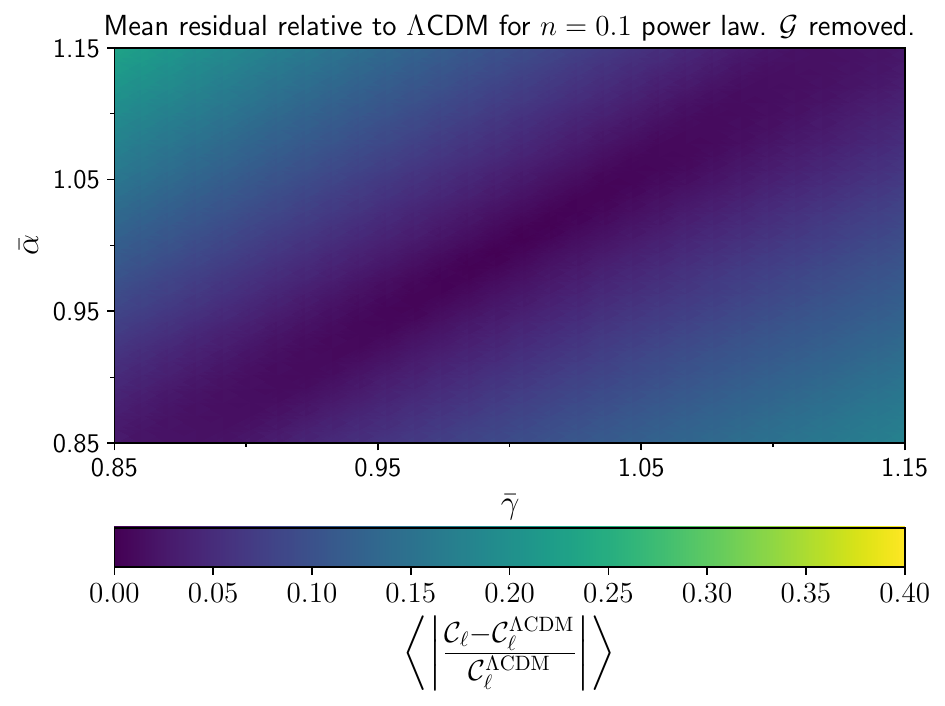}
    \caption{The same as Fig. \ref{fig_mean_residual_0pt1_with_G}, but where the term $\mathcal{G}\mathcal{H}\Psi$ has been artificially removed from the momentum constraint equation, in order to demonstrate the importance of the PPNC $\mathcal{G}$ function in driving the requirement that $\bar{\alpha} \approx \bar{\gamma}$\,.}
    \label{fig_mean_residual_0pt1_no_G}
\end{figure}

\begin{figure}
    \centering
    \includegraphics[width=\linewidth]{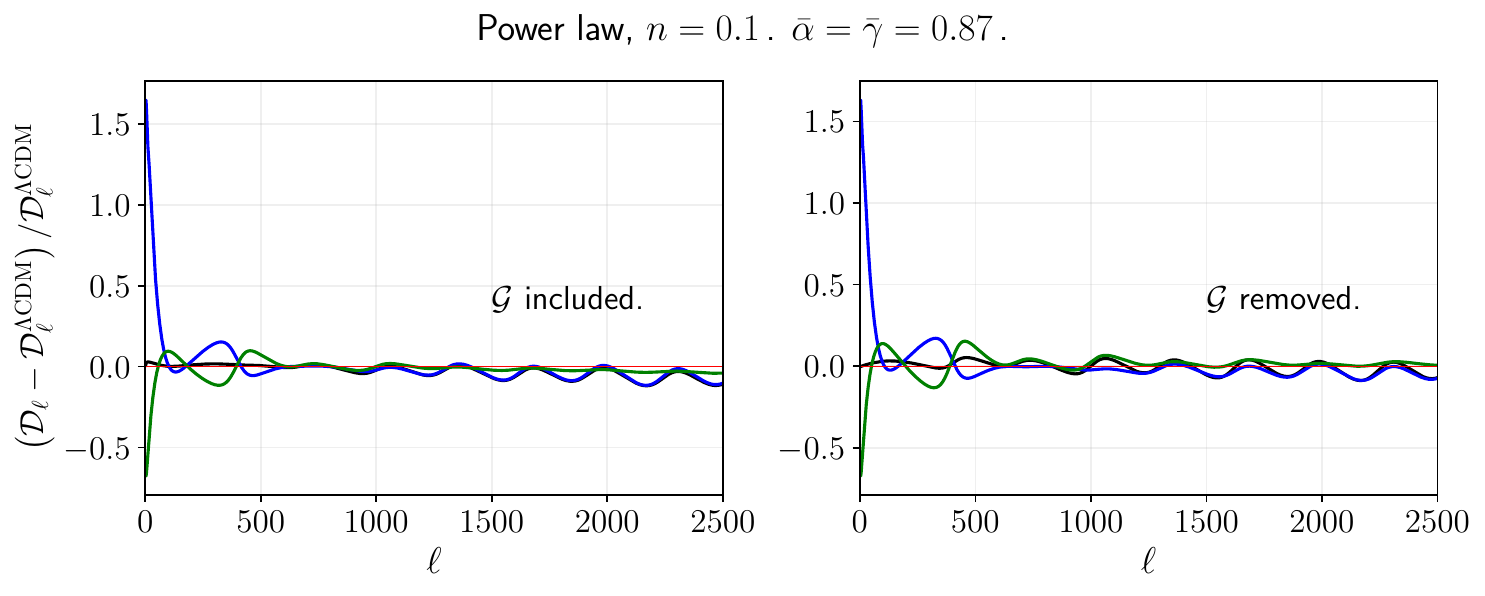}
    \includegraphics[width=\linewidth]{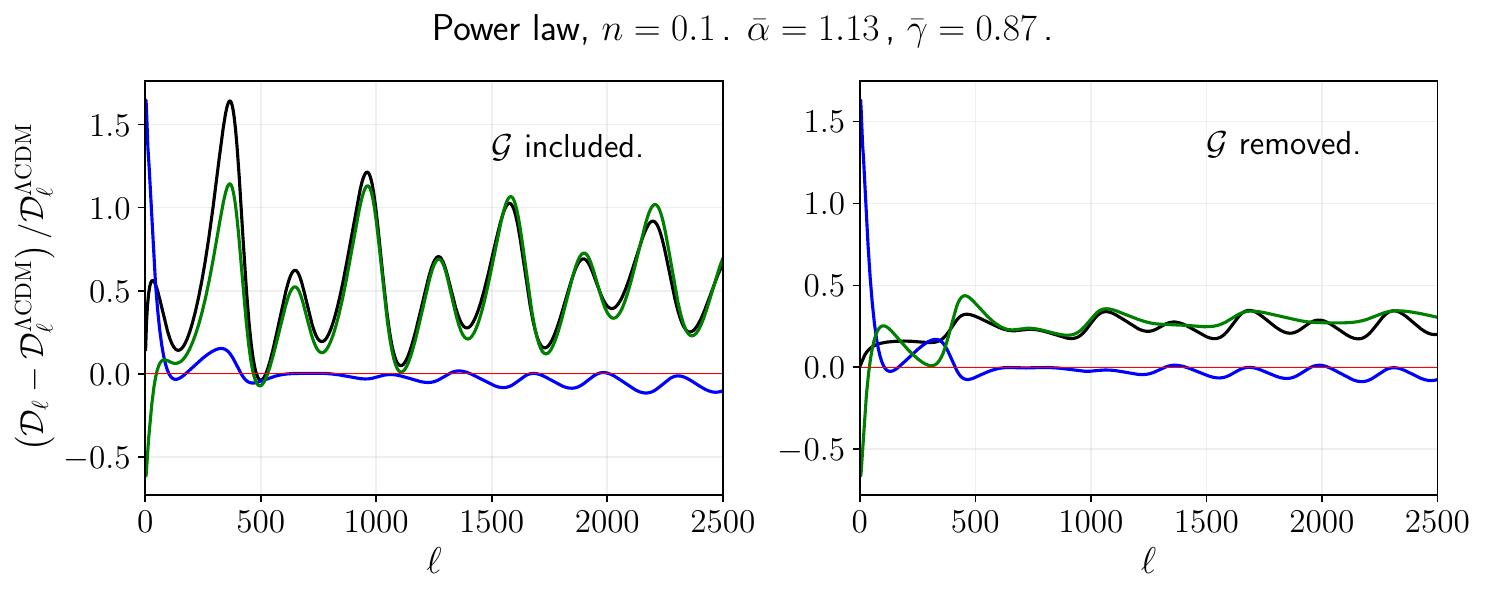}
    \caption{The relative effects of PPNC background and perturbation evolutions on the total angular power spectrum, compared to $\Lambda$CDM. In each case, we have computed results first using the full PPNC equations, and then with the $\mathcal{G}$ PPNC function artificially removed. 
    Results are shown for a power law with $n = 0.1$\,, where $\bar{\alpha} = \bar{\gamma} = 0.87\,$ in the top row, and in the bottom row $\bar{\alpha} = 1.13$ and $\bar{\gamma} = 0.87\,$. 
    Black curves show the results for a full PPNC evolution, green when the PPNC perturbation equations are used but the background expansion is $\Lambda$CDM, and blue when the background is evolved using Eq. (\ref{eq_ppnc_CLASS_Friedmann}), but perturbations are evolved according to $\Lambda$CDM equations.}
    \label{fig_total_background_vs_perturbations}
\end{figure}

\begin{figure}
    \centering
    \includegraphics[width=\linewidth]{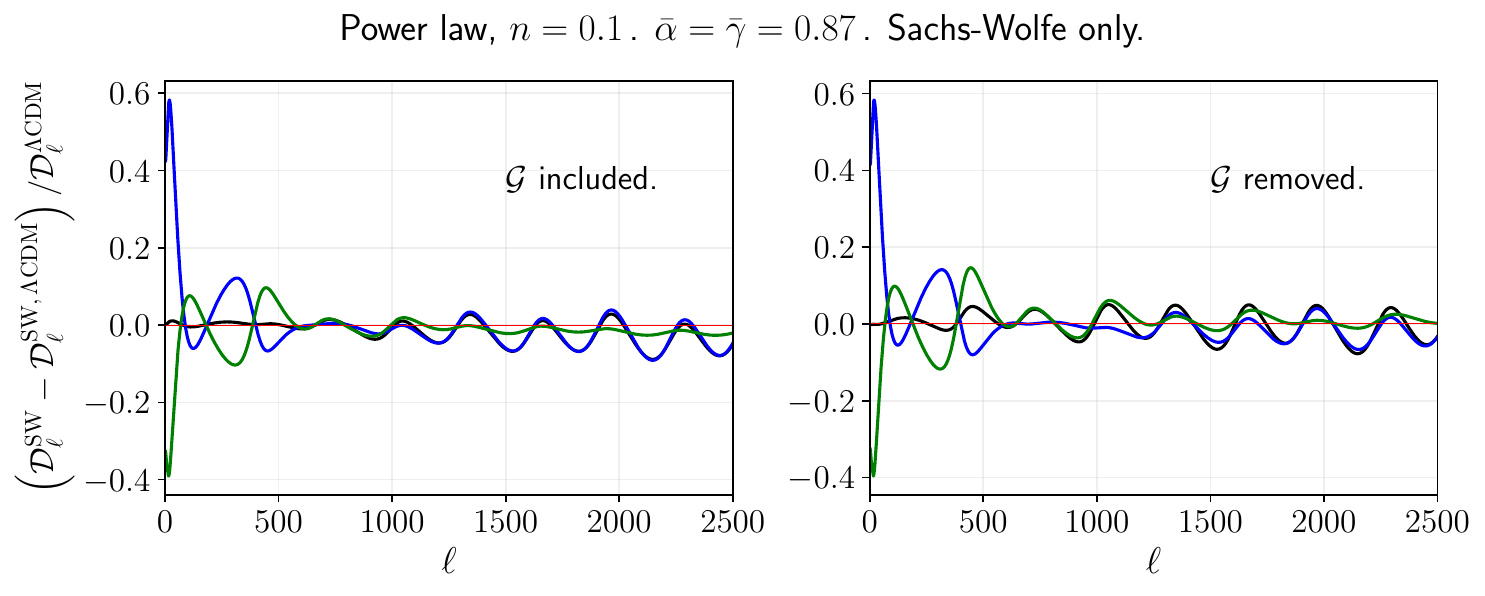}
    \includegraphics[width=\linewidth]{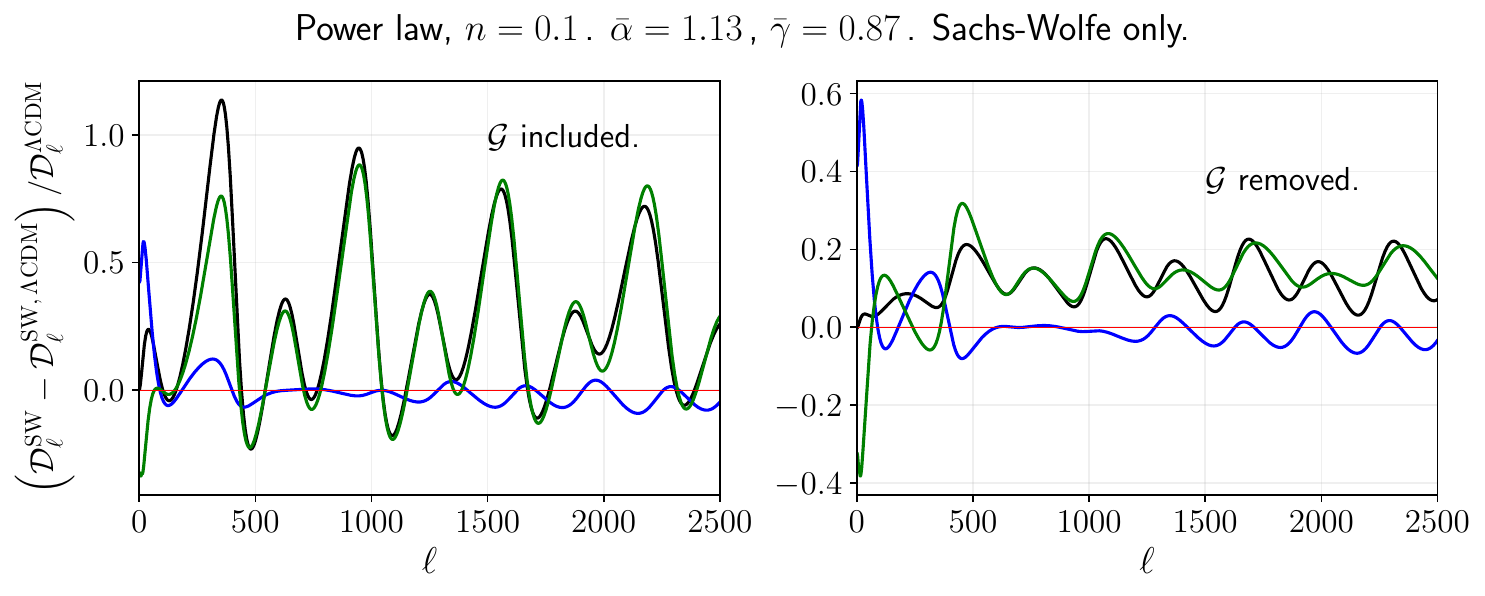}
    \caption{The same as Fig. \ref{fig_total_background_vs_perturbations}, but where instead of the total angular power spectrum $\mathcal{C}_l^{\rm tot}$, we have computed only the Sachs-Wolfe term $\mathcal{C}_l^{\rm SW}$\,.}
    \label{fig_SW_background_vs_perturbations}
\end{figure}


Fig. \ref{fig_mean_residual_0pt1_with_G} suggests that there ought to be an approximate equality between $\gamma(a)$ and $\alpha(a)\,$. Let us now try to identify what the physical reason for that is. 
First, we consider what happens if one artificially removes the $\mathcal{G}\mathcal{H}\Psi$ term from Eq. (\ref{eq_PPNC_CLASS_momentum}), and then re-runs the PPNC CLASS code without that term, in order to calculate the mean absolute residual in $\mathcal{C}_l$ relative to $\Lambda$CDM, as we previously calculated in Fig. \ref{fig_mean_residual_0pt1_with_G} for the true PPNC equations with this term retained. 
Fig. \ref{fig_mean_residual_0pt1_no_G} shows that the preference for $\bar{\alpha} = \bar{\gamma}$ is then substantially reduced, with a maximum value of $\left\langle \vert\left(\mathcal{C}_l - \mathcal{C}_l^{\Lambda{\rm CDM}}\right)/\mathcal{C}_l^{\Lambda{\rm CDM}}\vert\right\rangle$ of $24.0\%$ in the corner $\left(\bar{\alpha}, \bar{\gamma}\right) = \left(1.15, 0.85\right)\,$. 

This suggests that the PPNC $\mathcal{G}$ function, if it is large, can drive substantially different behaviour from $\Lambda$CDM at the level of the perturbations, leading to stark differences in the statistical properties of the temperature anisotropies. 
However, it does not tell us anything about which of the physical phenomena that produce $\mathcal{C}_l^{TT}$ (i.e. the CMB source functions) that term is affecting, or about how the modifications to the perturbations compare to the modifications to the background expansion.



To begin to investigate this, we consider the effects on $\mathcal{C}_l^{TT}$ of the PPNC background and perturbation evolution, in order to show that there is indeed a cancellation that occurs for $\bar{\alpha} = \bar{\gamma}$\,, and that this cancellation is worse for $\bar{\alpha} \neq \bar{\gamma}$ (driving the results much further from the Planck best-fit $\Lambda$CDM and making them disfavoured by the data). 
We will also show that the non-cancellation in the unequal case is driven primarily by the presence of the $\mathcal{G}$ PPNC function.
To do this, we calculate the angular power spectrum for a given choice of $n$, $\bar{\alpha}$ and $\bar{\gamma}$ three times: once where the entire system of PPNC equations is used, once where the perturbations are evolved according to the PPNC perturbation equations but the background is evolved with the $\Lambda$CDM Friedmann equations, and once where the perturbations follow $\Lambda$CDM equations but the background follows the PPNC Friedmann equations. 
We then do this again, but with the term proportional to $\mathcal{G}$ artificially removed from the momentum constraint equation.

We show the results of doing this first in Fig. \ref{fig_total_background_vs_perturbations} for the total $\mathcal{C}_l\,$, and then in Fig. \ref{fig_SW_background_vs_perturbations} for the Sachs-Wolfe source term.
The Sachs-Wolfe and Doppler terms are the dominant source functions driving the constraint between $\bar{\alpha}$ and $\bar{\gamma}$\,. This is because the power laws for the PPNC parameters can substantially shift their values from unity at $a_{LS}$, as displayed in Fig. \ref{fig_gamma_of_a_power_laws}. Therefore, they directly affect the metric and velocity perturbations at last scattering. For the sake of brevity, we do not display the Doppler term, as the phenomenological behaviour is the same as in the Sachs-Wolfe case in Fig. \ref{fig_SW_background_vs_perturbations}.

\begin{figure}
   \centering
   \includegraphics[width=\linewidth]{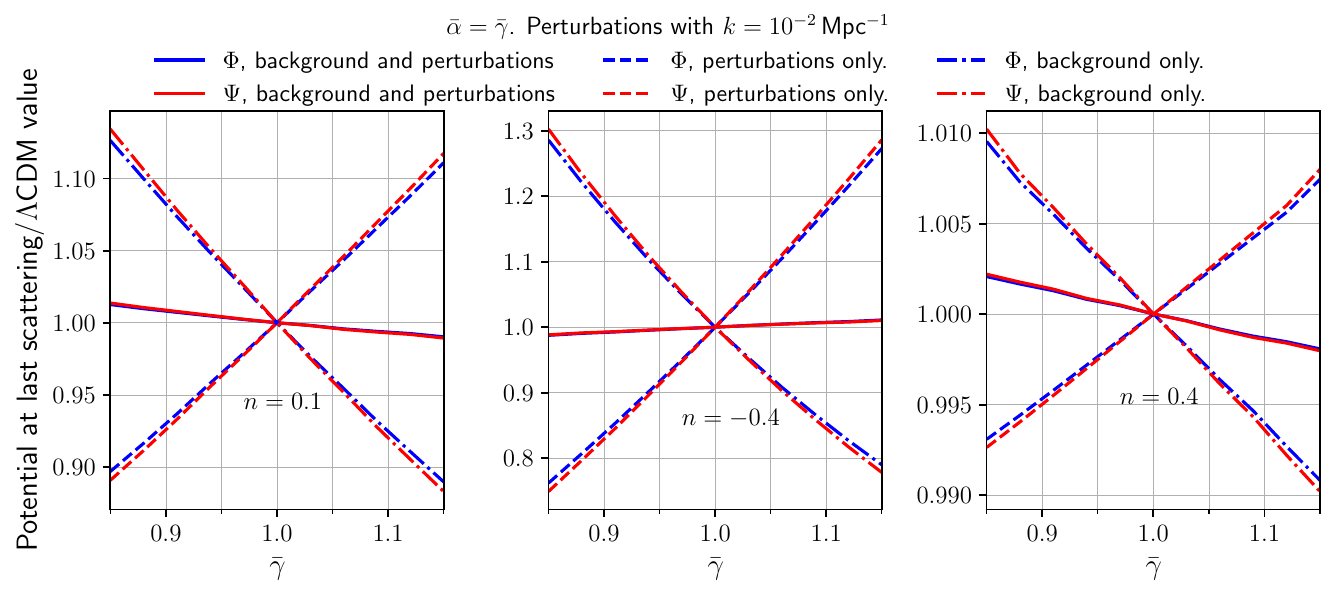}
   \caption{Metric perturbations $\Phi$ and $\Psi$ at last scattering, as a function of $\bar{\gamma}$, for $k = 10^{-2} \,{\rm Mpc}^{-1}$\,, which corresponds to a mode deep inside the horizon at last scattering. They are evaluated for the case where the full PPNC equations are used, for the case where only the perturbations are governed by PPNC equations (and the background by $\Lambda$CDM equations), and for the case where only the background is governed by PPNC equations (and the perturbations by $\Lambda$CDM equations). In these plots, we have set $\bar{\alpha} = \bar{\gamma}$, and have normalised each perturbation by its $\Lambda$CDM value at $a_{LS}$. Results are shown for the $n = 0.1$, $n = -0.4$ and $n = 0.4$ power laws.}
   \label{fig_Phi_Psi_LS_subhorizon}
\end{figure}

In the case $\left(\bar{\alpha}, \bar{\gamma}\right) = \left(1.13, 0.87\right)$, there is virtually no cancellation. Instead, the effects of the background and perturbations act on $\mathcal{C}_l^{\rm tot}$ and $\mathcal{C}_l^{\rm SW}$ (and indeed on $\mathcal{C}_l^{\rm Dop}\,$, although we have not displayed it here) in the same direction to further amplify the acoustic peaks. The $l \gtrsim 500$ region is dominated by the PPNC perturbations, which are highly deviant from $\Lambda$CDM.
However, we see that the non-cancellation in the case $\bar{\alpha} \neq \bar{\gamma}$ is much less severe if we artificially remove the $\mathcal{G}$ term from the PPNC momentum constraint equation, as shown in the right half of each quadrant. For $\bar{\alpha} = \bar{\gamma}\,$, removing $\mathcal{G}$ has very little effect, as expected.

The presence of the function $\mathcal{G}$ gives a good explanation as to why a parameterised post-Newtonian cosmology with $\alpha \neq \gamma$ should provide a poor fit to the CMB. However, we have not yet established the converse: why $\alpha \approx \gamma$ provides a good fit, even if that shared value can be rather deviant from its GR value of unity. 
To understand why this is the case, consider again the Sachs-Wolfe and Doppler source terms. These are sensitive to the values of both the scalar metric perturbations $\Phi$ and $\Psi$ themselves at last scattering, through Eqs. (\ref{eq_source_term_density}-\ref{eq_source_term_Doppler}), with the $v_b$ that enters the Doppler source function ultimately determined by $\Phi(a_{\rm LS})$ through the Euler equation (\ref{eq_CPT_Euler_eqn}), and $\Phi(a_{\rm LS})$ itself depending on $\Psi(a_{\rm LS})$ by the slip relation (\ref{eq_PPNC_CLASS_slip}).
Thus, deviations in $\mathcal{C}_l^{\rm SW}$ and $\mathcal{C}_l^{\rm Dop}$ from their $\Lambda$CDM predictions reflect the modifications to GR encoded in our equations of motion, evaluated at $a_{\rm LS}\,$.
This is seen directly by evaluating the metric perturbations $\Phi(a_{\rm LS})$ and $\Psi(a_{\rm LS})$ for a subhorizon mode\footnote{The horizon scale at last scattering $k_H^{\rm LS} = \mathcal{H}(a_{\rm LS})$ depends on the PPNC parameters and power-law index chosen, but it is typically around $k_H^{\rm LS} \sim 4 \ {\rm or} \ 5 \times 10^{-3}\,{\rm Mpc}^{-1}$ (e.g. $k_H^{\rm LS} = 4.81 \times 10^{-3}\,{\rm Mpc}^{-1}$ in $\Lambda$CDM, and $k_H^{\rm LS} = 4.51 \times 10^{-3}\,{\rm Mpc}^{-1}$ for a PPNC power law with $n = -0.4$ and $\bar{\gamma} = 0.85\,$).}. The ratio between them and their $\Lambda$CDM counterparts is shown in Fig. \ref{fig_Phi_Psi_LS_subhorizon}, as a function of $\bar{\gamma} = \bar{\alpha}$, for the $n = 0.1$, $-0.4$ and $0.4$ power laws\footnote{The cancellation between background and perturbations is phenomenologically very similar for superhorizon modes.}. 

If $\alpha = \gamma$, then the PPNC modifications to the background and perturbation equations have roughly equal and opposite effects on the gravitational potentials at last scattering. 
The reason why the effects are opposite is that if $\gamma > 1$, then the coupling $\mu$ to matter in the perturbation equation (\ref{eq_PPNC_CLASS_momentum}) is larger, and so the growth of perturbations is enhanced. 

However, $\gamma > 1$ means that for the same total matter density $\bar{\rho}_m$, the expansion rate $H$ is faster, according to Eq. (\ref{eq_parametrisedfriedmanneqns1}). The growth of perturbations is therefore suppressed.

\begin{figure}
    \centering
    \includegraphics[width=0.8\linewidth]{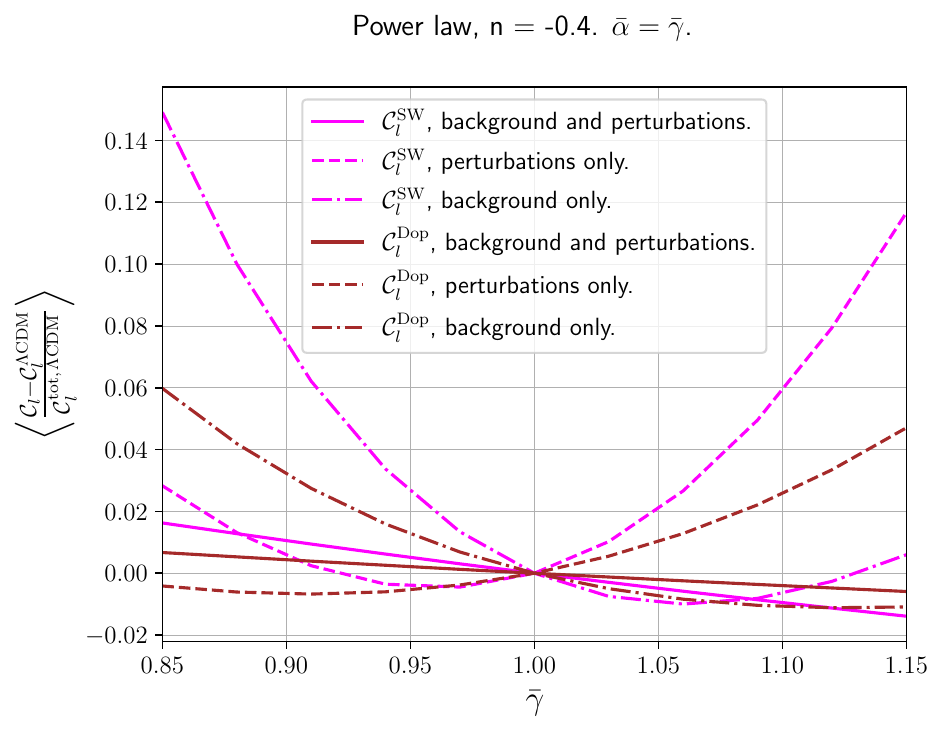}
    \caption{The average residual, relative to $\Lambda$CDM, in the Sachs-Wolfe and Doppler terms, for $l \in \left[2, 1001\right]$, as a function of $\bar{\gamma}$ (with $\bar{\alpha} = \bar{\gamma}$). They are evaluated for the case where the full PPNC equations are used, for the case where only the perturbations are governed by PPNC equations (and the background by $\Lambda$CDM equations), and for the case where only the background is governed by PPNC equations (and the perturbations by $\Lambda$CDM equations). Results are displayed for the $n = -0.4$ power law.}
    \label{fig_effect_on_SW_Doppler_m0pt4}
\end{figure}

\begin{figure}
    \centering
    \includegraphics[width=0.8\linewidth]{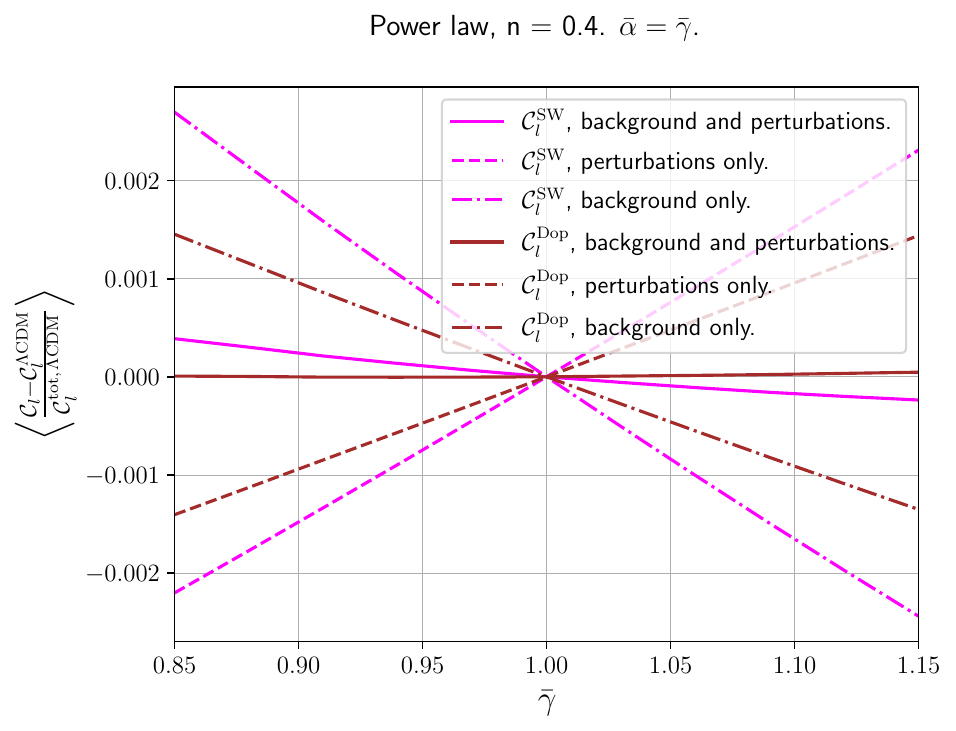}
    \caption{The same as Fig. \ref{fig_effect_on_SW_Doppler_m0pt4}, but for the $n = +0.4$ power law.}
    \label{fig_effect_on_SW_Doppler_0pt4}
\end{figure}

Figs. \ref{fig_effect_on_SW_Doppler_m0pt4} and \ref{fig_effect_on_SW_Doppler_0pt4} display the effects of the PPNC modifications to the background and perturbations in the mean absolute deviation of $\mathcal{C}_l$ from $\Lambda$CDM, in both the Sachs-Wolfe and Doppler terms.
They are shown as a function of $\bar{\gamma}$, and we have set $\bar{\alpha} = \bar{\gamma}$\,. Again, we consider the $-0.4$ and $0.4$ power laws, while noting that all departures from $\Lambda$CDM are far smaller for $n = 0.4$, for which even at the extremal cases $\vert \bar{\alpha} - 1 \vert = 0.15$ or $\vert \bar{\gamma} - 1 \vert = 0.15$ , both $\alpha$ and $\gamma$ are within $0.3\%$ of unity by last scattering. 

\begin{figure}
    \centering
    \includegraphics[width=0.65\linewidth]{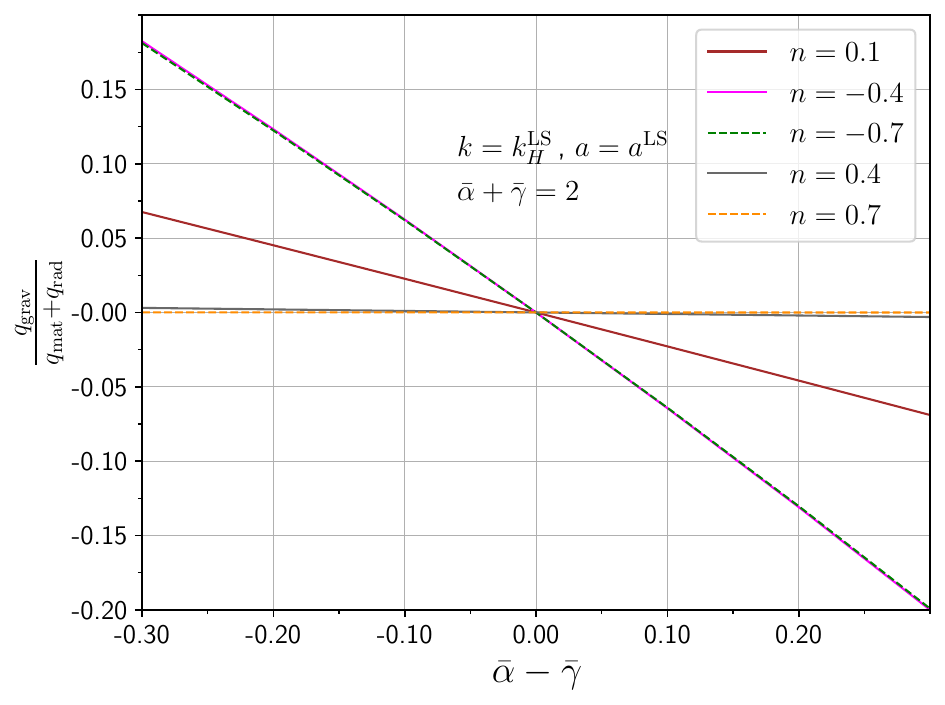}
    \caption{Ratio at last scattering of the purely gravitational term $q_{\rm grav} = \mathcal{G}\mathcal{H}\Psi$ in the momentum constraint equation to the sum of the conventional momentum-like terms arising from matter ($q_{\rm mat} = 4\pi G a^2  \mu \left[\bar{\rho}_c v_c + \bar{\rho}_b v_b\right]$) and radiation ($q_{\rm rad} = 4\pi G a^2 \left[\left(\bar{\rho}_{r}+\bar{p}_{r}\right) v_{r} + \left(\bar{\rho}_{\rm ur}+\bar{p}_{\rm ur}\right) v_{\rm ur}\right]$). The wavelength of the perturbations is set equal to the Hubble horizon at last scattering. Results are displayed as a function of the difference between the average PPNC parameters, $\bar{\alpha}-\bar{\gamma}$\,, for five different power law indices $n\,$. We have set $\bar{\alpha}+\bar{\gamma}$ equal to 2.}
    \label{fig_G_ppnc_last_scattering}
\end{figure}

Because of the cancellation between the non-GR effects of PPNC on the FLRW background and scalar perturbations in both the Sachs-Wolfe and Doppler terms, we find that $\bar{\gamma}$ itself is allowed to be quite substantially deviant from unity, as long as it is closely tracked by $\bar{\alpha}\,$.
We will see in Section \ref{sec:planck_results} that it is the difference between the PPNC parameters that is strongly constrained by the data, not their sum or average. 

In Fig. \ref{fig_G_ppnc_last_scattering}, we demonstrate explicitly that the cancellations seen in Figs. \ref{fig_effect_on_SW_Doppler_m0pt4} and \ref{fig_effect_on_SW_Doppler_0pt4} are primarily a result of the term $\mathcal{G}\mathcal{H}\Psi$ in the momentum constraint being small at $a_{\rm LS}$ when $\bar{\alpha} = \bar{\gamma}$\,. 
The contribution of $\mathcal{G}\mathcal{H}\Psi$ to the evolution of horizon-scale perturbations at last scattering is most sensitive to $\bar{\alpha} - \bar{\gamma}$ for $n < 0$, and very insensitive for $n \gtrsim 0.4\,$. 
This reiterates the point we have made, that the requirement $\alpha \approx \gamma$ is most stringent if the PPNC parameters are allowed to be noticeably deviant from GR until well after last scattering, whereas there is very little constraining power in the data if all the deviations from GR are happening in the very early Universe during radiation domination.

\begin{figure}
\centering
\hspace{-1cm}
\includegraphics[width=1.05\linewidth]{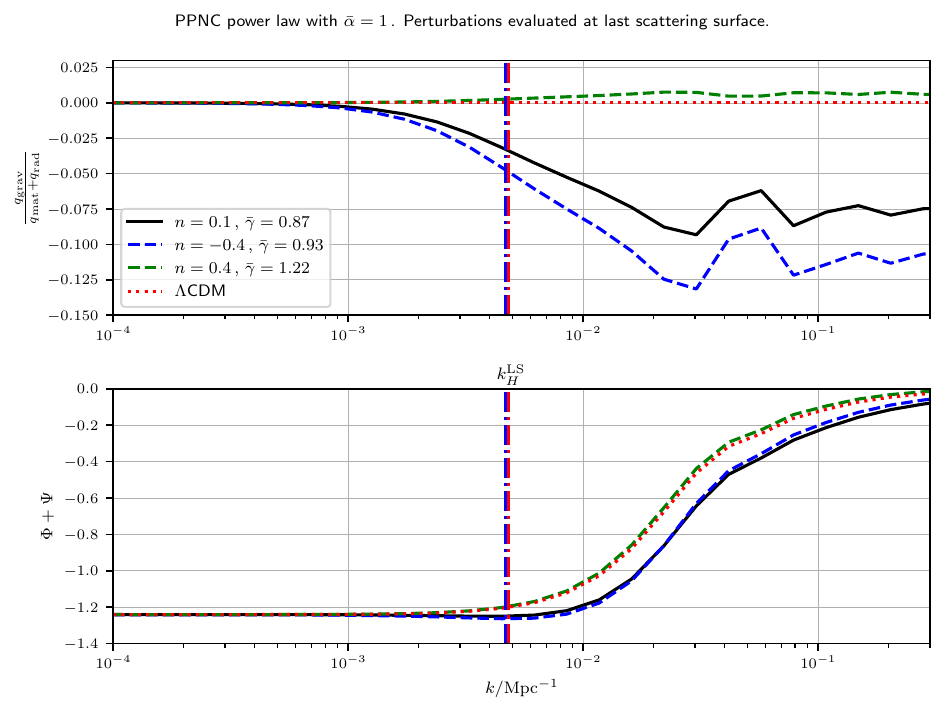}
    \caption{Upper panel: ratio $q_{\rm grav}/\left(q_{\rm mat} + q_{\rm rad}\right)$, as a function of $k$, evaluated at last scattering, where $q_{\rm grav}\,$, $q_{\rm mat}$ and $q_{\rm rad}$ are as defined in Fig. \ref{fig_G_ppnc_last_scattering}. Lower panel: $\Phi + \Psi$ as a function of $k$, also at last scattering. The results are shown for $\Lambda$CDM and for the $n = 0.1$, $n = -0.4$ and $n = 0.4$ PPNC power laws. For the PPNC cases, we have used $\bar{\alpha} =1\,$. The horizon scale at last scattering for each case is also displayed on both plots.}
    \label{fig_G_last_scattering_k}
\end{figure}

Finally, Fig. \ref{fig_G_last_scattering_k} shows the persistence of the effect of the purely non-GR term $\mathcal{G}\mathcal{H}\Psi$ across a wide range of scales at last scattering, for $\bar{\gamma} \neq \bar{\alpha} = 1\,$. It affects all $k \gtrsim k_H$, at the level of several percent in both the momentum constraint equation and the metric perturbations $\Phi$ and $\Psi$ themselves, as demonstrated by the upper and lower plots respectively. 
That effect on the metric perturbations is a roughly uniform amplification of the Weyl potential $\Phi + \Psi$ across sub-horizon $k$ scales during matter domination\footnote{The modified damping that $\mathcal{G}\mathcal{H}\Psi$ introduces in Eq. (\ref{eq_PPNC_CLASS_momentum}) becomes significant after matter-radiation equality, as $\alpha(a)$ and $\gamma(a)$ couple only to matter.}, when $\bar{\gamma} < \bar{\alpha} = 1\,$\,, as for $\bar{\gamma} < \bar{\alpha}$, we have $\mathcal{G}(a_{\rm LS}, k_H^{\rm LS}) > 0\,$. 

One can interpret the required (near-) equality between $\alpha(a)$ and $\gamma(a)$ as requiring that $\Phi$ and $\Psi$ are equal on all scales and at all times, except in the presence of anisotropic stress due to neutrinos, as per Eq. (\ref{eq_PPNC_CLASS_slip}), because we have already set the superhorizon limit of the slip $\Sigma$ to zero. Hence, if $\Sigma$ vanishes on both large and small scales, then it must vanish on all scales due to the form (\ref{eq_scale_dependent_coupling}) of its interpolation in $k\,$. 
The equality of the Newtonian gauge perturbations is a well-known feature of GR (see e.g. \cite{Clifton_2012,Bertschinger_2006}). It appears that one of our strongest constraints on deviations from GR, therefore, will be that they should approximately retain this property.

We can summarise our conclusions in this section as the following:
    \begin{enumerate}
        \item the $\mathcal{G}$ function drives the preference for $\bar{\alpha} = \bar{\gamma}$ as it becomes large if they are not equal.
        \item If $\bar{\alpha}$ and $\bar{\gamma}$ are equal, then there can be a cancellation between the effects of the background (which only involves $\alpha$) and the perturbations (which involve both $\gamma$ and $\alpha$), so that the overall CMB produced is consistent with the observed spectra.
        \item If $\bar{\alpha}$ and $\bar{\gamma}$ are not equal, then the background-perturbations cancellation does not happen, because the $\mathcal{G}$ function changes the evolution of the perturbations substantially. Hence, the perturbations can no longer cancel off the background.
    \end{enumerate}

Let us now focus on how parameterised post-Newtonian cosmology affects the FLRW background expansion.

\subsection{FLRW background expansion and the acoustic peaks}\label{subsec:gamma_h0_omc}

\begin{figure}
    \centering
    \includegraphics[width=0.75\linewidth]{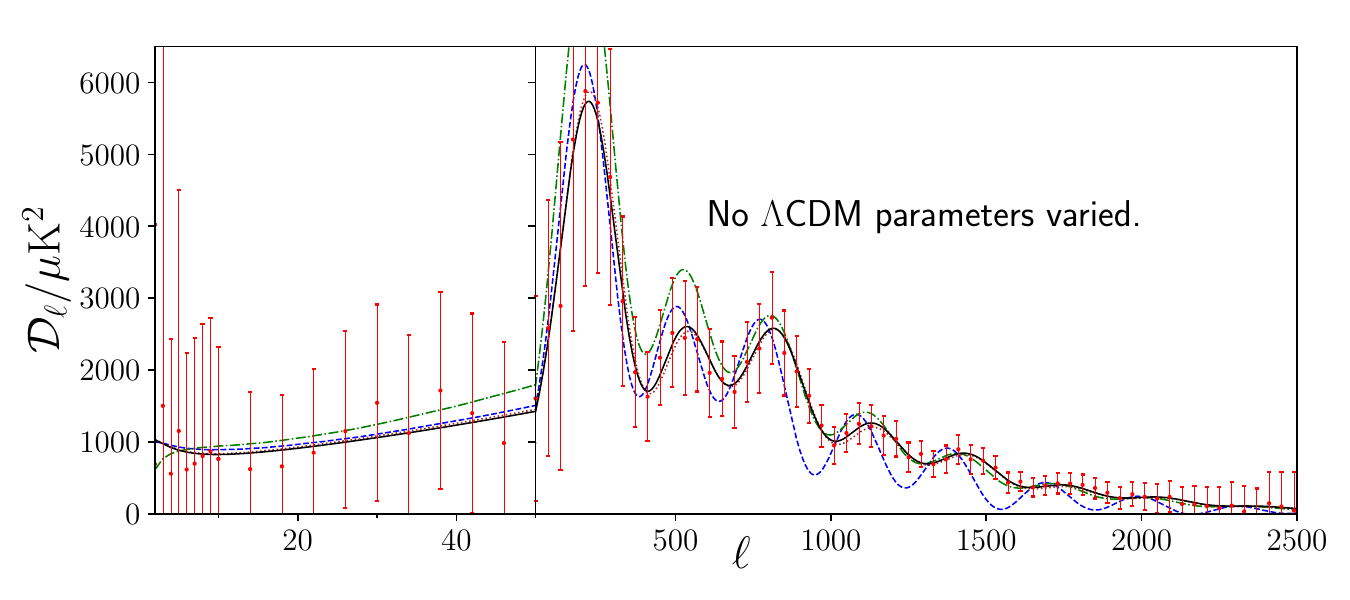}
    \includegraphics[width=0.75\linewidth]{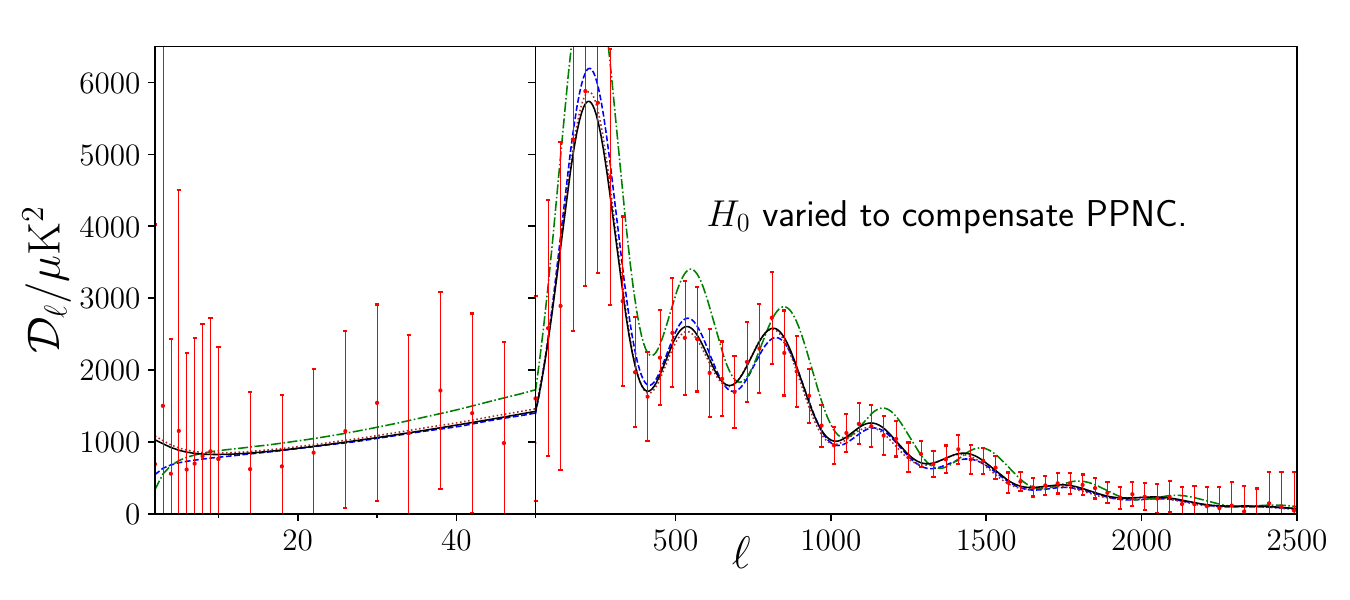}
    \includegraphics[width=0.75\linewidth]{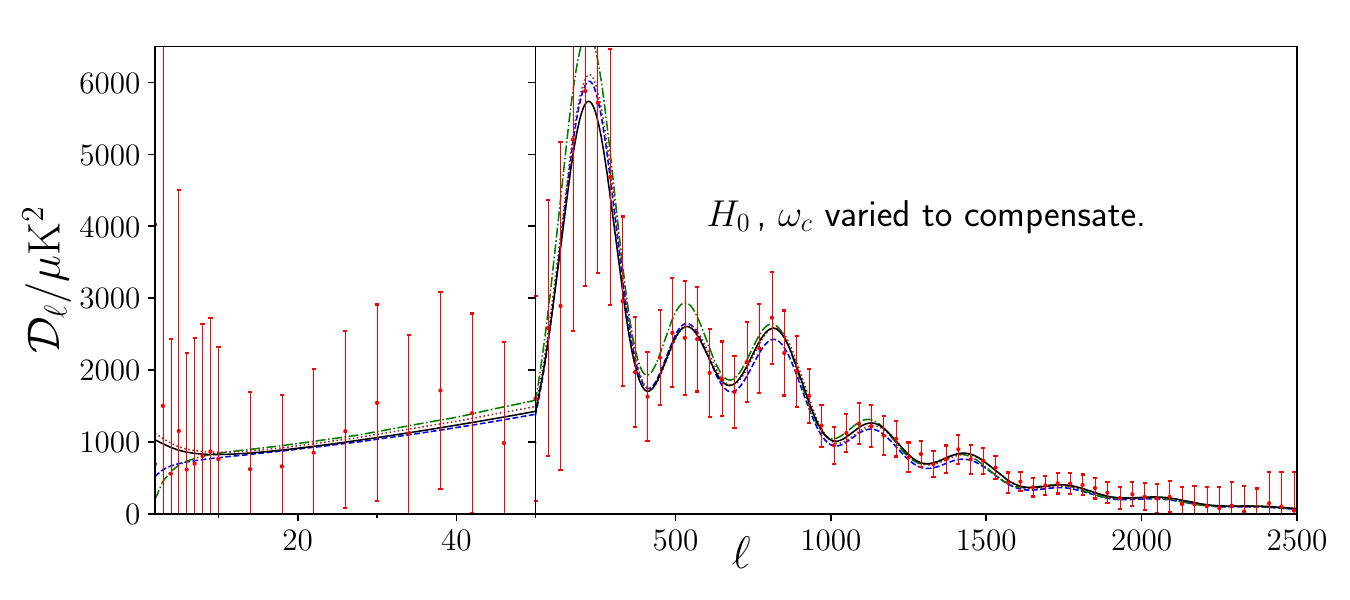}
    \includegraphics[width=0.75\linewidth]{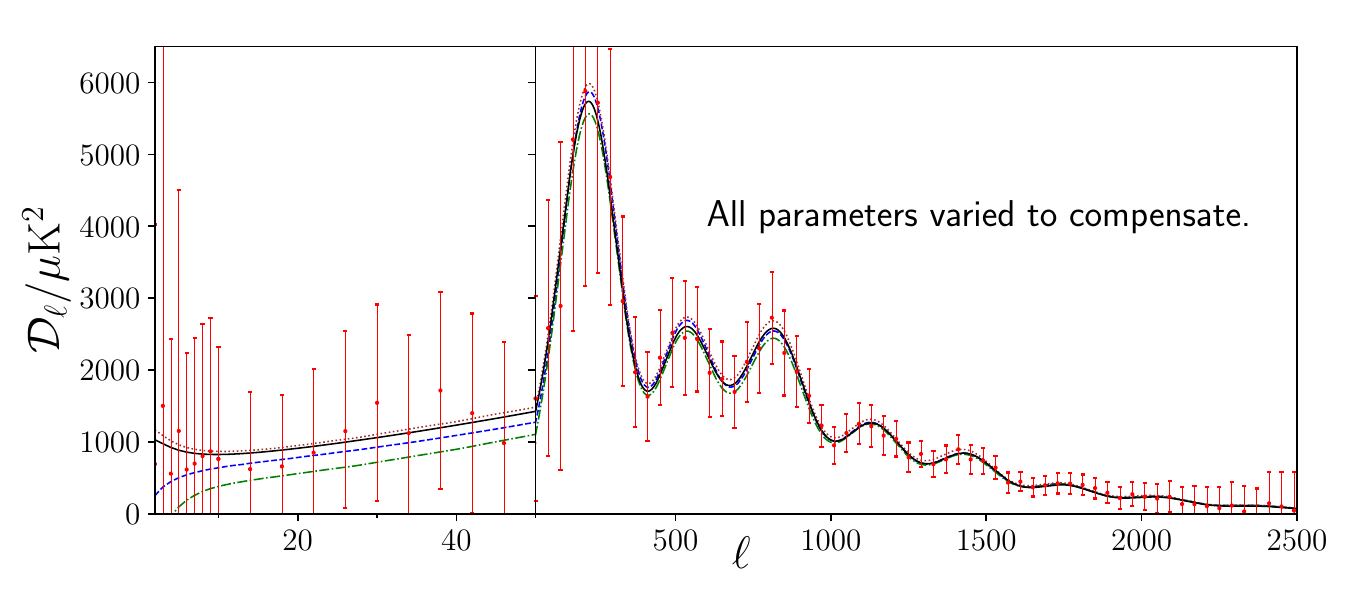}
    \caption{$\mathcal{D}_l^{TT}$ for the $n = 0.1$ (blue), $-0.4$ (green) and $0.4$ (brown) PPNC power laws, with their best-fit values of $\bar{\alpha}$ and $\bar{\gamma}$. 
    We have also shown the Planck 2018 data points (red), and the corresponding best-fit $\Lambda$CDM curve (black). In the first plot, all the basic $\Lambda$CDM cosmological parameters are fixed. Then in the remaining three, they are consecutively adjusted (first $H_0$\,, then $H_0$ and $\omega_c$\,, then all the other standard parameters.).
    The PPNC curves have all been shifted by $\mathcal{D}_l^{\rm PPNC} \longrightarrow \mathcal{D}_l^{\Lambda {\rm CDM}} + 10\left(\mathcal{D}_l^{\rm PPNC} - \mathcal{D}_l^{\Lambda {\rm CDM}}\right)$ in order to make the discrepancies visible, and the Planck error bars enlarged by a factor of 5.}
    \label{fig_vary_parameters}
\end{figure}

The most immediate effect of our CMB calculations is the requirement that $\bar{\alpha} \approx \bar{\gamma}$\,, for values of $n$ that might be considered more physically viable (i.e. those $\lesssim 0.25$ that do not push all the modifications to gravity deep into the radiation era).
There is still clearly some remaining constraining power: not all sets of cosmological parameters $\left\lbrace H_0, \omega_c, \omega_b, \tau, A_s, n_s, Y_{\rm P}\right\rbrace$ provide equally good fits to the data. 
In this section, we will show that we do not expect the constraints on all of those remaining parameters (the standard $\Lambda$CDM cosmological parameters plus $Y_{\rm P}$, which must be included, as mentioned in Section \ref{subsec:mcmc}) to be independent from $\bar{\alpha}$ and $\bar{\gamma}$.
We expect the most significant degeneracies to be between the PPNC parameters $\bar{\alpha}$ and $\bar{\gamma}$ and the parameters $H_0$ and $\omega_c\,$. This is because those are the two parameters in the $\Lambda$CDM concordance cosmology which most strongly affect the heights and locations of the acoustic peaks, as we explained in Section \ref{subsec:CMB}.

Before we go on, we stress again that the FLRW expansion is sensitive to $\gamma(a)$ but not $\alpha(a)$, in our implementation.
Therefore, what we are really dealing with are the competing effects on the cosmological background of $\bar{\gamma}$ and $H_0$, and the competing effects of $\bar{\gamma}$ and $\omega_c$. 
The requirement $\bar{\alpha} \approx \bar{\gamma}$, which is required by the evolution of the gravitational perturbations, as discussed in detail above, then means that any expected degeneracies between $\bar{\gamma}$ and $H_0$, and between $\bar{\gamma}$ and $\omega_c$, give rise to roughly the same degeneracies between $\bar{\alpha}$ and $H_0$, and between $\bar{\alpha}$ and $\omega_c$\,.
This can be seen through the effect of varying these parameters on the temperature anisotropies themselves, as displayed in Fig. \ref{fig_vary_parameters}. 
We show $\mathcal{C}_l^{\rm TT}$ for the $n = 0.1$, $n = -0.4$ and $n = 0.4$ PPNC power laws.
First, we keep all the standard cosmological parameters at their $\Lambda$CDM Planck best-fit values. Then we successively adjust the remaining cosmological parameters, in order to compensate the effect of varying $\gamma$ and hence improve the overall fit to the Planck data: in the second subplot, $H_0$ has been adjusted for the PPNC curves to its best-fit value for each power law, but all the other basic parameters have been held fixed. 
In the third, both $H_0$ and $\omega_c$ have been adjusted, in the fourth $H_0$, $\omega_c$ and $A_s$\,, and in the fifth $H_0$, $\omega_c$, $A_s$ and $n_s\,$. In the sixth, all the parameters have been adjusted (i.e. those above, plus $\omega_b$, $\tau_{\rm reio}$ and $Y_{\rm P}\,$). 

\begin{figure}
    \centering
    \hspace{-2cm}
    \includegraphics[width=1.1\linewidth]{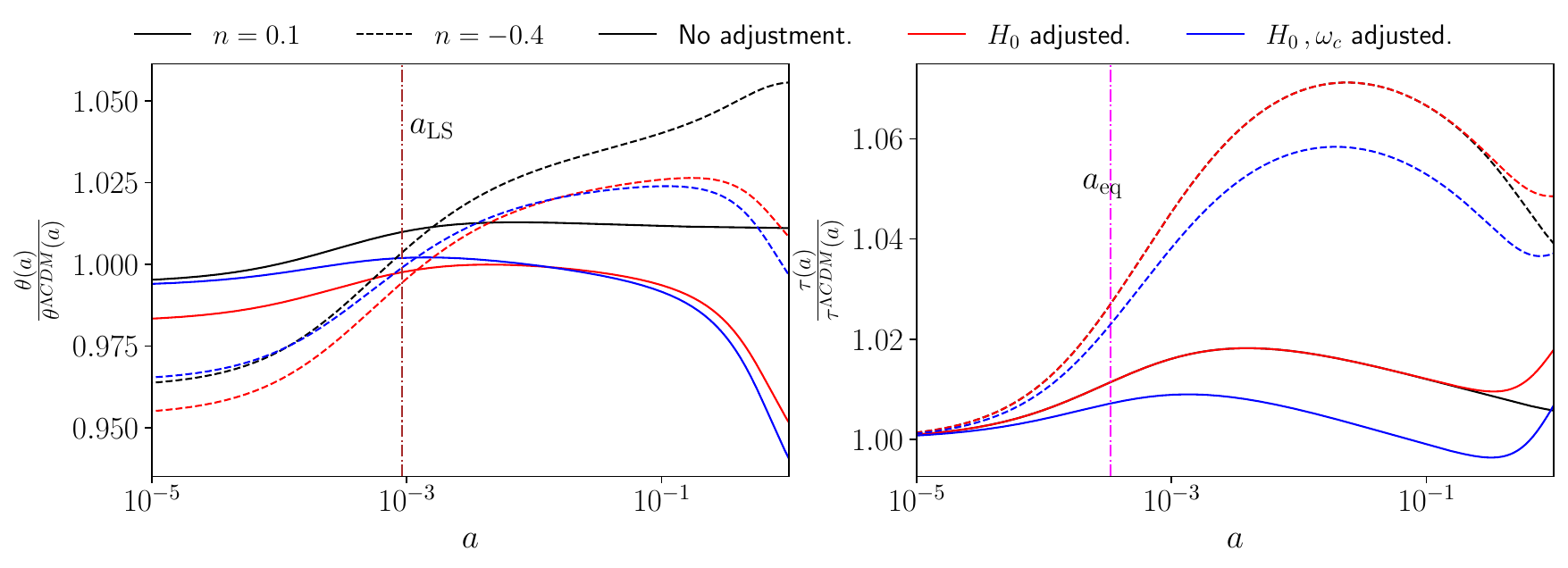}
    \caption{Comparison of the effect on the cosmological background expansion of varying $\bar{\gamma}$ (i.e. changing gravity) while holding all the standard $\Lambda$CDM parameters fixed (black), to the effect of varying $\bar{\gamma}$ by the same amount but adjusting $H_0$ (red), or both $H_0$ and $\omega_c$ (green), to compensate its effect on the FLRW expansion.
    We have displayed results for the $n = 0.1$ (solid) and $n = -0.4$ (dashed) power laws, with $\bar{\gamma} = 0.85$\,, so that the curves $\gamma(a)$ for each power law are as given in Fig. \ref{fig_gamma_of_a_power_laws}. We show the curves $\theta(a)$ and $\tau(a)$\,, for $a \in \left[10^{-5}, 1\right]\,$, where $\theta(a) = r_s(a)/d_A(a)\,$. 
    All curves have been normalised by their $\Lambda$CDM evolution. The quantities $\theta(a_{\rm LS})$ and $\tau(a_{\rm eq})$ respectively determine the locations and heights of the CMB acoustic peaks.}
    \label{fig_background_gamma_H0_omc}
\end{figure}

The change between the plot with all the standard parameters fixed and the plot with $H_0$ adjusted shows how $\gamma$ competes with $H_0$ to set the locations of the acoustic peaks. 
This is because, once $\bar{\alpha}$ has been fixed to be approximately equal to $\bar{\gamma}$, the most significant remaining difference in the CMB power spectrum between the PPNC case and $\Lambda$CDM is that the evolving $\gamma(a)$ modifies the angular diameter distance to last scattering, and the sound horizon at last scattering. Therefore, $\gamma(a)$ generically causes a shift in the peak locations, through the ratio $\theta(a_{\rm LS}) = r_s(a_{\rm LS})/d_A(a_{\rm LS})\,$, unless the changes to $d_A(a_{\rm LS})$ and $r_s(a_{\rm LS})$ happen to cancel, which does not typically occur.

The effect of the PPNC power law evolution of $\gamma(a)$ on the cosmological background is demonstrated by Fig. \ref{fig_background_gamma_H0_omc}.
The reason for the degeneracy between $\bar{\gamma}$ and $H_0$ is indicated by the upper set, where, of the standard cosmological parameters, only $H_0$ can be varied. 
Consider the first subplot, which shows $\theta(a)$ relative to $\Lambda$CDM\,. The black curves show the effect of varying $\bar{\gamma}$ while holding all the standard cosmological parameters fixed, whereas the red curves show the effect of varying $\bar{\gamma}$ while also compensating its effect by varying $H_0\,$. 
Without adjusting $H_0$\,, setting $\bar{\gamma} < 1$ shifts $\theta(a_{\rm LS})$ above its $\Lambda$CDM value, corresponding to a lower $l$ for the first acoustic peak. However, it can be compensated by reducing $H_0$ relative to $\Lambda$CDM, so that the peak locations return to their observed values. This shows that we expect to see a positive degeneracy between $\bar{\gamma}$ and $H_0$. 

Now, let us focus on what we expect will be the next-strongest degeneracy in our hierarchy, between $\bar{\gamma}$ and $\omega_c$ (and thus between $\bar{\alpha}$ and $\omega_c$ due to the approximate equality of $\bar{\alpha}$ and $\bar{\gamma}$). 
Fig. \ref{fig_vary_parameters} shows that the main effect of adjusting $\omega_c$ is to compensate the effect of the evolution of $\gamma$ on the heights of the acoustic peaks. This can be explained through the second subplot in Fig. \ref{fig_background_gamma_H0_omc}.

We focus in particular on the conformal time $\tau$ that has elapsed at a given $a$ since the beginning of the PPNC evolution at $a_1 = 10^{-10}\,$, which is shown in the third subplot as a ratio of its $\Lambda$CDM value. Of particular importance is $\tau_{\rm eq} = \tau(a_{\rm eq})\,$, the matter-radiation equality scale.
Setting $\bar{\gamma} < 1$ increases $\tau_{\rm eq}\,$, because weakening gravity at the level of the background means that the effective energy density of cold dark matter, $\gamma \rho_c\,$, that enters into the Friedmann equation (\ref{eq_parametrisedfriedmanneqns1}), is always lower than in $\Lambda$CDM.
Thus, the radiation era lasts longer, and so acoustic oscillations in the photon-baryon fluid have more time to drive acoustic anisotropies in the CMB, before dark matter takes over and suppresses the oscillations.
Such an amplification of the acoustic peaks is entirely equivalent to the effect of lowering $\omega_c$ within $\Lambda$CDM. Hence, the effect on $\tau_{\rm eq}$ of reducing $\bar{\gamma}$ from unity can be compensated, at least in part, by a suitable increase in $\omega_c\,$, as seen from the blue curves in the same subplot. This means that we predict a negative degeneracy between $\bar{\gamma}$ and $\omega_c\,$.

\section{Observational constraints from Planck}\label{sec:planck_results}

Our constraints can be split into two categories: the MCMC analyses with a fixed power law index $n$ for $\alpha(a)$ and $\gamma(a)\,$ (for which there are nine fitting parameters), and analyses with the power law index $n$ varied as an additional, tenth free parameter.

\subsection{Fixed power law results}\label{subsec:fixed_power_law}

\begin{figure}
    \centering
    \includegraphics[width=\linewidth]{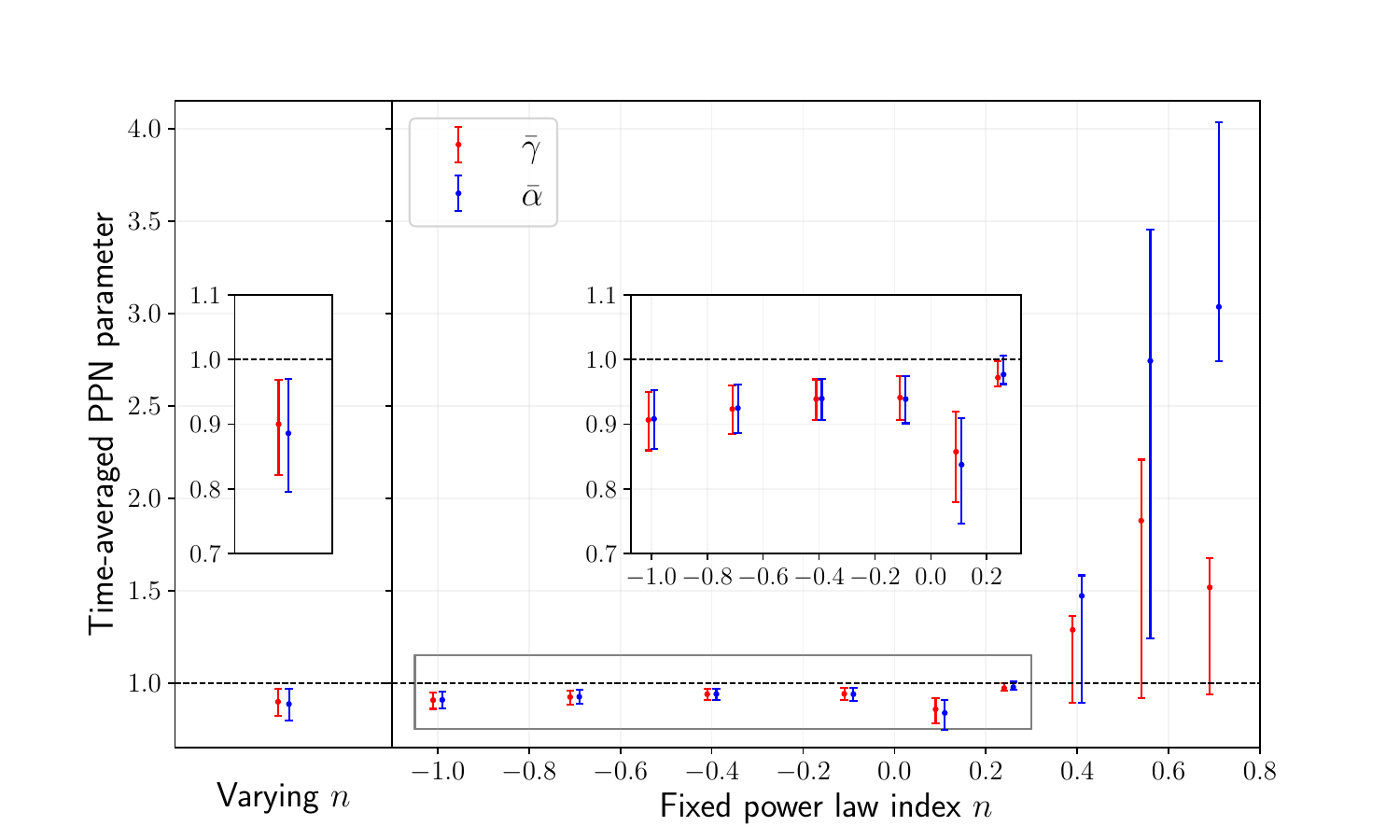}
    \caption{Constraints on $\bar{\gamma}$ and $\bar{\alpha}$ for various fixed power law indices $n$\,, and for the case where $n$ is varied (with a flat prior between $-15$ and $0.25$), as will be discussed in Section \ref{subsec:varying_power_law}. Results are shown with their $68\%$ confidence intervals.}
    \label{fig_alpha_gamma_fixed_n_results}
\end{figure}

For the fixed power law chains, we consider the following values of $n$: -0.7, -0.4, -0.1, 0.1, 0.25, 0.4, 0.55, 0.7. 
Let us focus first on the constraints on the averaged post-Newtonian parameters $\bar{\alpha}$ and $\bar{\gamma}\,$, that are obtained in parameterised post-Newtonian cosmologies governed by those specific power laws.
The results are displayed in Table \ref{table_alphagamma}, where we show the 68\% confidence intervals on $\bar{\alpha}$ and $\bar{\gamma}$, for fixed powerlaws of different values. This information is visualised in Figure \ref{fig_alpha_gamma_fixed_n_results}. 
For $n \leq 0.25\,$, the two PPNC parameters are broadly equally constrained, to within around 3-4\%, and $1$ to $2\sigma\,$, of their GR value. They are all slightly below GR (unity for both parameters), but consistent with it. 
The constraints on $\bar{\alpha}$ and $\bar{\gamma}$ are very similar, because the two parameters are highly degenerate with one another. This is a direct consequence of the requirement $\bar{\alpha} \approx \bar{\gamma}$\,, the physical origin of which we explored in detail in Section \ref{subsec:alpha_gamma}.

\begin{table}
\centering
\begin{mytabular}[1.2]{|c|c|c|c|c|c|c|c|c|c|} 
\hline 
Power law &  $-1$ & $-0.7$ & $-0.4$ &  $-0.1$ & $0.1$ \\
\hline 
$\bar{\gamma}$ & $0.91^{+0.04}_{-0.05}$ & $0.92^{+0.04}_{-0.04}$ & $0.94^{+0.03}_{-0.03}$ & $0.94^{+0.03}_{-0.03}$ & $0.86^{+0.06}_{-0.08}$ \\
\hline 
$\bar{\alpha}$  & $0.91^{+0.04}_{-0.05}$ & $0.92^{+0.04}_{-0.04}$ & $0.94^{+0.03}_{-0.03}$ & $0.94^{+0.04}_{-0.04}$ & $0.84^{+0.07}_{-0.09}$ \\
\hline 
Power law &  $0.25$ & $0.4$ & $0.55$ & $0.7$ & varying \\
\hline 
$\bar{\gamma}$ & $0.97^{+0.03}_{-0.01}$ & $1.29^{+0.08}_{-0.40}$ & $1.88^{+0.33}_{-0.96}$ & $1.52^{+0.16}_{-0.58}$ & $0.90^{+0.07}_{-0.08}$ \\
\hline 
$\bar{\alpha}$ & $0.98^{+0.03}_{-0.01}$ & $1.47^{+0.11}_{-0.58}$ & $2.74^{+0.71}_{-1.50}$ & $3.04^{+1.00}_{-0.29}$ & $0.89^{+0.08}_{-0.09}$ \\
\hline
\end{mytabular} 
\caption{Constraints on $\bar{\gamma}$ and $\bar{\alpha}$ for fixed power laws $n$ (and a varying power law with an $n_{\rm max} = 0.25$ upper bound on the prior), with their $68\%$ confidence intervals.}
\label{table_alphagamma}
\end{table}

It is notable that some of our fixed-$n$ MCMC analyses indicate a slight preference for ``weak gravity'', $\bar{\alpha} \approx \bar{\gamma} < 1\,$. This should not really be taken as evidence for deviations from General Relativity, not least because there is no uniquely well-motivated choice of power law $n$\,, and so any results for fixed $n$ are suggestive at most.
For $n \gtrsim 0.4\,$, the constraints are much weaker. We can explain this with direct reference to the issues we discussed in Section \ref{subsec:mcmc}: for these values of $n\,$, virtually all the modifications to gravity from deviant values of $\alpha$ and $\gamma$ are pushed deep into the radiation era. 
Therefore they have very little physical effect, so the data provide much less constraining power on $\bar{\alpha}$ and $\bar{\gamma}$\,, since by the time of last scattering the Universe is virtually indistinguishable from the concordance cosmology. Thus, it is not clear that these constraints have much physical meaning.

\begin{figure}
    \centering
    \includegraphics[width=\linewidth]{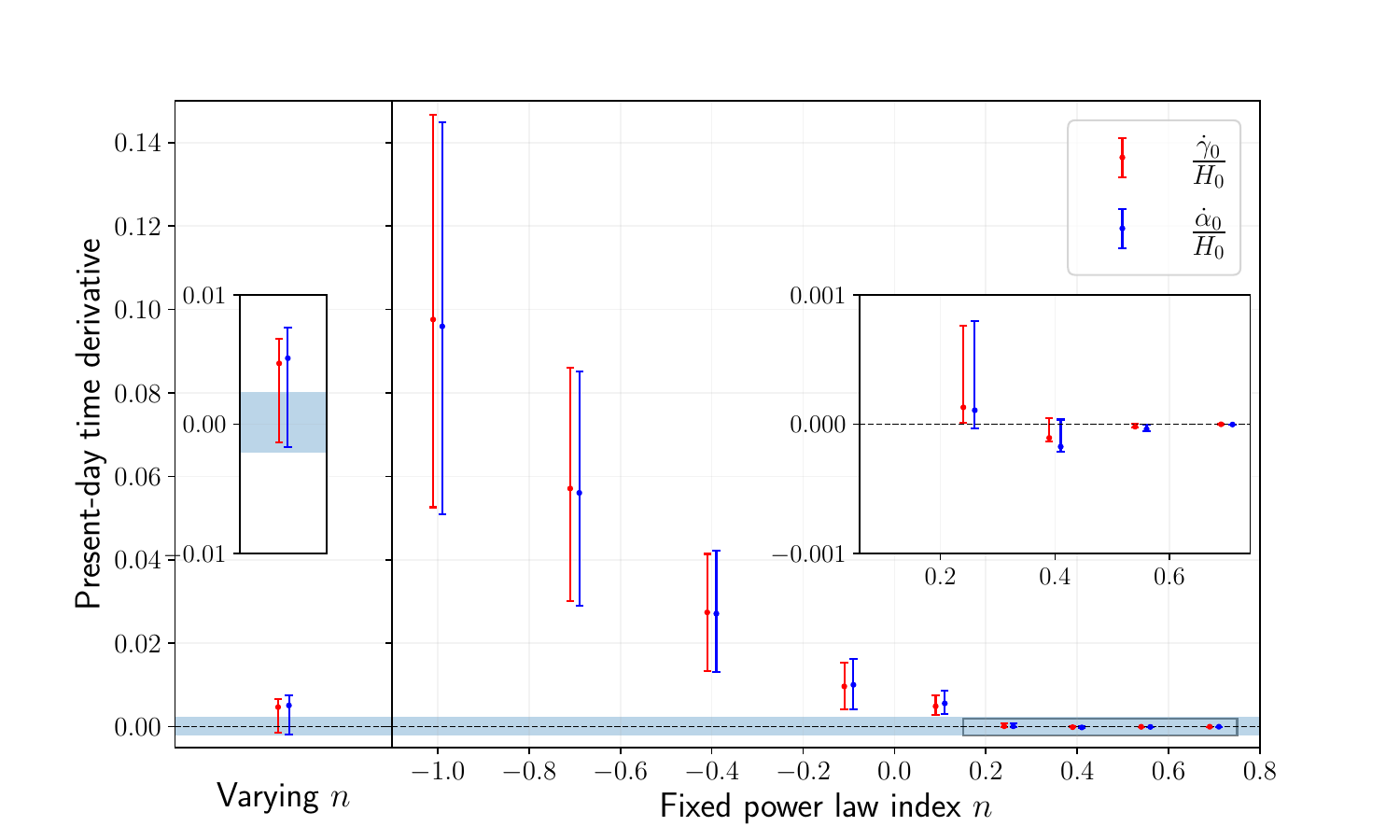}
    \caption{Constraints on the present-day time derivatives of $\gamma$ and $\alpha$, for various fixed power law indices $n\,$, and also for the varying-$n$ case. They are normalised by $H_0\,$, and shown along with their $68\%$ confidence intervals. The shaded region displays the tightest Solar System constraint on $\dot{\alpha}_0$\,, from the ephemeris of Mars \cite{konopliv2011mars}. This has similarly been normalised by $H_0\,$.}
    \label{fig_alpha_prime_gamma_prime_fixed_n_results}
\end{figure}

Let us now turn our attention to the present-day time derivatives of the post-Newtonian parameters, for which the constraints are displayed in Table \ref{table_derivatives_and_n} (with $95\%$ confidence intervals), and visualised in Fig. \ref{fig_alpha_prime_gamma_prime_fixed_n_results} (with $68\%$ intervals).
The constraints on $\dot{\gamma}_0$ and $\dot{\alpha}_0$ are similar to one another, as expected from our discussion of the gravitational phenomenology driving $\alpha$ and $\gamma$ to be equal to one another throughout cosmic history\footnote{Our constraints are actually on $\dfrac{\dot{\gamma}_0}{H_0}$ and $\dfrac{\dot{\alpha}_0}{H_0}\,$, due to the degeneracy between the post-Newtonian parameters and $H_0\,$. However, for simplicity we will just talk about these as being constraints on $\dot{\gamma}_0$ and $\dot{\alpha}_0\,$, because the 1D posterior distributions on them are much broader than that on $H_0\,$.}.
The best-fit values, and their confidence intervals, depend very strongly on the value of $n$\,, as can be understood with reference to Fig. \ref{fig_gamma_of_a_power_laws}.
For large positive $n$\,, there is simply no way to get a large derivative of a PPN parameter at $a = 1\,,$ unless the average of the PPN parameter is enormously, and probably unphysically, deviant from its GR value of unity. For the $n \geq 0.25$ power laws, the constraints on $\dot{\alpha}_0$ are better than those obtained in Solar System experiments (with the tightest such measurement coming from the ephemeris of Mars \cite{konopliv2011mars}).
For $n < 0$ the constraints on $\dot{\alpha}_0$ are not especially competitive with Solar System measurements.

It is important here to recall that Solar System tests do not constrain $\dot{\gamma}_0$ at all, because they are tests involving the motion of non-relativistic matter (such as the planets). The post-Newtonian geodesic equation for non-relativistic matter is described by $\alpha$ only, whereas $\alpha$ and $\gamma$ together describe trajectories of electromagnetic waves, and there are no Solar System experiments that directly study the time evolution of this. Thus, our constraints on $\dot{\gamma}$ are tighter than what have been obtained to date in the weak-field astrophysical regime.
They are still of course subject to an intrinsic dependence on the prescription for $n\,$, being derived parameters within the MCMC, rather than additional parameters to be varied independently. 

\begin{table}
\centering
\begin{mytabular}[1.2]{|l|c|c|} 
\hline 
Power law & $\dot{\gamma}_0/H_0$ & $\dot{\alpha}_0/H_0$ \\ 
\hline 
$-1$ & $0.098_{-0.045}^{+0.049}$ & $0.096_{-0.045}^{+0.049}$  \\
\hline 
$-0.7$ & $0.057_{-0.027}^{+0.029}$ & $0.056_{-0.027}^{+0.029}$  \\
\hline
$-0.4$ & $0.027_{-0.014}^{+0.014}$ & $0.027_{-0.014}^{+0.015}$  \\
\hline
$-0.1$ & $0.001_{-0.0055}^{+0.0057}$ & $0.010_{-0.0058}^{+0.0061}$ \\
\hline
$0.1$ & $0.0049_{-0.0021}^{+0.0027}$ & $0.0056_{-0.0025}^{+0.0031}$  \\
\hline
$0.25$ & $\left(1.3_{-1.2}^{+6.3}\right)\times 10^{-4}$ & $\left(1.1_{-1.4}^{+6.9}\right) \times 10^{-4}$  \\
\hline
$0.4$ & $\left(-1.1_{-0.3}^{+1.5}\right) \times 10^{-4}$ & $\left(-1.7_{-0.4}^{+2.1}\right)\times 10^{-4}$ \\
\hline
\end{mytabular}
\caption{$68\%$ constraints on the present-day time derivatives of the post-Newtonian parameters, as a ratio of $H_0$\,, for the fixed PPNC power laws with power law indices $n = \left\lbrace -1, -0.7, -0.4, -0.1, 0.1, 0.25, 0.4\right\rbrace$\,.}
\label{table_derivatives_and_n}
\end{table}

\begin{figure}
    \centering
    \includegraphics[width=\linewidth]{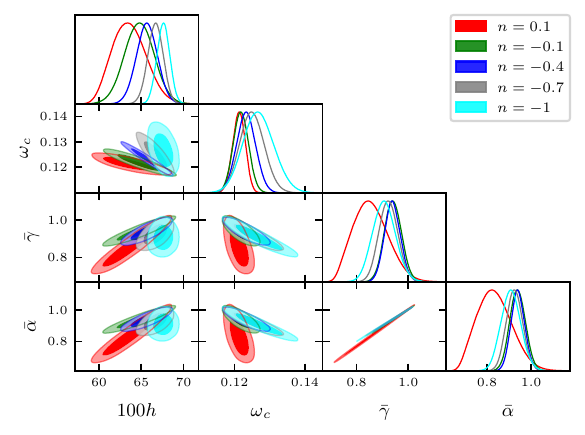}
    \caption{2D and 1D posteriors on $H_0\,$, $\omega_c$\,, $\bar{\gamma}$ and $\bar{\alpha}$\,, for the $n = 0.1$\,, $-0.1\,$, $-0.4$ and $-0.7$ power laws.}
    \label{fig_H0_omc_gamma_alpha_1}
\end{figure}

\begin{figure}
    \centering
    \includegraphics[width=\linewidth]{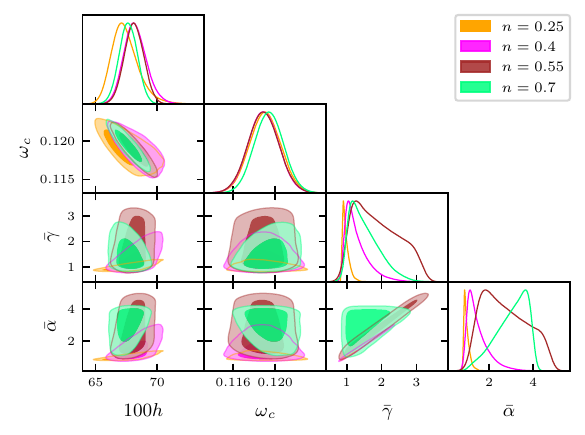}
    \caption{2D and 1D posteriors on $H_0\,$, $\omega_c$\,, $\bar{\gamma}$ and $\bar{\alpha}$\,, for the $n = 0.25$\,, $0.4$\,, $0.55$ and $0.7$ power laws. A smoothing scale of $0.3\sigma$ has been applied to the posteriors.}
    \label{fig_H0_omc_gamma_alpha_2}
\end{figure}

Let us now move on to the standard $\Lambda$CDM parameters, and their degeneracies with the time-averaged post-Newtonian parameters. As is to be expected when one expands the parameter space, we have a slight reduction in constraining power compared to the base $\Lambda$CDM cosmology. 
The key results are summarised by Figs. \ref{fig_H0_omc_gamma_alpha_1} and \ref{fig_H0_omc_gamma_alpha_2}, where we have shown the 1D and 2D posteriors for the most important reduced set of parameters, which is $H_0\,$, $\omega_c\,$, $\bar{\gamma}$ and $\bar{\alpha}\,$.
The full set of $95\%$ constraints on each standard cosmological parameter is displayed in Table \ref{table_all_powerlaws_summary}, where we have not included the $0.55$ and $0.7$ power laws, because their deviations from $\Lambda$CDM take place so deep into the radiation era that the constraints have little meaning.

\begin{table}
\hspace{-2cm}
\begin{mytabular}[1.2]{|l|c|c|c|c|c|c|c|c|c|c|} 
\hline 
Power law & $-1$ & $-0.7$ & $-0.4$ & $-0.1$ & $0.1$ & $0.25$ & $0.4$ & Varying \\ 
\hline 
$100\,\omega_b$ & $2.24_{-0.02}^{+0.02}$ & $2.24_{-0.03}^{+0.03}$ & $2.24_{-0.03}^{+0.03}$ & $2.24_{-0.03}^{+0.03}$ & $2.23_{-0.03}^{+0.03}$ & $2.21_{-0.02}^{+0.02}$ & $2.21_{-0.02}^{+0.02}$ & $2.23_{-0.03}^{+0.03}$ \\ 
\hline
$10\,\omega_c$ & $1.27_{-0.05}^{+0.04}$ & $1.25_{-0.04}^{+0.03}$ & $1.23_{-0.03}^{+0.03}$ & $1.22_{-0.02}^{+0.02}$ & $1.21_{-0.02}^{+0.02}$ & $1.19_{-0.02}^{+0.02}$ & $1.19_{-0.02}^{+0.02}$ & $1.21_{-0.02}^{+0.02}$ \\ 
\hline
$\ln{\left(10^{10}\,A_s\right)}$ & $3.02_{-0.02}^{+0.02}$ & $3.02_{-0.02}^{+0.02}$ & $3.02_{-0.02}^{+0.02}$ & $3.03_{-0.02}^{+0.02}$ & $3.04_{-0.02}^{+0.02}$ & $3.04_{-0.02}^{+0.02}$ & $3.03_{-0.02}^{+0.02}$ & $3.04_{-0.02}^{+0.02}$ \\ 
\hline
$10\, n_s$ & $9.72_{-0.09}^{+0.09}$ & $9.71_{-0.09}^{+0.09}$ & $9.69_{-0.08}^{+0.08}$ & $9.67_{-0.08}^{+0.08}$ & $9.69_{-0.08}^{+0.08}$ & $9.62_{-0.07}^{+0.07}$ & $9.59_{-0.07}^{+0.08}$ & $9.66_{-0.08}^{+0.08}$\\ 
\hline
$100\,\tau_{\rm reio}$ & $5.02_{-0.77}^{+0.85}$ & $4.94_{-0.79}^{+0.86}$ & $4.88_{-0.79}^{+0.84}$ & $5.00_{-0.80}^{+0.82}$ & $5.00_{-0.79}^{+0.80}$ & $5.41_{-0.80}^{+0.75}$ & $5.47_{-0.80}^{+0.73}$ &  $5.10_{-0.79}^{+0.77}$ \\ 
\hline
$10\,Y_{\rm P}$ & $2.32_{-0.16}^{+0.16}$ & $2.33_{-0.16}^{+0.16}$ & $2.34_{-0.16}^{+0.17}$ & $2.35_{-0.16}^{+0.17}$ & $2.32_{-0.17}^{+0.17}$ & $2.23_{-0.17}^{+0.18}$ & $2.21_{-0.18}^{+0.18}$ & $2.30_{-0.17}^{+0.18}$ \\ 
\hline
$100\,h$ & $67.6_{-0.8}^{+0.8}$ & $66.7_{-0.9}^{+1.0}$ & $65.7_{-1.3}^{+1.3}$ & $64.8_{-1.8}^{+1.8}$ & $63.7_{-2.3}^{+1.9}$ & $67.4_{-1.3}^{+0.9}$ & $68.2_{-1.0}^{+0.8}$ &  $65.1_{-1.8}^{+2.3}$ \\  
\hline
\end{mytabular}
\caption{$68\%$ constraints on the $\Lambda$CDM cosmological parameters, for the fixed PPNC power laws with power law indices $n = \left\lbrace -0.7, -0.4, -0.1, 0.1, 0.25, 0.4\right\rbrace$\,, and for a varying power law, with a flat prior on the power law index $n$ between $-15$ and $0.25\,$\,. Constraints on $H_0$ are expressed in terms of $100\,h = H_0/{\rm km}\,{\rm s}^{-1}\,{\rm Mpc}^{-1}\,$.}
\label{table_all_powerlaws_summary}
\end{table}

\begin{figure}
    \centering
    \includegraphics[width=\linewidth]{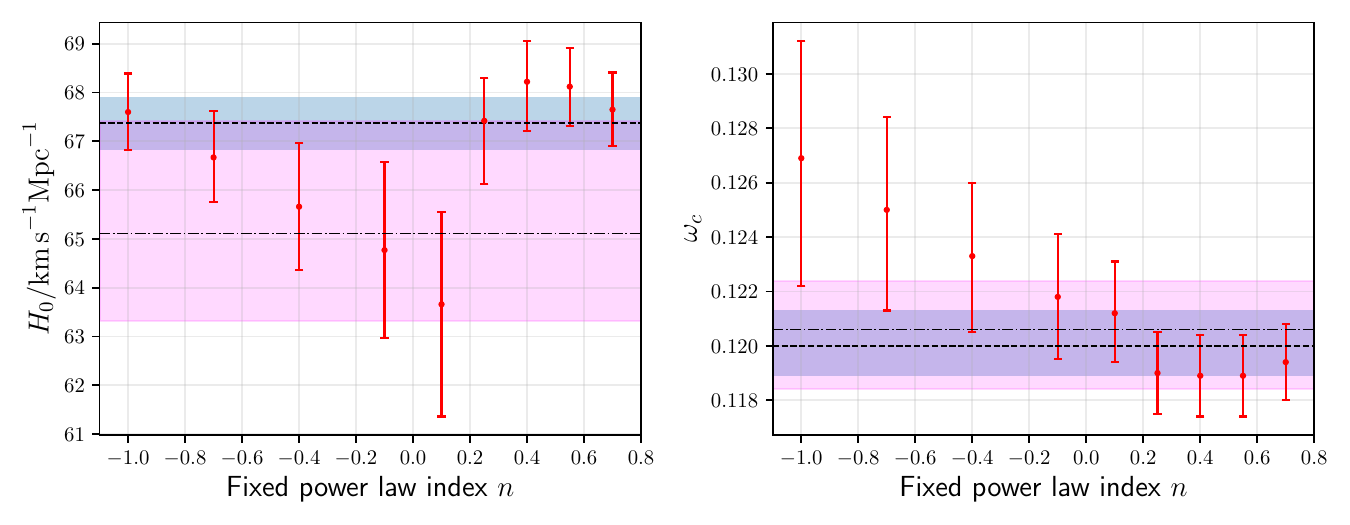}
    \caption{Constraints on $H_0$ and $\omega_c$ for the fixed power law PPNC analyses, along with their $68\%$ confidence intervals. The blue shaded regions display equivalent constraints (with the mean value given by the dashed black line) in the 7-parameter $\Lambda$CDM model - the standard 6-parameter $\Lambda$CDM, plus a varied primordial helium abundance $Y_{\rm P}$. The magenta shaded regions (with mean given by the dot-dashed black line) are the constraints for the varying power law case in Section \ref{subsec:varying_power_law}.}
    \label{fig_h0_omc_ns_As_fixed_n_results}
\end{figure}

\begin{figure}
    \centering
    \includegraphics[width=0.48\linewidth]{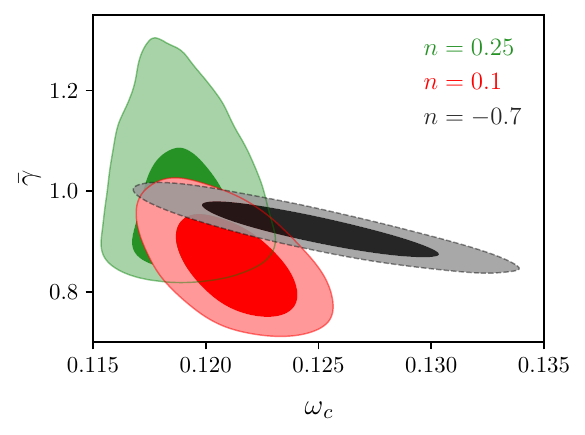}
    \includegraphics[width=0.48\linewidth]{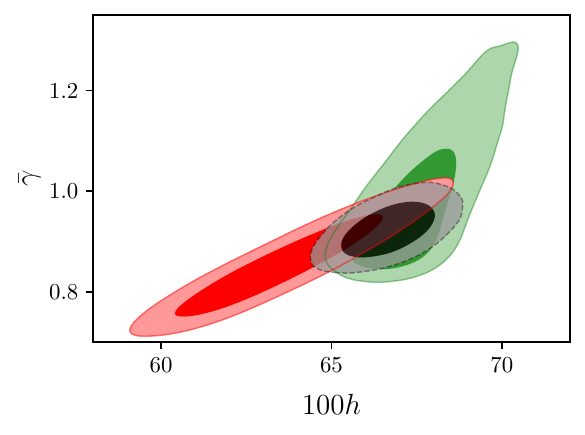}
    \includegraphics[width=0.6\linewidth]{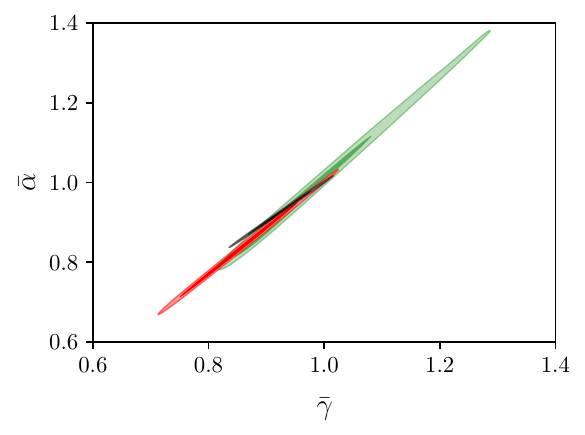}
    \caption{Depiction of how the 2D posteriors change for different fixed power law indices $n\,$. The plot on the top left shows the $\left(\omega_c,\bar{\gamma}\right)$ degeneracy, the top right $\left(H_0,\bar{\gamma}\right)$ and the bottom $\left(\bar{\gamma}, \bar{\alpha}\right)\,$. The power law indices shown have been chosen in order to demonstrate most clearly how the degeneracies change with $n$\,.}
    \label{fig_degeneracies_fixed_n_zoomed}
\end{figure}

We will focus here on the $\Lambda$CDM parameters whose constraint varies notably with power law index $n$\,, as it is in these cases that there is a substantial physical effect driving the degeneracy with $\bar{\gamma}$ and $\bar{\alpha}\,$.
The largest variations with $n$ appear in the parameters $H_0$ and $\omega_c$\,, the constraints on which are displayed in the top row of Fig. \ref{fig_h0_omc_ns_As_fixed_n_results}. We know that it must be $\bar{\gamma}$ where the fundamental degeneracy with these two parameters is occurring, not $\bar{\alpha}$\,, because the background evolution is independent of $\alpha$ in the PPNC implementation in this chapter, as we discussed in Section \ref{subsec:ppnc_class}. 
Then, the evolution of the cosmological perturbations enforces $\bar{\alpha} \approx \bar{\gamma}\,$ (as per Section \ref{subsec:alpha_gamma}), resulting in $\bar{\alpha}$ picking up roughly the same degeneracies with $H_0$ and $\omega_c$ that $\bar{\gamma}$ has. 
The expected strong degeneracy between $\bar{\alpha}$ and $\bar{\gamma}$ is very apparent for all but one of the power laws considered, as displayed in the bottom plot of Fig. \ref{fig_degeneracies_fixed_n_zoomed}, where we have focused on the 2D posterior on $\bar{\gamma}$ and $\bar{\alpha}$. 
The exception is the extremal case with $n = 0.7$\,, for which virtually all possible modifications to GR are pushed deep into the radiation era.

The constraint on $H_0$ is similar to the best-fit $\Lambda$CDM Planck value for power laws at either end of the range of $n$ considered, but is up to a factor of 3 worse for intermediate values of the power law index. This is because the degeneracy between $H_0$ and the residual effect of the PPNC parameters (after the cancellation that causes the $\left(\bar{\gamma}, \bar{\alpha}\right)$ degeneracy) is strongest within this region. 
The degeneracy between $H_0$ and $\bar{\gamma}$ is also displayed in the top-right plot of Fig. \ref{fig_degeneracies_fixed_n_zoomed}.
It has the additional effect of the central value of $H_0$ being (slightly, and not statistically significantly) pulled down for these power laws\footnote{$H_0$ moves in the wrong direction to help resolve the observed Hubble tension \cite{di2021realm}.}, due to the positive degeneracy between $H_0$ and $\bar{\gamma}$ and the data preferring $\bar{\gamma}$ below unity here.
The degeneracy results from the modifications $\gamma$ makes to the angular diameter distance to last scattering, and therefore to the acoustic peak scale $\theta_{*}\,$ (as shown by the adjusted peak locations in the second plot of Fig. \ref{fig_vary_parameters}). 

The constraint on $\omega_c$ is very close to its $\Lambda$CDM value for large positive $n$\,, because in that case the cosmology is virtually indistinguishable from $\Lambda$CDM except for deep in the radiation era. The constraint becomes worse, and the inferred value of $\omega_c$ larger, as $n$ decreases and becomes negative, such that the expansion history is more substantially modified by the evolution of $\gamma(a)\,$.
This affects $\omega_c$ through the $(\bar{\gamma}, \omega_c)$ degeneracy, displayed most clearly in the top-left plot of Fig. \ref{fig_degeneracies_fixed_n_zoomed}, which is due to the modification that $\gamma$ induces on the matter-radiation equality time $\tau_{\rm eq}$ that sets the heights of the acoustic peaks, as explained in Section \ref{subsec:gamma_h0_omc}, and displayed by the adjusted peak heights in the third plot of Fig. \ref{fig_vary_parameters}.

\subsection{Varying power law results}\label{subsec:varying_power_law}

\begin{figure}
    \centering
    \includegraphics[width=\linewidth]{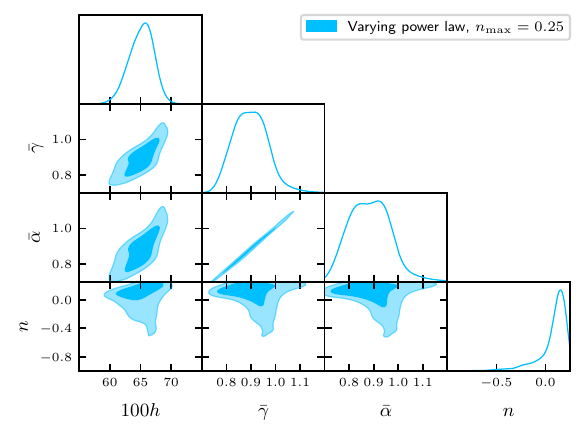}
    \caption{2D and 1D posteriors on $H_0\,$, $\bar{\gamma}$\,, $\bar{\alpha}$\,, and the power law index $n$\,. For $n$\,, we have a flat prior between $-15$ and $0.25\,$. A smoothing scale of $0.3\sigma$ has been applied to all posteriors.}
    \label{fig_varied_powerlaw_posteriors}
\end{figure}

Let us now present the constraints when the power law index $n$ is varied as an additional free parameter. As discussed in Section \ref{subsec:mcmc}, there is some inevitable dependence in this case on the choice of prior for $n\,$, especially its upper boundary which we set at $n = 0.25\,$. 
This is demonstrated clearly by the 1D posterior on $n$ in Fig. \ref{fig_varied_powerlaw_posteriors}, which is clearly stacked up against the upper boundary of the prior volume. Indeed, the $68\%$ confidence interval constraint on the power law is $n = 0.076_{-0.009}^{+0.17}$\,, so the upper end of the constraint is explicitly controlled by the prior selection.

In the 2D posteriors between $\left\lbrace \bar{\gamma}, \bar{\alpha}\right\rbrace$ and $n\,$, on the bottom line of Fig. \ref{fig_varied_powerlaw_posteriors}, there is a noticeable ``tail'' feature, where the $2\sigma$ confidence interval allows $n$ to extend to negative values $\sim -0.5\,$, for $\bar{\gamma}$ and $\bar{\alpha}$ near unity. 
This is reflective of the fact that the PPNC parametrisation adopted is a nested model: if $\bar{\gamma} = \bar{\alpha} = 1\,$, then we are back at $\Lambda$CDM in General Relativity. Thus, any value of $n$ is allowed, because the coefficient of $a^{-n}$ in the power laws for $\gamma(a)$ and $\alpha(a)$ vanishes. 
Likewise, if the power law index is close to the upper bound of the prior at $0.25\,$, then any deviations from GR are pushed further back towards radiation domination. Thus, there is much more freedom for $\bar{\gamma}$ and $\bar{\alpha}$ to deviate from unity, and so the 2D posterior spreads out in the horizontal direction. 
This combination, of a vertical tail for $\left\lbrace\bar{\gamma}, \bar{\alpha}\right\rbrace \approx 1\,$, and a horizontal spreading for $n \sim 0.25\,$, gives the $\left(\bar{\gamma}, n\right)$ and $\left(\bar{\alpha}, n\right)$ 2D posteriors a distinctive T shape.

Unsurprisingly, the constraints on the time-averaged post-Newtonian parameters are somewhat weaker than in the majority of the fixed-$n$ cases: we obtain $\bar{\gamma} = 0.898_{-0.078}^{+0.068}$ and $\bar{\alpha} = 0.886^{+0.084}_{-0.091}\,$, as shown on the left of Fig. \ref{fig_alpha_gamma_fixed_n_results}. 
Moreover, the 2D posterior in the $\left(\bar{\gamma},\bar{\alpha}\right)$ plane indicates that the requirement $\bar{\alpha} \approx \bar{\gamma}$ that was a recurrent feature of the fixed-$n$ analyses carries over to the varying-$n$ analysis. The degeneracies between the post-Newtonian parameters and $H_0$ are still present, but the degeneracies with $\omega_c$ are essentially lost compared to the fixed-$n$ results, so we have not shown them.

Finally, it is worth commenting that we get derived constraints, shown in Table \ref{table_alphadot_gammadot_vs_mars_ephemeris} (and visualised on the left of Fig. \ref{fig_alpha_prime_gamma_prime_fixed_n_results}), on the present-day time derivatives of $\alpha$ and $\gamma$ that is much more meaningful than in the fixed-$n$ case, because it is not so overwhelmingly driven by an {\it a priori} imposed choice of power law. 
The constraints on the standard cosmological parameters for the varying-$n$ MCMC analysis are shown in the final column of Table \ref{table_all_powerlaws_summary}.

\begin{table}[ht]
\centering
\begin{mytabular}[1.2]{|c|c|c|} 
\hline 
Parameter &  Solar System constraint & CMB constraint \\
\hline 
$\dot{\alpha}_0/H_0$ & $\left(0.15 \pm 2.3\right)\times 10^{-3}$ & $\left(5.1^{+2.4}_{-6.9}\right)\times 10^{-3}$ \\ 
\hline 
$\dot{\gamma}_0/H_0$ & --- & $\left(4.7^{+1.9}_{-6.1}\right)\times 10^{-3}$ \\ 
\hline
\end{mytabular} 
\caption{$68\%$ CMB constraints on $\dfrac{\dot{\alpha}_0}{H_0}$ and $\dfrac{\dot{\gamma}_0}{H_0}$ for a PPNC power law with varying $n$\,, compared to the tightest Solar System constraint on $\dot{\alpha}_0$\,, from the ephemeris of Mars \cite{konopliv2011mars} (which we have converted to a constraint on $\dfrac{\dot{\alpha}_0}{H_0}$ using the Planck $\Lambda$CDM best-fit $H_0$ value). There is no equivalent Solar System constraint on $\dot{\gamma}_0\,$.}
\label{table_alphadot_gammadot_vs_mars_ephemeris}
\end{table}

\subsection{Prospects for improved constraints}

The results we have presented in this section are the first constraints on the generalised PPN parameters, and the associated parameterised post-Newtonian cosmology, from a cosmological dataset. For several fixed power laws we could obtain fairly tight constraints, but in the more generic case where the index $n$ was allowed to vary, tight constraints were only possible with a rather aggressive prior selection.

In order to strengthen our constraints, it would be beneficial to combine our results with those from other implementations of the PPNC formalism that probe different times in cosmic history.
This is suggested by the fact that the results of the fixed power law chains produced much weaker constraints on the present-day time derivatives of $\gamma$ and $\alpha$ for $n < 0\,$, as per Fig. \ref{fig_alpha_prime_gamma_prime_fixed_n_results}. However, it is precisely these PPNC models that most strongly affect late time cosmology, because $\gamma(a)$ and $\alpha(a)$ can be more strongly deviant from unity at late times (see Fig. \ref{fig_gamma_of_a_power_laws}).

Therefore, we would hope to more tightly constrain the values of $\gamma$ and $\alpha$ in the late Universe, i.e. somewhere in between last scattering and the present day. 
This may improve the constraint on $\dot{\gamma}_0$ and $\dot{\alpha}_0$ for negative $n$ power laws, as well as the constraints on $\bar{\gamma}$\,, $\bar{\alpha}$ and $n$ for the varying power law investigation.
The most natural observational datasets to study would appear to be focused therefore around large scale structure, such as measurements of baryon acoustic oscillations from the Dark Energy Spectroscopic Instrument (DESI). These have recently found evidence at $\sim 3\sigma$ for a non-constant equation of state of dark energy \cite{adame2024desi}. 
It is certainly conceivable that data indicating an apparently evolving $w_{\rm DE}(a)$ could similarly be consistent with evolving PPNC parameters $\gamma(a)$\,, $\alpha(a)$\,, $\gamma_c(a)$ and $\alpha_c(a)\,$, although of course this idea must be thoroughly tested.
It may also be useful to study gravitational wave data, as this could give a handle on the values of $\alpha$ and $\gamma$ at redshifts $\lesssim 1$\,, although the use of these observations is complicated by the possibility of strong-field, fast-motion dynamics, which would violate the assumptions of our weak-field formalism. 

There are a plethora of other additional investigations one could carry out into the PPNC theory space, including:
\begin{itemize}

    \item Calculating constraints on PPN parameters using priors from astrophysical experiments. Imposing a prior of this kind may be particularly helpful for testing the evolution of $\alpha(a)$\,, for which the most precise constraint on its present-day time derivative remains the result from the ephemeris of Mars \cite{konopliv2011mars}.
    
    \item Constraining a more generalised form of $\gamma_c(a)\,$, rather than assuming it to be a constant as we have here. For example, one could study a power law for $\gamma_c$ in the same way as for $\gamma$ and $\alpha\,$. Note that we do not consider $\alpha_c(a)$ to be an independent function, as it can always be calculated from the other PPNC parameters according to the integrability condition (\ref{eq_integrabilitycondition}).
    
    \item Implementing a parameterised form of the gravitational slip $\Sigma(a, k)$ on all scales, in a more robust, physically well-motivated way. In this chapter we have just assumed that the slip vanishes in the superhorizon limit $k \longrightarrow 0\,$, but one ought instead to derive a form of the large-scale slip from the underlying principles of the PPNC formalism, to go with the small-scale slip which we know to be $\dfrac{\alpha - \gamma}{\gamma}\,$. The form of the slip in parameterised post-Newtonian cosmology is the subject of ongoing research \cite{Clifton_2024}. 

    \item Generalising the assumed FLRW cosmological background to include spatial curvature, rather than it being spatially flat, as we have assumed in the last two chapters. As long as $\Omega_K$ remains small at all points in cosmic history we are interested in, spatial curvature is not too difficult to implement in the PPNC framework, as it can be treated perturbatively \cite{Sanghai_2019}.

    \item Including the effects of frame-dragging vector perturbations $B_i$ to the FLRW metric. These have been ignored in this chapter, because they are not important for the temperature anisotropies in the cosmic microwave background, but they are a natural part of the post-Newtonian approach, as we saw in detail in Chapter 5, and introduce rich phenomenology such as preferred-frame effects. 
    
\end{itemize}

These investigations are beyond the remit of this thesis, but we include them as an indication of the potential utility of the PPNC framework in testing modifications to General Relativity on cosmological length and time scales.

\section{Discussion}

In this chapter, we have presented the first observational constraints on the parameter space of parameterised post-Newtonian cosmology \cite{Sanghai_2017,Sanghai_2019,anton2022momentum,Thomas_2023,Thomas_2024,Clifton_2024}, the theory-independent framework for testing gravity in cosmology whose theoretical origin we explained in detail in Chapter 5, using cosmic microwave background data from the Planck satellite \cite{Planck_2020}.
This is far from the first attempted theory-independent investigation into possible deviations from General Relativity on cosmological scales (see e.g. \cite{Clifton_2012,Ishak_2018} for reviews). 
The idea of a time-varying strength of gravity has, unsurprisingly, been studied previously using cosmological observables \cite{bbn}. However, we reiterate our earlier point that these are typically done with the $G$ in mind being the parameter that multiplies the total energy density $\rho$ in the Friedmann equation, so they do not really say anything about the $G$ that we have measured astrophysically and in table-top experiments \cite{ppnvscosmo}. 

This, in fact, is a generic issue with analyses of a ``time-varying Newton's constant'': most such tests take a specific equation in which $G$ appears, and consider the constraints on the deviation of the phenomenology in question from the assumption of a constant $G$\,. 
However, in general, modified theories of gravity provide different modifications to the effective coupling strength of non-relativistic matter to the metric tensor (which is what we should really think of as $G$) differently in different regimes and contexts, so constraints on the evolution of $G$ in a particular context do not necessarily carry over elsewhere. 
Thus, single-phenomenon tests of this kind may not always be internally consistent. Instead, one should use a generalised framework in which gravitational couplings can be described in the same mathematical language, across a variety of cosmological length and time scales, and for both the FLRW ``background'' expansion of the Universe and the evolution of perturbations.
This is the ultimate advantage of the PPNC constraints we have presented in this chapter: they are consistent with astrophysical tests of gravity, because they are built from the same parameters as the PPN formalism that is used for those tests.

Moreover, the extension of the PPN to cosmology allows for precision tests of parameters that cannot be constrained astrophysically, in particular $\alpha(t)$\,, which is unity by definition in Solar System observations, and $\dot{\gamma}\,$.
The constraints on $\dot{\alpha}$ and $\gamma$ from the Solar System are fundamentally valid only at the present day, with no indication of what the values of those parameters might have been at some earlier cosmological epoch.
Our CMB constraints extend these measurements to a much longer period of time (with the caveat that the constraints we obtain are highly dependent on the assumed form of the parameter as a function of scale factor). Constraining the full set of evolving post-Newtonian parameters, using data from almost the entirety of cosmic time, is a genuinely novel endeavour. 
At this stage the constraints we have obtained, coming from only one probe, are fairly weak. However, the vast array of existing and upcoming cosmological data should make it possible in the future to construct precise models for the allowed evolution of e.g. $\alpha(a)$ and $\gamma(a)$\,, and therefore to develop a strong understanding of what possible deviations from General Relativity might be allowed by cosmological observations.

\chapter{The emergence of cosmic anisotropy}

\lhead{\emph{Emergent anisotropy}}

For the remainder of the research-based part of the thesis, we turn our attention to alternatives to the concordance cosmology that are focused on deviations from the cosmological principle of homogeneity and isotropy, which is encoded in the Friedmann-Lema{\^ i}tre-Robertson-Walker metric.
We will develop and analyse a novel framework for studying anisotropy on large scales in the Universe, inspired in particular by the dipole \cite{Secrest_2021,Secrest_2022,Migkas_2021,Yeung_2022,antoniou2010searching,sorrenti2023dipole} and bulk flow \cite{Kashlinsky_2008, Magoulas_2014, hoffman2015cosmic} observational anomalies summarised by Fig. \ref{fig_dipole_tension}. A detailed exposition of those and other anisotropic anomalies is provided by Ref. \cite{Aluri_2023}.

In this chapter, we will motivate and construct our formalism, which we will refer to as ``emergent anisotropy''. In Section \ref{sec:anisotropy_covariant} we will introduce the key theoretical issues involved, including a detailed discussion of the problems of foliation and directional dependence. 
Then, in Section \ref{sec:anisotropic_averaging} we derive the full set of equations that govern the dynamics of such an anisotropic cosmological model, and discuss how the results can be interpreted physically. Finally, Section \ref{sec:backreaction_farnsworth} will be concerned with the application of the framework to an illustrative class of example cosmological spacetimes.
The contents of the chapter are based on Ref. \cite{Anton_2023}. Note that throughout this and the following chapter (i.e. the remainder of the thesis), we will choose our units so that in addition to $c$ being equal to unity, we also have $8\pi G = 1\,$.

\section{Modelling anisotropy covariantly}\label{sec:anisotropy_covariant}

The various observational anomalies that we discussed in Section \ref{subsec:observational_issues} have led some cosmologists to suggest that our Universe might be anisotropic on large scales. 
Although these anomalies are controversial, and we will not argue for or against their statistical significance or veracity, it is certainly worth considering whether a fundamental anisotropy in the Universe at late times is a viable possibility from a theoretical point of view.
This has been modelled in some quarters by considering the possibility that realistic observers might have worldlines that are tilted with respect to the preferred homogeneous foliation of the Universe, with potentially large consequences \cite{Tsagas_2009,Tsagas_2011,tsagas2015deceleration,tsagas2022deceleration,Santiago_2022}.
In fact, we recall from Chapter 4 that some authors have suggested that the presence tilted flows could imply that the apparent dynamics of the late Universe do not require the existence of dark energy to explain them \cite{colin2019evidence, Mohayaee_2020, Mohayaee_2021}. 

As we showed in Section \ref{sec:inhomogeneity_anisotropy}, however, there are a number of difficulties associated with modelling anisotropic universes, including in particular understanding how large-scale anisotropies might actually be supported in the late Universe.
Indeed, most of the proposals of anisotropic or tilted cosmologies have so far gone without a concrete mechanism by which such large-scale misalignments could be generated. 
We take this as further motivation for the development of a framework which could be used to evaluate the possibility of these scenarios being realised.

An important aspect in our approach is the concept of ``emergence'', which is intended to describe the possibility that the properties in which we are interested are realised on average, from a spacetime which is highly inhomogeneous on small scales. 
This concept is vital for the goal we have outlined so far, as anisotropy in exactly homogeneous cosmologies typically decays rapidly in time \cite{barrow1995universe}. Indeed, we calculated a concrete example of this in Section \ref{subsec:anisotropic_models}, for the Bianchi type {\it I} cosmologies.

The isotropisation of homogeneous cosmological spacetimes means that any anisotropy in the late Universe would mean enormous anisotropy at early times.
This is clearly not physically realistic, and so the only other possibility would be to have large-scale anisotropy realised as an emergent property of a more complicated inhomogeneous spacetime. 
We expect that anisotropy generated in this way should weaken the tight bounds that could otherwise be imposed from the CMB \cite{Pontzen_2007,Pontzen_2009,Saadeh_2016}, and might even allow for the intriguing fits of Bianchi ${\it VII}_h$ templates to the observational data to be reinterpreted \cite{jaffe2005evidence,jaffe2006viability,bridges2007markov,mcewen2013bayesian}.

The approach we will use in our construction will follow the general philosophy of the scalar averaging formalism developed by Buchert \cite{Buchert_2000}, which we discussed in detail in Section \ref{subsec:averaging_problem}. 
By explicitly including the consequences of the non-commutativity of averaging and evolution under Einstein's equations, we saw that in the isotropic case Buchert studied, the effective field equations that govern the behaviour of large-scale averages can give phenomenologically very different behaviour to the locally defined field equations of an exact FLRW spacetime, that are more frequently used in the standard approach to cosmological modelling.
Moreover, that different behaviour can arise without introducing any new physics, such as dark energy or a modification of gravity.
The novel phenomenology in Buchert's averaged, emergent description, could be attributed entirely to ``backreaction'' terms that account for the effects of nonlinear structure on the cosmological dynamics. 
This is exactly what we require for our current purpose, although we will need to extend and generalise Buchert's approach in a number of regards in order to make it suitable for the task at hand.

\subsection{Spacetime foliations}

First, let us consider how the possibility of a tilted cosmology, which must be considered in the general case of an anisotropic universe, can be accounted for covariantly. 
If we are to model situations in which matter is to have a bulk motion, then we will need to perform our averaging procedure on a set of hypersurfaces that are not necessarily orthogonal to the flow of matter \cite{Umeh_2011}, but just to some well-defined, irrotational normal vector $n^a$\,.
This will require generalising the standard approach, in which the averaging surfaces are usually chosen such that the matter 4-velocity $u^a$ is comoving with the normal to the surfaces. 

Second, we will need to introduce a preferred spatial direction $m^a$, and covariantly decompose all relevant quantities with respect to it, in the $1+1+2$ formalism described in Section \ref{subsec:1+1+2}. Unlike in that section, where both the preferred timelike and spacelike vectors were essentially arbitrary, in this chapter we will need to consider carefully how they can be picked out in the Universe.

By constructing these objects according to some clear physical principles, and then performing the $1+1+2$ decomposition, we can finally average all resultant scalar quantities on our generalised foliation. 
We will then identify backreaction terms by comparing the equations that result to the field equations of the locally rotationally symmetric (LRS) Bianchi or Kantowski-Sachs cosmologies. 
Our many backreaction terms will generalise the single term $\mathcal{Q}$ that occurs when considering only the isotropic part of the expansion, as in Eqs. (\ref{eq_Buchert_1}-\ref{eq_Buchert_2}).
The equations we will arrive at in Section \ref{sec:anisotropic_averaging} will allow us to study if and how anisotropic cosmological expansion and bulk flows could emerge from inhomogeneous spacetimes.

\subsubsection*{Changing foliation}\label{subsec:foliation_changing}

We have so far not specified anything about the timelike vector $n^a\,$, other than it being hypersurface-forming so that it is suitable for performing averaging according to Buchert's approach. 
Later in this section, we will consider specific choices of this vector, which are orthogonal to foliations with certain special properties. 
Here, we will simply note that it will be of interest to be able to transform between foliations, and present some of the mathematics that is required to do so.

In order to understand this, let us consider a boost from $n^a$\,, which defines the orthogonal projection tensor $f_{ab} = g_{ab} + n_a n_b$\,, to some second timelike vector $u^a$, which need not be irrotational in general. The components of these vectors are related by
\begin{equation}\label{eq_boost}
u_a = \gamma\left(n_a - v_a\right) \qquad {\rm and} \qquad n_a = \gamma\left(u_a + w_a\right) \, , 
\end{equation}
where $v_a = \gamma^{-1}f_a^{\ b} w_b$ and $\gamma = \left(1-v^2\right)^{-1/2} = \left(1-w^2\right)^{-1/2}$. Here the boost vectors $v^a$ and $w^a$ exist respectively in the surfaces orthogonal to $n^a$ and $u^a$, such that $n_a v^a = u_a w^a = 0$. 

\begin{figure}
    \centering
    \includegraphics[width=\linewidth]{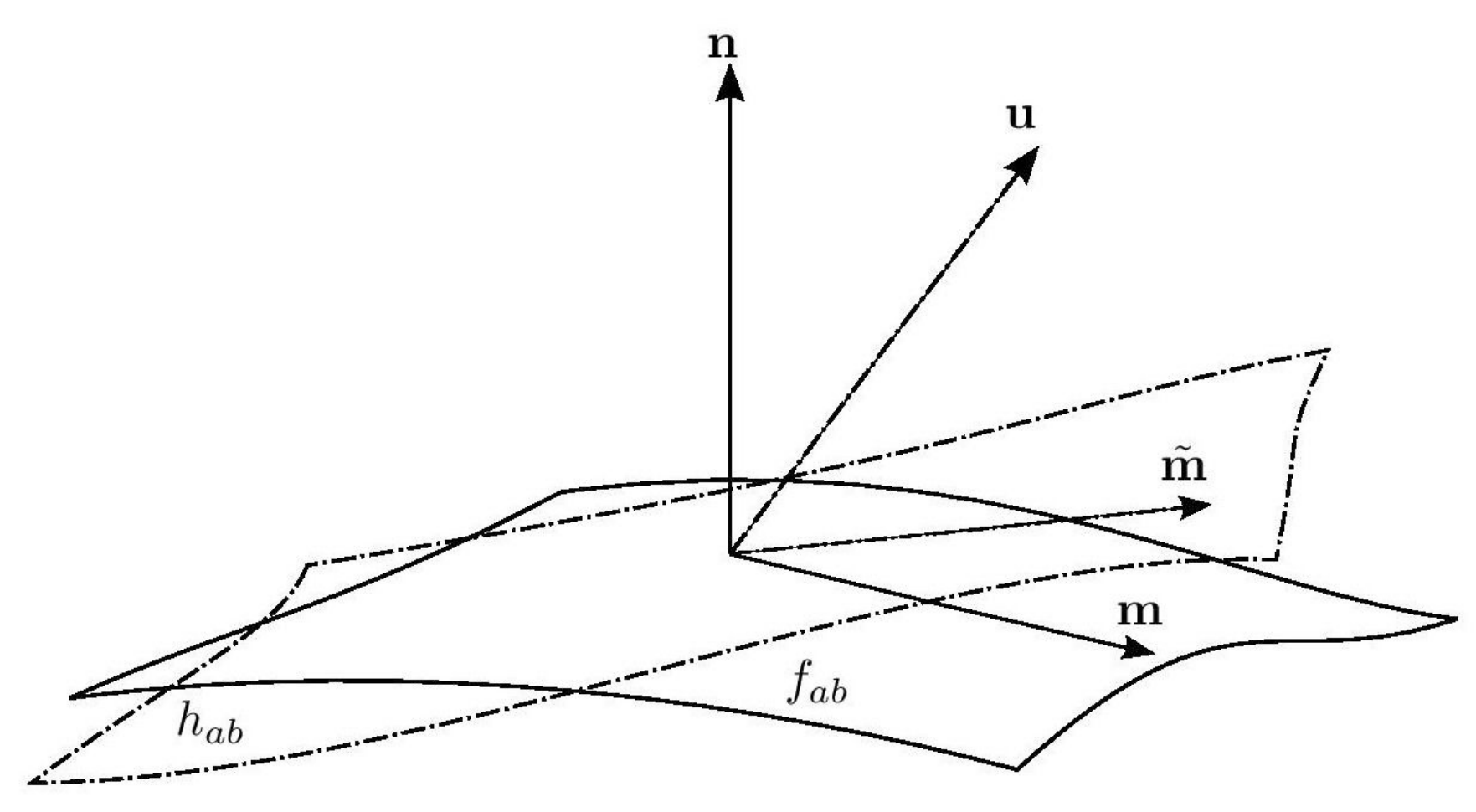}
    \caption{Visualisation of foliations induced by the time-like vectors $n^a$ and $u^a$, with the associated projection tensors $f_{ab}$ and $h_{ab}$, and the preferred space-like vectors $m^a$ and $\tilde{m}^a\,$.}
    \label{fig_tilted_foliations}
\end{figure}

The projection tensor into the instantaneous rest spaces orthogonal to $u^a$ can then be defined as $h_{ab} = g_{ab} + u_a u_b$, which allows us to write $w_a = \gamma^{-1} h_a^{\ b} v_b$. 
The same projection tensor also allows us to define a new vector $\tilde{m}^a$ with components
\begin{equation}\label{eq_projectm}
\tilde{m}_a = k \, h_a^{\ b} m_b \, , \qquad {\rm such \;\; that} \qquad m_a = \tilde{k} \, f_a^{\ b} \tilde{m}_b \,,
\end{equation}
where $k = \left[1+\left(\gamma m_c v^c\right)^2\right]^{-1/2}$ and $\tilde{k} = \left[1+\left(\gamma \tilde{m}_c w^c\right)^2\right]^{-1/2}$. These are the components of the preferred spacelike vector $m^a$ projected orthogonally to $u^a\,$, as illustrated in Figure \ref{fig_tilted_foliations}.

The whole system of $1+1+2$-equations from Section \ref{subsec:1+1+2} can now be set up with respect to either $n^a$ and $u^a\,$, with the preferred spacelike vector projected according to Eq. (\ref{eq_projectm}). 
After we boost from $n^a$ to $u^a\,$, we wish to express the scalars in the new foliation in terms of those in the old one, plus any terms arising from the boost $v^a$ that connects them. 

This can be done by defining the kinematic decomposition of the derivatives of $v^a$ in the usual way, with respective to the original irrotational timelike vector $n^a\,$:
\begin{eqnarray}
    \nabla_a v_b &=& f_a^{\ c}f_b^{\ d}\nabla_c v_d - n_a n^c \nabla_c v_b \\
    \nonumber &=& D_a v_b - n_a \dot{v}_b \\
    \nonumber &=& \frac{1}{3}\kappa \, f_{ab} + \beta_{ab} + W_{ab} - n_a \dot{v}_b\,,
\end{eqnarray}
where $\kappa = D_a n^a$\,, $\beta_{ab} = D_{\langle a} v_{b\rangle}$ and $W_{ab} = \eta_{abc} W^c = D_{[a}v_{b]}$ are respectively the expansion, shear and vorticity associated with the 3-velocity $v^a\,$, as measured by observers comoving with the irrotational congruence with tangent $n^a\,$\,, and we have defined from $W_{ab}$ the vorticity vector $W_a = \frac{1}{2}\eta_{abc}W^{bc}\,$ associated with $v_a\,$. 

If $\left\vert \mathbf{v} \right\vert \ll 1$, which is expected if, for example, $n^a\,$ corresponds to the canonical foliation of FLRW spacetime, and $u^a$ is the matter 4-velocity in that spacetime, then one can expand these expressions order by order in $v^a\,$, leading to vast simplifications. 
However, in the general case, one need not make this approximation. Performing the full calculations, one finds the following results.

The isotropic expansion $\tilde{\Theta} = h^{ab} \nabla_a u_b$ is related to $\Theta = f^{ab}\nabla_a n_b$ by 
\begin{equation}
\tilde{\Theta} = \gamma\left[\Theta - \kappa - v_a\left(\dot{n}^a + \gamma^2 \dot{v}^a\right) + \gamma^2\left(\frac{1}{3}\kappa v^2 + \beta_{ab} v^a v^b\right)\right]\,.
\end{equation}
This equation shows how the isotropic expansion rates of the rest spaces defined with respect to either $n^a$ or $u^a$ are related to each other, and explicitly provides the additional terms that contribute to the expansion after performing a boost. 

The shear scalar $\tilde{\Sigma} = \tilde{m}^a \tilde{m}^b \left(h_a^{\ c} h_b^{\ d} - \frac{1}{3}h_{ab}h^{cd}\right)\nabla_c u_d$ in the new frame is related to its counterpart $\Sigma = m^a m^b \left(f_a^{\ c} f_b^{\ d} - \frac{1}{3}f_{ab}f^{cd}\right)\nabla_c n_d$ by
\begin{eqnarray}
\tilde{\Sigma} &=& k^2\gamma \Sigma + \frac{\gamma}{3}\left(\Theta-\kappa\right)\left(k^2\gamma - 1\right) + \frac{\gamma}{3}v_a\left(\dot{n}^a+\gamma^2\dot{v}^a\right) - \frac{\gamma^3}{3}\left(\frac{\kappa}{3}v^2 + \beta_{ab}v^a v^b\right) \\
\nonumber && + k^2\gamma \Bigg[-\beta_{ab}m^am^b -\left(\gamma m_a v^a\right)^2 +\gamma^2 m^a v^b\left(m_c v^c - 1\right)\left(\beta_{ab}+W_{ab}\right) - \gamma^2 m_a v^a\mathcal{A} \\
\nonumber &&  + \gamma^2 m_b v^b\left(\frac{1}{3}\left(3\Sigma+\Theta-\kappa\right) m_a v^a + \Sigma_a v^a + \dot{v}_a \left(m^a + \gamma^2 m_c v^c v^a\right)\right) 
\\
\nonumber &&
 - \gamma^4 \left(m_c v^c\right)^2\left(\frac{1}{3}\kappa v^2 + \beta_{ab} v^a v^b\right) \Bigg]\,.
\end{eqnarray}
The vorticity and acceleration scalars $\tilde{\Omega} = \frac{1}{2}\tilde{m}^a \eta_{dabc}u^d \nabla^b u^c$ and $\tilde{\mathcal{A}} = \tilde{m}^a u^b \nabla_a u_b$ are respectively
\begin{equation}
\tilde{\Omega} = -k\gamma^2\left[W_a\left(m^a + \gamma^2 m_bv^b v^a\right) - \frac{1}{2}\gamma^2 m_a v^a \chi_{bc}W^{bc}\right]\,,
\end{equation}
and
\begin{eqnarray}
\tilde{\mathcal{A}} &=& k\gamma^2\Bigg[\mathcal{A} - m_a \dot{v}^a - \gamma^2 m_a v^a v_b \dot{v}^b - \frac{1}{3}m_a v^a\left(3\Sigma + \Theta - \kappa\right) \\
\nonumber && \quad - \gamma^2 m_a v^a \left(\frac{1}{3}\kappa v^2 + v^b v^c \beta_{bc}\right) - \Sigma_a v^a + v^a m^b\left(\beta_{ab} + W_{ab}\right)\Bigg]\,,
\end{eqnarray}
where we have used that $\Omega = \frac{1}{2} m^a \eta_{dabc}n^d \nabla^b n^c = 0$\,, because $n^a$ is irrotational by definition, and $\mathcal{A} = m^a n^b \nabla_a n_b\,$. 
We have also defined a novel alternating tensor, $\chi_{ab} = \eta_{dabc} v^c n^d\,$. 
One sees immediately that if the vorticity $W_a$ associated with the relative velocity $v^a$ vanishes, then $\tilde{\Omega} = 0$ as well, which means that $u^a$ can also be used to define a set of hypersurfaces upon which scalar averages can be covariantly calculated.

Meanwhile, the twist and expansion of the two-dimensional screen spaces along $\tilde{m}^a$\,, which are defined by $\tilde{\xi} = \frac{1}{2}\tilde{\epsilon}_{ab} \tilde{M}^{ac}\tilde{M}^{bd}\tilde{D}_c \tilde{m}_d$ and $\tilde{\phi} = \tilde{M}^{ab}\tilde{D}_a \tilde{m}_b$\,, are related to their equivalents in the $n^a$ foliation, $\xi = \frac{1}{2}\epsilon_{ab} M^{ac} M^{bd} D_c m_d$ and $\phi = M^{ab}D_a m_b$\,, by
\begin{equation}
\tilde{\xi} = k^2\gamma\left[\xi + \epsilon_{ab}v^a\left(\alpha^b + \Sigma^b\right)\right]\,,
\end{equation}
and
\begin{eqnarray}
\tilde{\phi} &=& k\phi + k\mathcal{A} - \tilde{A} - k\gamma m_a v^a \tilde{\Theta} + \left(m^a - \gamma^2 m_b v^b\left(n^a - v^a\right)\right)\nabla_a k \\
\nonumber && + k\gamma^2 \Bigg[-\alpha_a v^a + m_a v^a\left(a_b v^b + \frac{\kappa}{3}\right) - m_a\dot{v}^a + \left(\frac{\phi}{2} M_{ab} + \zeta_{ab}\right)v^a v^b \\
\nonumber && + \left(\beta_{ab} + W_{ab}\right) v^a m^b \Bigg] + k\gamma^4 m_c v^c\left(-v_a\dot{v}^a + \frac{\kappa}{3}v^2 + \beta_{ab}v^a v^b\right)\,,
\end{eqnarray}
\begin{eqnarray}
\nonumber {\rm where} \quad \nabla_a k &=& - \gamma^2 k^3 m_c v^c \Bigg[\gamma^2 m_b v^b v^d\left(-n_a \dot{v}_d + \beta_{ad} + W_{ad}\right) + \frac{\gamma^2}{3}m_b v^b \kappa v_a - \alpha_b v^b n_a \\
\nonumber && + v^b\left(\frac{\phi}{2}M_{ab} + \zeta_{ab} + \xi \epsilon_{ab} + m_a a_b\right) - \dot{v}_b m^b n_a + \frac{\kappa}{3}m_a + m^b\left(\beta_{ab} + W_{ab}\right)\Bigg]\,.
\end{eqnarray}

For the matter variables, the energy density $\tilde{\rho} = T_{ab}u^a u^b$, isotropic pressure $\tilde{p} = \frac{1}{3}T_{ab}h^{ab}$, and scalar momentum density $\tilde{Q} = -T_{bc}u^b h^c_{\ a} \tilde{m}^a$ are given by
\begin{eqnarray}
\tilde{\rho} &=& \gamma^2\left(\rho + pv^2 + 2q_a v^a + \pi_{ab}v^av^b\right)\,, \\
\tilde{p} &=& \left(1+\frac{1}{3}\gamma^2 v^2\right) p + \frac{1}{3}\gamma^2 v^2 \rho + \frac{1}{3}\gamma^2\left(2q_a v^a + \pi_{ab}v^a v^b\right)\,, \\
\tilde{Q} &=& k\gamma\left[Q + m_c v^c\left(\gamma^2\left(\rho +p+2q_a v^a + \pi_{ab}v^a v^b\right) + \Pi\right) + \Pi_a v^a \right]\,,
\end{eqnarray}
and the scalar anisotropic stress $\tilde{\Pi} = \tilde{m}^a \tilde{m}^b\left(h_a^{\ c} h_b^{\ d} - \frac{1}{3}h_{ab}h^{cd}\right)T_{cd}$ is given by
\begin{eqnarray}
\tilde{\Pi} &=& \gamma^2 \rho\left(\left(k\gamma m_a v^a\right)^2 - \frac{1}{3}v^2\right) \\
\nonumber && + p \left[ k^2 - 1 + 2\left(k\gamma m_a v^a\right)^2   + \frac{1}{3}\gamma^2 v^2\left(3\left(k\gamma m_a v^a\right)^2 - 1\right)\right] \\
\nonumber && + 2k^2\gamma^2 m_a v^a Q + k^2 \Pi\left(1+2\left(\gamma m_a v^a\right)^2\right) \\
\nonumber && + \frac{\gamma^2}{3}\left(3\left(k\gamma m_c v^c\right)^2-1\right)\left[2q_a v^a + \pi_{ab} v^a v^b\right]\,,
\end{eqnarray}
where $\rho = T_{ab} n^a n^b\,$, $p = \frac{1}{3}T_{ab}f^{ab}\,$, $Q = -T_{bc} n^c f^c_{\ a} m^a$\,, and $\Pi = m^a m^b \left(f_a^{\ c}f_b^{\ d} - \frac{1}{3}f_{ab}f^{cd}\right)T_{cd}\,$.

Finally, one can compute the scalar parts $\tilde{\mathcal{E}} = \tilde{E}_{ab} \tilde{m}^a \tilde{m}^b$ and $\tilde{\mathcal{H}} = \tilde{H}_{ab} \tilde{m}^a \tilde{m}^b$ of the Weyl curvature for the new congruence, compared to $\mathcal{E} = E_{ab} m^a m^b$ and $\mathcal{H} = H_{ab} m^a m^b\,$. 
This gives \cite{ellis2012relativistic} 
\begin{eqnarray*}
\tilde{\mathcal{E}} &=& k^2\gamma^2\left[\left(1+v^2 - \left(m_a v^a\right)^2 - \frac{1}{2}M_{ab}v^a v^b\right)\mathcal{E} + 2\chi^{ab}v_c m_a\mathcal{H}_b + \mathcal{E}_{ab}v^a v^b\right]\,, \\
\tilde{\mathcal{H}} &=& k^2\gamma^2\left[\left(1+v^2 - \left(m_a v^a\right)^2 - \frac{1}{2}M_{ab}v^a v^b\right)\mathcal{H} - 2\chi^{ab}v_c m_a\mathcal{E}_b + \mathcal{H}_{ab}v^a v^b\right]\,.
\end{eqnarray*}
This completes the transformation rules for all of our covariantly defined scalars. The vector and tensor quantities in the $1+1+2$ decomposition can also be transformed between foliations, but only appear in our averaged description via backreaction terms. We therefore omit presenting their transformation rules here. 
In the non-relativistic limit $\left\vert\mathbf{v}\right\vert \ll 1\,$, the above expressions simplify dramatically, to the following linearised set:
\begin{eqnarray*}
    \tilde{\Theta} &=& \Theta - \kappa - \dot{n}_a v^a \\
    \tilde{\Sigma} &=& \Sigma + \frac{1}{3}\dot{n}_a v^a - m^a m^b \beta_{ab} - \mathcal{A}m_a v^a \\
    \tilde{\Omega} &=& - W_a m^a \\
    \tilde{\mathcal{A}} &=& \mathcal{A} - m_a \dot{v}^a - \frac{1}{3}m_a v^a\left(\Theta + 3\Sigma\right) - \Sigma_a v^a \\
    \tilde{\xi} &=& \xi + \epsilon_{ab}v^a\left(\alpha^b + \Sigma^b\right) \\
    \tilde{\phi} &=& \phi + m_a \dot{v}^a - m_a v^a\left(\frac{2}{3}\Theta - \Sigma\right) \\
    \tilde{\rho} &=& \rho + 2q_a v^a \\ 
    \tilde{p} &=& p + \frac{2}{3}q_a v^a \\
    \tilde{Q} &=& Q + \Pi_a v^a + m_a v^a\left(\rho + p + \Pi\right) \\
    \tilde{\Pi} &=& \Pi + 2 v^a\left(Q m_a - \frac{1}{3}q_a\right) \\
    \tilde{\mathcal{E}} &=& \mathcal{E} \\
    \tilde{\mathcal{H}} &=& \mathcal{H}
\end{eqnarray*}
This is often the situation one is dealing with in standard cosmological models, where different possible timelike congruences are usually related to one another by a perturbatively small $v^a$ that can be interpreted as a peculiar velocity relative to the chosen canonical foliation.

In what follows, we will use often use $u^a$ to refer to the flow lines of an irrotational dust fluid. In the presence of such matter, this vector is uniquely defined and therefore also provides a unique foliation on which to perform averaging.
This is the foliation that is most often used in the literature on this subject, and is the one chosen for the standard approach to formulating Buchert's equations \cite{Buchert_2000}. 
We will then use $n^a$ to refer to any other well-defined spacetime foliation, that might be chosen, for example, for observational or geometric reasons. Such freedom is required if we are to allow for the possibility of bulk flows, as the 3-velocity of matter vanishes by construction, if we choose the foliation to be induced by the flow of dust $u^a\,$. 
It also allows us the freedom to refoliate in situations in which the description of spacetime might break down if it were specified by the fluid flow.
This would happen in perturbed FLRW cosmologies that contain nonlinear structures \cite{Clifton_2020}, or if one wished to consider fluids with non-zero vorticity or caustics in their flow lines \cite{umeh2023vorticity}.

\subsubsection*{Choice of foliation}\label{subsec:foliation_choices}

An application of the Buchert averaging procedure requires a choice of foliation, or equivalently a choice of the irrotational timelike vector $n^a = - N \nabla_a t$\,, whose orthogonal spaces define the constant-$t$ leaves of the foliation. 
The averages obtained using Buchert's scheme (\ref{eq_Buchert_averaging_procedure}) will then correspond to the large-scale properties of the 3-dimensional spaces that constitute the leaves. Different foliations will mean that one is considering different 3-dimensional spaces, and hence different averages will be obtained. 
It is therefore necessary to make sure an appropriate choice of foliation is made, for the situation being considered. This is given further importance by the fact that observers in different frames will infer different cosmological parameters from the Hubble flow around them \cite{Tsagas_2009,Tsagas_2011,tsagas2015deceleration,tsagas2022deceleration,Santiago_2022,colin2019evidence,Mohayaee_2020,Mohayaee_2021}.

In general, one might be interested in choosing a foliation that is expected to give results that can be associated with a particular observable \cite{rasanen2009light}, that has a particular mathematical or physical meaning associated with it \cite{rendall1996constant}, or that is perhaps convenient in some other way \cite{Umeh_2011}. For example, one may wish to construct Hubble diagrams in a frame comoving with the flow of matter \cite{Buchert_2000}, or in a frame of ``most uniform Hubble flow'' \cite{Wiltshire_2013, McKay_2015, Kraljic_2016}. 
Of course, in order to relate observables corresponding to quantities calculated on different foliations one will need to be able to transform between frames\footnote{We recall that foliation dependence should not be confused with gauge dependence, as choice of foliation is in general a covariant and non-perturbative process, meaning that no background manifold is defined and that there is therefore no gauge issue \cite{Buchert_2012}.}, as we considered above.

Let us now consider some specific choices of foliation that one might use.

\begin{itemize}

    \item {\it The comoving foliation} exists when the Universe is filled with irrotational dust, and $n^a$ is chosen to be coincident with the 4-velocity $u^a$ of that dust. This choice shares some properties with the comoving synchronous gauge that is often used in cosmological perturbation theory about an FLRW background. 
    The comoving foliation has the distinct benefit of being tied to a physical quantity, the large-scale flow of matter. It therefore takes a particularly privileged position in the pantheon of possible choices, but does require the matter flow to be vorticity-free, which is not expected to hold for realistic astrophysical structures. 
    It also cannot account for the existence of any bulk flow, as it corresponds to the choice of frame in which matter is at rest. Nevertheless, this is the standard choice of foliation in much of the literature on mathematical cosmology \cite{Ellis_1999}, as well as in many studies of perturbation theory. 
    Indeed, Einstein-Boltzmann solvers are typically run in synchronous gauge as a default \cite{lesgourgues2011cosmic,zumalacarregui2017hi_class}, which corresponds at the perturbative level to a comoving foliation of spacetime.

    \item {\it The gravitational rest frame} was defined in Ref. \cite{Umeh_2011} to be that in which the magnetic part of the Weyl tensor vanishes, i.e. 
    \begin{equation*}
        H_{ab} = 0\,,
    \end{equation*}
    so that the emergent cosmological model is as Newtonian in character as possible \cite{Maartens_1998}, and is described as silent \cite{bruni1994dynamics,van_Elst_1997}. 
    Up to second order in CPT this choice corresponds to purely scalar perturbations in Newtonian gauge \cite{Umeh_2011}.
    However, when applied non-perturbatively, the condition of silence ends up being a strong restriction on the allowed solutions \cite{Barnes_1989, van_Elst_1997}. 
    It therefore appears to be particularly useful for studies of perturbed FLRW models, as the Newtonian gauge is one of the few that is expected to be valid into the regime of nonlinear structure formation, where post-Newtonian expansions are required to describe weak gravitational fields \cite{Clifton_2020}, but it is likely to be overly restrictive in a more general setting, particularly as it explicitly discounts the existence of gravitational radiation \cite{Dunsby_1997}.

    \item {\it The gravitational wave frame} is a more conservative choice defined by the demand that $H_{ab}$ is divergence-free, such that
    \begin{equation*}
        D^b H_{ab} = 0\,.
    \end{equation*}
    This is a covariant way of stating that the only gravitomagnetic contributions to the curvature measured by an observer comoving with  $n^a$ would be those coming from gravitational waves \cite{Hawking_1966, Dunsby_1997}. 
    It is therefore less restrictive than the condition $H_{ab} = 0$, and typically corresponds to a congruence which is distinct from the flow of dust \cite{Sopuerta_1999}. It is well-suited to numerical-relativistic cosmological simulations \cite{Heinesen_2022}, but in a general spacetime it is not guaranteed that the frames it picks out are hypersurface-forming (though one may be able to take the irrotational part of $n^a$\,, which would be hypersurface-forming by construction).

    \item {\it The constant mean curvature foliation} has a long history \cite{bartnik1988remarks,rendall1996constant}, and is defined by the condition that the spatial gradient of the expansion scalar (also known as the ``mean curvature'') vanishes. That is,
    \begin{equation*}
        D_a \Theta = 0\,.
    \end{equation*}
    This foliation has the property of being unique (under certain circumstances), and having a monotonic variation in the expansion scalar between leaves \cite{rendall1996constant}. 
    It therefore provides a plausible candidate for the provision of a universal arrow of time. The literature on this particular choice of foliation has been largely restricted to mathematical cosmology, but it also exists in perturbed FLRW models under the name of uniform expansion gauge \cite{Malik_2009}.

    \item {\it The zero-shear foliation} is specified by the condition 
    \begin{equation*}
       \sigma_{ab} = 0\,.
    \end{equation*}
    It is closely related to the gravitational rest frame, in that the Ricci identities provide the constraint (\ref{eq_Hweyl_vs_shear_vorticity}) that sets $H_{ab}$ equal to the covariant curl of the shear tensor, for an irrotational and geodesic timelike congruence. 
    Vanishing of the shear therefore corresponds to the vanishing of $H_{ab}$, which explains why the Newtonian gauge in CPT, which is also sometimes called the zero-shear gauge \cite{noh2012cosmological}, corresponds to the gravitational rest frame. 

    \item {\it The constant density foliation} is defined by the requirement that the spatial gradient of the energy density vanishes,
    \begin{equation*}
        D_a \rho = 0\,.
    \end{equation*}
    As with the comoving foliation, this has the benefit of being directly tied to the matter content of the spacetime. However, it also shares the weakness that it will be highly problematic in the presence of nonlinear structures \cite{Clifton_2020}.
    
\end{itemize}

The list of choices presented above is by no means exhaustive, and what one might consider to be an appropriate choice of foliation for a given situation is generally dependent on the physical context, as well as practical considerations, such as whether one is performing analytical or computational calculations.

\subsubsection*{Choice of preferred spatial direction}\label{subsec:preferred_direction_choices}

In order to extract scalars from anisotropic quantities, we also need to choose a preferred spacelike direction $m^a$\,. All vector and tensor quantities can be decomposed with respect to this direction, as prescribed by Eqs. (\ref{eq_vdef}-\ref{eq_tdef}). Then, the scalar parts of the decomposed quantities can be averaged according to Eq. (\ref{eq_Buchert_averaging_procedure}). 
In order for the procedure to make sense, this vector must (in some sense) point in the same direction at every point in an averaging domain. Just as in the case of the timelike vector $n^a\,$, this presents a choice. 
There is in general not going to be any single uniquely preferred spacelike vector $m^a\,$, and we will need to select a suitable way to define this direction based on the situation at hand. This choice will be important for the outcome of our averaging process.

As in the case of choosing $n^a\,$, we may wish to apply geometric or observational considerations when selecting a preferred direction $m^a\,$. In spacetimes with symmetry, it may be that a preferred direction is selected by the existence of some Killing vectors. 
For example, in LRS geometries $m^a$ can unambiguously be taken to correspond to the spatial axis around which the rotational symmetry exists. 
This can be identified by searching for the non-degenerate eigenvector of any non-vanishing $1+3$-covariant tensor, such as $\sigma_{ab}$ or $E_{ab}$ \cite{Clarkson_2007, vanElst_1996}. 
Alternatively, in algebraically special geometries it may be possible to pick out a unique direction from the projections of the canonical null tetrad into our chosen foliation. These will be specified by the properties of the Weyl tensor, through the five complex Newman-Penrose scalars that describe different physical effects encoded in $C_{abcd}$ \cite{Newman_1961,pirani1961geometrical,Szekeres_1965, Stephani_2003}.

An alternative method for selecting a preferred spatial direction would be to use observational methods at each point within an averaging domain. This method would be better adapted to situations in which anisotropy in a particular observable is being considered, or where the spacetime has no explicit symmetries or special algebraic properties (as is the case for the real Universe).
For example, one might choose to take $m^a$ to correspond to the axis of CMB parity asymmetry \cite{Zhao_2014,Cheng_2016}, the direction of greatest asymmetry in the galaxy distribution \cite{Migkas_2021,Siewert_2021}, the quasar or Type Ia supernovae dipole \cite{Secrest_2021,Singal_2022_a}, or the direction of greatest variation of a cosmological parameter or coupling constant such as $H_0$ or the fine structure constant $\alpha$ \cite{Webb_2011, King_2012, Yeung_2022}. 
As long as these directions line up at different points in space, as one would expect in an anisotropic universe, then they should provide a well-motivated choice of preferred direction.

\section{Averaging in anisotropic universes}\label{sec:anisotropic_averaging}

Let us now turn to the question of how to construct an emergent cosmological model for an anisotropic universe. To do this, we will make use of Buchert's scalar averaging procedure that we discussed in Chapter 4. 
It is characterised by the spatial averaging rule (\ref{eq_Buchert_averaging_procedure}).
In Buchert's approach, which is focused on averaging to an isotropic emergent model that satisfies Friedmann-like equations, the spacetime is foliated into hypersurfaces $\Sigma_t$ which are orthogonal to the irrotational and geodesic flow of pressureless dust. 
This means that the lapse function $N$, between the leaves of the chosen foliation, is independent of the spatial coordinates $x^i$ on $\Sigma_t$\,. Each leaf in the foliation can then be identified with a single value of the proper time $t$ along the worldlines of the dust fluid, and will be the same at all points in each spatial hypersurface.
The lapse may therefore be set to unity \cite{Buchert_2018}, and the shift vector $N^i$ set to zero, which leads to the commutation rule (\ref{eq_Buchert_commutation_rule}) between spatial averaging and differentiation with respect to coordinate time that we introduced earlier in the thesis.

However, with a more general matter distribution, it is no longer typically the case that we can construct a well-defined congruence $n^a$ with unit lapse, or even that we would wish to. 
Rather, we will have 
\begin{equation}
    \dot{n}_a = n^b \nabla_b n_a = - D_a \ln{N} \, \neq 0\,,
\end{equation}
meaning that the proper time of a timelike congruence of observers following some $n^a$ is not constant across an averaging domain $\mathcal{D}\,$.

In principle, one could simply proceed with the Buchert commutation rule (\ref{eq_Buchert_commutation_rule}), but just upgrade the coordinate time derivative to covariant time derivatives $n^a \nabla_a\,$, and accept that the surfaces orthogonal to $n^a$ are not labelled by a single parameter $t\,$. 
However, for the purposes of calculating cosmological averages and interpreting them physically, this is clearly far from desirable. 
Instead, we hope that we can construct quantities which are defined over a spatial averaging domain $\mathcal{D}$ to have associated with them a single value for the time coordinate $t$\,. 
This can be achieved by writing $n_a = - N \nabla_a t$ where $N (t, x^i) = \left(\dot{t}\right)^{-1} = \left(n^b \nabla_b t\right)^{-1}$\,, such that the partial derivative of a scalar with respect to $t$ can be written as $\partial_t S = N \dot{S}$. 
The commutation rule (\ref{eq_Buchert_commutation_rule}) then becomes
\begin{equation}\label{eq_commutation_rule_lapseshift}
\partial_t \avg{S} - \avg{\partial_t S} = \avg{N\Theta S} - \avg{N\Theta}\avg{S} \equiv {\rm Cov}\left(N\Theta, S\right)\,,   
\end{equation}
which allows us to calculate the evolution of scalar averages as a function of the coordinate time $t$\,, once the lapse function $N$ is determined from the acceleration of $n^a\,$.

Hence, we can construct averaged equations by applying the averaging rule (\ref{eq_Buchert_averaging_procedure}) and commutation rule (\ref{eq_commutation_rule_lapseshift}) to any covariantly defined scalars we wish. 
Following the same ethos as Buchert, we can then obtain from that non-locally averaged scalar equation a new one that is written entirely in terms of averages and their derivatives with respect to time. 
Equations derived in this way can be thought of as governing the large-scale, emergent, behaviour of averages, rather than the behaviour of locally-defined quantities (as usually occurs in theories of gravity). 
Finally, we can write the emergent equations that result from this procedure in a form identical to that of the local gravitational equations of an homogeneous cosmology. Any extra terms that occur can be explicitly tied to the phenomenon of cosmological backreaction from inhomogeneous structures that exist below the averaging scale.

What we need to do, however, is identify a suitable set of scalars that can be used to study the emergence of anisotropic quantities. Fortunately, we have already explained in this thesis that such a set of scalars exists: it is provided by the $1+1+2$-covariant decomposition of General Relativity, as described in Section \ref{subsec:1+1+2}.
Here, we will work out how an anisotropic cosmological model can be obtained from the averages of $1+1+2$-covariant scalars. 
It should be noted that an interesting step in this direction was already taken by Barrow and Tsagas, who extended the Buchert formalism to include an evolution equation for the shear scalar in the $1+3$ formalism, $\sigma^2 = \frac{1}{2} \sigma_{ab}\sigma^{ab}$ \cite{Barrow_2007}. 
However, there are several more objects that describe anisotropic cosmologies that we ought to consider, in order to tackle the problem in full generality.
Therefore, we will extend the result in Ref. \cite{Barrow_2007} further by deriving the full set of averaged $1+1+2$-covariant scalar equations, and the backreaction scalars that describe how the evolution of those large-scale averages is affected by small-scale inhomogeneities.

\subsection{Interpretation as an LRS Bianchi cosmology}\label{subsec:interpretation_Bianchi_LRS}

Let us now consider what the model that emerges from averaging the scalars in the 1+1+2 decomposition will represent. These will be precisely the set of scalars presented in Eq. (\ref{eq_112scalars}), except for $\Omega$\,, which vanishes because we consider only irrotational timelike vectors $n^a\,$. 

The averaged scalars that result from the 1+3-decomposition, as provided in Buchert's ground-breaking approach \cite{Buchert_2000}, are relatively straightforward to interpret. 
They result in the equations (\ref{eq_Buchert_1}) and (\ref{eq_Buchert_2})\footnote{Plus an averaged energy conservation equation and an integrability condition on the time evolution of $\avg{^{(3)}R}$ \cite{Buchert_2012}.} that govern the averages of the expansion scalar $\avg{\Theta}\,$, the energy density of dust $\avg{\rho}\,$, and the curvature scalar of the 3-spaces $\avg{^{(3)}R}\,$, with everything else being collected together into backreaction terms.
The Buchert equations (\ref{eq_Buchert_1}) and (\ref{eq_Buchert_2}) bear a striking resemblance to the dust-dominated Friedmann equations. This is no accident, as the three averaged scalars that arise are precisely those which are required to fully characterise a dust-dominated, homogeneous and isotropic universe in the $1+3$ formulation of General Relativity.
All other $1+3$ covariantly defined quantities will be vectors or tensors, but these must vanish in geometries that are isotropic around every point in space \cite{Ellis_1999}. So, after discarding such quantities in the scalar averaging approach, we are left with exactly the set that is required to describe an FLRW model. One could say that the geometry has been averaged to an FLRW cosmology.

If we now consider the scalars that result from the 1+1+2-decomposition, as given in Eq. (\ref{eq_112scalars}), then we have a more complicated situation. It can, however, be understood in terms of the locally rotationally symmetric (LRS) cosmologies \cite{Ellis_1967, Stewart_1968} that we mentioned in Section \ref{subsec:anisotropic_models}.
We remind the reader that a locally rotationally symmetric spacetime is one in which every point has associated with it a single preferred spacelike direction, about which local $U(1)$ rotations leave the geometry fixed. They do not need to be homogeneous in general: for example, the inhomogeneous, spherically symmetric Lema{\^ i}tre-Tolman-Bondi models (including the Schwarzschild metric as its vacuum case) fall under the LRS umbrella. 
For the remainder of this thesis, we refer to homogeneous LRS Bianchi spacetimes as LRS cosmologies.
The one-dimensional continuous isotropy group of an LRS cosmology essentially corresponds in the $1+1+2$ formalism to rotations about the local symmetry axis which is defined precisely by the preferred spacelike vector $m^a\,$. 
The existence of such a symmetry means that all $1+3$ covariant vectors are proportional to $m^a\,$, and all projected symmetric and trace-free $1+3$-covariant tensors are proportional to $m_a m_b - \frac{1}{2}M_{ab}$, where $M_{ab}$ is the 2-space projection tensor defined by Eq. (\ref{eq_bigm}). In the language of the $1+1+2$-decomposition described in Section \ref{subsec:1+1+2}, this means that all $1+1+2$-covariantly defined vectors and tensors must vanish. 

The entire dynamics of an LRS spacetime is therefore described purely by the $1+1+2$-covariant scalars. 
For this reason, we expect to recover a set of equations for our averaged scalars that can be written in a form similar to the equations that govern LRS cosmologies, with additional backreaction terms due to the averaging of small-scale structure. We will therefore say that we are averaging to an LRS Bianchi cosmology.
Due to the homogeneity of Bianchi spacetimes, the only non-zero derivative of any of the scalars $S$ within these models will be their coordinate time derivative. 

Of course, this assumes that we are foliating the spacetime into successive surfaces of transitivity of the Bianchi spacetime's group of isometries, with the fundamental timelike vector in the $1+3$ and $1+1+2$ decompositions being the normal $n^a$ to those surfaces.
An implication of this is that, because the Bianchi spacetimes can be tilted, with the matter 4-velocity $u^a \neq n^a\,$, the energy-momentum tensor of a perfect fluid $T_{ab} = \tilde{\rho} \,u_a u_b + \tilde{p}\, h_{ab}$ will give rise to an apparent imperfect fluid description in the homogeneous hypersurfaces, $T_{ab} = \rho n_a n_b + p f_{ab} + 2 q_{(a}n_{b)} + \pi_{ab}\,$, with $q_a \neq 0$ and $\pi_{ab} \neq 0$ in general. 
This, however, will be a necessary complication, because we do not want to have any spatial derivatives in our final results. After all, the fundamental utility of averaged cosmological equations is that one deals only with a set of ordinary differential equations in some time coordinate $t\,$, rather than the full set of PDEs that arise in any formulation of General Relativity, including the $1+3$ and $1+1+2$-covariant formalisms.

All the governing equations will then necessarily have the form
\begin{equation}\label{eq_general_LRS_scalar}
a_i \, N^{-1}\,\partial_t S_i + \sum_j b_{ij} S_j +\sum_{j,k} c_{ijk} S_j  S_k = 0\,.
\end{equation}
Here $S_i$ label the allowed scalars from the $1+1+2$ decomposition. The objects $b_{ij}$ and $c_{ijk}$ are constants, and $a_i = 0$ or $1$ depending on whether the equation is a constraint or an evolution equation. Note that the subscripts in this expression are just labels, not tensor indices, and so repeated indices should not be summed over unless explicitly stated. 
The complete set of equations is the following 15, consisting of 8 ODEs in time and 7 algebraic constraints:
\begin{eqnarray*}
N^{-1}\,\partial_t \Theta - \mathcal{A}\left(\mathcal{A}+\phi\right) + \frac{1}{3}\Theta^2 + \frac{3}{2}\Sigma^2 + \frac{1}{2}\left(\rho + 3p\right) - \Lambda &=& 0 \,,\\
Q + \frac{3}{2}\phi\Sigma &=& 0\,, \\
N^{-1}\,\partial_t \Sigma + \frac{2}{3}\Theta\Sigma + \frac{1}{2}\Sigma^2 + \mathcal{E} - \frac{1}{2}\Pi - \frac{1}{3}\left(\mathcal{A}-\phi\right)\mathcal{A} &=& 0 \,, \\
\mathcal{A}\xi &=& 0\,, \\
\frac{2}{9}\Theta^2 - \frac{1}{2}\phi^2 + \frac{1}{3}\Theta\Sigma - \Sigma^2 - \frac{2}{3}\rho - \frac{2}{3}\Lambda - \frac{1}{2}\Pi - \mathcal{E} + 2\xi^2 &=& 0\,, \\
N^{-1}\,\partial_t \phi - Q - \left(\frac{2}{3}\Theta - \Sigma\right)\left(\mathcal{A}-\frac{1}{2}\phi\right) &=& 0\,, \\
\mathcal{H} - 3\xi\Sigma &=& 0\,, \\
N^{-1}\,\partial_t \xi + \frac{1}{2}\left(\frac{2}{3}\Theta-\Sigma\right)\xi - \frac{1}{2}\mathcal{H} &=& 0\,, \\
\phi\xi &=& 0\,, \\
N^{-1}\,\partial_t \rho + \Theta\left(\rho + p\right) + \frac{3}{2}\Sigma\Pi + \left(\phi + 2\mathcal{A}\right)Q &=& 0\,, \\
N^{-1}\,\partial_t Q + \left(\frac{4}{3}\Theta + \Sigma\right)Q - \mathcal{A}\left(\rho + p + \Pi\right) + \frac{3}{2}\Pi\phi &=& 0\,, \\
N^{-1}\,\partial_t \mathcal{E} + \frac{1}{2}N^{-1}\,\partial_t \Pi + \left(\Theta - \frac{3}{2}\Sigma\right)\mathcal{E} \qquad \qquad \qquad \qquad \qquad && \\
+ \frac{1}{2}\left(\frac{1}{3}\Theta + \frac{1}{2}\Sigma\right)\Pi - \frac{1}{3}\left(\frac{1}{2}\phi - 2\mathcal{A}\right)Q + \frac{1}{2}\left(\rho + p\right)\Sigma - 3\xi\mathcal{H} &=& 0\,, \\
N^{-1}\,\partial_t \mathcal{H} + \left(\Theta - \frac{3}{2}\Sigma\right)\mathcal{H} + 3\xi\left(\mathcal{E}-\frac{1}{2}\Pi\right) &=& 0\,, \\
\frac{3}{2}\left(\mathcal{E}+\frac{1}{2}\Pi\right)\phi + \left(\frac{1}{3}\Theta - \frac{1}{2}\Sigma\right)Q &=& 0\,, \\
{\rm and} \qquad \frac{3}{2}\phi\mathcal{H} + Q\xi &=& 0\,.
\end{eqnarray*}

These equations are a significant simplification of the full set of equations of motion, and will be the form we expect for the equations that govern the dynamics of the spatially averaged $1+1+2$-covariant scalars, with the caveat that we will need to multiply the scalars by appropriate factors of the lapse function, in order to correctly account for the presence of a non-unit lapse $N$ in the commutation rule (\ref{eq_commutation_rule_lapseshift}).

Specifically, those averaged equations will be of the form
\begin{equation}
    a_i \,\partial_t\avg{N^{p_i} S_i} + \sum_j b_{ij}\,\avg{N^{p_i + 1}S_j} + \sum_{j,k} c_{ijk} \,\avg{N^{p_j} S_j}\avg{N^{p_k} S_k} = \mathcal{B}_i\,,
\end{equation}
where the exponent $p_i$ of the lapse function $N$ depends on the scalar being considered: for the kinematic scalars $\left\lbrace \Theta,\Sigma,\mathcal{A},\phi,\xi\right\rbrace$, $p_i = 1$, and for the matter and Weyl curvature scalars, $\left\lbrace \rho, p, Q, \Pi, \mathcal{E}, \mathcal{H}\right\rbrace$, $p_i = 2\,$\,, and where $p_j + p_k = p_i + 1\,$.

One can see that the only difference between the local equations in the exact LRS case, and the non-local averaged equations in the effective, emergent, LRS case, will be that we will find scalar backreaction terms on the right-hand side of each equation, instead of zero. 
These can be interpreted just like Buchert's backreaction scalar $\mathcal{Q}\,$: they describe the integrated effects of small-scale inhomogeneities on the large-scale cosmic expansion.
Let us now derive the emergent equations, and the backreaction scalars that source them.

\subsection{Emergent equations of motion}\label{subsec:emergent_equations}

In this section we will present the explicit form of the equations governing the averaged scalars from the $1+1+2$ formalism. These equations are valid for any hypersurface-orthogonal $n^a\,$, and any choice of spacelike vector $m^a$ in the orthogonal hypersurfaces, although they are of course most useful if the chosen $n^a$ and $m^a$ have well-motivated physical meaning.
They are obtained through applying the averaging rule (\ref{eq_Buchert_averaging_procedure}), and commutation relation (\ref{eq_commutation_rule_lapseshift}), to the scalar equations in Section \ref{subsec:1+1+2}, and finally making sure that all time derivatives are with respect to coordinate time. 
This is required so that they can be more easily interpreted physically, as $t$ takes the same value at every point in any spatial averaging domain $\mathcal{D}$ of a given $\Sigma_t\,$.
We will start with the averaged scalar equations that come from the Ricci identities for $n^a$, then the equations that come from the Ricci identities for $m^a\,$, and finally the Bianchi identities.

Let us start with the $1+1+2$ formulation of the Raychaudhuri equation, Eq. (\ref{eq_Raychaudhuri_112}), for the evolution of the expansion. After averaging, we get
\begin{eqnarray}\label{eq_averaged_Raychaudhuri}
\partial_t \avg{N\Theta} - \avg{N\mathcal{A}}\left(\avg{N\mathcal{A}} + \avg{N\phi}\right) + \frac{1}{3}\avg{N\Theta}^2 + \frac{3}{2}\avg{N\Sigma}^2 && \\
\nonumber 
+\frac{1}{2}\avg{N^2\left(\rho + 3p\right)} - \avg{N^2}\Lambda &=& \mathcal{B}_1\,.
\end{eqnarray}
This equation is identical to the corresponding equation from LRS Bianchi cosmologies up to the backreaction term $\mathcal{B}_1$, and the presence of the lapse function $N$, which cannot in general be set to unity.
The backreaction term is
\begin{eqnarray*}\label{eq_backreaction_scalar_1}
\mathcal{B}_1 &=& \frac{2}{3}{\rm Var}\, N\Theta  - \frac{3}{2}{\rm Var}\,N\Sigma  + {\rm Var}\,N\mathcal{A} + {\rm Cov}\left(N\mathcal{A},N\phi\right)  -2\avg{N^2\Sigma_a \Sigma^a} - \avg{N^2 \Sigma_{ab}\Sigma^{ab}} \\
&& + \avg{N^2 m^a D_a \mathcal{A}} + \avg{N^2 M^{ab}D_a\mathcal{A}_b} + \avg{N^2\left(\mathcal{A}_a-a_a\right)\mathcal{A}^a} + \avg{N\Theta \partial_t \ln{N}}\,,
\end{eqnarray*}
which encodes all information about the influence of small-scale inhomogeneities on the acceleration of the expansion of space. Here, we recall the definitions of covariance and variance, which we introduced in Section \ref{subsec:averaging_problem} as being ${\rm Cov}\left(S_1, S_2\right) \equiv \avg{S_1 S_2}-\avg{S_1}\avg{S_2}$ and ${\rm Var}\,S \equiv {\rm Cov}\left(S, S\right)\,$. 
This is the only equation that can be derived from $S_{abc}$ that has a counterpart in the standard approach, pioneered by Buchert, of averaging to isotropic cosmology. 
As with $\mathcal{Q}$ in Eq. (\ref{eq_Buchert_2}), a sufficiently large (and positive) $\mathcal{B}_1$ would lead to an accelerating universe, at least within the spatial domain being considered.

The scalar part of the momentum constraint in the $1+1+2$ formalism, Eq. (\ref{eq_momentum_constraint_112}), gives
\begin{eqnarray}\label{eq_mom_constraint_avg}
\avg{N^2 Q} + \frac{3}{2}\avg{N\phi}\avg{N\Sigma} = \mathcal{B}_2 \,,
\end{eqnarray}
where the backreaction term is given by
\begin{eqnarray*}\label{eq_backreaction_scalar_2}
\mathcal{B}_2 &=& \frac{2}{3}\avg{N^2 m^a D_a \Theta} - \avg{N^2 m^a D_a \Sigma} - \frac{3}{2}{\rm Cov}\left(N\phi,N\Sigma\right)  - \avg{N^2 M^{ab} D_a \Sigma_b} \\
&& + 2\avg{N^2 a_b \Sigma^b} + \avg{ N^2\Sigma_{ab}\zeta^{ab}}\,. 
\end{eqnarray*}
This term describes the direct contribution from inhomogeneity to the large-scale momentum density, projected along the preferred spacelike direction picked out by $m^a\,$.

Next, we can average the $1+1+2$-scalar projection (\ref{eq_shear_evol_eqn_112}) of the shear evolution equation, to obtain
\begin{eqnarray}\label{eq_ shear_evol_avg}
\partial_t \avg{N\Sigma} + \frac{2}{3}\avg{N\Theta}\avg{N\Sigma} + \frac{1}{2}\avg{N\Sigma}^2 + \avg{N^2 \mathcal{E}}  - \frac{1}{2}\avg{N^2\Pi} \\
\nonumber - \frac{1}{3}\left(\avg{N\mathcal{A}}-\avg{N\phi}\right)\avg{N\mathcal{A}} &=& \mathcal{B}_3\,,
\end{eqnarray}
where the backreaction term is given by
\begin{eqnarray*}\label{eq_backreaction_scalar_3}
\hspace{-2.5cm}
\mathcal{B}_3 = \frac{1}{3}{\rm Cov}\left(N\Theta,N\Sigma\right) + \frac{2}{3}{\rm Var}\, N\mathcal{A} - \frac{1}{3}{\rm Cov}\left(N\phi, N\mathcal{A}\right) + \frac{2}{3}\avg{N^2 m^a D_a \mathcal{A}}  && \\
\nonumber - \frac{1}{2}{\rm Var}\,N\Sigma - \frac{1}{3}\avg{N^2 M^{ab} D_a \mathcal{A}_b} - \frac{1}{3}\avg{N^2\Sigma_a\Sigma^a} + \frac{1}{3}\avg{ N^2 \mathcal{A}_a\mathcal{A}^a} && \\
\nonumber  + \frac{1}{3}\avg{ N^2\Sigma_{ab}\Sigma^{ab}}  + 2\avg{ N^2\alpha_a\Sigma^a} - \frac{2}{3}\avg{ N^2 a_a\mathcal{A}^a} + \avg{ N\Sigma\, \partial_t \ln{N}} \, . &&
\end{eqnarray*}
This drives the generation of anisotropy in the expansion of space.

The $1+1+2$-scalar projection (\ref{eq_vorticity_evolution_112}) of the vorticity evolution equation tells us that for an exactly LRS homogeneous cosmology where the preferred timelike vector is irrotational, we must have $\mathcal{A}\xi = 0\,$\,.
Thus, either the acceleration $\mathcal{A}$ associated with $n^a$ or the twist $\xi$ of the 2-dimensional spaces orthogonal to both $n^a$ and $m^a$ must be set to zero.
However, for a spacetime which is only an LRS Bianchi model on average, not locally, the result of averaging Eq. (\ref{eq_vorticity_evolution_112}) is instead
\begin{eqnarray}\label{eq_A_xi_constraint_avg}
    \avg{N\mathcal{A}}\avg{N\xi} = \mathcal{B}_4\,,
\end{eqnarray}
where the backreaction term is given by 
$\mathcal{B}_{4} = -\frac{1}{2}\avg{N^2 \epsilon_{ab}D^a \mathcal{A}^b} - {\rm Cov}\left(N\mathcal{A},N\xi\right)$\,.
In this case we see that any non-zero effect from the small-scale inhomogeneity will lead to a more complicated result, with $ \avg{N\mathcal{A}} \neq 0 \neq \avg{N\xi}\,$.

Let us now turn our attention to the equations that come from the Ricci identities for the preferred spacelike vector $m^a\,$.
The first, which has a counterpart in isotropic cosmological modelling, comes from the average of Eq. (\ref{eq_hamiltonian_constraint_112}). 
As we are considering an irrotational timelike congruence, this equation is entirely equivalent to the Hamiltonian constraint. 
Performing Buchert's averaging procedure on it yields
\begin{eqnarray}\label{eq_Hamiltonian_avg}
\frac{2}{9}\avg{N\Theta}^2 - \frac{1}{2}\avg{N\phi}^2 + \frac{1}{3}\avg{N\Theta}\avg{N\Sigma} - \avg{N\Sigma}^2 - \frac{2}{3}\avg{N^2\rho} - \frac{2}{3}\avg{N^2}\Lambda && \\
\nonumber -\frac{1}{2}\avg{N^2\Pi} - \avg{N^2\mathcal{E}} + 2\avg{N\xi}^2 &=& \mathcal{B}_5\,.
\end{eqnarray}
The backreaction term in this case is given by the following expression:
\begin{eqnarray*}\label{eq_backreaction_scalar_5}
\mathcal{B}_5 &=& \avg{N^2 m^a D_a \phi} + \frac{1}{2}{\rm Var}\,N\phi - \frac{2}{9}{\rm Var}\,N\Theta + {\rm Var}\,N\Sigma - \frac{1}{3}{\rm Cov}\left(N\Theta,N\Sigma\right) - 2\,{\rm Var}\,N\xi \\
\nonumber &&  + \avg{N^2\zeta_{ab}\zeta^{ab}} + \avg{N^2 \Sigma_a\Sigma^a} + \avg{ N^2 a_b a^b} - \avg{N^2 M^{ab} D_a a_b}\,. 
\end{eqnarray*}
This term acts in the same way as an additional effective contribution to the overall energy density, in what would reduce to the Friedmann equation in isotropic cosmology. 
We can relate the unfamiliar-looking terms on the left-hand side of the Hamiltonian constraint (\ref{eq_Hamiltonian_avg}) to the more familiar average Ricci curvature $\avg{^{(3)}R}$ of the spacelike hypersurfaces, by averaging Eq. (\ref{eq_ricci_3_112}). Doing this, we get
\begin{eqnarray*}\label{meanRicci3}
\avg{N^2 \ {^{(3)} R}} &=& -\frac{3}{2}\avg{N\phi}^2 - \frac{3}{2}\avg{N\Sigma}^2 + \avg{N\Theta}\avg{N\Sigma} - 3\avg{N^2 \mathcal{E}} - \frac{3}{2}\avg{N^2 \Pi} + 6\avg{N\xi}^2 \\
\nonumber && - 3\avg{N^2 m^a D_a\phi} - \frac{3}{2}\,{\rm Var}\,N\phi - \frac{3}{2}\,{\rm Var}\,N\Sigma + {\rm Cov}\left(N\Theta,N\Sigma\right) + 6\,{\rm Var}\,N\xi \\
\nonumber && - \avg{N^2\Sigma_a\Sigma^a} - 3\avg{N^2 a_b a^b} + 3\avg{N^2 M^{ab}D_a a_b} \\
\nonumber && - 3\avg{N^2\zeta_{ab}\zeta^{ab}} + \avg{N^2 \Sigma_{ab}\Sigma^{ab}}\,.
\end{eqnarray*}

The evolution equation for the large-scale expansion $\avg{\phi}$ of the two-dimensional screen spaces, along the preferred direction $m^a\,$, is given by averaging Eq. (\ref{eq_phi_evolution_eqn_112}):
\begin{eqnarray}
&& \partial_t \left\langle N\phi\right\rangle - \left\langle N^2 Q\right\rangle -\left(\frac{2}{3}\left\langle N\Theta\right\rangle - \left\langle N \Sigma\right\rangle\right)\left(\left\langle N\mathcal{A}\right\rangle - \frac{1}{2}\left\langle N\phi\right\rangle\right) = \mathcal{B}_{6} \, ,
\end{eqnarray}
where we find
\begin{eqnarray*}\label{eq_backreaction_scalar_6}
\mathcal{B}_{6} &=& \frac{2}{3}{\rm Cov}\left(N\Theta,N\phi\right) + \frac{1}{2}{\rm Cov}\left(N\Sigma,N\phi\right) + \frac{2}{3}{\rm Cov}\left(N\Theta,N\mathcal{A}\right)-{\rm Cov}\left(N\Sigma,N\mathcal{A}\right) \\
\nonumber && + \left\langle N^2 M^{ab} D_a \Sigma_b\right\rangle  - \left\langle N^2 \zeta_{ab}\Sigma^{ab}\right\rangle + \left\langle N^2 \mathcal{A}^a\left(\alpha_a - \Sigma_a\right)\right\rangle \\
\nonumber && + \left\langle N^2 a^a \left(\Sigma_a - \mathcal{A}_a\right)\right\rangle + \left\langle N\phi \, \partial_t \ln{N}\right\rangle.
\end{eqnarray*}

The next non-trivial averaged scalar equation we can construct comes from the scalar projection (\ref{eq_Hweyl_vs_vorticity_112}) of the constraint (\ref{eq_Hweyl_vs_shear_vorticity}) that relates the magnetic part $H_{ab}$ of the Weyl tensor to the shear of the irrotational congruence tangent to $n^a\,$. After averaging, we have
\begin{eqnarray}\label{eq_H_weyl_constraint_avg}
\left\langle N^2 \mathcal{H}\right\rangle - 3\left\langle N\xi\right\rangle\left\langle N\Sigma\right\rangle = \mathcal{B}_{7} \, ,
\end{eqnarray}
where the backreaction scalar in this case is 
\begin{equation*}
\mathcal{B}_{7} = 3\,{\rm Cov}\left(N\xi,N\Sigma\right) + \left\langle N^2 \epsilon_{ab}\left(D^a\Sigma^b - \zeta^{ac}\Sigma^b_{\:c}\right)\right\rangle\,.
\end{equation*}
For our averaged cosmology to be silent, with $H_{ab} = \avg{\mathcal{H}}\left(m_a m_b - \frac{1}{2}M_{ab}\right) = 0\,$, we therefore require $\mathcal{B}_{7}$ to vanish along with either $\left\langle N\xi\right\rangle$ or $\left\langle N\Sigma\right\rangle$, or to conspire to cancel $3\left\langle N\xi\right\rangle\left\langle N\Sigma\right\rangle$\,, which seems unlikely in a generic spacetime without special symmetries.

The evolution equation for the average twist $\avg{\xi}$ is
\begin{eqnarray}
&& \partial_t\left\langle N\xi\right\rangle + \frac{1}{2}\left(\frac{2}{3}\left\langle N\Theta\right\rangle - \left\langle N\Sigma\right\rangle\right)\left\langle N\xi\right\rangle - \frac{1}{2}\left\langle N^2\mathcal{H}\right\rangle = \mathcal{B}_{8} \,,
\end{eqnarray}
with backreaction term 
\begin{eqnarray*}\label{eq_backreaction_scalar_8}
\mathcal{B}_{8} &=& \frac{2}{3}{\rm Cov}\left( N\Theta, N\xi\right) + \frac{1}{2}{\rm Cov}\left(N\Sigma,N\xi\right)   + \frac{1}{2}\left\langle N^2 \epsilon_{ab}\left(a^a + \mathcal{A}^a\right)\left(\alpha^b+\Sigma^b\right)\right\rangle 
\\ &&  + \frac{1}{2}\left\langle N^2 \epsilon_{ab}D^a\alpha^b\right\rangle + \frac{1}{2}\left\langle N^2 \epsilon_{ac}\zeta_b^{\ c}\Sigma^{ab}\right\rangle + \left\langle N\xi\,\partial_t \ln{N}\right\rangle\,.
\end{eqnarray*}

For an exact LRS Bianchi cosmology, the constraint (\ref{eq_phi_xi_constraint_112}) on the projected derivative of $\xi$ gives that $\phi \xi = 0\,$, so at least one of these two variables must vanish. However, in the emergent case that comes from our explicit averaging procedure, we have instead
\begin{eqnarray}\label{eq_phi_xi_constraint_avg}
&& \left\langle N\phi\right\rangle\left\langle N\xi\right\rangle = \mathcal{B}_{9} \,,
\end{eqnarray}
where the backreaction scalar is
\begin{eqnarray*}\label{eq_backreaction_scalar_9}
\mathcal{B}_{9} = -\left\langle N^2 m^a D_a \xi \right\rangle - {\rm Cov}\left(N\phi,N\xi\right) + \frac{1}{2}\left\langle N^2 \epsilon_{ab}\left(D^a a^b + \Sigma^a a^b\right)\right\rangle\,.
\end{eqnarray*}
As in Eq. (\ref{eq_A_xi_constraint_avg}), the backreaction term in this last equation prevents one obtaining the usual result that either $\left\langle N\phi\right\rangle$ or $\left\langle N\xi\right\rangle$ vanishes, with $\left\langle N\phi\right\rangle \neq 0 \neq \left\langle N\xi\right\rangle$ being required if $\mathcal{B}_{9}\neq 0$\,.

Finally, let us move on to the final six averaged equations, which all come from the Bianchi identities. They tell us about the emergent Ricci curvature (i.e. the energy-momentum tensor) and Weyl curvature on large scales.

The only equation that could be derived from these identities in the case of isotropic cosmologies would be the local energy conservation equation, which in our case takes the form
\begin{eqnarray}\label{eq_Bianchi_energy_avg}
\partial_t \left\langle N^2 \rho\right\rangle + \left\langle N\Theta\right\rangle\left(\left\langle N^2\rho\right\rangle+\left\langle N^2 p \right\rangle \right) + \frac{3}{2}\left\langle N\Sigma\right\rangle\left\langle N^2 \Pi\right\rangle && \\
\nonumber +\left(\left\langle N\phi\right\rangle + 2\left\langle N\mathcal{A}\right\rangle\right)\left\langle N^2 Q \right\rangle &=& \mathcal{B}_{10} \, ,
\end{eqnarray}
where the backreaction term is given by
\begin{eqnarray*}\label{eq_backreaction_scalar_10}
\mathcal{B}_{10} &=& -{\rm Cov}\left(N^2 p,N\Theta\right) - \left\langle N^3 M^{ab} D_a Q_b \right\rangle - \left\langle N^3 m^a D_a Q \right\rangle \\
\nonumber && - {\rm Cov}\left(N\left(\phi+2\mathcal{A}\right),N^2Q\right) - \frac{3}{2}{\rm Cov}\left(N\Sigma,N^2\Pi\right) - 2\left\langle N^3 \mathcal{A}_a Q^a\right\rangle  \\
\nonumber && - 2\left\langle N^3\Sigma_a\Pi^a\right\rangle - \left\langle N^3\Sigma_{ab}\Pi^{ab}\right\rangle + \left\langle N^3 a_a Q^a\right\rangle + 2\left\langle N^2 \rho\,\partial_t \ln{N}\right\rangle\,,
\end{eqnarray*}
and where we have obtained the large-scale equation by averaging Eq. (\ref{eq_energy_conservation_112}).
The backreaction term here provides an additional source that drives the rate of change of the average energy density, $\avg{N^2 \rho}\,$. 
All other equations correspond to anisotropic cosmologies only, and vanish in the isotropic limit. It can be seen that if one chooses the foliation such that $n^a$ is the 4-velocity of pressureless dust, then $\mathcal{B}_{10}$ vanishes identically. 
For a generic foliation, however, this need not be the case, as an observer that is not comoving with the dust will typically measure non-zero pressure, momentum density and anisotropic stress.

The first of the remaining equations comes from averaging the $1+1+2$-covariant scalar projection (\ref{eq_momentum_conservation_112}) of the local equation of momentum conservation,
\begin{eqnarray} \label{eq_momentum_conservation_avg}
\partial_t\left\langle N^2 Q\right\rangle + \left(\frac{4}{3}\left\langle N\Theta\right\rangle + \left\langle N\Sigma\right\rangle\right) \left\langle N^2 Q\right\rangle  - \left\langle N\mathcal{A}\right\rangle\left\langle N^2\left(\rho + p + \Pi\right)\right\rangle && \\
\nonumber + \frac{3}{2}\left\langle N^2 \Pi\right\rangle\left\langle N\phi\right\rangle &=& \mathcal{B}_{11}\,.
\end{eqnarray}
This equation contains the backreaction scalar
\begin{eqnarray*}\label{eq_backreaction_scalar_11}
\mathcal{B}_{11} &=& -\frac{1}{3}{\rm Cov}\left(N\Theta, N^2 Q\right) - \left\langle N^3 m^a D_a \left(p + \Pi\right) \right\rangle - \left\langle N^3 M^{ab} D_a \Pi_b\right\rangle \\ 
\nonumber && - {\rm Cov}\left(\frac{3}{2}N\phi+N\mathcal{A},N^2\Pi\right) - {\rm Cov}\left(N\Sigma, N^2 Q\right) - {\rm Cov}\left(N^2\left(\rho+p\right), N \mathcal{A}\right) \\
\nonumber &&  - \left\langle N^3 \mathcal{A}_a\Pi^a\right\rangle + \left\langle N^3 \zeta_{ab}\Pi^{ab}\right\rangle + \left\langle N^3 \left(\alpha_a-\Sigma_a\right)Q^a\right\rangle \\
\nonumber && + 2\left\langle N^3 a_a \Pi^a\right\rangle + 2\left\langle N^2 Q \,\partial_t \ln{N}\right\rangle\,. 
\end{eqnarray*}
The backreaction scalar $\mathcal{B}_{11}$ drives the evolution of the large-scale momentum density, and if non-zero can therefore source a bulk flow in the averaged cosmology. This has obvious utility for modelling how the anomalously large bulk flows reported in e.g. Refs. \cite{Magoulas_2014,hoffman2015cosmic} might be sourced by smaller-scale inhomogeneities.

Finally, we can determine averaged equations for the scalar projections $\mathcal{E}$ and $\mathcal{H}$ of the Weyl tensor.
The first is the evolution equation for $\avg{\mathcal{E}}\,$, which is
\begin{eqnarray}\label{eq_E_weyl_evol_avg}
\partial_t\left\langle N^2 \mathcal{E}\right\rangle + \frac{1}{2}\partial_t\left\langle N^2 \Pi\right\rangle + \left(\left\langle N\Theta\right\rangle - \frac{3}{2}\left\langle N\Sigma\right\rangle \right)\left\langle N^2\mathcal{E}\right\rangle && \\
\nonumber + \frac{1}{2}\left(\frac{1}{3}\left\langle N\Theta\right\rangle + \frac{1}{2}\left\langle N\Sigma\right\rangle \right)\left\langle N^2\Pi\right\rangle
- \frac{1}{3}\left(\frac{1}{2}\left\langle N\phi\right\rangle-2\left\langle N\mathcal{A}\right\rangle\right)\left\langle N^2 Q \right\rangle && \\
\nonumber + \frac{1}{2}\left(\left\langle N^2 \rho\right\rangle+\left\langle N^2 p \right\rangle \right) \left\langle N\Sigma\right\rangle - 3\left\langle N\xi \right\rangle\left\langle N^2 \mathcal{H}\right\rangle &=& \mathcal{B}_{12} \, ,
\end{eqnarray}
where
\begin{eqnarray*}\label{eq_backreaction_scalar_12}
\mathcal{B}_{12} &=& -\frac{1}{3}\left\langle N^3 m^a D_a Q \right\rangle + \frac{1}{3}{\rm Cov}\left(N\Theta,N^2\Pi\right) + \left\langle N^3\epsilon_{ab}D^a \mathcal{H}^b\right\rangle + \frac{1}{6}\left\langle N^3 M^{ab} D_a Q_b\right\rangle \\
\nonumber &&  + \frac{3}{2}{\rm Cov}\left(N\Sigma,N^2\mathcal{E}\right) + \left\langle N^3\epsilon_{ab}\mathcal{H}^{bc}\zeta^a_{\ c}\right\rangle -\frac{1}{2}{\rm Cov}\left(N\Sigma,N^2\left(\rho+p+\frac{1}{2}\Pi\right)\right) \\
\nonumber &&+ \frac{1}{3}{\rm Cov}\left(N^2 Q,\frac{1}{2}N\phi-2N\mathcal{A}\right)   - \frac{1}{6}\left\langle N^3 \Sigma_a\Pi^a\right\rangle + \frac{1}{3}\left\langle N^3 \left(\mathcal{A}_a + a_a\right)Q^a\right\rangle \\
\nonumber && + 2\left\langle N^3 \epsilon_{ab}\mathcal{A}^a\mathcal{H}^b\right\rangle - \left\langle N^3 \Sigma_{ab}\left(\mathcal{E}^{ab}+\frac{1}{2}\Pi^{ab}\right)\right\rangle  + 3\,{\rm Cov}\left(N\xi, N^2\mathcal{H}\right)\\
\nonumber &&  + \left\langle N^3\alpha_a\Pi^a\right\rangle + \left\langle N^3\left(2\alpha_a+\Sigma_a\right)\mathcal{E}^a\right\rangle + 2\left\langle N^2 \mathcal{E}\,\partial_t \ln{N}\right\rangle\,.
\end{eqnarray*}

Likewise, the averaged evolution equation for $\avg{\mathcal{H}}$ is 
\begin{eqnarray}\label{eq_H_weyl_evol}
\hspace{-2cm} && \partial_t\left\langle N^2\mathcal{H}\right\rangle + \left(\left\langle N\Theta\right\rangle - \frac{3}{2}\left\langle N\Sigma\right\rangle\right)\left\langle N^2\mathcal{H}\right\rangle + 3\left\langle N\xi\right\rangle\left(\left\langle N^2\mathcal{E}\right\rangle-\frac{1}{2}\left\langle N^2\Pi\right\rangle\right) = \mathcal{B}_{13} \, ,
\end{eqnarray}
which has backreaction 
\begin{eqnarray*}\label{eq_backreaction_scalar_13}
\mathcal{B}_{13} &=& -\left\langle N^3\epsilon_{ab}D^a\left(\mathcal{E}^b-\frac{1}{2}\Pi^b\right)\right\rangle + \frac{3}{2}{\rm Cov}\left(N\Sigma,N^2\mathcal{H}\right) + 2\left\langle N^3\epsilon_{ab}\mathcal{E}^a \mathcal{A}^b\right\rangle  \\
\nonumber && + \left\langle N^3\left(2\alpha_a + \Sigma_a\right)\mathcal{H}^a\right\rangle  + \frac{1}{2}\left\langle N^3\epsilon_{ab}\Sigma^a Q^b\right\rangle - \left\langle N^3\Sigma_{ab}\mathcal{H}^{ab}\right\rangle   \\
\nonumber && + \frac{1}{2}\left\langle N^3\epsilon_{ab}\mathcal{E}^{ac}\zeta^b_{\ c}\right\rangle - 3\,{\rm Cov}\left(N\xi,N^2\mathcal{E}-\frac{1}{2}N^2\Pi\right) + \left\langle N^2 \mathcal{H} \, \partial_t \ln{N}\right\rangle\,.
\end{eqnarray*}

Finally, the averaged constraint equations for the $1+1+2$-covariant scalar projections of $E_{ab}$ and $H_{ab}$ are respectively
\begin{eqnarray}\label{eq_div_E_weyl_constraint_avg}
&& \frac{3}{2}\left(\left\langle N^2\mathcal{E}\right\rangle + \frac{1}{2}\left\langle N^2\Pi\right\rangle\right)\left\langle N\phi\right\rangle + \left(\frac{1}{3}\left\langle N\Theta\right\rangle-\frac{1}{2}\left\langle N\Sigma\right\rangle\right)\left\langle N^2 Q\right\rangle = \mathcal{B}_{14}\,,
\end{eqnarray}
which we find to have a backreaction term given by
\begin{eqnarray*}\label{eq_backreaction_scalar_14}
\mathcal{B}_{14} &=& -\left\langle N^3 m^a D_a \mathcal{E}\right\rangle + \frac{1}{3}\left\langle N^3 m^a D_a \rho \right\rangle - \frac{1}{2}\left\langle N^3 m^a D_a \Pi\right\rangle  \\
\nonumber && - \left\langle N^3 M^{ab} D_a \left(\mathcal{E}_b +\frac{1}{2}\Pi_b\right)\right\rangle - \frac{3}{2}{\rm Cov}\left(N\phi,N^2\mathcal{E}+\frac{1}{2}N^2\Pi\right) \\
\nonumber &&  + \frac{1}{2}{\rm Cov}\left(N^2 Q,N\Sigma-\frac{2}{3}N\Theta\right) +  \frac{1}{2}\left\langle N^3\Sigma_a Q^a\right\rangle \\
\nonumber && + \left\langle N^3 \epsilon_{ab}\Sigma^{ac}\mathcal{H}_c^{\ b}\right\rangle + \left\langle N^3\left(\mathcal{E}_{ab}+\frac{1}{2}\Pi_{ab}\right)\zeta^{ab}\right\rangle + \left\langle N^3\left(2\mathcal{E}_a+\Pi_a\right)a^a\right\rangle\,,
\end{eqnarray*}
and 
\begin{eqnarray}\label{eq_div_H_weyl_constraint_avg}
&& \frac{3}{2}\left\langle N\phi\right\rangle\left\langle N^2\mathcal{H}\right\rangle + \left\langle N^2 Q \right\rangle \left\langle N\xi\right\rangle = \mathcal{B}_{15}\,
\end{eqnarray}
where
\begin{eqnarray*}\label{eq_backreaction_scalar_15}
\mathcal{B}_{15} &=& \left\langle N^3 m^a D_a \mathcal{H} \right\rangle - \frac{3}{2}{\rm Cov}\left(N\phi,N^2\mathcal{H}\right) - {\rm Cov}\left(N^2 Q,N\xi\right) \\
\nonumber &&  - \frac{1}{2}\left\langle N^3\epsilon_{ab}D^a Q^b\right\rangle - \left\langle N^3 M^{ab} D_a \mathcal{H}_b\right\rangle + 2\left\langle N^3 a_b \mathcal{H}^b\right\rangle \\
\nonumber && + \left\langle N^3 \zeta_{ab}\mathcal{H}^{ab}\right\rangle - \left\langle N^3\epsilon_{ab}\Sigma^a_{\ c}\left(\mathcal{E}^{bc}+\frac{1}{2}\Pi^{bc}\right)\right\rangle\,.
\end{eqnarray*}

Eqs. (\ref{eq_averaged_Raychaudhuri}-\ref{eq_div_H_weyl_constraint_avg}) provide a complete set that can be used to describe the large-scale properties of an anisotropic cosmological model after averaging. 
These equations are a considerable complication on those that govern exact LRS cosmologies, due to the presence of the backreaction scalars $\mathcal{B}_i$\,. If all of these terms are sufficiently small when calculated for a domain $\mathcal{D}$, then the expansion of $\mathcal{D}$ should be expected to be well approximated by an exact LRS Bianchi model. 
If this is not the case, then we will be in a situation where backreaction from small-scale inhomogeneous structures on the large-scale properties of the cosmology is no longer negligible. In Section \ref{sec:backreaction_farnsworth} we will consider a family of exact cosmological models, to help develop our understanding of these terms by explicitly calculating them for an example geometry.

Let us briefly summarise and comment on what we have done here.
On a mathematical level, the equations we have derived constitute a full, novel, approach for modelling anisotropy in the Universe, where we have modelled the possibility of cosmic anisotropy emerging from inhomogeneous spacetimes. 
This has been achieved by performing a $1+1+2$-decomposition of all relevant fields \cite{Clarkson_2007}, and averaging the covariantly defined scalars that result \cite{Buchert_2000}. 
The equations that result describe a locally rotationally symmetric Bianchi cosmological model, with additional source terms due to backreaction from inhomogeneity. 

Our approach is very general. It allows for:
\begin{enumerate}
    \item The underlying spacetime to be arbitrarily inhomogeneous, with no Killing symmetries in general.
    \item The averaging surfaces, determined by the choice of foliation, to be specified freely. The foliation might be chosen by one or more of the means discussed in Section \ref{subsec:foliation_choices}. We can convert freely between different foliations, according to the transformation rules in Section \ref{subsec:foliation_changing}, with obvious utility for near-FLRW geometries with small tilt \cite{Tsagas_2009,Tsagas_2011,tsagas2022deceleration,Santiago_2022}.
    \item The direction of anisotropy to be freely chosen, which might be done according the observational or mathematical criteria laid down in Section \ref{subsec:preferred_direction_choices}.
    \item A completely general matter content $T_{ab}\,$, with no special properties such as requiring it to be pressureless dust.
\end{enumerate}

This is a new way to model deviations from FLRW, motivated by anisotropic and inhomogeneous observational anomalies, such as the dipole anomaly we explored in Chapter 4. 
Helpfully, the framework is fully relativistic and covariant. It does not introduce any new physics.

However, it remains to be seen whether any particular model will be able to produce an emergent anisotropy that would be compatible with any of the anomalies claimed in the literature, or whether any of the additional degrees of freedom in this approach would be sufficient to alleviate any observational tensions. 
Our framework provides a mechanism by which such questions can be investigated, but doing this in general is very difficult, because of the substantial mathematical complexity and nonlinearity of the backreaction problem. 

What we would really like, of course, is to have some understanding of how large the backreaction terms can be, and under what conditions they can drive anisotropy.
The magnitude of backreaction terms will depend on the geometry of the underlying spacetime. In the case of nearly-FLRW cosmologies, which have been extensively studied, most authors find the relative size of backreaction terms to be small in the Buchert equations of isotropic cosmology, such that $\mathcal{Q}/H^2 \sim 10^{-5}$ (see e.g. Ref. \cite{Adamek_2018}). 
While such contributions are indeed small, they are not as small as one may have na{\" i}vely  assumed, given that $\mathcal{Q}$ itself has leading-order contributions at second-order in cosmological perturbation theory. 
The reason why $\mathcal{Q}$ is not even smaller is that it contains terms $\sim \Phi \, \nabla^2 \Phi \sim \Phi \, \delta$, where $\delta$ is a density contrast that can become of order unity (or larger) in the presence of nonlinear structures. 

In the present case we have backreaction terms appearing not only in the averaged versions of the Friedmann equations, but also (for example) in the momentum conservation equation (\ref{eq_momentum_conservation_avg}). Using the same logic, the presence of nonlinear structure might be expected to result in backreaction terms of order $\mathcal{B}_{11}/H^2 \sim 10^{-5}$. 
In this case, however, such a contribution would be of the same size as the usual terms on the left-hand side of Eq. (\ref{eq_momentum_conservation_avg}), meaning that the backreaction terms could potentially be more significant overall. Whether these expectations are realised in typical cosmological spacetimes will require a detailed study.

In the next section, we will carry out a first step towards such a study, by performing the first full calculation of the anisotropic backreaction scalars $\left\lbrace\mathcal{B}_1,...,\mathcal{B}_{15}\right\rbrace\,$, in an exact cosmological spacetime.
The geometry we will study possesses symmetries and an idealised matter distribution, and is really a toy model rather than a valid description of our Universe. However, it will still provide us with some very valuable insight, particularly regarding the issue of foliation dependence.

\section{Application to Farnsworth models}\label{sec:backreaction_farnsworth}

In order to understand our formalism, it is illustrative to apply it to a class of exact cosmological models. For this we choose the anisotropic cosmologies found by Farnsworth \cite{Farnsworth_1967}. These are exact solutions to Einstein's equations. 
They are of Bianchi type $V$, admitting a four-parameter group of isometries, including local rotational symmetry. The three-dimensional spacelike surfaces of transitivity of this isometry group are in general not coincident with the hypersurfaces orthogonal to the dust 4-velocity: these are tilted cosmologies.
The $k=-1$ FLRW metric is contained within this wider class of metrics as a special case, and although they do not contain anything that could be considered as nonlinear structure, they do provide us with a precise example geometry to illustrate the application of our formalism.

The metric for the Farnworth solutions can be written as \cite{Farnsworth_1967}
\begin{equation}\label{eq_Farnsworth_metric}
\mathrm{d}s^2 = -\mathrm{d}t^2 + X^2(t+Cr)\,\mathrm{d}r^2 + e^{-2\alpha r}Y^2(t+Cr)\,\left(\mathrm{d}y^2+\mathrm{d}z^2\right)\,,
\end{equation}
where the functional dependence of the metric functions $X$ and $Y$ has been fixed by the presence of the Killing vectors
\begin{equation}\label{eq_Farnsworth_Killing_vectors}
\mathbf{X}_1 = \partial_y, \quad \mathbf{X}_2 = \partial_z, \quad \mathbf{X}_3 = -C\partial_t + \partial_r + y\partial_y + z\partial_z, \quad {\rm and} \quad \mathbf{X}_4 = -z\partial_y + y\partial_z \,.
\end{equation} 
Here $C$ is a constant, and the one-parameter isotropy group is generated by $\mathbf{X}_4$.
The parameter $\alpha$ is related to the curvature scale of the surfaces of transitivity, which we reiterate are not in general the matter-orthogonal surfaces. It can be set to unity by a choice of units for $r\,$, which we will now do.
The dust 4-velocity can be written in these coordinates as $u^a = \delta^a_{\ t}\,$, for which the corresponding axis of rotational symmetry is given by $\tilde{m}^a = X^{-1} \delta^a_{\ r}\,$. 

Substituting the metric from Eq.  (\ref{eq_Farnsworth_metric}) into Einstein's equations, one obtains a relationship between $X$ and $Y$,
\begin{equation}\label{eq_Farnsworth_XY_relation}
X = k^{-1}\left(CY'-Y\right)
\end{equation}
where a prime denotes a derivative with respect to $t+Cr$ and $k$ is a positive constant, as well as the following Friedmann-like equation:
\begin{equation}\label{eq_Farnsworth_Friedmann}
Y'^2 = \frac{D}{3Y} + k^2,
\end{equation}
where $D$ is a positive constant related to the energy density. Note that we have set the cosmological constant to zero.

The solution to Eq. (\ref{eq_Farnsworth_Friedmann}) can be written in parametric form as
\begin{eqnarray}\label{eq_Farnsworth_Fried_solution}
Y(t+Cr) &=& \frac{Wk}{2}\left(\cosh{\eta}-1\right)\,, \quad{\rm and} \\
t + Cr &=& \frac{W}{2}\left(\sinh{\eta}-\eta\right), 
\end{eqnarray}
where we have defined $W \equiv \dfrac{D}{3k^3}$\,. 
The constant $C$ controls the size of the bulk flow, and in the limit $C \longrightarrow 0$ we recover FLRW, and $\eta$ becomes the usual conformal time coordinate $\tau\,$. In what follows, when required, we will choose $C=2$ and $W=125$ in order to display numerical results.

Let us now consider refoliating this spacetime by boosting along the symmetry axis from the matter frame $u^a$ into the frame defined by some new irrotational timelike vector $n^a$, such that
\begin{eqnarray}\label{eq_Farnsworth_boost}
n_a = \gamma\left(u_a + w_a\right) \,, \qquad {\rm where} \qquad w_a = v(t+Cr) X(t+Cr) \delta_a^{\ r} \,,
\end{eqnarray}
and $\gamma= \left(1-v^2\right)^{-1/2}$ takes its usual form in terms of the magnitude $v$ of the 3-velocity $w_a\,$. The functional dependence of $v$ is motivated by the symmetries of the spacetime. 
This choice of boost, staying parallel to the local rotational symmetry axis of the spacetime, means that all $1+1+2$-covariant vectors and tensors vanish in any new frame. This will greatly simplify the calculation of our backreaction scalars in the $1+1+2$ formalism.

In order to calculate spatial Buchert averages in our anisotropic framework, it is of course necessary to consider the non-zero $1+1+2$-covariant scalars.
These can be calculated with respect to either $\left\lbrace u^a, \tilde{m}^a\right\rbrace$ or $\left\lbrace n^a, m^a\right\rbrace\,$. The scalars are related between the frames according to the equations in Section \ref{subsec:foliation_changing}.
After a boost of the kind in Eq. (\ref{eq_Farnsworth_boost}), the kinematic scalars can be written as
\begin{eqnarray}
\Theta &=& \frac{\gamma^3}{XY}\left[Y\left\lbrace Cv'+X'+v\left(2v^2-2+Xv'-vX'\right)\right\rbrace +2\gamma^{-2}Y'\left(X+Cv\right)\right] \label{eq_boost_theta} \\
\mathcal{A} &=& \frac{\gamma^3}{X}\left[Xv'-v^3 X'+ v\left(X'+Cv'\right)\right] \label{eq_boost_A} \\
\Sigma &=& \frac{2\gamma^3}{3XY}\left[Y\left\lbrace Cv'+X'+v\left(1+Xv'-v^2 -vX'\right)\right\rbrace - \gamma^{-2}Y'\left(X+Cv\right)\right] \label{eq_boost_sigma}\\
\phi &=& \frac{2\gamma}{XY}\left[Y'\left(C+Xv\right) - Y\right]\,, \quad {\rm and} \label{eq_boost_phi} \\
\Omega &=& \xi \ = \ 0\,, \label{eq_boost_omega}
\end{eqnarray}
whence we see that vorticity vanishes in all cases, so that any $n^a$ does indeed define a foliation of spacetime. The form of the kinematic scalars in the matter-comoving frame is obtained from the above set simply by setting $v \overset{!}{=} 0\,$.

The Ricci curvature, or equivalently the energy-momentum tensor $T_{ab}$\,, can be entirely characterised by the following four scalars describing the matter content,
\begin{eqnarray}\label{eq_Farnsworth_matter_scalars}
\rho = \gamma^2 \rho^{(u)}\left(t+Cr\right) \, ,\qquad p = \frac{1}{2}\Pi = \frac{1}{3}\gamma^2 v^2 \rho^{(u)} \qquad {\rm and} \qquad Q = -\gamma^2 v \rho^{(u)}\,, 
\end{eqnarray}
where the rest mass density is given by $\rho^{(u)} = T_{ab} u^a u^b = G_{ab}u^a u^b\,$\,:
\begin{equation*}
\hspace{-0.5cm}
\rho^{(u)} = \frac{2X^2YX'Y' + X^3 {Y'}^2 + 2CYX'\left(CY'-Y\right) - X\left(3Y^2 + C^2{Y'}^2 + 2CY\left(CY'' - 3Y'\right)\right)}{X^3 Y^2}\,.
\end{equation*}

Finally, we need the Weyl curvature. As the spacetime is of Petrov type D \cite{Petrov_1964,Stephani_2003}, the electric Weyl scalar $\mathcal{E}$ is invariant under boosts. For all possible $v$\,, it takes the form
\begin{equation}\label{eq_Farnsworth_Eweyl}
\hspace{-0cm}
\mathcal{E} = \frac{X\left(Y'^2 - YY''\right)\left(C^2-X^2\right) + YY'X'\left(C^2+X^2\right) - Y^2\left(CX'+X^2 X''\right)}{3 X^3 Y^2} \,.
\end{equation} 
The magnetic part of the Weyl tensor vanishes.
With these results in place, we can now consider refoliations of the Farnsworth cosmology, and calculate the $1+1+2$-covariant scalars and their associated averages and backreaction terms for any of them. 

\subsection{Homogeneous foliation}\label{subsec:farnsworth_homogeneous_foliation}

Let us first consider the foliation composed of leaves that coincide with the surfaces of transitivity of the Killing vectors from Eq. (\ref{eq_Farnsworth_Killing_vectors}). These are spatially homogeneous 3-dimensional spaces with a timelike normal $n^a$ that must satisfy $n_a X_3^a = 0$.
In terms of the function $v(t+Cr)$ that relates $n_a$ to $u_a$ through Eq. (\ref{eq_Farnsworth_boost}), this immediately implies $v = -C/X$\,, and results in a set of spaces of constant $\eta$\,.
These spaces are spanned by the induced metric $f_{ab} = g_{ab} + n_a n_b$\,.

The set of non-vanishing scalars in this foliation is larger than in the foliation orthogonal to the matter flow, as it contains a bulk flow that gives rise to non-zero pressure, momentum density and anisotropic stress. 
The full set of scalars is $\left\lbrace \Theta, \Sigma, \mathcal{E}, \phi, \rho, p, Q,\Pi \right\rbrace$, which can immediately be obtained from Eqs. (\ref{eq_boost_theta}-\ref{eq_Farnsworth_matter_scalars}). For each of these scalars $S$ we also know that their projected covariant derivative $D_a S$ must vanish, due to homogeneity. This immediately implies that this foliation is one of constant density, constant mean curvature and constant spatial curvature, i.e. $D_a \rho = D_a \Theta = D_a ^{(3)}R = 0\,$.

The line element in Eq. (\ref{eq_Farnsworth_metric}) can be recast into coordinates adapted to this foliation by making the transformation \cite{Petrov_1964}
\begin{eqnarray}\label{eq_Farnsworth_coord_transformation}
t &=& Cx + \int \mathrm{d}T \, \beta(T) \quad {\rm and} \quad r = - x + \int \mathrm{d}T\, \frac{\alpha(T)}{\beta(T)}\,, \\
{\rm where} \quad \alpha &=& \frac{C}{X^2 - C^2} \quad {\rm and} \quad \beta^2 = 1 - \frac{C^2}{X^2 - C^2}\,.
\end{eqnarray}
This results in
\begin{equation}
\mathrm{d}s^2 = -\mathrm{d}T^2 + A(T)\,\mathrm{d}x^2 + B(T)\,e^{2x}\,\left(\mathrm{d}y^2+\mathrm{d}z^2\right)\,,
\end{equation}
where 
\begin{equation}
    X^2 = C^2 + A\,, \quad {\rm and} \quad Y^2 = B e^{4x} \exp{\left\{2\int dT \, \dfrac{\alpha}{\beta}\right\}}\,.
\end{equation}
All quantities that were functions of $t+Cr$ are now purely functions of $T$, and moreover we have that $n^a = \delta^a_{\ T}$ and $\mathbf{X}_3 = -\partial_x + y\partial_y + z\partial_z$\,. 
This makes clear the orthogonality of $n^a$ to the surfaces of transitivity spanned by the Killing vectors from Eq. (\ref{eq_Farnsworth_Killing_vectors}). 
The lapse function is again equal to unity in these coordinates, which further simplifies the calculation of the emergent equations of motion.

\begin{figure}
\centering
\includegraphics[width=\linewidth]{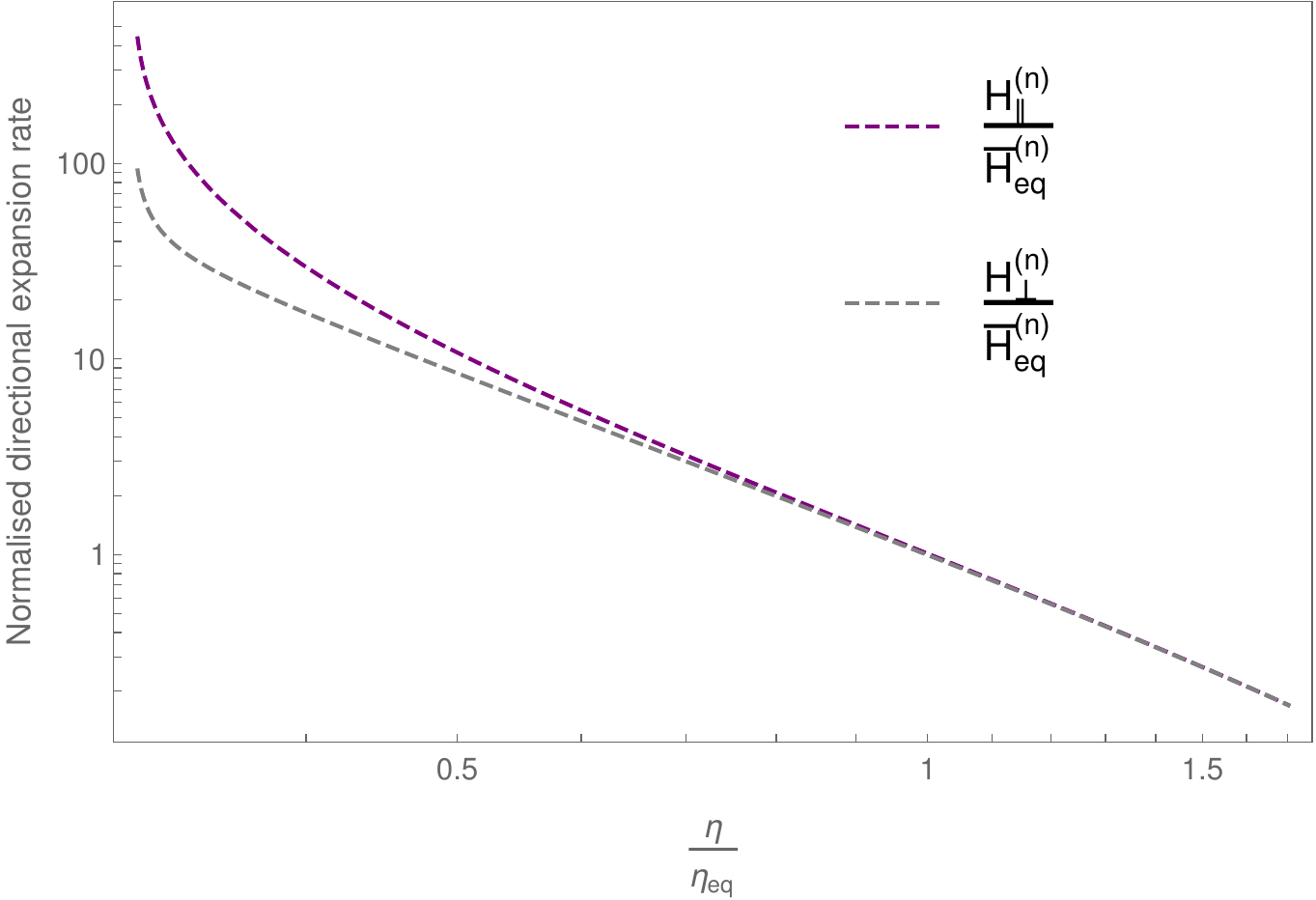}
\caption{The normalised rate of spatial expansion in the directions parallel, $H^{(n)}_{\parallel} = \dfrac{1}{3}\Theta^{(n)} + \Sigma^{(n)}$, and orthogonal, $H^{(n)}_{\perp} = \dfrac{1}{3}\Theta^{(n)} - \frac{1}{2}\Sigma^{(n)}$, to the axis of rotational symmetry, in the homogeneous foliation of the Farnsworth solution.}
\label{fig_Farnsworth_directional_rates_n}
\end{figure}

\begin{figure}
\centering
\includegraphics[width=\linewidth]{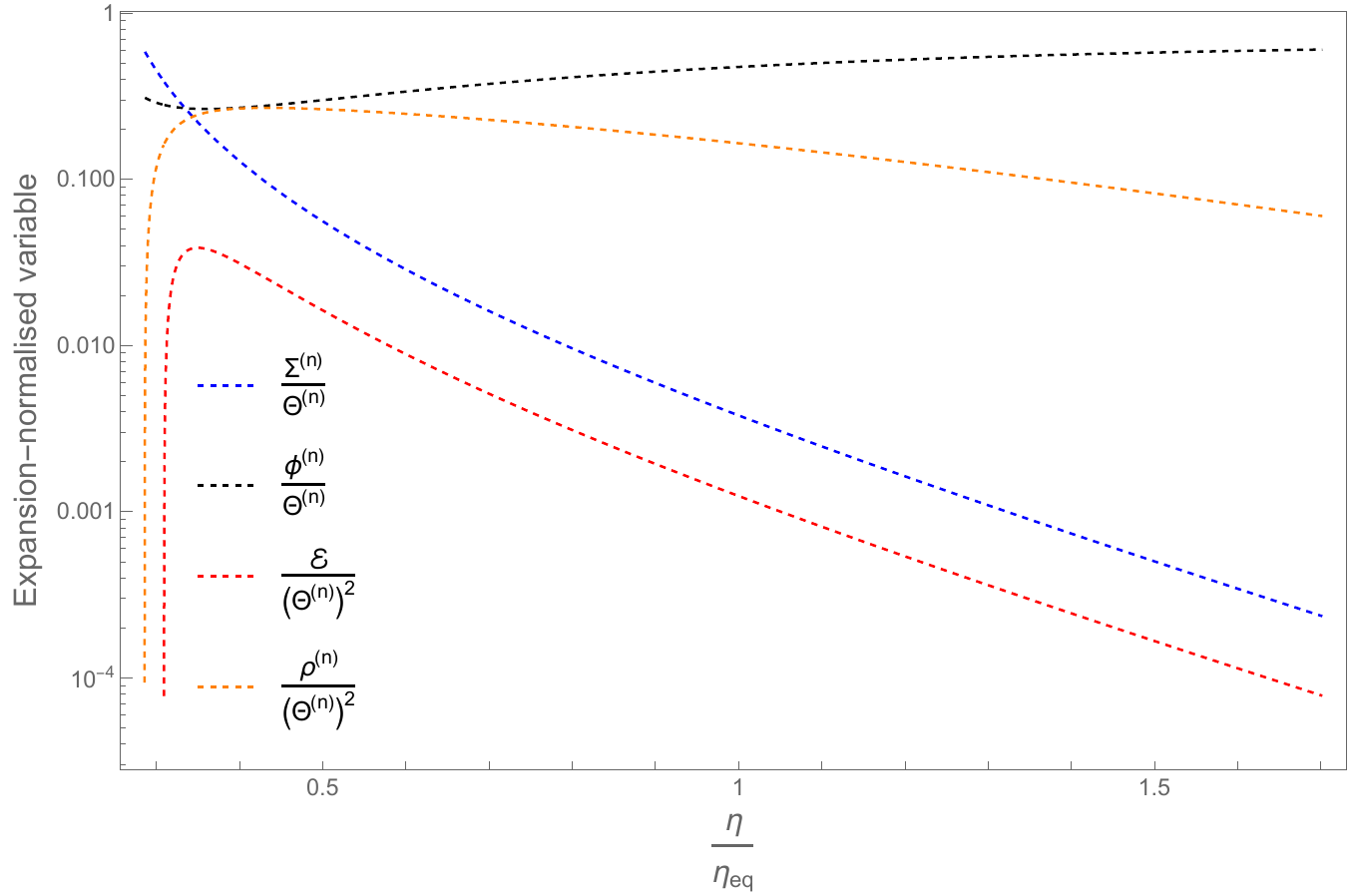}
\caption{Expansion-normalised variables $\left\lbrace \Sigma^{(n)}, \phi^{(n)}, \mathcal{E}, \rho^{(n)} \right\rbrace$ in the homogeneous foliation of the Farnsworth solution.}
\label{fig_Farnsworth_homogeneous_functions}
\end{figure}

The behaviour of the direction-dependent expansion rates of space, and the expansion-normalised variables $\left\lbrace \Sigma^{(n)}, \phi^{(n)}, \mathcal{E}, \rho^{(n)} \right\rbrace$, are displayed in Figs. \ref{fig_Farnsworth_directional_rates_n} and \ref{fig_Farnsworth_homogeneous_functions}. 
Fig.\ref{fig_Farnsworth_directional_rates_n} shows the expansion rates parallel and orthogonal to the direction picked out by the bulk flow, and displays significant anisotropy at early times (small $\eta$). As the bulk flow decays at late times (large $\eta$), the difference between the expansion rates in different directions also decays, demonstrating that space isotropises in this foliation. 
Fig. \ref{fig_Farnsworth_homogeneous_functions} shows other relevant scalars as a function of $\eta$, normalised by the appropriate power of the expansion rate $\Theta$ to make the resultant quantity dimensionless. One sees that shear $\Sigma$ and electric Weyl curvature $\mathcal{E}$ both increase at early times, consistent with growing anisotropy in this limit.
In both plots $\eta$ is made dimensionless by normalising it relative to $\eta_{\rm eq}$, its value when $Y=\dfrac{D}{3k^2}\,$, such that the two terms on the right-hand side of Eq. (\ref{eq_Farnsworth_Friedmann}), which are roughly analogous to the matter and curvature terms in the FLRW Friedmann equation, are equal. 
In Fig. \ref{fig_Farnsworth_directional_rates_n} the expansion rates are made dimensionless by normalising them with respect to the isotropic expansion rate $\bar{H}^{(n)} =\frac{1}{3} \Theta^{(n)}$ at the equality time $\eta_{\rm eq}\,$. 

In this foliation it can be seen that all of backreaction scalars $\mathcal{B}_i$ from Section \ref{subsec:emergent_equations} must be equal to zero. 
This happens because 
\begin{itemize}
    \item All variances and covariances must vanish, as $\int_{\mathcal{D}} d^3 x\,\sqrt{f}\, S(T) = S(T) V_{\mathcal{D}}$\,.
    \item All terms containing projected covariant derivatives, such as $m^a D_a S$\,, must also vanish, because $D_a$ acts on manifestly homogeneous hypersurfaces.
\end{itemize} 
The cosmology one observes in this foliation can therefore be entirely described by the locally defined scalar quanitites, without any need for averaging. This is exactly as expected for a spacetime that admits a homogeneous foliation. 
In most geometries such a simplifying set of symmetries will not exist, and in more realistic cosmologies one would need to perform explicit averages in order to gain a set of quantities that could be used to describe the large-scale properties of the cosmological spacetime. 

Let us now consider an inhomogeneous foliation of the Farnsworth solutions, to show how this would work in this simple example of a tilted Bianchi cosmology.

\subsection{Matter-rest-space foliation}\label{subsec:farnsworth_matter_foliation}

Let us now foliate the exact same spacetime into hypersurfaces orthogonal to the matter flow $u^a$\,, with induced metric $h_{ab}=g_{ab} + u_a u_b$\,. 
This corresponds to $v=0$ in Eqs. (\ref{eq_boost_theta}--\ref{eq_Farnsworth_matter_scalars}), and is a valid choice for averaging due to its irrotationality.
Indeed, if we were performing our averaging procedure according to Buchert's original approach \cite{Buchert_2000}, this is precisely the foliation we would use. 
The leaves of the matter foliation are level surfaces of the coordinate time $t$, as per the metric (\ref{eq_Farnsworth_metric}). 
The lapse is again equal to unity, so covariant and coordinate time derivatives are equivalent. 

The only non-zero quantities in this case are the scalars $\left\lbrace \Theta, \Sigma, \phi, \rho, \mathcal{E} \right\rbrace$, which in terms of the parameters from the solution in Eq. (\ref{eq_Farnsworth_Fried_solution}) are
\begin{eqnarray}
\Theta &=& \frac{2\left[3W\sinh{\eta}\left(\cosh{\eta}-1\right)+2C\left(2\cosh{\eta}+1\right)\right]}{W\left(\cosh{\eta}-1\right)\left[W\left(\cosh{\eta}-1\right)^2-2C\sinh{\eta}\right]} \\
\Sigma &=& \frac{8C\left(\sinh^2{\eta}+\cosh{\eta}-1\right)}{3W\left(\cosh{\eta}-1\right)^2\left[W\left(\cosh{\eta}-1\right)^2-2C\sinh{\eta}\right]} \\
\phi &=& \frac{4}{W\left(\cosh{\eta}-1\right)}\,,
\label{eq_rho_Farnsworth}
\rho = \frac{24}{W\left(\cosh{\eta}-1\right)\left[W\left(\cosh{\eta}-1\right)^2-2C\sinh{\eta}\right]}\,,
\end{eqnarray}
and the electric Weyl curvature, which is given by Eq. (\ref{eq_Farnsworth_Eweyl}). 

\begin{figure}
\centering
\includegraphics[width=\linewidth]{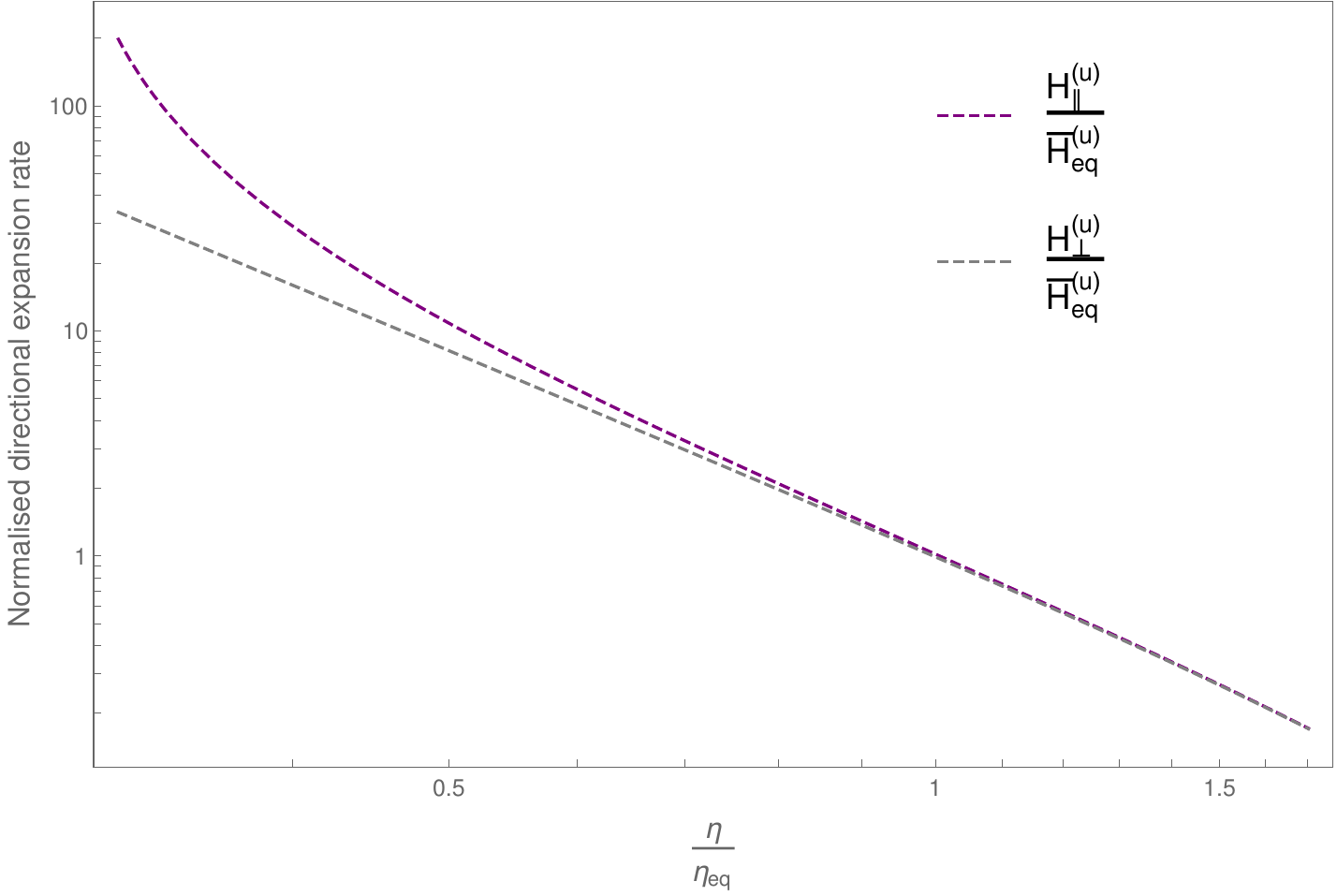}
\caption{The normalised rate of spatial expansion in the directions parallel, $H^{(u)}_{\parallel} = \dfrac{1}{3}\Theta^{(u)} + \Sigma^{(u)}$, and orthogonal, $H^{(u)}_{\perp} = \frac{1}{3}\Theta^{(u)} - \dfrac{1}{2}\Sigma^{(u)}$, to the axis of rotational symmetry, in the matter-rest-space foliation of the Farnsworth solution.}
\label{fig_Farnsworth_directional_rates_u}
\end{figure}

\begin{figure}
\centering
\includegraphics[width=\linewidth]{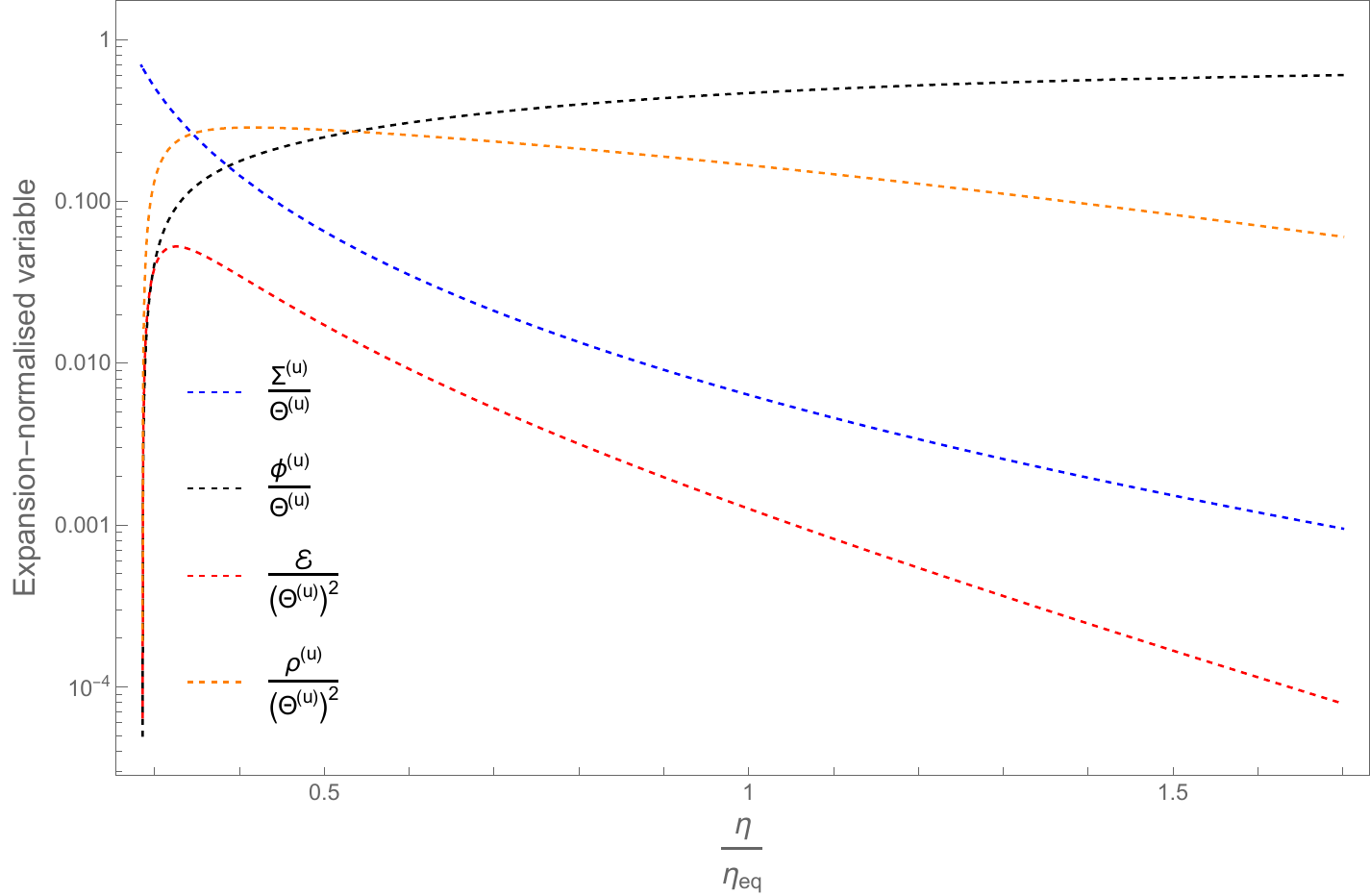}
\caption{Expansion-normalised variables $\left\lbrace \Sigma^{(u)},\phi^{(u)},\mathcal{E},\rho^{(u)} \right\rbrace$ in the matter-rest-space foliation of the Farnsworth solution.}
\label{fig_Farnsworth_functions_eta}
\end{figure}

The normalised expansion rates as a function of $\eta$ are shown in Fig.\ref{fig_Farnsworth_directional_rates_u}, and the other expansion-normalised scalars are shown in Fig. \ref{fig_Farnsworth_functions_eta}. 
These plots, displaying quantities calculated in the foliation generated by the matter flow $u^a$\,, may be compared directly with Figs. \ref{fig_Farnsworth_directional_rates_n} and \ref{fig_Farnsworth_homogeneous_functions}.

\begin{figure}
\centering
\includegraphics[width=\linewidth]{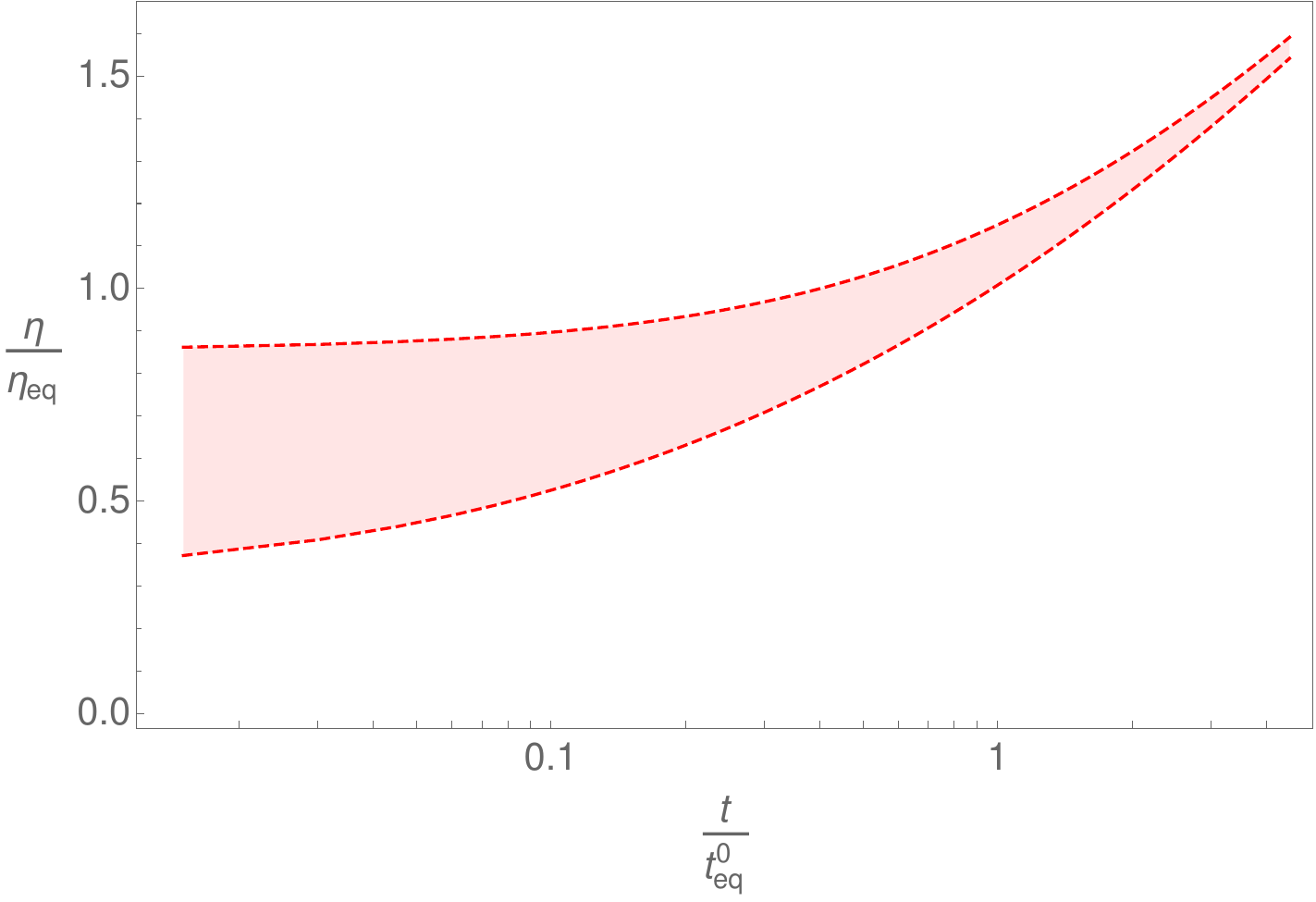}
\caption{The integration domain from $r_{\rm min}=1$ to $r_{\rm max}=20$ (shaded region) in the matter-rest-frame foliation as a function of time $t$, normalised relative to $t^0_{\rm eq}=W \left(\sinh{\eta_{\rm eq}}-\eta_{\rm eq}\right)/2$\,.}
\label{fig_Farnsworth_integrationlimits}
\end{figure}

In order to average these quantities we need an averaging domain $\mathcal{D}$ on each of the surfaces orthogonal to $u^a\,$. 
If we take this domain to have coordinate extents $\left\lbrace \Delta r, \Delta y, \Delta z\right\rbrace\,$, then the spatial volume is given by
\begin{eqnarray}
V_{\mathcal{D}}(t) &=& \int_{\mathcal{D}} \mathrm{d}y\mathrm{d}z\mathrm{d}r \sqrt{h(\eta)} = \frac{\Delta y \Delta z W^3 k^2}{16 C} \,e^{2t/C} \, \int_{\eta(r_{\rm min},t)}^{\eta(r_{\rm max}, t)} d\eta \, J(\eta)\,,
\end{eqnarray}
where 
\begin{equation}
    J(\eta) = \left(\cosh{\eta}-1\right)^2 \left[W\left(\cosh{\eta}-1\right)^2 - 2C\sinh{\eta}\right]\exp{\left\{-\frac{W}{C}\left(\sinh{\eta} - \eta\right)\right\}}\,,
\end{equation}
and $\Delta r = r_{\rm max} - r_{\rm min}\,$.

We can then compute all the desired scalar averages as \footnote{In the isotropic case $C \longrightarrow 0$, the integration limits $\eta(r_{\rm min},t)$ and $\eta(r_{\rm max}, t)$ coincide. However, in that case $S(\eta) \rightarrow S(t)$ can be brought outside of the integral, such that one safely obtains $\left\langle S(t)\right\rangle = S$\,.}
\begin{equation}
\avg{S}(t) = \frac{\int_{\eta(r_{\rm min},t)}^{\eta(r_{\rm max}, t)} d\eta \, J(\eta) S(\eta)}{\int_{\eta(r_{\rm min},t)}^{\eta(r_{\rm max}, t)} d\eta \, J(\eta)}\, .
\end{equation}

In choosing the integration domain (i.e. $r_{\rm min / max}$), it is important to avoid the bang time singularity that is located at $r_{\rm B}(t)$ such that $t + C r_{\rm B}(t) = 0$\,. 
For any $C>0$ this is satisfied by $r_{\rm min} >0\,$. The value of $r_{\rm max}$ needs only to be sufficiently large that $\eta_{\rm min}$ and $\eta_{\rm max}$ are relatively far apart, but once this is ensured to be true the backreaction effects are only weakly dependent on its value. 
If we choose $r_{\rm min} = 1$, $r_{\rm max} = 20$ and $C=2$ as an example, then one obtains the integration domain shown in Fig. \ref{fig_Farnsworth_integrationlimits}. 
At early times the distance from $r_{\rm min}$ to $r_{\rm max}$ corresponds to a large separation in $\eta$, while at late times the integration domain shrinks, as the spacelike hypersurfaces isotropise.

\begin{figure}
\centering
\includegraphics[width=\linewidth]{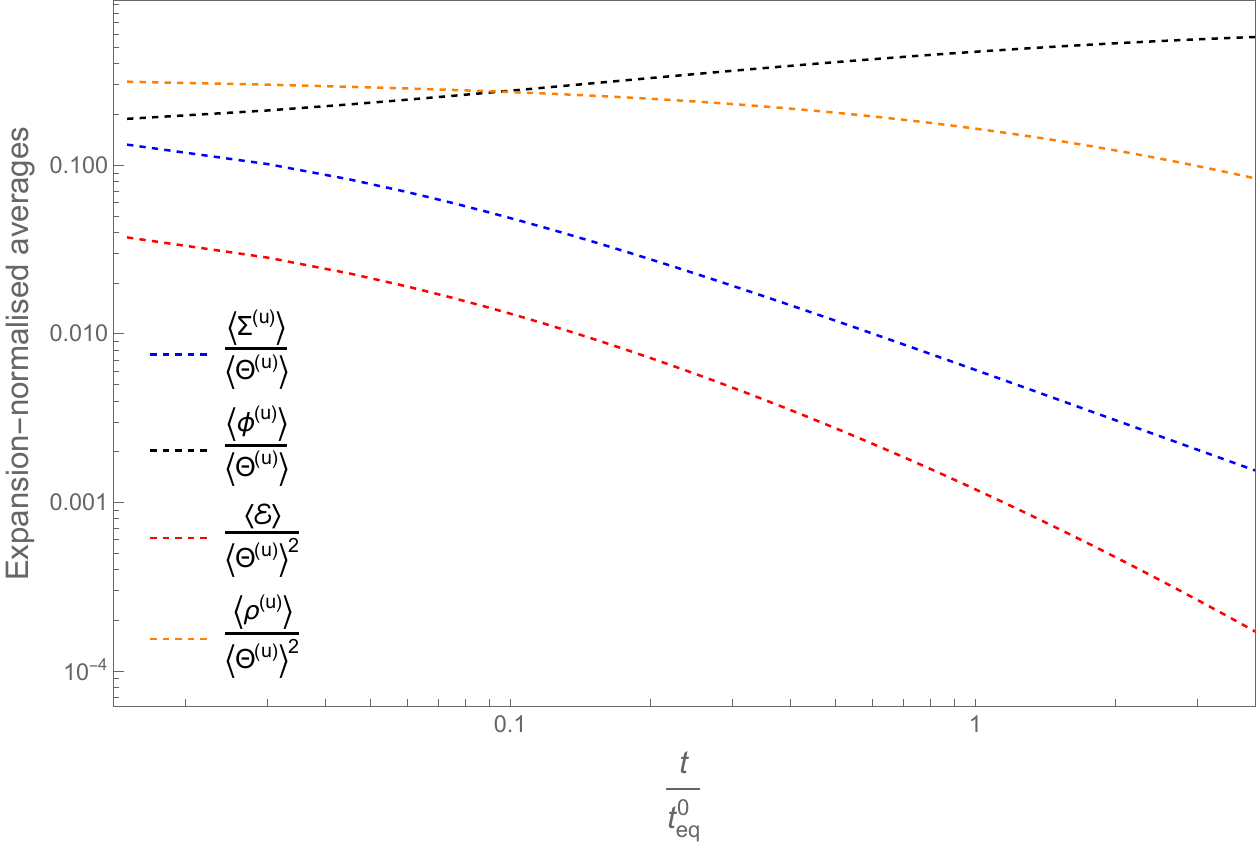}
\caption{Evolution of $\left\lbrace \avg{\phi^{(u)}}, \avg{\Sigma^{(u)}}, \avg{\mathcal{E}}, \avg{\rho^{(u)}} \right\rbrace$ with $t$, normalised by the relevant power of $\avg{\Theta^{(u)}}\,$. 
The time coordinate $t$ is normalised by $t^0_{\rm eq}$. The averaging domain is from Figure \ref{fig_Farnsworth_integrationlimits}.}
\label{fig_Farnsworth_means_normalised}
\end{figure}

\begin{figure}
\centering
\includegraphics[width=\linewidth]{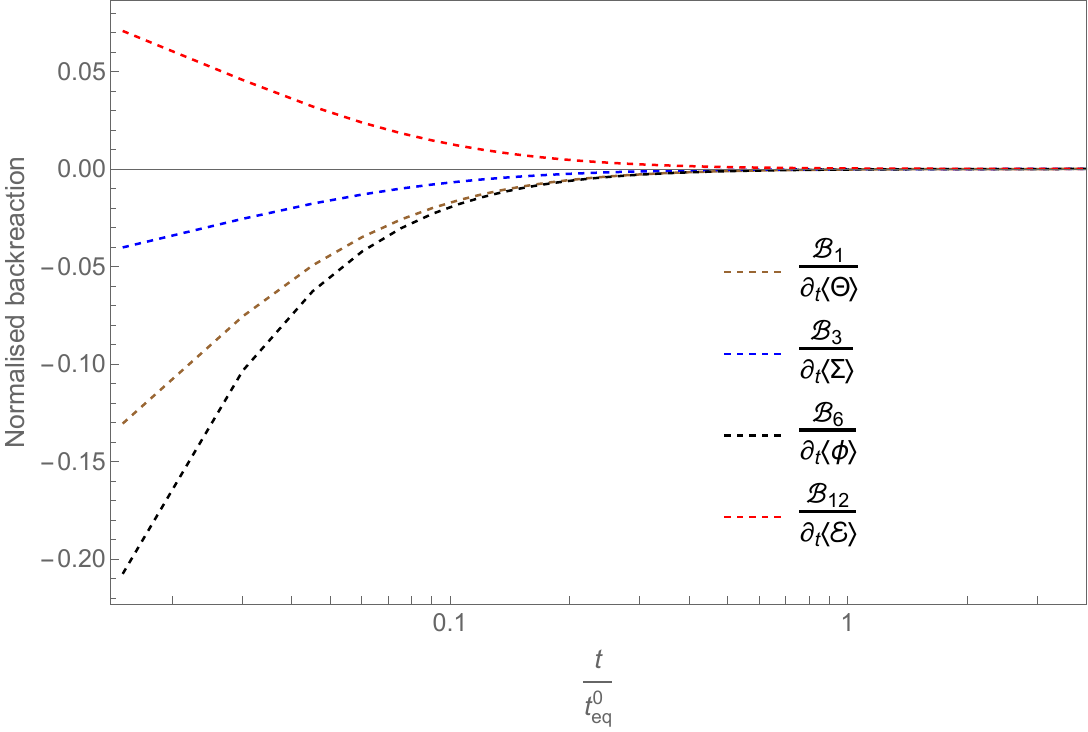}
\caption{Backreaction scalars $\left\lbrace \mathcal{B}_1,\mathcal{B}_3,\mathcal{B}_6,\mathcal{B}_{12}\right\rbrace$ as a function of $t/t^0_{\rm eq}\,$. Each scalar has been normalised by the time derivative of the averaged quantity whose evolution equation they appear in.}
\label{fig_Farnsworth_backreaction_scalars}
\end{figure}

The evolution with $t$ of our averaged scalars, normalised by the averaged expansion $\left\langle\Theta\right\rangle$, is shown in Fig. \ref{fig_Farnsworth_means_normalised}. The inhomogeneity in these spaces has been averaged out to give a set of purely time-dependent scalar functions. 
These averages provide a cosmological description of the large-scale properties of the spaces orthogonal to the matter flow $u^a$\,, including anisotropic properties which are communicated by $\avg{\Sigma}$ and $\avg{\mathcal{E}}$\,. 

The magnitudes of the backreaction scalars that source the evolution of the large-scale averaged cosmology are displayed in Fig. \ref{fig_Farnsworth_backreaction_scalars}, where we focus on the subset $\left\lbrace\mathcal{B}_1,\mathcal{B}_3,\mathcal{B}_6,\mathcal{B}_{12}\right\rbrace$\,, as they appear in evolution rather than constraint equations. 
These quantities are shown as ratios of the time derivative of the relevant average, as per the equations in Section \ref{subsec:emergent_equations}, in order to show the fractional contribution they make to their evolution. 
In each case, the averaging domain used is the one displayed in Fig. \ref{fig_Farnsworth_integrationlimits}, and we have again used $C = 2$ and $W = 125\,$.

It can be seen from Fig. \ref{fig_Farnsworth_backreaction_scalars} that the backreaction effect becomes large at early times, and dissipates to zero at late times. In the Farnsworth spacetime, this happens because the anisotropy (and hence inhomogeneity in the matter-rest-space foliation) is entirely due to the bulk flow $v=-C/X$\,, which becomes small as the directional scale factor $X$ grows and the spacelike hypersurfaces isotropise. 

This example demonstrates that the large-scale averaged properties of a cosmological model depend on the chosen foliation, and that one requires in addition a choice of averaging domain (and in general a spatial direction $m^a\,$). 
The kinematic and matter quantities that result can display different behaviours when calculated on different foliations, and so can the magnitude of the backreaction scalars. This clearly demonstrates the need for these choices to be made carefully, and for the interpretation of scalar averaging results in cosmology to be understood in terms of these choices.
The consequences of constructing an average cosmological model in a spacetime with more than one well-motivated possible choice of foliation (such as a tilted Bianchi cosmology, or indeed our Universe which has no Killing vectors in reality) will be explored further in Chapter 8, where we will again use the Farnsworth model as one of our canonical examples.

\section{Discussion}

In this chapter, we have developed a way to model anisotropic universes on cosmological scales, where the large-scale anisotropic description emerges from explicitly averaging over small-scale structures. 
The hope is that as long as any given cosmological spacetime has a well-defined homogeneity scale, then averages taken over spatial domains of that size or larger will produce a sensible description of the averaged properties of the spacetime. This, after all, is the implicit assumption behind all standard cosmological modelling, where the averaging procedure is essentially considered to be trivial, irrespective of whether the large-scale model in question is FLRW or some other spatially homogeneous cosmology.

The averaging formalism developed by Buchert is designed, among other reasons, to quantify to what extent the implicit assumption of trivial averaging is reasonable in the context of an isotropic Universe at late times, through the backreaction scalar $\mathcal{Q}\,$.
Here, we have derived the equivalent quantities $\mathcal{B}_i$ that would enable one to ask this question in an anisotropic universe, with the envisaged goal in particular of testing whether an emergent anisotropic cosmology would isotropise in the same way as its exact Bianchi counterpart, or whether the presence of the backreaction scalars would complicate that situation and potentially sustain anisotropic signatures on large scales in the late Universe. 
This is a very difficult question to answer in general, but we now have the mathematical tools in place to approach it. The exposition at the end of Section \ref{subsec:emergent_equations} suggests that the backreaction present in our formalism might have a sizeable effect on observational signatures, even if the scalars $\mathcal{B}_i$ are small enough to be analysed using perturbative methods.
To work out whether this is the case it is necessary to build an understanding of relevant late-time observables (the most obvious being the Hubble diagram) in the context of our formalism. 

If we are dealing with a spacetime displaying a well-defined homogeneity scale $L_{\rm hom}\,$, then it ought to follow that cosmological observables over scales $> L_{\rm hom}$ should be described well by the average model, with local effects due to inhomogeneities washed out in all measurements that are made over sufficiently large distances. If they are not, then an homogeneous averaged description cannot be considered a viable model for the observational properties of the Universe on the largest scales.
In the next chapter, we will study how our ``emergent anisotropy'' framework can be used to model luminosity distance and redshift observations in inhomogeneous and anisotropic universes.

\chapter{Hubble diagrams in statistically homogeneous, anisotropic universes}

\lhead{\emph{Hubble diagrams}}

Hubble diagrams describe the relationship between the redshift of light received from distant sources, and the luminosity distance to them. 
They are of fundamental importance in cosmology, having played a crucial historical role in the development of the standard cosmological model, and underpin a wide array of cosmological observables. 
Of course, Hubble diagrams are typically interpreted within the homogeneous and isotropic FLRW models of the Universe, and within this class of models can be used to determine the isotropic Hubble rate, $H_0$, and deceleration parameter $q_0$\,, as we explained in Section \ref{subsec:Hubble_diagram}. 
However, they can also be constructed in anisotropic cosmological models. In such cases, the luminosity distance as a function of redshift depends on the direction of observation in space, and one could imagine constructing Hubble diagrams along certain lines of sight, so that $H_0$ and $q_0$ become functions on the sky. 
This issue has been brought to the fore by the apparent directional dependence of dipole asymmetries in the CMB and matter distribution. It has led some to ask whether the late Universe might in fact be best described as being anisotropic on large scales \cite{Aluri_2023}.

In Chapter 7, we considered how such an anisotropy might potentially emerge from the formation of nonlinear structures \cite{Anton_2023}. In this chapter, we consider what the observational consequences of such a scenario might be.
We will focus in particular on how Hubble diagrams might be constructed in cosmologies that have emergent large-scale anisotropy in their expansion, and how the properties of those diagrams might be related to the average expansion and shear of space itself. The contents of this chapter are based on Ref. \cite{anton2024hubble}.

In order to investigate the possibility of emergent anisotropy, we will use the Sachs formalism for propagating bundles of rays of light in general spacetimes, which we introduced back in Section \ref{sec:light_propagation}. 
We will combine these ideas with our general framework, whose full set of equations we derived in the previous chapter, for incorporating the direction-dependent backreaction of inhomogeneities on the large-scale properties of space.
Our approach is entirely relativistic and non-perturbative, and will therefore allow us to test explicitly whether the optical properties of inhomogeneous cosmological spacetime can be described by an anisotropic average model, even in situations where the inhomogeneities are of very large scale and/or amplitude.

The problem of calculating luminosity distances in inhomogeneous cosmologies is far from a new one \cite{Dyer_1973}, with numerous studies having been performed in (for example) Swiss cheese models \cite{Koksbang_2020_a,Koksbang_2020_b,Brouzakis:2006dj,Brouzakis:2007zi,Biswas:2007gi,Vanderveld:2008vi}, Lema\^{i}tre-Tolman-Bondi and Szekeres cosmologies \cite{Marra:2007pm,Marra:2007gc,Bolejko_2008,clifton2009hubble,Bolejko:2008xh,Bolejko:2010eb,Peel:2014qaa,Koksbang:2017arw}, Lindquist-Wheeler models \cite{Clifton:2011mt,Liu:2015bya}, post-Newtonian cosmologies \cite{Sanghai:2017yyn}, N-body simulations \cite{Koksbang:2015ima,Koksbang:2023wez,adamek2014distance,lepori2020weak,lepori2021cosmological,lepori2023halo,adamek2024cosmography}, models using numerical relativity \cite{Heinesen_2022,macpherson2021luminosity,Macpherson:2022eve}, and cosmographic analyses that construct the low-$z$ Hubble diagram in a generalised way \cite{Heinesen_2021, kalbouneh2024cosmography, maartens2024covariant}.

In this chapter, we will extend this field by investigating the extent to which Hubble diagrams in inhomogeneous universes can be accurately described by a large-scale averaged model that is anisotropic. We will study not only the all-sky average of the Hubble diagram (i.e. the monopole), but also the full variation of the luminosity distance function across the skies of many observers. 
This is made possible by the framework we built in Chapter 7, as per Ref. \cite{Anton_2023}, which is an extended version of the spatial averaging procedure of Buchert \cite{Buchert_2000}. 
We find that our formalism can account well for that variation, as long as the average model is allowed to be anisotropic, and as long as the spatial averaging is done on an appropriate foliation of the spacetime.

The family of spacetimes we choose to consider for our study are dust-filled and plane symmetric. 
These solutions are discussed in detail in Section \ref{sec:plane_symmetric}, after we discuss our set of target models on to which averages can be mapped in Section \ref{subsec:emergent_dust}, and the techniques we use to calculate distance measures in Section \ref{subsec:ray_tracing_theory}. 
In Section \ref{sec:sinusoidal} we bring these techniques together in the context of inhomogeneous and anisotropic spacetimes with zero backreaction. We show that in that case there is a unique choice of homogeneous model corresponding to the scalar averages, and that within this class of models the averaged geometry permits a very good understanding of the Hubble diagrams of observers within it. 
We follow the same procedure in Sections \ref{sec:farnsworth} and \ref{sec:linear}, for spacetimes with non-zero backreaction. Section \ref{sec:farnsworth} deals with the tilted Farnsworth cosmologies we introduced already in Section \ref{sec:backreaction_farnsworth}. 
In contrast, Section \ref{sec:linear} considers an inhomogeneous and anisotropic spacetime where backreaction is small but non-zero. 

\section{Observables in anisotropic cosmologies}

Let us discuss how luminosity distance and redshift measurements can be interpreted in the context of anisotropic universes, while retaining the concept of emergence that is crucial to our entire framework. 
This will require us to introduce two key ingredients: (i) a tractable set of emergent cosmological models on to which the averages of covariantly defined scalars can be mapped, which we will take to be dust-dominated locally rotationally symmetric Bianchi cosmologies, and (ii) a method for calculating null geodesics in general curved spacetimes, and then using their properties to construct distances and redshifts using the null congruence formalism from Section \ref{sec:light_propagation}.

\subsection{Emergent LRS dust models}\label{subsec:emergent_dust}

When analysing large-scale cosmological observations, one typically has a model geometry in mind, whose parameters are to be fit to the data. Here, we are interested in anisotropic cosmologies, whose emergent behaviour is calculated using the Buchert-like averaging scheme we discussed at length in Chapter 7.
We expect that the large-scale dynamics of an anisotropic universe can be modelled as a locally rotationally symmetric (LRS) Bianchi spacetime, described by a set of averaged $1+1+2$-covariant scalars that obey the equations of motion (\ref{eq_averaged_Raychaudhuri}--\ref{eq_div_H_weyl_constraint_avg}). These equations are rather unwieldy, but they can be simplified significantly if we make two more additional suppositions:
\begin{itemize}
    \item The spacetime geometry is plane-symmetric, i.e. there exist two-dimensional planes of homogeneity orthogonal to the preferred spacelike direction $m^a$ through which the $1+1+2$-covariant scalars are constructed. We will explain the properties of plane-symmetric cosmological models in detail in Section \ref{sec:plane_symmetric}.
    \item The matter content of the universe is described by pressureless dust, with an irrotational 4-velocity that is accordingly the normal vector $n^a$ to the constant-$t$ leaves of a spacetime foliation $\Sigma_t\,$.
\end{itemize}
Under these assumptions, we can model the large-scale cosmology in terms of the Buchert averages $\avg{S}$ of only five covariantly defined scalars $S\,$. These are
\begin{eqnarray*}
    \Theta = \nabla_a n^a\,, \quad \rho = T_{ab}n^a n^b\,, \quad \phi = D_a m^a\,, \quad \Sigma = \sigma_{ab}m^a m^b \,, \quad \mathcal{E} = E_{ab}m^a m^b \,,
\end{eqnarray*}
which we recall from Section \ref{subsec:1+1+2} are scalars describing respectively the timelike expansion of $n^a\,$, the energy density measured by comoving observers, the spacelike expansion of $m^a\,$, the shear of $n^a\,$, and the electric part of the Weyl tensor.

After averaging, the LRS scalars $\left\lbrace \Theta, \rho, \phi, \Sigma, \mathcal{E}\right\rbrace$ should therefore be thought of as non-local, large-scale quantities that evolve according to
\begin{eqnarray}
\partial_t \avg{\Theta} + \frac{1}{3}\avg{\Theta}^2 + \frac{3}{2}\avg{\Sigma}^2 + \frac{1}{2}\avg{\rho} &=& \mathcal{B}_1 \label{eq_avg_LRS_first} \\
\partial_t \avg{\Sigma} + \frac{2}{3}\avg{\Theta}\avg{\Sigma} + \frac{1}{2}\avg{\Sigma}^2 + \avg{\mathcal{E}} &=& \mathcal{B}_3\\
\partial_t \avg{\phi} + \frac{1}{2}\left(\frac{2}{3}\avg{\Theta} - \avg{\Sigma}\right)\avg{\phi} &=& \mathcal{B}_6\\
\partial_t \avg{\mathcal{E}} + \left(\avg{\Theta} - \frac{3}{2}\avg{\Sigma}\right)\avg{\mathcal{E}} + \frac{1}{2}\avg{\rho}\avg{\Sigma} &=& \mathcal{B}_{12}\\
\partial_t\avg{\rho} + \avg{\Theta}\avg{\rho} &=& 0  \, ,\phantom{\frac{1}{2}} \label{eq_conservation_equation_avg}
\end{eqnarray} 
and which obey the constraints
\begin{eqnarray}
\frac{3}{2}\avg{\phi}\avg{\Sigma} &=& \mathcal{B}_2 \\
\frac{2}{9}\avg{\Theta}^2 - \frac{1}{2}\avg{\phi}^2 + \frac{1}{3}\avg{\Theta}\avg{\Sigma} -\avg{\Sigma}^2 - \frac{2}{3}\avg{\rho} - \avg{\mathcal{E}} &=& \mathcal{B}_5 \\
\frac{3}{2}\avg{\mathcal{E}}\avg{\phi} &=& \mathcal{B}_{14} \, . \label{eq_avg_LRS_last}
\end{eqnarray} 
These are a reduced and simplified set of equations compared to Eqs. (\ref{eq_averaged_Raychaudhuri}--\ref{eq_div_H_weyl_constraint_avg}). They have been tailored to the case of plane-symmetric dust-filled spacetimes, and we have also chosen the lapse function to be $N=1\,$, which we are free to do for pressureless matter \cite{Buchert_2018}.
Notably, the plane symmetry of the spacetimes we are dealing with here means that of the 15 emergent equations (\ref{eq_averaged_Raychaudhuri}--\ref{eq_div_H_weyl_constraint_avg}) in the general case, the ones containing $\left\lbrace \mathcal{B}_4, \mathcal{B}_7, \mathcal{B}_8, \mathcal{B}_9, \mathcal{B}_{11}, \mathcal{B}_{13}, \mathcal{B}_{15}\right\rbrace$ are absent. The backreaction term $\mathcal{B}_{10}$ would enter on the right hand side of Eq. (\ref{eq_conservation_equation_avg}) in general, but it vanishes for a foliation comoving with the dust fluid, which is exactly the situation that we are dealing with here.

The backreaction scalars $\mathcal{B}_i$ in Eqs. (\ref{eq_avg_LRS_first}--\ref{eq_avg_LRS_last}) encode the contribution of inhomogeneities to the averaged equations. They are given in our current setting by
\begin{eqnarray*}
\mathcal{B}_1 &=& \frac{2}{3}\,\var{\Theta} - \frac{3}{2}\,\var{\Sigma} \\
\mathcal{B}_2 &=& \frac{2}{3}\avg{m^a D_a \Theta} - \avg{m^a D_a \Sigma} - \frac{3}{2}\,\cov{\phi}{\Sigma} \\
\mathcal{B}_3 &=& \frac{1}{3}\,\cov{\Theta}{\Sigma} - \frac{1}{2}\,\var{\Sigma} \\
\mathcal{B}_5 &=& \frac{1}{2}\,\var{\phi} - \frac{2}{9}\,\var{\Theta} + \var{\Sigma} - \frac{1}{3}\,\cov{\Theta}{\Sigma} \\
\mathcal{B}_6 &=& \frac{2}{3}\,\cov{\Theta}{\phi} + \frac{1}{2}\,\cov{\phi}{\Sigma} \\
\mathcal{B}_{12} &=& \frac{3}{2}\,\cov{\Sigma}{\mathcal{E}} - \frac{1}{2}\cov{\Sigma}{\rho} \\
\mathcal{B}_{14} &=& - \avg{m^a D_a \mathcal{E}} + \frac{1}{3}\,\avg{m^a D_a \rho} - \frac{3}{2}\cov{\phi}{\mathcal{E}} \,, 
\end{eqnarray*}
which are far simpler than their general forms presented in Section \ref{subsec:emergent_equations}.
It can be seen that these scalars are composed by averaging spatial gradients of scalars along the preferred direction $m^a$\,, as well from the variances (Var) and covariances (Cov) of various of the covariantly-defined scalars.

Given a set of averaged scalars, one may wish to construct an homogeneous but anisotropic LRS cosmology out of them, in order to interpret the large-scale behaviour of the averaged spacetime. 
We can then calculate null geodesics on these emergent Bianchi spacetimes, and thereby construct the Hubble diagrams that observers in such idealised universes would measure. These can be compared to the Hubble diagrams constructed by observers in the underlying, non-averaged inhomogeneous spacetime.
In such a scenario, the existence of any non-zero backreaction scalars would indicate that the inhomogeneities that have been averaged away are having an influence on the dynamics of the large-scale cosmology. 

For the simplest possible example, the averaged scalars $\avg{\Theta}$ and $\avg{\Sigma}$ can be used to specify an emergent Bianchi type {\it I} line element of the form (\ref{eq_LRS_Bianchi_I}). 
The directional scale factors $A(t)$ and $B(t)$ are related to our averaged scalars by
\begin{eqnarray}\label{eq_Bianchi_Theta_Sigma}
\avg{\Theta} =    \frac{\dot{A}}{A} + \frac{2\dot{B}}{B}  \qquad \mathrm{and} \qquad \avg{\Sigma}= \frac{2}{3}\left(\frac{\dot{A}}{A} - \frac{\dot{B}}{B}\right) \,.
\end{eqnarray}
Similarly, one can write down a Bianchi type {\it V} line element of the form (\ref{eq_LRS_Bianchi_V}). In this case, $\avg{\Theta}$ and $\avg{\Sigma}$ are related to $A(t)$ and $B(t)$ as in the equations above, and the spatial curvature parameter $\beta$ is given according to 
\begin{equation}\label{eq_Bianchi_V_beta}
    -\frac{6\beta}{A^2} - \frac{\dot{A}^2}{2A^2} - \frac{\dot{B}^2}{2B^2} + \frac{\dot{A}\dot{B}}{AB} = \big\langle {}^{(3)}R \big\rangle = 2\avg{\rho} - \frac{2}{3}\avg{\Theta}^2 + \frac{3}{2}\avg{\Sigma}^2\,,
\end{equation}
where $^{(3)}R$ is the Ricci scalar curvature of the hypersurfaces orthogonal to $n_a$. 
In the final equality here we have made use of the Buchert average of the Hamiltonian constraint equation (\ref{eq_Hamiltonian_constraint_1+3}), in order to relate $\beta$ to $\avg{\rho}$, $\avg{\Theta}$ and $\avg{\Sigma}\,$. 

In both cases we recover an FLRW model when $A(t) = B (t)$\,, as can be seen from the vanishing of the shear in this case. For the type {\it I} spacetime that FLRW model is spatially flat, whereas for the type {\it V} model it is spatially curved. 
In principle, one could also construct average LRS models of Bianchi types {\it II, III, VII} or {\it IX}\,, but they will not be required in what follows.

\subsection{Calculating the Hubble diagram}\label{subsec:ray_tracing_theory}

In order to calculate redshifts and luminosity distances we first need to know the trajectories of rays of light in a given spacetime. 
As discussed in Section \ref{sec:light_propagation}, under the eikonal approximation these will be null geodesics with $k^b \nabla_b k_a = 0$, where $k^a$ is the tangent vector to the ray. 
Finding these paths can be achieved straightforwardly by constructing the Hamiltonian $\mathcal{H} = g^{ab}k_a k_b/2$, subject to the constraint $\mathcal{H}=0$\,, and using Hamilton's equations:
\begin{eqnarray}\label{eq_geodesic_equation_Hamilton}
    \frac{\mathrm{d}x^a}{\mathrm{d}\lambda} &=& \frac{\partial \mathcal{H}}{\partial k_a} = g^{ab}k_b\,, \\
    \nonumber \frac{\mathrm{d}k_a}{\mathrm{d}\lambda} &=& -\frac{\partial \mathcal{H}}{\partial x^a} = -\frac{1}{2}g^{bc}_{\ \ ,a}k_b k_c \,.
\end{eqnarray}
These equations will be integrated backwards in time from the observer to the source, by choosing $k^a$ to be past-directed. For the initial conditions, one requires a spacetime location $x^a_{\rm obs}$, and a propagation direction $e_a^{\rm obs}$ along which rays arrive at the observer.  
In defining $e_a^{\rm obs}$, we have decomposed the tangent vector $k^a$ with respect to the timelike vector $n^a$, such that $k^a = -E\left(n^a - e^a\right)$, where $e^a n_a = 0$ and $e^a e_a = 1\,$. 
The photon energy is then $E = k^a n_a$, and $e^a$ gives the direction of the ray in the spacelike hypersurfaces orthogonal to $n_a$\,.

The initial direction of propagation $e_a^{\rm obs}$ is chosen by specifying the angles $\left(\theta_c, \phi_c\right)\,$ on the observer's celestial sphere. 
These angles pick out a spacelike unit vector
\begin{equation}
\epsilon^i = -\left(\cos{\theta_c}, \sin{\theta_c}\cos{\phi_c}, \sin{\theta_c}\sin{\phi_c}\right)\, .
\end{equation}
Aligning $e_a^{\rm obs}$ with this unit vector by writing $e_a^{\rm obs} = g_{ai}(x^c_{\rm obs})\, \epsilon^i$\,, and using the null condition $k^a k_a=0$\,, is then sufficient to determine $k^a$ at the observer (up to the specific value of $E$). 
By varying the observing angles $\left(\theta_c, \phi_c\right)\,$, and the observer's spacetime location $x^a_{\rm obs}$\,, we can then calculate the path of any null geodesic. We can of course also calculate the redshift along any particular geodesic, using the usual definition
\begin{equation}\label{eq_redshift_def}
    1 + z := \frac{E(\lambda)}{E(0)} = \frac{k^a n_a\vert_{\lambda}}{k^a n_a \vert_{0}}\,,
\end{equation}
where we have specified to the case of a past-directed null geodesic that we are concerned with in this chapter.

Once the full set of null geodesic curves in the spacetime has been calculated, one must then solve Sachs' optical equations (\ref{eq_sachs_1}-\ref{eq_sachs_2}) for the expansion $\h{\theta}$ and shear $\h{\sigma}$ of a congruence of null geodesics. 
As explained in Section \ref{subsec:distance_measures}, these equations are more straightforwardly solved and interpreted by recasting them in terms of the angular diameter distance $d_A$\,, so that the Sachs equations  can be written in the simpler form (\ref{eq_sachs_dA_1}-\ref{eq_sachs_dA_2}). 

For the sake of clarity, we reproduce here all the equations that will be used in the remainder of this chapter to calculate the Hubble diagram given a set of null geodesics:
\begin{eqnarray}
    \frac{\mathrm{d}^2}{\mathrm{d} \lambda^2}\, d_A &=& \left(\Phi_{00} - \bar{\hat{\sigma}}\hat{\sigma}\right) d_A \, \label{eq_sachs_repeat_1}\\
    \frac{\mathrm{d}}{\mathrm{d}\lambda}\, \left(\hat{\sigma}\, d_A^2\right) &=& \Psi_0\, d_A^2 \,, \label{eq_sachs_repeat_2} \\
    \label{eq_dz_dlambda_2} \frac{\mathrm{d}z}{\mathrm{d}\lambda} \ &=& \: \left(1+z\right)^2 H^{\parallel}(z)\,, \\
    \label{eq_reciprocity_theorem_2} d_L &=& \left(1+z\right)^2\, d_A\,, \quad {\rm and} \\
    \label{eq_distance_modulus_2} \mu &=& 5\log{\left(\frac{d_L}{\rm Mpc}\right)} + 25\,,
\end{eqnarray}
where $\Phi_{00} = -R_{ab}k^a k^b/2$ and $\Psi_0 = C_{ab}s^a k^b s^c k^d\,$, with $s^a$ the complex null vector spanning the screen space orthogonal to $k^a\,$.
The null energy condition implies that $\Phi_{00} \leq 0$, and we recall that the quantity $H^{\parallel} = \dfrac{1}{E^2}k^a k^b \nabla_a n_b = \frac{1}{3}\Theta + \sigma_{ab} e^a e^b$ is the rate of expansion of space in the direction of the ray of light.
The fourth equation is Etherington's reciprocity relation \cite{Etherington_1933,ellis2009republication,ellis2012relativistic} whose physical origin was earlier demonstrated by Fig. \ref{fig_reciprocity_theorem}, and the final equation defines the distance modulus $\mu\,$.

\section{Plane-symmetric cosmological models}\label{sec:plane_symmetric}

The next ingredient that we need in order to implement our formalism is a set of inhomogeneous cosmological models. Our intent is to average these models using the equations from Section \ref{subsec:emergent_dust}, and to calculate observables in both the averaged and un-averaged spacetimes using the approach outlined in Section \ref{subsec:ray_tracing_theory}, which itself is based on the theory of light propagation in curved spacetime which we introduced in Section \ref{sec:light_propagation}. 
An appropriate choice is provided by the family of plane-symmetric dust-filled cosmologies, which exhibit a single preferred spacelike direction orthogonal to the planes of symmetry, and which can exhibit arbitrary amounts of inhomogeneity along this direction.

It is important to make clear at this point that plane-symmetric spacetimes are a subset of the locally rotationally symmetric spacetimes. In an LRS spacetime every point has associated with it a preferred spatial direction, but these directions are not typically aligned within a plane, with an obvious counterexample being the case of spherical symmetry, so there are LRS models that are not plane-symmetric. 
However, all plane-symmetric spacetimes are locally rotationally symmetric, because every point in each plane of symmetry has associated with it a 1-parameter isotropy group consisting of rotations in the plane \cite{Stephani_2003}.
 
The metric for plane-symmetric spacetimes can be written in the general form
\begin{equation}\label{eq_plane_symmetric_general_metric}
\mathrm{d}s^2 = -e^{2\nu(t)}\mathrm{d}t^2 + e^{2\lambda(t,r)}\mathrm{d}r^2 + R^2(t,r)\left(\mathrm{d}y^2 + \mathrm{d}z^2\right)\,.
\end{equation}
This general class of metrics includes the spatially-flat and negatively-curved FLRW, degenerate Kasner, and vacuum Taub solutions as special cases. 
Solutions to Einstein's equations for the metric (\ref{eq_plane_symmetric_general_metric}) can be split into two distinct classes: (i) those with $R' = \partial R/\partial r = 0\,$\footnote{We will follow the convention throughout this chapter that a prime denotes a partial derivative with respect to the preferred spatial coordinate $r\,$.}, and (ii) those with $R' \neq 0$. Both of these classes allow for significant inhomogeneity, but only the second will turn out to have non-zero backreaction scalars, $\mathcal{B}_i\,$.

The $y$ and $z$ coordinates in Eq. (\ref{eq_plane_symmetric_general_metric}) label points in the planes of symmetry, while $t$ and $r$ correspond to time and space directions orthogonal to those planes. 
Note that despite the notation, the coordinate $r$ should not be considered radial, but rather just a parameter that labels successive homogeneous constant-$r$ planes within a constant-$t$ hypersurface. The convention that it is labelled by $r$ is a result of the study of plane-symmetric spacetimes having been developed in analogy with pre-existing studies of spherically symmetric spacetimes, such as the LTB cosmologies \cite{Stephani_2003}.

For the plane-symmetric models, a natural choice for unit vectors in the preferred timelike and spacelike directions is $n_a = -\delta_a^{\ t}$ and $m_a = \exp \{ \lambda(t,r) \}\,\delta_a^{\ r}\,$\,. Then, the scalars $\left \lbrace \Theta,\rho,\phi,\Sigma,\mathcal{E}\right\rbrace\,$ are naturally all functions of $t$ and $r$ only.
Plane-symmetric spacetimes are therefore very well-suited for study within our formalism, and for the remainder of this section we will study the form that the metric functions $\nu(t)$, $\lambda(t,r)$ and $R(t,r)$ must take, in order to provide solutions to Einstein's equations.

\subsection{Dust-filled solutions with $R' = 0$}

When $R' = 0$\,, one can redefine the time coordinate such that $R(t) = t$\,. The metric functions can be expressed as \cite{Stephani_2003, Clifton_2019}
\begin{equation}
    e^{2\nu(t)} = \frac{t}{t_0} \quad {\rm and} \quad e^{2\lambda(t,r)} = \frac{t_0}{t}\left[c_1(r) \left(\frac{t}{t_0}\right)^{3/2} + c_2(r) \right]\,,
\end{equation}
where $c_1(r)$ and $c_2(r)$ are arbitrary functions of $r$, and $t_0$ is a constant with units of time.
By a further redefinition of the time coordinate, $t \longrightarrow t_0 \left({3t}/{2t_0}\right)^{2/3}$, we can set $\nu=0$. Finally, the factors of $3/2$ and $t_0$ can be absorbed into $c_1(r)$ and $c_2(r)$, to end up with the general solution in the form
\begin{equation}\label{eq_Rprime=0_metric}
\mathrm{d}s^2 = -\mathrm{d}t^2 + \left[c_1(r) \left(\frac{t}{t_0}\right)^{2/3} + c_2(r)\left(\frac{t}{t_0}\right)^{-1/3}\right]^2 \mathrm{d}r^2 + \left(\frac{t}{t_0}\right)^{4/3}\left(\mathrm{d}y^2 + \mathrm{d}z^2\right)\,.
\end{equation} 
The line element above corresponds to the Einstein-de Sitter solution when $c_2$ vanishes, and the degenerate Kasner solution when $c_1$ vanishes. 
The $1$+$1$+$2$-covariantly defined scalars in this case are given by
\begin{eqnarray}\label{eq_R'=0_scalars}
\Theta &=& \frac{2tc_1+ t_0 c_2}{t\left(tc_1+ t_0 c_2\right)}\,, \\
\nonumber \rho &=& \frac{4c_1}{3t\left(tc_1+ t_0 c_2\right)}\,,\\
\nonumber \Sigma &=&  -\frac{2t_0 c_2}{3t\left(tc_1+ t_0 c_2\right)}\,, \\
\nonumber \mathcal{E} &=&  -\frac{4t_0 c_2}{9t^2\left(tc_1+ t_0 c_2\right)}\,,
\end{eqnarray}
and $\phi = 0$.

Due to the plane symmetry, averages over spacelike domains reduce to ratios of one-dimensional integrals over the specified range of the $r$ coordinate, such that
\begin{equation}\label{eq_R'=0_average_definition}
\avg{S}(t) \ \equiv \ \frac{\int_{\mathcal{D}} \mathrm{d}^3 x \, \sqrt{^{(3)}g(t,r)} \, S(t,r)}{\int_{\mathcal{D}} \mathrm{d}^3 x \, \sqrt{^{(3)}g(t,r)}} \ = \ \frac{\int_{r_{\mathrm{min}}}^{r_{\mathrm{max}}} \mathrm{d}r \, \left(t c_1(r) + t_0 c_2(r)\right)\,S(t,r)}{\int_{r_{\mathrm{min}}}^{r_{\mathrm{max}}}\mathrm{d}r \, \left(t c_1(r) + t_0 c_2(r)\right)}\, ,
\end{equation}
for any scalar $S(t,r)\,$. Evaluating the simplified set of backreaction scalars in Section \ref{subsec:emergent_dust}, one finds that for any choice of the functions $c_1(r)$, $c_2(r)$, and for any interval $\left(r_{\mathrm{min}},r_{\mathrm{max}}\right)$, all $\mathcal{B}_i=0\,$, as long as $c_1(r)$ and $c_2(r)$ are integrable. 

As an example, to illustrate how this works, consider the scalar $\mathcal{B}_1=\dfrac{2}{3}\,\var{\Theta} - \dfrac{3}{2}\,\var{\Sigma}$, for which we have
\begin{equation}
    \nonumber {\rm Var}\,\Theta = \frac{\int \mathrm{d}r \, \left(t c_1(r) + t_0 c_2(r)\right)\int \mathrm{d}r' \, \frac{\left(2tc_1(r')+t_0 c_2(r')\right)^2}{tc_1(r')+ t_0 c_2(r')} - \left(\int\mathrm{d}r \,\left(2tc_1(r)+ t_0 c_2(r)\right)\right)^2}{t^2\left(\int \mathrm{d}r \, \left(t c_1(r) + t_0 c_2(r)\right)\right)^2}\,,
\end{equation}
and
\begin{equation}
    \nonumber {\rm Var}\,\Sigma = \frac{4}{9t^2}\,\frac{\int \mathrm{d}r \, \left(t c_1(r) + t_0 c_2(r)\right)\int \mathrm{d}r' \,\frac{t_0^2 c_2(r')^2}{tc_1(r')+ t_0 c_2(r')} - \left(\int \mathrm{d}r \, t_0 c_2(r)\right)^2 }{\left(\int \mathrm{d}r \,\left(tc_1(r)+ t_0 c_2(r)\right)\right)^2}\,,
\end{equation}
where all integrals should be understood to be between $r_{\mathrm{min}}$ and $r_{\mathrm{max}}\,$. This clearly demonstrates that $\mathcal{B}_1$ vanishes, as long as all the integrals in the two above expressions are well-defined. Similar calculations show that the other $\mathcal{B}_i$ also vanish. 

This pleasing result can be understood by thinking about the averages of $c_{1}(r)$ and $c_{2}(r)$: 
\begin{equation}
\left\langle c_{1,2}\right\rangle(t) := \frac{\int\mathrm{d}r\,\left(tc_1(r)+t_0 c_2(r)\right)c_{1,2}(r)}{\int \mathrm{d}r\,\left(tc_1(r)+ t_0 c_2(r)\right)}\,,
\end{equation}
where the averages pick up a time dependence due to the presence of $t$ in the integrands. From these, we can define an effective line element
\begin{equation}
    \mathrm{d}s^2_{\rm eff} = -\mathrm{d}t^2 + A^2(t)\,\mathrm{d}r^2 + B^2(t)\,\left(\mathrm{d}y^2 + \mathrm{d}z^2\right),
\end{equation}
where 
\begin{equation}\label{eq_Bianchi_I_Aeff}
    A(t) = \left\langle c_1\right\rangle(t) \,\left(\frac{t}{t_0}\right)^{2/3} + \left\langle c_2\right\rangle(t)\,\left(\frac{t}{t_0}\right)^{-1/3}\qquad {\rm and} \qquad B(t) = \left(\frac{t}{t_0}\right)^{2/3} \, .
\end{equation}
This means that the averaged geometry behaves precisely like a degenerate Bianchi type {\it I} cosmology, with directionally-dependent scale factors $a_r(t) = A(t)$ and $a_y(t) = a_z(t) = B(t)\,$.
As this is a member of the target space of solutions in our averaging formalism (i.e. it is an LRS Bianchi spacetime), there is no backreaction.
However, if one were to take the target space for one's averaging procedure to be FLRW, as in most approaches to scalar averaging in cosmology, then this effective line element cannot be mapped exactly onto that space. Hence, performing averages that map this class of spacetimes onto an FLRW cosmology must necessarily involve some non-zero amount of backreaction. 
This exemplifies the usefulness of our approach, which is designed specifically for understanding spacetimes with large-scale anisotropy.
It also provides a way to understand the result that the average square of the shear, $\avg{\sigma^2}$, need not be small \cite{marozzi2012late}, which would usually be interpreted as a contribution to Buchert's backreaction scalar $\mathcal{B}$, but in the present case would be accounted for by the emergent large-scale anisotropy.
The result in Eq. (\ref{eq_Bianchi_I_Aeff}) means that we can always identify a unique homogeneous model that describes the large-scale dynamics, as long as we are prepared for that model to be anisotropic.

Finally, let us write down the quantities required to solve the Sachs equations (\ref{eq_sachs_dA_1}) and (\ref{eq_sachs_dA_2}), plus the equation (\ref{eq_dz_dlambda}) that allows us to write our results in terms of redshifts rather than the affine parameter along a null ray
In the $R' = 0$ class of plane-symmetric spacetimes, these quantities are
\begin{eqnarray*}\label{eq:H_parallel_R'=0}
    \hspace{-1cm}
    H^{\parallel} &=& \frac{2tc_1 + t_0 c_2}{3t\left(tc_1 + t_0 c_2\right)} 
    + \frac{\left(\frac{t}{t_0}\right)^{2/3} t_0^3 c_2 \left\lbrace -2k_r^2 t^2 + \left(k_y^2 + k_z^2\right)\left(t c_1 + t_0 c_2\right)^2\right\rbrace}{3k_t^2 t^3 \left(tc_1 + t_0 c_2\right)^3}\,, \\
\label{eq:ricci_R'=0}
    \hspace{-1cm} \, \Phi_{00} &=& -\frac{c_1}{3\left(tc_1 + t_0 c_2\right)}\left[\frac{k_r^2 t_0}{\left(\frac{t}{t_0}\right)^{1/3}\left(tc_1+t_0c_2\right)^2} + \frac{k_t^2 t^2 + \left(k_y^2 + k_z\right)^2 t_0^2 \left(\frac{t}{t_0}\right)^{2/3}}{t^3}\right]\,,\\
\label{eq:weyl_R'=0}
\Psi_0 &=& \frac{\left(k_y^2 + k_z^2\right) t_0 c_2 \left[k_r^2 t^2 + \left(k_t^2 \left(\frac{t}{t_0}\right)^{4/3} + \left(k_y^2 + k_z^2\right)\right)\left(tc_1 + t_0 c_2\right)^2 \right]}{3t^2 \left(\frac{t}{t_0}\right)^{4/3}\left(tc_1 + t_0c_2\right)\left[k_r^2 t^2 + \left(k_y^2 + k_z^2\right)\left(tc_1 + t_0c_2\right)^2\right]}\,,
\end{eqnarray*}
where $k_a$ is the tangent vector to the ray of light.
With these quantities calculated as a function of affine distance along every null ray arriving at the observer, one can solve Sachs' optical equations directly.

\subsection{Dust-filled solutions with $R' \neq 0$}

Let us now consider the class of plane-symmetric dust-filled cosmologies with $R' \neq 0$. In this case, one has
\begin{equation}
    G_{tr} = \frac{2}{R}\left[R' \dot{\lambda} - \dot{R'}\right] \, \overset{!}{=} \, 0\,,
\end{equation}
where dots denote partial derivatives with respect to $t$, and where we have assumed $T_{tr}=0$ by aligning $\partial_t$ with the flow lines of the dust. This equation is solved by $\lambda = \ln{R'} - \ln{f}$, where $f \equiv f(r)$ is any arbitrary function of $r$\,. We can therefore write the metric as
\begin{equation}\label{eq_R'_neq_0_metric_general}
\mathrm{d}s^2 = -\mathrm{d}t^2 + \frac{R'^2(t,r)}{f^2(r)}\,\mathrm{d}r^2 + R^2(t,r)\left(\mathrm{d}y^2+\mathrm{d}z^2\right)\,,
\end{equation}
where we have assumed that the matter is dust and chosen the time coordinate to set $\nu = 0\,$. 

The kinematic $1$+$1$+$2$-scalars are then given by
\begin{eqnarray}\label{eq_R'_neq_0_scalars}
\Theta &=& \frac{2\dot{R}}{R} + \frac{\dot{R}'}{R'} \qquad {\rm and} \qquad
\Sigma = \frac{2}{3}\left(\frac{\dot{R}'}{R'} - \frac{\dot{R}}{R}\right) \, ,
\end{eqnarray}
while the scalars $\rho$ and $\mathcal{E}$ that fully characterise the Ricci and Weyl curvature tensors are respectively
\begin{eqnarray}
\rho &=& \frac{-2Rff' - f^2 R' + \dot{R}\left(R'\dot{R} + 2R\dot{R}'\right)}{R^2 R'}
\\
\mathcal{E} &=& \frac{-Rff' + R'\left(f^2 + R\ddot{R}-\dot{R}^2\right) + R\left(\dot{R}\dot{R}' - R\ddot{R}'\right)}{3R^2 R'}\, ,
\end{eqnarray}
and we also have $\phi = \dfrac{2f}{R}\,$. 
Note that in the $R'\neq 0$ class we have $\phi\neq 0\,$, in contrast to what happens in the case $R' = 0\,$.

Within this class, the rest of Einstein's equations are solved completely if we write the following constraint equation \cite{Stephani_2003}:
\begin{equation}\label{eq_constraint_rdotsq}
    \dot{R}^2 - f^2(r) = \frac{2m(r)}{R(t,r)}\,,
\end{equation}
where $m(r)$ is another arbitrary function, which can be related to the energy density of the dust according to
\begin{equation}\label{eq_mu_mprime}
    \rho(t,r) = \frac{2m'(r)}{R^2 R'(t,r)}\,.
\end{equation}
Eq. (\ref{eq_constraint_rdotsq}) is solved in parametric form by
\begin{eqnarray}\label{eq_R'_neq_0_solution_parametric}
    R(t,r) &=& \frac{m(r)}{f^2(r)}\left(\cosh{\eta}-1\right)\,, \\
    \nonumber t - t_0(r) &=& \frac{m(r)}{f^3(r)}\left(\sinh{\eta}-\eta\right)\,,
\end{eqnarray}
where $t_0(r)$ is a third free function, which can thought of as setting the bang time at each value of $r$\,, such that the coordinate extent of the spacetime is bounded by the curve $t = t_0(r)\,$. A choice of the three free functions $f(r)$, $m(r)$ and $t_0(r)$ specifies a solution to Einstein's equations of the form given in Eq. (\ref{eq_R'_neq_0_metric_general}). 
These constitute two independent functional degrees of freedom, as there remains a freedom in reparametrising the $r$ coordinate.

In the present case, the plane symmetry of the spacetime means that calculating the Buchert averages reduces to computing a set of one-dimensional integrals of the form
\begin{equation}
\avg{S}(t) = \frac{\int_{r_{\rm min}}^{r_{\rm max}} \mathrm{d}r\, R^2(t,r) \, \left\vert \frac{R'(t,r)}{f(r)}\right\vert \, S(t,r)}{\int_{r_{\rm min}}^{r_{\rm max}} \mathrm{d}r\, R^2(t,r) \, \left\vert \frac{R'(t,r)}{f(r)}\right\vert }\: .
\end{equation}
Aside from a small number of special cases, the backreaction scalars are generically non-zero for these solutions, as will be verified numerically in Sections \ref{sec:farnsworth} and \ref{sec:linear}.

For any set of functions $f(r)$, $m(r)$ and $t_0(r)$, the quantities required to solve Sachs' optical equations in the $R' \neq 0$ class are
\begin{eqnarray*}
\hspace{-2cm}H^{\parallel} &=& \frac{2k_r^2 f^2 R^2\left(\dot{R}' R - R' \dot{R}\right) + R'^2\left\lbrace \left(k_y^2 + k_z^2 + 2k_t^2 R^2\right)R'\dot{R}-\left(k_y^2 + k_z^2 - k_t^2 R^2\right)R\dot{R}'\right\rbrace}{3k_t^2 R^3 R'^3}\,
\\
\hspace{-2cm}\Phi_{00} &=& \frac{1}{2R^4 R'^3}\Big[2k_r^2 f^3 f' R^3 + \left(k_y^2 + k_z^2\right) ff'RR'^2 + f^2\left\lbrace \left(k_y^2 + k_z^2\right)R'^3 - k_r^2 R^3 \left(2\dot{R}\dot{R}' + R\ddot{R}'\right)\right\rbrace \\
\hspace{-2cm}&&\hspace{0.5cm}  - R'^2 \left\lbrace \left(k_y^2 + k_z^2\right)R\dot{R}\dot{R}' + R'\left(\left(k_y^2 + k_z^2\right)\dot{R}^2 + \left(k_y^2 + k_z^2 - 2k_t^2 R^2 \right)R\ddot{R}\right) - k_t^2 R^4 \ddot{R}'\right\rbrace\Big]\,,
\\
\hspace{-2cm}\Psi_0 &=& \frac{\left(k_y^2 + k_z^2\right)\left(k_r^2 f^2 R^2 + \left(k_y^2 + k_z^2 + k_t^2 R^2\right)R'^2 \right)\left[Rff' + R'\left(\dot{R}^2 - R\ddot{R} - f^2\right) + R\left(R\ddot{R}' - \dot{R}\dot{R}'\right)\right]}{4R^4 R' \left[k_r^2 f^2 R^2 + \left(k_y^2 + k_z^2\right)R'^2\right]}\,.
\end{eqnarray*}
We will use these equations to create Hubble diagrams in tilted and inhomogeneous spacetimes in Sections \ref{sec:farnsworth} and \ref{sec:linear}.

\section{An $R' = 0$ universe with inhomogeneity}\label{sec:sinusoidal}

In this section we consider Hubble diagrams constructed in plane-symmetric dust-dominated cosmologies with $R' = 0\,$, in which all $\mathcal{B}_i=0$\,. This means that the average evolution of the cosmology is exactly equivalent to that of a Bianchi model, and the metric can be written as in Eq. (\ref{eq_Rprime=0_metric}). 
In these cases, Buchert's backreaction scalar $\mathcal{Q}$, as introduced in Eqs. (\ref{eq_Buchert_1}-\ref{eq_Buchert_2}), does not vanish \cite{Clifton_2019}, even though the scalars $\mathcal{B}_i$ from our anisotropic formalism are all zero. 

\begin{figure}
    \centering
    \includegraphics[width=0.8\linewidth]{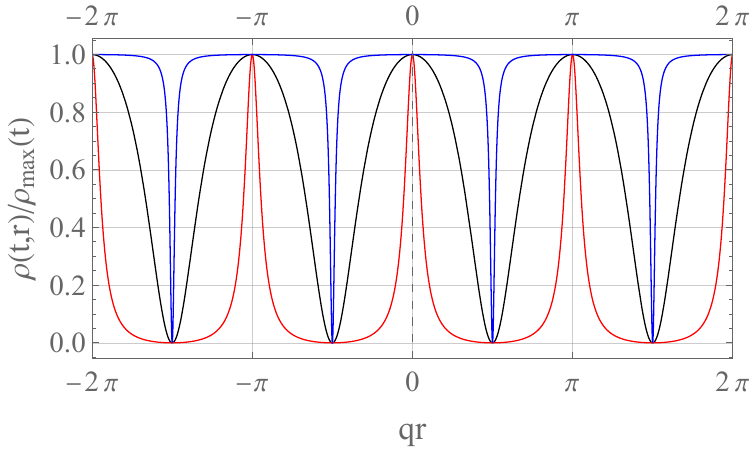}
    \caption{Energy density profile on constant-time hypersurfaces, normalised by its maximum value, in the $R' = 0$ geometry with $c_1(r) = 2\cos^2{qr}$, $c_2(r) = 2\eta\sin^2{qr}\,$, and $\eta = 1/3\,$. Curves correspond to $t=t_0$ (black), $t=t_0/100$ (red) and $t=100 t_0$ (blue).}
    \label{fig_sinusoidal_model_density}
\end{figure}

Let us construct a spacetime within this class that exhibits non-perturbative inhomogeneity in the matter distribution. This can be achieved by choosing the free functions $c_1(r)$ and $c_2(r)$ to be oscillatory, such that $c_1(r) = 2 \cos^2{qr}$ and $c_2(r) = 2\eta \sin^2{qr}$. 
The energy density of the dust, as measured by comoving observers, is then
\begin{equation}
\rho(t,r) = \frac{8\cos^2{qr}}{3t\left(t+\eta + \left(t-\eta\right)\cos{2qr}\right)}\ ,
\end{equation}
where for simplicity we have normalised the time coordinate so that $t_0 = 1\,$. 
For $t \ll \eta$, $\rho \longrightarrow \dfrac{4\tan^2{qr}}{3t\eta}\,$, so the density profile is dominated by Kasner-like vacuum at early times, with small regions of very high density, whereas for $t \gg \eta$\,, $\rho \longrightarrow \dfrac{4}{3t^2}\,$, as the density profile tends towards that of an homogeneous Einstein-de Sitter (EdS) universe. These features are indicated by the red and blue curves respectively in Fig. \ref{fig_sinusoidal_model_density}.

Correspondingly, we have from Eq. (\ref{eq_R'=0_scalars}) that the expansion, shear and electric Weyl scalars are
\begin{eqnarray}
\Theta(t,r) &=& \frac{2t + \eta + \left(2t-\eta\right) \cos{2qr}}{t\left[t + \eta + \left(t-\eta\right)\cos{2qr}\right]}\: ,\\
\Sigma(t,r) &=& \frac{-4\eta \sin^2{qr}}{3t\left[t + \eta + \left(t-\eta\right)\cos{2qr}\right]}\: ,\\
\mathcal{E}(t,r) &=& \frac{-8\eta \sin^2{qr}}{9t^2\left[t + \eta + \left(t-\eta\right)\cos{2qr}\right]}\: = \frac{2}{3t}\,\Sigma(t,r) \, .
\end{eqnarray}
From these one may reconstruct the shear tensor as $\sigma_{ab} = \Sigma \left(3m_a m_b - n_a n_b - g_{ab}\right)/2\,$, and the electric part of the Weyl tensor as $E_{ab} = \mathcal{E} \left(3m_a m_b - n_a n_b - g_{ab}\right)/2\,$.  
The expansion scalar $\Theta$ behaves in a similar fashion to the plots of $\rho$ in Fig. \ref{fig_sinusoidal_model_density}, while the scalars $\mathcal{E}$ and $\Sigma$ have the opposite behaviour. At early (vacuum-dominated) times $\mathcal{E}$ and $\Sigma$ are mostly large and non-zero (as in Kasner), whereas at late (matter-dominated) times they are mostly zero (as in EdS), except for spikes in the vacuum regions. 

\subsection{Ray tracing}

We solve the geodesic equations (\ref{eq_geodesic_equation_Hamilton}) for a large number of observing directions $\theta_c$, and observer positions $r_{\rm obs}\,$, for observers at time $t = t_0$. Because of the plane symmetry of the spacetime, the initial coordinates $y_0$ and $z_0$ are irrelevant, as is the azimuthal angle $\phi_c$ on the observer's celestial sphere (we choose $\phi_c = 5\pi/4\,$, for the sake of numerical simplicity). 
Moreover, the symmetry of the sinusoidal metric profile means that one only need consider $\theta_c$ in the range $[0,\pi/2]\,$. We can then numerically integrate the geodesic and Sachs equations.  The free parameters in the metric are set to $\eta = 1/3$ and $q = 100\,$. 
We normalise the time coordinate so that the observing time $t_{\rm obs} = t_0$ is equal to 1, and consider $100$ observers at even $r$-coordinate spacings between $r = 0$ and $r = \pi/q$\,. The angular range is split up into discrete intervals of $\Delta \,\theta_c = \pi/200\,$.

\begin{figure}
    \centering
    \includegraphics[width=\linewidth]{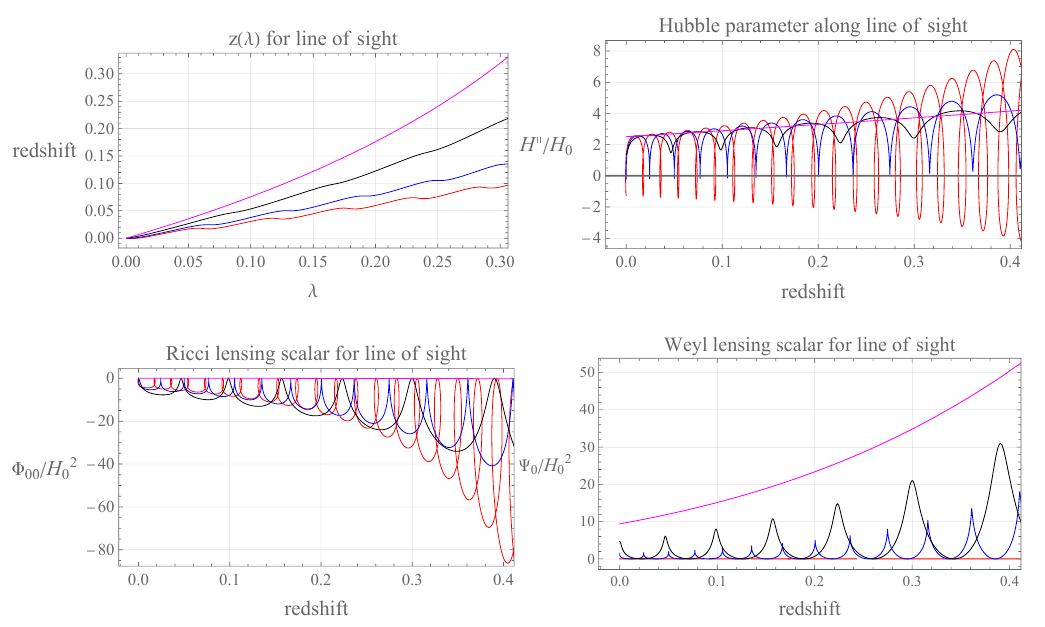}
    \caption{Top left panel: photon redshift $z$ as a function of affine parameter $\lambda$, for observing angles $\theta_c = 0$ (red), $\pi/8$ (blue), $\pi/4$ (black) and $\pi/2$ (magenta) in the $R'=0$ geometry. Top right: line-of-sight Hubble parameter $H^{\parallel} = \Theta/3 + \sigma_{ab}e^a e^b$ (normalised by its monopole $H_0$ at $z = 0$). Bottom left and right: Ricci and Weyl lensing scalars $\Phi_{00} = -R_{ab}k^a k^b/2$ and $\Psi_0 = C_{abcd} s^a k^b s^c k^d$ as functions of redshift for the same observer (both made dimensionless by normalising with respect to $H_0^2\,$).}
    \label{fig_sinusoidal_ray_tracing}
\end{figure}

Some key results of carrying out the ray tracing are shown in Fig. \ref{fig_sinusoidal_ray_tracing}. In these plots, we have considered an observer at time $t_0$ who is placed at a point $r_{\rm obs} = \pi/2q$, which corresponds to the centre of an underdensity, i.e. $\rho(t_0, r_{\rm obs}) = 0\,$.
For each initial direction $\theta_c$, we can first calculate the redshift $z$ as a function of the affine parameter in the geodesic equation (\ref{eq_geodesic_equation_Hamilton}). 
For small $\theta_c$ (e.g. the red and blue curves in the top left plot of Fig. \ref{fig_sinusoidal_ray_tracing}), the bumpy nature of the function $z(\lambda)$ indicates the strong oscillatory inhomogeneities in the metric as light rays propagate in that direction. 
For those directions, the function $z(\lambda)$ is not monotonically increasing. When a future-directed null ray passes through the vacuum-dominated regions, it can gain energy as it moves forward in time (and vice versa for past-directed null rays), if it is directed along or close to the spatially contracting symmetry axis of the rotationally symmetric Kasner-like geometry.
For large $\theta_c$ (e.g. the black and pink curves), $z(\lambda)$ is much smoother, because those observers are looking in directions along which the spacetime is much closer to homogeneous.

With $z(\lambda)$ obtained, one can then calculate any quantity as a function of the observable redshift, rather than the non-observable $\lambda$. In the top right plot of Fig. \ref{fig_sinusoidal_ray_tracing} we show the expansion rate $H^{\parallel}$ parallel to the corresponding null geodesic, as a function of the redshift along that curve. 
As the light ray passes through the underdense regions, the function $H^{\parallel}$ dips and can become negative if the effect of the inhomogeneity is sufficiently strong, as can happen for small $\theta_c$\,. 

The loops in the $H^{\parallel}$ curve for $\theta_c = 0$ reflect the non-monotonicity of $z(\lambda)$, wherein a past-directed null ray moving from an overdense to an underdense region initially has $H^{\parallel}$ decreasing but $\dfrac{\mathrm{d}z}{\mathrm{d}\lambda}$ remaining positive. This produces the leading, right-hand side of each loop. 
Then as the light ray is moving towards the centre of the underdensity, $\dfrac{\mathrm{d}z}{\mathrm{d}\lambda}$ becomes negative as $H^{\parallel}$ crosses zero, as per Eq. (\ref{eq_dz_dlambda_2}). 
After the null ray passes the centre of the underdensity, $H^{\parallel}$ begins to increase once more. Finally $H^{\parallel}$ passes back through zero, and therefore the redshift begins to increase again. For null rays travelling in directions where $H^{\parallel}$ is always positive, these loops do not exist. Moreover, as seen in the pink curve, for $\theta_c = \pi/2$ the geometry that the ray moves through appears spatially homogeneous, and $H^{\parallel}$ increases monotonically.

Because we are displaying the results for observers situated within a vacuum-dominated region, the expansion rate $H^{\parallel}_0$ at redshift zero is large and negative for $\theta_c = 0$, with $\vert H_0^{\parallel} \vert$ similar to the magnitude of the all-sky average of $H_0$. 
The expansion rate of space along the line of sight then monotonically increases as a function of $\theta_c$, to over twice the all-sky average at $\theta_c = \pi/2\,$. This is indicative of the very strong anisotropy in the spacetime.

One sees a similar set of effects for the Ricci and Weyl curvature terms $\Phi_{00}$ and $\Psi_0$, in the bottom left and bottom right plots of Fig. \ref{fig_sinusoidal_ray_tracing} respectively. 
The scalar $\Psi_0$ is small in matter-dominated regions, where the EdS-like dynamics dominate, and rises sharply in the Kasner-like underdensities. For null rays with $\theta_c$ near $\pi/2$, $\Psi_0$ increases monotonically, and $\Psi_0 \gg \vert \Phi_{00} \vert$ at all redshifts. The first of these facts is explained, like for the $H^{\parallel}$ curve in the top right plot of this figure, by the apparent homogeneity along that line of sight. 
For the latter fact, we recall that $\Phi_{00} = -R_{ab}k^a k^b/2 = -T_{ab}k^a k^b/2$. Therefore, for a light ray that always resides in regions of zero (or very low) matter density, the energy-momentum tensor remains zero (or near zero), and so the spacetime curvature is dominated instead by the free gravitational field encoded in the Weyl curvature.

For null rays with $\theta_c = 0$, $\Phi_{00}$ is large (and negative, as it must be due to the null energy condition) at most points where the density is sufficiently large that the dynamics are EdS-like, and then drops sharply to zero in vacuum/near-vacuum regions. 
Note, however, that the Weyl scalar $\Psi_0$ remains zero throughout, because a null congruence that travels directly along the rotational symmetry axis of the spacetime cannot be sheared (it is a principal null direction). 
The loops in $\Phi_{00}(z)$ have the same origin as those in $H(z)\,$. They are also just about visible in the blue curve ($\theta_c = \pi/8$) in the bottom right plot. For $\theta_c = \pi/8\,$, the Weyl term is non-zero, but is strongly suppressed relative to the Ricci curvature.

With past-directed solutions to the geodesic equation (\ref{eq_geodesic_equation_Hamilton}) known, we can now solve the Sachs equations (\ref{eq_sachs_repeat_1}) and (\ref{eq_sachs_repeat_2}) to obtain the angular diameter distance $d_A(z)$\,, and hence the luminosity distance, along individual lines of sight. This will be done for each possible $\theta_c$, at multiple observer locations $r_{\rm obs}$ on the $t = t_0$ constant-time hypersurface. 

\subsection{Hubble diagrams}

A typical approach used in cosmology is to consider the Hubble diagrams that would result in some average cosmology, which is expected to reproduce the large-scale properties of the Universe. This process is left somewhat implicit in the standard FLRW framework, but is done explicitly in the case of the Buchert averaging procedure. In doing this, one obtains a ``Hubble diagram of the average''. 
As we are currently dealing with an $R' = 0$ metric, it follows that for any choice of averaging domain the backreaction scalars will vanish, and the large-scale model is obtained by simply replacing the sinusoidal functions $c_1(r)$ and $c_2(r)$ by their average values. 
In the present case, the average model is therefore described by a Bianchi type {\it I} metric (\ref{eq_LRS_Bianchi_I}), with scale factors $A(t) = \left(\dfrac{t}{t_0}\right)^{2/3} + \eta \left(\dfrac{t}{t_0}\right)^{-1/3}$ and $B(t) = \left(\dfrac{t}{t_0}\right)^{2/3}\,$. 
We can now perform ray tracing in that averaged model universe, and compare the results to those of the inhomogeneous model that existed before averaging. The averaged model in this case is homogeneous but anisotropic, meaning that the spatial location of the observer is irrelevant, but that we still expect angular variation in the Hubble diagram. 

By integrating the Sachs equations using the averaged Bianchi {\it I} metric, we finally calculate $d_L^{\rm model}(z)$ in different directions on the observer's sky. This will let us determine how accurately that homogeneous averaged model fits the distance-redshift relation for observers in the actual inhomogeneous spacetime.

\begin{figure}
    \centering
    \includegraphics[width=\linewidth]{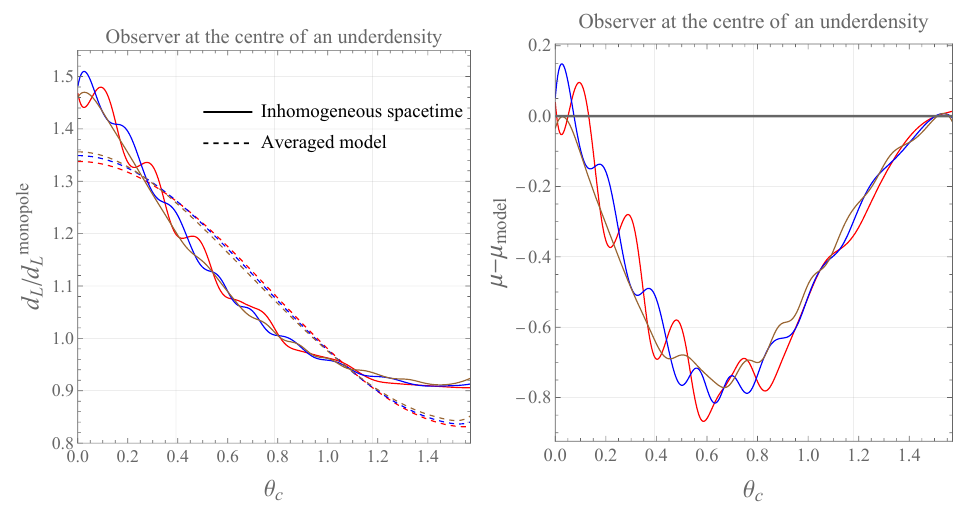}
    \caption{Left: luminosity distance $d_L$ at a given redshift as a function of observing angle $\theta_c$, relative to the monopole, for an observer at the centre of an underdensity in the $R'=0$ geometry (as per Eq. (\ref{eq_Rprime=0_metric}), with $c_1(r) = 2\cos^2{qr}$ and $c_2(r) = 2\eta\sin^2{qr}$\,, where $\eta = 1/3$\,). The curves are for $z = 0.1$ (red), $z = 0.2$ (blue), and $z = 0.3$ (brown). The solid lines are obtained by performing ray tracing in the inhomogeneous spacetime, and the dashed lines come from ray-tracing in the averaged models. 
    Right: difference between distance modulus $\mu$ obtained by from ray-tracing in the inhomogeneous spacetime and averaged model, as a function of $\theta_c$. Colours are as given in the left-hand plot.}
    \label{fig_sinusoidal_dL_theta}
\end{figure}

First, we consider the angular variation of $d_L$, at a constant specified redshift. Such a variation is entirely absent in an isotropic universe, but is non-trivial, and dependent on the observer location, in an inhomogeneous spacetime. Our results are shown in the left-hand plot of Fig. \ref{fig_sinusoidal_dL_theta}. 
The function $d_L(\theta_c)$ is rather complicated in the inhomogeneous spacetime, as each light ray will have passed through a continually oscillating geometry, leading to an oscillatory pattern within the function, arising from the $\cos^2{qr}$ and $\sin^2{qr}$ terms in the metric. 
That pattern is itself contained in a quadrupolar envelope, which, rather than coming from any oscillations, is due to the overall discrepancy between the expansion rates in the $r$ coordinate direction and each of the $y$ and $z$ directions, producing a shear $\sigma_{ab}$ which is naturally quadrupolar and affects the Hubble diagram through e.g. the line-of-sight Hubble parameter $H_0^{\parallel}$ at $\mathcal{O}(z)\,$, the line-of-sight deceleration parameter $q_0^{\parallel}$ at $\mathcal{O}(z^2)\,$, and so on.
Thus, the envelope is a signature of the global anisotropy that arises due to the presence of Kasner-like regions, which are contracting along $\theta_c = 0$ and therefore reducing redshifts for a given luminosity distance in that direction. 
Therefore, to reach the target redshift requires the past-directed null geodesic to travel ``further'', meaning that $d_L$ exceeds the all-sky average (i.e. the monopole). 

Conversely, observing at $\theta_c = \pi/2$ means that one sees the lowest possible $d_L$ for that redshift, as there is only EdS-like expansion along that line of sight. 
The averaged Bianchi {\it I} model captures something of the quadrupolar envelope, which in that context can be interpreted in terms of the shear anisotropy $\sigma_{ab}$, but loses all information about any higher-order multipoles (with $l \geq 2$).

For observers residing in underdense, roughly Kasner-like, regions, $d_L$ is overestimated at the axes and underestimated in between, because the local geometry is less isotropic than the average. 
The extent of the average model's success (or lack thereof) in predicting the Hubble diagrams of an individual observer is displayed in the right plot of Fig. \ref{fig_sinusoidal_dL_theta}, wherein we see that the angular variation of $\mu - \mu_{\rm model}$ is roughly maintained as $z$ increases. 

\begin{figure}
    \centering
    \includegraphics[width=0.68\linewidth]{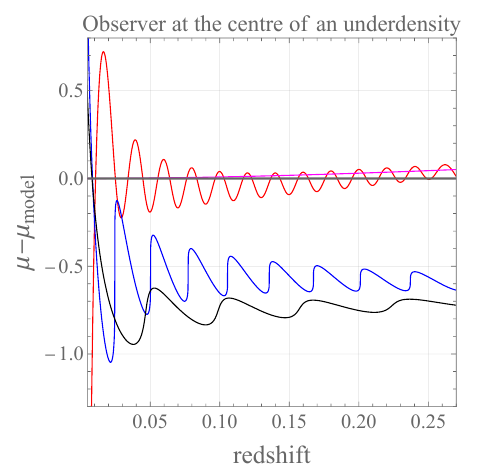}
    \caption{Difference in distance modulus from the averaged model, $\mu - \mu_{\mathrm{model}}\,$, as a function of redshift, shown for an observer at the centre of an underdense region in the $R'=0$ geometry. The curves are for $\theta_c = 0$ (red), $\theta_c = \pi/8$ (blue), $\theta_c = \pi/4$ (black), and $\theta_c = \pi/2$ (magenta).}
    \label{fig_sinusoidal_magnitude_vs_model}
\end{figure}

Fig. \ref{fig_sinusoidal_magnitude_vs_model} shows the performance of the averaged homogeneous model as a function of redshift, for the same observer.
The magnitude difference can be very large at low redshifts, as one expects: on very small scales, and in the presence of very large inhomogeneities, observations in one's immediate vicinity do not produce an accurate Hubble diagram of the universe at large. 
However, as null congruences travel through many cycles of the oscillatory geometry, $d_L$ becomes much closer to the average. This is seen most clearly in the red curve, where observing along the axis of inhomogeneity $\theta_c = 0$ means that the light ray ultimately samples equal numbers of overdensities and underdensities. 
It therefore converges, with an oscillatory pattern, towards the homogeneous model obtained through our averaging procedure, as it corresponds to one of the axes of the quadrupole in $d_L$ for that model. 

Similarly, for the pink curve, which corresponds to $\theta_c = \pi/2$, the quadrupolar nature of the average model means that $d_L$ is accurately reproduced, as also shown by the return of $\mu - \mu_{\rm model}$ to zero as a function of $\theta_c$ in the right plot of Fig. \ref{fig_sinusoidal_dL_theta}. 
For observing angles $\theta_c$ that are off the quadrupole axes, the curve $\mu(z) - \mu_{\rm model}(z)$ does not always converge back to zero, but to a constant offset value. This is to be expected for single observers, as they cannot be said to be measuring a fair sample of the universe in all directions.

\begin{figure}
    \centering
    \includegraphics[width=0.68\linewidth]{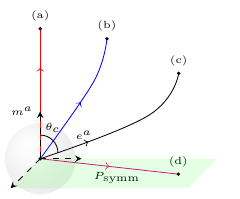}
    \caption{An illustration of the ray tracing procedure used to calculate the sky map of $d_L(z)$ for each observer. Past-directed null geodesics emanating from the observer are distinguished by their value of $\theta_c$\,. The spacelike vector $m^a$ is indicated as being normal to the planes of symmetry $P_{\rm symm}\,$, which here are in the horizontal plane.}
    \label{fig_ray_tracing_picture}
\end{figure}

\begin{figure}
    \centering
    \includegraphics[width=\linewidth]{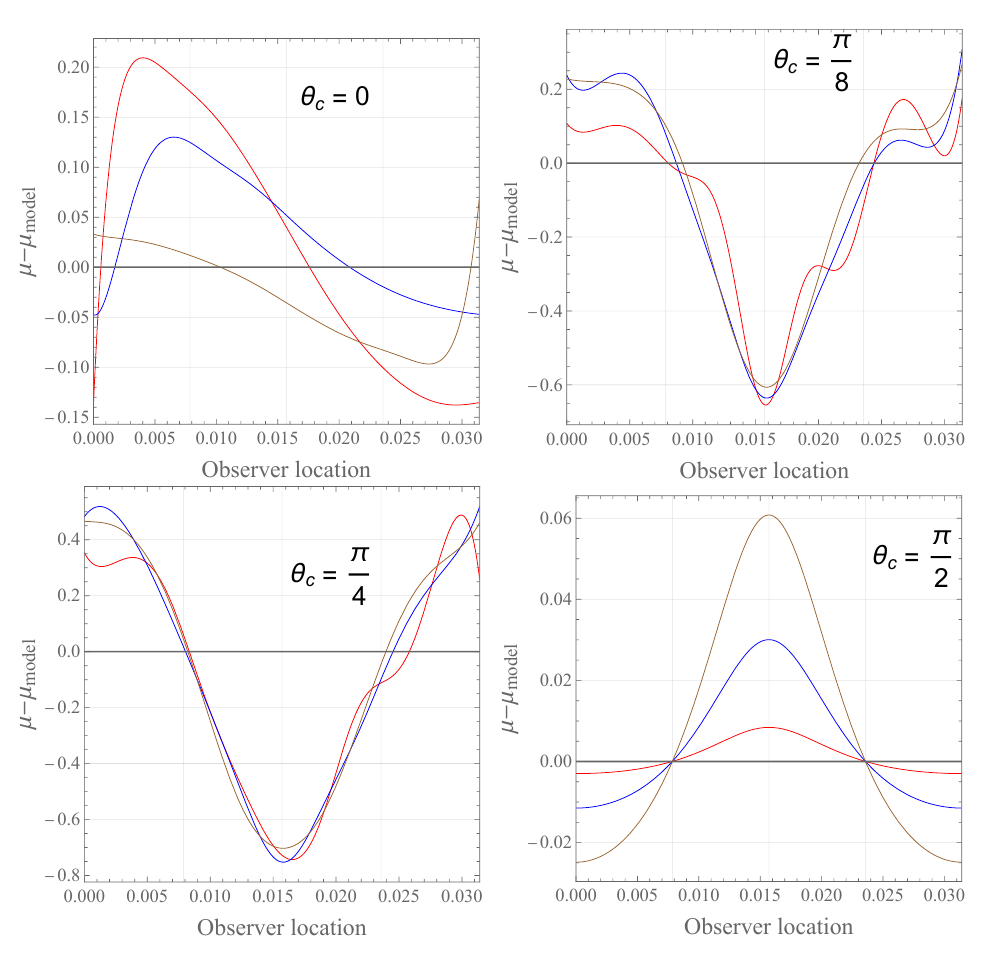}
    \caption{$\mu - \mu_{\mathrm{model}}$ for a given line of sight $\theta_c$, as a function of observer location within a single averaging domain in the $R'=0$ geometry. The left and right edges of the plot correspond to overdense regions, whereas the middle corresponds to an underdensity. The  curves are for $z = 0.1$ (red), $z = 0.2$ (blue), and $z = 0.3$ (brown).}
    \label{fig_sinusoidal_mu-mu_model_vs_observer_location}
\end{figure}

While any averaged model could not be expected to reproduce the Hubble diagrams of every individual observer, we might expect it to produce a good representation of the average Hubble diagram that would be obtained by combining, and subsequently averaging, the results from many different observers. 
To investigate this possibility, we pick an observing direction, $\theta_c$, and consider the $d_L(z)$ from $100$ different observers across an averaging domain, separated from one another by equal intervals of their $r$ coordinate\footnote{As the spacetime geometry is curved, it is not possible to define the ``distance'' between different points in a spacelike hypersurface covariantly, and so we restrict ourselves to referring to the coordinate interval between such observers, rather than a physical distance.} $r_{\rm obs}$. They each view in that direction $\theta_c$ on their local celestial sphere, as envisaged in Fig. \ref{fig_ray_tracing_picture}. 
Comparing their measurements gives rise to a situation as in Fig. \ref{fig_sinusoidal_mu-mu_model_vs_observer_location}, where we consider the difference $\mu - \mu_{\rm model}$ for a given $\theta_c$ and $z\,$, as a function of the observer's location within the averaging domain.

\begin{figure}
    \centering
    \includegraphics[width=\linewidth]{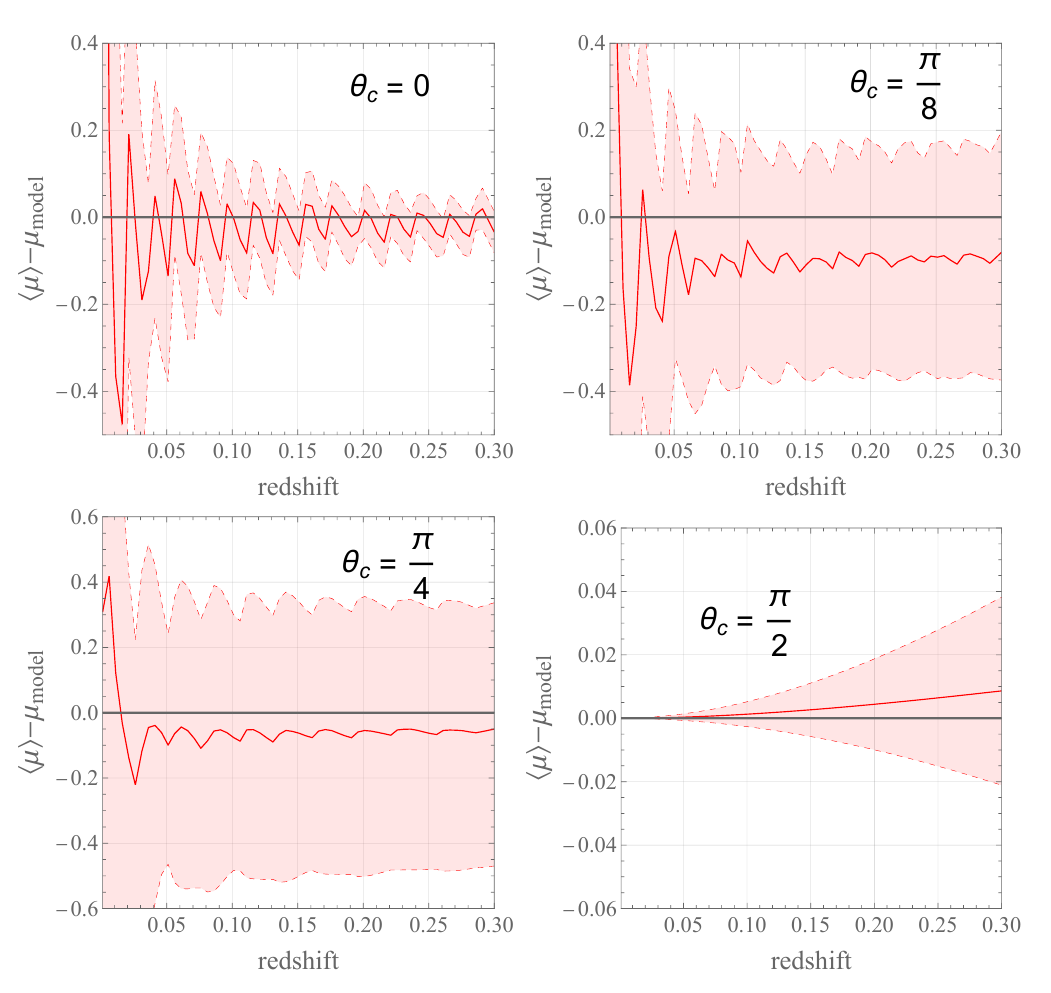}
    \caption{Averaged distance moduli, $\avg{\mu} - \mu_{\mathrm{model}}\,$, as a function of redshift, in the geometry with $R'=0\,$. These averages are obtained from the mean of $100$ observers, at evenly distributed $r$ coordinate intervals throughout the averaging domain. The shaded red regions indicate the $1\sigma$ confidence intervals. Results are displayed for sets of observers all viewing in the directions $\theta_c = \left\lbrace 0, \pi/8, \pi/4, \pi/2\right\rbrace$.}
    \label{fig_sinusoidal_mean_mu_vs_model}
\end{figure}

By choosing a large range of redshift intervals, we can then calculate the inferred $d_L(z)$ for each point in the discretised parameter space spanned by $\left(r_{\rm obs}, \theta_c\right)$, and finally average this over $r_{\rm obs}$ to obtain the mean and variance of $d_L(z)$ for all possible observing directions.
The averaged distance modulus $\avg{\mu}$ that results is displayed in Fig. \ref{fig_sinusoidal_mean_mu_vs_model}, where we have calculated $\avg{\mu} - \mu_{\rm model}$ for a variety of directions in the redshift range $z \in \left[0, 0.30\right)\,$. Although at low redshift the homogeneous large-scale averaged model describes the distance modulus poorly, the curve $\avg{\mu}(z)$ rapidly converges to $\mu_{\rm model}(z)$. 
This convergence is particularly strong for a collection of observers viewing along the axis of symmetry, $\theta_c = 0$, but even for the collection of observers who are off-axis the average distance modulus can be seen to settle down to have only a small offset from the averaged model (comfortably within one standard deviation). 
This result reflects the capacity of our formalism to account for large-scale anisotropy.

\section{An $R' \neq 0$ universe with tilt}\label{sec:farnsworth}

While the backreaction scalars, $\mathcal{B}_i$, vanish in the $R' = 0$ cosmologies, the situation with $R' \neq 0$ cosmologies is more complicated. For example, the presence of a non-zero $R'$ means that the dust that sources the spacetime curvature does not need to be moving along integral curves of $n_a$. 
This is true even if the spacetime is homogeneous, and constitutes the tilted class of anisotropic cosmological models \cite{King_1973}. As in Chapter 7, we wish to study this possibility using the anisotropic cosmologies found by Farnsworth \cite{Farnsworth_1967}.

\begin{figure}
    \centering
    \includegraphics[width=0.63\linewidth]{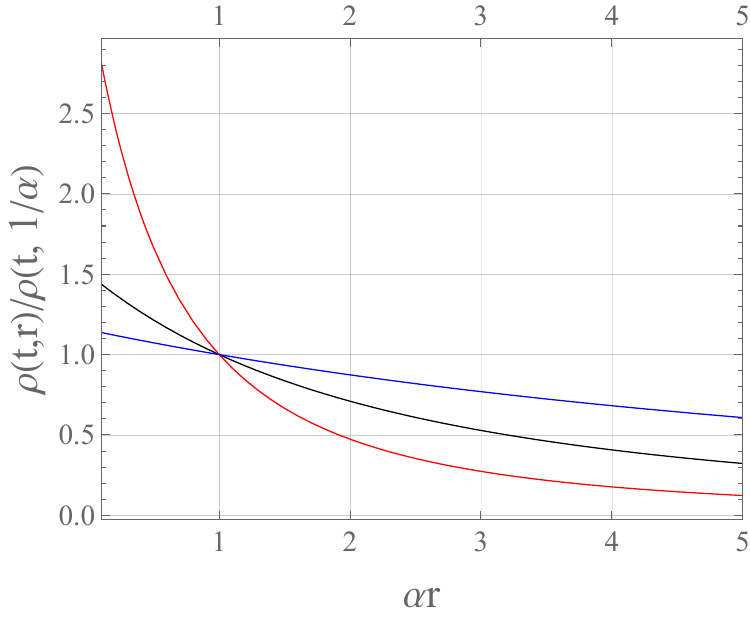}
    \caption{Energy density profile on constant-time surfaces of the Farnsworth cosmologies, normalised by its value at $r=\alpha^{-1}$. The curves shows $\rho(r)$ at the observing time $t=t_{\rm obs}$ (black), $t=t_{\rm obs}/3$ (red), and $t=3 t_{\rm obs}$ (blue). Parameter values are $\left\lbrace \alpha, k, W, C, t_{\rm obs}\right\rbrace = \left\lbrace 1, 5, 125, 2, 40\right\rbrace \,$.}
    \label{fig_farnsworth_scalars}
\end{figure}

We recall that Farnsworth's cosmologies are exact homogeneous solutions of Einstein's equations of Bianchi type {\it V}. 
They are locally rotationally symmetric, but tilted.
We will describe them in slightly different terms to Section \ref{sec:backreaction_farnsworth}, because we wish to make their origin within the plane-symmetric class of cosmological models explicit.
In terms of the metric functions from Eqs. (\ref{eq_R'_neq_0_metric_general}) and (\ref{eq_R'_neq_0_solution_parametric}), the Farnsworth solutions are given as
\begin{eqnarray}\label{eq_Farnsworth_definition}
    m(r) &=& \frac{Wk^3}{2}\,e^{-3\alpha r}\,,\qquad
    f(r) = k\, e^{-\alpha r}\,,\qquad {\rm and} \qquad
    t_0(r) = -C\,r \,,
\end{eqnarray}
where $W$, $k$ $\alpha$ and $C$ are the same constants that we defined in Section \ref{sec:backreaction_farnsworth}.

We remind the reader that the tilted nature of the spacetime means that a set of observers comoving with the matter flow should measure global inhomogeneity in the hypersurface that is spanned by their rest spaces, so the ``homogeneous'' cosmological model that they would construct by averaging over domains of those hypersurfaces would appear to exhibit backreaction.
For the exposition in this section, the key consequence of that fact is that the Hubble diagram they would infer within that model may well be a poor fit to observations of distance measures \footnote{The exception to this is the special case $C = 0$, in which case the solution (\ref{eq_Farnsworth_definition}) is just an FLRW geometry with negative spatial curvature.}.
Fig. \ref{fig_farnsworth_scalars} shows explicitly that observers comoving with the dust would have orthogonal rest spaces that are inhomogeneous, provided $C \neq 0$\,. This effect is entirely due to the tilt. 

\begin{figure}
    \centering
    \includegraphics[width=0.68\linewidth]{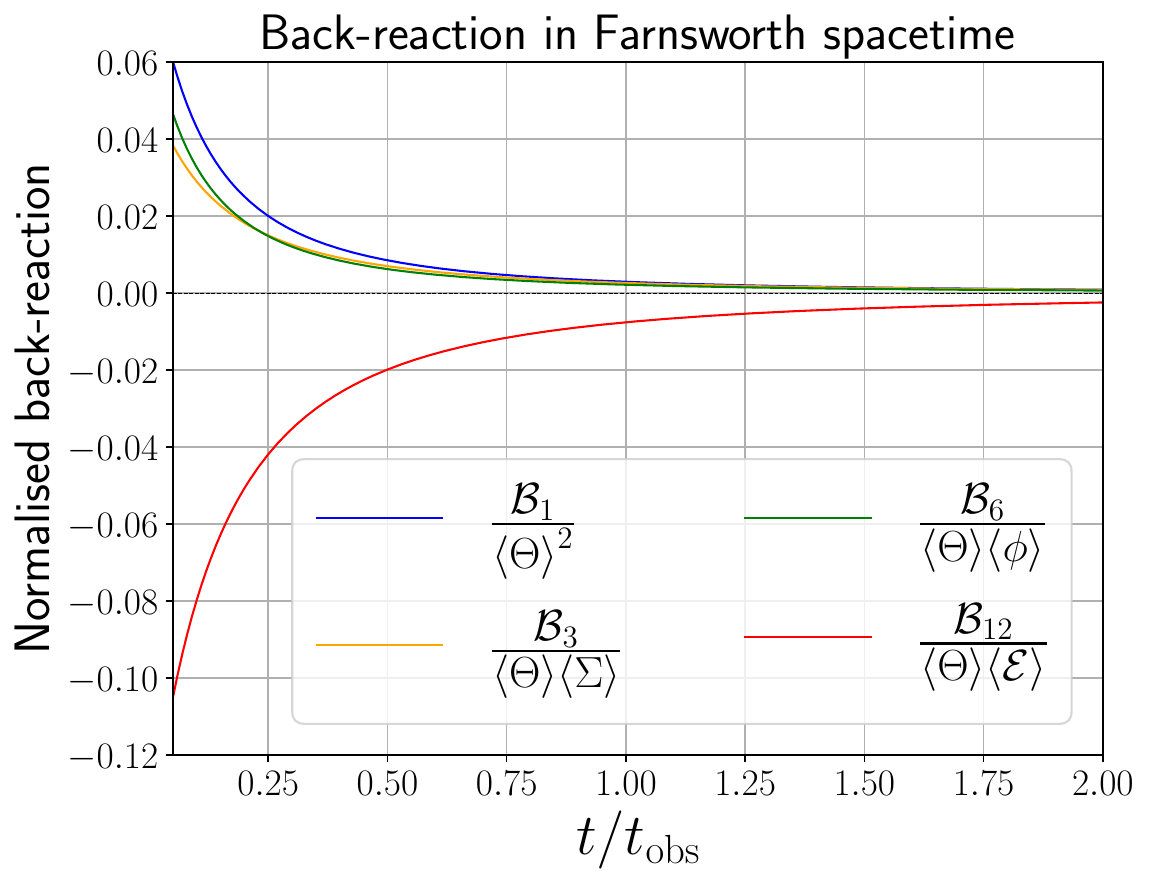}
    \caption{All non-vanishing backreaction scalars (that appear in evolution equations) in constant-time hypersurfaces of the Farnsworth geometry (\ref{eq_Farnsworth_definition}). The scalars $\mathcal{B}_i$ have all been made dimensionless, by normalising them with respect to the largest term from the equation in which they appear. This plot is very closely related to Fig. \ref{fig_Farnsworth_backreaction_scalars}.}
    \label{fig_farnsworth_backreaction_2}
\end{figure}

We now wish to carry out numerical integrations for rays of light in this spacetime, for which we make the following choices for parameter values: $\alpha=1$ for the characteristic inverse length scale, and $k = 5$, $W = 125$ and $C = 2$ for the curvature, density, and tilt parameters, respectively (exactly as in Section \ref{sec:backreaction_farnsworth}). 

Because of the nature of the apparent inhomogeneity induced by the tilt, the rest spaces of the observers in this case are not statistically homogeneous. 
This means that there is no natural homogeneity scale that can be used to define our averaging domain, $\mathcal{D}$. We therefore choose to average between $\left\{r_{\rm min}, r_{\rm max}\right\} = \left\{\alpha^{-1}, 3 \alpha^{-1} \right\}\,$, which in the absence of an homogeneity scale is made purely out of computational convenience. Note that this a different choice to what was made in Section \ref{sec:backreaction_farnsworth}, although the conclusions are qualitatively unchanged. 
We find once again that backreaction scalars in this case can be non-zero, with relative sizes of up to $10\%\,$ (as shown in Fig. \ref{fig_farnsworth_backreaction_2}), though they decay at late times. 

\subsection{Ray tracing}

\begin{figure}
    \centering
    \includegraphics[width=\linewidth]{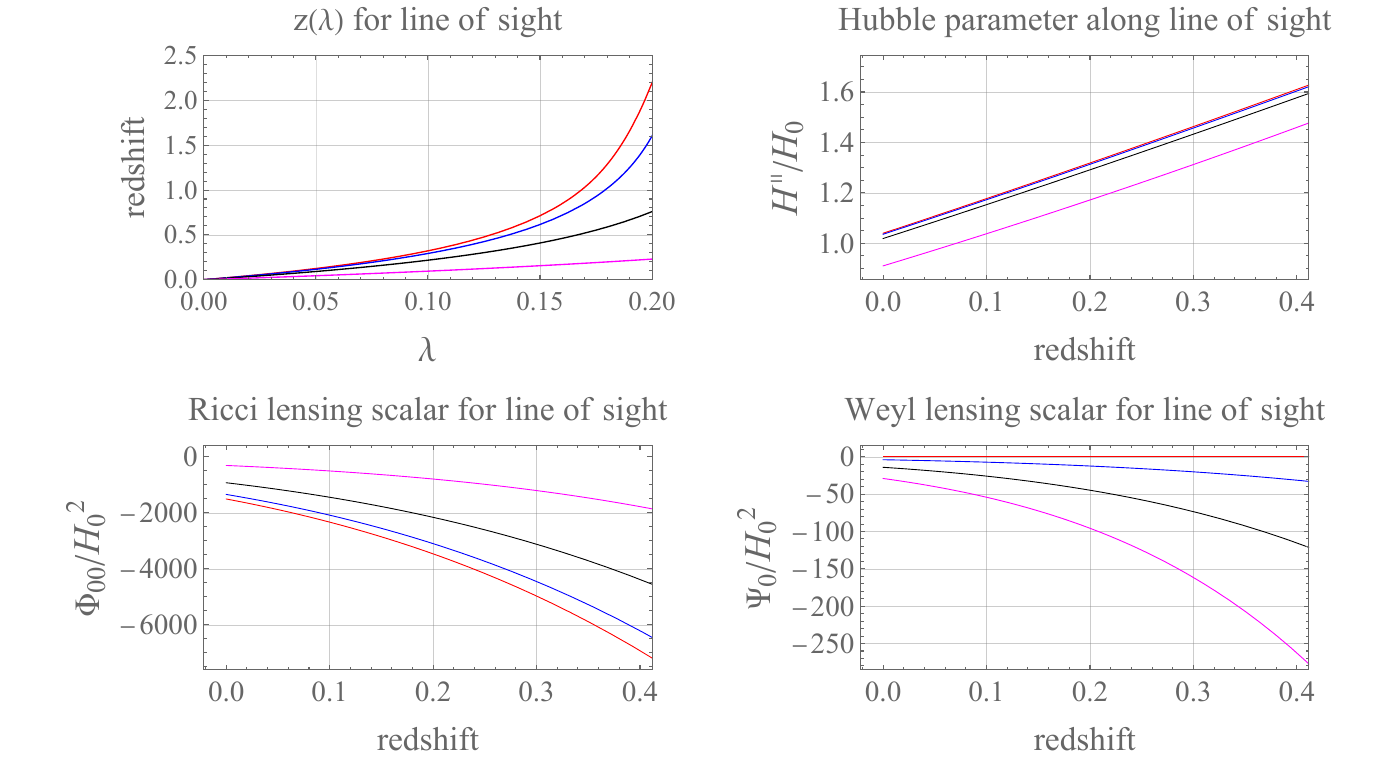}
    \caption{Top-left: redshift as a function of affine distance $\lambda$ in the Farnsworth cosmology, for observing angles $\theta_c = 0$ (red), $\pi/8$ (blue), $\pi/4$ (black) and $\pi/2$ (magenta). Top-right: line-of-sight Hubble parameter $H^{\parallel}$ as a function of redshift. Bottom-left and right: Ricci and Weyl lensing scalars $\Phi_{00}$ and $\Psi_0$, for the same observing directions.}
    \label{fig_farnsworth_ray_tracing}
\end{figure}

We now calculate the paths of null geodesics arriving at observers on a constant-time hypersurface, by repeatedly solving the geodesic equation in this spacetime. Once again, the plane symmetry of the spacetime restricts the initial conditions we need to vary to $r_{\rm obs}$ and $\theta_c$. 
Although the spacetime has homogeneous surfaces, these are not the constant-time surfaces of observers comoving with the dust, and so we would expect different observers on any given $t = t_{\rm obs}$ hypersurface to construct different Hubble diagrams, even if they observe in the same direction, as they will not exist on the same hypersurface of homogeneity. 
By contrast, if we picked observers at different $t_{\rm obs}$, but the same value of the combination $t_{\rm obs} + Cr_{\rm obs}$, they would all be on the same homogeneous hypersurface, and would therefore construct identical Hubble diagrams.

We consider $100$ different observers at regular $r$ coordinate separations throughout our averaging domain, on a hypersurface of constant  time, $t = t_{\rm obs} = 10\,$. In Fig. \ref{fig_farnsworth_ray_tracing} we show results for an observer located at $r_{\rm obs} = 2.5/\alpha\,$. 
As expected, there are no bumps in the functions $z(\lambda)$, $H^{\parallel}(z)$, $\Phi_{00}(z)$ and $\Psi_0(z)$\,. The effect of the Weyl curvature on the light ray's propagation, given by $\Psi_0$, is in all directions substantially smaller than the effect of Ricci curvature, $\Phi_{00}$. 
This is indicative of the late-time isotropisation of the Farnsworth metric, wherein it tends towards an (open) FLRW universe, which has no Weyl curvature.

\subsection{Hubble diagrams}

\begin{figure}
    \centering
    \includegraphics[width=\linewidth]{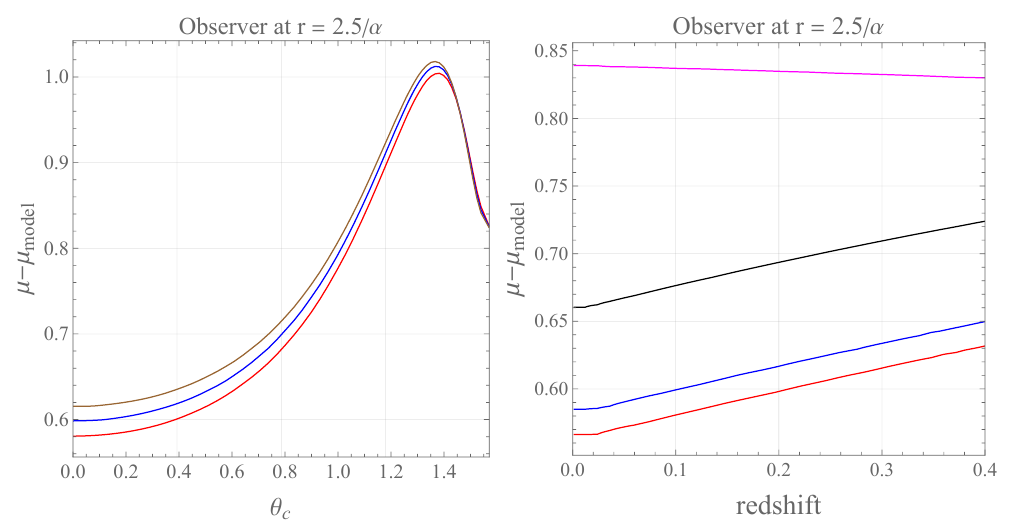}
    \caption{Left: $\mu - \mu_{\rm model}$ as a function of $\theta_c$, for at observer at $r_{\rm obs} = 2.5/\alpha$ in the Farnsworth cosmology. Curves are for redshifts $0.1$ (red), $0.2$ (blue) and $0.3$ (brown). The solid lines are obtained by performing ray tracing in the tilted spacetime, while the dashed lines come from ray tracing in the averaged LRS Bianchi type {\it V} model.
    Right: $\mu - \mu_{\rm model}$ as function of redshift, for the observer at $r = 2.5/\alpha$ considered in the last set of plots. Curves are for observing angles $\theta_c = 0$ (red), $\pi/8$ (blue), $\pi/4$ (black) and $\pi/2$ (magenta). Again, $\mu_{\rm model}$ refers to a Bianchi type {\it V} model.}
    \label{fig_farnsworth_magnitude_vs_model}
\end{figure}

\begin{figure}
    \centering
    \includegraphics[width=\linewidth]{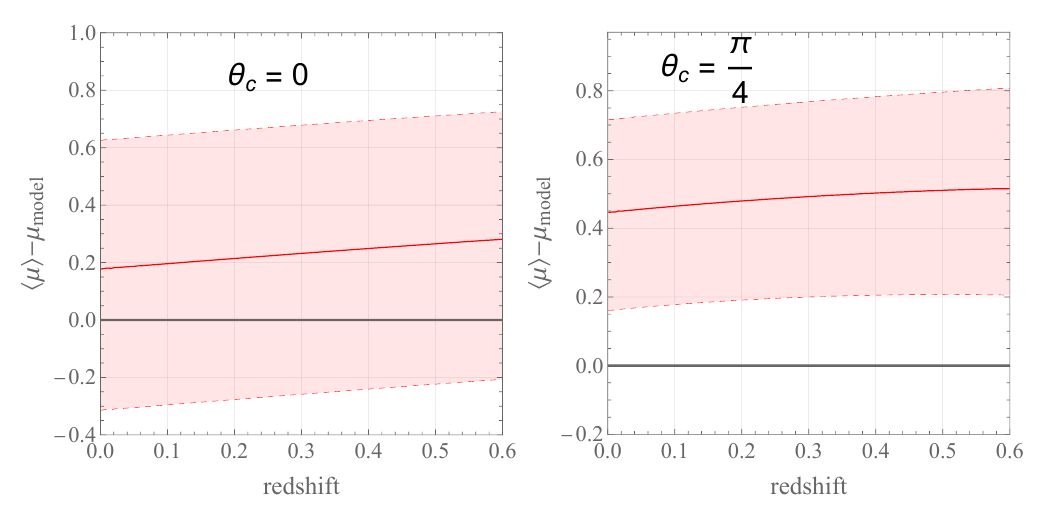}
    \caption{The ensemble mean of $\mu$ over 100 observers evenly spaced throughout the averaging domain in the Farnsworth geometry, compared to $\mu_{\rm model}$ obtained by mapping the scalar averages from that domain onto an LRS Bianchi type $V$ model, for $\theta_c = 0$ (left) and $\pi/4$ (right).}
    \label{fig_farnsworth_mean_mu_vs_model}
\end{figure}

The Farnsworth cosmologies are of Bianchi type {\it V}, with homogeneous surfaces tilted by a 3-velocity $Cf(r)/R'(t,r)$ with respect to the matter-comoving surfaces of constant time. 
It therefore seems natural to map the averages calculated on $t = {\rm cst.}$ surfaces to the associated quantities in Bianchi type {\it V} cosmologies. We now show the key results of that comparison.
The angular variation of $\mu$ at specified redshift shells, compared to this averaged Bianchi type {\it V} cosmology, is shown in the left plot of Fig. \ref{fig_farnsworth_magnitude_vs_model}, where it can be seen that the difference between the curves increases with redshift at all angles.
This suggests that the model will fail to capture the correct Hubble diagram at intermediate and high redshifts, as verified in the right plot of Fig. \ref{fig_farnsworth_magnitude_vs_model}. 
The result here is in contrast to the case studied in Section \ref{sec:sinusoidal}, and we interpret it as being due to the lack of a homogeneity scale in the tilted hypersurfaces.

As expected, averaging over a large ensemble of observers does not alleviate the discrepancy, as displayed in Fig. \ref{fig_farnsworth_mean_mu_vs_model}. Not only does $\avg{\mu} - \mu_{\rm model}$ not return to zero, but the standard deviation remains large in all cases. 
This indicates, in addition to the lack of homogeneity scale making the averaged Bianchi model perform poorly, that averaging in the constant-$t$ surfaces may not be a sensible procedure in the first place, because the tilt gives rise to a type of global inhomogeneity.
Hence, imposing an homogeneous model gives rise to a Hubble diagram that bears no relation to one that observers in the spacetime would record. 

\section{An $R' \neq 0$ universe with inhomogeneity}\label{sec:linear}

Finally, let us consider an inhomogeneous plane-symmetric universe with $R' \neq 0\,$. In this case the backreaction scalars are not {\it a priori} restricted to be zero. 
An homogeneous, tilt-free solution to the plane-symmetric metric with $R' \neq 0$ is provided by the metric functions in Eqs. (\ref{eq_R'_neq_0_metric_general}) and (\ref{eq_R'_neq_0_solution_parametric}) taking the form $f(r) = kr$, $m(r) = A f^3(r)$ and $t_0(r) = 0$, for some positive constants $k$ and $A$. 
To introduce inhomogeneity, we can therefore simply modify these functions, such that
\begin{eqnarray}\label{eq_linear_model_definition}
    m(r) = A k^3 \,r^3\,,
    \qquad f(r) = k\left(r-b\sin^2{qr}\right)\,,\qquad {\rm and} \qquad 
    t_0(r) = 0\,,
\end{eqnarray}
where $b$ and $q$ are free parameters controlling the amplitude and frequency of oscillations in the inhomogeneities. 

\begin{figure}
    \centering
    \includegraphics[width=\linewidth]{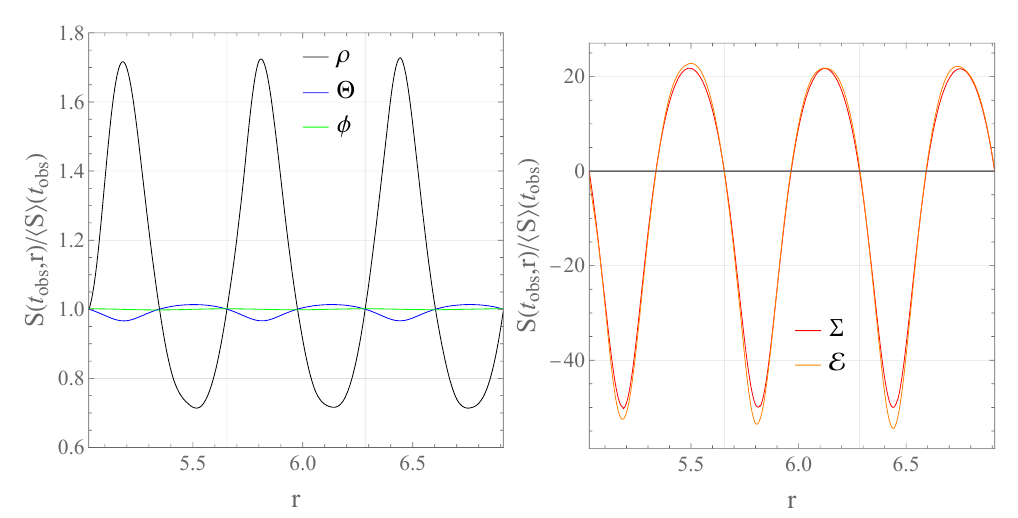}
    \caption{Left panel: matter density $\rho\,$, isotropic expansion $\Theta\,,$ and spacelike expansion $\phi\,$, at the observing time $t_{\rm obs}\,$, displayed as a function of the $r$ coordinate in the $R' \neq 0$ models. Right panel: shear $\Sigma$ and electric Weyl curvature $\mathcal{E}$\,, also evaluated at $t_{\rm obs}$\,. All quantities are normalised by their Buchert averages, which have been calculated over the domain $\left[r_{\rm min},r_{\rm max}\right)\,$.}
    \label{fig_linear_scalars}
\end{figure}

The $1$+$1$+$2$-scalars, evaluated at the observing time $t_{\rm obs} = 40$, are displayed in Fig. \ref{fig_linear_scalars} for this case, where we have chosen $k = 5$, $A = 1$, and $q = 5$\,. We have separated the scalars out into the set $\left\lbrace \rho, \Theta, \phi\right\rbrace$ which are non-vanishing in the isotropic limit, and the pair $\left\lbrace \Sigma,\mathcal{E}\right\rbrace$ which are intrinsically anisotropic in nature.

Importantly, we restrict the amplitude $b$ of the oscillatory part of $f(r)$\,, so that the matter density undergoes fluctuations of order unity, but is always non-negative. 
We find that $b = 0.1$ produces a density that is never less than $30\%$, or more than $180\%\,$, of its average value at $t=t_{\rm obs}\,$. This means that the model never reaches perfect matter domination nor vacuum domination (unlike in the $R' = 0$ case from Section \ref{sec:sinusoidal}), but that the density variations are still large. 

\begin{figure}
\centering
\includegraphics[width=0.9\linewidth]{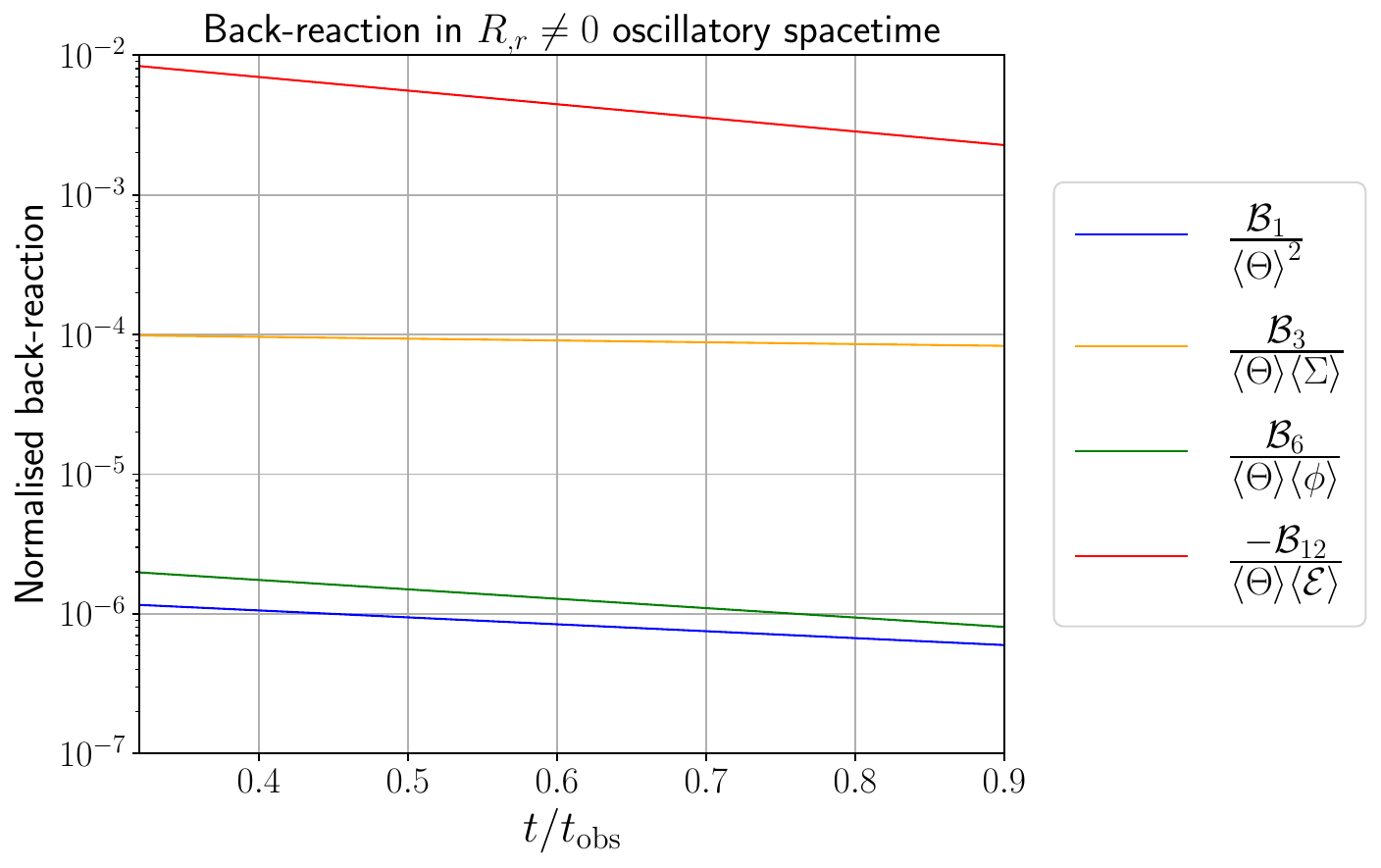}
\caption{Non-vanishing backreaction scalars $\mathcal{B}_i$ for constant-time surfaces of the $R' \neq 0$ geometry. These scalars have all been made dimensionless, by normalising them with respect to the largest term from the evolution equation in which they appear.}
\label{fig_backreaction_linear}
\end{figure}

We then calculate the full set of scalar averages, and the backreaction scalars $\mathcal{B}_i$ that source them, according to the emergent equations of motion, Eqs. (\ref{eq_avg_LRS_first}--\ref{eq_avg_LRS_last}). 
By inspection of Fig. \ref{fig_linear_scalars}, one sees that $\Delta r = \pi/q$ defines the statistical homogeneity scale, meaning that one may choose any such oscillation cycle as constituting the extent of $r$-coordinate in a well-motivated averaging domain (the $y$ and $z$ coordinates again being irrelevant, due to the plane symmetry of the spacetime). 
We choose $r_{\rm min} = 9\pi/q$\,, so $r_{\rm max} = 10\pi/q\,$, and show our results in Fig. \ref{fig_backreaction_linear}. 

Fig. \ref{fig_backreaction_linear} shows that backreaction scalars all make very small contributions to the evolution equations in this case. Their contributions are roughly constant in time, relative to the overall scale of the quantities in the equation, and only $\mathcal{B}_{12}$ (the backreaction term in the evolution equation for $\avg{\mathcal{E}}\,$) is larger than $0.01\%$ of the dominant term in its evolution equation. 
Furthermore, $\mathcal{B}_{12}$ has the opposite sign to $\avg{\Theta}\avg{\mathcal{E}}\,$, so it is not causing the average Weyl curvature to grow, but rather is suppressing it. The other $\mathcal{B}_i$ are all of positive sign, but are negligibly small. 
Overall, one sees that although the backreaction scalars are allowed to be non-zero in the $R' \neq 0$ plane-symmetric geometries, they are still highly restricted by the Killing symmetries of the spacetime. This is true even though the matter density fluctuations are of order unity, and seems to be independent of our precise choice of parameter values and functions.

\subsection{Ray tracing}

\begin{figure}
    \centering
    \includegraphics[width=\linewidth]{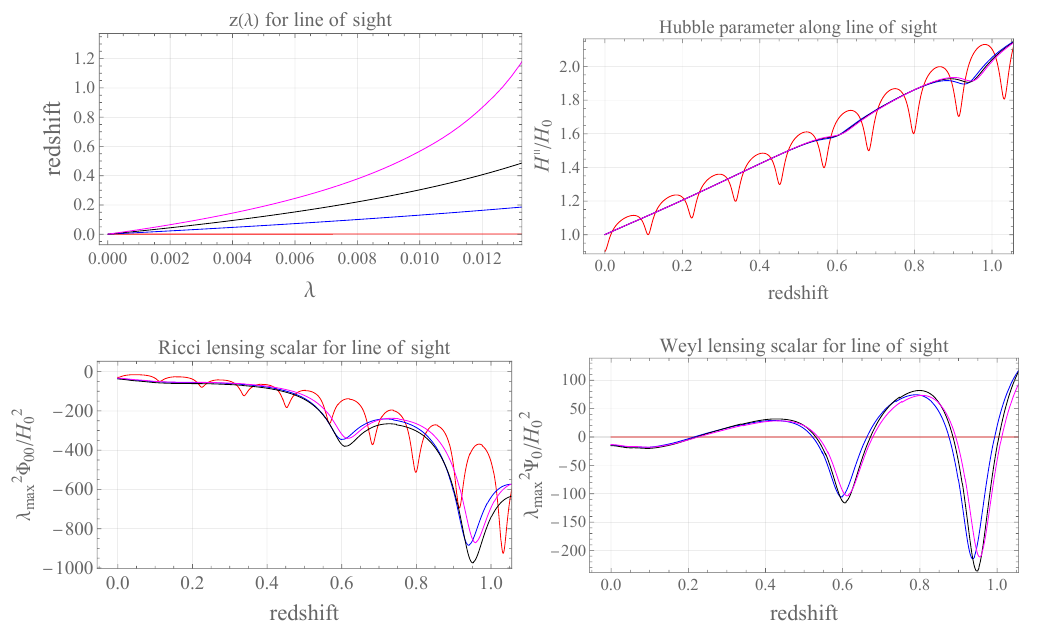}
    \caption{Top left: redshift as a function of affine distance in models with $R'\neq0$, for observing angles $\theta_c = 0$ (red), $\pi/4$ (blue), $\pi/2$ (black) and $\pi/2$ (magenta) . Top right: line-of-sight Hubble parameter $H^{\parallel}$ as a function of redshift, normalised by its monopole at $z=0\,$. 
    Bottom left and right: Ricci and Weyl lensing scalars $\Phi_{00}$ and $\Psi_0$ as functions of redshift. Each of the Ricci and Weyl terms have been normalised by the maximum affine parameter value $\lambda_{\rm max}^{-2}$\,.}
    \label{fig_linear_ray_tracing}
\end{figure}

The results of our ray tracing procedure are summarised in Fig. \ref{fig_linear_ray_tracing}, where we have displayed $z(\lambda)$, $H^{\parallel}(z)$, $\Phi_{00}(z)$ and $\Psi_0(z)$ for null geodesics arriving at an observer at time $t=t_{\rm obs}$ and location $r_{\rm obs} = r_{\rm min} + \dfrac{\pi}{4q}$\,. 
The top left plot of the figure shows that the redshift $z$ is monotonic in affine parameter $\lambda$ for all $\theta_c$ considered. 
The effect of inhomogeneities on the null rays is clear, however, once one considers the inferred line-of-sight Hubble parameter, which is displayed in the top right plot as a function of redshift. 
Here the oscillations in the metric function $f(r)$ are of amplitude $10\%$\,, so the effects of inhomogeneity on $H^{\parallel}$ are less drastic than in the $R'=0$ case, where the metric oscillated entirely between dust-dominated and vacuum regions, as displayed in Fig. \ref{fig_sinusoidal_model_density}.

\begin{figure}
    \centering
    \includegraphics[width=0.63\linewidth]{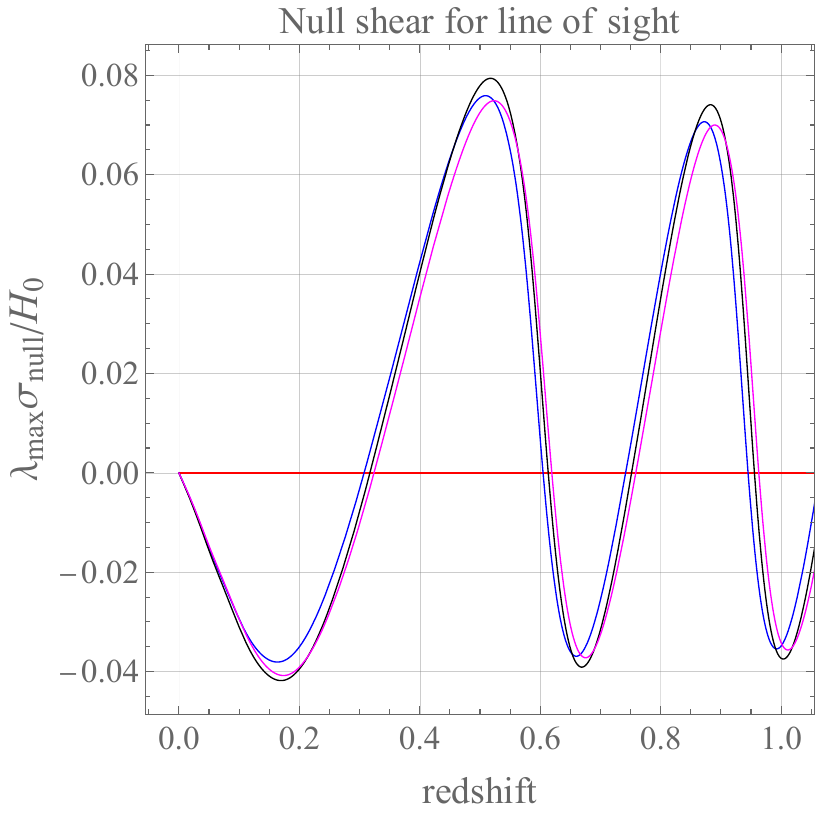}
    \caption{Shear $\hat{\sigma}$ (labelled $\sigma_{\rm null}$) of the null congruence, as a function of redshift, for models with $R'\neq 0\,$. Results are displayed for observing angles $\theta_c = 0$ (red), $\pi/8$ (blue), $\pi/4$ (black) and $\pi/2$ (magenta), and normalised by $\lambda_{\rm max}^{-1}$ in each case. We have considered an observer at $r_{\rm obs} = r_{\rm min} + \pi/4q\,$, and at time $t=t_{\rm obs}$\,.}
    \label{fig_linear_model_null_shear}
\end{figure}

The Ricci and Weyl curvature terms have the same oscillatory pattern as $H^{\parallel}$. 
Unless one is observing directly along the symmetry axis ($\theta_c = 0$\,), in which case $\Psi_0$ vanishes identically, then the Weyl curvature contribution $\Psi_0$ is typically of the same order of magnitude as the Ricci contribution $\Phi_{00}$, which is displayed in the bottom left plot of Fig. \ref{fig_linear_ray_tracing}. The bottom right plot of the figure shows that $\Psi_0$ has a changing sign in each case.
This means that for observations at high redshift, for which null rays will typically have to travel through many oscillations in the geometry, the effect of Weyl curvature will be suppressed relative to Ricci curvature, even though the two terms $\Psi_0$ and $\Phi_{00}$ are typically of comparable magnitude. 
This can be understood by studying Sachs' equation for the null shear (\ref{eq_sachs_repeat_2}), which integrates to
\begin{equation}\label{eq_solution_to_sachs2}
    \hat{\sigma}(z) = \frac{1}{d_A^2(z)}\int_0^z \frac{\mathrm{d}\tilde{z} \, \Psi_0(\tilde{z})\, d_A^2(\tilde{z})}{\left(1+\tilde{z}\right)^2 H^{\parallel}(\tilde{z})}\,.
\end{equation}
As $H^{\parallel}$ is always positive in this case, the oscillations in the sign of $\Psi_0$ cause the right-hand side of Eq. (\ref{eq_solution_to_sachs2}) to remain bounded, when integrated to intermediate and high redshifts, as can be seen in Fig \ref{fig_linear_model_null_shear}. 

We reminder the reader that although $\h{\sigma}$ is in general a complex scalar, the plane symmetry of the spacetime being considered reduces the number of degrees of freedom in the null shear from 2 to 1. Hence, $\h{\sigma}$ can be written as a single real scalar in this case.
The effect of Weyl curvature on $d_A$ is communicated through the presence of $\bar{\hat{\sigma}}\hat{\sigma}$ in Sachs' equation for $d_A$\,. By $z = 1$\,, $\bar{\hat{\sigma}}\hat{\sigma}$ is four orders of magnitude smaller than $\Phi_{00}$, and hence the integrated effects of Weyl curvature are small.

\subsection{Hubble diagrams}

In this section we consider both the Bianchi type {\it I} and {\it V} geometries as candidate models, by mapping the scalar averages $\avg{\Theta}$, $\avg{\Sigma}$ and $\avg{^{(3)}R}$ onto these classes as appropriate. 
Our averages are calculated over the domains described above, for time coordinate values in the range $\left[\dfrac{t_{\rm obs}}{4}, t_{\rm obs}\right]\,$. 
As in the previous sections, we then perform ray tracing in those homogeneous averaged models by solving the geodesic equation (\ref{eq_geodesic_equation_Hamilton}) for past-directed null geodesics emanating from observers at $t_{\rm obs}$\,. Finally, we compare the results of that procedure to the results of ray tracing in the true inhomogeneous spacetime. 

\begin{figure}
    \centering
    \includegraphics[width=0.7\linewidth]{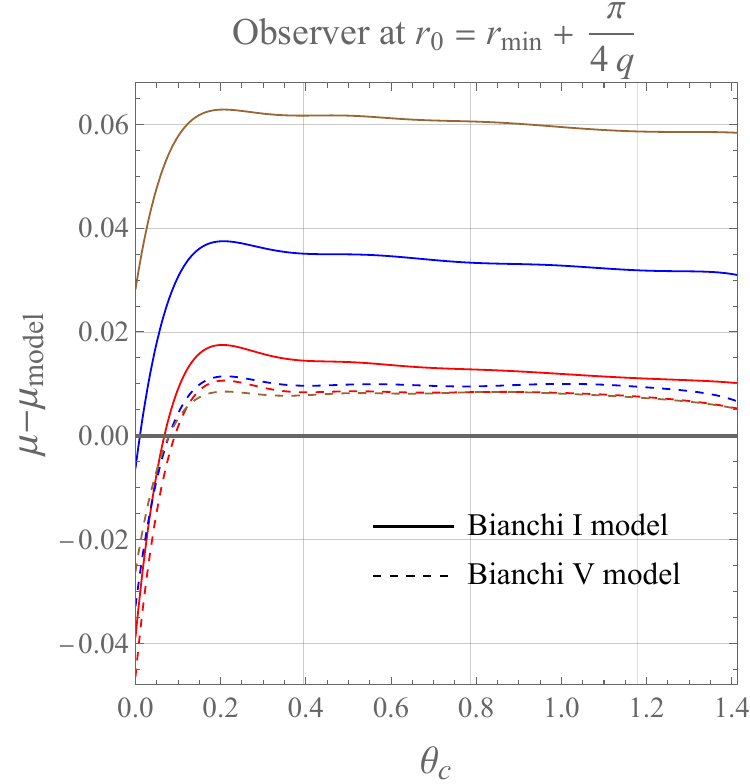}
    \caption{$\mu - \mu_{\rm model}$ as a function of $\theta_c$\,, for an observer at $r_{\rm min} + \dfrac{\pi}{4q}$ in the $R' \neq 0$ model. Curves correspond to $z = 0.1$ (red), $z = 0.2$ (blue) and $z = 0.3$ (brown). The function $\mu_{\rm model}$ refers to an averaged Bianchi type {\it I} cosmology for the solid curves, and to an averaged Bianchi type {\it V} cosmology for the dashed curves.}
    \label{fig_linear_model_mu-mu_model_theta}
\end{figure}

Fig. \ref{fig_linear_model_mu-mu_model_theta} shows a comparison of the distance moduli to the average models, as a function of $\theta_c$ for a given redshift. It demonstrates that the Bianchi type {\it V} average model appears more appropriate than type {\it I}, which is apparent once one considers observing directions $\theta_c$ that are sufficiently far from the rotational symmetry axis. 
Not only are the differences $\mu - \mu_{\rm model}$ significantly smaller for type {\it V} than type {\it I}, but they also decay with redshift. 
However, at this point it is not yet clear that observations of Type Ia supernovae would actually indicate a preference for either average model, as any overall shift in $\mu$ can be removed by a recalibration of the intrinsic magnitude of the supernovae, as discussed in Section \ref{subsec:Hubble_diagram} and demonstrated explicitly by Eq. (\ref{eq_SNE1a_nuisance_parameters}).
All that would remain in that case would be the angular profile, which is not easily explained by either choice of homogeneous large-scale geometry. If this were all of the information available then one would not be able to select which average model would be more appropriate, which could lead to serious errors in inferring cosmological parameters from the properties of the Hubble diagram, such as the best-fit Taylor series coefficients in the function $d_L(z)\,$, which provide information in the standard FLRW cosmology about $H_0\,$, $\Omega_{m0}$ and so on.

\begin{figure}
    \centering
    \includegraphics[width=0.7\linewidth]{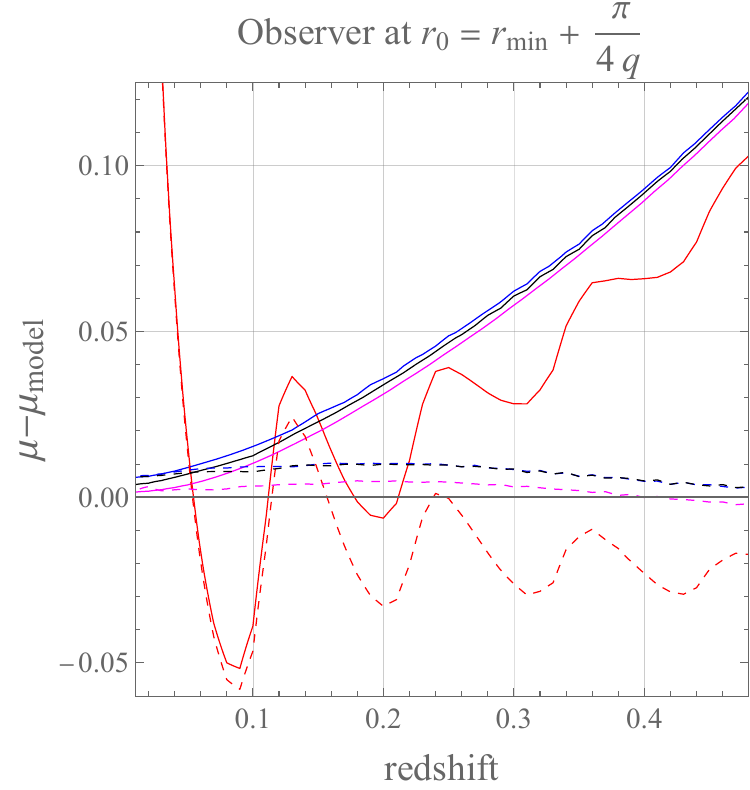}
    \caption{$\mu - \mu_{\rm model}$ as a function of redshift, for the same observer at $r_{\rm min} + \dfrac{\pi}{4q}$ in the $R' \neq 0$ model. Curves are for $\theta_c = 0$ (red), $\pi/8$ (blue), $\pi/4$ (black) and $\pi/2$ (magenta). The function $\mu_{\mathrm{model}}$ refers to averaged Bianchi type {\it I} (solid) and {\it V} (dashed) cosmologies.}
    \label{fig_linear_magnitude_vs_model}
\end{figure}

Figure \ref{fig_linear_magnitude_vs_model} shows the Hubble diagrams $\mu(z)$ that would be constructed for observing directions $\theta_c = \left\lbrace 0, \pi/8, \pi/4, \pi/2\right\rbrace\,$, for redshifts up to $0.48\,$. 
One sees that at low redshifts there is little distinction between the models, but at higher redshifts the Bianchi type {\it V} interpretation of the scalar averages is substantially preferred over the type {\it I} interpretation. 
In the case of the type {\it V} models, $\mu - \mu_{\rm model}$ remains very close to zero at all redshifts for $\theta_c = \pi/8$ and larger. This is a direct result of the angular profile of $\mu$ displayed in Fig. \ref{fig_linear_model_mu-mu_model_theta}: the function $\mu(\theta_c)$ is steepest near $\theta_c = 0$ and then rapidly becomes a nearly flat line just above the monopole. Therefore, $\mu - \mu_{\rm model}$ for each $\theta_c \gtrsim \pi/8$ is nearly flat.
This explains the lack of features in Fig. \ref{fig_linear_magnitude_vs_model}, for both observing locations and all the observing directions shown (except directly along the symmetry axis). On the other hand, the effect of inhomogeneities is clearest for $\theta_c = 0\,$, but there is still a relatively rapid trend towards zero, with $\vert \mu - \mu_{\rm model} \vert < 0.05$ for $z > 0.2$ in both plots, when the average model is chosen to be an emergent Bianchi type {\it V} cosmology. 
This shows the success of our averaging procedure at describing the large-scale Hubble diagram of an anisotropic universe, if the large-scale averaged model is chosen appropriately.

\begin{figure}
    \centering
    \includegraphics[width=\linewidth]{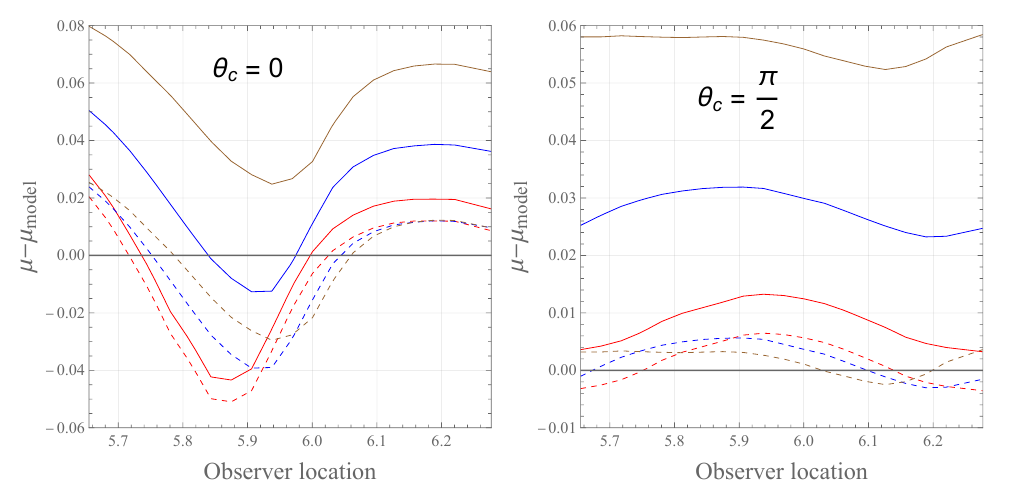}
    \caption{The same set of plots as in Fig. \ref{fig_sinusoidal_mu-mu_model_vs_observer_location}, but for the $R'\neq 0$ model, and where $\mu_{\mathrm{model}}$ refers to averaged Bianchi type {\it I} (solid) and type {\it V} (dashed) cosmologies. Curves are for $z = 0.1$ (red), $z = 0.2$ (blue), and $z = 0.3$ (brown). The averaging domain extends from $r_{\rm min} = 9\pi/q$ to $r_{\rm max} = 10\pi/q\,$.}
    \label{fig_linear_mu-mu_model_vs_observer_location}
\end{figure}

Let us now suppose that we had access to information from $100$ such observers, with their $r$ coordinate values evenly spaced throughout our averaging domain (on the same $t = t_{\rm obs}$ constant-time hypersurface), i.e. at coordinate positions $r_{\rm obs} \in \left[9\pi/q, 10\pi/q\right)\,$. The effect of the inhomogeneity on distance measures in this case is displayed in Fig. \ref{fig_linear_mu-mu_model_vs_observer_location}, where it can be seen that the preference for the Bianchi type {\it V} model only becomes clear at redshifts $z \gtrsim 0.2\,$. 

\begin{figure}
    \centering
    \includegraphics[width=\linewidth]{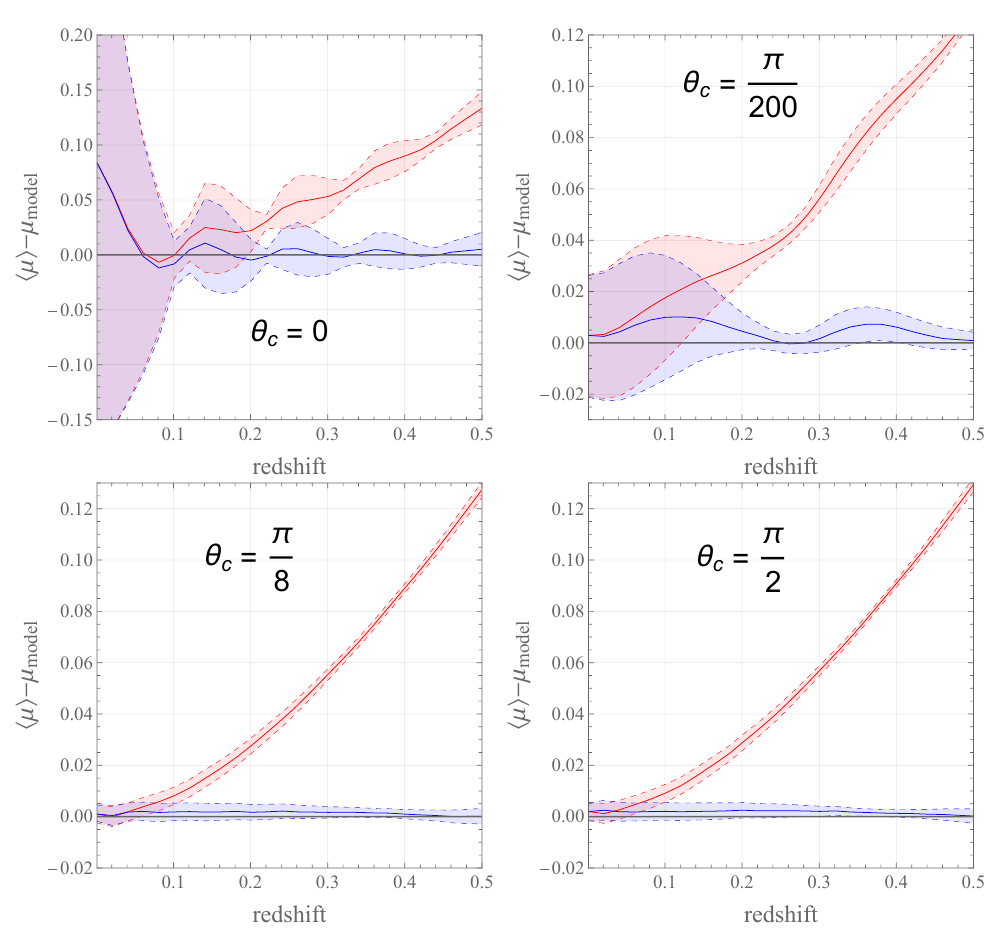}
    \caption{$\avg{\mu} - \mu_{\rm model}$, calculated as a function of redshift in the $R'\neq 0$ models. We have displayed results for $\theta_c = 0$, $\pi/200$, $\pi/8$ and $\pi/2$, and constructed $\avg{\mu}$ by taking an ensemble average over 100 observers. The function $\mu_{\mathrm{model}}$ refers to averaged Bianchi type {\it I} (red), and type {\it V} (blue).}
    \label{fig_linear_mean_mu_vs_model}
\end{figure}

Performing an ensemble average over this set of observers gives the results displayed in Fig. \ref{fig_linear_mean_mu_vs_model}, which shows that there are small effects in all directions in the redshift range $z \lesssim 0.1$\,. 
This is due to some observers receiving photons that have just moved through a region of high density, and so accordingly a region with a large negative Ricci curvature term $\Phi_{00}$\,, and others receiving photons coming through underdensities (and thus small $\left\vert \Phi_{00} \right\vert\,$). These local effects are, however, only pronounced for $\theta_c$ close to zero, as can be seen from the results for $\theta_c = {\pi}/{200}$\,, which show a rapid transition in behaviour as the observing angle $\theta_c$ is increased from zero. 

One may also note that oscillatory features in the average $\avg{\mu}(z)$ of the Hubble diagram are highly suppressed if one is observing away from the symmetry axis. This is in obvious contrast to the $R' = 0$ models considered in Section \ref{sec:sinusoidal}, where Fig. \ref{fig_sinusoidal_mean_mu_vs_model} showed that for the $R' = 0$ model, the averaged Hubble diagram has distinct features of inhomogeneity for $\theta_c = \pi/8$ and $\pi/4$\,, as well as just $\theta_c = 0$\,. 

While the $\theta_c = 0$ and $\theta_c = {\pi}/{200}$ cases demonstrate significant effects from the inhomogeneities, it remains true that $\avg{\mu}$ is always consistent with $\mu_{\rm model}$ to within $1\sigma$\,, at all redshifts considered. 
It can also be seen that the size of the oscillations, and their associated confidence intervals, is clearly decaying with increased redshift. 

Finally let us note that the offset at $z = 0$ can be explained with reference to the Hubble parameter, since for $z \ll 1$ a Taylor series expansion of $d_L(z)$ shows that the leading term at low redshifts, calculated in a generic inhomogeneous spacetime, is given by $d_L \simeq \dfrac{z}{H_0^{\parallel}}$ \cite{Heinesen_2021}. 
In our diagrams, a difference in $H_0^{\parallel}$ leads to a vertical displacement in $\mu \sim \log{d_L}$\,. At the same time, the ensemble average of the line-of-sight Hubble parameters $H^{\parallel}_0$ that are measured by each of the observers is not always equal to the line-of-sight Hubble parameter in the averaged homogeneous model, which means that a local effect on $\mu$ is entirely expected. 
In the $\theta_c = \pi/8$ and $\pi/2$ diagrams we have subtracted off the small offset. Thus, the nearly flat line $\avg{\mu} - \mu_{\rm model}$ that appears for $z \gtrsim 0.1$ is consistent with zero, which reflects the fact that a flat vertical displacement in $\mu$ would not be observable in a Type Ia supernova.  

This concludes our study of the Hubble diagrams constructed by observers in statistically homogeneous, anisotropic cosmologies, and their ability to be fit by the models that result from the averaging formalism that we developed in Chapter 7.

\section{Discussion}

Let us briefly recap what we have learnt from the analysis carried out in this chapter. Our aim was to study the validity of emergent, anisotropic, cosmological models in describing the fundamental cosmological observables of redshifts and luminosity distances. 

With that goal in mind, we constructed Hubble diagrams in universes that are inhomogeneous, and anisotropic on large scales. These diagrams depend on the line-of-sight of the particular observer, as well as their position in spacetime. We have then compared these diagrams to those that would be created by considering the large-scale average spacetime, using the emergent anisotropy formalism we developed in Chapter 7 \cite{Anton_2023}.
In order to carry out this comparison, we have focused on three families of cosmological models within the plane-symmetric class of dust-dominated solutions of Einstein's equations. These solutions admit closed-form exact solutions, and allow for arbitrary amounts of inhomogeneity to be introduced in the directions orthogonal to the surfaces of symmetry (though care is needed to avoid situations that may involve shell-crossings and singularities). 
The homogeneous subclasses of this set of solutions belong to the LRS Bianchi type {\it I} and {\it V} cosmologies, which are therefore considered to be the ``target spaces'' for the averaged cosmological models, through which large-scale measurements might be interpreted.

While the observations made by any one observer in these spacetimes are not necessarily reproduced well by the averaged cosmological models, we find that the observations made by many observers have an average that can be described accurately. 
In particular, we found in Sections \ref{sec:sinusoidal} and \ref{sec:linear} that there exist averaged anisotropic cosmological models that can accurately predict the Hubble diagram for average observations made in all directions to within $1\sigma$\,.
This is a non-trivial result, as the Hubble diagrams can take very different shapes in different directions in universes that exhibit large-scale anisotropy. In particular, the work presented in this chapter extends previous studies that focused only on observations along special lines of sight, that are aligned with principal null directions of the spacetime \cite{Bull_2012}.
Furthermore, the $1\sigma$ confidence interval generically shrinks as redshift increases, meaning that the average model becomes a better approximation for typical observers who make measurements over larger distances. This is precisely what one would hope for a useful cosmological model.

While our study has shown some successes for the anisotropic cosmologies that result from the application of our averaging formalism, it has also shown some clear warning signs. In particular, it is clear that the choice of foliation on which the averaging is performed must be made with care, as we discussed previously in Section \ref{subsec:foliation_choices}. 
This is exemplified by the tilted Farnsworth cosmologies studied in Section \ref{sec:farnsworth}, where we found that an anisotropic homogeneous model constructed from our averaging procedure could not fit the Hubble diagrams of observers accurately, even in an average sense. 
Although that result was not unexpected, after the calculation of cosmological backreaction for that model that we had already undertaken in Section \ref{sec:backreaction_farnsworth}, the analysis in this chapter showed that the foliation problem carries over from unobservable non-locally averaged scalars to Hubble diagram observables.
This is despite the fact that the Farnsworth spacetime is genuinely spatially homogeneous, and clearly indicates the importance of suitably foliating the spacetime. In general, therefore, one should not expect averaged cosmological models to reproduce the average of observables if there is no statistical homogeneity scale. If a foliation is such that no scale of this type exists, then that choice should be expected to fail in general.

Another area which requires care, in order to get a sensible result for the averaged cosmology, is the choice of target space for the symmetries of the averaged model. In particular, if the averaged model does not allow for all aspects of the averaged covariant scalars to be accounted for, then it is unlikely to reproduce the average of observations made within the spacetime. 
We believe this to be the reason why the Bianchi type {\it I} models, described by Eq. (\ref{eq_LRS_Bianchi_I}), failed to reproduce the average Hubble diagrams for the observers considered in Section \ref{sec:linear}. In that case the average spatial curvature is non-zero, and so it needs an averaged cosmology that allows for this possibility to exist. 
This is the case for Bianchi type {\it V} models, characterised by Eq. (\ref{eq_LRS_Bianchi_V}), which reproduced observables well, but not for Bianchi type {\it I}, which did not, at high $z$\,. 
We suspect the same will be true for cosmological models with large-scale tilt, which is only allowed in a restricted set of Bianchi classes \cite{King_1973}.

Let us comment briefly on the feasibility of making accurate sky maps on our local celestial sphere of $d_L(z)\,$, particularly with a view to estimating the multipoles of $d_L$ at a variety of redshift shells.
To do so is a rather speculative notion at present. However, as the number of SNEIa, quasars, radio galaxies and other distant sources we observe rises in the coming years, it will be possible to construct an increasingly precise sky map of $d_L(z)$, to $z \sim 1$ and beyond. 
If these observations continue to support the existence of large-scale anisotropies in the Universe, then we will need cosmological models that can include that freedom. 

The analysis presented in this work constitutes a step towards understanding the Hubble diagram in such spacetimes. 
On an immediate level, it suggests that the emergence-based approach we presented in Chapter 7 should be of use for understanding the theoretical problem of modelling anisotropy in the late Universe. 
More importantly, it may provide a framework within which anomalous anisotropic signals in the real Universe can be understood and interpreted, if those observational signatures become sufficiently mature in the near future, and cannot be explained by survey systematics or mere statistical noise. 

One of the main drawbacks of the situations considered in this chapter was the degree of symmetry present in the models, which is not a realistic assumption in our Universe. A consequence of the plane symmetry was that the backreaction scalars remained small in both the cases (Sections \ref{sec:sinusoidal} and \ref{sec:linear}) where there was an homogeneity scale present in the problem, which is likely to be closer to the situation we face in the Universe than the situation studied in Section \ref{sec:farnsworth} where a tilted flow was present globally. 
Hence, a natural extension of the study made here would be to investigate situations in which our backreaction scalars can be large, while maintaining statistical homogeneity within the averaging domains. This could potentially be found in more general situations, in which one would not expect there to be spacetime symmetries on small scales. 
The use of relativistic simulations \cite{macpherson2021luminosity, Macpherson:2022eve, Adamek_2018, lepori2020weak, lepori2021cosmological} may well prove to be fruitful in this regard, because they can allow inhomogeneities to be studied nonperturbatively, but do not make the Newtonian assumption of standard N-body simulations, which is incompatible with the intrinsically relativistic idea of cosmological backreaction.

\chapter{Conclusions}

\lhead{\emph{Conclusions}}

In this thesis, we have explored the use of relativistic frameworks for understanding cosmological gravity. We have focused our efforts on developing and testing two such frameworks, one of which is designed to test the overall landscape of gravitational theories, and the other of which is concerned with the emergence of anisotropy in the Universe from averaging inhomogeneous spacetimes. 
We have thus been building general approaches to modifying and testing the $\Lambda$CDM concordance model, rather than attempting to constrain individual alternative models with the set of techniques developed for the standard model.

Of course the simpler, less truly relativistic methods that are standard practice in the cosmological community are standard practice for good reason: they are mathematically tractable, interpretable and readily applicable to observations. Moreover, they have had enormous success in describing the majority of cosmological observations, especially the cosmic microwave background.
In contrast, covariant approaches are typically more mathematically complicated than standard ones based on the coordinate picture of FLRW cosmology and linear perturbation theory on large scales, and Newtonian gravity on small scales, and their observational application can be rather difficult and opaque. 

However, with the various observational tensions and anomalies pointing to a possible breakdown of the concordance model, it may be that a paradigm shift is needed. 
While the results of recent observations, and indeed upcoming surveys such as Euclid \cite{nesseris2022euclid} and the SKA \cite{maartens2015overview}, may point to inconsistencies in our present understanding, it seems unlikely that they will point to any one particular model, amongst the vast zoo of alternatives that have been proposed (see e.g. Refs. \cite{Clifton_2012,di2021realm} for reviews). 
It seems, therefore, that if we do not truly understand our Universe as much as it may have previously appeared, then we should test whether the problem might lie in our failure to appreciate all the many consequences of gravity being a fundamentally relativistic phenomenon. 

Furthermore, given that there may exist some theory, yet to be devised, that successfully accounts for all our observations, resolves the tensions and solves fundamental theoretical problems such as the cosmological constant and averaging problems\footnote{As far fetched as such a theory may seem at present!}, it may be safer not to attach ourselves to any specific theory content (known unknowns), but rather to use generalised approaches that, by avoiding any such content, implicitly include not just the known unknowns but also the unknown unknowns. 
There remains the possibility that certain cosmological models might appear equivalent to $\Lambda$CDM on the scales, and at the precision, accessible to current experiments, but make novel predictions in other contexts, such as on nonlinear cosmological scales or in the regime of strong gravity.
Thus, it is important to make sure that we do not accidentally bias our measurements by na{\" i}vely applying conclusions obtained in the FLRW cosmology or in Newtonian gravity to complex relativistic phenomena in inhomogeneous curved spacetime. 
If, for example, the anomalous dipole measurements are truly due to fundamental anisotropy in the Universe, then to attribute them to our kinematic motion with respect to some supposed cosmic rest frame would represent a serious theoretical source of systematic error.

Through our development and analysis of these frameworks, we have reached the following key conclusions. 
\begin{enumerate}

    \item We showed that the formalism of parameterised post-Newtonian cosmology can be used to describe the relativistic gravitational fields sourced not only by the energy density and isotropic pressure of matter fields, but also their momentum densities. Accordingly, we can model not only the scalar sector of gravitational perturbations in a theory-agnostic way, but also the divergenceless vector perturbations.
    Furthermore, the formalism can equally well be applied to either conservative or non-conservative theories of gravity (such as scalar-tensor or vector-tensor theories respectively).
    
    \item We demonstrated that the PPNC framework correctly accounts for the evolution of the cosmological background and linear scalar perturbations in a canonical class of example theories. Thus, it can be used to make viable observational predictions, using simple interpolating functions to describe the time and scale dependence of the coupling functions that describe deviations from GR in the scalar sector.

    \item We obtained novel constraints on the time-dependent PPN parameters $\alpha(a)$ and $\gamma(a)\,$, using observations of the anisotropies in the CMB. The data exhibit a strong degeneracy between their weighted time averages $\bar{\alpha}$ and $\bar{\gamma}$\,, due primarily to a novel term $\mathcal{G}\mathcal{H}\Psi$ in the momentum constraint equation we derived for the PPNC formalism. 
    If the functional form of $\alpha(a)$ and $\gamma(a)$ is prescribed to be a power law with a fixed index $n$\,, then for $n \leq 0.25$ we found a mild preference in the Planck data for $\bar{\alpha}$ and $\bar{\gamma}$ to be below their GR values of unity. However, these values are consistent with GR to within $\sim 2\sigma\,$.

    \item We also identified degeneracies between the PPN parameters, and the standard cosmological parameters $H_0$ and $\omega_c\,$, through their impact on the acoustic peaks in the cosmic microwave background anisotropies. This demonstrated the importance of correctly modelling the FLRW background, not just the perturbations, as is sometimes done in other general approaches to testing gravity in cosmology.

    \item By considering power laws for $\alpha(a)$ and $\gamma(a)$ with a varying power-law index $n\,$, and applying a fairly aggressive cut $n \leq 0.25$ on the prior volume, we obtained constraints on the present-day time variation of $\alpha$ and $\gamma$\,, that are competitive with Solar System bounds in the case of $\alpha\,$, and better than can be obtained in the Solar System for $\gamma\,$. 

    \item We developed a new framework whereby large-scale anisotropy in the Universe may emerge from the growth of nonlinear structures, using a scalar averaging procedure based on Buchert's approach. We derived the full set of effective equations of motion for any emergent large-scale anisotropic universe, which can be interpreted as a backreaction-dependent LRS Bianchi cosmology. 

    \item Using the Farnsworth model spacetimes, we demonstrated the importance of correctly accounting for foliation dependence in cosmological averaging, especially in the presence of anisotropy, because an anisotropic cosmology can be tilted. We showed that tilted flows can give rise to substantial backreaction, and that the Hubble diagrams constructed by observers in a tilted Universe are poorly reproduced by an average model.

    \item Conversely, we showed that in anisotropic, inhomogeneous spacetimes that nonetheless possess an homogeneity scale, an LRS Bianchi average model, if chosen appropriately, can provide an excellent fit to Hubble diagram observables on large scales. In those cases, most high-redshift deviations between the real and averaged spacetime, in the inferred magnitudes $\mu$ of distant sources, take the form of small constant offsets, which are not observable for measurements of Type Ia supernovae.
    
\end{enumerate}

These investigations have indicated that the two main relativistic approaches to cosmology we have studied in this thesis - parameterised post-Newtonian cosmology, and the emergent anisotropy framework - are insightful ways to model and test deviations from the standard cosmology. 
At this stage, it is not possible to state definitively whether there is any evidence for cosmological PPN parameters being inconsistent with General Relativity, or whether the backreaction mechanism we developed produces a large enough effect to account for observed signatures of anisotropy in the late Universe.
However, with a plethora of observational probes at our disposal, and with relativistic simulations \cite{macpherson2017inhomogeneous,macpherson2019einstein,adamek2014distance,Adamek_2015,Adamek_2018} providing a concrete way to make predictions from fundamentally relativistic approaches such as ours, there is substantial scope to further constrain the allowed possibilities in each framework in the near future.

Let us conclude with some final remarks on the validity of the concordance model, consisting of General Relativity, the FLRW geometry, and the $\Lambda$CDM model for the Universe's energy-momentum content. 
This model is extraordinarily successful, and although it is possible that the $H_0$ and $\sigma_8$ tensions, the anomalous cosmic dipole measurements, or other tensions and anomalies yet to be identified, will eventually force it to be discarded, it is certainly not out of the question at this point that it will survive the next generation of observational tests.

Even if this were to happen, though, it should not spell the end for relativistic cosmological modelling outside the paradigms of GR and FLRW. Instead, we should seek to understand why such a simple model is, on the face of it, so unreasonably effective. If one simply assumes the concordance model, and takes the perspective that one should simply calculate observables within it to ever higher precision, then we would still be left in the dark as to why such an approach works so well.
It may be that the answer to either this question, or to the question of replacing the concordance model, if it happens to be falsified in the coming years, lies in developing a deeper understanding of relativistic gravity. 
I aim to continue to develop and test the covariant frameworks we have discussed in this thesis in the coming years, using a mixture of further theoretical work, observational constraints, and numerical simulations. I hope that they will provide valuable tools into appreciating the full implications of relativity for cosmological physics.


\addtocontents{toc}{\vspace{2em}} 

\appendix 

\chapter{The Einstein-Boltzmann equations}\label{subsec:einstein_boltzmann}

\lhead{\emph{Appendix: the Einstein-Boltzmann equations}} 

Einstein-Boltzmann solvers are a key computational method in cosmology. They make it possible to calculate quantities associated with linear cosmological perturbations in the context of a realistic set of matter, radiation and dark energy species in the Universe.

The equations of linear cosmological perturbations must be solved in order to calculate key observables associated with linear perturbations to the FLRW cosmology, especially the temperature, polarisation and lensing anisotropies in the cosmic microwave background and the power spectrum of linear fluctuations in the matter density field. 

In order to calculate the metric perturbations, it is necessary to understand the form of $T_{ab}$\,. The energy-momentum tensor is ultimately a description of all the particles (photons, baryons, dark matter, neutrinos etc.) that make up the matter and radiation fields. 
These must be described using kinetic theory. That is, $T_{ab}$ is calculated by integrating over all the possible conjugate momenta $P_i$ of particles in phase space. The number of particles in an infinitesimal phase-space volume $dN$ is given by a probability distribution $f(\tau, x^i, P_i)$: $dN = \mathrm{d}^3\mathbf{x}\,\mathrm{d}^3\mathbf{P}\, f(\tau, \mathbf{x}, \mathbf{P})\,$. Then we can write the energy-momentum tensor as \cite{Ma_1995}
\begin{equation}
    T_{ab} = \int \mathrm{d}^3\mathbf{P}\,\sqrt{-g}\,\frac{P_a\,P_b}{P_0}\,f(\tau, \mathbf{x},\mathbf{P})\,.
    \end{equation}
The distribution $f(\tau, x^i, P_i)$ evolves according to Boltzmann's equation,
\begin{equation}
    \frac{Df}{\mathrm{d}\tau} = \frac{\partial f}{\partial \tau} + \frac{\mathrm{d}x^i}{\mathrm{d}\tau}\frac{\partial f}{\partial x^i} + \frac{\mathrm{d}P_i}{\mathrm{d}\tau}\frac{\partial f}{\partial P_i} = \left(\frac{\partial f}{\partial \tau}\right)_C\,,
\end{equation}
where $\left(\frac{\partial f}{\partial \tau}\right)_C$ is a collision operator, and we have implicitly summed over all the particles of the species we are interested in. For the cosmic microwave background, $f$ is the phase-distribution for photons.
By entering into $T_{ab}$, the phase-space distribution $f$ sources the metric perturbations via Einstein's equations. The perturbations themselves determine the evolution of $f$ via the Boltzmann equation. Hence, one must calculate the coupled Einstein-Boltzmann system of equations.

Because of the isotropy of the FLRW background, it is advantageous to solve these equations in Fourier space. As an example, consider the Newtonian gauge perturbation to the lapse function, $\Phi(\tau, \mathbf{x})$. Its Fourier transform $\Phi(\tau, \mathbf{k})$ is defined
\begin{equation}
    \Phi(\tau, \mathbf{k}) = \int \mathrm{d}^3 x\, e^{-i k_i x^i}\, \Phi(\tau, \mathbf{x})\,.
\end{equation}
Under the Fourier transform, partial derivatives with respect to spatial coordinates transform as $\partial_i \longrightarrow -i k_i$\,. Performing this mapping to all the perturbation theory equations in Section \ref{subsec:newtonian_gauge} gives the equivalent Fourier-space equations trivially.

Crucially, the isotropy of the FLRW background means that perturbations depend only on their scale, not on their orientation, i.e. $\Phi(\tau, \mathbf{k}) = \Phi(\tau, \vert \mathbf{k} \vert) = \Phi(\tau, k)$. Likewise, the background value $f_0$ of $f$ can depend only on the magnitudes of particle momenta, and not their directions or locations, so that $f_0 = f_0(a p)\,$, where we have used that $p$ redshifts as $a^{-1}\,$.
Transforming to Fourier space ($\mathbf{x} \longrightarrow \mathbf{k}$), one can write $f(\tau, \mathbf{k},\mathbf{P}) = f_0(ap)\left(1 + \Upsilon(\tau, \mathbf{k}, ap, \theta)\right)$, where $\cos{\theta} = \frac{k_i p^i}{kp}\,$.
Then, the Boltzmann equation in Fourier space is
\begin{equation}
    \frac{\partial \Upsilon}{\partial \tau} - i\Upsilon\cos{\theta} + \left(\frac{\mathrm{d}\ln{f_0}}{\mathrm{d}\ln{ap}}\right)\left[i\Phi\cos{\theta} - \Psi'\right] = \frac{1}{f_0}\left(\frac{\partial f}{\partial \tau}\right)_C\,,
\end{equation}
where for the sake of simplicity we have specialised to the case of a massless species. An obvious example is photons, for which the collision operator on the right hand side is due to photon-baryon Thomson scattering before decoupling, and is zero afterwards.
Although this equation looks daunting, it is actually fairly straightforward to solve numerically, because the isotropy of the background has come to our rescue by making the phase-space distribution dependent only on the particles' directions through the scalar product $k_i p^i = kp \cos{\theta}\,$.
Thus, $\Upsilon$ can be decomposed into a series of Legendre polynomials $P_l(\cos{\theta})$, and the coupled Einstein-Boltzmann equations are solved successively in multipole moments $l\,$.

The primary tools for solving the coupled linear Einstein-Boltzmann system are the well-established cosmological Boltzmann codes. In the present day, these are usually CAMB (Code for Anisotropies in the Microwave Background) \cite{lewis2011camb} and CLASS (Cosmic Linear Anisotropy Solving System) \cite{lesgourgues2011cosmic,Diego_Blas_2011}.
In Chapters 4 and 6, we make extensive use of the CLASS code, which is set up to allow for modifications to either the matter/radiation/dark energy sectors \cite{von2019cosmological}, or the laws of gravity themselves \cite{zumalacarregui2017hi_class,renk2016gravity}.



\addtocontents{toc}{\vspace{2em}}  
\backmatter

\label{Bibliography}
\lhead{\emph{Bibliography}}  
\begingroup
\setstretch{1.0}
\footnotesize
\bibliographystyle{unsrt} 

\endgroup

\end{document}